%% file: panda_pb.tex
\documentclass[pdftex]{panda_book}
\markpanda{Strong interaction studies with antiprotons}{Physics Book}
\begin{document}
%
%
\include{panda_pb_tit}
%
%
\include{panda_pb_int}
%
%
\include{panda_pb_exp}
\include{panda_pb_soft}
\include{panda_pb_phys}
\include{panda_pb_sum}
%
\include{panda_pb_end}
%
\end{document}

%% file: panda_pb_tit.tex
\pagenumbering{roman}
\onecolumn
%
%
\begin{center}
{\bfseries \sffamily \huge Physics Performance Report for:\\ \ \\ \Panda{} \\
{\sffamily \small (Anti\underline{P}roton \underline{An}nihilations at \underline{Da}rmstadt)}\\
\ \\ Strong Interaction Studies with Antiprotons}
\vskip 1cm
{\large \Panda{} Collaboration}
%
%
\end{center}
\vskip 1cm
To study  fundamental questions of hadron and nuclear physics
in interactions of antiprotons with nucleons and nuclei, 
the universal \Panda{} detector will be build. Gluonic
excitations, the physics of strange and charm quarks and nucleon structure
studies will be performed with unprecedented accuracy thereby allowing high-precision
tests of the strong interaction. The proposed \Panda{} detector is a state-of-the-art
internal target detector at the \HESR{} at \FAIR{} allowing the detection
and identification of neutral
and charged particles generated within the relevant
angular and energy range.\\
This report presents a summary of the physics accessible at \Panda{} and
what performance can be expected.
\vskip 1cm
\begin{center}
\includegraphics[width=1.7\swidth]{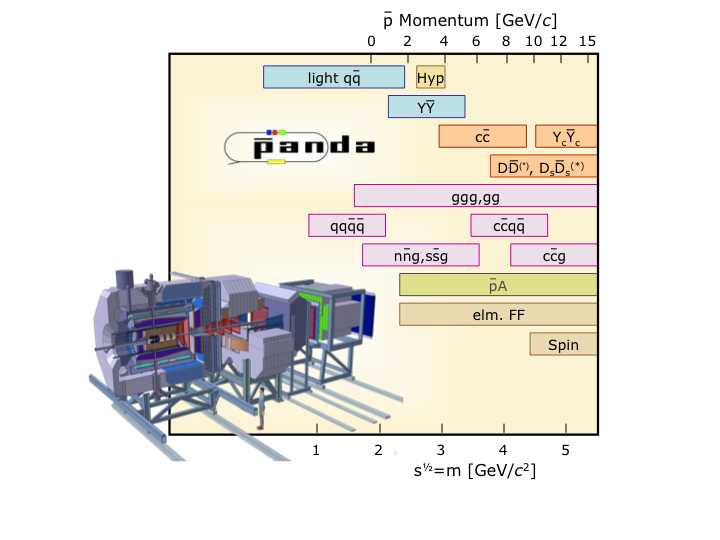}
\end{center}
\vfill
%
%
\newpage
\begin{center}
\vspace*{3mm }
{\LARGE \bfseries \sffamily The \Panda{} Collaboration}
\vskip 7mm
\input{./main/authors}
\end{center}
%
%
\begin{center}
\vspace*{1.5cm }
{\Large \bfseries \sffamily AND}
\vskip 7mm
\input{./main/additionalAuthors}
\end{center}

%
%
\vfill
\hrulefill\\
\begin{tabbing}
Editors:  \hspace{1.cm} \= Diego Bettoni (chief editor)  \hspace{0.5cm}  \= Email: \verb$bettoni@fe.infn.it$ \\
   	\> Rob Timmermans (chief editor)  \hspace{1cm}   \> Email: \verb$timmermans@kvi.nl$ \\
   	\> Maria Pia Bussa  \hspace{1cm}   \> Email: \verb$bussa@to.infn.it$ \\
   	\> Michael Dueren  \hspace{1cm}   \> Email: \verb$Michael.Dueren@exp2.physik.uni-giessen.de$ \\
   	\> Alessandro Feliciello  \hspace{1cm}   \> Email: \verb$Alessandro.Feliciello@to.infn.it$ \\
        \> Albrecht Gillitzer  \hspace{1cm}   \> Email: \verb$a.gillitzer@fz-juelich.de$ \\
        \> Felice Iazzi  \hspace{1cm}   \> Email: \verb$felice.iazzi@polito.it$ \\
        \> Tord Johansson  \hspace{1cm}   \> Email: \verb$tord.johansson@tsl.uu.se$ \\
        \> Bertram Kopf   \hspace{1cm}   \> Email: \verb$bertram@ep1.rub.de$ \\
        \> Andreas Lehrach  \hspace{1cm}   \> Email: \verb$a.lehrach@fz-juelich.de$ \\
        \> Matthias F.M. Lutz  \hspace{1cm}   \> Email: \verb$m.lutz@gsi.de$ \\
        \> Frank Maas  \hspace{1cm}   \> Email: \verb$maas@KPH.UNI-MAINZ.DE$ \\
        \> Marco Maggiora  \hspace{1cm}   \> Email: \verb$marco.maggiora@to.infn.it$ \\
        \> Matteo Negrini  \hspace{1cm}   \> Email: \verb$negrini@fe.infn.it$ \\
        \> Klaus Peters       \hspace{1cm}   \> Email: \verb$K.Peters@gsi.de$ \\
        \> Josef Pochodzalla  \hspace{1cm}    \> Email: \verb$pochodza@KPH.UNI-MAINZ.DE$ \\
	\> Lars Schmitt  \hspace{1cm}   \> Email: \verb$L.Schmitt@gsi.de$ \\ 
	\> Olaf Scholten  \hspace{1cm}   \> Email: \verb$scholten@kvi.nl$ \\ 
	\> Giulio Stancari  \hspace{1cm}   \> Email: \verb$stancari@fe.infn.it$ \\ \ \\
 
Spokesperson: \>  Ulrich Wiedner \> Email: \verb$ulrich.wiedner@ruhr-uni-bochum.de$ \\
Deputy:  \> Paola Gianotti  \> Email: \verb$paola.gianotti@lnf.infn.it$ \\
\end{tabbing}
\hrulefill\\
%
%
\vfill
%
%
\cleardoublepage
\input{./main/preamble}
%
%
\cleardoublepage
\tableofcontents
%
%

%% file: main/authors.tex
%
%
\institem{Universit\"at {\bf Basel}, Switzerland}
\authitem{W.~Erni},
\authitem{I.~Keshelashvili},
\authitem{B.~Krusche},
\authitem{M.~Steinacher}\lastitem
\institem{Institute of High Energy Physics, Chinese Academy of Sciences, {\bf Beijing}, China}
\authitem{Y.~Heng},
\authitem{Z.~Liu},
\authitem{H.~Liu},
\authitem{X.~Shen},
\authitem{O.~Wang},
\authitem{H.~Xu}\lastitem
% ...
%
%
%

%% file: main/additionalAuthors.tex
%
%
\institem{GSI Helmholtzzentrum  f\"ur Schwerionenforschung GmbH, {\bf Darmstadt}, Germany}
\authitem{M.F.M.~Lutz}
\institem{CPhT, Ecole Polytechnique, CNRS, {\bf Palaiseau}, France}
\authitem{B.~Pire}
\institem{Kernfysisch Versneller Instituut, University of {\bf Groningen}, Netherlands}\lastitem
\authitem{O.~Scholten},
\authitem{R.~Timmermans}
%
%

%% file: main/preamble.tex
%
\begin{center}
\vspace*{1.5cm}
{\Large \bfseries \sffamily Preface}\addcontentsline{toc}{chapter}{Preface}
\end{center}
\vskip 1.5cm

\PANDA is one of the major projects of the \FAIR-Facility in Darmstadt. \FAIR is an extension of the existing 
Heavy Ion Research Lab (GSI) and is expected to start its operation in 2014. 
\PANDA studies interactions between antiprotons and fixed target protons and nuclei in the momentum range 
of 1.5-15 GeV/c using the high energy storage ring \HESR. The antiproton project was initiated by a large 
community of scientists outside GSI, who had worked very successfully with antiprotons at LEAR/CERN and the 
Fermilab antiproton accumulator. Many of the physics ideas of \PANDA were already described in a 
Letter of Intent (Construction of a GLUE/CHARM-Factory at GSI, Ruhr-University Bochum, 1999) and were extended 
afterwards in the \FAIR Conceptual Design Report (GSI, 2001), the Technical Progress Report (\FAIR, 2005) and 
further \PANDA specific reports. 
After the approval of \FAIR further projects involving antiprotons were proposed 
(experiments with low energy and polarized antiprotons) which are now in the preparatory phase.

The \PANDA scientific program includes several measurements, which address fundamental questions of QCD, 
mostly in the non-perturbative regime:
\begin{itemize}
\item Hadron spectroscopy up to the region of charm quarks. Here the search for exotic states like glueballs,
hybrids and multiquark states in the light quark domain and in the hidden and open charm region is in the 
focus of interest. The recently found XYZ states will be further explored.
\item Study of properties of hadrons inside nuclear matter. Mass and width modifications have been reported 
and will be investigated also in the charm region.
\item Study of nonperturbative dynamics, also including spin degrees of freedom.
\item Antiproton induced reactions are a very effective tool to implant strange baryons in nuclei.
\PANDA will particularly study double $\Lambda$ hypernuclei, which are of great importance for nuclear 
structure studies and the $\Lambda\Lambda$ interaction.
\item Hard exclusive antiproton-proton reactions can be used to study the structure of nucleons
(time-like form factors) and the relevance of certain models, like the Hand Bag approach.
Interesting aspects of Transverse Parton Distributions will be studied in Drell-Yan production.
\item In a later stage of the project, when all systematic effects are well studied, also contributions 
to electroweak physics can be expected, like direct CP violation in hyperon decays and CP violation and 
mixing in the charm sector.  
\end{itemize}

All measurements will profit from the high yield of antiproton induced reactions and from the fact that, 
in contrast to $\ee$ reactions, all non-exotic quantum number combinations for directly formed states are 
allowed, whereas states with exotic quantum numbers can be observed in production.
The achievable precision, 
as far mass and width measurements are concerned, is very high as was successfully demonstrated by the 
Fermilab experiments.
It is independent of the mass resolution of the detector and only limited by the 
tiny energy spread of the primary cooled antiproton beam.
It will allow a measurement of the widths of the recently discovered very narrow states.
The international \PANDA collaboration was established in 2002. 
More than 400 scientists from 16 countries and 53 institutions are involved in
R\&D hardware and software projects. The most recent achievement is the definition of 
the final setup of the electromagnetic calorimeter. The delivery of a first tranche of 
PWO crystals for the calorimeter has already started. An overview of all studies and results 
achieved in the last years can be found on the \PANDA Web site 
(http://www\--panda.gsi.de/auto/\_home.htm).

This \PANDA Physics Book describes in detail the physics topics envisioned. 
The first chapter gives a comprehensive overview of the challenges of QCD; 
the \PANDA detector and the high energy storage ring HESR are described in detail in chapter 2;  
in chapter 3 the status of the software is discussed. Chapter 4 shows very detailed 
simulations of selected benchmark reactions. 
They take into account the event generation, digitization,
reconstruction, event selection and background estimations. 
A refinement of the analysis is achieved by using kinematical fitting and neural network tools. 
A summary and outlook concludes the Physics Book.
The setup used in the simulations described in this book is not final: some detector components are 
still undergoing R\&D and will be finalised in the coming months. At the same time a new software 
framework is being developed. In the next two years we also expect advances in background simulations 
and better estimates of presently unknown cross-sections. A new version of the physics book is planned, 
which will reflect the progress described above and which will feature 
a more complete list of benchmark channels.

\vspace*{2cm}
\centerline{\includegraphics[width=0.8\dwidth]{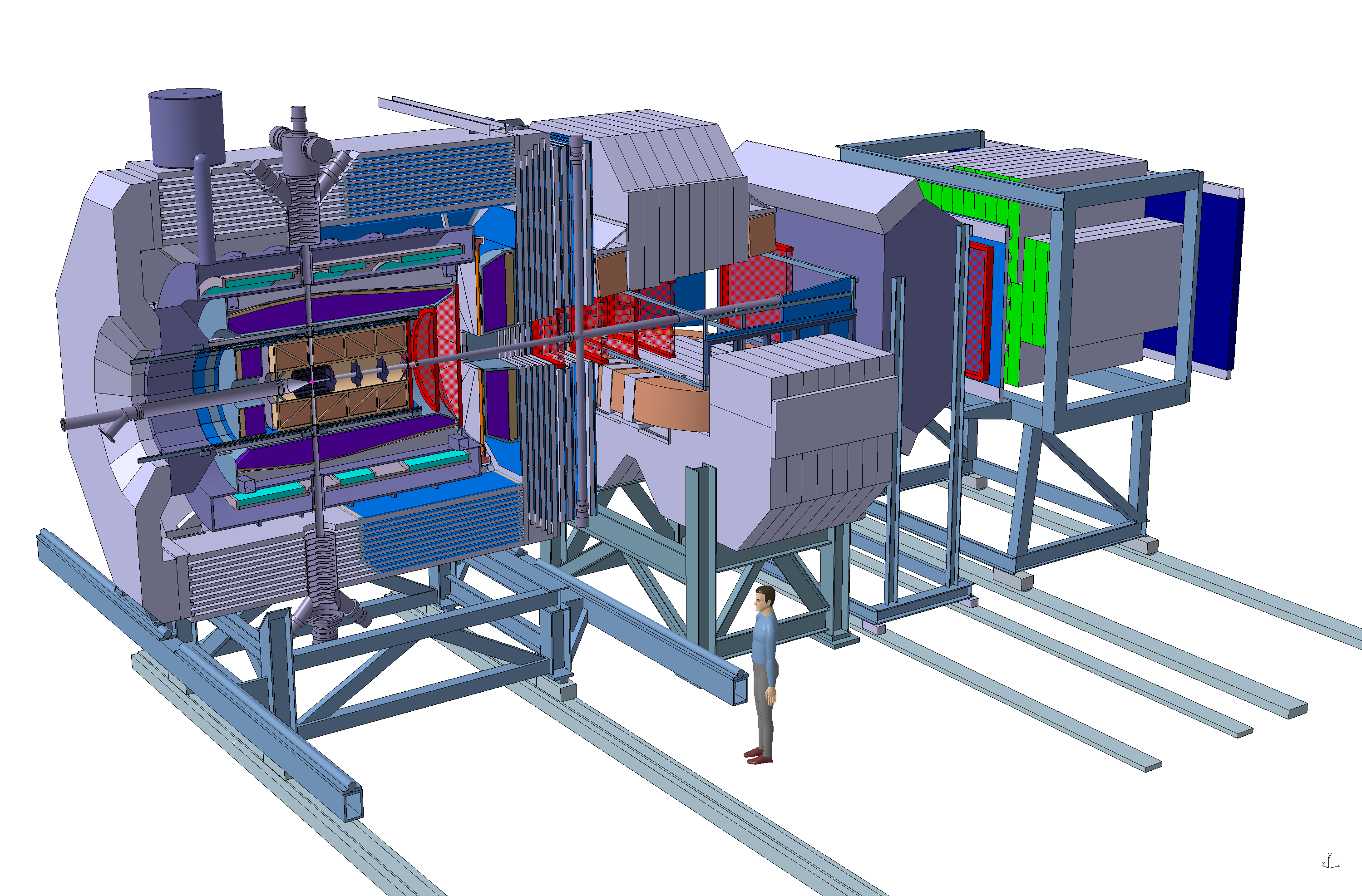}}
\vfill
\clearpage
\vspace*{18cm}
\hrulefill\\
\vspace*{2cm}\\
\begin{minipage}[t]{10cm}
\sloppy
The use of registered names, trademarks, \etc in this publication does not
imply, even in the absence of specific statement, that such names are exempt
from the relevant laws and regulations and therefore free for general use.
\end{minipage}
\vfill
%

%% file: panda_pb_int.tex
%
%
\cleardoublepage
\pagenumbering{arabic}
\setcounter{page}{1}
\chapter{Introduction}
\label{sec:int}
\COM{Author(s): R. Timmermans, D. Bettoni, K. Peters}

\input{./int/intro}
\newpage
\bibliographystyle{panda_pb_lit}
\bibliography{./int/lit_int,./main/lit_main}
%

%% file: int/intro.tex
%
\section{The Challenge of QCD}
The modern theory of the strong interactions is Quantum Chromodynamics (QCD), the
quantum field theory of quarks and gluons based on the non-abelian gauge group
SU(3). Together with the SU(2)$\times$U(1) electroweak theory, QCD is part of the
Standard Model of particle physics. QCD is well tested at high energies, where
the strong coupling constant becomes small and perturbation theory applies. In
the low-energy regime, however, QCD becomes a strongly-coupled theory, many
aspects of which are not understood. The thriving questions are: How can we bring
order into the rich phenomena of low energy QCD? Are there effective degrees of
freedom in terms of which we can understand the resonances and bound states of
QCD efficiently and systematically? Does QCD generate exotic structures so far
undiscovered? \PANDA will be in a unique position to provide answers to such
important questions about non-perturbative QCD. A major part of the physics
programme of \PANDA is designed to collect high-quality data that allow a clean
interpretation in terms of the predictions of non-perturbative QCD. In this
introductory chapter, we first summarize the basics of QCD and then review the
theoretical approaches that can be justified rigorously within QCD and provide
testable predictions for experiments like \PANDA.
\subsection{Quantum Chromodynamics}
The development of QCD as the theory of strong interactions is a success story.
Its quantitative predictions at high energies, in the perturbative regime, are
such that it is beyond serious doubt that QCD is the correct theory of the strong
interactions. Nevertheless, in the non-perturbative low-energy regime, it remains
very hard to make quantitative predictions starting from first principles, {\it
i.e.} from the QCD Lagrangian. Conceptually, QCD is simple: it is a relativistic
quantum field theory of quarks and gluons interacting according to the laws of
non-abelian forces between colour charges. The starting point of all considerations
is the celebrated QCD Lagrangian density:
\begin{eqnarray}
&&{\cal L}_{QCD} = -\frac{1}{4}\, G^{\mu\nu}_a \,G_{\mu\nu}^a \nonumber \\
 &&\qquad +\,\sum_{f} \bar q_f \left[ i\,\gamma^\mu D_\mu - m_f \right] q_f \, ,
\label{def-QCD-Lag}
\end{eqnarray}
where
\begin{equation}
G^{\mu\nu}_a = \partial^\mu A^\nu_a - \partial^\nu A^\mu_a
              + g \,f_a^{\phantom{a}bc}\, A^\mu_b\, A^\nu_c\,,
\label{def-Gmunu}
\end{equation}
is the gluon field strength tensor, and
\begin{equation}
          D^\mu = \partial^\mu - i\, \frac{g}{2}\, A^\mu_a\, \lambda^a \,,
\label{def-D}
\end{equation}
the gauge covariant derivative involving the gluon field $A^\mu_a$; $g$ is the
strong coupling constant, $\alpha_S=g^2/4\pi$; $f$ denotes the quark flavour,
where for the energy regime of \PANDA, the relevant quark flavours are $u$, $d$,
$s$, $c$: up, down, strange, and charm. We take $\hbar=1=c$.
\par
This deceptively simple looking QCD Lagrangian is at the basis of the rich and
complex phenomena of nuclear and hadronic physics. How this complexity arises in
a theory with quarks and gluons as fundamental degrees of freedom is only
qualitatively understood. The QCD field equations are non-linear, since the gluons
that mediate the interaction carry colour charge, and hence interact among
themselves. This makes every strongly-interacting system intrinsically a
many-body problem, wherein apart from the valence quarks many quark-antiquark
pairs and many gluons are always involved. These non-abelian features of QCD are
believed to lead to the phenomenon that the basic degrees of freedom, the quarks
and the gluons, cannot be observed in the QCD spectrum: the confinement of colour
charge is the reason behind the complex world of nuclear and hadronic physics.
\par
The process of renormalization in quantum field theory generates an intrinsic QCD
scale $\Lambda_{QCD}$ through the mechanism of dimensional transmutation;
$\Lambda_{QCD}$ is, loosely speaking, the scale below which the coupling constant
becomes so large that standard perturbation theory no longer applies. All hadron
masses are in principle calculable within QCD in terms of $\Lambda_{QCD}$. This
dynamical generation of the mass scale of the strong interactions is the famous
QCD gap phenomenon: the proton mass is non-zero because of the energy of the
confined quarks and gluons. Although a mathematical proof of colour confinement is
lacking, qualitatively this is thought to be linked to the fact that the quark
and gluon bilinears $\overline{q}_a q^a$ and $G^a_{\mu\nu}G_a^{\mu\nu}$ acquire
non-zero vacuum expectation values.
\par
Now, some 35 years after the discovery of QCD, it is fair to say that strong
interactions are understood in principle, but a long list of unresolved questions
about low-energy QCD remains. Our present understanding of QCD thereby serves as
the basis to set priorities for theoretical and experimental research. Clearly,
not all phenomena in nuclear physics need to be understood in detail from QCD.
Many areas of nuclear physics will be happily described in terms of
well-established phenomenology with its own degrees of freedom, just like many
complex phenomena in atomic physics and chemistry do not have to be understood
directly in terms of QED. Likewise, while not all experiments in nuclear and
hadronic physics should be guided by QCD, dedicated experiments that test QCD in
the non-perturbative regime and to improve our limited understanding of these
aspects of QCD are crucial. In its choice of topics, the \PANDA physics programme
aims to achieve precisely this.
\subsection{The QCD Coupling Constant}
The qualitative understanding of QCD as outlined above is to a large extent based
on the classical calculation of the renormalization scale dependence of the QCD
coupling constant $\alpha_S$ as given by the $\beta$-function at an energy scale
$\mu$,
\begin{equation}
   \beta(\alpha_S) \equiv \frac{\mu}{2}\,\frac{\partial\alpha_S}{\partial\mu} =
   -\frac{\beta_0}{4\pi}\,\alpha_S^2 -\frac{\beta_1}{8\pi^2}\,\alpha_S^3 - \dots
\end{equation}
where
\begin{eqnarray}
  \beta_0 & = & 11 - \frac{2}{3}\,n_f \ , \\
  \beta_1 & = & 51 - \frac{19}{3}\,n_f \ ,
\end{eqnarray}
where $n_f$ is the number of quarks with mass less than $\mu$; expressions for
$\beta_2$ and $\beta_3$ exist. In solving this differential equation for
$\alpha_S(\mu)$, one introduces the scale $\Lambda$ to provide the $\mu$
dependence of $\alpha_S$. The solution then demonstrates the famous properties of
asymptotic freedom, $\alpha_S\rightarrow0$ when $\mu\rightarrow\infty$, and of
strong coupling at scales below $\mu\sim\Lambda$.
\begin{figure}[ht]
\centerline{\includegraphics*[width=\swidth]{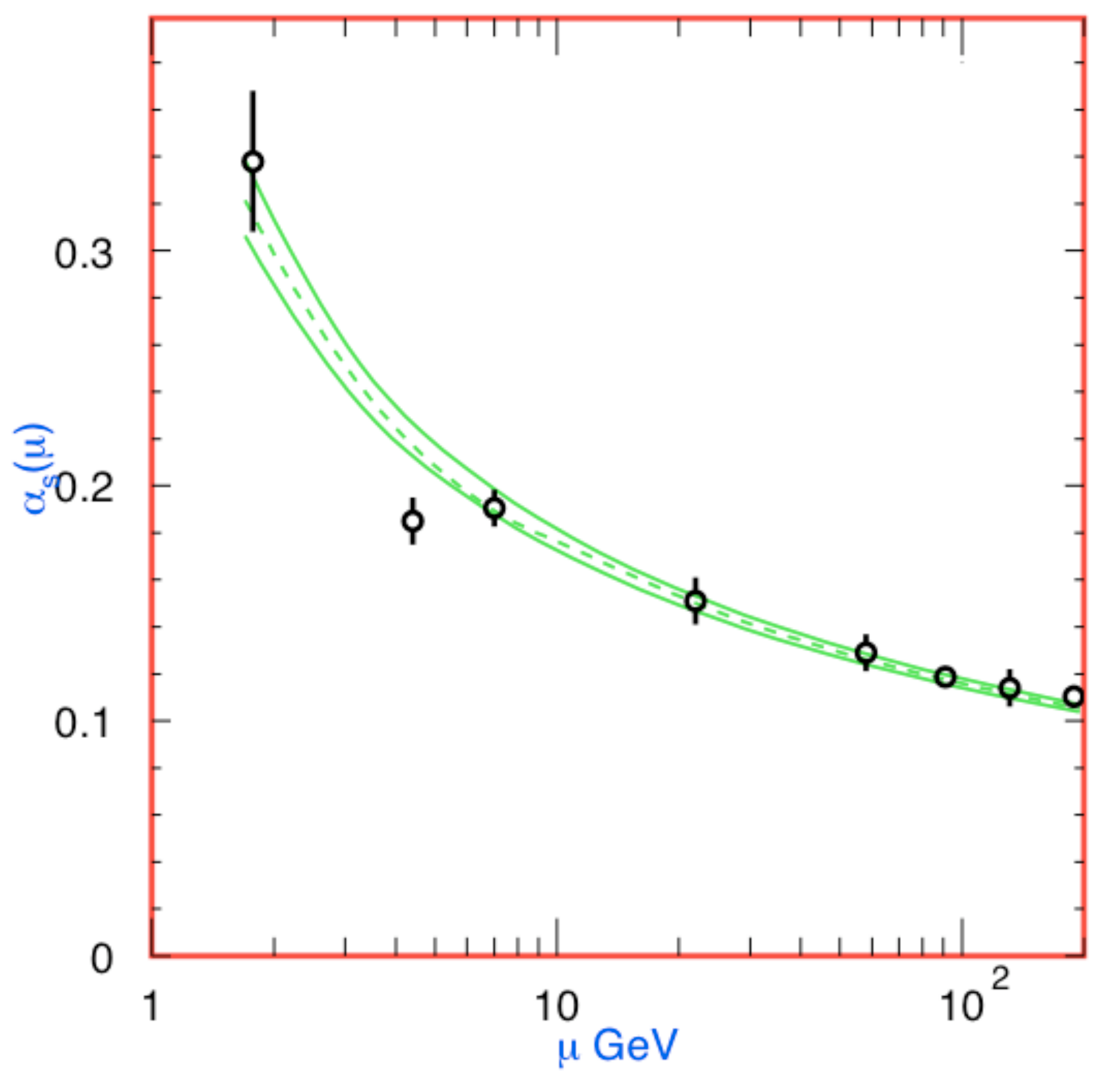}}
\caption{The running of the strong coupling constant as function of
          the scale $\mu$~\cite{PDG06}.}
\end{figure}
Based on this result for the scale dependence of the QCD coupling constant, one
may roughly divide the field of strong interaction physics into the areas of
perturbative QCD (pQCD) and of non-perturbative QCD. QCD has been very successful
in quantitatively describing phenomena where perturbation theory with its
standard machinery of Feynman rules applies. An important example is $e^+e^-$
annihilation in the area of the $Z^0$ boson, where the multi-particle hadronic
final-state system reveals the perturbative QCD physics in the form of the quark
and gluon jets. In this perturbative regime predictions can be made on the basis
of the magnitude of the QCD coupling constant. Its value as a function of energy
determines a host of phenomena, such as scaling violations in deep inelastic
scattering, the $\tau$ lifetime, high-energy hadron collisions, heavy-quarkonium
(in particular bottomonium) decay, $e^+e^-$ collisions, and jet rates in $ep$
collisions. The coupling constants derived from these processes are consistent
and lead to an average value~\cite{PDG06}
\begin{equation}
   \alpha_S(M_Z) = 0.1176 \pm 0.0002 \ .
\end{equation}
The non-perturbative regime is the area of strong nuclear forces and hadronic
resonances, which is quantitatively much less well understood and where important
questions still have to be addressed. In between are areas like deep inelastic
lepton-hadron scattering where perturbation theory is used, however, with
non-perturbative input.
\subsection{The Symmetries of QCD}
It has been said that QCD is a most elegant theory in physics, since its
structure is solely determined by symmetry principles: QCD is the most general
renormalisable quantum field theory based on the gauge group SU(3). In addition
to exact Lorentz invariance and SU(3) colour gauge invariance, it has several
other important symmetry properties. The QCD Lagrangian as given above has a
number of ``accidental'' symmetries, {\it i.e.} symmetries that are an automatic
consequence of the assumed gauge invariance. The discrete symmetries parity and
charge conjugation are such accidental symmetries (we ignore here the mysterious
$\theta$-term that results in the still unsolved strong CP-problem of QCD).
Flavour conservation is another: the number of quarks (minus antiquarks) of each
flavour ({\it e.g.} strangeness) is conserved, corresponding to an automatic
invariance of the QCD Lagrangian under phase rotations of the quark fields of
each flavour separately.
\par
Additional symmetries result from the consideration that the masses of the up,
down, and strange quarks can be considered small compared to the typical hadronic
scale $\Lambda_{QCD}$. To the extent that these masses can be ignored, the QCD
Lagrangian is invariant under unitary transformations of the quark fields of the
form $q^\prime_i = U_{ij}\,q_j$. This accounts for the rather accurate
SU(2)-isospin and the approximate SU(3)-flavour symmetries of nuclear and hadronic
physics. Moreover, when the $u$, $d$, and $s$ masses can be ignored, QCD is
invariant under separate unitary transformations among the left- and right-handed
quarks, $q^{L\prime}_i = U^L_{ij}\,q^L_j$ and $q^{R\prime}_i = U^R_{ij}\,q^R_j$,
resulting in the chiral symmetry group U(3)$_L$$\times$U(3)$_R$. The diagonal
subgroup ($U_L=U_R$) corresponds to the SU(3)-flavour (and baryon number) symmetry
mentioned. The remaining chiral SU(3) symmetry ($U_L^{-1}=U_R$) is believed to be
spontaneously broken by the vacuum state of QCD, resulting in the existence of an
octet of Goldstone bosons identified with the pseudoscalar mesons $\pi$, $K$,
$\eta$.
\par
These approximate flavour and chiral symmetries due to the smallness of the $u$,
$d$, and $s$ quark masses, are important, since they can be exploited to
formulate effective field theories that are equivalent to QCD in a certain energy
range. A classic example is chiral perturbation theory for the interaction of
baryons with the octet of the pseudoscalar mesons, which results in an expansion
of matrix elements in terms of small momenta or light-quark masses~\cite{Ber08}.
On similar footing is heavy-quark effective theory (HQET) for hadrons containing
a quark ($c$, $b$, $t$) with mass $m_Q\gg\Lambda_{QCD}$. In the limit
$m_Q\rightarrow\infty$, the heavy quark becomes on-shell and the dynamics becomes
independent of its mass. The hadronic matrix elements can be expanded as a power
series in $1/m_Q$, resulting in symmetry relations between various matrix
elements~\cite{Man00}.
\par
Generalizing QCD to an SU($N_c$) gauge theory, the inverse of the number of
colours, $1/N_c$, is a hidden expansion parameter~\cite{Hoo74}. This theory,
wherein the coupling is decreased such that $g^2N_c$ is constant, is
``large-$N_c$ QCD''. Diagrammatic considerations suggest that large-$N_c$ QCD is
a weakly-coupled theory of mesons and baryons, wherein baryons are heavy
semiclassical objects. Significant, mostly qualitative, insight into QCD can be
obtained from considering the large-$N_c$ limit, especially when combined with
the techniques of effective field theory.
\subsection{Theoretical Approaches to non-Perturbative QCD}
In this brief introduction, we focus on theoretical frameworks that have a
rigorous justification in QCD and that, with allowance for further progress in
the coming years and with a reasonable extrapolation of available computing
resources, can be expected to provide a direct confrontation of the data from
experiments like \PANDA  with the predictions of non-perturbative QCD. Among these
theoretical approaches the best established are ($i$) lattice QCD, which attempts
a direct attack to solve QCD non-perturbatively by numerical simulation, and
($ii$) effective field theories, which exploit the symmetries of QCD and the
existence of hierarchies of scales to provide predictions from effective
Lagrangians that are equivalent to QCD; among the latter we distinguish
systematic effective field theories formulated in terms of quark-gluon and in
terms of hadronic degrees of freedom.
\par
It should be kept in mind that the theoretical approaches will, in many cases,
calculate quantities that require an additional step to be compared with measured
data. This is because a full computation of the cross section as measured by
experimentalists in antiproton-proton collisions demands significantly more
effort. Such an additional step may involve a partial-wave analysis of the
measured data, in order to, for instance determine the quantum numbers of a
resonance. Effective field theory with hadronic degrees of freedom offers some
advantages in this respect. In the case of \PANDA, an example is the associated
production of hyperon-antihyperon pairs in the reactions
$\pbarp\rightarrow\overline{Y}Y$, where the spin observables in the final
state can be precisely measured~\cite{Pas06}. Therefore, the possibility exists
in this case to perform a full partial-wave analysis of the data to study in
detail the contribution of resonances.
\section{Lattice QCD: Status and Prospects}
Lattice QCD (LQCD) is an {\it ab initio} approach to deal with QCD in the
non-perturbative low-energy regime. The equations of motion of QCD are discretised
on a 4-dimensional space-time lattice and solved by large-scale numerical
simulations on big computers. For numerical reasons the QCD action is Wick
rotated into Euclidean space-time. The lattice spacing, $a$, acts as the
ultraviolet regulator of the theory. By letting $a\rightarrow0$, the regulator is
removed and continuum results are obtained. LQCD (originally proposed by Wilson
in 1974) has made enormous progress over the last decades. In the past, the
accuracy of LQCD results were limited by the use of the ``quenched
approximation'' ({\it i.e.} the neglect of sea quarks), by unrealistic heavy up
and down quarks, and by the use of only two instead of three light quark flavours.
These deviations from ``real QCD'' were partly mandated by the limited
availability of CPU-time. In recent years, all these limiting aspects (finite
volume effects, lattice artefacts, unrealistic quark masses, exclusion of sea
quarks) are being improved upon gradually. Thus, there is every reason to expect
that progress in LQCD will continue in the future, to the extent that precise
LQCD predictions will be available when the \PANDA experiment starts.
\begin{figure}[tb]
\centerline{\includegraphics*[width=\swidth]{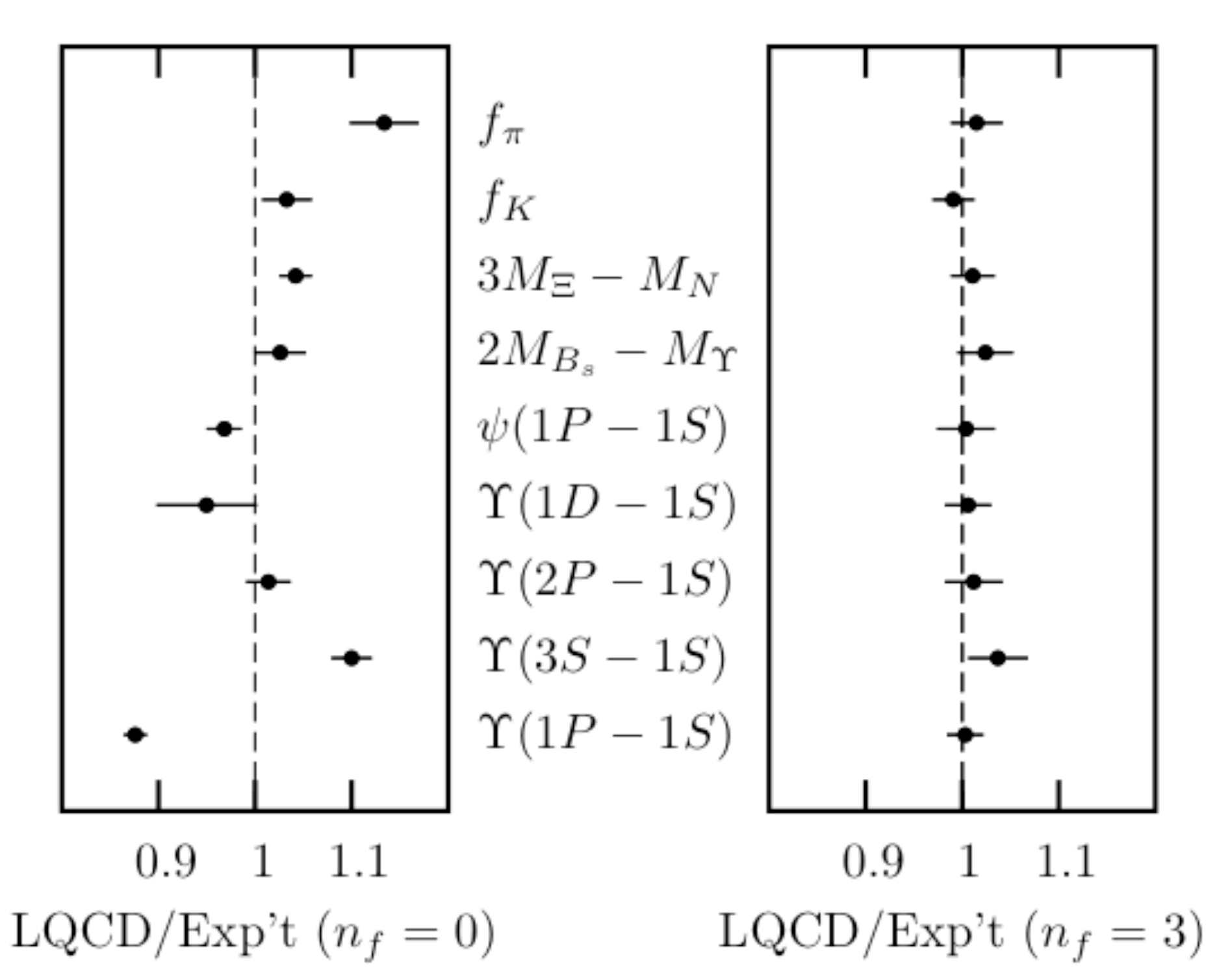}}
\caption[LQCD results divided by experimental values for selected quantities
         in hadronic physics.]{On the left the quenched calculations are shown,
         on the right the corresponding unquenched ones. While the former exhibit
         systematic deviations of some 10\percent from experiment, the latter are
         in impressive agreement with experiment.}
         \label{fig:3}
\end{figure}
\par
In LQCD, the SU(3) group elements $U_{x,\mu}$ are 3$\times$3 matrices defined on
the links that connect the neighbouring sites $x$ and $x+a\hat\mu$ on the lattice.
Traces of products of such matrices along closed paths, so-called Wilson loops,
are gauge invariant. The elementary building block is the ``plaquette,'' the
1$\times$1 lattice square. The simplest discretised action is then the Wilson
action, which is proportional to the (gauge-invariant) trace of the sum over all
plaquettes,
\begin{equation}
   S_{W} = -\frac{6}{g^2}\,{\rm Re}\sum_{x,\mu>\nu}{\rm Tr}\,\Pi_{x,\mu,\nu} \ ,
\end{equation}
where
$\Pi_{x,\mu\nu}=U_{x,\mu}\,U_{x+a\hat\mu,\nu}\,U^\dagger_{x+a\hat\nu,\mu}\,U^\dagger_{x,\nu}$.
In the continuum limit, this action agrees with the (Yang-Mills part of the) QCD
action (in 4D Euclidean space) to order ${\cal O}(a^2)$,
\begin{eqnarray}
   S_{YM} &=& -\frac{1}{4\,g^2}\int {\rm d}^4x \, G^{\mu\nu}_a(x) \, G_{\mu\nu}^a(x) \nonumber \\
          &=& S_W + \mbox{constant} + {\cal O}(a^2) \ .
\end{eqnarray}
The simplest discretised version for the fermionic quark part of the QCD action
reads
\begin{eqnarray}
   &&S_{f} =
	\sum_{x}\, \Big( \frac{1}{2}\,\gamma_\mu\,\overline{\psi}_x\,
   	\Big[U_{x+a\hat\mu,\mu}\,\psi_{x+a\hat\mu}
   \nonumber \\ && \quad  -\,U^\dagger_{x-a\hat\mu,\mu}\,\psi_{x-a\hat\mu} \Big]
               + m \,a\,\overline{\psi}_x\,\psi_x \Big) \, ,
\end{eqnarray}
which again corresponds to the continuum action up to order ${\cal O}(a^2)$
terms. A major part of the progress in LQCD has consisted in developing improved
versions of these naive discretised actions for the gluon and the quark fields,
for instance to remove the order ${\cal O}(a^2)$ artefacts when taking the
continuum limit. Improved fermionic actions have been developed for instance to
implement an exact chiral symmetry on the lattice that in the continuum limit
corresponds to the usual continuum chiral symmetry.
\par
In LQCD the expectation values of $n$-point Green functions are calculated in the
path integral sense, by evaluating their averages over all possible gauge field
configurations on the lattice, weighted with the exponent of the action. For
instance, for a hadron mass a zero-momentum two-point Green function with the
desired quantum numbers is calculated that creates the hadron at Euclidean time
zero and annihilates it at time $t$. For large $t$, this quantity will decay as
$\exp(-m\,t)$, from which the mass can be extracted. In such a LQCD calculation
it is the production of the gauge-field configurations that consumes the bulk of
the CPU-time, especially in (``unquenched'') calculations that include sea
quarks.
\par
\begin{figure}[ht]
\centerline{\includegraphics*[width=\swidth]{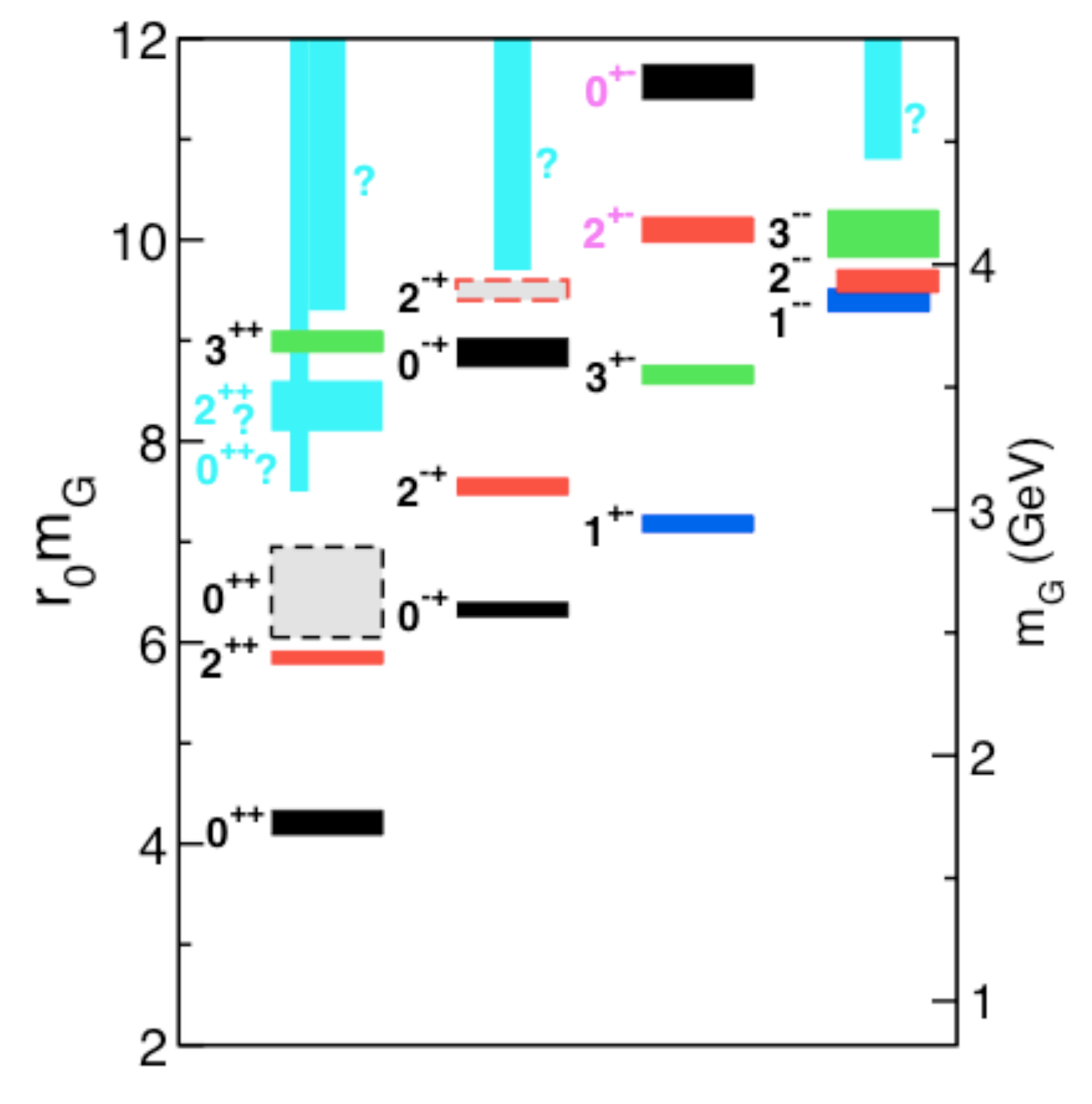}}
\caption{The LQCD glueball spectrum in pure SU(3) gauge theory~\cite{Mor99}.}
\label{fig:LQCD-glueball}
\end{figure}

Many impressive results have been obtained for hadron spectroscopy within LQCD.
As an example of great relevance to the \PANDA programme, we discuss briefly
(quenched) lattice calculations for the glueball spectrum. One first chooses an
interpolating operator for a glueball with specific quantum numbers, {\it e.g.}
for a scalar glueball one can take
\begin{equation}
O(\vec{x},t) = \sum_{i<j=x,y,z} {\rm Re}\,{\rm Tr}\,U_{ij}(\vec{x}) \ ,
\end{equation}

where $U_{ij}$ is the plaquette in the $ij$-plane. The glueball masses are then
obtained from the asymptotic behaviour of the time correlator
\begin{eqnarray}
C(t) &=& \sum_{x,x',t}\,\langle O(\vec{x},t)\,O^\dagger(\vec{x}',0) \rangle \nonumber \\
     &=& \sum_i |\langle 0|O|i\rangle|^2 \exp(-m_i\,t) \ .
\end{eqnarray}
However, because the scalar glueball has vacuum quantum numbers, it provides from
the simulation point of view a particular ``noisy'' signal.
The mass must be obtained from a fit to a
function of the type $C(t)\simeq C_0+C_1\exp(-m\,t)$. Special techniques have
been developed to improve the signal-to-noise quality of the glueball signal on
the lattice. For instance, one can use an anisotropic lattice with a smaller
lattice spacing in the temporal direction. The glueball spectrum obtained in this
manner~\cite{Mor99} is shown in Fig.~\ref{fig:LQCD-glueball}.
\par
While this LQCD result for the glueball spectrum is already very impressive,
further improvement is clearly needed in order to compare ultimately to
spectroscopic results obtained from a partial-wave analysis of the experimental
data. The glueballs from a quenched calculations, for instance, decay only to
lighter glueballs and the $\eta'$ meson. Moreover, significant limitations will have
to be overcome for LQCD to become as predictive for dynamical (scattering)
observables as it is for spectroscopy. Lattice QCD applications in the charm
sector constitute an important test for applications to hadronic corrections
required in B physics.
\par
From lattice calculation of various quantities the QCD coupling constant can be
determined. For instance, from the bottomonium spectrum, the value $\alpha_S(M_Z)
= 0.1170 \pm 0.0012$ was extracted. Also the running of $\alpha_S$ was studied
and the results are in good agreement with the two-loop result calculated within
pQCD.
\par
An area that so far has proven challenging for LQCD, but where it is foreseeable
that significant progress will be made in the coming years, is the (ground-state)
structure of the nucleon. The nucleon electromagnetic form factors, measured with
increasing precision in electron scattering, are classical observables in this
respect, but recent surprising results obtained at \INST{JLAB} demonstrate that our
understanding is far from complete. The proton charge form factor falls off more
rapidly than the standard dipole form supported by pQCD, and also the neutron
data are not consistent with pQCD. LQCD results are becoming more accurate, but
at present only the space-like region, in a limited $q^2$-range, can be handled.
Significant progress has been achieved in recent years by applying dispersion
relation techniques to relate the space-like and the time-like
regime~\cite{Pac07}. The latter can be addressed by \PANDA in a wide $q^2$-range
via the reaction $\pbarp\rightarrow e^+e^-$.
\par
Certain hard exclusive processes in electron scattering have offered more
detailed probes of the transition from the non-perturbative to the perturbative
regime in QCD. The success of the theoretical framework of generalized parton
distributions (GPDs) to describe the ``soft'' part of deeply virtual Compton
scattering is a case in point. With \PANDA the opportunity exists to measure the
time-like counterparts of such processes, {\it viz.\/} antiproton-proton
annihilation with crossed kinematics, as in
$\pbarp\rightarrow\gamma\gamma$. A framework analogous to the GPDs for the
space-like case has been developed: the amplitudes that encode the soft part of
these annihilation reactions are the generalized distribution amplitudes (GDAs).
The theoretical description within QCD of such time-like dynamical processes
needs to be developed.
\section{EFT with Quark and Gluon Degrees of Freedom}
{\it Ab initio\/} calculations from QCD, be it pQCD or LQCD, are and will remain
very difficult, especially in situations where several dynamical scales are
involved. Effective field theory (EFT) techniques in many such cases can provide
a solution. For instance, LQCD is particularly powerful when it is combined with
EFT. A variety of EFTs with quark and gluon degrees of freedom have been
developed in recent years. Exploiting a scale separation a simpler theory is
obtained that is equivalent to full QCD in the energy region considered. The
degrees of freedom above a chosen energy scale are integrated out (in a
path-integral sense) and the resulting field theory is organized as a power
series of operators containing the low-energy degrees of freedom over the heavy
scales. The high-energy physics is encoded in the coupling constants multiplying
these operators, which are calculated by ``matching'' selected observables in the
EFT and in full QCD. In the process, the symmetries of QCD need to be obeyed.
\subsection{Non-Relativistic QCD}
Applications of QCD to systems involving charm, beauty or top quarks, can be
simplified significantly by the use of effective field theory methods. It is not
always necessary to solve the dynamics of such systems based on the QCD
Lagrangian (\Refeq{def-QCD-Lag}) directly. One may integrate out fast modes of
the heavy quarks by a systematic non-relativistic expansion~\cite{God95}. The
role of the four-component Dirac spinor fields $q_f$ is taken over by
two-component Pauli spinor fields $\psi_f$ and $\chi_f$ describing heavy quarks
and antiquarks. For the purpose of charm physics as studied with \PANDA at FAIR it
is useful to integrate out the $b$ and $t$ quarks and keep $u$, $d$, $s$, and $c$
quarks as active degrees of freedom only. The resulting effective Lagrangian of
QCD reads
\begin{eqnarray}
  && {\cal L}^{\rm eff}_{QCD}  =  -\frac{1}{4}\, G^{\mu\nu}_a G_{\mu\nu} \nonumber\\
	& & +\,\sum_{f=u,d,s} \bar q_f \left[ i\,\gamma^\mu D_\mu - m_f \right] q_f \,
\nonumber\\
&& +\, \psi^\dagger \left[ i\,D_0 \nonumber + \frac{1}{2\,m_c}{\boldsymbol{D}}^2 \right] \psi
\nonumber\\
   & &+ \,\chi^\dagger \left[ i\,D_0 - \frac{1}{2\,m_c}{\boldsymbol{D}}^2 \right] \chi
+ \sum_{n=0}^\infty \,{\mathcal L}_n ,
\label{def-NRQCD}
\end{eqnarray}
where the non-trivial terms ${\mathcal L}_n $ are expanded in inverse powers of
the charm-quark mass. The index $n$ denotes the number of heavy-quark and
antiquark fields involved. Terms with $n\geq 4$ arise only if part of the gluon
dynamics is integrated out, as will be discussed below. For a given heavy-quark
species we illustrate the form of typical one-body terms,
\begin{eqnarray}
&&{\cal L}_{2}
= \frac{c_1}{8\,m_f^3}
\,\Big( \psi^\dagger ({\bf D}^2)^2 \psi \;-\; \chi^\dagger ({\bf D}^2)^2 \chi
\Big) \nonumber \\
&&\quad +\, \frac{c_2}{8\,m_f^2}\,
\Big( \psi^\dagger ({\bf D} \cdot g\, {\bf E} - g \,{\bf E} \cdot {\bf D})\, \psi \nonumber \\
&& \qquad +\, \chi^\dagger ({\bf D} \cdot g \,{\bf E} - g \,{\bf E} \cdot {\bf D})\, \chi\Big) \nonumber \\
&&\quad +\, \frac{c_3}{8\,m_f^2}\,
\Big( \psi^\dagger (i \,{\bf D} \times g\, {\bf E} - g\, {\bf E} \times i \,{\bf D})
\cdot \mbox{\boldmath $\sigma$} \,\psi \nonumber \\
&& \qquad +\, \chi^\dagger (i\, {\bf D} \times g \,{\bf E} - g \,{\bf E} \times i \,{\bf D})
\cdot \mbox{\boldmath $\sigma$} \,\chi \Big) \nonumber \\
&&\quad +\, \frac{c_4}{2\,m_f}
\,\Big( \psi^\dagger (g\, {\bf B} \cdot \mbox{\boldmath $\sigma$})\, \psi
\;-\; \chi^\dagger (g\, {\bf B} \cdot \mbox{\boldmath $\sigma$}) \,\chi \Big) \nonumber \\
&&\quad +\, \dots \;,
\label{def-NRQCD-1-body}
\end{eqnarray}
where $E^i = G^{0i}$ and $B^i = \mbox{$\frac{1}{2}$} \epsilon^i_{\phantom{i}jk}
G^{jk}$ are the electric and magnetic components of the gluon field strength
tensor $G^{\mu \nu}$. The coefficients $c_i$ in (\Refeq{def-NRQCD-1-body}) are
calculable in perturbation theory with $c_i =1 + {\mathcal O} \left(
\alpha_s\right)$. The relevance of the various terms in
(\Refeq{def-NRQCD-1-body}) is predicted by power-counting rules~\cite{God95}.
\par
The characteristic quantity that controls the relative importance of the infinite
number of terms in ${\cal L}_n$ is the typical velocity $v $ of the heavy quark.
By assumption it must hold $v \ll 1$, as to justify the use of the
non-relativistic fields in (\Refeq{def-NRQCD}). The precise realization of the
power-counting rules depends on the system, whether for instance heavy-light or
heavy-heavy systems are studied. We focus on heavy-quarkonium
systems~\cite{Bra04} for which the two terms $D_0$ and ${\boldsymbol{D}}^2/(2\,
m_f)$ in (\Refeq{def-NRQCD}) are of equal importance. This is reflected in the
counting rules with $D_0 \sim m_f\,v^2$ and ${\boldsymbol{D}} \sim m_f \,v$. If
supplemented by the identification $\alpha_S(m_c) \sim v$ and $g\,{\bf E } \sim
m_f^2 \,v^3$ and $g\,{\bf B } \sim m_f^2 \,v^4$ all terms displayed in
(\Refeq{def-NRQCD-1-body}) are of equal relevance~\cite{God95}.
\par
The effective Lagrangian (\Refeq{def-NRQCD}) defines still a quite complicated
theory, since it is not always justified to treat the gluon dynamics in an
expansion in powers of $\alpha_S $. Nevertheless, it is a powerful tool since it
can be and is used for simulations on the lattice. Depending on the typical
velocity $v$ of the heavy quark it is possible to integrate out the gluon
dynamics at least in part. This is desirable since it would make contact with the
phenomenological quark-potential model~\cite{Swa06}. The one-gluon exchange part
of such models follows by the assumption of perturbative gluon dynamics. A
corresponding contribution in ${\cal L}_4$ of (\Refeq{def-NRQCD}) would arise.
\begin{figure}[tb]
\centerline{\includegraphics*[width=1.1\swidth]{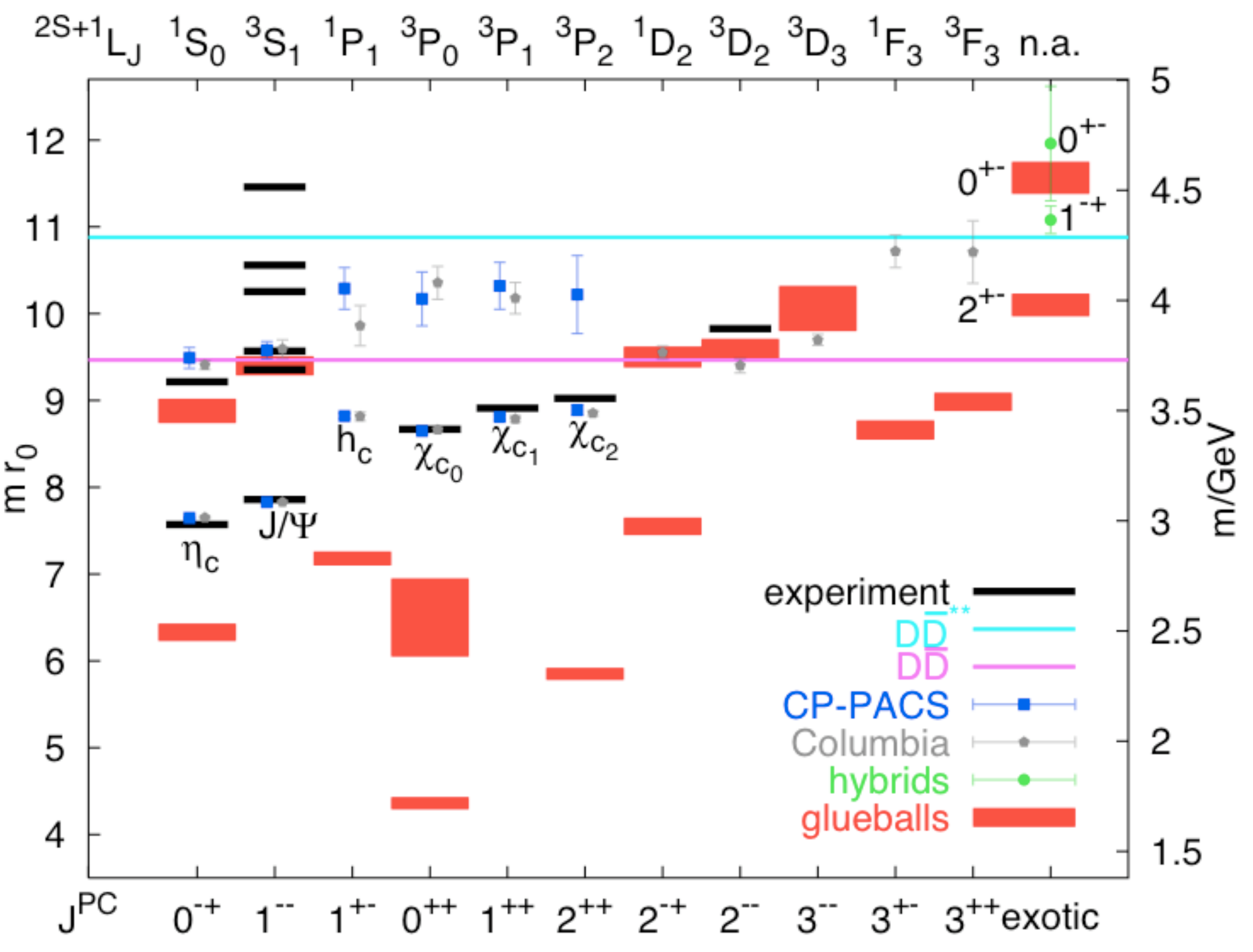}}
\caption{LQCD predictions for the charmonium, the glueball and the spin-exotic
         $\overline{c}c$-glue hybrids spectrum in quenched lattice QCD.}
         \label{fig:4}
\end{figure}
For charmonium systems there are three relevant scales: the mass $m_c$ (the
``hard scale''), the momentum transfer $m_c\,v$ (the ``soft scale'', proportional
to the inverse of the typical size of the system), and the binding energy
$m_c\,v^2$ (the ``ultra-soft scale'', proportional to the inverse of the typical
time of the system). In charmonium $v^2 \simeq 0.3$. The crucial question is how
$\Lambda_{QCD}$ relates to the three scales discussed above. If $\Lambda_{QCD} >
m_c\,v^2$ there must be non-perturbative physics involved when integrating out
gluon dynamics. Since one expects $m_c > m_c \,v \sim \Lambda_{QCD} $, part of
the gluon dynamics is perturbative~\cite{Luk97,Bra05}.
\par
One can integrate out the soft scale $m_c\,v$ by employing a matching procedure
keeping only dynamical ultra-soft degrees. This results in potential
Non-relativistic QCD (pNRQCD). In lowest order the problem reduces to solving a
Schr\"odinger equation for the $c\, \bar c$ states. The quark-potential model is
recovered from pNRQCD, with potentials calculated from QCD following a formal
non-perturbative procedure. An actual evaluation of the low-energy part requires
a calculation on a lattice or QCD vacuum models. Some actual examples are shown
in Fig.~\ref{fig:4}. The charmonium ground-state hyperfine splitting has been
calculated at NLO.
\section{EFT with Hadronic Degrees of Freedom}
We discuss the concepts of EFT with hadronic degrees of freedom at hand of
open-charm systems in some detail. This is a sector most relevant for the \PANDA
experiment with its goal to study properties of the spectrum of the $D$ mesons in
free-space and in nuclear matter. Analogous developments are possible in the
light sector of QCD with up, down and strange quarks only. The description of
baryon resonances with double strangeness, another important topic of the \PANDA
experiment, will profit from such developments. The study of hypernuclei with
\PANDA is motivated in part as a mean to learn on the interaction of hyperons. For
the latter chiral effective field theories are being developed \cite{Pol06}.  In
the charmonium sector, the construction of an EFT with hadronic degrees of
freedom is in its infancy, though further developments would be important for the
charmonium programme at \PANDA.
\par
There are three important steps in the course of constructing an EFT. The first
step is the choice of degrees of freedom. In most cases this can not be derived
from QCD, but must be conjectured and then falsified by explicit computations to
be confronted with QCD lattice results or experiments. Most predictive are EFTs
involving Goldstone boson fields, since the interaction of the latter with matter
fields is strongly constrained by the spontaneously broken chiral SU(3) symmetry
QCD. The second step is of more technical nature, the construction of the
effective Lagrangian, in a manner consistent with the symmetries of QCD. Though
the second step is least controversial, it involves an infinite number of unknown
parameters. This leads to the third crucial step: the identification of a
suitable approximation scheme for the effective Lagrangian based on
power-counting rules. This programme is quite rewarding since the leading order
Lagrangian involves typically a few unknown parameters only, in terms of which a
lot of physics can be understood.
\par
A quite radical approach is the hadrogenesis conjecture~\cite{Lut02}, which
postulates the spectrum of QCD to be generated by the interaction of a few
``quasi-fundamental'' hadronic degrees of freedom, the selection of which is
guided by symmetry properties of QCD. Clearly, alternative or complementary
assumptions may be used at this stage. It is an important goal of the \PANDA
project to give answers to the fundamental question how QCD manifests itself in
the hadronic spectrum.
\subsection{Chiral Symmetry and Open Charm Meson Systems}
In the open charm sector consider the flavour octet of Goldstone bosons
\begin{eqnarray}
\Phi =\left(\begin{array}{ccc}
\piz+\frac{1}{\sqrt{3}}\,\eta &\sqrt{2}\,\pi^+&\sqrt{2}\,K^+\\
\sqrt{2}\,\pi^-&-\piz+\frac{1}{\sqrt{3}}\,\eta&\sqrt{2}\,K^0\\
\sqrt{2}\,K^- &\sqrt{2}\,\bar{K}^0&-\frac{2}{\sqrt{3}}\,\eta
\end{array}\right) \,,
\label{def-Phi}
\end{eqnarray}
and the flavor anti-triplets of pseudo-scalar and vector open-charm mesons
\begin{eqnarray}
P&=&(D^0, -D^+, D_s^+) \,, \nonumber \\
P_{\mu \nu}&=&(D^{*0}_{\mu \nu}, -D^{*+}_{\mu \nu}, D_{s, \mu \nu}^{*+}) \,,
\label{}
\end{eqnarray}
where one may represent the vector states by antisymmetric tensor fields. In the
limit of an infinite charm-quark mass the properties of pseudo-scalar and vector
$D$ mesons are closely related, in particular they are mass degenerate. In order
to construct the effective Lagrangian describing the interaction of Goldstone
bosons and the $D$ mesons it is useful to identify building blocks that have
convenient transformation properties under chiral transformations
\begin{eqnarray}
U_\mu &=& \frac{1}{2}\, u^\dagger \,\Big(\Big(\partial_\mu e^{i\,\frac{\Phi}{f}}\Big) + i\,e\,A_\mu\,
\Big[Q,e^{i\,\frac{\Phi}{f}}\Big] \Big)\,u^\dagger \,, \nonumber \\
u &=& \exp \left( \frac{ i \,\Phi }{2\,f} \right)\,, \nonumber \\
\chi_\pm &=& \frac{1}{2}\,u\,\chi_0 \,u \pm \frac{1}{2}\,u^\dagger\,\chi_0 \,u^\dagger \,, \nonumber \\
\chi_0 &=& 2\,B_0 \, \left(
\begin{array}{ccc}
m_u & 0 & 0\\
0 & m_d & 0 \\
0 & 0 & m_s
\end{array}
\right) \,,
\label{building-blocks}
\end{eqnarray}
where the parameter $f \simeq f_\pi$ may be identified with the pion-decay
constant, $f_\pi =92.4$\,MeV, at leading order. A precise determination of $f$
requires a chiral SU(3) extrapolation of some data set. The field $U_\mu$
involves the photon field $A_\mu$ and the electromagnetic coupling constant $e
\simeq 0.303$ and the quark charge matrix $Q = {\rm diag}(2,-1,-1)/3$. The
building blocks in (\Refeq{building-blocks}) illustrate two important aspects
of EFTs: first the Goldstone boson field $\Phi$ enters in a non-linear fashion
and second the chiral symmetry breaking fields $\chi_\pm $ are proportional to
the quark-mass matrix of QCD. The parameter $B_0$ is related to the chiral quark
condensate.
\par
The chiral Ward identities of QCD are transported into the EFT by the use of a
covariant derivative, $D_\mu$,
\begin{eqnarray}
D_\mu \,U_\nu &=&
\partial_\mu\,U_\nu +\Big[\Gamma_\mu ,\,U_\nu \Big]
+i\,e\,A_\mu\,\Big[Q,\, U_\nu \Big] \,, \nonumber\\
D_\mu \,P &=& \partial_\mu\,P - P\,\Gamma_\mu +i\,e\,P\,Q'\,A_\mu \,, \nonumber\\
\Gamma_\mu &=& \frac{1}{2}\,\Big( u^\dagger \,\partial_\mu\,u +u
\,\partial_\mu\,u^\dagger \Big) \,, \qquad
\label{def-Umu}
\end{eqnarray}
with $Q'={\rm }(0,1,1)$. Given the building blocks Eqs.~(\ref{building-blocks},
\ref{def-Umu}) it is straightforward to write down interaction terms that are
compatible with the chiral constraints of QCD. This is because a covariant
derivative $D_\mu$ acting on the fields $U_\mu$, $P$ , $P_{\mu \nu}$ or $\chi_\pm
$ does not alter their transformation properties under chiral
transformation~\cite{Kra90}.
\par
We display the leading order Lagrangian constructed in terms of the building
blocks Eqs.~(\ref{building-blocks}, \ref{def-Umu})
\begin{eqnarray}
&& {\mathcal L}= f^2\, {\rm tr \,} \Big\{ U^\mu\,U^\dagger_\mu \Big\}
+\frac{1}{2}\,{\rm tr \,} (\chi_+) \nonumber\\
&&\quad +\, (D_\mu P) \, (D^\mu \bar P)- P\,M^2_{0^-} \,\bar P \nonumber\\
&&\quad -\,(D_\mu P^{\mu \alpha}) \,(D^\nu \bar P_{\nu \alpha})
+ \frac{1}{2}\,P^{\mu \alpha}\,M^2_{1^-} \,\bar P_{\mu \alpha} \nonumber\\
&&\quad+ \,2\,g_P\,\Big\{P_{\mu \nu}\,U^\mu\,(D^\nu \bar P)
- (D^\nu P )\,U^\mu \,\bar P_{\mu \nu} \Big\}
\nonumber\\
&& \quad - \,i\,\frac{\tilde g_P}{2}\,\epsilon^{\mu \nu \alpha \beta}\,\Big\{
P_{\mu \nu}\,U_\alpha \,(D^\tau \bar P_{\tau \beta} ) \nonumber\\
&& \quad \qquad +\, (D^\tau P_{\tau \beta})\,U_\alpha\,\bar P_{\mu \nu} \Big\} \,,
\label{def-kin-Goldstone}
\end{eqnarray}
where we use the notation $\bar P = P^\dagger$. The decay of the charged
$\Dx$-mesons implies $|g_P| = 0.57 \pm 0.07 $. The parameter $\tilde g_P$ in
(\Refeq{def-kin-Goldstone}) can not be extracted from empirical data directly.
In the absence of an accurate evaluation within unquenched lattice QCD, the size
of $\tilde g_P$ can be estimated using the heavy-quark symmetry, one expects
$\tilde g_P = g_P \,$ at leading order~\cite{Yan92}. In that limit it holds also
that $M_{0^-}=M_{1^-}$.
\par
If we admit isospin-breaking effects, {\it i.e.} $m_u \neq m_d$, there is a term in
\Refeq{def-kin-Goldstone}) after insertion of \Refeq{def-Phi} proportional to
$(m_u-m_d)\,\piz\,\eta$. A unitary transformation is required such that the
transformed fields $\tilde \piz$ and $\tilde \eta$ decouple. The Lagrangian
density (\Refeq{def-kin-Goldstone}), when written in terms of the new fields
and a mixing angle $\epsilon$, does not show a $\tilde\pi^0\,\tilde\eta$ term if
and only if
\begin{eqnarray}
\frac{\sin (2\,\epsilon)}{\cos (2 \,\epsilon)} =
\sqrt{3}\,\frac{m_d-m_u}{2\,m_s-m_u-m_d} \,,
\label{eps-def}
\end{eqnarray}
where we recalled the value for the mixing angle $\epsilon = 0.010 \pm 0.001$ as
determined in Ref.~\cite{Gas85}.
\par
We give a brief discussion of the power-counting rules underlying
(\Refeq{def-kin-Goldstone}). Such counting rules are based on naive dimensional
counting supplemented by a naturalness assumption. Dimensional counting rules are
realized in a computation of a given Feynman diagram if the theory is treated in
dimensional regularization. It is an important issue to devise renormalisation
schemes that are compatible with a given counting scheme. To be specific we
collect some counting rules
\begin{eqnarray}
U_\mu &\sim& Q\,, \nonumber\\
D_\mu \,U_\mu &\sim& Q^2 \,, \nonumber\\
D_\mu \,P &\sim& Q^0 \,, \nonumber\\
\qquad \chi_\pm &\sim& Q^2 \,.
\end{eqnarray}
The fact that a covariant derivative acting on $U_\mu$ must be counted $Q$
reflects the ``lightness'' of the Goldstone boson fields: the squared mass of the
Goldstone boson is proportional to the current quark mass of QCD. On the other
hand the mass of a $D$ meson is much larger than the masses of the Goldstone
bosons. Thus, at a formal level one must assign a covariant derivative acting on
a $D$-meson field the order $Q^0$. A systematic approach arises if one orders the
interaction terms of the chiral Lagrangian according to inverse powers in the
charm-quark mass. The leading-order term is then linear in the charm quark mass,
like the QCD action is linear in that parameter.
\par
The central guiding rule of EFTs is the derivation of most general approximations
compatible with the symmetries of the underlying theory but also the fundamental
concepts of local quantum field theory like micro-causality, covariance,
unitarity, and crossing symmetry. As emphasized by Weinberg for the case of
nucleon-nucleon scattering a systematic approximation scheme can be devised if
one evaluates a two-body potential based on power-counting rules in a
perturbative manner~\cite{Wei90}. The latter should then be used in a
Schr\"odinger-type equation as to arrive at scattering amplitude compatible with
the unitarity constraint. It is important to realize that within this type of EFT
one does not apply the counting rules to the scattering amplitude directly,
rather to the subset or irreducible diagrams. The rationale behind this approach
is the observation that the ``naive'' counting rules are spoiled for a Feynman
diagram in phase-space regions close to the opening of thresholds: an appropriate
summation is required.
\par
We illustrate the power of the effective Lagrangian (\Refeq{def-kin-Goldstone})
and extract the leading order two-body interaction terms of the Goldstone bosons
with the $D$ mesons
\begin{eqnarray}
\!\!\!\!\!&&\!\!\!\!\! \!\!\!\!\!{\mathcal L}_{WT} = \frac{1}{8\,f^2}\,\Big\{ (\partial^\mu P)\,
\,[\Phi , (\partial_\mu \Phi)]\,\bar P \nonumber\\
&& \quad \qquad -\,P\,\,[\Phi , (\partial_\mu \Phi)]\,(\partial^\mu \bar P ) \Big\} \,, \nonumber\\
&&\quad -\, \frac{1}{8\,f^2}\,\Big\{ (\partial^\nu P_{\nu \alpha })\,
\,[\Phi , (\partial_\mu \Phi)]\,\bar P^{\mu \alpha } \nonumber\\
&& \quad \qquad -\, P_{\nu \alpha }\,
\,[\Phi , (\partial_\nu \Phi)]\,(\partial_\mu \bar P^{\mu \alpha } ) \Big\} \,.
 \label{WT-term}
\end{eqnarray}
\par
The Weinberg-Tomozawa interaction (\Refeq{WT-term}) is a direct consequence of
the chiral SU(3) symmetry of QCD and illustrates the predictive power of chiral
EFT: the interaction is determined by one parameter $f \simeq 90$\,MeV only that
is known from the pion-decay process. Analogous interactions of the Goldstone
bosons with other matter fields were derived and studied in detail in the recent
literature~\cite{Kai97,Oll98,Gar04,Kol04b,Lut04}.
It was shown that the chiral interaction (\Refeq{WT-term}) is relevant for the
study of open-charm resonances with $J^P=0^+$ or $1^+$ quantum number: it implies
unambiguously the existence of two resonances with masses below the $D K$ and
$\Dx K$ thresholds and astonishingly close to the empirical masses of the
$D_{s0}^\ast(2317)$ and $D_{s1}^\ast(2460)$~\cite{Kol04}. Such states manifest
themselves as poles in the $S$-wave scattering amplitude of Goldstone bosons with
the pseudo-scalar or vector open-charm ground states. For the formation of the
scalar $D_{s0}^\ast(2317)$ resonance the four isospin states $\langle K\,D, I |,
\langle \pi \,D_s, 1 | $ and $\langle \eta \,D_s, 0 |$ are relevant with their
coupled-channel interactions determined by (\Refeq{WT-term}) at leading order.
In the presence of isospin mixing all channels couple. The mixing of the two
isospin sectors is of order $\epsilon$. It predicts the leading isospin-violating
hadronic decay $D_{s0}^\ast(2317) \to \piz\,D_s$. Analogous statements hold for the
axial-vector state.
\par
The detailed consequence of the EFT approach for the properties of scalar and
axial vector open-charm states has been addressed in the recent literature where
chiral correction terms were considered systematically~\cite{Lut07,Guo08}. In
particular their electromagnetic and isospin-violating decay properties have been
computed and confronted with experimental bounds successfully~\cite{Lut07}. The
leading order interaction predicts sizeable attraction in channels only that have
an interpretation as $c \,\bar q$ states. Exotic channels, which would require
tetra-quark configurations, are sensitive to chiral correction terms. Since the
latter involve further a priori unknown parameters, the predictions in exotic
channel are more uncertain. Nevertheless, it was shown that if further
constraints of QCD, which arise in the limit of large number of colours $N_c$, are
considered one would expect exotic signals in the invariant mass distribution of
the $\pi D$ and $\eta \Dx$ channels. It would be an important step towards a
better understanding of the physics of open-charm meson systems to measure such
mass distributions with high accuracy~\cite{Lut07}.
\par
Further studies of open-charm systems that involve light vector mesons as active
degrees of freedom are necessary. Within the hadrogenesis conjecture one would
expect further resonance states, like tensor states, to be formed.
\subsection{Phenomenology of Open Charm Baryon Systems}
While EFT is a rigorous and systematic approach to solve QCD, there are many
systems for which such tools are not yet available. In such cases it is useful to
develop schematic models to pave the way towards more systematic approaches. A
good example is the interaction of $D$ mesons with nucleons. The theoretical
efforts to describe such systems with hadronic degrees of freedom are few so far.
Since the experimental study of the open-charm baryon spectrum and properties of
$D$ mesons in cold nuclear matter may be feasible with \PANDA, it is important to
work out the relevance of this physics for the better understanding of strong
QCD.
One may model the interaction of $D$ mesons with matter fields by a $t$-channel
exchange of universally coupled light vector mesons. If written down for all
two-body channels that are implied by the presence of up, down, strange and charm
quarks such an interaction is compatible with the leading order chiral
interaction in the case where the initial and final state involves a Goldstone
boson~\cite{Hof05}. For instance, the chiral interaction of Goldstone bosons with
the open-charm baryon ground state, that is analogous to (\Refeq{WT-term}), is
recovered. The open-charm baryon systems are considerably more complicated than
the open-charm meson systems since the charm quark may be exchanged from a meson
to baryon and vice versa. For instance such processes are implied by the
$t$-channel exchanges of $D$ mesons. This complicates the construction of an EFT
considerably. On the other hand, coupled-channel models based on that simple
$t$-channel exchange force predict a surprising rich phenomenology. Thus a
dedicated theoretical and experimental study appears quite rewarding and should
be undertaken.

%% file: panda_pb_exp.tex
%
%
\cleardoublepage
\chapter{Experimental Setup}
\label{sec:exp}
\COM{Author(s): A. Lehrach, L. Schmitt, G. Stancari}
%
%
\input{./exp/exp_overview}
\input{./exp/exp_detector}
\input{./exp/exp_hesr}
\input{./exp/exp_scan}
%
%
\newpage
\bibliographystyle{panda_pb_lit}
\bibliography{./exp/lit_exp_detector,./exp/lit_exp_hesr,./exp/lit_exp_scan,./main/lit_main}
%

%% file: exp/exp_overview.tex
%
\section{Overview}
\COM{Author(s): K. Peters}
%

%% file: exp/exp_detector.tex
%
\begin{figure*}[hbt]
\begin{center}
\includegraphics[width=\dwidth]{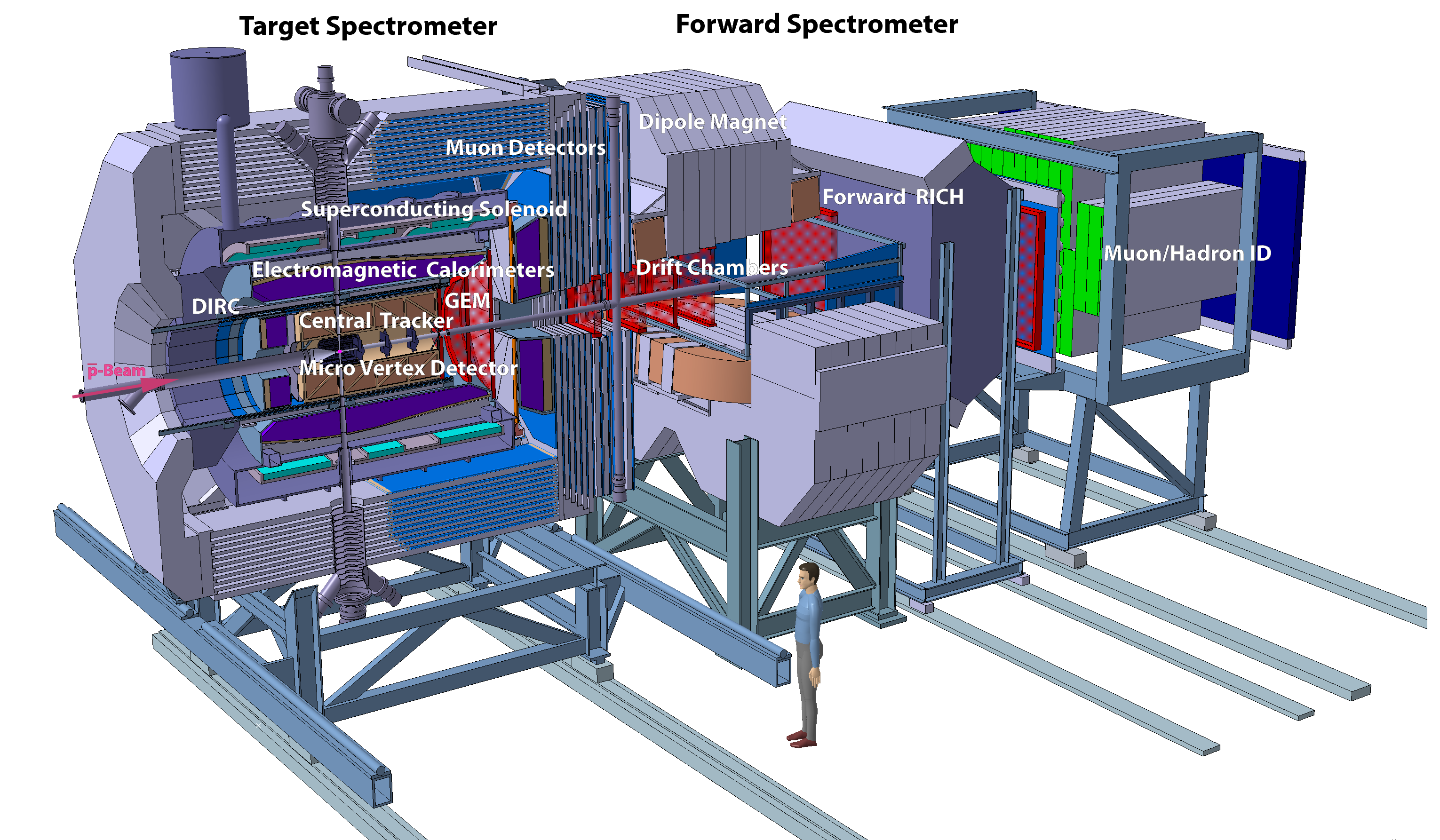}
\caption[Artistic view of the \PANDA Detector]
{Artistic view of the \PANDA Detector}
\label{fig:det:exp}
\end{center}
\end{figure*}
\section{The \PANDA Detector}
\label{exp:detector}
\COM{Author(s): L. Schmitt}
The main objectives of the design of the \PANDA experiment pictured in
\Reffig{fig:det:exp} are to achieve $4\pi$ acceptance, high resolution
for tracking, particle identification and calorimetry, high rate
capabilities and a versatile readout and event selection. To obtain a
good momentum resolution the detector is split into a {\em target
  spectrometer} based on a superconducting solenoid magnet surrounding
the interaction point and measuring at high angles and a {\em forward
  spectrometer} based on a dipole magnet for small angle tracks. A
silicon vertex detector surrounds the interaction point. In both
spectrometer parts tracking, charged particle identification,
electromagnetic calorimetry and muon identification are available to
allow to detect the complete spectrum of final states relevant for the
\PANDA physics objectives.
\par
In the following paragraphs the components
of all detector subsystems are briefly described.
\subsection{Target Spectrometer}
The target spectrometer surrounds the interaction point and measures
charged tracks in a solenoidal field of 2\,T. It consists of detector
layers arranged in an onion shell configuration. Pipes for the
injection of target material have to cross the spectrometer
perpendicular to the beam pipe.
\par
\begin{figure*}[hbt]
\begin{center}
\includegraphics[width=0.8\dwidth]{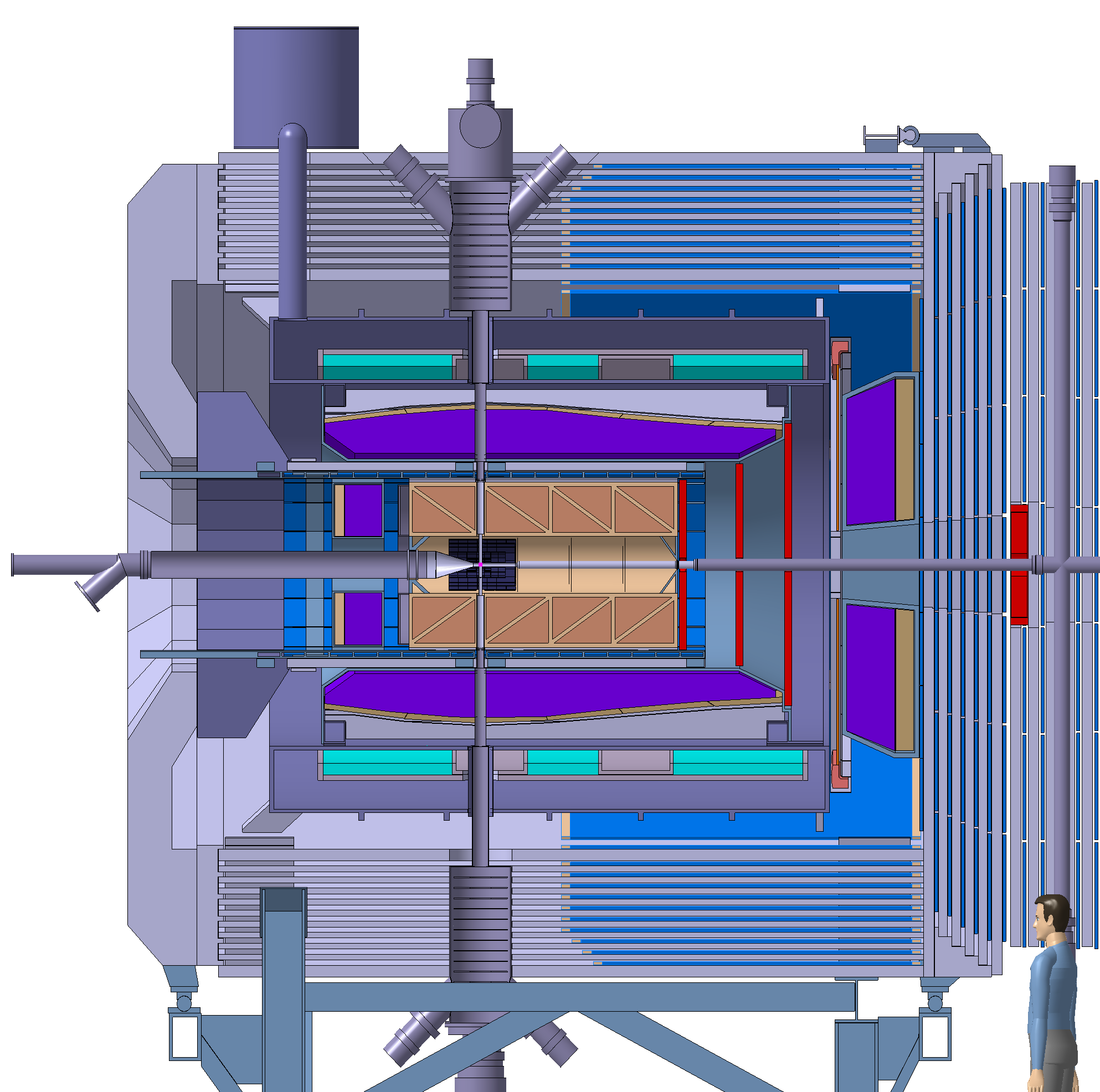}
\caption[Side view of the target spectrometer]
{Side view of the target spectrometer}
\label{fig:det:ts}
\end{center}
\end{figure*}

The target spectrometer is arranged in a barrel part for angles
larger than 22$\degrees$ and an end-cap part for the forward range down
to 5$\degrees$ in the vertical and 10$\degrees$ in the horizontal plane.
A side view of the target spectrometer is shown in \Reffig{fig:det:ts}.
\par
One of the main design requirements is compactness to avoid a too large
and a too costly magnet and crystal calorimeter.
\subsubsection{Target}
\label{sec:det:ts:tgt}
\begin{figure*}[hbt]
\begin{center}
\includegraphics[angle=90,width=\dwidth]{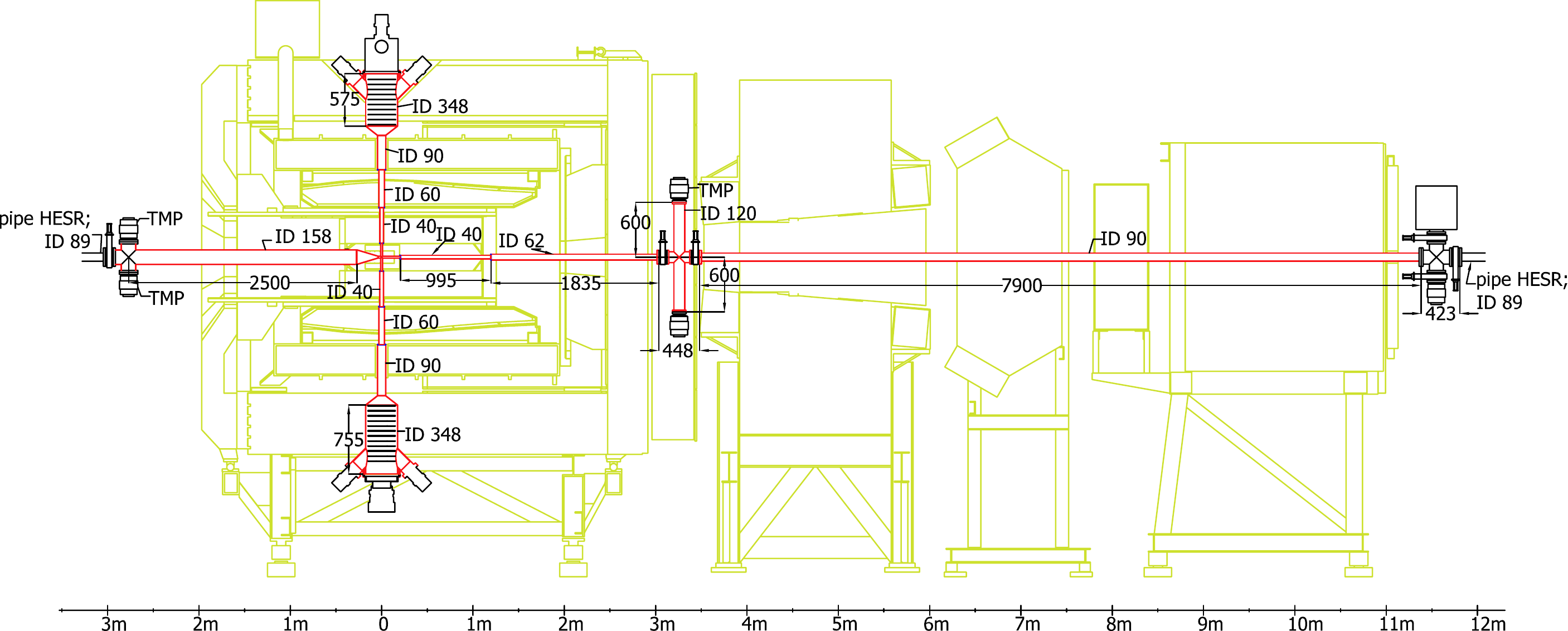}
\caption[Schematic of the target and beam pipe setup with pumps.]
{Schematic of the target and beam pipe setup with pumps.}
\label{fig:det:target}
\end{center}
\end{figure*}

The compact geometry of the detector layers nested inside the
solenoidal magnetic field combined with the request of minimal
distance from the interaction point to the vertex tracker leaves very
restricted space for the target installations.  The situation is
displayed in \Reffig{fig:det:target}, showing the intersection
between the antiproton beam pipe and the target pipe being gauged to
the available space.  In order to reach the design luminosity of
$2\EE{32}$\,cm$^{-2}$s$^{-1}$ a target thickness of about
$4\EE{15}$ hydrogen atoms per cm$^2$ is required assuming
$10^{11}$ stored antiprotons in the \INST{HESR} ring.
\par
These are conditions posing a real challenge for an internal target
inside a storage ring. At present, two different, complementary
techniques for the internal target are being developed: the
cluster-jet target and the pellet target. Both techniques are capable
of providing sufficient densities for hydrogen at the interaction
point, but exhibit different properties concerning their effect
on the beam quality and the definition of the interaction point. In
addition, internal targets also of heavier gases, like deuterium,
nitrogen or argon can be made available.
\par
For non-gaseous nuclear targets the situation is different in
particular in case of the planned hypernuclear experiment. In these
studies the whole upstream end cap and part of the inner detector
geometry will be modified.
\paragraph*{Cluster-Jet Target}
The expansion of pressurized cold hydrogen gas into vacuum through a
Laval-type nozzle leads to a condensation of hydrogen molecules
forming a narrow jet of hydrogen clusters. The cluster size varies
from $\EE{3}$ to $\EE{6}$ hydrogen molecules tending to become larger at higher
inlet pressure and lower nozzle temperatures.  Such a cluster-jet
with density of $\EE{15}$\,atoms/cm$^3$ acts as a very diluted target since it
may be seen as a localized and homogeneous mono-layer of hydrogen atoms
being passed by the antiprotons once per revolution.
\par
Fulfilling the luminosity demand for \PANDA still requires a density
increase compared to current applications. Additionally, due to
detector constraints, the distance between the cluster-jet nozzle and
the target will be larger. The size of the target region will be given
by the lateral spread of hydrogen clusters. This width should stay
smaller than 10 mm when optimized with skimmers and collimators both
for maximum cluster flux as well as for minimum gas load in the
adjacent beam pipes. The great advantage of cluster targets is the
homogeneous density profile and the possibility to focus the
antiproton beam at highest phase space density. Hence, the interaction
point is defined transversely but has to be reconstructed
longitudinally in beam direction. In addition the low $\beta$-function
of the antiproton beam keeps the transverse beam target heating effects
at the minimum. The possibility of adjusting the target density along
with the gradual consumption of antiprotons for running at constant
luminosity will be an important feature.
\paragraph*{Pellet Target}
The pellet target features a stream of frozen hydrogen micro-spheres,
called pellets, traversing the antiproton beam perpendicularly.
Pellet targets are presently in use at the \INST{WASA} at \INST{COSY} experiment
\cite{bib:tgt:pe:Buescher:2006} and were previously developed at \INST{TSL}
\cite{bib:tgt:pe:Ekstroem:1996jt}. Typical parameters for pellets at
the interaction point are the rate of 1.0 - 1.5\,$\EE{4}$\,s$^{-1}$, the
pellet size of 25 - 40\,$\umu$m, and the velocity of about\,60 m/s. At
the interaction point the pellet train has a lateral spread of
$\sigma\approx$\,1\,mm and an inter spacing of pellets that varies
between 0.5 to 5\,mm. With proper adjustment of the $\beta$-function of
the coasting antiproton beam at the target position, the design
luminosity for \PANDA can be reached.  The present R\&D is
concentrating on minimizing the luminosity variations such that the
instantaneous interaction rate does not exceed the acceptance of the
detector systems.  Since a single pellet becomes the vertex for more
than hundred nuclear interactions with antiprotons during the time a
pellet traverses the beam, it will be possible to determine the
position of individual pellets with the resolution of the micro-vertex
detector averaged over many events. R\&D is going on to devise an
optical pellet tracking system. Such a device could determine the
vertex position to about 50\,$\umu$m precision for each individual
event independently of the detector. It remains to be seen if this
device can later be implemented in \PANDA.
\par
The production of deuterium pellets is also well established, the use
of other gases like N$_2$, Ar or Xe as pellet target material does not
pose problems \cite{Boukharov:2008vh}.
\paragraph*{Other Targets}
are under consideration for the
hypernuclear studies where a separate target station upstream will
comprise primary and secondary target and detectors. Moreover, current
R\&D is undertaken for the development of a liquid helium target and a
polarized $^3$He target. A wire target may be employed to study
antiproton-nucleus interactions.
\subsubsection{Solenoid Magnet}
The magnetic field in the target spectrometer is provided by a
superconducting solenoid coil with an inner radius of 90$\,$cm and a
length of 2.8$\,$m. The maximum magnetic field is 2$\,$T. The field
homogeneity is foreseen to be better than 2\percent over the volume of the
vertex detector and central tracker.  In addition the transverse
component of the solenoid field should be as small as possible, in
order to allow a uniform drift of charges in the time projection
chamber. This is expressed by a limit of $\int B_r/B_z dz < 2$\,mm for
the normalized integral of the radial field component.
\par
In order to minimize the amount of material in front of the
electromagnetic calorimeter, the latter is placed inside the magnetic
coil. The tracking devices in the solenoid cover angles down to
5\degrees{}/10\degrees{} where momentum resolution is still
acceptable. The dipole magnet with a gap height of 1.4 m provides a
continuation of the angular coverage to smaller polar angles.
\par
The cryostat for the solenoid coils has two warm bores of 100$\,$mm
diameter, one above and one below the target position, to
allow for insertion of internal targets.
\begin{figure*}[hbt]
\begin{center}
\includegraphics[width=0.8\dwidth]{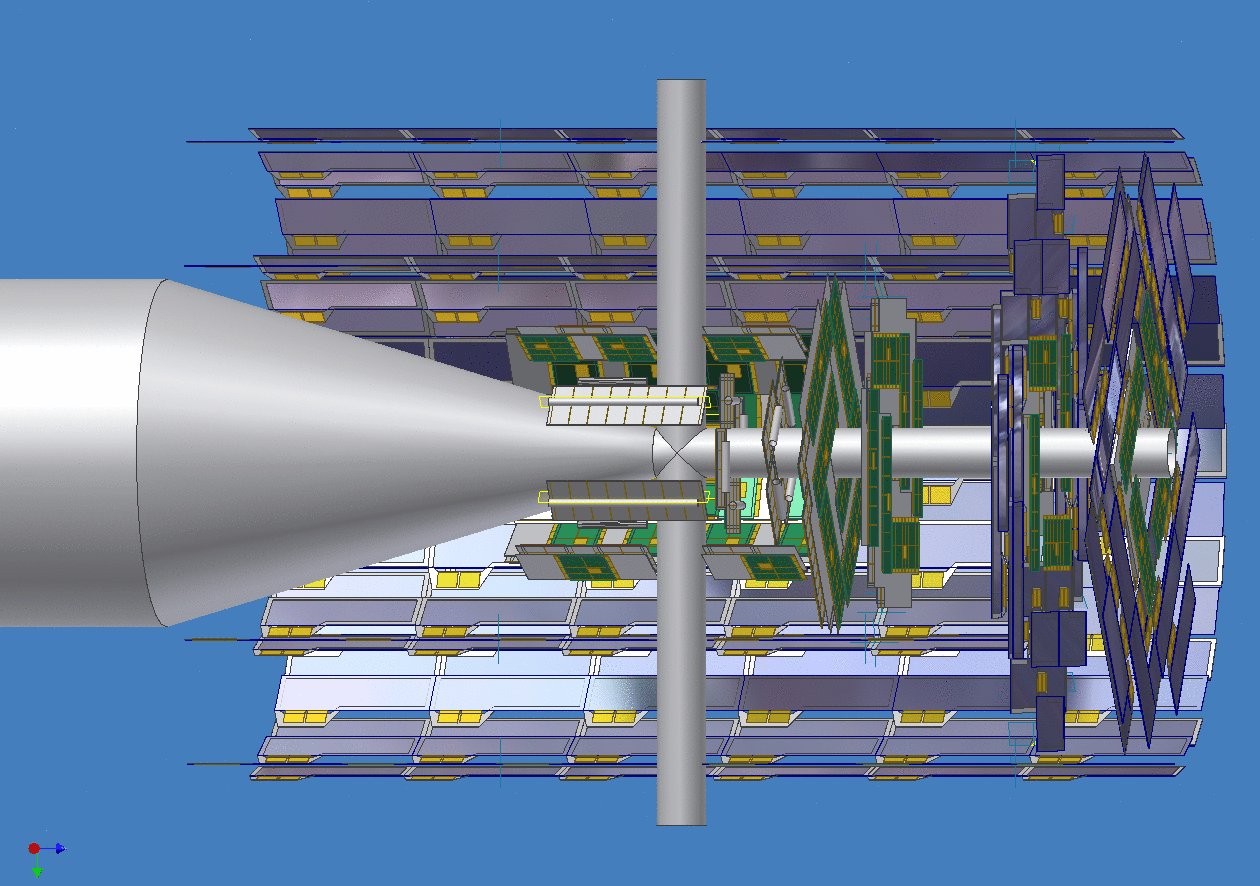}
\caption[The Micro-vertex detector of \PANDA]
{The Micro-vertex detector of \PANDA}
\label{fig:det:mvd}
\end{center}
\end{figure*}
\subsubsection{Microvertex Detector}
The design of the micro-vertex detector (\Mvd) for the target
spectrometer is optimized for the detection of secondary vertices from
\D and hyperon decays and maximum acceptance close to the interaction
point. It will also strongly improve the transverse momentum
resolution. The setup is depicted in \Reffig{fig:det:mvd}.
\par
The concept of the \Mvd is based on radiation hard silicon pixel
detectors with fast individual pixel readout circuits and silicon
strip detectors. The layout foresees a four layer barrel detector with
an inner radius of 2.5\,cm and an outer radius of 13\,cm. The two
innermost layers will consist of pixel detectors while the outer two
layers are considered to consist of double sided silicon strip
detectors.
\par
Eight detector wheels arranged perpendicular to the beam will achieve
the best acceptance for the forward part of the particle spectrum.
Here again, the inner two layers are made entirely of pixel detectors,
the following four are a combination of strip detectors on the outer
radius and pixel detectors closer to the beam pipe. Finally the last
two wheels, made entirely of silicon strip detectors, are placed
further downstream to achieve a better acceptance of hyperon cascades.
\par
The present design of the pixel detectors comprises detector wafers
which are 200$\,\umu$m thick (0.25\%$\,X_0$).  The readout via
bump-bonded wafers with ASICs as it is used in {\INST{ATLAS}} and
{\INST{CMS}} \cite{Atlas:Tdr:11,Cms:Tdr:5} is foreseen as the default
solution.  It is highly parallelised and allows zero suppression as
well as the transfer of analogue information at the same time. The
readout wafer has a thickness of 300 $\umu$m (0.37\%$\,X_0$). A pixel
readout chip based on a 0.13\,$\umu$m CMOS technology is under
development for \PANDA. This chip allows smaller pixels,
lower power consumption and a continuously sampling readout without
external trigger.
\subsubsection{Central Tracker}
The charged particle tracking devices must handle the high particle
fluxes that are anticipated for a luminosity of up to several
$10^{32}\,$cm$^{-2}$s$^{-1}$.  The momentum resolution $\delta p/p$
has to be on the percent level.  The detectors should have good
detection efficiency for secondary vertices which can occur outside
the inner vertex detector (e.g.\ $\Ks$ or $\Lambda$).  This is
achieved by the combination of the silicon vertex detectors close to
the interaction point (\Mvd) with two outer systems. One system is
covering a large area and is designed as a barrel around the
\Mvd. This will be either a stack of straw tubes (\Stt) or a
time-projection chamber (\Tpc). The forward angles will be covered
using three sets of GEM trackers similar to those developed for the
\INST{COMPASS} experiment \cite{Ketzer:2004jk} at \INST{CERN}. The
two options for the central tracker are explained briefly in the
following.
\paragraph*{Straw Tube Tracker (\Stt)}
\label{sec:det:ts:stt}
\begin{figure*}[htb]
\begin{center}
\includegraphics[width=0.6\dwidth]{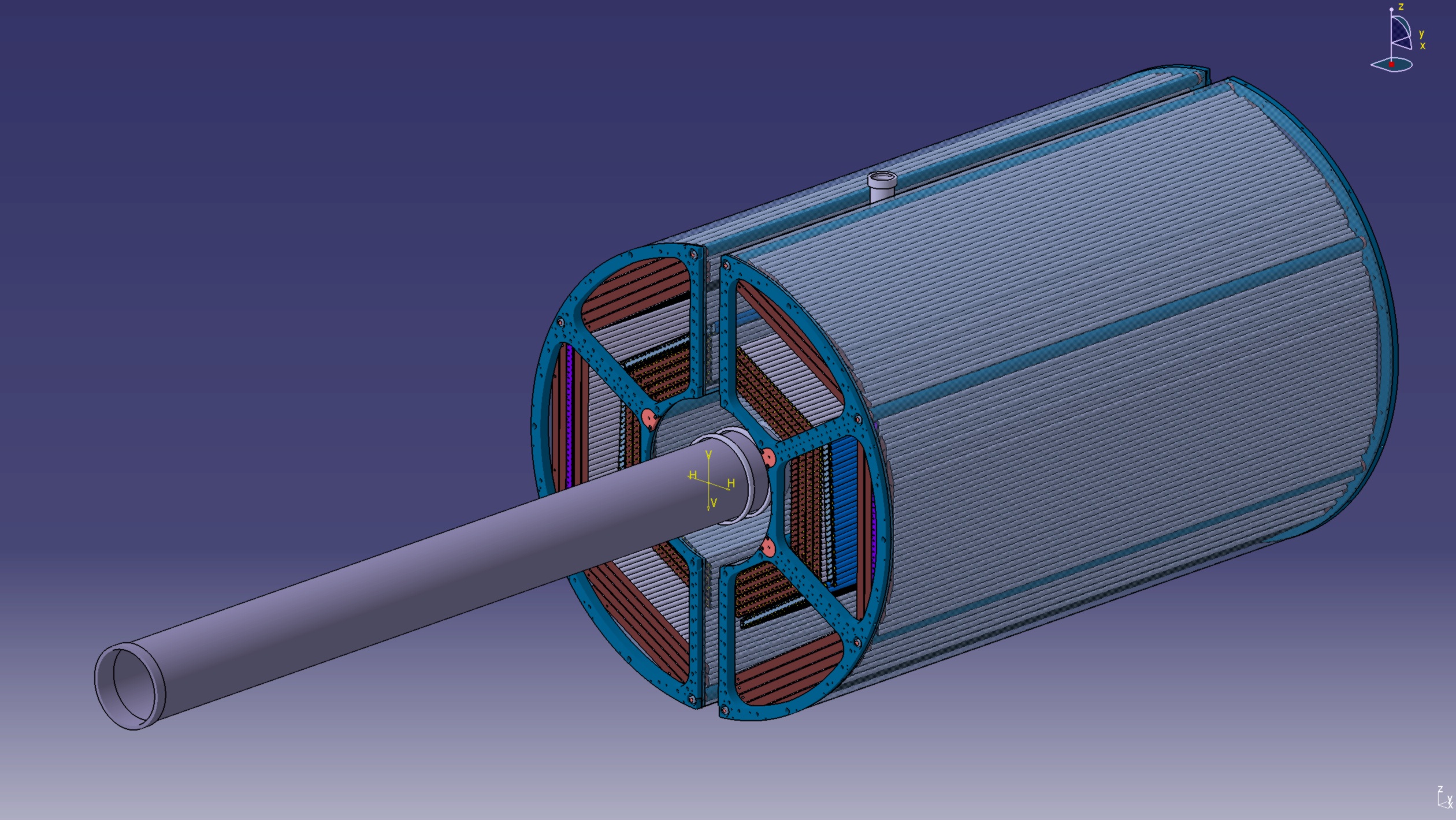}
\caption[Straw Tube Tracker in the Target Spectrometer]
{Straw Tube Tracker in the Target Spectrometer.}
\label{fig:exp:ts:stt}
\includegraphics[width=0.6\dwidth]{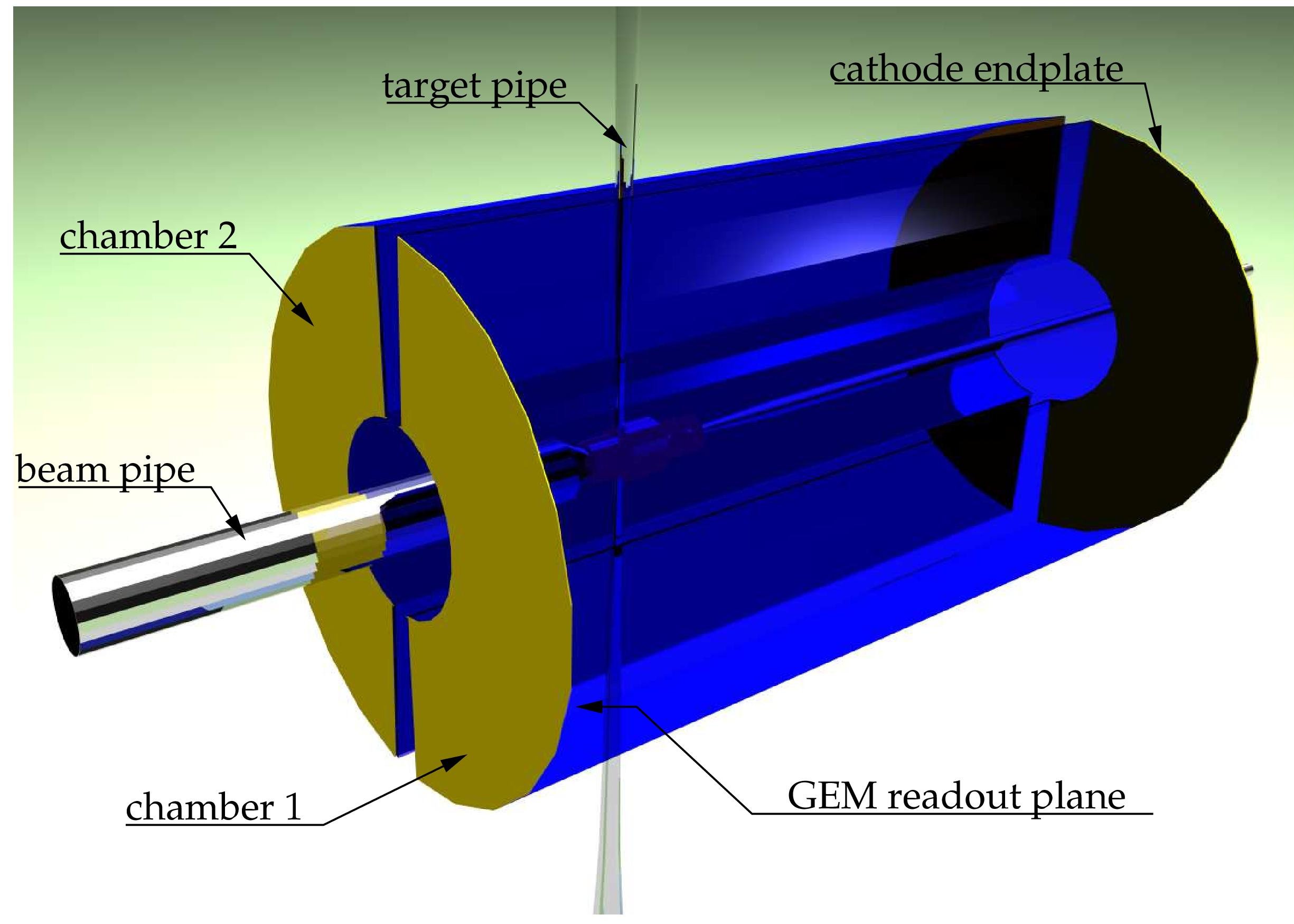}
\caption[GEM Time Projection Chamber in the Target Spectrometer]
{GEM Time Projection Chamber in the Target Spectrometer.}
\label{fig:exp:ts:tpc}
\end{center}
\end{figure*}
This detector consists of aluminized mylar tubes called {\em straws},
which are self supporting by the operation at 1\,bar overpressure. The
straws are arranged in planar layers which are mounted in a hexagonal
shape around the \Mvd as shown in \Reffig{fig:exp:ts:stt}. In total
there are 24 layers of which the 8 central ones are tilted to achieve
an acceptable resolution of 3\,mm also in $z$ (parallel to the beam). The
gap to the surrounding detectors is filled with further individual
straws. In total there are 4200 straws around the beam pipe at radial
distances between 15$\,$cm and 42$\,$cm with an overall length of
150$\,$cm. All straws have a diameter of 10\,mm. A thin and light space
frame will hold the straws in place, the force of the wire however is
kept solely by the straw itself. The mylar foil is 30\,$\umu$m thick,
the wire is made of 20$\,\umu$m thick gold plated tungsten. This
design results in a material budget of 1.3\percent of a radiation length.
\par
The gas mixture used will be Argon based with CO$_2$ as quencher. It
is foreseen to have a gas gain no greater than 10$^5$ in order to
warrant long term operation. With these parameters, a resolution in
$x$ and $y$ coordinates of about 150$\,\umu$m is expected.
\paragraph*{Time Projection Chamber (\Tpc)}
\label{sec:det:ts:tpc}
A challenging but advantageous alternative to the \Stt is a \Tpc, which
would combine superior track resolution with a low material budget and
additional particle identification capabilities through energy loss
measurements.
\par
The \Tpc depicted in a schematic view in \Reffig{fig:exp:ts:tpc}
consists of two large gas-filled half-cylinders enclosing the target
and beam pipe and surrounding the \Mvd. An electric field along the
cylinder axis separates positive gas ions from electrons created by
ionizing particles traversing the gas volume. The electrons drift with
constant velocity towards the anode at the upstream end face and
create an avalanche detected by a pad readout plane yielding
information on two coordinates.  The third coordinate of the track
comes from the measurement of the drift time of each primary electron
cluster. In common TPCs the amplification stage typically occurs in
multi-wire proportional chambers. These are gated by an external
trigger to avoid a continuous backflow of ions in the drift volume
which would distort the electric drift field and jeopardize the
principle of operation.
\par
In \PANDA the interaction rate is too high and there is no fast
external trigger to allow such an operation. Therefore a novel readout
scheme is employed which is based on GEM foils as amplification stage.
These foils have a strong suppression of ion backflow, since the ions
produced in the avalanches within the holes are mostly caught on the
backside of the foil. Nevertheless about two ions per primary
electron are drifting back into the ionization volume even at moderate
gains. The deformation of the drift field can be measured by a laser
calibration system and the resulting drift can be corrected
accordingly. In addition a very good homogeneity of the solenoid field
with a low radial component is required.
\par
A further challenge is the large number of tracks accumulating in the
drift volume because of the high rate and slow drift. While the TPC is
capable of storing a lot of tracks at the same time, their assignment
to specific interactions has to be done by time correlations with
other detectors in the target spectrometer. To achieve this, first a
tracklet reconstruction has to take place. The tracklets are then
matched against other detector signals or are pointed to the
interaction.  This requires either high computing power close to the
readout electronics or a very high bandwidth at the full interaction
rate.
\paragraph*{Forward GEM Detectors}
Particles emitted at angles below 22\degrees{} which are not covered
fully by the Straw Tube Tracker or \Tpc will be tracked by three
stations of GEM detectors placed 1.1\,m, 1.4\,m and 1.9\,m downstream of
the target. The chambers have to sustain a high counting rate of
particles peaked at the most forward angles due to the relativistic
boost of the reaction products as well as due to the small angle
$\pbarp$ elastic scattering. With the envisaged luminosity, the
expected particle flux in the first chamber in the vicinity of the 5\,cm
diameter beam pipe is about 3$\cdot10^{4}\,$cm$^{-2}$s$^{-1}$. In
addition it is required that the chambers work in the 2$\,$T magnetic
field produced by the solenoid.  Drift chambers cannot fulfil the
requirements here since they would suffer from aging and the occupancy
would be too high. Therefore gaseous micropattern detectors based on
GEM foils as amplification stages are chosen. These detectors have
rate capabilities three orders of magnitude higher than drift
chambers.
\par
In the current layout there are three double planes with two
projections per plane. The readout plane is subdivided in an outer ring
with longer and an inner ring with shorter strips. The strips are
arranged in two orthogonal projections per readout plane. Owing to the
charge sharing between strip layers a strong correlation between the
orthogonal strips can be found giving an almost 2D information rather
than just two projections.
\par
The readout is performed by the same front-end chips as are used for
the silicon microstrips. The first chamber has a diameter of 90 cm,
the last one of 150\,cm. The readout boards carrying the ASICs are
placed at the outer rim of the detectors.
\subsubsection{Cherenkov Detectors and Time-of-Flight}
Charged particle identification of hadrons and leptons over a large
range of angles and momenta is an essential requirement for
meeting the physics objectives of \PANDA. There will be several
dedicated systems which, complementary to the other detectors, will
provide means to identify particles. The main part of the momentum
spectrum above 1\,\gevc will be covered by Cherenkov detectors.
Below the Cherenkov threshold of kaons several other processes have
to be employed for particle identification: the tracking detectors
are able to provide energy loss measurements. Here in particular the
TPC with its large number of measurements along each track excels.
In addition a time-of-flight barrel can identify slow particles.
\paragraph*{Barrel DIRC}
Charged particles in a medium with index of refraction $n$,
propagating with velocity $\beta c\,<\,\mbox{1}/n$, emit radiation at
an angle $\Theta_C\,=\,\arccos(\mbox{1}/n\beta$).  Thus, the mass of
the detected particle can be determined by combining the velocity
information determined from $\Theta_C$ with the momentum information
from the tracking detectors.
\par
A very good choice as radiator material for these detectors is fused
silica ({\it i.e.} artificial quartz) with a refractive index of 1.47. This
provides pion-kaon-separation from rather low momenta of 800\,\mevc
up to about 5\,\gevc and fits well to the compact design of the
target spectrometer. In this way the loss of photons converting in the
radiator material can be reduced by placing the conversion point as
close as possible to the electromagnetic calorimeter.
\par
At polar angles between 22\degrees{} and 140\degrees{}, particle
identification will be performed by the detection of internally
reflected Cherenkov (\Dirc) light as realized in the {\INST{BaBar}}
detector~\cite{Staengle:1997xp}.  It will consist of 1.7$\,$cm thick
quartz slabs surrounding the beam line at a radial distance of
45 - 54 $\,$cm. At {\INST{BaBar}} the light was imaged across a large
stand-off volume filled with water onto 11\,000 photomultiplier
tubes. At \PANDA{}, it is intended to focus the images by lenses onto
micro-channel plate photomultiplier tubes (MCP PMTs) which are
insensitive to magnetic fields. This fast light detector type allows a
more compact design and the readout of two spatial coordinates. In
addition MCP PMTs provide good time resolution to measure the time of
light propagation for dispersion correction and background suppression.
\paragraph*{Forward Endcap DIRC}
A similar concept can be employed in the forward direction for
particles between 5\degrees{} and 22\degrees{}. The same radiator,
fused silica, is to be employed however in shape of a disk. At the rim
around the disk focussing will be done by mirroring quartz elements
reflecting onto MCP PMTs. Once again two spatial coordinates plus the
propagation time for corrections will be read. The disk will be 2 cm
thick and will have a radius of 110 cm. It will be placed directly
upstream of the forward end-cap calorimeter.
\paragraph*{Barrel Time-of-Flight}
For slow particles at large polar angles particle identification will
be provided by a time-of-flight detector. In the target spectrometer
the flight path is only of the order of 50 - 100\,cm. Therefore the
detector must have a very good time resolution between 50 and 100\,ps.
\par
Implementing an additional start detector would introduce too much
material close to the interaction point deteriorating considerably the
resolution of the electromagnetic crystal calorimeter. In the absence
of a start detector relative timing of a minimum of two particles has
to be employed.
\par
As detector candidates scintillator bars and strips or pads of
multi-gap resistive plate chambers are considered. In both cases a
compromise between time resolution and material budget has to be
found. The detectors will cover angles between 22\degrees{} and
140\degrees{} using a barrel arrangement around the \Stt/\Tpc at
42 - 45$\,$cm radial distance.
\begin{figure*}[hbt]
\includegraphics[width=\dwidth]{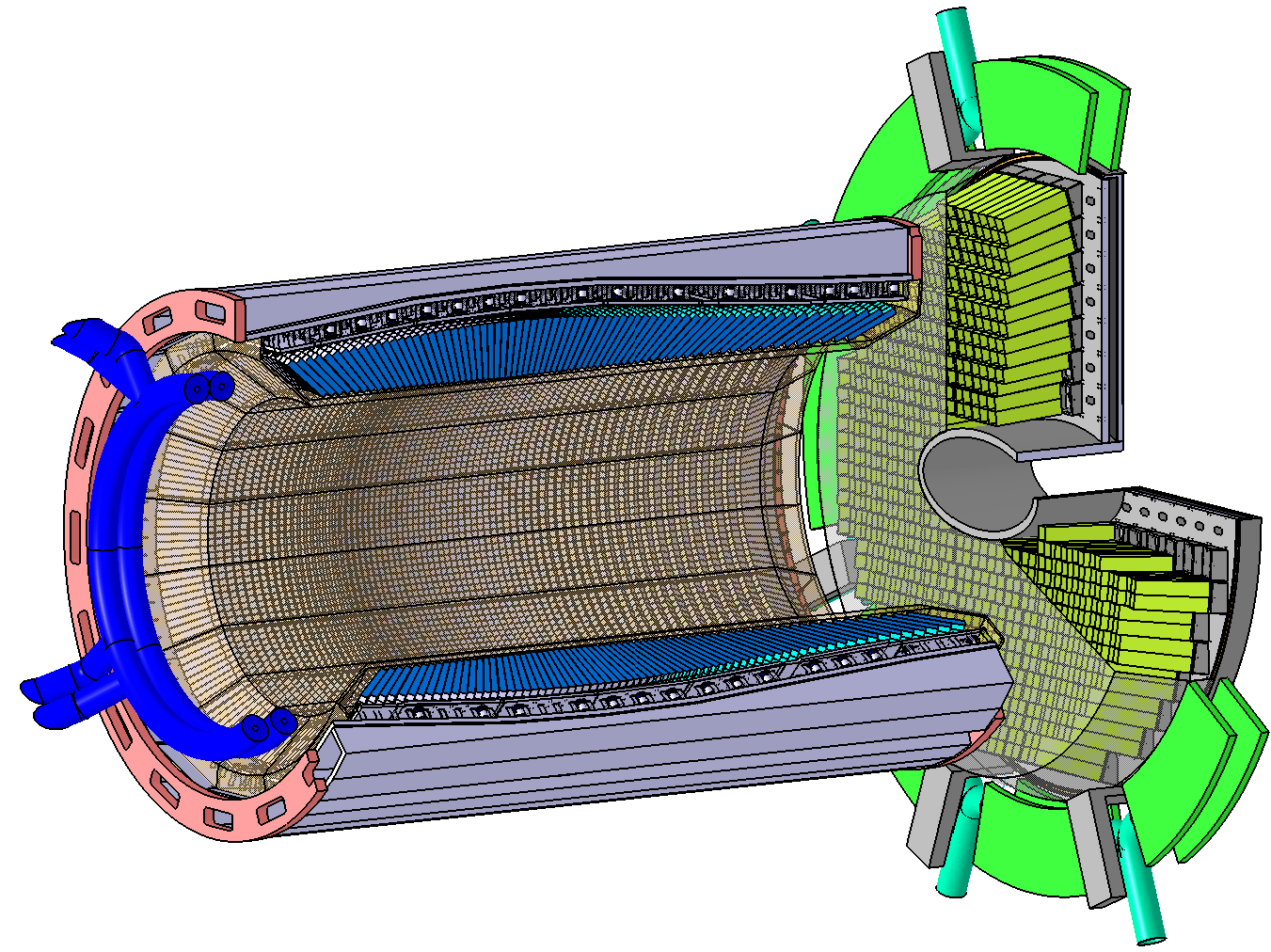}
\caption[The \PANDA barrel and end-cap EMC]
{The \PANDA barrel and forward end-cap EMC}
\label{fig:det:emc}
\end{figure*}
\subsubsection{Electromagnetic Calorimeters}
Expected high count rates and a geometrically compact design of the
target spectrometer require a fast scintillator material with a short
radiation length and Moli\`ere radius for the construction of the
electromagnetic calorimeter (\Emc). Lead tungstate (PbWO$_4$) is a
high density inorganic scintillator with sufficient energy and time
resolution for photon, electron, and hadron detection even at
intermediate energies \cite{Mengel:1998si,Novotny:2000zg,Hoek:2002ss}.
For high energy physics PbWO$_4$ has been chosen by the {\INST{CMS}}
and \INST{ALICE} collaborations at \INST{CERN}
\cite{Alice:Tp,Cms:Tp:1994} and optimized for large scale production.
Apart from a short decay time of less than 10$\,$ns good radiation
hardness has been achieved \cite{Auffray:1999}.  Recent developments
indicate a significant increase of light yield due to crystal
perfection and appropriate doping to enable photon detection down to a
few MeV with sufficient resolution. The light yield can be increased
by a factor of about 4 compared to room temperature by cooling the
crystals down to -25$\degC$.
\par
The crystals will be 20\,cm long, {\it i.e.} approximately 22$\,X_0$, in
order to achieve an energy resolution below 2\percent{} at 1$\,\gev$
\cite{Mengel:1998si,Novotny:2000zg,Hoek:2002ss} at a tolerable energy
loss due to longitudinal leakage of the shower.  Tapered crystals with
a front size of $2.1\times 2.1\,\cm^2$ will be mounted with an inner
radius of 57$\,$cm.  This implies 11\,360 crystals for the barrel part
of the calorimeter.  The forward end-cap calorimeter will have 3600
tapered crystals, the backward end-cap calorimeter 592.  The readout of
the crystals will be accomplished by large area avalanche photo diodes
in the barrel and vacuum photo-triodes in the forward and backward
endcaps.
\par
The barrel part and the forward endcap of the target spectrometer EMC
are depicted in \Reffig{fig:det:emc}.
\subsubsection{Muon Detectors}
Muons are an important probe for, among others, $\jpsi$ decays,
semi-leptonic $D$-meson decays and the Drell-Yan process. The
strongest background are pions and their decay daughter muons.
However at the low momenta of \PANDA the signature is less clean than
in high energy physics experiments. To allow nevertheless a proper
separation of primary muons from pions and decay muons a range
tracking system will be implemented in the yoke of the solenoid
magnet. Here a fine segmentation of the yoke as absorber with
interleaved tracking detectors allows the distinction of energy loss
processes of muons and pions and kinks from pion decays. Only in this
way a high separation of primary muons from the background can be
achieved.
\par
In the barrel region the yoke is segmented in a first layer of 6\,cm
iron followed by 12 layers of 3\,cm thickness. The gaps for the
detectors are 3\,cm wide. This is enough material for the absorption of
pions in the momentum range in \PANDA at these angles. In the forward
end-cap more material is needed. Since the downstream door of the
return yoke has to fulfil constraints for space and accessibility,
the muon system is split in several layers.  Six detection layers are
placed around five iron layers of 6\,cm each within the door, and a
removable muon filter with additional five layers of 6\,cm iron is
located in the space between the solenoid and the dipole. This filter
has to provide cut-outs for forward detectors and pump lines and has
to be built in a way that it can be removed with few crane operations
to allow easy access to these parts.
\par
As detector within the absorber layers rectangular aluminium drift
tubes are used as they were constructed for the \INST{COMPASS} muon detection
system \cite{Abbon:2007pq}. They are essentially drift tubes with
additional capacitively coupled strips read out on both ends to obtain
the longitudinal coordinate.
\subsubsection{Hypernuclear Detector}
\begin{figure}[!h]
\begin{center}
  \includegraphics[width=\swidth]{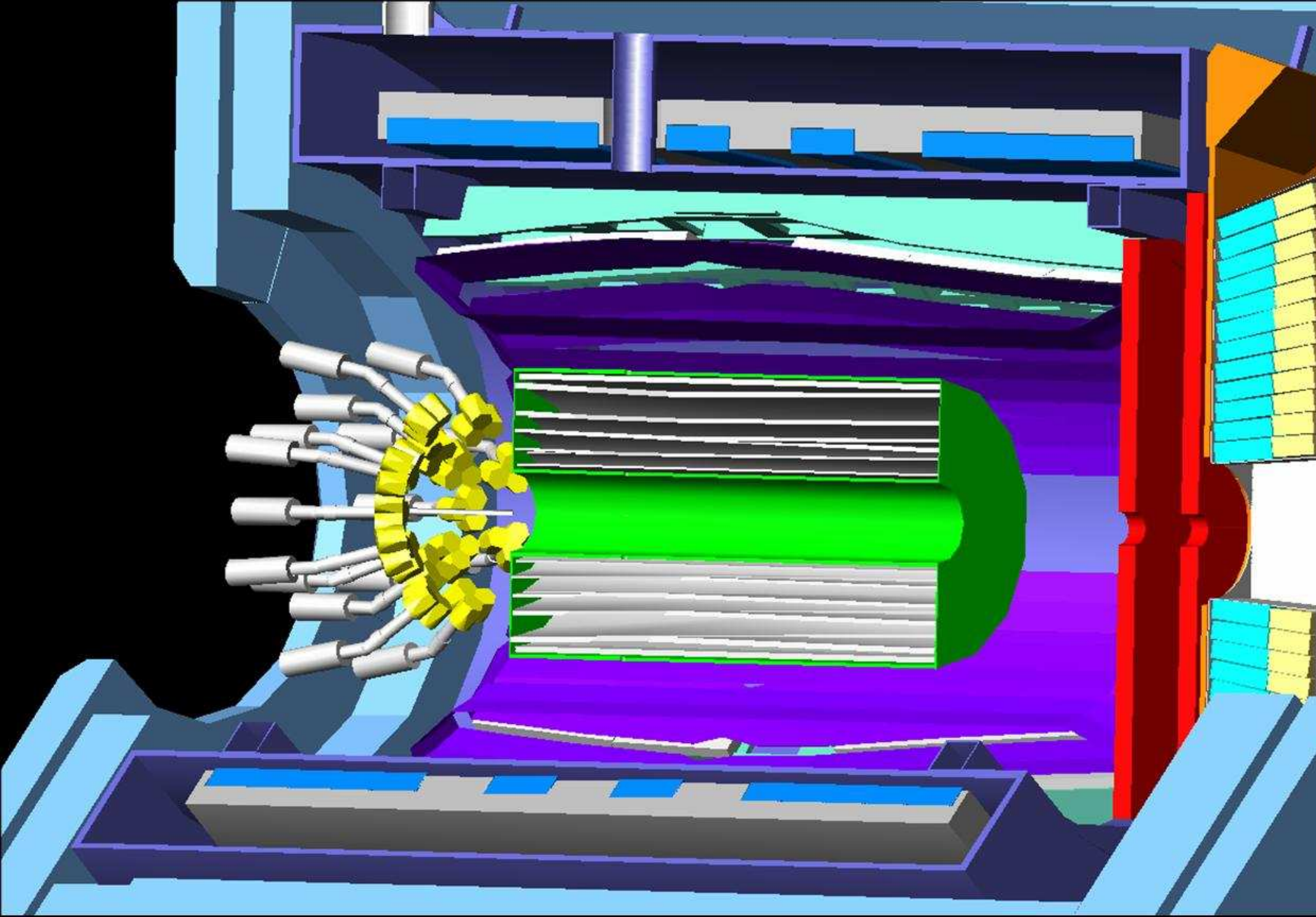}
  \caption[Integration of the secondary target and the germanium
  Cluster--array in the \PANDA detector.]{The beam enters from left.}
 \label{fig:phys:hyp:fig_PandaGeSetup}
\end{center}
\end{figure}

The hypernuclei study will make use of the modular structure of
\PANDA{}. Removing the backward end-cap calorimeter will allow to
add a dedicated nuclear target station and the required additional
detectors for $\gamma$ spectroscopy (see \Reffig{fig:phys:hyp:fig_PandaGeSetup})
close to the entrance of \PANDA{}. While the detection of anti-hyperons
and low momentum $K^+$
can be ensured by the universal detector and its PID system, a
specific target system and a $\gamma$-detector are additional
components required for the hypernuclear studies.
\paragraph*{Active Secondary Target}
%
The production of hypernuclei proceeds as a two-stage process. First
hyperons, in particular $\Xi\bar{\Xi}$, are produced on a primary
nuclear target. The slowing down of the $\Xi$ proceeds through a
sequence of nuclear elastic scattering processes inside the residual
nucleus in which the antiproton annihilation has occurred and by
energy loss during the passage through an active absorber. If
decelerated to rest before decaying, the particle can be captured in
a nucleus, eventually releasing two $\Lambda$ hyperons and forming a
double hypernucleus.
\par
\begin{figure}[!h]
\begin{center}
  \includegraphics[width=\swidth]{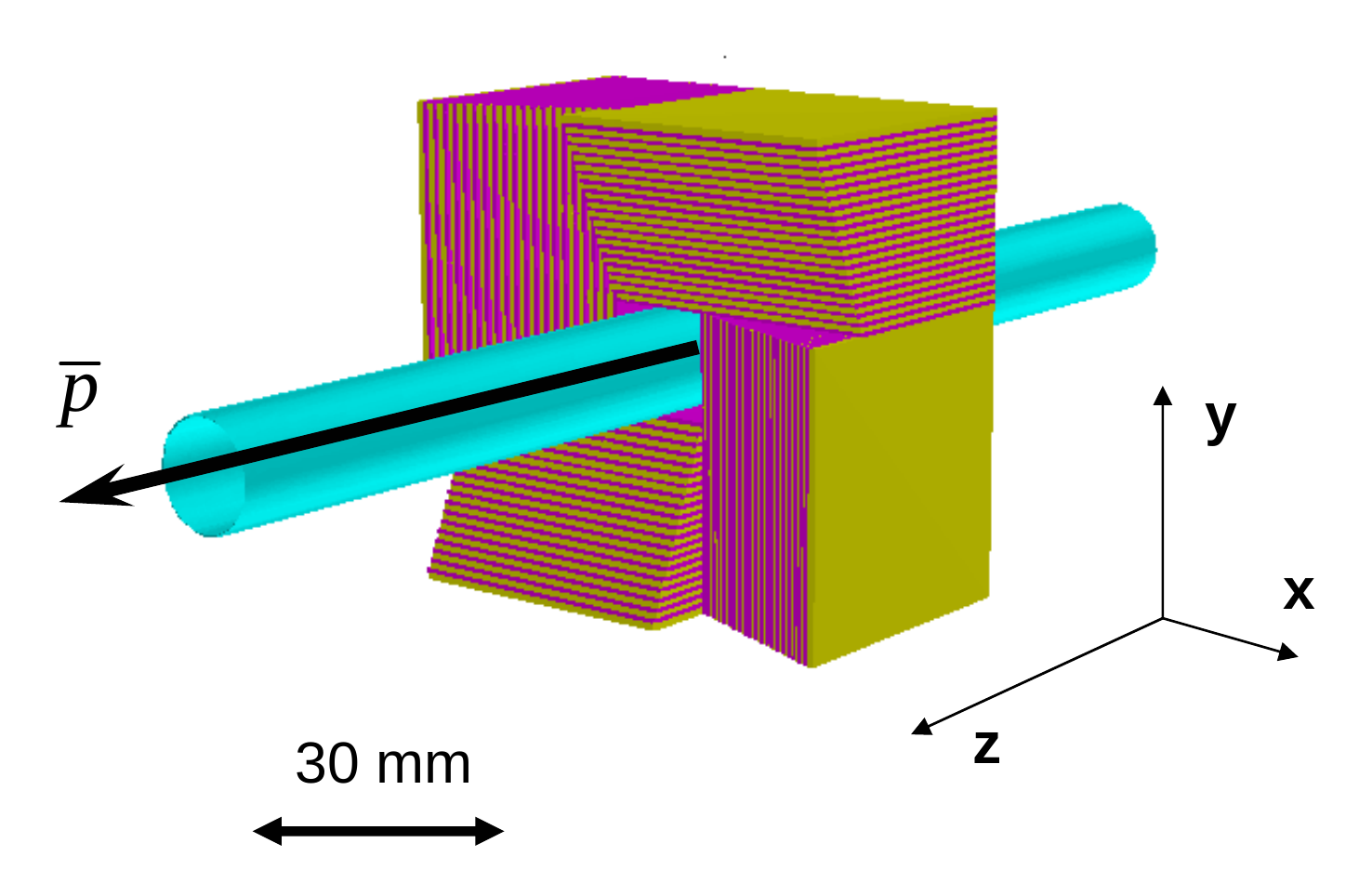}
  \caption[Layout out the secondary sandwich target.]{ The active
silicon layers provide also tracking information on
the emitted weak decay products of the produced hypernuclei.}
 \label{fig:phys:hyp:fig_sectarg1}
\end{center}
\end{figure}

The geometry of this secondary target is essentially determined by
the short mean life of the $\Xi^-$ of only 0.164$\,$ns and its
stopping time in solid material. This limits the required thickness
of the active secondary target to about 25--30$\,$mm. The present
layout shows a compact structure of 26mm thickness, consisting
of 20 layers of silicon strip detectors with alternating layers of
absorber material (\Reffig{fig:phys:hyp:fig_sectarg1}). The active
silicon layers provide also tracking information on the emitted weak
decay products of the produced hypernuclei (see
\Reffig{fig:phys:hyp:fig_sectarg2} in \Refsec{sec:hyp:exp}).

\begin{figure}[!h]
\begin{center}
  \includegraphics[width=\swidth]{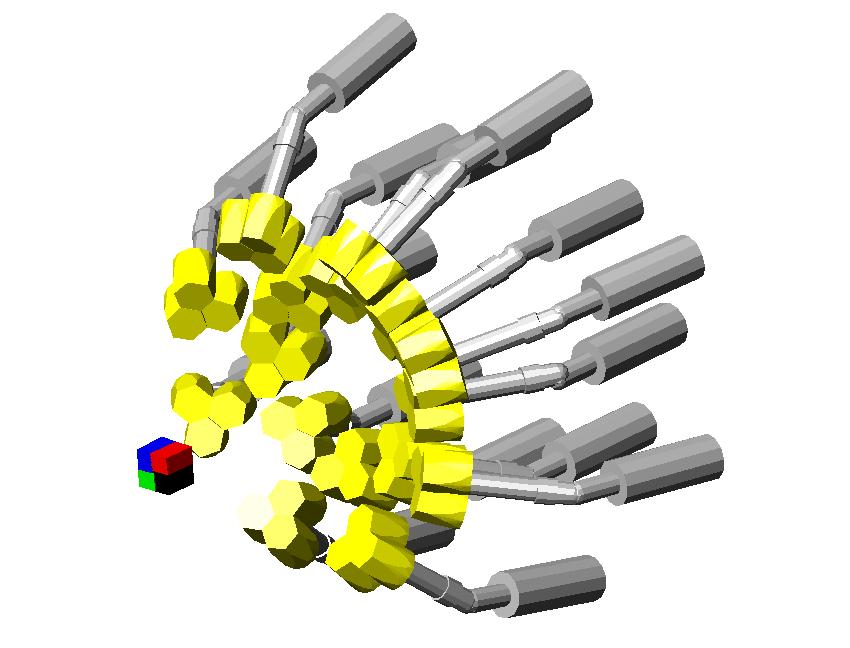}
  \caption{ Design of
the $\gamma$--ray spectroscopy setup with 15 germanium cluster
detector, each comprising 3 germanium crystals.}
 \label{fig:phys:hyp:fig_triGeclust}
\end{center}
\end{figure}
\paragraph*{Germanium Array}
An existing germanium-array with refurbished readout is planned to
be used for the $\gamma$-spectroscopy of the nuclear decay cascades
of hypernuclei. The main limitation will be the load due to neutral
or charged particles traversing the germanium detectors. Therefore,
readout schemes and tracking algorithms are presently being
developed which will enable high resolution $\gamma$-spectroscopy in
an environment of high particle flux. The germanium-array crystals
will be grouped asymmetrically by forming triple clusters. Each
cluster consists of three encapsulated n--type Germanium crystals of
the \INST{Euroball} type. The total $\gamma$--array set-up consists of 15
triple Germanium clusters positioned at backward axial angle around
the target region as shown in \Reffig{fig:phys:hyp:fig_triGeclust}.

\subsection{Forward Spectrometer}
\subsubsection{Dipole Magnet}
A dipole magnet with a window frame, a 1$\,$m gap, and more than
2$\,$m aperture will be used for the momentum analysis of charged
particles in the forward spectrometer.  In the current planning, the
magnet yoke will occupy about 2.5$\,$m in beam direction starting from
3.5$\,$m downstream of the target.  Thus, it covers the entire angular
acceptance of the forward spectrometer of $\pm$10\degrees{} and
$\pm$5\degrees{} in the horizontal and in the vertical direction,
respectively.  The maximum bending power of the magnet will be 2$\,$Tm
and the resulting deflection of the antiproton beam at the maximum
momentum of 15$\,\gevc$ will be 2.2\degrees{}.  The design acceptance
for charged particles covers a dynamic range of a factor 15 with the
detectors downstream of the magnet.  For particles with lower momenta,
detectors will be placed inside the yoke opening. The beam deflection
will be compensated by two correcting dipole magnets, placed around
the \PANDA{} detection system.
\subsubsection{Forward Trackers}
\label{sec:det:fs:trk}
The deflection of particle trajectories in the field of the dipole
magnet will be measured with a set of wire chambers (either small cell
size drift chambers or straw tubes), two placed in front, two within
and two behind the dipole magnet.  This will allow to track particles
with highest momenta as well as very low momentum particles where
tracks will curl up inside the magnetic field.
\par
The chambers will contain drift cells of 1$\,$cm width. Each chamber
will contain three pairs of detection planes, one pair with vertical
wires and two pairs with wires inclined by +10\degrees{} and
-10\degrees{}.  This configuration will allow to reconstruct tracks in
each chamber separately, also in case of multi-track events. The beam
pipe will pass through central holes in the chambers.  The most
central wires will be separately mounted on insulating rings
surrounding the beam pipe. The expected momentum resolution of the
system for 3$\,\gevc$ protons is $\delta p/p\,=\,0.2$\percent{} and is
limited by the small angle scattering on the chamber wires and gas.
\subsubsection{Forward Particle Identification}
\paragraph*{RICH Detector}
To enable the $\pi$/$K$ and $K$/p separation also at the very highest
momenta a \Rich detector is proposed. The favoured design is a dual
radiator \Rich detector similar to the one used at
\INST{Hermes}~\cite{Akopov:2000qi}. Using two radiators, silica
aerogel and C$_4$F$_{10}$ gas, provides $\pi$/$K$/p separation in a
broad momentum range from 2--15$\,\gevc$.  The two different indices
of refraction are 1.0304 and 1.00137, respectively.  The total
thickness of the detector is reduced to the freon gas radiator
\mbox{(5\percent$\,X_0$),} the aerogel radiator (2.8\percent$\,X_0$), and the
aluminium window (3\percent$\,X_0$) by using a lightweight mirror focusing
the Cherenkov light on an array of photo tubes placed outside the
active volume. It has been studied to reuse components of the
\INST{HERMES} \Rich{}.
\paragraph*{Time-of-Flight Wall}
A wall of slabs made of plastic scintillator and read out on both ends
by fast photo tubes will serve as time-of-flight stop counter placed at
about 7$\,$m from the target. In addition, similar detectors will be
placed inside the dipole magnet opening, to detect low momentum
particles which do not exit the dipole magnet. The relative time of
flight between two charged tracks reaching any of the time-of-flight
detectors in the experiment (including the barrel TOF)
will be measured. The wall in front of the
forward spectrometer \Emc will consist of vertical strips varying in
width from 5 to 10$\,$cm to account for the differences in count
rate. With the expected time resolution of $\sigma\,=\,50\,$ps
$\pi$-$K$ and $K$/p separation on a 3$\,\sigma$ level will be possible
up to momenta of 2.8$\,\gevc$ and 4.7$\,\gevc$, respectively.
\subsubsection{Forward Electromagnetic Calorimeter}
For the detection of photons and electrons a {Shashlyk}-type
calorimeter with high resolution and efficiency will be employed. The
detection is based on lead-scintillator sandwiches read out with
wave-length shifting fibres passing through the block and coupled to
photomultipliers. The technique has already been successfully used in
the \INST{E865} experiment~\cite{bib:emc:E865}.  It has been adopted
for various other experiments~\cite{bib:emc:PHENIX, bib:emc:HERAB,
  bib:emc:LHCb, bib:emc:KOP99, bib:emc:KOP1, bib:emc:KOP04}.  An
energy resolution of $4\percent/\sqrt{E}$~\cite{bib:emc:KOP99} has been
achieved.  To cover the forward acceptance, 26 rows and 54 columns are
required with a cell size of 55 mm, {\it i.e.} 1404 modules in total,
which will be placed at a distance of 7--8$\,$m from the target.
\subsubsection{Forward Muon Detectors}
For the very forward part of the muon spectrum a further range
tracking system consisting of interleaved absorber layers and
rectangular aluminium drift-tubes is being designed, similar to the
muon system of the target spectrometer, but laid out for higher
momenta. The system allows discrimination of pions from muons,
detection of pion decays and, with moderate resolution, also the
energy determination of neutrons and antineutrons.
\subsection{Luminosity Monitor}
In order to determine the cross section for physical processes, it is
essential to determine the time integrated luminosity $L$ for
reactions at the \PANDA interaction point that was available while
collecting a given data sample. Typically the precision for a relative
measurement is higher than for an absolute measurement. For many
observables connected to narrow resonance scans a relative measurement
might be sufficient for \PANDA, but for other observables an absolute
determination of $L$ is required.  The absolute cross section can be
determined from the measured count rate of a specific process with
known cross section. In the following we concentrate on elastic
antiproton-proton scattering as the reference channel.  For most other
hadronic processes that will be measured concurrently in \PANDA the
precision with which the cross section is known is poor.
\par
The optical theorem connects the forward elastic scattering amplitude
to the total cross section.  The total reaction rate and the
differential elastic reaction rate as a function of the
4-momentum transfer {\it t} can be used to determine the total cross
section.
\par
The differential cross section $d\sigma_{el}/dt$ becomes dominated by
Coulomb scattering at very low values of $t$. Since the
electromagnetic amplitude can be precisely calculated, Coulomb elastic
scattering allows both the luminosity and total cross section to be
determined without measuring the inelastic rate
\cite{Armstrong:1996np}.
\par
Due to the 2\,T solenoid field and the existence of the MVD it appears
most feasible to measure the forward going antiproton in \PANDA. The
Coulomb-nuclear interference region corresponds to 4-momentum
transfers of $-t\approx 0.001$\,GeV$^2$ at the beam momentum range of
interest to \PANDA. At a beam momentum of 6\,\gevc this momentum
transfer corresponds to a scattering angle of the antiproton of about
5\,mrad.
\par
The basic concept of the luminosity monitor is to reconstruct the
angle (and thus $t$) of the scattered antiprotons in the polar angle
range of 3-8\,mrad with respect to the beam axis.  Due to the large
transverse dimensions of the interaction region when using the pellet
target, there is only a weak correlation of the position of the
antiproton at e.g. $z=$+10.0\,m to the recoil angle. Therefore, it is
necessary to reconstruct the angle of the antiproton at the luminosity
monitor. As a result the luminosity monitor will consist of a sequence
of four planes of double-sided silicon strip detectors located as far
downstream and as close to the beam axis as possible. The planes are
separated by 20\,cm along the beam direction. Each plane consists of 4
wafers (e.g. 2\,cm $\times$ 5\,cm $\times\ 200\,\umu$m, with 50\,$\umu$m
pitch) arranged radially to the beam axis. Four planes are required
for sufficient redundancy and background suppression. The use of 4
wafers (up, down, right, left) in each plane allows systematic errors
to be strongly suppressed.
\par
\begin{figure}
  \begin{center}
    \includegraphics[width=0.85\linewidth]{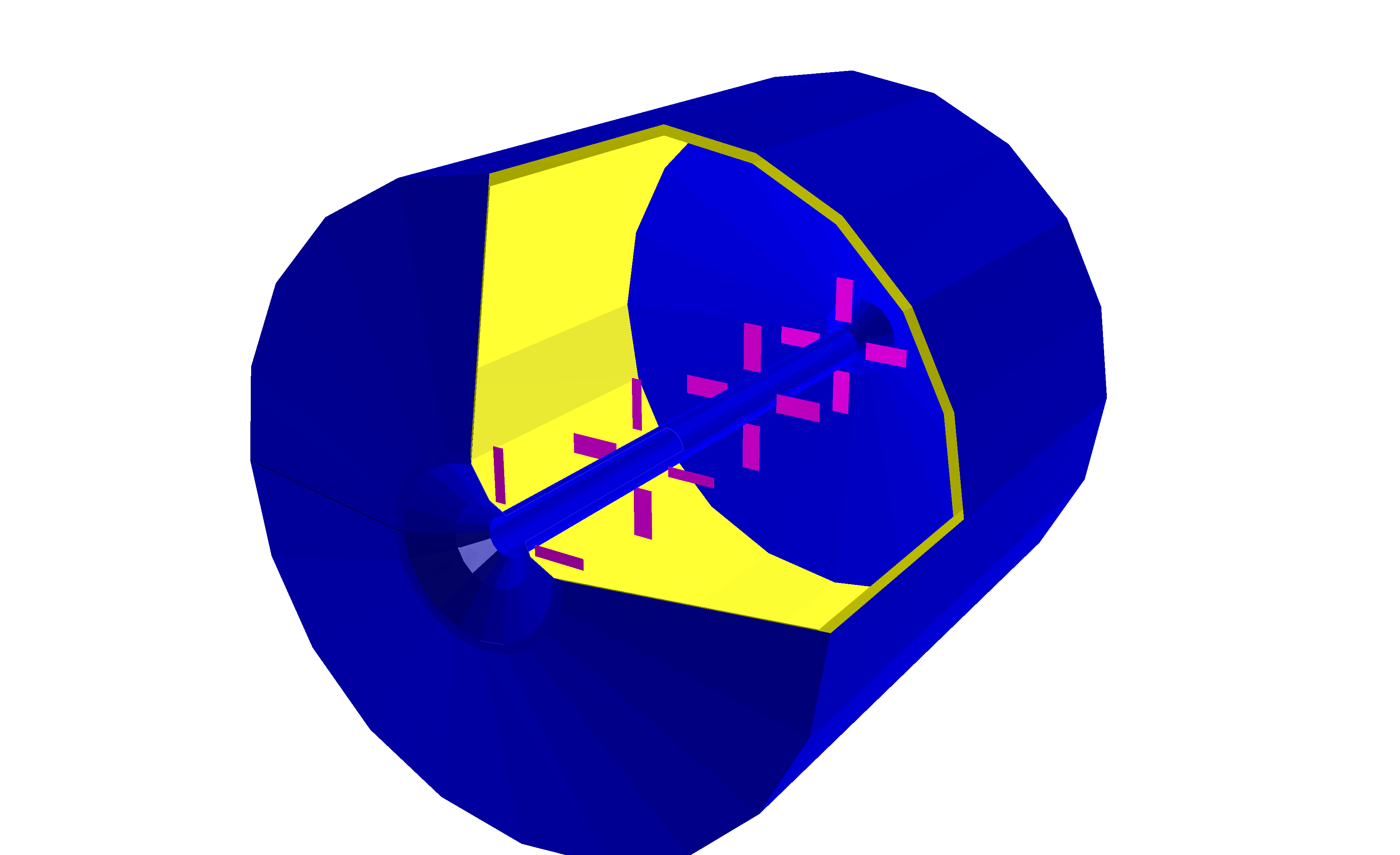}
    \caption{Schematic overview of the luminosity monitor concept.}
    \label{fig:LumMon}
  \end{center}
\end{figure}

The silicon wafers are located inside a vacuum chamber to minimize
scattering of the antiprotons before traversing the 4 tracking
planes. The acceptance for the antiproton beam in the HESR is $\pm$3\,mrad, 
corresponding to the 89 mm inner diameter of the beam pipe at
the quadrupoles located at about 15 m downstream of the interaction
point.  The luminosity monitor can be located in the space between the
downstream side of the forward spectrometer hadronic calorimeter and
the HESR dipole needed to redirect the antiproton beam out of the
\PANDA chicane back into the direction of the HESR straight stretch
({\it i.e.} between $z=$+10.0\,m and $z=$+12.0\,m downstream of the target). At
this distance from the target the luminosity monitor needs to measure
particles at a radial distance of between 3 and 8\,cm from the beam
axis. A sketch of the detector concept is given in \Reffig{fig:LumMon}.
As pilot simulations show, at a beam momentum of 6.2\,\gevc the
proposed detector measures antiprotons elastically scattered in the
range $0.0006\,$GeV$^2< -t < 0.0035$\,GeV$^2$, which spans the Coulomb-nuclear
interference region.  Based upon the granularity of the readout the
resolution of $t$ could reach $\sigma_t \approx 0.0001$\,GeV$^2$. In
reality this value is expected to degrade to $\sigma_t \approx 0.0005$\, GeV$^2$
when taking small-angle scattering into account. At the
nominal \PANDA interaction rate of $2\EE{7}$/s there will be an
average of 10\,kHz/cm$^2$ in the sensors. In comparison with
other experiments an absolute precision of about 3\percent is considered
feasible for this detector concept at \PANDA, which will be verified
by more detailed simulations.
\subsection{Data Acquisition}
In many contemporary experiments the trigger and data acquisition (DAQ)
system is based on a two layer hierarchical approach. A subset of
specially instrumented detectors is used to evaluate a first level
trigger condition. For the accepted events, the full information of
all detectors is then transported to the next higher trigger level or
to storage. The available time for the first level decision is usually
limited by the buffering capabilities of the front-end electronics.
Furthermore, the hard-wired detector connectivity severely constrains
both the complexity and the flexibility of the possible trigger
schemes.
\par
In \PANDA, a data acquisition concept is being developed which is
better matched to the high data rates, to the complexity of the
experiment and the diversity of physics objectives and the rate
capability of at least $2\EE{7}$\,events/s.
\par
In our approach, every sub-detector system is a self-triggering
entity.  Signals are detected autonomously by the sub-systems and are
preprocessed.  Only the physically relevant information is extracted
and transmitted.  This requires hit-detection, noise-suppression and
clustering at the readout level.  The data related to a particle hit,
with a substantially reduced rate in the preprocessing step, is marked
by a precise time stamp and buffered for further processing.  The
trigger selection finally occurs in computing nodes which access the
buffers via a high-bandwidth network fabric. The new concept provides
a high degree of flexibility in the choice of trigger algorithms. It
makes trigger conditions available which are outside the capabilities
of the standard approach. One obvious example is displaced vertex
triggering.
\par
In this scheme, sub-detectors can contribute to the trigger decision
on the same footing without restrictions due to hard-wired
connectivity. Different physics can be accessed either in parallel or
via software reconfiguration of the system.
\par
High speed serial (10$\,$Gb/s per link and beyond) and high-density FPGA
(field programmable gate arrays) with large numbers of programmable
gates as well as more advanced embedded features
are key technologies to be exploited within the DAQ framework.
\par
The basic building blocks of the hardware infrastructure which
  can be combined in a flexible way to cope with varying demands, are
  the following:
  \begin{itemize}
  \item Intelligent front-end modules capable of autonomous hit
    detection and data preprocessing (e.g.\ clustering, hit time
    reconstruction, and pattern recognition) are needed.
  \item A  precise time distribution system is mandatory to
    provide a clock norm from which all time stamps can be derived.
    Without this, data from subsystems cannot be correlated.
  \item Data concentrators provide point-to-point communication,
    typically via optical links, buffering and online data
    manipulation.
  \item Compute nodes aggregate large amounts of computing power in a
    specialized architecture rather than through commodity PC hardware.
    They may employ fast FPGAs (Fast Programmable Gate Arrays),
    DSPs (Digital Signal Processors), or other computing units.
    The nodes have to deal with feature extraction, association of data
    fragments to events, and, finally, event selection.
  \end{itemize}
\par
A major component providing the link for all building blocks is the
network fabric. Here, special emphasis is put on embedded switches
which can be cascaded and reconfigured to reroute traffic for
different physics selection topologies. Alternatively, with an even
higher aggregate bandwidth of the network, which according to
projections of network speed evolution will be available by the time
the experiment will start, a flat network topology where all data is
transferred directly to processing nodes may be feasible as well.
This requires a higher total bandwidth but would have a simpler
architecture and allow event selection in a single environment. The
bandwidth required in this case would be at least 200 GB/s. After
event selection in the order of 100-200\,MB/s will be saved to mass
storage.
\par
An important requirement for this scheme is that all detectors perform
a continuous online calibration with data. The normal data taking is
interleaved with special calibration runs. For the monitoring of the
quality of data, calibration constants and event selection a small
fraction of unfiltered raw data is transmitted to mass storage.
\par
To facilitate the association of data fragments to events the beam
structure of the accelerator is exploited: Every 1.8\,$\umu$s there is
a gap of about 400\,ns needed for the compensation of energy loss with
a bucket barrier cavity. This gap provides a clean division between
consistent data blocks which can be processed coherently by one
processing unit.
\subsection{Infrastructure}
\begin{figure*}[htb]
\centerline{\includegraphics[width=\dwidth]{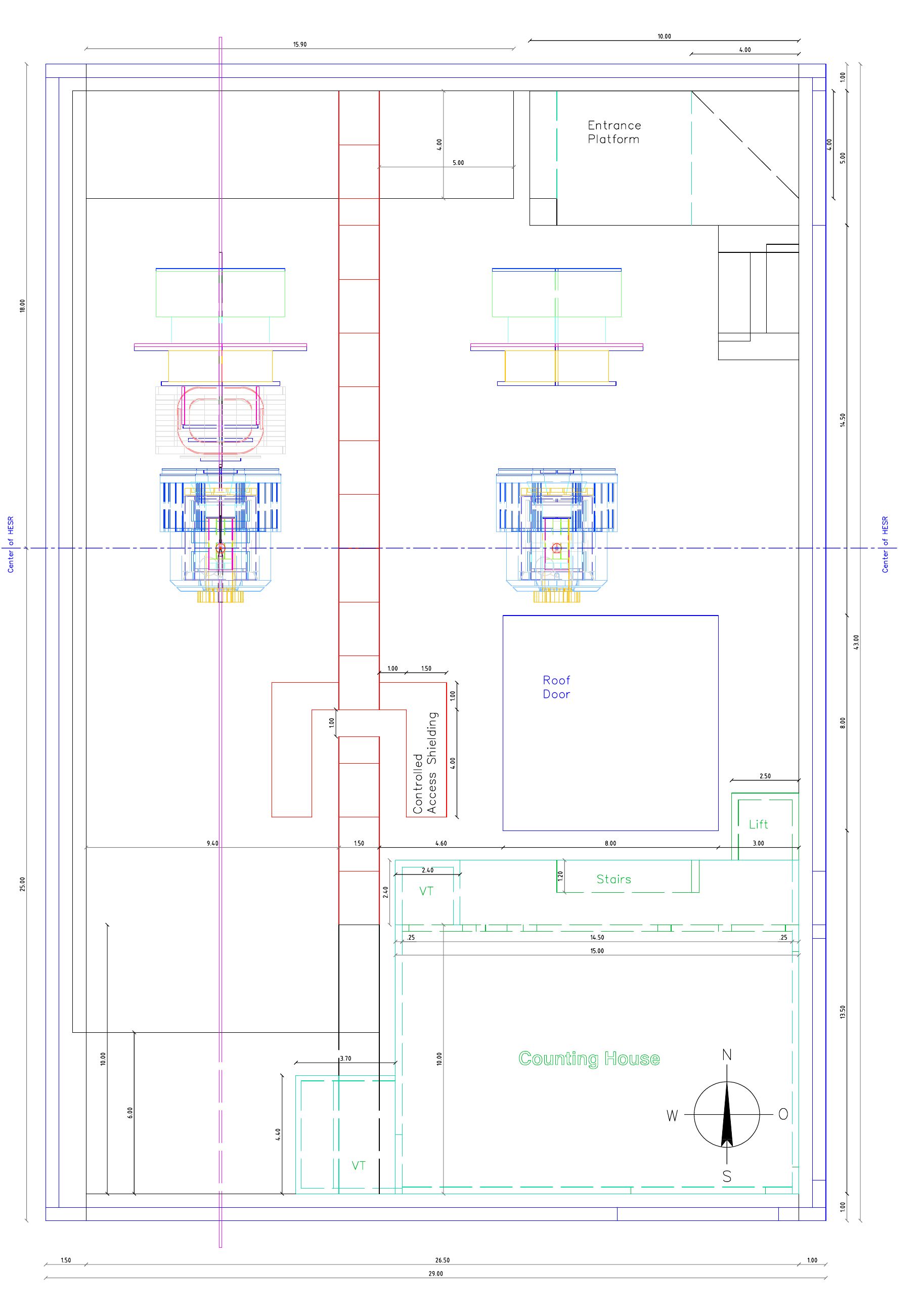}}
\caption[Top view of the experimental area]{Top view of the
  experimental area indicating the location of \PANDA in the \Hesr
  beam line. The target centre is at the centre of \Hesr and is
  indicated as vertical dash-dotted line. North is to the right and
  the beam comes in from the left. The roll-out position of the
  detector will be on the east side of the Hall.}
\label{fig:exp:inf:exphall}
\end{figure*}

The target for antiproton physics is located in the straight section
at the east side of the \HESR.  At this location an experimental hall
of 43\,m $\times$ 29\,m floor space and 14.5\,m height is planned (see
\Reffig{fig:exp:inf:exphall}). A concrete radiation shield of 2$\,$m
thickness on both sides along the beam line is covered by concrete
bars of 1$\,$m thickness to suppress the neutron sky shine. Within the
elongated concrete cave the \PANDA{} detector together with auxiliary
equipment, beam steering, and focusing elements will be housed. The
roof of the cave can be opened and heavy components hoisted by crane.
The shielded beam line area for the \PANDA experiment including
dipoles and focusing elements is foreseen to have 37\,m $\times$
9.4\,m floor space and a height of 8.5\,m with the beam line at a height
of 3.5\,m.  The general floor level of the \HESR is 2\,m higher.
This level will be kept for a length of 4\,m in the north of the
hall (right part in \Reffig{fig:exp:inf:exphall}), to facilitate
transport of heavy equipment into the \HESR tunnel.
\par
The target spectrometer with electronics and supplies will be mounted
on rails which makes it retractable to a parking position outside
the \HESR{} beam line {\it i.e.} into the lower part of the hall in
\Reffig{fig:exp:inf:exphall}). The experimental hall provides
additional space for delivery of components and assembly of the
detector parts. In the south corner of the hall, a counting house
complex with five floors is foreseen. The lowest floor will contain
various supplies for power, high voltage, cooling water, gases
{\it etc.}\,. The next level is planned for readout electronics including data
concentrators. The third level will house the online computing farm.
The fourth floor is at level with the surrounding ground and will house
the control room, a meeting room and social rooms for the shift crew.
Above this floor, hall electricity supplies and ventilation is placed.
A crane (15$\,$t) spans the whole area with a hook at a height of
about 10$\,$m. Sufficient (300$\,$kW) electric power will be
available.
\par
Liquid helium coolant may come from the main cryogenic liquefier for
the \INST{SIS} rings.  Alternatively, a separate small liquefier
(50$\,$W cooling power at 4$\,$K) would be mounted. The temperature of
the building will be moderately controlled. The more stringent
requirements with respect to temperature and humidity for the
detectors have to be maintained locally. To facilitate cooling and
avoid condensation the target spectrometer will be kept in a tent
with dry air at a controlled temperature.
\clearpage
%

%% file: exp/exp_hesr.tex
%
\input{./exp/exp_commands.tex}
\section{The HESR}
\COM{Author(s): A. Lehrach}

\subsection{Introduction}
\label{sec:exp:hesr}

\begin{figure*}[bhtp]
  \centering
    \includegraphics[width=0.9\dwidth]{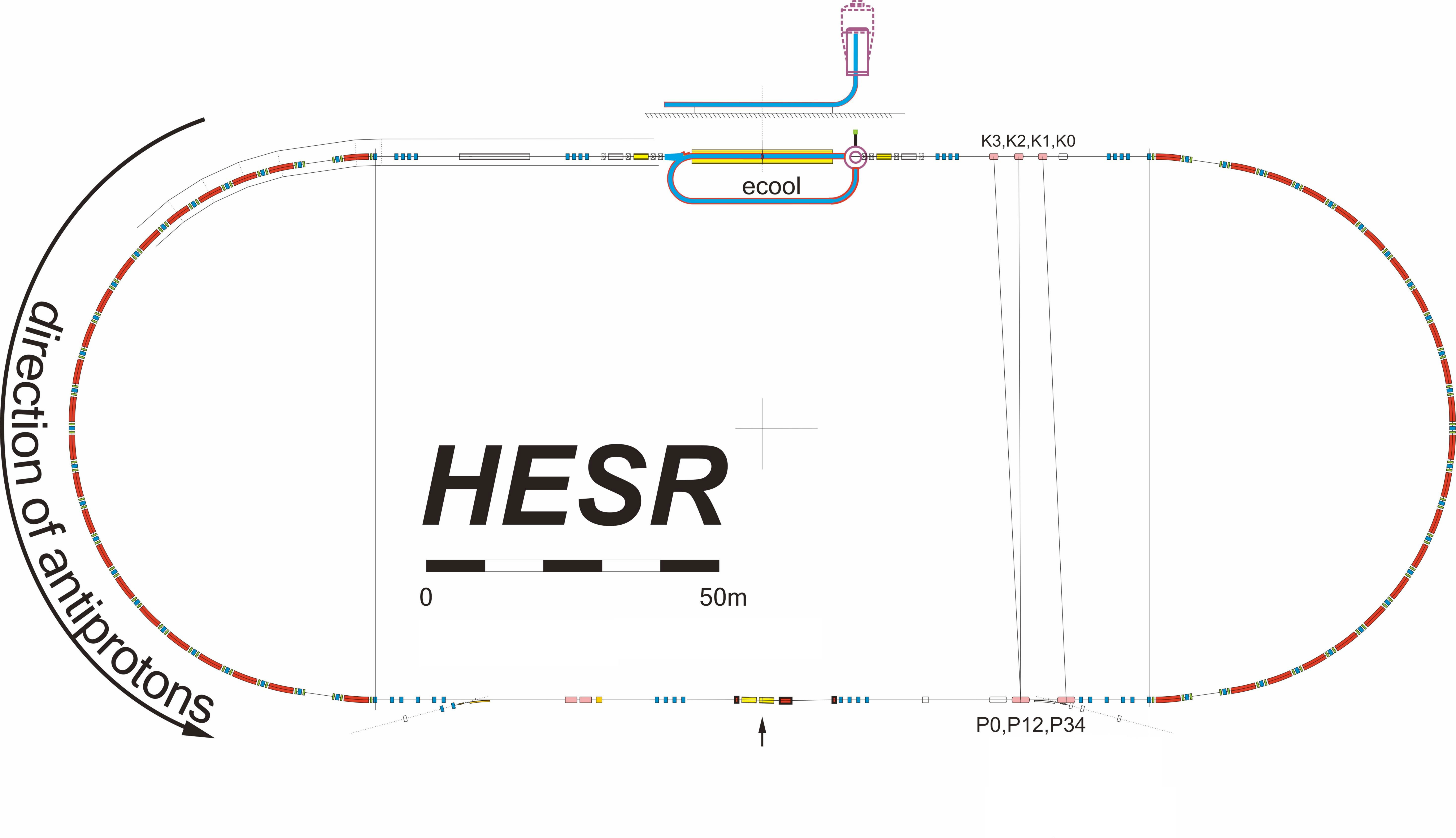}
  \caption[Schematic view of the HESR.]{ Tentative positions for
    injection, cooling devices and experimental installations are
    indicated.}
  \label{fig:HESR}
\end{figure*}

\begin{table*}
\begin{center}
{\small
\begin{tabular}{ll}
\multicolumn{2}{c}{\textbf{Injection Parameters}}\\ \hline\hline
Transverse emittance &
\q{0.25}{mm\cdot mrad} (normalized, rms) for $3.5\cdot 10^{10}$ particles, \\
 & scaling with number of accumulated particles: \(\eperp \sim N^{4/5}\) \\
Relative momentum spread &
$3.3\cdot 10^{-4}$ (normalized, rms) for $3.5\cdot10^{10}$ particles, \\
 & scaling with number of accumulated particles: $\sigma_p/p \sim
N^{2/5}$ \\
Bunch length & 150\,m \\
Injection Momentum & \q{3.8}{GeV/c} \\
Injection & Kicker injection using multi-harmonic RF cavities \\
 & \\
\multicolumn{2}{c}{\textbf{Experimental Requirements}}\\ \hline
Ion species & Antiprotons \\
$\pbar$ production rate &
\q{2\cdot 10^7}{/s} ($1.2\cdot 10^{10}$ per 10~min) \\
Momentum / Kinetic energy range &
1.5 to \q{15}{GeV/c} / 0.83 to \q{14.1}{GeV} \\
Number of particles &
$10^{10}$ to $10^{11}$ \\
Target thickness &
\q{4\cdot 10^{15}}{atoms/cm^2} (H$_2$ pellets) \\
Transverse emittance &
\q{< 1}{mm\cdot mrad} \\
Betatron amplitude E-Cooler &
25--200\,m \\
Betatron amplitude at IP &
1--15\,m \\
 & \\
\multicolumn{2}{c}{\textbf{Operation Modes}}\\ \hline
High resolution (HR) &
Luminosity of \q{2\cdot 10^{31}}{\mbox{cm}^{-2} s^{-1}} for $10^{10}\
\pbar$ \\
 & rms momentum spread $\sigma_p / p \leq 2\cdot 10^{-5}$, \\
 & 1.5 to \q{9}{\gevc}, electron cooling  up to \q{9}{\gevc} \\
High luminosity (HL) &
Luminosity of \q{2\cdot 10^{32}}{\mbox{cm}^{-2} s^{-1}} for $10^{11}\
\pbar$ \\
 & rms momentum spread $\sigma_p / p \sim 10^{-4}$, \\
 & 1.5 to \q{15}{\gevc}, stochastic cooling  above \q{3.8}{\gevc}\\ \hline\hline
\end{tabular}
}
\end{center}
\caption{Injection parameters, experimental requirements and operation
  modes.}
\label{tab:param}
\end{table*}

\begin{table*}
\begin{center}
{\small
\begin{tabular}{lccc}\hline\hline
 & \multicolumn{3}{c}{$(\tauloss^{-1})$ / s$^{-1}$} \\
Heating process & \q{1.5}{\gevc} & \q{9}{\gevc} & \q{15}{\gevc} \\ \hline
Hadronic Interaction &
$1.8\cdot 10^{-4}$ &
$1.2\cdot 10^{-4}$ &
$1.1\cdot 10^{-4}$ \\
Single Coulomb &
$2.9\cdot 10^{-4}$ &
$6.8\cdot 10^{-6}$ &
$2.4\cdot 10^{-6}$ \\
Energy Straggling &
$1.3\cdot 10^{-4}$ &
$4.1\cdot 10^{-5}$ &
$2.8\cdot 10^{-5}$ \\
Touschek Effect &
$4.9\cdot 10^{-5}$ &
$2.3\cdot 10^{-7}$ &
$4.9\cdot 10^{-8}$ \\ \hline
Total relative loss rate &
$6.5\cdot 10^{-4}$ &
$1.7\cdot 10^{-4}$ &
$1.4\cdot 10^{-4}$ \\
$1/e$ beam lifetime \tpbar\ / s &
$\sim 1540$ &
$\sim 6000$ &
$\sim 7100$ \\ \hline
\Lmax\ / \q{10^{32}}{\mbox{cm}^{-2} \mbox{s}^{-1}} &
0.82 &
3.22 &
3.93 \\ \hline\hline
\end{tabular}
}
\end{center}
\caption{Upper limit for relative beam loss rate, $1/e$ beam lifetime
  \tpbar, and maximum average luminosity {\Lmax} for a H$_2$ pellet target.}
\label{tab:lifetimes}
\end{table*}

\begin{figure*}[hbtp]
  \centering
  \includegraphics[width=0.9\dwidth]{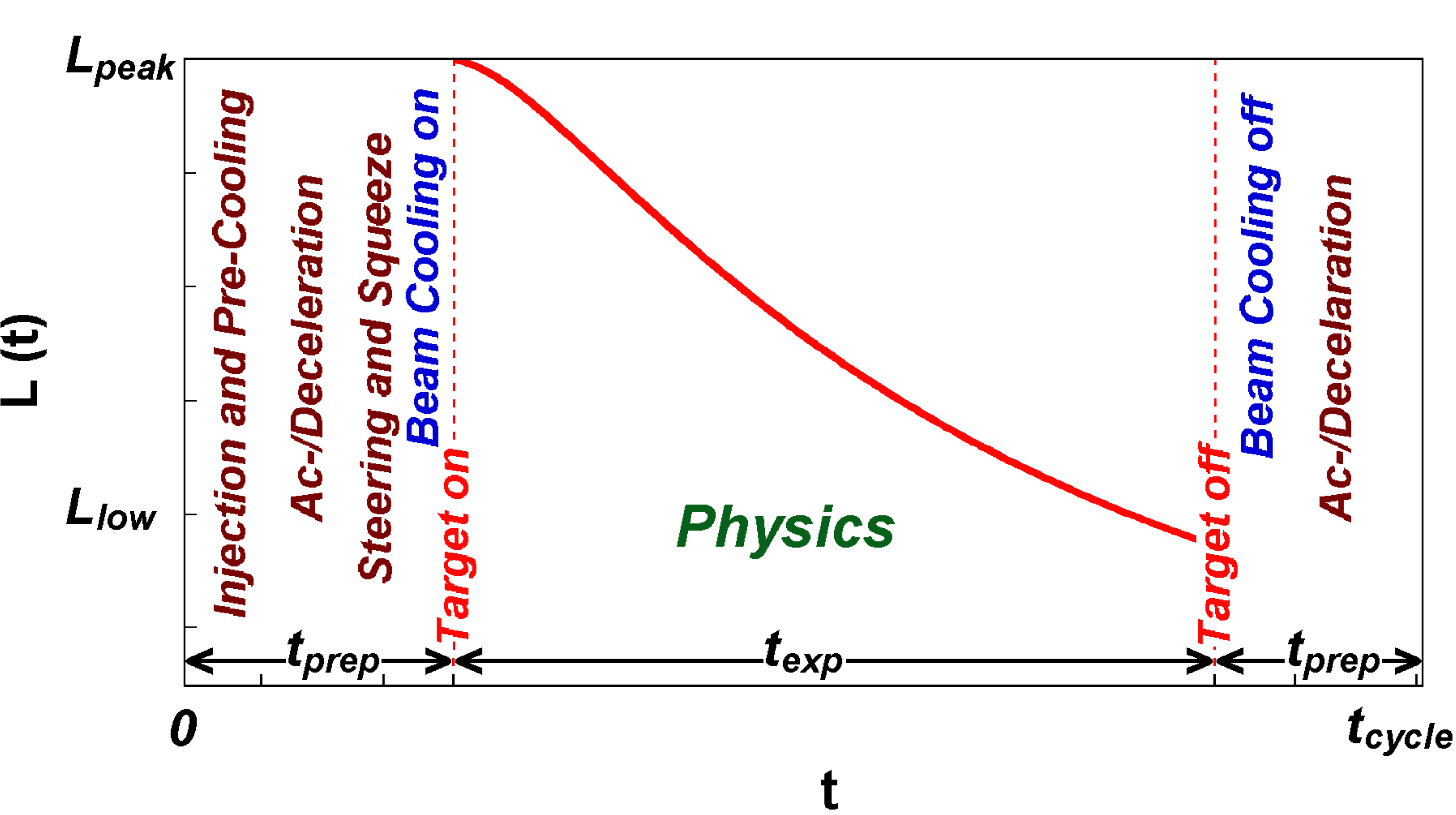}
  \caption{Time dependent luminosity during the cycle~$L(t)$ versus the time
in the cycle. Different measures for beam preparation are
indicated.}
  \label{fig:lumcycle}
\end{figure*}

\begin{figure*}[hbtp]
  \centering
    \includegraphics[width=0.45\dwidth]{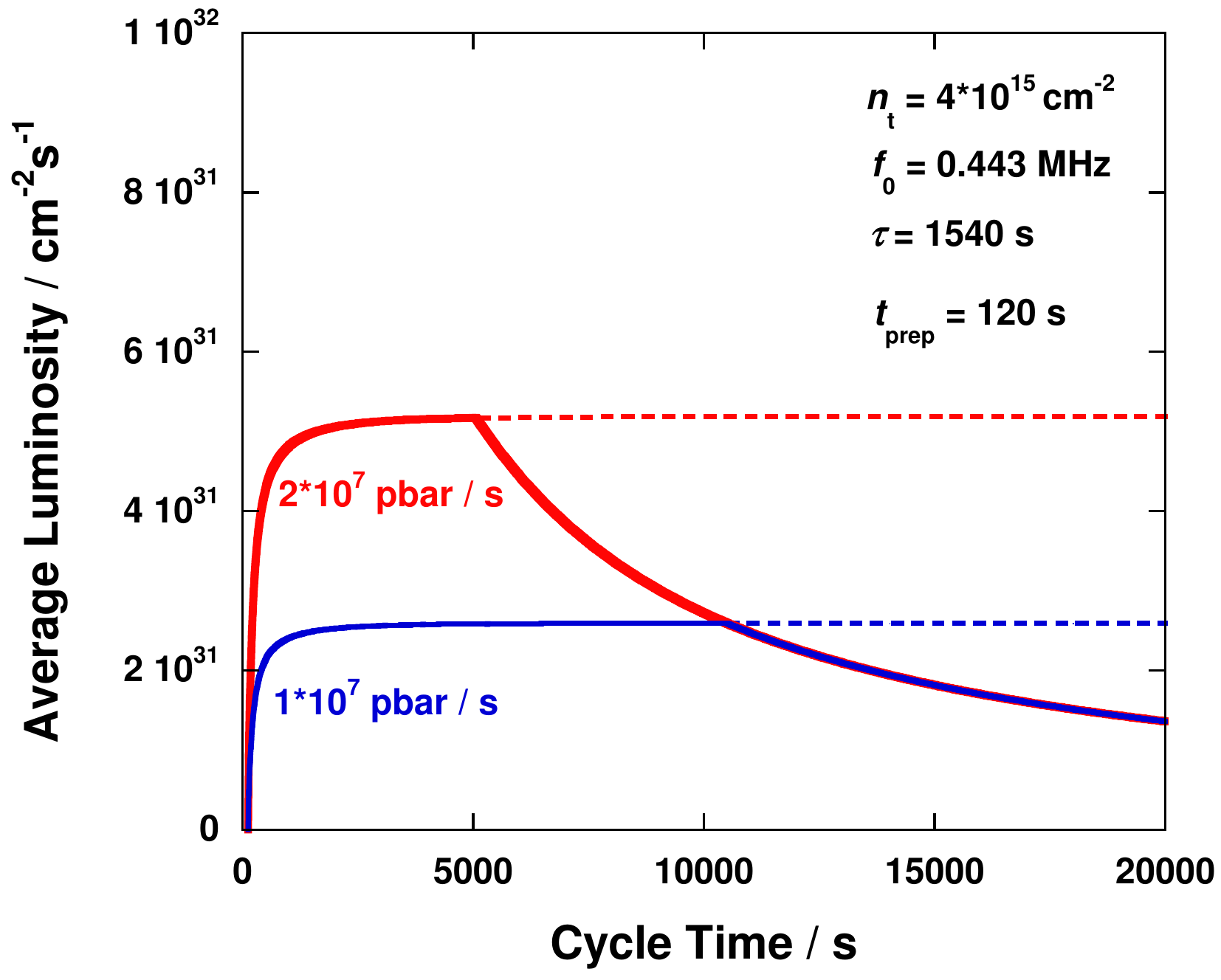} \hfill
    \includegraphics[width=0.45\dwidth]{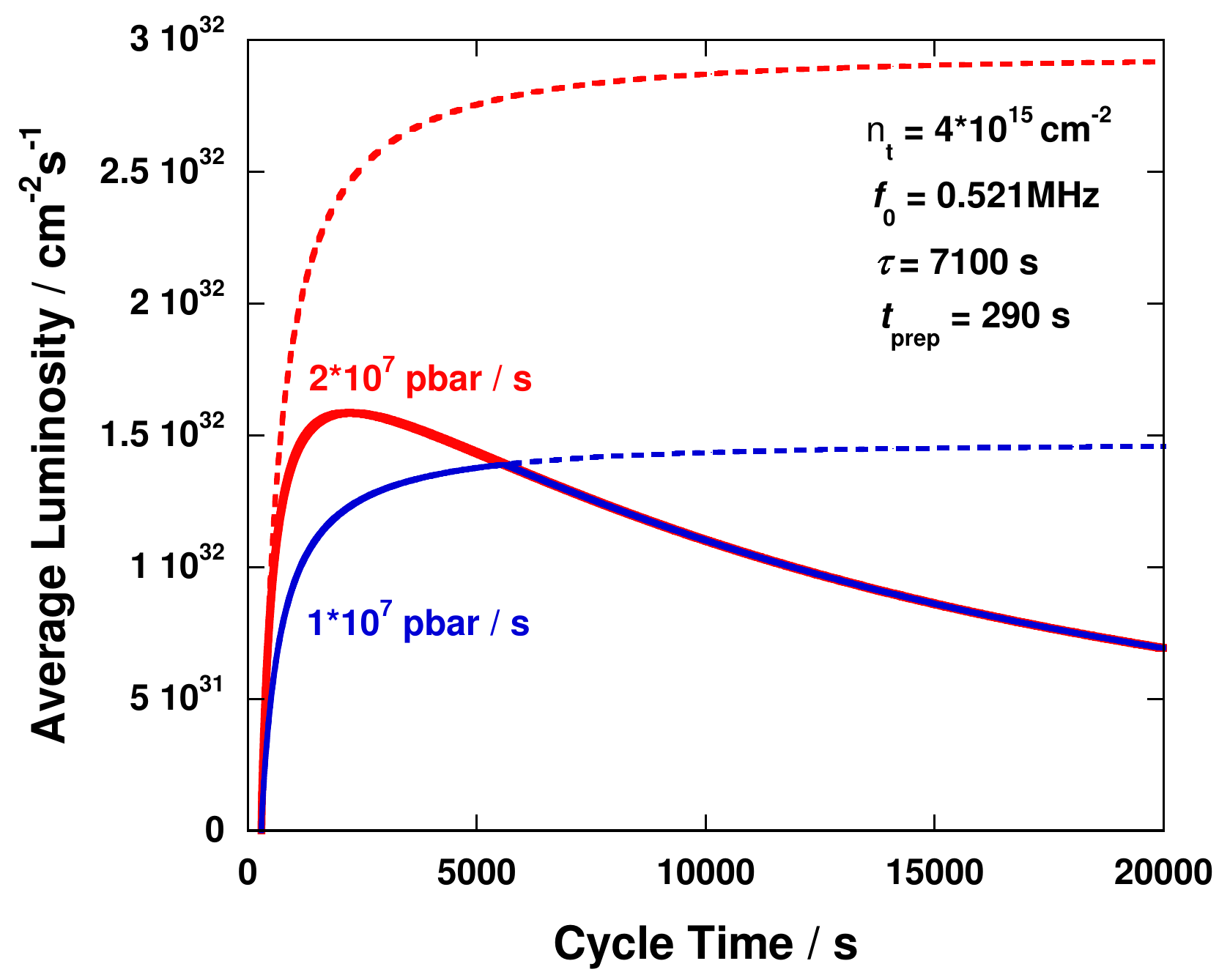}
  \caption[Cycle average luminosity vs. cycle time at 1.5 (left) and
    \q{15}{\gevc} (right).]{ The maximum number of particles is limited
    to $10^{11}$ (solid line), and unlimited (dashed lines).}
  \label{fig:avglum}
\end{figure*}

\begin{figure}[hbtp]
  \centering
    \includegraphics[width=\swidth]{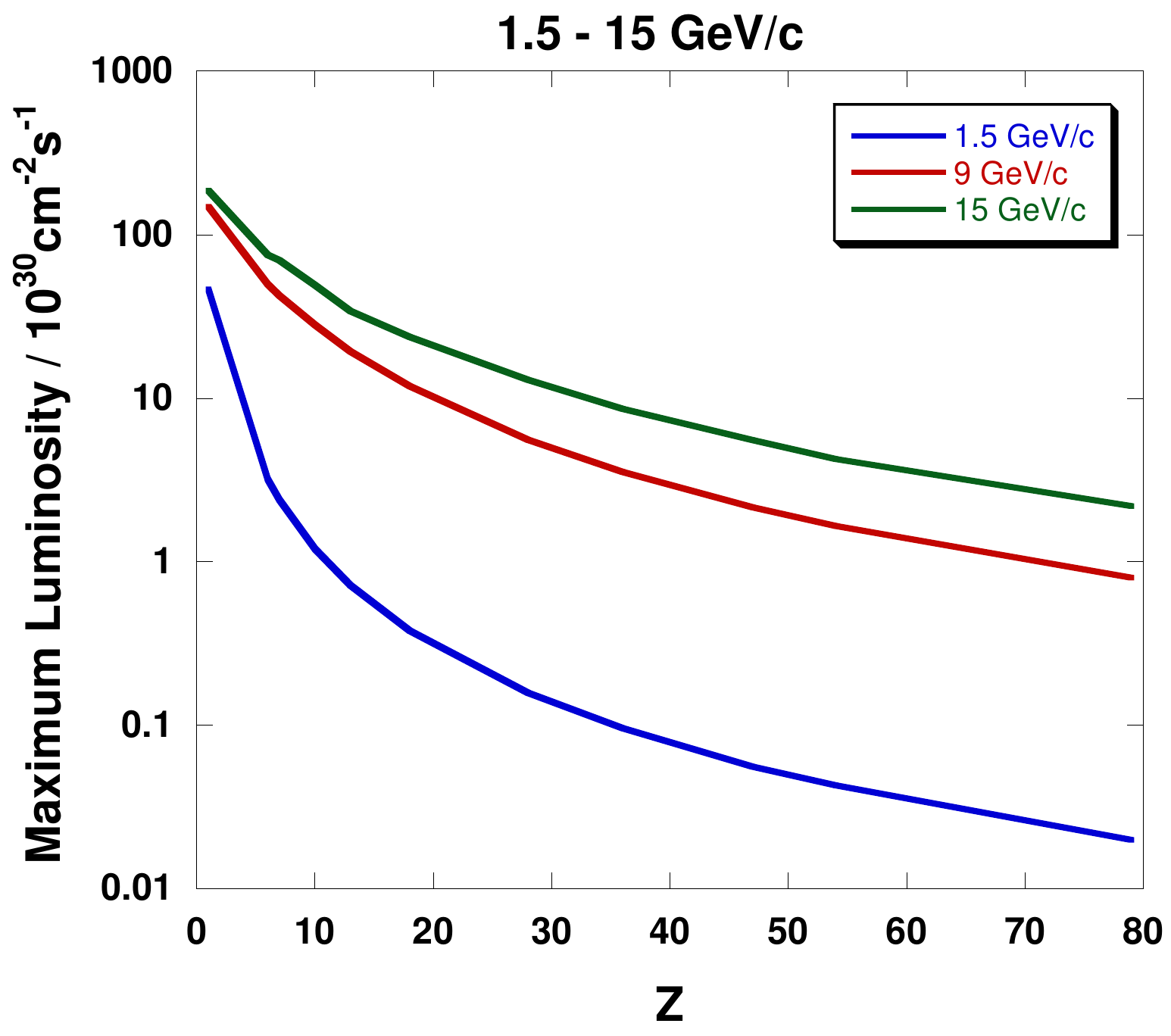}
    \caption{Maximum average luminosity \Lmax vs. atomic charge $Z$
      for three different beam momenta.}
  \label{fig:nucleart}
\end{figure}

The HESR is being realized by a consortium consisting of IKP at
Forschungszentrum J\"ulich, TSL at Uppsala University, and GSI
Darmstadt~\cite{FAIR:2006}. An important feature of this new
facility is the combination of phase-space cooled beams and dense
internal targets, comprising challenging beam parameters in two
operation modes: high-luminosity mode with beam intensities up to
10$^{11}$, and high-resolution mode with a momentum spread down to
a few times 10$^{-5}$, respectively. Powerful electron and
stochastic cooling systems are necessary to meet the experimental
requirements.

The HESR lattice is designed as a racetrack shaped ring,
consisting of two 180$^\circ$ arc sections connected by two long
straight sections. One straight section will mainly be occupied by
the electron cooler. The other section will host the experimental
installation with internal H$_2$ pellet target, RF cavities,
injection kickers and septa (see \Reffig{fig:HESR}). For
stochastic cooling pickup (P) and kicker (K) tanks are also
located in the straight sections, opposite to each other. Special
requirements for the lattice are dispersion free straight sections
and small betatron amplitudes in the range of a few metres at the
internal interaction point. In addition the betatron amplitudes at
the electron cooler are adjustable within a large range.

\Reftbl{tab:param} summarizes the specified injection parameters,
experimental requirements and operation modes.

\subsection{Beam Equilibria and Luminosity Estimates}

The equilibrium beam parameters are most important for the
high-resolution mode. Calculations of beam equilibria for beam
cooling, intra-beam scattering and beam-target interaction are
being performed utilizing different simulation codes like BETACOOL
(JINR, Dubna), MOCAC (ITEP, Moscow), and PTARGET (GSI, Darmstadt).
Cooled beam equilibria calculations including special features of
pellet targets have been carried out with a simulation code based
on PTARGET.

\subsubsection{Beam Equilibria with Electron Cooling}

An electron beam with up to 1~A current, accelerated in special
accelerator columns to energies in the range of 0.4 to 4.5\,MeV is
proposed for the HESR. Since the design is modular it facilitates
future increase of the high voltage to 8\,MV. The 22\,m long
solenoidal field in the cooler section has a longitudinal field
strength of 0.2\,T with a magnetic field straightness on the order
of $10^{-5}$~\cite{FAIR:2006}. This arrangement allows beam
cooling for beam momentum between \q{1.5}{\gevc} and
\q{8.9}{\gevc}.

To simulate the dynamics of the core particles, an analytic
``rms'' model was applied~\cite{boine:2006}. The empirical
magnetized cooling force formula by V.~V.~Parkhomchuk for electron
cooling~\cite{parkhomchuk:2000} and an analytical description for
intra-beam scattering~\cite{sorensen:1987} was used. Beam heating
by beam-target interaction is described by transverse and
longitudinal emittance growth due to Coulomb scattering and energy
straggling,
respectively~\cite{hinterberger:1989a,hinterberger:1989b}. In the
HR mode, rms relative momentum spreads are
$7.9\times 10^{-6}$ at \q{1.5}{\gevc},
$2.7\times 10^{-5}$ at \q{8.9}{\gevc}, and
$1.2\times 10^{-4}$ at \q{15}{\gevc}~\cite{reistad:2007}.

\subsubsection{Beam Equilibria with Stochastic Cooling}

The main stochastic cooling parameters were determined for a
cooling system utilizing pickups and kickers with a band-width of
2 -- 4~GHz and the option for an extension to 4 -- 6~GHz.
Stochastic cooling is presently specified above
\q{3.8}{\gevc}~\cite{stockhorst:2008}. Beam equilibria have been
simulated based on a Fokker-Planck approach. Applying stochastic
cooling, one can achieve rms relative momentum spreads of
$5.1\times 10^{-5}$ at \q{3.8}{\gevc},
$5.4\times 10^{-5}$ at \q{8.9}{\gevc}, and
$3.9\times 10^{-5}$ at \q{15}{\gevc}
for the HR mode.
With a combination of electron and stochastic cooling the beam
equilibria can be further improved.
In the HL mode, rms relative
momentum spread of roughly $10^{-4}$ can be expected. Transverse
stochastic cooling can be adjusted independently to ensure
sufficient beam-target overlap.

\subsubsection{Beam Losses and Luminosity Estimates}

Beam losses are the main restriction for high luminosities, since
the antiproton production rate is limited. Three dominating
contributions of beam-target interaction have been identified:
hadronic interaction, single Coulomb scattering and energy
straggling of the circulating beam in the target. In addition,
single intra-beam scattering due to the Touschek effect has also
to be considered for beam lifetime estimates. Beam losses due to
residual gas scattering can be neglected compared to beam-target
interaction, if the vacuum is better than \q{10^{-9}}{mbar}. A
detailed analysis of all beam loss processes can be found
in~\cite{lehrach:2006,hinterberger:2006}.

\subsubsection{Beam lifetime}

The relative beam loss rate for the total cross section~\stot\ is
given by the expression
\begin{equation}
(\tauloss^{-1}) = n_t \stot f_0
\end{equation}
where $(\tauloss^{-1})$ is the relative beam loss rate, $n_t$ the
target thickness and $f_0$ the reference particle's revolution
frequency. In \Reftbl{tab:lifetimes} the upper limit for beam
losses and corresponding lifetimes are listed for a transverse beam
emittance of \q{1}{mm\cdot mrad}, a longitudinal ring acceptance of
$\Delta p/p = \pm 10^{-3}$ and $10^{11}$ circulating particles in the ring.

For beam-target interaction, the beam lifetime is independent of
the beam intensity, whereas for the Touschek effect it depends on
the beam equilibria and beam intensity. Beam lifetimes are ranging
from 1540\,s to 7100\,s. Beam lifetimes at low momenta strongly
depend on the beam cooling scenario and the ring acceptance. Beam
losses corresponding to beam lifetimes below half an hour
obviously cannot be compensated by the antiproton production rate.

\subsubsection{Luminosity Considerations for Hydrogen-Pellet Targets}

The maximum average luminosity depends on the antiproton
production rate \( d\Npbar / dt = \q{2\cdot 10^7}{/s} \) and loss
rate
\begin{equation}
\Lmax = \frac{d\Npbar / dt}{\stot}
\end{equation}
and is also given in \Reftbl{tab:lifetimes} for different beam
momenta. The maximum average luminosity for \q{1.5}{\gevc} is
below the specified value for the HL mode.

To calculate the cycle average luminosity, machine cycles and beam
preparation times have to be specified. After injection, the beam
is pre-cooled to equilibrium (with target off) at \q{3.8}{\gevc}.
The beam is then accelerated or decelerated to the desired beam
momentum. A maximum ramp rate of 25~mT/s is specified. After
reaching the final momentum, beam steering and focusing in the
target and beam cooler region takes place. The total beam
preparation time \tprep\ ranges from 120\,s for \q{1.5}{\gevc} to
290\,s for \q{15}{\gevc}. A typical example for the evolution of
the luminosity during a cycle is plotted in
\Reffig{fig:lumcycle} versus the time in the cycle.

In the high-luminosity mode, particles should be re-used in the
next cycle. Therefore the used beam is transferred back to the
injection momentum and merged with the newly injected beam. A
bucket scheme utilizing broad-band cavities is foreseen for beam
injection and the refill procedure. During acceleration 1\percent and
during deceleration 5\percent beam losses are assumed.  The cycle
average luminosity reads
\begin{equation}
\bar{L} = f_0 N_{i,0} n_t
  \frac{\tau \left[ 1 - e^{-\texp/\tau} \right]}{\texp + \tprep}
\end{equation}
where~$\tau$ is the $1/e$ beam lifetime, \texp\ the experimental
time (beam on target time), and \tcycle\ the total time of the
cycle, with \( \tcycle = \texp + \tprep \). The dependence of the
cycle average luminosity on the cycle time is shown for different
antiproton production rates in \Reffig{fig:avglum}.

With limited number of antiprotons of $10^{11}$, as specified for
the high-luminosity mode, cycle average luminosities of up to
\q{1.6\cdot 10^{32}}{\mbox{cm}^{-2} \mbox{s}^{-1}} can be achieved at
\q{15}{\gevc} for cycle times of less than one beam lifetime. If
one does not restrict the number of available particles, cycle
times should be longer to reach maximum average luminosities close
to \q{3\cdot 10^{32}}{\mbox{cm}^{-2} \mbox{s}^{-1}}. This is a theoretical upper
limit, since the larger momentum spread of the injected beam would
lead to higher beam losses during injection due to the limited
longitudinal ring acceptance. For the lowest momentum, more than
$10^{11}$ particles can not be provided in average, due to very
short beam lifetimes. As expected, cycle average luminosities are
below \q{10^{32}}{\mbox{cm}^{-2} \mbox{s}^{-1}}.

\subsubsection{Luminosity Considerations for Nuclear Targets}

The hadronic cross section for the interaction of antiprotons with
nuclear targets can be estimated from geometric considerations.
Starting from the antiproton-proton hadronic cross section
$\sigma^\mathrm{\pbarp}_\mathrm{H}$ for 1.5\,\gevc
\begin{equation*}
\sigma^\mathrm{\pbarp}_\mathrm{H} \approx 100\,\mathrm{mbarn} :=
\pi r_\mathrm{p}^2
\end{equation*}
with the proton radius of $r_\mathrm{p} = 0.9$\,fm, the
antiproton-nucleus hadronic cross section can be deduced to be
\begin{equation}
\label{Formel:HadronischeWW} \sigma^\mathrm{\pbarA}_\mathrm{H} =
\pi (R_\mathrm{A} + r_\mathrm{p})^2.
\end{equation}
The radius of a spherical nucleus as a first approximation reads
$R_\mathrm{A} = r_0 A^{1/3}$, with $r_0 = 1.2$\,fm and the mass
number $A$. The total hadronic cross section decreases with the
beam momentum from 100, 60 to 50\,mb for hydrogen targets at 1.5,
9, and 15\,\gevc, respectively. The cross sections for nuclear
targets is scaled with beam momentum accordingly
\cite{lehrach:pdb}. To evaluate beam losses also single Coulomb
scattering and energy straggling of the circulating beam in the
target have been calculated \cite{lehrach:ar}.
\Reffig{fig:nucleart} shows maximum average luminosities for
nuclear targets under the same conditions as for hydrogen targets.

For the specified antiproton production rate maximum average
luminosities of $5 \cdot 10^{31}$, $4 \cdot 10^{29}$ to $4 \cdot
10^{28}$\,cm$^{-2}$ $\cdot$ s$^{-1}$ (deuterium, argon to gold)
are achieved at 1.5\,\gevc beam momentum. For maximum beam momentum
of 15\,\gevc the maximum average luminosities increase by more than
one order of magnitude to $1.9 \cdot 10^{32}$, $2.4 \cdot 10^{31}$
to $2.2 \cdot 10^{30}\,$cm$^{-2}\cdot\,$s$^{-1}$ (deuterium, argon
to gold) for $10^{11}$ circulating antiprotons.

In order to reach these values an effective target thickness of
3.6 $\cdot 10^{15}$\,atoms/cm$^2$ for a deuterium target, 4.6
$\cdot 10^{14}$\,atoms/cm$^2$ for an argon target to 4.1 $\cdot
10^{13}$\,atoms/cm$^2$ for gold at $10^{11}$ $\pbar$ is
required. The optimum effective target thickness can be adjusted
by proper beam focussing and steering onto the target.


%% file: exp/exp_commands.tex



\newcommand{\q}[2]{\ensuremath{#1\ \mathrm{#2}}}
\newcommand{\psip}{\ensuremath{\psi (2S)}}
\newcommand{\psinc}{\ensuremath{\jpsi + X}}
\newcommand{\eeX}{\ensuremath{e^+ e^- X}}
\newcommand{\Gin}{\ensuremath{\Gamma_{\mathrm{in}}}}
\newcommand{\Gout}{\ensuremath{\Gamma_{\mathrm{out}}}}
\newcommand{\Gee}{\ensuremath{\Gamma_{\ee}}}
\newcommand{\Gpp}{\ensuremath{\Gamma_{\pbarp}}}
\newcommand{\GGG}{\ensuremath{\Gee \Gpp / \Gamma}}
\newcommand{\GiGoG}{\ensuremath{\Gin \Gout / \Gamma}}
\newcommand{\Lumn}{\ensuremath{\mathcal{L}}}
\newcommand{\lik}{\ensuremath{\Lambda}}
\newcommand{\sBW}{\ensuremath{\sigma_\mathrm{BW}}}
\newcommand{\sBWr}{\ensuremath{\sigma_\mathrm{BWr}}}
\newcommand{\scf}{\ensuremath{b}}
\newcommand{\sbkg}{\ensuremath{\sigma_\mathrm{bkg}}}
\newcommand{\Bi}{\ensuremath{B_\mathrm{in}}}
\newcommand{\Bo}{\ensuremath{B_\mathrm{out}}}
\newcommand{\fcav}{\ensuremath{f^\mathrm{cav}}}
\newcommand{\frf}{\ensuremath{f^\mathrm{rf}}}
\newcommand{\Vrf}{\ensuremath{V^\mathrm{rf}}}
\newcommand{\vrf}{\ensuremath{v^\mathrm{rf}}}
\newcommand{\betarf}{\ensuremath{\beta^\mathrm{rf}}}
\newcommand{\gammarf}{\ensuremath{\gamma^\mathrm{rf}}}
\newcommand{\wrf}{\ensuremath{w^\mathrm{rf}}}
\newcommand{\frfz}{\ensuremath{f^\mathrm{rf}_0}}
\newcommand{\frfi}{\ensuremath{f^\mathrm{rf}_i}}
\newcommand{\dL}{\ensuremath{\Delta L}}
\newcommand{\reff}{\ensuremath{\varepsilon^X_\mathrm{co}/\varepsilon^X_\mathrm{cf}}}
\newcommand{\muee}{\ensuremath{\mu^{ee}}}
\newcommand{\Nee}{\ensuremath{N^{ee}}}
\newcommand{\effee}{\ensuremath{\varepsilon^{ee}}}
\newcommand{\muX}{\ensuremath{\mu^{X}}}
\newcommand{\NX}{\ensuremath{N^{X}}}
\newcommand{\effXcf}{\ensuremath{\varepsilon^{X}_\mathrm{cf}}}
\newcommand{\resG}{\q{290 \pm 25 \mathrm{(sta)} \pm 4 \mathrm{(sys)}}{keV}}
\newcommand{\resA}{\q{579 \pm 38 \mathrm{(sta)} \pm 36 \mathrm{(sys)}}{meV}}
\newcommand{\eperp}{\ensuremath{\varepsilon_\perp}}
\newcommand{\stot}{\ensuremath{\sigma_\mathrm{tot}}}
\newcommand{\tauloss}{\ensuremath{\tau_\mathrm{loss}}}
\newcommand{\tpbar}{\ensuremath{t_{\bar{p}}}}
\newcommand{\Lmax}{\ensuremath{L_\mathrm{max}}}
\newcommand{\Npbar}{\ensuremath{N_{\bar{p}}}}
\newcommand{\tprep}{\ensuremath{t_\mathrm{prep}}}
\newcommand{\texp}{\ensuremath{t_\mathrm{exp}}}
\newcommand{\tcycle}{\ensuremath{t_\mathrm{cycle}}}

\newcommand{\code}[1]{\textsc{#1}}



%% file: exp/exp_scan.tex
%
\newcommand{\sigmapk}{\ensuremath{\sigma_p}}
\newcommand{\sphys}{\ensuremath{\sigma_\mathrm{phys}}}
\newcommand{\sphyspeak}{\ensuremath{\sigma_\mathrm{phys}^\mathrm{peak}}}
\newcommand{\sobs}{\ensuremath{\sigma_\mathrm{obs}}}
\newcommand{\dstn}{\ensuremath{d}}

\section{Precision Measurements of Resonance Parameters}
\COM{Author(s): G. Stancari}

The study of resonances is an important part of the \PANDA physics
programme. Masses, widths and decay fractions are measured by scanning
the beam energy across the resonance under study.  In
antiproton-proton annihilations, there are two main advantages over
inclusive production: (a)~resonances can be formed directly; (b)~the
detector is used as an event counter ($y$ axis of the excitation
curve), while the energy determination ($x$ axis) relies entirely on
the precisely-calibrated and cooled antiproton beam.

This is an area where a close interplay between machine and detector
is needed, and this is why this discussion is included here.  In this
section, we describe the technique of resonance scans for precise
determination of resonance parameters.  A discussion of the main
sources of uncertainty is given, as it is an important input for the
following physics chapters.  The issue of determining line shapes is
also briefly addressed.

Much of the discussion is based upon E835 experience with charmonium
resonances~\cite{bib:exp:mcginnis:2003,bib:exp:andreotti:2007}.  We
assume Breit-Wigner resonant shapes and constant backgrounds, but the
analysis can be easily extended to more general cases.

\begin{figure*}[btp]
  \centering
    \resizebox{0.67\dwidth}{!}{\includegraphics{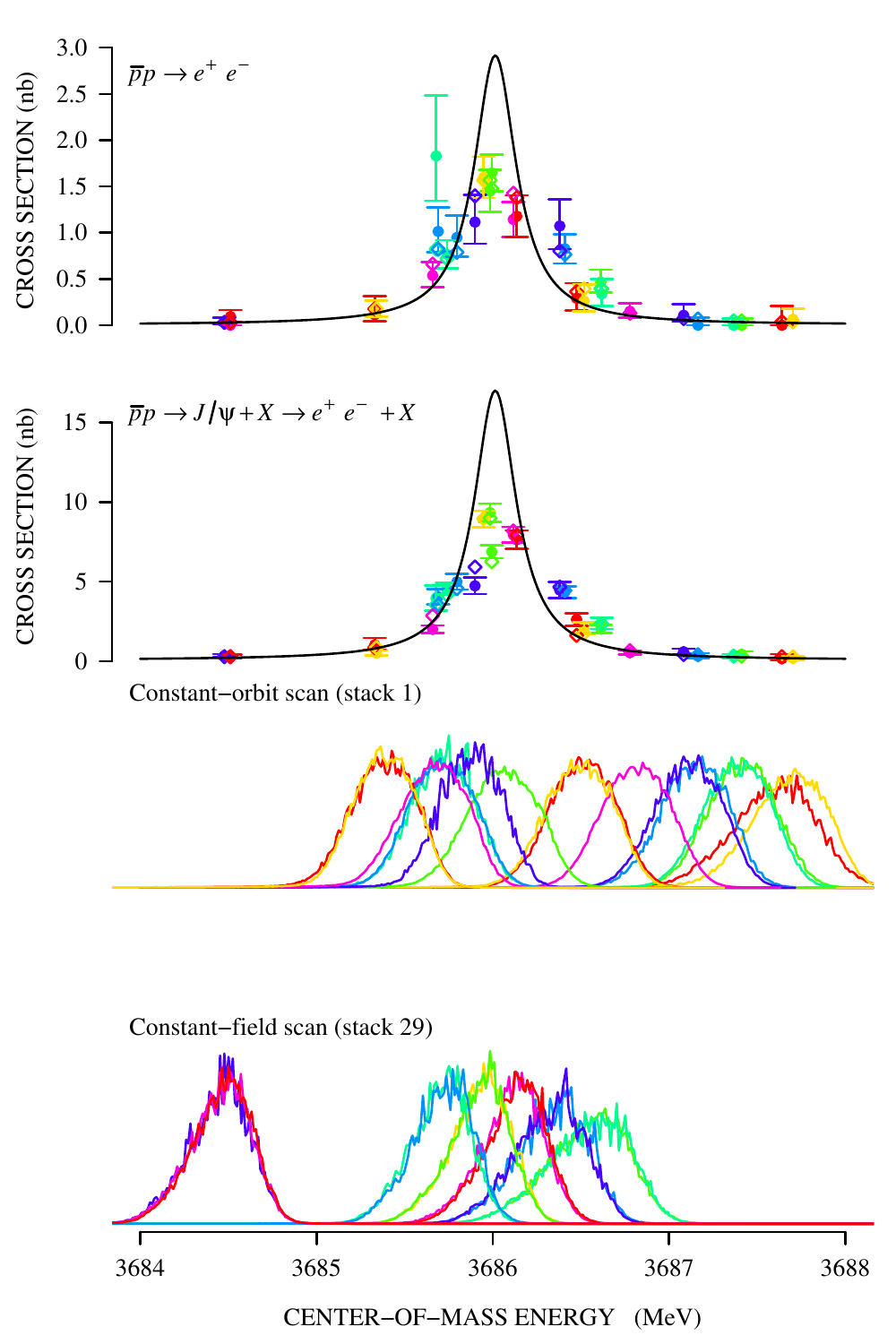}}
  \caption[\psip\ resonance scans.]{
    The observed cross section (excitation curve)
    for each channel (filled dots);
    the expected cross section from the fit (open diamonds);
    the `bare' resonance curves~\sBW\ from the fit (solid lines).
    The two bottom plots show the normalized energy distributions~$B_i$
    (from Ref.~\cite{bib:exp:andreotti:2007}).}
  \label{fig:exp:scan}
\end{figure*}

\begin{figure}[btp]
  \centering
    \resizebox{\swidth}{!}{\includegraphics{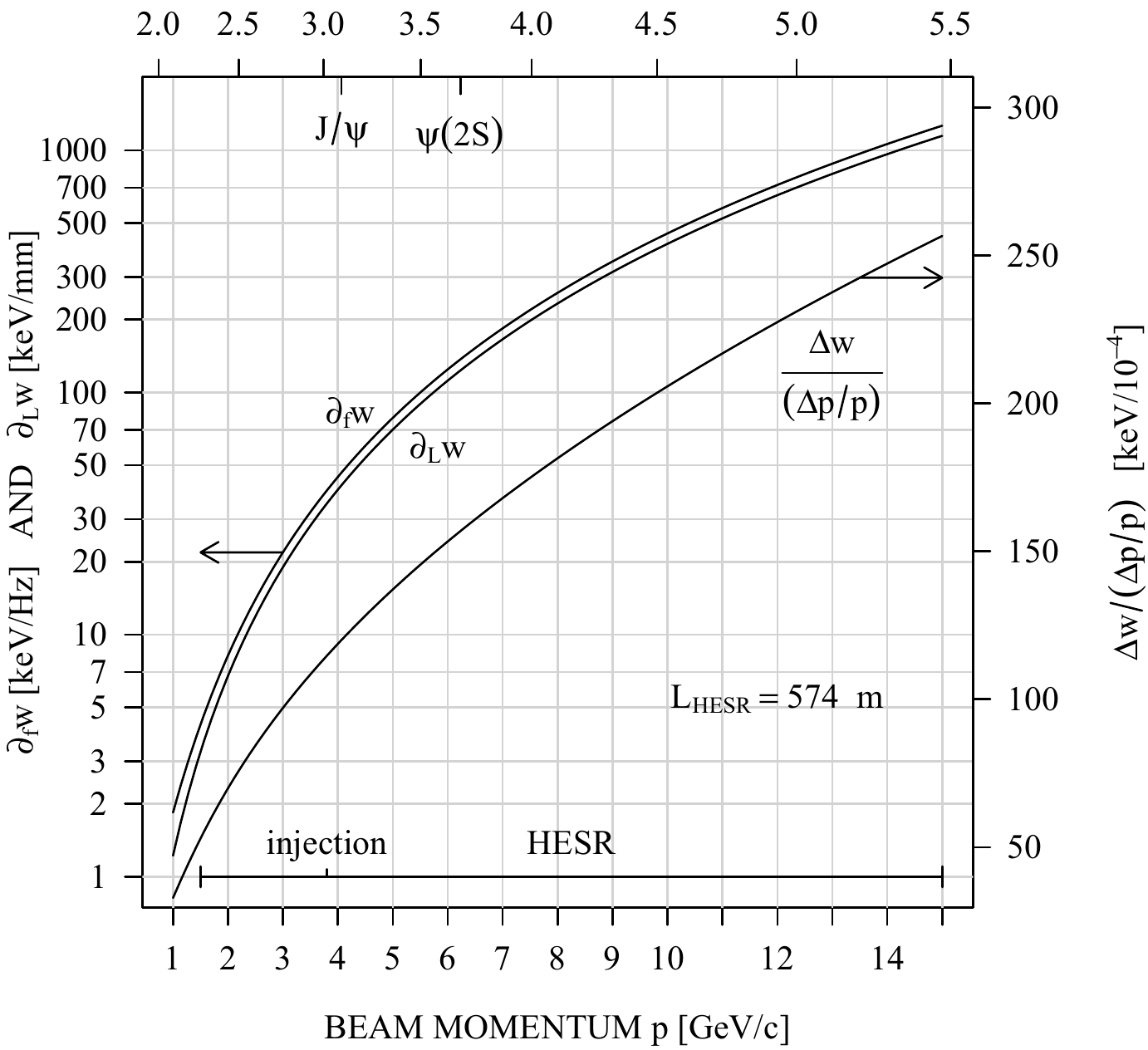}}
  \caption[Kinematical factors used in the determination of the
    centre-of-mass energy from the velocity of antiprotons in the rf
    bucket.]{
    Left vertical axis: partial derivatives of the
    centre-of-mass energy~$w$ with respect to revolution frequency~$f$
    and orbit length~$L$.
    Right vertical axis: energy spread in the centre of mass~$\Delta w$ for a
    given momentum spread~$\Delta p / p$.}
  \label{fig:energymeas}
\end{figure}

\newlength{\unctable}
\setlength{\unctable}{0.6\dwidth}
\newcommand{\Msyst}{\begin{minipage}[t]{\unctable}
\begin{flushleft}
-- \jpsi\ or \psip\ mass from resonant depolarization: 10~keV\\
-- \dL: 5~keV (single scan), 100~keV (scans at different energies)
\end{flushleft}\end{minipage}}
\newcommand{\Gsyst}{\begin{minipage}[t]{\unctable}
\begin{flushleft}
-- $\eta$: $\sim$5\percent \\
-- \dL\ (BPM noise): \q{\sim 5}{keV}
\end{flushleft}\end{minipage}}
\newcommand{\Asyst}{\begin{minipage}[t]{\unctable}
\begin{flushleft}
-- efficiency: a few \% \\
-- luminosity: a few \%
\end{flushleft}\end{minipage}}

\begin{table*}[btp]
  \centering
  \begin{tabular}{ccc}\hline\hline\\
  & statistical & systematic \\ \hline
  mass~$M$ & 3~keV & \Msyst \\
  width~$\Gamma$ & 1.9\percent & \Gsyst \\
  `area'~$\GiGoG$ & 1.5\percent & \Asyst \\ \hline\hline
  \end{tabular}
  \caption[Summary of the sources of uncertainty in the resonance parameters]{Summary of the sources of uncertainty in the resonance parameters.
    Statistical errors refer to
    $\mathcal{N} \equiv \varepsilon \Lumn \sigma = 10^4$ and scale as
    $1/\sqrt{\mathcal{N}}$.
    It is assumed that the luminosity distribution is optimized
    and that the beam width is negligible, and so they represent lower
    limits for a given $\mathcal{N}$.} 
\label{tab:exp:uncertainties}
\end{table*}

\begin{figure}[btp]
  \centering
   \resizebox{\swidth}{!}{\includegraphics{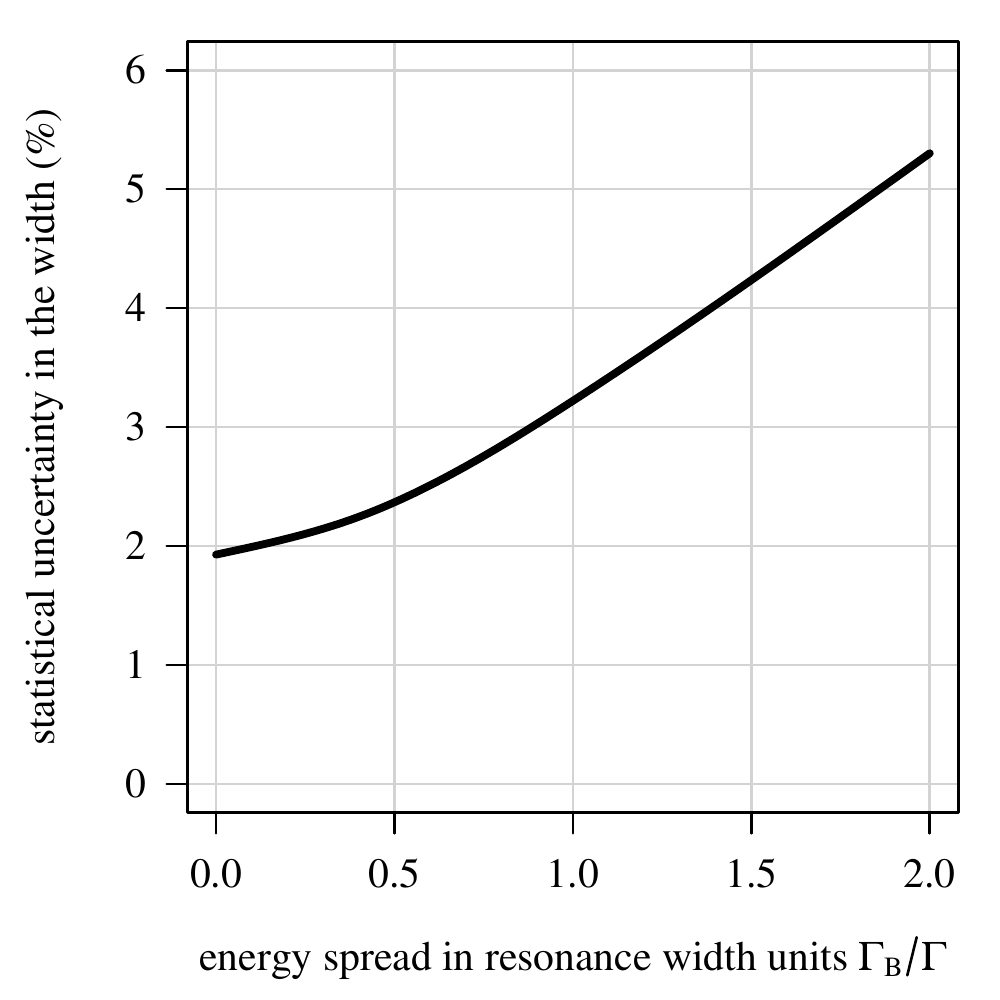}}
   \caption[Statistical uncertainty in the resonance width as a
     function of the ratio between energy spread (Gaussian
     FWHM~$\Gamma_B$) and resonance width~$\Gamma$.]{ The amount of data
     is $\mathcal{N} \equiv \varepsilon \Lumn \sigmapk = 10^4$ in this
     example. An
     optimal luminosity distribution is assumed.}
  \label{fig:exp:dGvsG}
\end{figure}

\begin{figure}[btp]
  \centering
   \resizebox{\swidth}{!}{\includegraphics{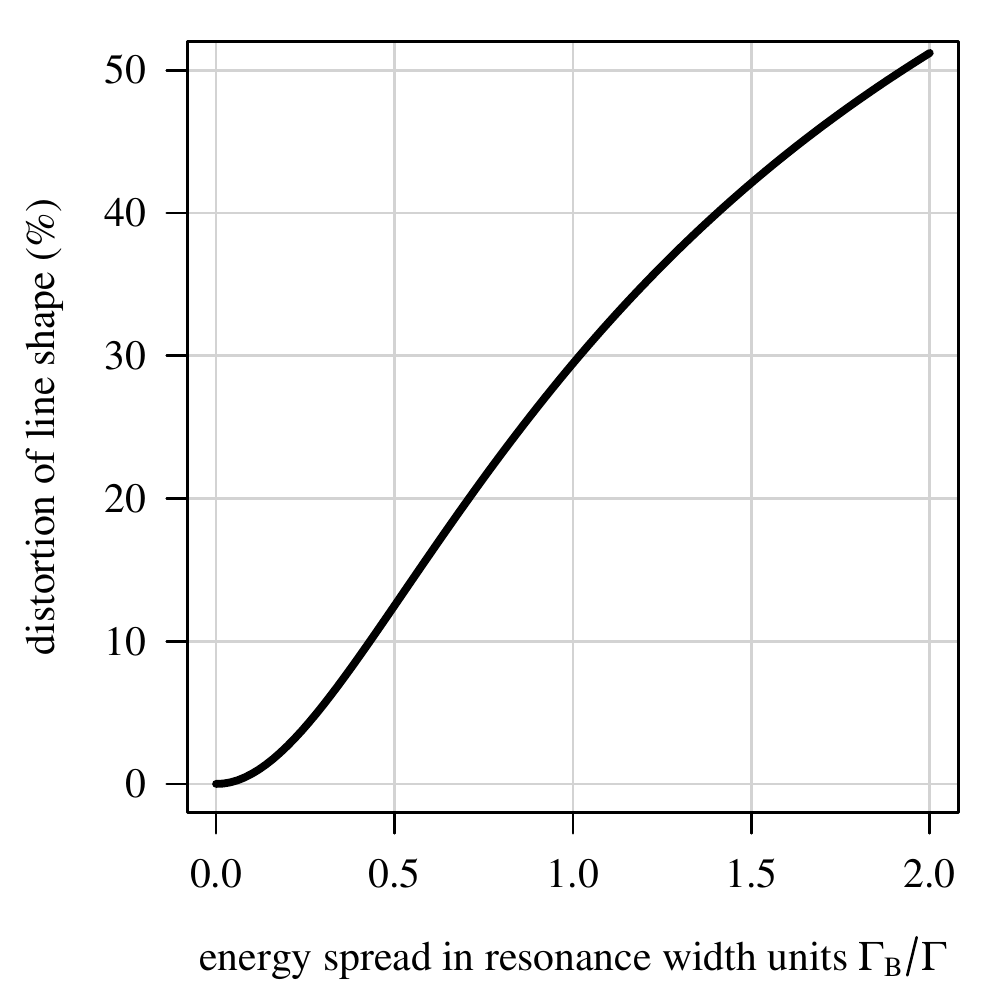}}
   \caption{Distortion of the resonance shape as a
     function of the ratio between energy spread (Gaussian
     FWHM~$\Gamma_B$) and resonance width~$\Gamma$.}
  \label{fig:exp:distortion}
\end{figure}

\subsection{Experimental Technique}

The resonance parameters are determined from a maximum-likelihood fit
to the excitation curve (\Reffig{fig:exp:scan}).  For each
data-taking run (subscript~$i$), we assume that the average number of
observed events~$\mu_i$ in each channel depends on a Breit-Wigner
cross section \sBWr\ and on the centre-of-mass energy distribution,
$B_i$, as follows:
\begin{equation}
  \mu_i = \varepsilon_i \Lumn_i \left[ \int \sBWr(w) \, B_i(w) \, dw
    + \sbkg
  \right],
  \label{eq:mu}
\end{equation}
where~$w$ is the centre-of-mass energy, $\varepsilon_i$ is the
detector efficiency, $\Lumn_i$ is the integrated luminosity, and
\sbkg\ is a constant background cross section.  The integral is
extended over the energy acceptance of the machine.  The spin-averaged
Breit-Wigner cross section for a spin-$J$ resonance of mass~$M$ and
width~$\Gamma$ formed in \pbarp\ annihilations is
\begin{equation}
  \sBW(w) = \frac{(2J+1)}{(2S+1)^2}
  \frac{16\pi}{w^2-4m^2}
  \frac{(\Gin \Gout / \Gamma) \cdot \Gamma}
  {\Gamma^2 + 4 (w-M)^2};
\end{equation}
$m$ and $S$ are the (anti)proton mass and spin, while~$\Gin$
and~$\Gout$ are the partial resonance widths for the entrance (\pbarp,
in our case) and exit channels.  The Breit-Wigner cross section is
corrected for initial-state radiation to obtain
\sBWr~\cite{bib:exp:armstrong:1993,bib:exp:kennedy:1992}:
\begin{eqnarray*}
  \lefteqn{\sBWr(w) =} \\
  & & \scf \int_0^{w/2}
  \frac{dk}{k}
  \left(\frac{2k}{w}\right)^\scf
  \sBW(\sqrt{w^2-2kw}) = \\
  & & \left(2/w\right)^\scf
  \int_0^{(w/2)^\scf} dt \, \sBW(\sqrt{w^2-2t^{1/\scf}w}),
\end{eqnarray*}
where the second form is more suitable for numerical integration and
$\scf(w)$ is the semiclassical collinearity
factor~\cite{bib:exp:kennedy:1992}, equal to 0.00753 at the \psip.

The resonance mass~$M$, width~$\Gamma$, `area'~$A \equiv (\GiGoG)$ and
the background cross section~\sbkg\ are left as free parameters in the
maximization of the log-likelihood function \( \log(\lik) = \sum_i
\log P(\mu_i,N_i) \), where~$P(\mu,N)$ are Poisson probabilities of
observing~$N$ events when the mean is~$\mu$.  $N_i$ are the observed
number of events, and $\sigma_i \equiv N_i / (\varepsilon_i \Lumn_i)$
is the observed cross section.  The area parameter is usually chosen
in the parameterization of the resonance shape because it is
proportional to the total number of events in each channel.  It is
less correlated with the width than the product of branching
fractions.

The detector determines the cross section ($y$-axis) measurements and
their uncertainty. Here we focus on energy measurements ($x$-axis),
their uncertainty and their impact on the determination of resonance
parameters through~$B_i$.

\subsection{Mass Measurements}
\label{sec:beam_energy}

The centre-of-mass energy distribution $B_i(w)$ of the \pbarp\ system
can be determined from the velocity of the antiproton beam, since
revolution frequencies and orbit lengths can be measured very
precisely.

The revolution-frequency distribution of the antiprotons can be
measured by detecting the Schottky noise signal generated by the
coasting beam.  The signal is sensed by a longitudinal Schottky pickup
and recorded on a spectrum analyser.  An accuracy of 0.05~Hz can be
achieved on a revolution frequency of 0.52~MHz, over a wide dynamic
range in intensity (60~dBm), using commercial spectrum analysers.

The beam is slightly bunched by an rf cavity operating at $\fcav \sim
\q{0.52}{MHz}$, the first harmonic ($h=1$) of the revolution
frequency.  The beam is bunched both for stability (ion clearing) and
for making the beam position monitors (BPMs) sensitive to a portion of
the beam.  Therefore, recorded orbits refer to particles bunched by
the rf system, and their revolution frequency is \( \frf = \fcav / h
\).  The bunched-beam revolution frequency~\frf\ is usually close to
the average revolution frequency of the beam.  Each orbit consists of
several horizontal and vertical readings.  BPM noise in an important
consideration for precision energy measurements.

From the BPM readings and the HESR lattice model, differences \dL\ in
the length of one orbit and another can be calculated accurately.  The
main systematic uncertainties come from lattice differences, from BPM
calibrations, from bend-field drifts, and from neglecting second-order
terms in the orbit length. In the Fermilab Antiproton Accumulator, the
systematic uncertainty in \dL\ was estimated to be 0.05\,mm out of
474\,m during the \INST{E835} run within a single \psip\
scan~\cite{bib:exp:andreotti:2007}, and about 1\,mm over the entire
run.

The absolute length~$L$ of an orbit can be calculated from a reference
orbit of length $L_0$: \( L = L_0 + \dL \).  The calibration of~$L_0$
is done by scanning a resonance (the \psip, for instance) the mass of
which is precisely known from the resonant-depolarization method in
\ee\ experiments~\cite{bib:exp:aulchenko:2003}.

For particles in the bunched portion of the beam (rf bucket), the
relativistic parameters~\betarf\ and~\gammarf\ are calculated from
their velocity \( \vrf = \frf \cdot L \), from which the
centre-of-mass energy~$w$ of the \pbarp\ system is calculated: \( \wrf
= w(\frf,L) \equiv m \sqrt{2 \left( 1+\gammarf \right)} \).  (The
superscript rf is omitted from orbit lengths because they always refer
to particles in the rf bucket.)  In the charmonium region, this method
yields good accuracies on~$w$.  Fig.~\ref{fig:energymeas} shows the
magnitude of the partial derivatives of~$w$ with respect to~$\frf$
and~$L$, $\partial_f w$ and $\partial_L w$.  Being based upon velocity
measurements, the precision is quickly degraded as the beam energy
increases.  At constant~$L$ and relativistic energies, the partial
derivatives scale as the fifth power of the centre-of-mass energy:
\( \partial_{f,L} w \sim w^5\).  From the uncertainties in the orbit
length and in the revolution frequency, it is reasonable to expect
absolute measurements of the beam energy with an uncertainty of the
order of \q{0.1}{MeV} in the charmonium region.

\subsection{Total and Partial Widths}

For width and area determinations, energy differences are crucial, and
they must be determined precisely.  During normal data taking, the
beam is kept near the central orbit of the machine.  A particular run
is chosen as the reference (subscript~0).  Energy differences between
the reference run and other runs in the scan (subscript~$i$), for
particles in the rf bucket, are simply
\begin{equation}
  \wrf_i - \wrf_0 =
  w(\frfi,L_0+\dL_i) - w(\frfz,L_0).
  \label{eq:dw_dL}
\end{equation}
Within the energy range of a resonance scan, these differences are
largely independent of the choice of~$L_0$.  For this reason, the
absolute energy calibration is irrelevant for width and area
measurements.  Only uncertainties coming from~\dL\ are to be considered.

Once the energy~$\wrf_i$ for particles in the rf bucket is known, the
complete energy distribution is obtained from the Schottky
spectrum using the relation between frequency differences and momentum
differences at constant magnetic field:
\begin{equation}
  \frac{\Delta p}{p} = - \frac{1}{\eta} \frac{\Delta f}{f},
  \label{eq:dpp}
\end{equation}
where~$\eta$ is the energy-dependent phase-slip factor of the machine,
which is one of the parameters governing synchrotron oscillations.
(The dependence of~$\eta$ on beam energy is chosen during lattice
design; the variation of~$\eta$ within a scan can usually be
neglected.)  In terms of the centre-of-mass energy,
\begin{equation}
  w - \wrf_i = - \frac{1}{\eta}
    \frac{(\betarf_i)^2 (\gammarf_i) m^2}{\wrf_i}
    \frac{f - \frf_i}{\frf_i}.
  \label{eq:dw}
\end{equation}
Within a run, rf frequencies, beam-frequency spectra, and BPM readings
are to be updated frequently with respect to expected variations in
energy or luminosity.  Frequency spectra can then translated into
centre-of-mass energy through \Refeq{eq:dw}, weighted by luminosity
and summed, to obtain the luminosity-weighted normalized energy
spectra~$B_i(w)$ for each data-taking run.

The phase-slip factor is usually determined from the slope of the
measured synchrotron frequency~$f_s$ as a function of rf voltage
settings~\Vrf:
\begin{equation}
f_s^2 = - \frac{\eta \cos{\phi_s} (\frf)^2 q \Vrf}{2 \pi h \beta^2 E_s},
\end{equation}
where~$\phi_s$ is the synchronous phase, $q$ is the particles' charge,
$E_s$ their energy and~$\beta^2$ is the relativistic factor.  The main
uncertainty comes from the absolute calibration of the rf voltage and
it is usually of the order of a few percent.

The resonance width and area are affected by a systematic error due to
the uncertainty in~$\eta$.  This effect is most noticeable when the
beam width is not negligible compared to the resonance width.
Usually, the resonance width and area are positively correlated with
the phase-slip factor.  A larger~$\eta$ implies a narrower energy
spectrum, as described in \Refeq{eq:dw}.  As a consequence, the
fitted resonance will more closely resemble the measured excitation
curve, yielding a larger resonance width.

For precision measurements, one needs a better estimate of the
phase-slip factor or determinations that are independent of~$\eta$, or
both.  In E760, the `double scan' technique was
used~\cite{bib:exp:armstrong:1993}.  It yielded~$\eta$ with an
uncertainty of 6\percent at the \psip\ and width determinations largely
independent of the phase-slip factor, but it had the disadvantage of
being operationally complex.  For E835, a new method of `complementary
scans' was developed~\cite{bib:exp:andreotti:2007}.  It resulted in a
similar precision on~$\eta$ and arbitrarily small correlations between
resonance parameters and phase-slip factor; the technique is also
operationally simpler.

The resonance is scanned once on the central orbit, as described
above.  A second scan is then performed at constant magnetic bend
field.  The energy of the beam is changed by moving the longitudinal
stochastic-cooling pickups.  The beam moves away from the central
orbit, and the range of energies is limited but appropriate for narrow
resonances.

Since the magnetic field is constant, beam-energy differences can be
calculated independently of~\dL, directly from the
revolution-frequency spectra and the phase-slip factor, according to
\Refeq{eq:dw}.  A pivot run is chosen (subscript~$p$).  The rf
frequency of this run is used as a reference to calculate the energy
for particles in the rf bucket in other runs.  These particles have
revolution frequency~$\frf_i$ and the energy is calculated as follows:
\begin{equation}
 \wrf_i - \wrf_p = -\frac{1}{\eta}
 \frac{(\betarf_p)^2 (\gammarf_p) m^2}{\wrf_p}
 \frac{\frf_i - \frf_p}{\frf_p}.
 \label{eq:dw_eta}
\end{equation}
For the scan at constant magnetic field, this relation is used instead
of \Refeq{eq:dw_dL}.  Once the energy for particles at $\frf_i$ is
known, the full energy spectrum within each run is obtained from
\Refeq{eq:dw}, as usual. For the constant-field scan, the energy
distributions may be obtained directly from the pivot energy by
calculating \( w - \wrf_p \), instead of using \Refeq{eq:dw_eta}
first and then \Refeq{eq:dw}.  The two-step procedure is chosen
because it is faster to rescale the energy spectra than to
re-calculate them from the frequency spectra when fitting for~$\eta$.
Numerically, the difference between the two calculations is
negligible.  Moreover, the two-step procedure exposes how the width
depends on~$\eta$.

Using this alternative energy measurement, the width and area
determined from scans at constant magnetic field are negatively
correlated with~$\eta$.  The increasing width with increasing~$\eta$
is still present, as it is in scans at nearly constant orbit.  But the
dominant effect is that a larger~$\eta$ brings the energy points in
the excitation curve closer to the pivot point, making the width
smaller.

The constant-orbit and the constant-field scan can be combined.  The
resulting width has a dependence on~$\eta$ that is intermediate
between the two.  An appropriate luminosity distribution can make the
resulting curve practically horizontal.  The combined measurement is
dominated by the statistical uncertainty.  Moreover, thanks to this
complementary behavior, the width, area and phase-slip factor can be
determined in a maximum-likelihood fit where~$\eta$ is also a free
parameter.  Errors and correlations are then obtained directly from
the fit.

\subsection{Line Shapes}

The discussion so far was focused on the determination of Breit-Wigner
resonance parameters. One might also wish to determine the line shape
of resonances near the open charm threshold.  The question then
arises: how narrow should the beam be in order to distort the line
shape by less than a given amount?

The distortion of the line shape can be characterized by the maximum
difference~\dstn\ between the physical cross section~$\sphys(w)$ and the
observed cross section~$\sobs(w)$ (arising from the convolution of the
physical cross section with the \pbarp\ energy distribution in the
centre of mass~$B(w)$),
divided by the physical cross section at the peak~\sphyspeak:
\begin{equation}
\dstn \equiv \max_w{\left| \sphys(w) - \sobs(w) \right|}\ /\ \sphyspeak
\end{equation}

Fig.~\ref{fig:exp:distortion} shows the distortion~\dstn\ as a
function of the ratio between the FWHM of the energy distribution
(assumed to be Gaussian) and the FWHM of the resonance (a
Breit-Wigner, in this example).  For instance, if a distortion of less
than 10\percent is needed, than the FWHM ratio needs to be smaller than
0.43.  If the FWHM of the resonance is \( \Gamma = \q{1}{MeV} \), the
rms of the energy distribution in the centre of mass needs to be
smaller than \q{0.18}{MeV}, corresponding to a momentum spread
of~$0.8\times 10^{-4}$ at \q{6}{GeV/c} (see also
\Reffig{fig:energymeas}).

\subsection{Achievable Precision}

From the above discussion, it is clear that some features of the
machine are essential for precision measurements. Here is a list of
the most important requirements:
\begin{itemize}
\item longitudinal Schottky pickups;
\item low-noise horizontal and vertical BPMs, with fluctuations
  corresponding to less than 0.1\,mm in the orbit length;
\item small lattice differences between the energy of the calibration
  resonance and the energies of interest;
\item absolute rf voltage calibration to within a few percent;
\item motorized longitudinal cooling pickups, for active feedback on
  energy drifts and for constant-field scans.
\end{itemize}

\Reftbl{tab:exp:uncertainties} summarizes the sources of
statistical and systematic uncertainty in the resonance parameters.
Statistical errors were normalized to $\mathcal{N}$, the product of
detector efficiency~$\varepsilon$, total integrated luminosity \(
\Lumn \equiv \sum \Lumn_i \), and peak cross section~$\sigmapk \equiv
\sBW(M)$: \( \mathcal{N} \equiv \varepsilon \Lumn \sigmapk = 10^4 \),
in this example. Statistical uncertainties are affected by how the
total integrated luminosity~$\Lumn$ is spent.  Here we assume that the
optimal distribution is used~\cite{bib:exp:pesce:2007}.  For the
numbers in the table, it is also assumed that the beam width is
negligible, and so they represent lower limits for a given
$\mathcal{N}$.  The uncertainty in the width is the one that is most
affected by a larger beam energy spread. Its dependence on the ratio
between energy spread and resonance width is shown in
\Reffig{fig:exp:dGvsG}.


%% file: panda_pb_soft.tex
%
%
\cleardoublepage
\chapter{Software}
\label{sec:soft}
%
%
\input{./soft/soft_intro}
\input{./soft/soft_evtgen}
\input{./soft/soft_simu}
\input{./soft/soft_reco}
\input{./soft/soft_ana}
\input{./soft/soft_prod}
\input{./soft/soft_pandaroot}
%
%
\newpage
\bibliographystyle{panda_pb_lit}
\bibliography{./soft/lit_soft,./main/lit_main}
%

%% file: soft/soft_intro.tex
%
The offline software which has been devised for the \PANDA Physics Book benchmark studies 
follows an object oriented approach, and most of the code
has been written in C++. Several well-tested software tools and packages from other 
HEP experiments are used and have been adapted to the \PANDA needs. The software contains
\begin{itemize}
\item event generators with accurate decay models for the individual physics channels as
well as for the relevant background channels,
\item particle tracking through the complete \PANDA detector by using the GEANT4 transport
code,
\item the digitization which models the signals of the individual detectors
and their processing in the front-end-electronics,
\item the reconstruction and identification of charged and neutral particles,
providing lists of particle candidates for the physics analysis and
\item user friendly high level analysis tools which allow to make use of vertex
and kinematic fits and to reconstruct extensive decay trees very easily.
\end{itemize}

%

%% file: soft/soft_evtgen.tex
%
\section{Event Generation}
For generating events representing the benchmark reactions, {\tt EvtGen} \cite{bib:soft:evtgen}
was used. It allows to generate the resonances of interest, taking into
account the known decay properties, including angular distributions, polarization,
etc., and allows user defined decay models.

For the simulation of the generic annihilation background, the Dual Parton 
Model based generator, DPM,
was used in the case of \pbarp, and the  Ultra-relativistic Quantum Molecular Dynamic 
model, UrQMD, in case of \pbarN, which are described in more detail in the following.

\subsection{{\tt EvtGen} Generator}
{\tt EvtGen} was first developed within the \INST{BaBar} collaboration, and originally it was designed
for the needs of studies at B-meson factories. The modular design allows an easy extension
to other physics channels, and
meanwhile {\tt EvtGen} has been adapted and used for \INST{ATLAS} \cite{bib:soft:EvtGenLHC}
and \Panda studies, too.

{\tt EvtGen}makes use of the formalism of spin density matrices. This enables the
inclusion of spin effects into the simulation, and allows the user to study 
angular distributions of particles in the final state.

The input data for each decay process are passed to the code as a complex amplitude. In
cases where more than one complex amplitude are involved for the same process, these
are added before the decay probabilities are calculated. Consequently, interference terms,
which are of significant importance in many channels studied with \Panda, are included.

The package also uses a novel {\it{nodal}} decay algorithm, where each decay step is
treated independently, addressing the problem of cascade decays. In a conventional
Monte Carlo generator, kinematics for the whole chain would be generated at once,
and the accept-reject decision applied on the result. This method is inefficient
as a rejection leads to the whole chain being regenerated from scratch. The node-wise
method of {\tt EvtGen} avoids this by generating kinematics for each step separately.
This approach does lead to increasingly complex spin density matrices being
attached to the amplitudes for each node, but the computation time required to
calculate these is very much less than what would be needed to continually re-generate
kinematics for the whole decay tree.

{\tt EvtGen} is controlled by means of a decay table, which lists all possible decay
processes, their branching ratios and the decay model.
A user decay table can be written to override the default table
and, thereby, exclude unwanted processes.

\subsection{Dual Parton Model}
The Dual Parton Model \cite{bib:soft:DPM} is a synthesis of the Regge
theory, topological expansions of QCD $1/N_f$ or $1/N_c$,
and ideas from the parton model. The Regge theory gives the energy
dependence of hadron-hadron cross sections assuming various exchanges
of particles between projectile and target in the $t$-channel. The cross
sections are in a correspondence with diagrams of $1/N$ expansion. The
diagrams describe creation of unstable intermediate $s$-channel states
-- quark-gluon strings or colour tubes.

The main objects of the model are constituent quarks having masses
$\sim\,$300--350$\,\mev$, strings, and string junctions for baryons. It is
assumed that mesons consist of a quark and an antiquark which are
coupled by colour forces. The vortex lines of the field are concentrated
in a small space region forming a string-like configuration. So, mesons
are considered as strings with small masses.

Baryons are assumed to consist of three quarks. Two possible string
configurations in baryons were considered: triangle and Mercedes
star configurations
\cite{bib:soft:Artru,bib:soft:Rossi,bib:soft:Montanet,bib:soft:Takahashi,bib:soft:Bali,bib:soft:Kuzmenko1,bib:soft:Kuzmenko2}.
Three strings are joined in the central point in the Mercedes star case
and give the string junction.

\begin{figure}[cbth]
  \begin{center}
    \includegraphics[width=\swidth]{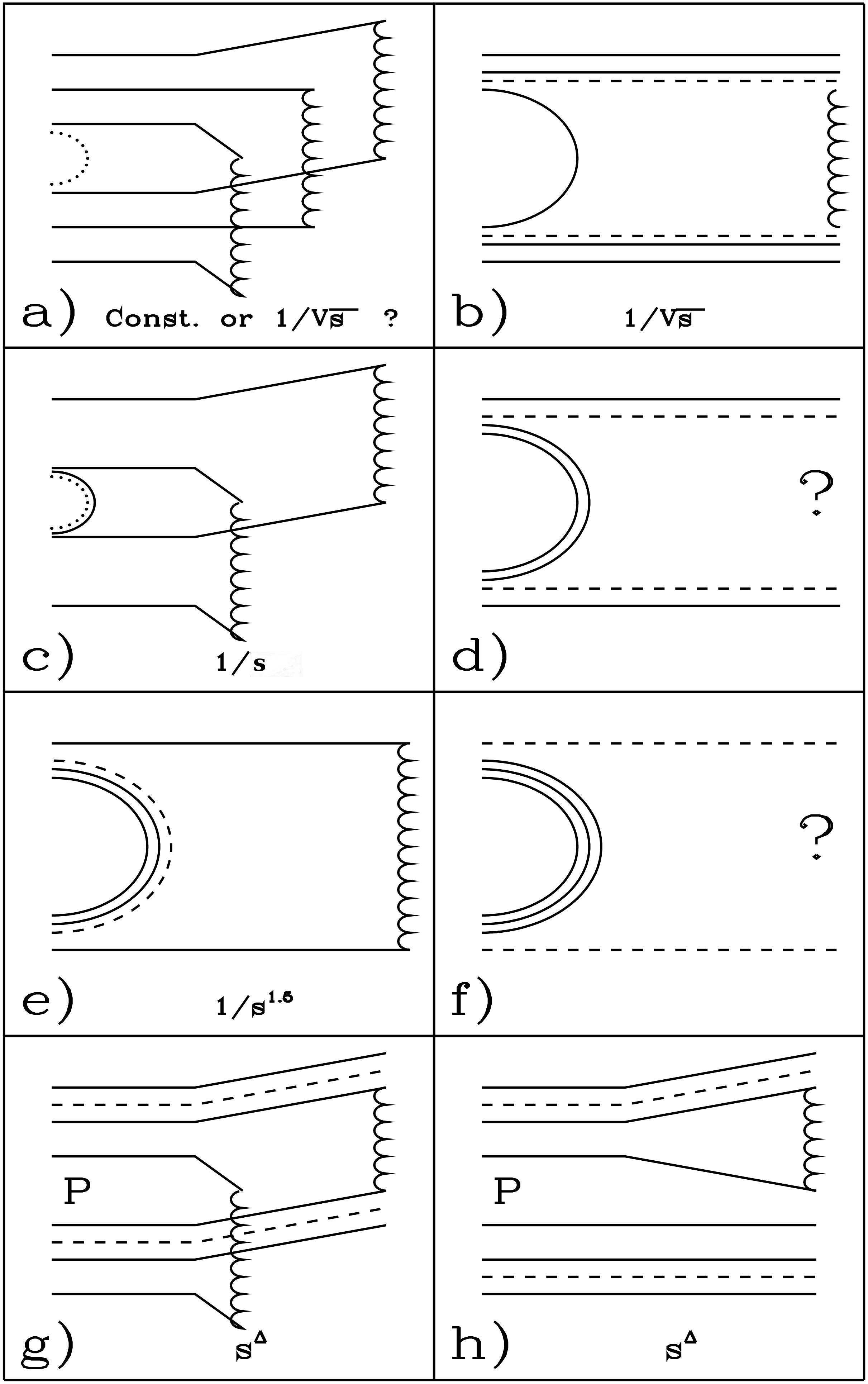}
  \end{center}
\caption[Possible processes in $\pbarp$-interactions and their estimated energy dependencies.]{
The question marks mean that the corresponding estimations are absent.}
\label{fig:soft:dpm_diagrams}
\end{figure}

Various processes are possible in baryon-antibaryon interactions. Some
of them are shown in \Reffig{fig:soft:dpm_diagrams} where string
junctions are presented by dashed lines. The diagram of
\ReffigX{fig:soft:dpm_diagrams}a represents a process with string
junction annihilation and creation of three strings. The diagram
\ref{fig:soft:dpm_diagrams}b describes quark-antiquark annihilation and
string creation between diquark and anti-diquark. Quark-antiquark
and string junctions annihilation is shown in \ReffigX{fig:soft:dpm_diagrams}c. 
Finally, one string is created in the
process of \Reffig{fig:soft:dpm_diagrams}e. After fragmentation of the
strings, hadrons appear in the same way as in $\ee$-annihilation.
One can assume that excited strings with complicated configuration are
created in processes \ref{fig:soft:dpm_diagrams}d and \ref{fig:soft:dpm_diagrams}f.
If the collision energy is sufficiently
small, glueballs can be formed in the process \ref{fig:soft:dpm_diagrams}f.
Mesons with constituent gluons can be created in the process
\ref{fig:soft:dpm_diagrams}d.

The pomeron exchange is responsible for 2 strings formation and
diffraction dissociation (\ReffigX{fig:soft:dpm_diagrams}g and
\ref{fig:soft:dpm_diagrams}h).  They are dominant at high energies.

In the simplest approach it is assumed that the cross sections of the
processes have an energy dependence given in
\Reffig{fig:soft:dpm_diagrams} where $s$ is the square of the total energy
in the centre-of-mass system (\INST{CMS}). 
Some of the processes (\ref{fig:soft:dpm_diagrams}d, 
\ref{fig:soft:dpm_diagrams}f) have not a well-defined energy dependence
of the cross sections. Since it is usually assumed that their cross sections are 
small, we have neglected them.

A calculation of the cross sections is a
rather complex procedure (see \cite{bib:soft:Kaidalov_ZPC,bib:soft:Uzhinsky_Galoyan}) because
there are interactions in initial and final states. 
Here we follow the approach developed in Ref. \cite{bib:soft:Uzhinsky_Galoyan}. 
The cross sections calculated by us have a complicated energy dependence.
To reproduce it we have
parametrised the cross sections of \Reffig{fig:soft:dpm_diagrams} in
the following form:
\begin{eqnarray}
  \sigma_a &=& 51.6/s^{0.5}-58.8/s+16.4/s^{1.5},   \nonumber \\
  \sigma_b &=& 77.4/s^{0.5}-88.2/s+24.6/s^{1.5},   \nonumber \\
  \sigma_c &=& 93/s-106/s^{1.5}+30/s^2,            \nonumber \\
  \sigma_d &=& \sigma_e \, =\, \sigma_f \,= \, 0,  \nonumber \\
  \sigma_g &=& 18.6/s^{0.08}-33.5/s^{0.5}+30.8/s,  \nonumber \\
  \sigma_h &=& 0,                                  \nonumber 
\end{eqnarray}
where only the leading terms correspond to that shown in \Reffig{fig:soft:dpm_diagrams}.
All cross sections are given in mb with $s$ in \gev$^2$.

String masses are determined by kinematic properties of quarks and
antiquarks at their ends. According to Ref. \cite{bib:soft:DPM} we 
assume the following probability distributions, $dW_i$, in the processes 
\ref{fig:soft:dpm_diagrams}a, \ref{fig:soft:dpm_diagrams}c and \ref{fig:soft:dpm_diagrams}g:
\begin{eqnarray}
  dW_a & \propto & x_{+}^{-\alpha_R(0)}dx_+,                                     \nonumber \\
  dW_c & \propto & x_{+}^{-\alpha_R(0)}(1-x_{+})^{-\alpha_R(0)}dx_+,             \nonumber \\
  dW_{g}& \propto &x_{+}^{-\alpha_R(0)}(1-x_{+})^{\alpha_R()-2\alpha_N(0)}dx_+,\nonumber
\end{eqnarray}
where $x_+$ is the light-cone momentum fraction of quarks, and $\alpha_R(0)$
is the intercept of the non-vacuum reggeon trajectory.

The transverse momentum distribution of quarks has been chosen in the form:
\begin{equation}
d^2W = B^2 e^{-B p_T} d^2 p_T, \qquad B=4.5\ (\gevc)^{-1}, \nonumber
\end{equation}
where $p_T$ is the transverse momentum, and $B$ is the adjusted parameter.

The strings fragment into hadrons. The mechanism of the fragmentation
is like the one applied in the LUND
model~\cite{bib:soft:Andersson1,bib:soft:Andersson2}. Here we use a code
proposed by S.~Ritter \cite{bib:soft:Ritter1} with fragmentation
functions (hadron distributions on light-cone momentum) taken from
Refs. \cite{bib:soft:Kaidalov1,bib:soft:Kaidalov2}. Strings with small
masses are considered as hadrons, and we put them on the mass-shell.

After the string fragmentation all unstable hadrons decay. We simulate
the processes with the help of the code DECAY~\cite{bib:soft:Ritter2}.

The corresponding event generator has been developed and tested successfully. It gives
a possibility to simulate the inelastic interactions as well as the elastic
\pbarp-scattering.

\subsection{UrQMD}
The Ultra-relativistic Quantum Molecular Dynamic model (UrQMD)
\cite{bib:soft:UrQMD1,bib:soft:UrQMD2} is a microscopic model based on a
phase space description of nuclear reactions. It describes the
phenomenology of hadronic interactions at low and intermediate
energies ($\sqrt s\,<\,5$\,\gev) in terms of interactions between known
hadrons and their resonances. At higher energies, $\sqrt s\,>\,5$\,\gev,
the excitation of colour strings and their subsequent fragmentation
into hadrons are taken into account in the UrQMD model.

The model is based on the covariant propagation of all hadrons
considered on the (quasi-)particle level on classical trajectories in
combination with stochastic binary scattering, colour string
formation and resonance decay. It represents a Monte Carlo solution of a
large set of coupled integro-differential equations for the time
evolution of the various phase space densities of particle species
$i\,=\,N$, $\Delta$ , $\Lambda$, etc.. The main ingredients of the model
are the cross sections of binary reactions, the two-body potentials and
decay widths of resonances.

In the UrQMD model, the total cross section $\sigma_{\rm tot}$ depends
on the isospins of colliding particles, their flavor and the c.m. energy.
The total and elastic proton-proton and proton-neutron cross sections
are well known \cite{bib:soft:PDG96}. 
Since their functional dependence on $\sqrt{s}$ shows a complicated
shape at low energies, UrQMD uses a lookup-table for those cross
sections. The neutron-neutron cross section is assumed to be equal to the
proton-proton cross section (isospin-symmetry).  In the high energy
limit ($\sqrt{s}\,\ge\,5$~\gev) the \INST{CERN}/\INST{HERA} parametrization for the
proton-proton cross section is used \cite{bib:soft:PDG96}.

Baryon resonances are produced in two different ways, namely:
{\it hard production\/} -- $N+N\,\to\,\Delta \,N$,
$\Delta\Delta$, $N^\ast N$, etc.
and
{\it soft production\/} -- $\pi^-+$p$\,\to\,\Delta^0$,
K$^-$+p$\,\to\,\Lambda^\ast$ $\ldots$

The cross sections of $s$-channel resonances formation are fitted to measured data.
Partial cross sections are used to calculate the relative weights for
the different channels.

There are six channels for the excitation of
non-strange resonances in the UrQMD model, namely $NN \to N
\Delta_{1232}$, $NN^{\ast}$, $N\Delta^{\ast}$, $\Delta_{1232}
\Delta_{1232}$, $\Delta_{1232}N^{\ast}$,  and $\Delta_{1232}
\Delta^{\ast}$.
The $\Delta_{1232}$ is explicitly listed,
whereas higher excitations of the $\Delta$ resonance
have been denoted as $\Delta^{\ast}$.
For each of these 6 channels specific assumptions have been made
with respect to the form of the matrix element, and the free
parameters have been adjusted to the available experimental data.

Meson-baryon  (MB) cross sections are dominated by the formation of 
$s$-channel
resonances, {\it i.e.}\ the formation of a transient state of mass
$m\,=\,\sqrt{s}$, containing the total c.m. energy of the two incoming
hadrons. On the quark level such a process implies that a quark from
the baryon annihilates an antiquark from the incoming meson. At c.m. energies below 2.2\,\gev,
intermediate resonance states get excited. At higher energies
the quark-antiquark annihilation processes become less important. There,
$t$-channel excitations of the hadrons dominate, where the exchange of
mesons and Pomeron exchange determine the total cross section of the
MB interaction \cite{bib:soft:donna}.

To describe the total meson-meson  (MM) reaction cross sections,
the additive quark model and the principle of detailed
balance, which assumes the reversibility of the particle interactions,
are used.

Resonance formation cross sections from the measured decay properties
of the possible resonances up to c.m. energies of 2.25~\gev for
baryon resonance and 1.7~\gev in the case of MM and MB reactions
have been calculated based on the principle. Above these energies
collisions are modelled by the formation of an $s$-channel string or, at
higher energies (beginning at $\sqrt s\,=\,3$\,\gev), by one or two
$t$-channel strings. In the strangeness channel elastic collisions are
possible for those meson-baryon combinations which can not
form a resonance, while the creation of $t$-channel strings is always
possible at sufficiently large energies. At high collision energies
both cross sections become equal due to quark counting rules.

A parametrization proposed by Koch and Dover \cite{bib:soft:kochP89a}
is used in the UrQMD model for the baryon-antibaryon annihilation
cross section.  It is assumed that the antiproton-neutron annihilation
cross section is identical to the antiproton-proton annihilation
cross section.

The potential interaction is based on a non-relativistic density-dependent
Skyrme-type equation of state with additional Yukawa- and Coulomb potentials.
The Skyrme potential consists of a
sum of two- and a three-body interaction terms. The two-body
term, which has a linear density dependence models the long range attractive
component of the nucleon-nucleon interaction, whereas the three-body
term with its quadratic density dependence is responsible for the
short range repulsive part of the interaction. The parameters of the
components are connected with the nuclear equation of state. Only the
hard equation of state has been implemented into the current
UrQMD model.

At the beginning of an event generation a target nucleus is modelled 
according to the Fermi-gas ansatz.
The wave-function of the nucleus is defined as the product
of nucleon wave-functions. A nucleon wave-function is represented by the Gauss function 
in the configuration and momentum space. In configuration space, the centroids
of the Gaussians are randomly distributed within a sphere with radius $R(A)$,
where $R(A)$ is the nucleus radius.
The initial momenta of the nucleons are randomly chosen between 0 and
the local Thomas-Fermi momentum.

The impact parameter of a collision is sampled according to the
quadratic measure ($dW\,\sim\,bdb$). At given impact parameter, $b$, the
centres of projectile and target are placed along the collision axis
in such a manner that the distance between the surfaces of the projectile
and the target is equal to 3$\,\fm$. Momenta of nucleons are
transformed in the system where the projectile and target have equal
velocities directed in different directions of the axis. After that
the time propagation starts.  During the calculation, 
at the beginning of each time step each particle is
checked  whether it will collide within that time step.  
A collision between two hadrons will occur if
$d<\sqrt{\sigma_{\rm tot}/\pi}$, where $d$ and $\sigma_{\rm tot}$ are
the impact parameter and the total cross section of the
two hadrons, respectively. After each binary collision or decay the
outgoing particles are checked for further collisions within the
respective time step.

The hadron-hadron interactions at high energies are simulated in 3 stages.
According to the cross sections, probabilities of interaction are defined
(elastic, inelastic, antibaryon-baryon annihilation etc.), and a type of 
interaction is sampled. 
In the case of
inelastic collisions with string excitation, the kinematic characteristics
of strings are determined. The strings between quark and diquark
(antiquark) from the same hadron are produced. The strings have the continuous
mass distribution $f(M)\,\propto\,1/M$ with the masses $M$, limited by
the total collision energy $\sqrt{s}$: $M_1+M_2\,\le\,\sqrt{s}$.
The remaining energy is equally distributed
between the longitudinal momenta of two produced strings.

The second stage of hadron-hadron interactions is related to string fragmentation. 
The fragmentation functions used in the UrQMD model are different from the ones 
in the well-known LUND model \cite{bib:soft:andersson83a}.

The formation time of created hadrons is taken into account in 
hadron-nucleus and nucleus-nucleus interactions.

The decay of the resonances proceeds according to the
branching ratios compiled by the Particle Data Group \cite{PDBook}.
The resonance decay products have isotropic distributions in the rest
frame of the resonance. If a resonance is among the outgoing particles,
its mass must first be determined  according to a Breit-Wigner
mass-distribution. If the resonance decays into $N\,>\,2$ particles,
then the corresponding $N$-body phase space is used to calculate the 
momenta of the final particles.

The Pauli principle is applied to hadronic collisions or decays by
blocking the final state if the outgoing phase space is occupied.

The final state of a baryon-antibaryon annihilation is generated
via the formation of two meson-strings. The available c.m. energy
of the reaction is distributed in equal parts to the two strings
which decay in the rest frame of the reaction.
On the quark level this procedure implies the annihilation of a
quark-antiquark pair and the reordering of the remaining
constituent quarks into newly produced hadrons (additionally taking
sea-quarks into account). This model for the baryon-antibaryon
annihilation thus follows the topology of a rearrangement graph.

The collision term in the UrQMD model contains 55 different baryon species
(including nucleon, delta and hyperon resonances with masses up
to 2.25\,\gev) and 32 different meson species (including strange
meson resonances with masses up to 1.9\,\gev), which are supplemented 
by their corresponding
antiparticle and all isospin-projected states. The states can either
be produced in string decays, s-channel collisions or resonance decays.
For excitations with masses larger than 2\,\gev, a string picture is
used. Full baryon/antibaryon symmetry is included:
the number of implemented baryons therefore defines the number
of antibaryons in the model, and the antibaryon-antibaryon interaction
is defined via the baryon-baryon interaction cross sections.

A very important improvement of the UrQMD model was proposed in
Ref.~\cite{bib:soft:Khaled} where the model was coupled with the
Statistical Multi-fragmentation (SM) Model~\cite{bib:soft:Botvina}.
According to Ref.~\cite{bib:soft:Khaled}, the UrQMD calculation is
carried out up to a time scale referred to as the transition time
$t_{tr}\,\sim\,100\fm/c$.  The positions of the nucleons are then used to
calculate the distribution of mass and charge numbers of
pre-fragments. In determining the mass and charge numbers of the
pre-fragments, the minimum spanning tree method \cite{bib:soft:Cluster}
is employed. A pre-fragment is formed if the distances between nucleons are
lower than $3\,\fm$. The total energy of each
pre-fragment is determined in its rest frame by the Lorentz
transformation. The excitation energy of a hot
pre-fragment is calculated as the difference between the binding
energy of the hot pre-fragment and the binding energies of this
pre-fragment in the ground state. The decay of the pre-fragments 
is described by the SM model. 

We use the combination of the models for estimation of neutron production
in $\pbar$-interactions with nuclear targets. 

%
%

%% file: soft/soft_simu.tex
%
\section{Particle Tracking and Detector Simulation}
The detector simulation is subdivided into two steps. The first step is the propagation
of the generated particles through the \Panda detector by using the GEANT4 transport code
\cite{bib:soft:Allison, bib:soft:GEANT4}. It
takes into account the full variety of interactions and decays that the different kinds of particles 
may undergo. The output of this first step is a collection of hits, which contain mainly the 
intersection points and energy losses of all particles in the individual detector parts.    
Based on this information the digitization step follows, which models the signals and their processing 
in the front-end-electronics of the individual detectors, producing a digitised detector response as 
similar as possible to beam data. This design ensures that in the future the {\it{same}} reconstruction
code can be used for Monte Carlo and for beam data. For performance reasons for some detectors an 
effective smearing was used, which was derived from Monte Carlo calculations using a full digitization. 

\subsection{Detector Setup}
The geometry description of the \Panda detector is based on the detector description database 
developed by the \INST{CMS} experiment \cite{bib:soft:DDDcms1}. It provides interfaces between the geometry 
information 
stored in Extensible Markup Language (XML) files and the corresponding 
transient objects of the individual applications. The used XML schema consists of type-safe XML 
constructs, abstract types, and hierarchical inheritance structures. This
allows an easy integration into a C++ environment. 

Tools have been developed which are able to convert technical drawings created by widely used 
Computer Aided Design (CAD) programs directly into the XML based detector description. This
makes it very easy to add and remove geometries or to update modifications of specific detector parts 
in the simulation.   

The simulations have been done with the complete setup which was already described in detail 
in \Refsec{exp:detector}. The still not finally established Time-Of-Flight (TOF) and the Forward RICH 
detectors have not been considered, and the Straw Tube option has been used for the central tracker 
device. The Muon detector
consists only of two scintillator layers instead of a multilayer scenario within the iron yoke
which was the most favoured option during the implementation phase of the geometry into 
the software. For the \pbarp reactions the pellet scenario as an internal target has
been chosen taking into account the material budget of the target pipe and the pumping stations.
The interaction point has been considered with a spread of $\sigma$=0.275\,mm in each direction. 

\begin{figure}[htb]
\begin{center}
\includegraphics[width=\swidth]{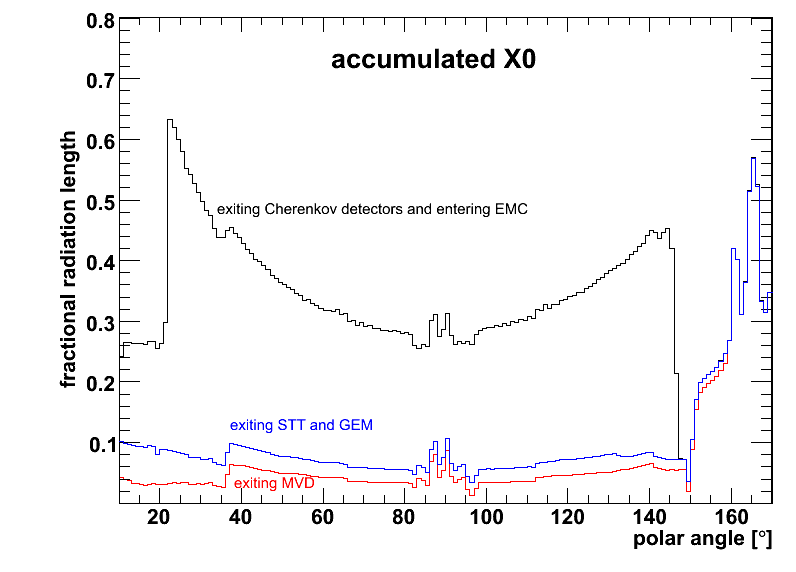}
\caption[Material budget in front of the \Emc]
{Contributions of the subdetectors to the material budget in front of the \Emc
in units of a radiation length X$_0$ as a function of the polar angle $\theta$.}
\label{fig:soft:x0}
\end{center}
\end{figure}

\Reffig{fig:soft:x0} shows the contributions of the subdetectors to the material which a particle
has to traverse to reach the \Emc as a function of $\theta$. The material thicknesses are presented in
units of the corresponding radiation lengths X$_0$.

\subsection{Digitization}

\subsubsection{Readout of the Tracking Devices}

\paragraph{Silicon Readout of the MVD}

The Micro Vertex Detector (MVD) makes use of two different silicon detector types, 
silicon strip and pixel detectors.
The readout of the silicon devices is for both types different and is treated differently in
the digitization scheme. 
The signal in the sensor is formed by using the local trajectory within the detector material
to calculate the corresponding channel relative to the readout matrix of the sensor. 
The hit position on the sensor surface defines the channel number and the deposited 
energy the charge collected by the electronics. The channel mapping of the trajectory 
is done on both sides of the sensor. In the case of the pixel detectors, the trajectory is 
projected to the surface and 
depending on its relative orientation, all excited pixel cells are calculated and the charge 
signal is shared among all pixel cells depending on the fraction of the local track.  
Strip sensors will be sensitive on both sides and the formation of digitised channels is done
independently on both sides of the sensor. The procedure is similar to the pixel case 
but done only in one dimension.

The size of the readout structures are defined by the size of the pixel cell or the spacing of 
the strips. These parameters can be changed interactively in order to test various settings 
and have to meet the dimensions of the sensor. The channel is assigned to a frontend and a
common number of 128 channels per frontend have been chosen.
In the case of the pixel detector the size of the \INST{ATLAS} frontend chip was used as basis to 
assign a frontend number to a certain channel. Since the electronics chip is bump bonded onto 
the surface of the detector the number of pixel cells per frontend is defined by the dimension
of the frontend chip. A threshold for the electronics signal can be set and was chosen 
as standard to an equivalent of 300 electrons which is a reasonable value for pixel detectors.

\paragraph{Straw Tube Tracker and Drift Chambers}
The digitization for the Straw Tube Tracker (STT) and the Drift Chambers (DCH) have been treated in a 
similar way. Both devices consist of wires inside an ArCO$_{2}$ gas mixture volume. 
If a charged particle traverses this gas volume, the local helix trajectory is derived from the
corresponding GEANT4 intersection points. The drift time of the ionization electrons is estimated
from the smallest distance of this helix trajectory to the wire $d_{poca}$. The uncertainty of 
the drift time is taken into account by smearing $d_{poca}$ with a Gaussian distribution 
with a standard deviation of 
$\sigma = 150 \, \mum$ for the STT, and $\sigma = 200 \, \mum$ for the DCH devices. 

The average number of primary ionization electrons is the total deposited 
energy in the gas  volume divided by the ionization energy of 27 eV for ArCO$_{2}$. The 
energy signal of a straw tube is finally calculated by taking into account
Poisson statistics.
 
\paragraph{GEMs Readout}
Each GEM station consists of two detector planes.
The distance between the detector planes is 1 cm.
It has been assumed that each detector plane has two strip detector layers with
perpendicular orientation to each other. The gas amplification process
and the response of the strip detector has not been simulated in detail.
Instead, the entry point of a charged track into the detector plane 
has been taken directly from GEANT4 and smeared with a Gaussian distribution of $70\, \mum$ 
width in each strip orientation direction.

\subsubsection{Readout of the DIRC Detectors}
The light propagation in the Cerenkov radiators, the signal processing in the 
front-end-electronics, and the reconstruction of the Cerenkov angle have been 
modelled in a single effective step.

\begin{figure}[htb]
\begin{center}
\includegraphics[width=\swidth]{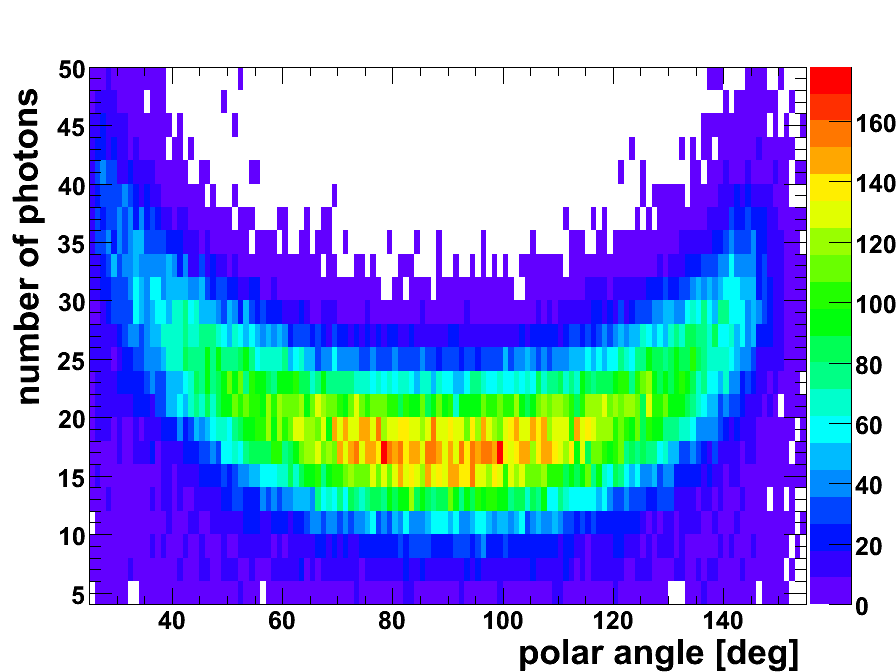}
\caption[The number of detected Cerenkov photons versus the polar angle of pions]
{The number of detected Cerenkov photons versus the polar angle of pions with
momenta of 1\,\gevc.}
\label{fig:soft:nCherPhotons}
\end{center}
\end{figure}

The resolution of the reconstructed Cerenkov angle $\sigma_{C}$ is mainly driven 
by the uncertainty of the single photon angle $\sigma_{C,\gamma}$ and the statistics
of relatively small numbers of detected Cerenkov photons $N_{ph}$:
\begin{equation}
\sigma_{C} = \frac{\sigma_{C,\gamma}}{\sqrt{N_{ph}}} \nonumber
\end{equation}     
A single photon resolution of $\sigma_{C,\gamma}\,=\,10$~mrad was used, corresponding to
the experience with existing DIRC detectors. The number of detected photons was
calculated from the velocity, $\beta$, of charged particles passing through the quartz radiators and 
the path length $L$ within the radiator, via:
\begin{equation}
N_{ph} = \epsilon\,2\pi\alpha\,L\,(1-\frac{1}{\beta^2 n^2_{quartz}})
        ( \frac{1}{\lambda_{min}}-\frac{1}{\lambda_{max}})\,, \nonumber
\end{equation}
where $\alpha$ is the fine structure constant and $n_{quartz}=1.473$ the refraction index of
quartz. The sensitive wavelength interval, $\lambda_{min}$--$\lambda_{max}$, 
was chosen to $[280\,\nm\, , \, 350\,\nm]$,
and a total efficiency of $\epsilon=$~7.5\percent was used to take into account the transmission and 
reflectivity losses as well as the quantum efficiency of the photo detectors. 
\Reffig{fig:soft:nCherPhotons} shows the simulated number of photons for the Barrel DIRC 
as a function of the polar angle for 1\,\gevc pions. The average number of detected photons
at a polar angle of 90$^\circ$, 20, increases by a factor of 2 for very forward and backward 
directions. 
    
\begin{figure}[htb]
\begin{center}
\includegraphics[width=\swidth]{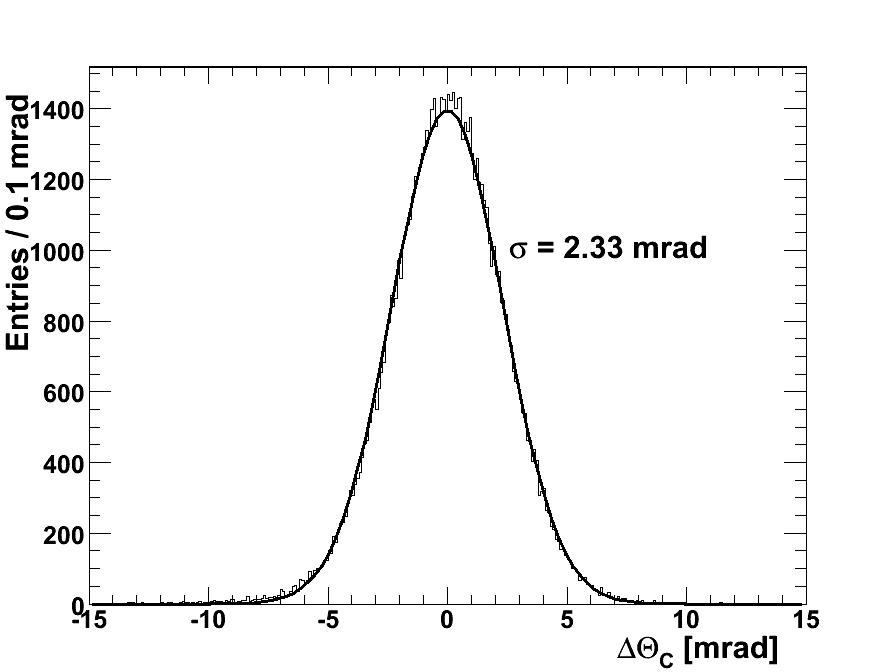}
\caption[Difference between the reconstructed and the expected Cerenkov angle]
{Difference between the reconstructed and expected Cerenkov angle $\Delta \Theta_C$ for single 1\,\gevc 
pions.}
\label{fig:soft:dThetaC}
\end{center}
\end{figure}

\Reffig{fig:soft:dThetaC} shows the precision of the measured Cerenkov angle obtained
with the digitization and reconstruction procedure described above. A fit with a 
Gaussian distribution yields to a resolution of $\sigma\,=\,2.33\,\mrad$. 

\subsubsection{\Emc Scintillator Readout}
\label{sec:soft:emcDigi}
For the \Emc in the target spectrometer (TS \Emc) reasonable properties of PbWO$_4$ crystals at the 
operational temperature of -25$\degC$ have been considered. 
A Gaussian distribution with a $\sigma$ of 1$\,\mev$ has been used for the constant electronics 
noise. The statistical fluctuations were estimated by 80 photo electrons per \mev produced in the 
Large Area Avalanche Photo Diode (\LAAPD).
An excess noise factor of 1.38 has been used, corresponding to the measurements with
the first \LAAPD prototype at an internal gain of $M=50$ (see \cite{bib:soft:PandaEmcTdr}). 
This results in a photo statistics noise term of 0.41\percent/$\sqrt{E / \gev}$.

For the forward calorimeter (FS \Emc) a Shashlyk detector consisting of 
lead-scintillator sandwiches is foreseen. Therefore, only a fraction of roughly 30\percent of the energy
are deposited in the scintillator material. Based on this energy deposit the electronics noise
with $\sigma\,=\,3\,\mev$ has been considered which yields to a statistic noise term of
0.8\percent/$\sqrt{E / \gev}$.

\subsubsection{Muon Detector Readout}
\label{sec:soft:muoDigi}
Because the layout of the muon detector and its readout is still under investigation,
a parametrised digitization was used. The intersections of the particle tracks with the
scintillators of the muon detectors were derived from the detector hits provided by the
GEANT4 transport code. In the process of matching muon detector hits with reconstructed
charged tracks, the uncertainties are dominated by the errors from the extrapolation of
the reconstructed tracks to the muon detector, and therefore the finite position resolution
of the muon detector was neglected. The algorithm that was used to deduce PID probabilities 
from the muon detector hits will be described in \Refsec{sec:soft:muoPID}.

%


%% file: soft/soft_reco.tex
%
%
\section{Reconstruction}

\subsection{Charged Particle Track Reconstruction}

\subsubsection{MVD Cluster Reconstruction}

\begin{figure}[htb]
\begin{center}
\includegraphics[width=\swidth]{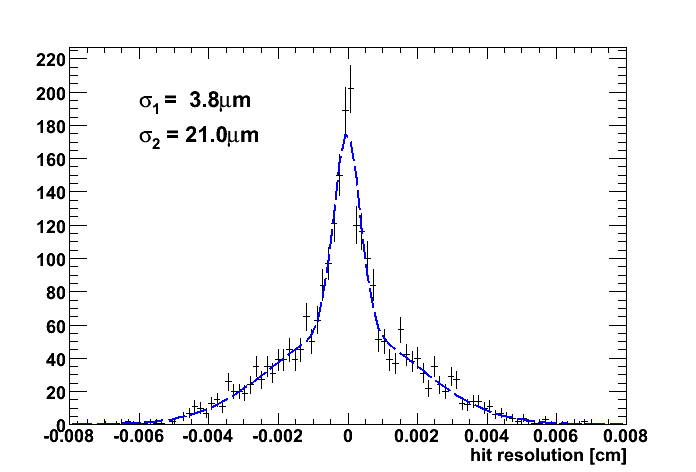}
\caption[Pixel hit resolution on sensor coordinates]
{The resolution of the reconstructed hit position after clustering with respect to the simulated value.}
\label{fig:soft:mvdHitReco}
\end{center}
\end{figure}

The MVD provides very precise space point measurements as a basis for the track and vertex
reconstruction. The hit resolution of individual MVD measurements is shown in 
\Reffig{fig:soft:mvdHitReco} in the case of the pixel detector. 

The distribution in \Reffig{fig:soft:mvdHitReco} shows the difference between the reconstructed
position on the sensor and the generated Monte Carlo value. The broad contribution comes
from hits where only one pixel is contributing to the hit cluster and the narrow contribution
from multi hit clusters. In the latter case, charge weighting between the pixel cells in the 
cluster can be used to calculate the mean position of the hit. Without using the energy 
information, the resolution of the position measurement would be 
\begin{equation*}
      \sigma_{geom} = p/\sqrt{12}
\end{equation*}
where $p$ is the size of the readout structure. Using $100\times100\,\mu$m large pixel 
cells, the geometric resolution is $\sigma_{geom}=28\,\mu$m, which is the uncertainty of the 
broader distribution caused by single pixel clusters.
For high momentum particles the hit resolution limits the overall track resolution whereas for
low momentum particles the small angle scattering is the limiting factor.

\subsubsection{Global Track Reconstruction}
\label{sec:soft:globTracking}
The track object provides information about a charged particle path through space.
It contains a collection of hits in the individual
tracking subdetectors.
Each hit knows about the residual of the hit to a given reference trajectory and
the precision of the measurement. For example, the hit residual of the STT is
defined as the closest approach of the reference trajectory to the wire of
the straw minus the actual drift distance. In case of hits in the MVD or in a 
GEM station, the residual is defined as the distance between 
the sensor and the reference track in the detector plane.
An idealised pattern recognition has been used for track building based on 
Monte Carlo information to assign reconstructed hits to their original tracks.
Tracks which contain less than eight detector hits were rejected.

The tracks in the target spectrometer are fitted with the Kalman Filter algorithm,
which considers not only the measurements and their corresponding resolutions but also the
effect of the interaction with the detector material, {\it i.e.} multiple scattering and energy
loss.
A detailed description of the implementation of the algorithm, which has been adapted from
the reconstruction software of the \INST{BaBar} Collaboration can be found in \cite{bib:soft:BaBarTracking}.
For simplification a constant magnetic field parallel to the z-axis has been assumed in 
the target spectrometer region. 
A typical choice for the parametrization of the track in a solenoidal field is a
five-parameter helix along the principle field direction z.
The following parameter vector P is used:
\begin{equation*}
   P=(d_0, \phi_{0}, \omega, z_{0}, \tan\lambda)
\end{equation*}
where $d_{0}$ is the distance of closest approach to the origin in the x-y plane,
signed by the angular momentum $\vec{r} \times \vec{p}$ at that point, 
$\phi_{0}$ is the angle in the x-y plane at
closest approach, and $z_{0}$ the distance of closest approach to the origin in the z projection.
The parameter $\omega$ gives the curvature of the track in the x-y plane, and $tan \lambda$
is the tangent of the track dip angle in the projection plane defined by the cylindrical coordinates 
$\rho$ and $z$.
The position of the particle as a function of the x-y plane projection
of the flight length from the point of closest approach $l$ is given by:
\begin{eqnarray}
      x &=&   \sin(\phi_{0}+\omega \cdot l)/\omega - (1/\omega + d_{0}) \sin \phi_{0} \nonumber \\
      y &=&  -\cos(\phi_{0}+\omega \cdot l)/\omega - (1/\omega + d_{0}) \cos \phi_{0} \nonumber \\
      z &=&  z_{0}+l \cdot \tan \lambda \nonumber 
\end{eqnarray}
The momentum of the track for a given magnetic field (0,0,B) is then:
\begin{eqnarray}
     p_{x} &=& q \cdot c \cdot B/\omega \cdot \cos(\phi_{0}+\omega \cdot l)  \nonumber \\
     p_{y} &=& q \cdot c \cdot B/\omega \cdot \sin(\phi_{0}+\omega \cdot l)  \nonumber \\
     p_{z} &=& q \cdot c \cdot B/\omega \cdot \tan \lambda  \nonumber 
\end{eqnarray}
The task of the track-fitting algorithm is to determine the optimal parameter vector and its
covariance matrix as a function of the flight length $l$ in order to create a
representation of the track as a piecewise helix. In the physics analysis
the particle position and momentum can then be accessed through this piecewise
helix representation.

Tracks which have hits in the target spectrometer as well as
in the drift chambers of the forward spectrometer are treated in the following way.
From the hit residuals of the drift chambers a $\chi^{2}$ is calculated, where the
propagation of the track through space is done with a Runge-Kutta integration method.
For minimising the $\chi^{2}$ the package MINUIT~\cite{bib:soft:Minuit} is used. 
The result of the fit
is a five-parameter helix and its covariant matrix at $z = 1.9 m$.
This information serves as a constraint for the Kalman Filter fit,
where also the hits in the target spectrometer are considered.

Multiple scattering and energy loss effects depend on the mass 
of the particle. Therefore, the tracks have been refitted separately for all five particle hypotheses 
(e, $\mu$, $\pi$, K and p). This results in an adequate accuracy of the reconstructed 
kinematics assigned to the individual particle species.

\subsubsection{Tracking Performance}
In the following the track reconstruction efficiency, the momentum resolution, and the 
spatial resolution for reconstructed vertexes are shown, which were achieved for the chosen
detector setup using the reconstruction software described above.

\begin{figure}[htb]
\begin{center}
\includegraphics[width=\swidth]{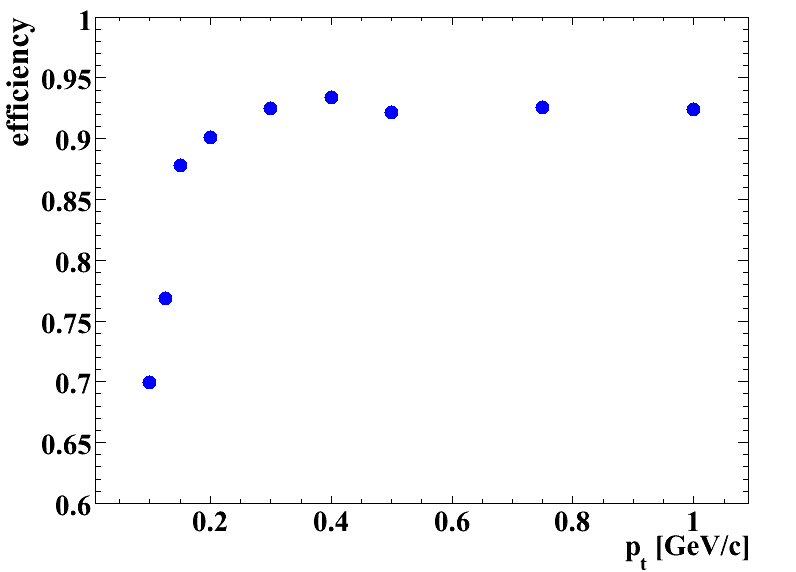}
\includegraphics[width=\swidth]{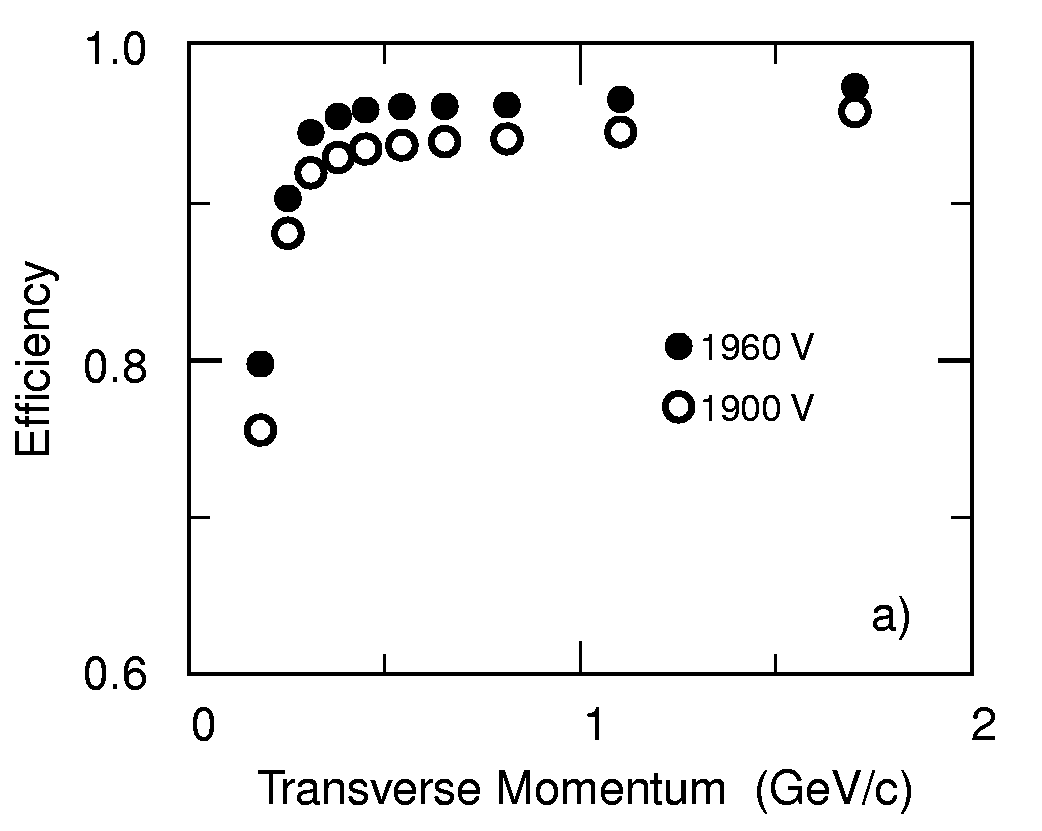}
\caption[Tracking reconstruction efficiency]
{The track reconstruction efficiency as a function of the transverse momentum for pions at a 
polar angle of 60$^\circ$ (upper histogram).\\
The lower histogram was taken from \cite{bib:soft:BaBarDetector} and illustrates the track reconstruction 
efficiency achieved at \INST{BaBar} for two different operation modes of the central drift chamber.  
}
\label{fig:soft:trkEff}
\end{center}
\end{figure}

The upper histogram of \Reffig{fig:soft:trkEff} shows the obtained track efficiency as a function of 
the transverse 
momentum for single pions generated at a polar angle of 60$^\circ$. The efficiency is here defined as the ratio 
between the number of tracks which fulfil only a low criterion on the accuracy of the reconstructed transverse 
momentum to the number of generated tracks. It is required that the difference between the reconstructed to the 
generated momentum is less than 3~$\sigma$ of its resolution. While above p$_t>0.2$~\gevc more than 90\percent
of the tracks are well measured, the efficiency decreases to 70\percent for a transverse momentum of 0.1~\gevc.
These results are comparable to the track efficiency for real data obtained with the \INST{BaBar} detector
(\Reffig{fig:soft:trkEff} and \cite{bib:soft:BaBarDetector}).

\begin{figure}[htb]
\begin{center}
\includegraphics[width=\swidth]{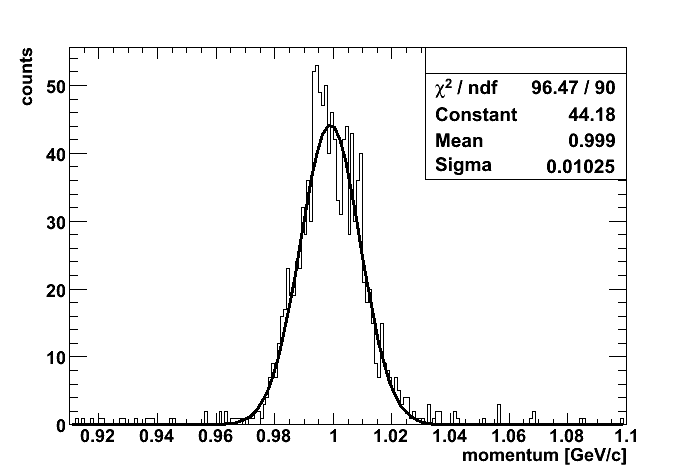}
\caption[The momentum resolution for pions momentum of 1~\gevc momentum at a polar angle of $20^\circ$]
{The momentum resolution for pions of 1~\gevc momentum at a polar angle of $20^\circ$.}
\label{fig:soft:piRes1gev}
\end{center}
\end{figure}

\Reffig{fig:soft:piRes1gev} illustrates the reconstructed momentum for pions
of 1~\gevc momentum at a polar angle of $20^\circ$. This example 
yields to a momentum resolution of $\sigma_p$/$p$~=~1\percent.

\begin{figure}[htb]
\begin{center}
\includegraphics[width=0.9\swidth]{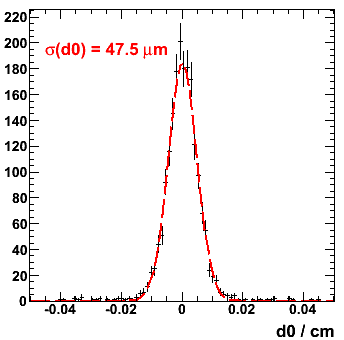}
\includegraphics[width=0.9\swidth]{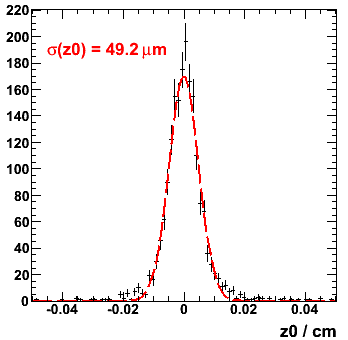}
\caption[The vertex resolution for pions of $p=3$\,\gevc momentum.]
{The achieved vertex resolution for pions of $3$\,\gevc momentum.}
\label{fig:soft:trkVertex}
\end{center}
\end{figure}

The achieved vertex resolution is shown in \Reffig{fig:soft:trkVertex} for 
pions of 3\,\gevc momentum. Here all tracking detectors were taken into account.

\subsection{Photon Reconstruction}
\subsubsection{Reconstruction Algorithm} 
\label{sec:soft:recoalgo}
A photon entering one scintillator module of the EMC develops an electromagnetic shower which,
in general, extends over several modules. A contiguous area of such modules is called a cluster.

The energy deposits and the positions of all scintillator modules in a cluster allow a
determination of the four vector of the initial photon. Most of the EMC reconstruction
code used in the offline software is based on the cluster finding and bump-splitting
algorithms which were developed and successfully applied by the \INST{BaBar} experiment
\cite{bib:soft:BaBarDetector, bib:soft:Strother}.

The first step of the cluster reconstruction is the finding of a contiguous area of
scintillator modules with energy deposit. The algorithm starts at the module exhibiting
the largest energy deposit.
Its neighbours are then added to the list of modules  if the
energy deposit is above a certain threshold $E_{xtl}$. The same procedure is continued on
the neighbours of newly added modules  until no module fulfils the threshold criterion.
Finally a cluster gets accepted if the total energy deposit in the contiguous area is above
a second threshold $E_{cl}$.

The next step is the search for bumps within each reconstructed cluster. 
A cluster can be formed by more than one particle if the angular
distances of the particles are small. In this case the cluster has to be subdivided
into regions which can be associated with the individual particles. This procedure
is called the {\it{bump splitting}}. A bump is defined by a local maximum inside the cluster:
The energy deposit of one scintillator module $E_{\mbox{local}}$ must be above $E_{max}$, while 
all neighbour
modules have smaller energies. In addition the highest energy $E_{\mbox{Nmax}}$ of any of the
$N$ neighbouring modules must fulfil the following requirement:
\begin{eqnarray}
 0.5\,(N-2.5) \, > \, E_{\mbox{Nmax}} \, / \,  E_{\mbox{local}}
\end{eqnarray}
The total cluster energy is then shared between the bumps, taking into account the shower shape of the
cluster. For this step an iterative algorithm is used, which assigns a weight $w_i$
to each scintillator module, so that the bump energy is defined as
$E_b= \sum_{i} \, w_i \, E_i$. $E_i$ represents the energy deposit in the iTH module
and the sum runs over all modules within the cluster.
The module weight for each bump is calculated by
\begin{eqnarray}
w_i = \frac{E_i \, exp(-2.5 \, r_i\, / \,r_m)}
            {\sum_{j} E_j \, exp(-2.5 \, r_j\, / \,r_m)}
\end{eqnarray}
with
\begin{itemize}
\item $r_m$ = Moli\`ere radius of the scintillator material,
\item  $r_i$,$r_j$ = distance of the iTH and jTH module to the centre of the bump, respectively, and
\item  index $j$ runs over all modules.
\end{itemize}
The procedure is iterated until convergence. The centre position is always determined from the weights
of the previous iteration and convergence is reached when the bump centre stays stable within a tolerance 
of 1$\,\mm$.

The spatial position of a bump is calculated via a centre-of-gravity method. 
The radial energy distribution, originating from a photon, decreases mainly exponentially.
Therefore, a logarithmic weighting with
\begin{eqnarray}
  W_i \, = \, max(0, A(E_b) \, + \, \ln(E_i / E_b))
\end{eqnarray}
 was chosen, where only modules with positive weights are used. The energy dependent factor
$A(E_b)$ varies between 2.1 for the lowest and 3.6 for the highest photon energies.

\subsubsection{Reconstruction Thresholds}
The optimal choice for the three photon reconstruction thresholds \--- as already explained in 
\ref{sec:soft:recoalgo} \--- depends strongly on the light yield of the scintillator material and the 
electronics noise. To detect low energetic photons and to 
achieve a good energy resolution, the thresholds should be set as low as possible. On the other 
hand, the thresholds must be sufficiently high for a  suppression of misleadingly reconstructed 
photons originating from the noise of the readout and from statistical fluctuations of the 
electromagnetic showers. 
The single crystal threshold was set to 3\,\mev of deposited energy, corresponding to the energy equivalent of 3 
$\sigma$ of the electronics noise (see \Refsec{sec:soft:emcDigi}). All reconstruction
thresholds for the TS EMC as well as for the FW EMC are listed in \Reftbl{tab:soft:emc_reco_thresh}.

\begin{table}
\begin{center}
\begin{tabular}{lcc}
  \hline\hline
           & TS EMC & FW EMC\\
    \hline
   $E_{xtl}$ & 3\,$\mev$   & 8\,$\mev$ \\
   $E_{cl}$ & 10\,$\mev$   & 15\,$\mev$\\
   $E_{max}$ & 20\,$\mev$  & 10\,$\mev$  \\
  \hline\hline
\end{tabular}
\caption[Reconstruction thresholds for the PbWO$_4$ and Shashlyk calorimeter]{Reconstruction 
thresholds for the PbWO$_4$ and Shashlyk calorimeter.}
\label{tab:soft:emc_reco_thresh}
\end{center}
\end{table}

\subsubsection{Leakage Corrections}
The sum of the energy deposited in the scintillator material of the calorimeters is in general less
than the energy of the incident photon. While only a few percent is lost in the TS EMC, which 
mainly originates from energy losses in the material between the individual crystals, a fraction of 
roughly 70\percent of the energy is deposited in the absorber material for the Shashlyk calorimeter.   

The reconstructed energy of the photon in the TS EMC is expressed as a product
of the measured total energy deposit and a correction function which depends 
logarithmically on the energy and \--- due to the layout \--- also on 
the polar angle. Monte Carlo simulations using single photons have been carried out to determine the
corrected photon energy $E_{\gamma,cor} = E * f(ln E,\theta)$ with the correction function
\begin{eqnarray}
f(ln E,\theta) = exp( a_0 + a_1\,ln E + a_2 \,ln^2 E + a_3 \,ln^3 E \nonumber\\
                   + a_4 \, \cos(\theta)+ a_5 \, \cos^2(\theta) + a_6 \, \cos^3(\theta) \nonumber\\
                  + a_7 \, \cos^4(\theta) + a_8 \, \cos^5(\theta) \nonumber\\
                  + a_9 \, ln E \, \cos(\theta) ) \nonumber 
\end{eqnarray} 

\begin{figure}[htb]
\begin{center}
\includegraphics[width=\swidth]{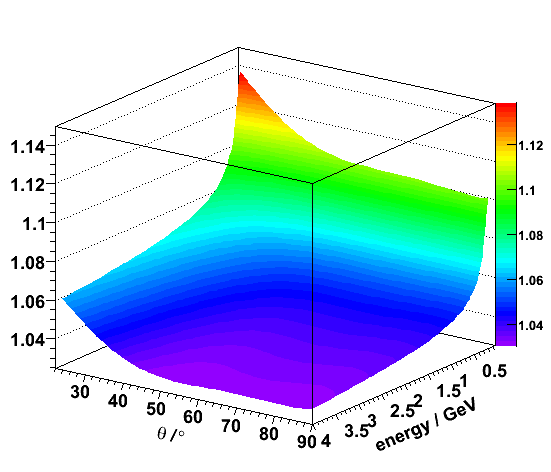}
\caption[Leakage correction function for the barrel EMC in the $\theta$ range 
between 22$^\circ$ and 90$^\circ$ ]
{The leakage correction function for the barrel EMC in the $\theta$ range 
between 22$^\circ$ and 90$^\circ$.}
\label{fig:soft:leakCorr}
\end{center}
\end{figure}

\Reffig{fig:soft:leakCorr} shows the result for the barrel part in the $\theta$ range 
between 22$^\circ$ and 90$^\circ$.  

\begin{figure}[htb]
\begin{center}
\includegraphics[width=\swidth]{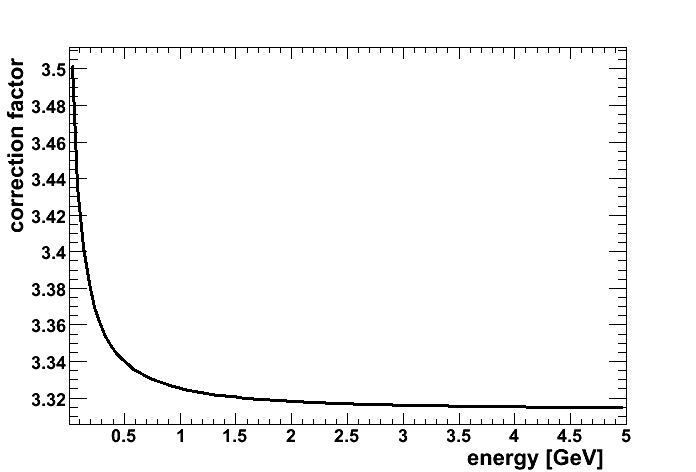}
\caption[Leakage correction function depending on the deposited energy for the Shashlyk calorimeter]
{The leakage correction function depending on the deposited energy for the Shashlyk calorimeter.}
\label{fig:soft:leakCorrShash}
\end{center}
\end{figure}

The energy leakage in the FW EMC is mainly driven by the huge amount of absorber material
and is just slightly caused by the geometry. Therefore, a correction has been considered which only
depends on the collected energy. \Reffig{fig:soft:leakCorrShash} shows the resulting correction
function for energies up to 5~\gev.
 
\subsection{Charged Particle Identification}
\label{sec:soft:recochargedPID}
Good particle identification for charged hadrons and leptons plays an essential role for \Panda
and must be guaranteed over a large momentum range from 200\,\mevc up to approximately 10\,\gevc.
Several subdetectors provide useful PID information for specific particle species and momenta.
While energy loss measurements within the
trackers obtain good criteria for the distinction between the different particle types below
1\,\gevc, the DIRC detector is the most suitable device for the identification of particles
with momenta above the Cerenkov threshold. Moreover, in combination with the tracking detectors,
the EMC is the most powerful detector for
an efficient and clean electron identification, and the Muon detector is designed for the separation
of muons from the other particle species. The best PID performance however can be obtained by 
taking into consideration all available information of all subdetectors.

The PID software is divided in two different parts. In the first stage the recognition is done 
for each detector individually, so that finally probabilities for all five particle hypothesis 
(e, $\mu$, $\pi$, K and p) are provided. The probabilities are normalised uniquely by assuming same 
fluxes for each particle species.\\ 
In the second stage the global PID combines this information
by applying a standard likelihood method. Based thereon, flexible tools can be 
used which allow an optimization of efficiency and purity, depending on the requirements of the 
particular physics channel.

\subsubsection{Subdetector PID}

\begin{figure}[htb]
\begin{center}
\includegraphics[width=\swidth]{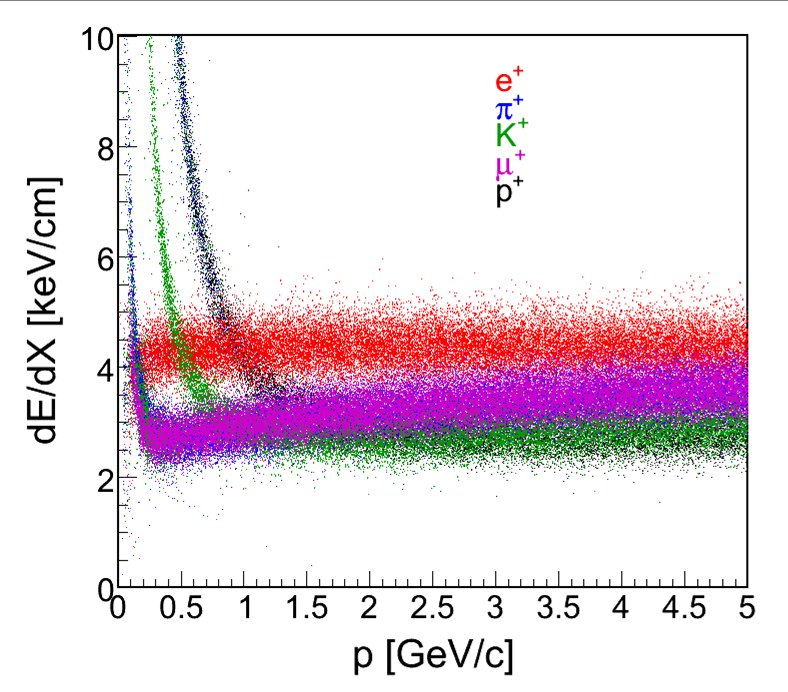}
\caption[Typical truncated $dE/dx$ plot]
{Typical truncated $dE/dx$ plot as a function of momentum for the 5 particle types}
\label{fig:soft:trunc_dedx_STT}
\end{center}
\end{figure}

\paragraph{$\mathbf{dE/dx}$ Measurements}
 The energy loss of particles in thin layers of material directly provides an access to the $dE/dx$.
As can be seen directly from the Bethe-Bloch formula, for a given momentum particles of different
types have different specific energy losses, $dE/dx$. This property can be used
for particle identification, as illustrated in \Reffig{fig:soft:trunc_dedx_STT}.
The method however suffers from two limitations. First of all, at the crossing points, there is no
possibility to disentangle particles. Secondly, the distribution of the specific energy loss
displays a long tail which constitutes a limitation to the separation, especially when
large differences exist between the different particle yields. Various methods have been
worked out to circumvent this second limitation. In \Panda, two detectors will give access to a $dE/dx$
measurement,
the MVD detector setup, and the central spectrometer tracking system. In the present status of the 
simulation,
only the Straw Tube Tracker option has been investigated in detail. Although not discussed here, the
TPC will also be able to give very good identification capabilities through $dE/dx$ measurements.

\subparagraph{MVD}

Although the number of reconstructed MVD hit points per track is limited to 4 in  the barrel
section and 5-6 in the forward domain the energy loss information provided by the readout
electronics can be used as part of the particle identification decision. 
\begin{figure}[htb]
\begin{center}
\includegraphics[width=\swidth]{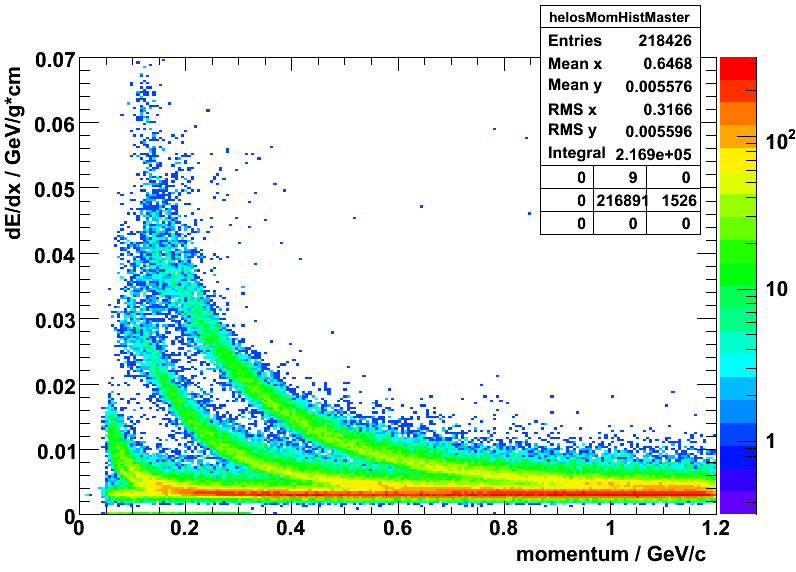}
\caption[$dE/dx$ versus track momentum in the MVD for different particle species]
{$dE/dx$ information from the MVD versus track momentum for protons (upper band), kaons (middle)
and pions/muons/electrons (lower).}
\label{fig:soft:dEdxMvd}
\end{center}
\end{figure}      
The ability of 
separating different particle species relies on an accurate energy loss information and a good
knowledge of the track position with respect to the sensors. Contrary to the usual method of summing
the individual hit measurements $dE_i/dx_i$ to an energy loss information
 all hit measurements can be combined to a total quantity
\begin{equation*}
      S = \frac{\sum dE_i}{\sum dx_i} n_e^{-1}
\end{equation*}
where $n_e^{-1}$ is the electron density in silicon. The $dE_i$ are the energy information
from the reconstructed hit and the $dx_i$ are obtained by calculating the length of the 
reconstructed track traversing the sensor material. This method gives a slightly smaller 
spread of the Landau-smeared energy loss.
\Reffig{fig:soft:dEdxMvd} shows the 
calculated energy loss versus particle momentum for different particle types. Separation is
possible only for protons (upper band) and kaons (middle) from the lowest band which is a
superposition of pions, muons and electrons.

\begin{figure}[htb]
\begin{center}
\includegraphics[width=\swidth]{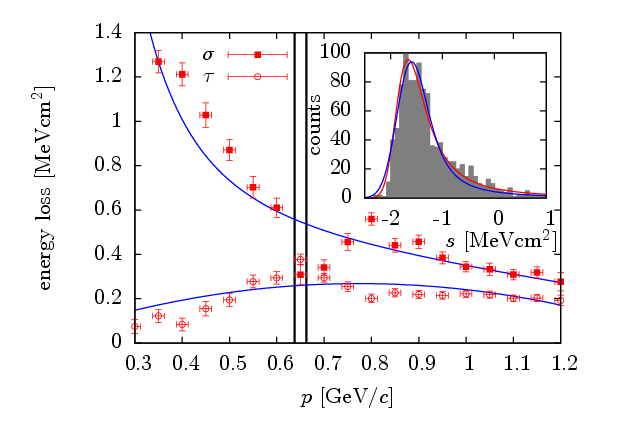}
\caption[Width parameters of the energy loss information from the MVD]
{The change of the width parameters $\sigma$ and $\tau$ for protons.}
\label{fig:soft:mvdPidParams}
\end{center}
\end{figure}

The width of the individual bands depends on the energy-loss distribution which varies with 
momentum. To reproduce the distribution all uncertainties are merged
into a single Gaussian-distributed error which is added to the
already Landau-distributed energy loss. 
The distribution has to be calculated using numerical integration of the convolution integral
\begin{equation*}
  w(s)=\int L(x) \, G(s-x) dx
\end{equation*}
with the parameters  $\sigma$ for the Gauss width, $\tau$ respectively for the 
Landau width and  $s$, which is the most probable value of the Landau 
distribution. 
The used parametrizations for the distributions are 
\begin{equation*}
  G_\sigma(x)=\frac{1}{\sqrt{2\pi}\sigma} e^{-x^2/\sigma^2}
\end{equation*}
for the Gauss distribution and
\begin{equation*}
L_\tau(x)=\frac{1}{\pi\tau}\int_0^\infty e^{-t(\ln{t}- x/\tau)} \sin(\pi t) dt
\end{equation*}
for the scaled Landau distribution.
For each particle type the parameters were obtained independently by generating
$3\times10^5$ single particle events each over a momentum range from $50\,$\mevc to $1.5\,$\gevc.
The energy loss distribution was fitted for $25$\,\mevc bins to get the parameters $\hat s_i$, 
$\sigma_i$, $\tau_i$ for a given $p_i$.
The width parameters $\sigma$ and $\tau$ for protons are shown in \Reffig{fig:soft:mvdPidParams}
and the small inset shows the energy loss distribution for the momentum bin 
$\Delta{p} = 0.65 \ldots 0.675$\,\gevc. For all particle types the evolution of the parameters
$\hat s(p)$, $\sigma(p)$ and $\tau (p)$ are fitted with polynomials and used as a basis for the 
calculation of the particle hypothesis.

\subparagraph{STT}
The energy loss of particles through thin layers of gas provides an opportunity for
particle identification in a large domain of momentum.  However, when thicknesses as low as a
few mm of gas are considered, the fluctuations resulting from the relatively low number
of primary collisions (less than 100 on a 1 cm Ar pathlength) give rise to an extended 
Landau tail in the energy loss distribution. To circumvent this phenomenon which could result
into a dramatic reduction of PID capabilities, truncation methods associated to different
averaging procedures are often used. The truncated arithmetic mean has been used
in the framework of \Panda. Moreover, as straw tubes are cylindrical detectors, a path
length determination is necessary to calculate a $dE/dx$. A transverse resolution of 150  $\mu$m
was assumed. With the used hexagonal STT setup, particles pass 
24 straw tubes on average in the angular range from $22^\circ$ to $140^\circ$. 
This number decreases rapidly to 8 straw tubes at $14^\circ$.

A truncation parameter of 70, corresponding to keeping 70\percent of the smallest energy loss values,
was found to be the best compromise between the resolution defined by the Gaussian fit and
the tail still remaining after truncation.
Parameters of the Gaussians ($\mu _i$ and $\sigma _i$) were obtained and tabulated for all particle 
types over the
angular range [$14^\circ,140^\circ$] covered by the STT and for momenta ranging from a
minimum of 400\,\mevc up to the maximum allowed by kinematics in \pbarp collisions at 15\,\gevc.
The dependence on the angle $\phi$ was not included in the simulation. However, the effect is expected to
be important only in the two $6^\circ$ wide $\phi$ angular regions at $90^\circ$ and $270^\circ$.
Outside the above mentioned limits, a priori probabilities are shared
equally between all particle types $ e,\mu ,\pi ,K$ and $p$.
For a given triplet ($p$, $\theta$, $dE/dx_{trunc}$), five
a priori probabilities are calculated using the parameters of the Gaussians
($\mu _i$,$\sigma _i$, $i=$, $ e,\mu ,\pi ,K$ and $p$), properly normalised.
To take into account the role of
possible non Gaussian tails, a lower limit on the likelihood was set to 1\percent.
 These likelihoods can then be directly combined with the ones from the other detectors to 
calculate a global likelihood.

\begin{figure}[htb]
\begin{center}
\includegraphics[width=\swidth]{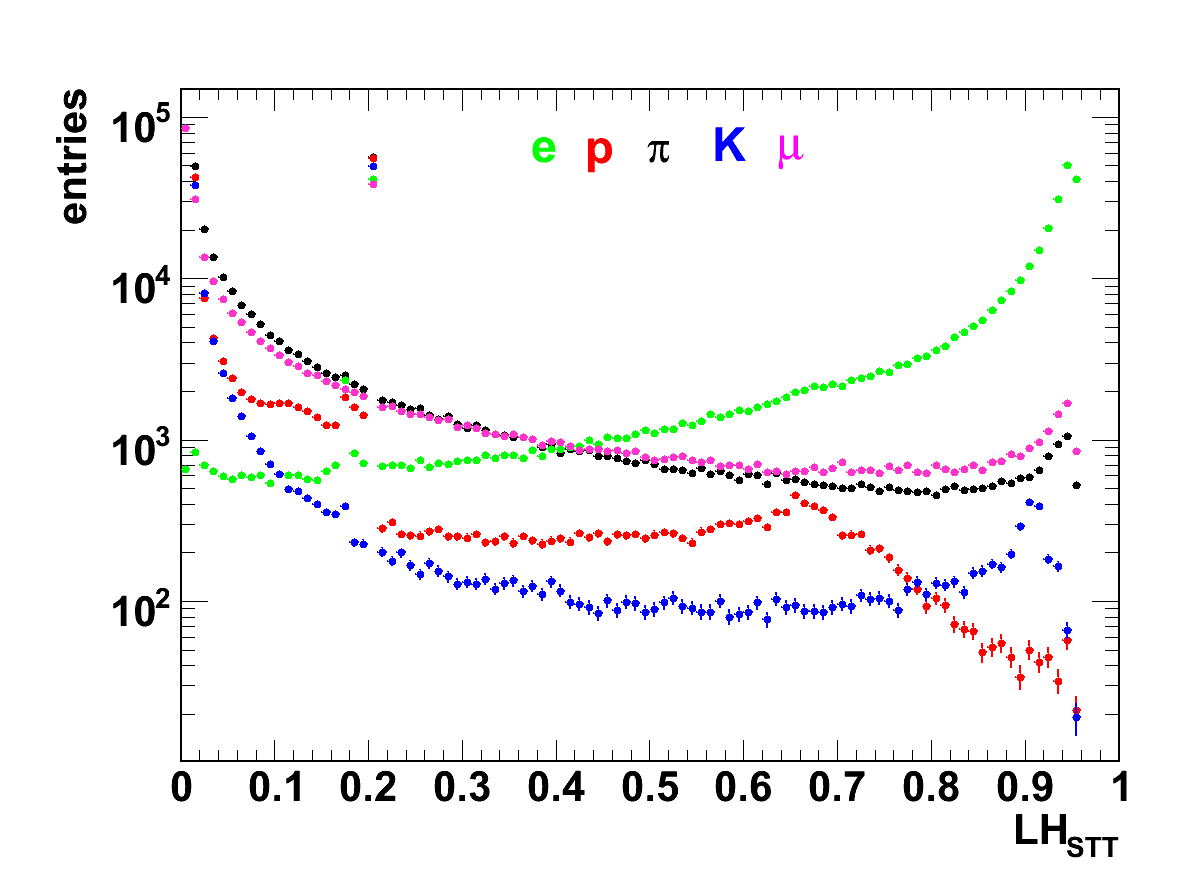}
\caption[Distribution of the likelihood for electron identification.]
{The distribution of the likelihoods for electron identification, averaged over the whole STT angular 
range in the [0.2,10.]\,\gevc momentum range. The peak at 0.2 corresponds to cases when all
particle likelihoods are finally set to the same value of 0.2: this happens when all 5 
first-guess likelihoods are below 1\percent. }
\label{fig:soft:LH_SttPID}
\end{center}
\end{figure}

\Reffig{fig:soft:LH_SttPID} shows the likelihoods for being identified as electrons for particles
with momenta between 0.2\,\gevc and 10\,\gevc and polar angles between $14^\circ$ and $140^\circ$.

\begin{figure}[htb]
\begin{center}
\includegraphics[width=\swidth]{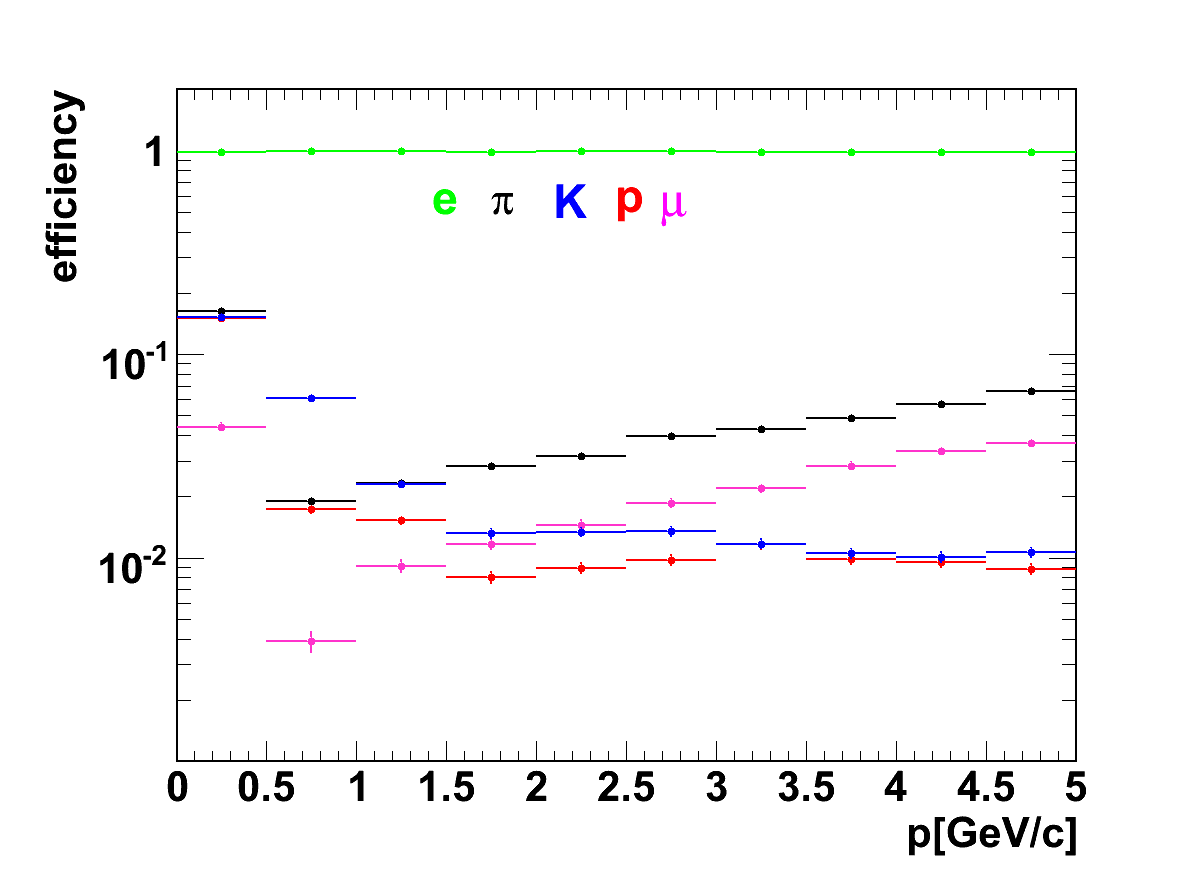}
\caption[Efficiency for electrons and contamination rate for the 4 other particle types.]
{The efficiency for electrons in the momentum range between 0.2\,\gevc and 10\,\gevc 
and the contamination rates for the four other particle types, averaged over the whole STT angular range.}
\label{fig:soft:eff_SttPID}
\end{center}
\end{figure}

Whereas an efficiency above 98\percent
is observed for electrons (see figure \Reffig{fig:soft:eff_SttPID}), the contamination rate,
that is the probability for another type
of particle ($\mu ,\pi ,K$ and $p$) to be identified as an electron varies between 1\percent and 16\percent.
One should. however, note that these values are averaged over the polar angle: at forward angles,
the contamination increases as the result of the strong decrease of the number of hits
in the straw tubes.

\paragraph{PID with DIRC}

\begin{figure}[htb]
\begin{center}
\includegraphics[width=\swidth]{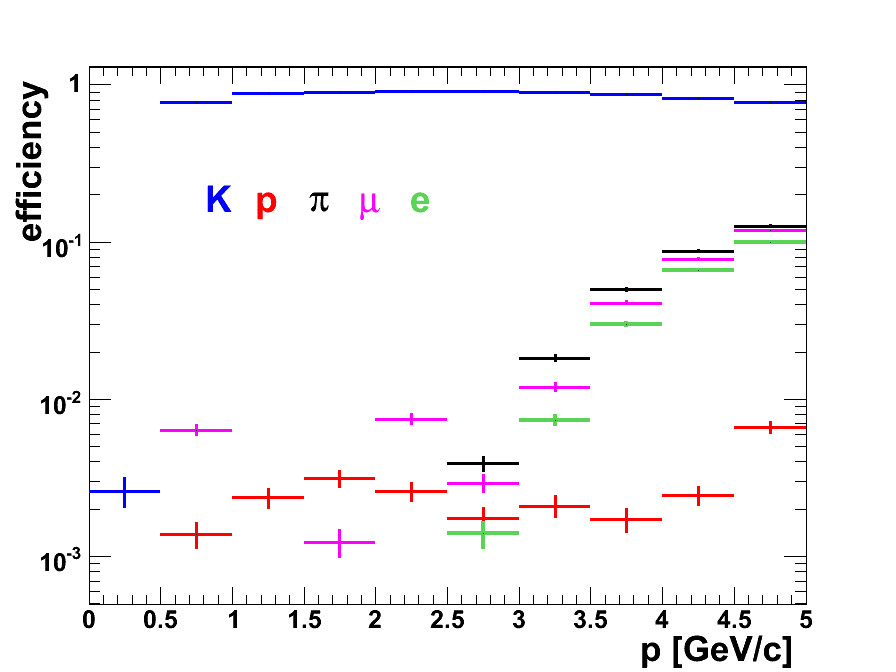}
\caption[Kaon efficiency and contamination rate of the remaining particle species in different momentum 
ranges by using the DIRC information]
{The kaon efficiency and contamination rate of the remaining particle species in different momentum ranges
by using the DIRC information.}
\label{fig:soft:kEffDIRC}
\end{center}
\end{figure}

Charged tracks are considered if they can be associated with the production of Cerenkov light in
the DIRC detector. Based on the reconstructed momentum, the reconstructed path length of the particle in 
the quartz radiator and the particle hypotheses the expected Cerenkov
angles and its errors are estimated. Compared with the measured Cerenkov angle the likelihood and
significance level  for each particle species are calculated. As an example for the DIRC performance
\Reffig{fig:soft:kEffDIRC} shows the obtained 
kaon efficiency and contamination rate by applying a \loo kaon  criterion on the DIRC PID. The loose 
criterion corresponds to the kaon probability of 30\percent. While below the Cerenkov threshold of approximately
500\,\mevc 
almost no kaon can be identified the efficiency above the threshold is more than 80\percent over the whole 
momentum range up to 5\,\gevc. The fraction of pions misidentified as kaons is substantially less than 
10$^{-3}$ for momenta below 3~\gevc and increases up to 10\percent for 5~\gevc.

\begin{figure}[htb]
\begin{center}
\includegraphics[width=\swidth]{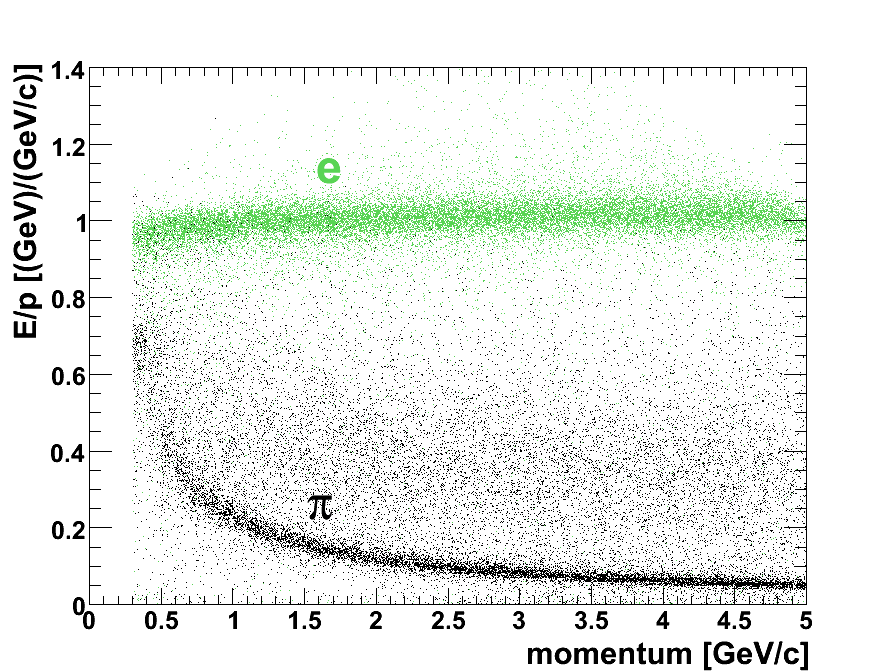}
\caption[E/p versus track momentum for electrons and pions]
{E/p versus track momentum for electrons (green) and pions (black) in the momentum 
range between 0.3\,\gevc and 5\,\gevc.}
\label{fig:soft:eOpElectronsPions}
\end{center}
\end{figure}

\paragraph{Electron Identification with the EMC}
The footprints of deposited energy in the calorimeter differ distinctively for
electrons, muons and hadrons. The most suitable
property is the deposited energy in the calorimeter. While muons and hadrons in 
general lose only
a certain fraction of their kinetic energy by ionization processes, electrons deposit
their complete energy in an electromagnetic shower. The 
ratio of the measured energy deposit in the calorimeter to the reconstructed track momentum 
($E/p$) will be approximately unity. Due to the fact that hadronic 
interactions can take place, hadrons can also have a higher $E/p$ ratio than expected from 
ionization.
Figure \Reffig{fig:soft:eOpElectronsPions} shows the reconstructed $E/p$ fraction for electrons and pions
as a function of the momentum.

\begin{figure}[htb]
\begin{center}
\includegraphics[width=\swidth]{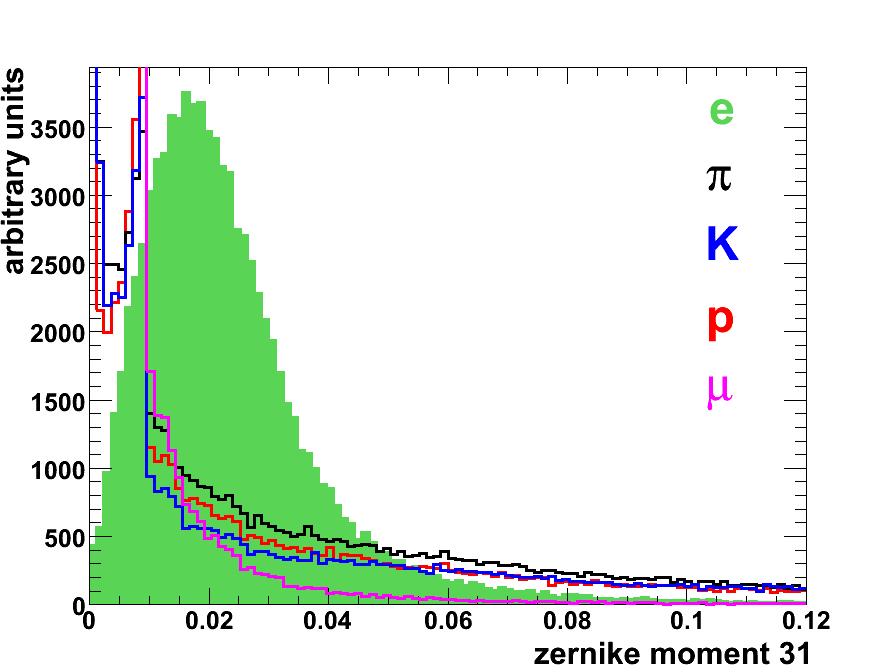}
\caption[Zernike moment 31 for electrons, muons and hadrons.]
{Zernike moment for electrons, muons and hadrons.}
\label{fig:soft:zernike31}
\end{center}
\end{figure}

Furthermore, the shower shape of a cluster is helpful to 
distinguish between electrons, muons and hadrons. Since the chosen size of the scintillator modules 
corresponds to the Moli\`ere radius of the material, the largest fraction of
an electromagnetic shower originating from an electron is contained in just a few 
modules. Instead, an hadronic shower with a similar energy deposit is less concentrated.
These differences are reflected in the shower shape of the cluster, which can be 
characterised by the following properties:
\begin{itemize}
\item $E_1/E_9$ which is the ratio of the energy deposited in the central scintillator module and in 
the 3$\times$3 module array containing the central module and the first innermost ring.
Also the ratio between $E_9$ and the energy deposit in the 
5$\times$5 module array  $E_{25}$ is useful for electron identification.    
 
\item The lateral moment of the cluster defined by
\begin{equation*}
      mom_{LAT} = \sum_{i=3}^n E_i r_i^2 / (\sum_{i=3}^n E_i r_i^2 + E_1 r_0^2 +  E_2 r_0^2) 
\end{equation*}
with
\begin{itemize}
\item $n$: number of modules associated to the shower
\item $E_i$: deposited energy in the iTH module with $E_1 \geq E_2 \geq ... \geq E_n$
\item $r_i$: lateral distance between the central and the iTH module
\item $r_0$: the average distance between two modules. 
\end{itemize}
 
\item A set of Zernike moments which describe the energy distribution within a cluster by
radial and angular dependent polynomials. An example is given in \Reffig{fig:soft:zernike31}, 
where the Zernike moment 31 is depicted for each particle type.  
\end{itemize}

\begin{figure}[htb]
\begin{center}
\includegraphics[width=\swidth]{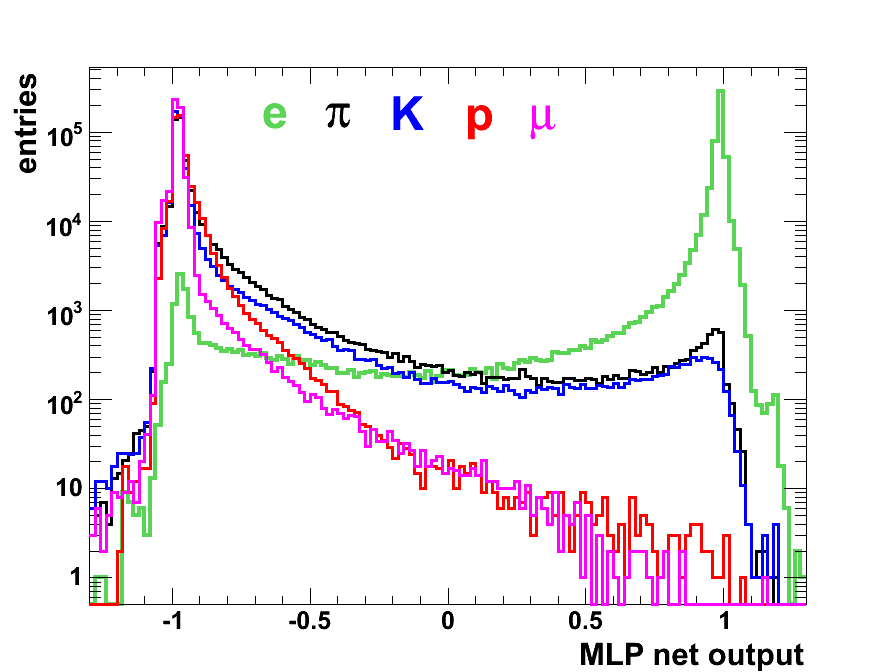}
\caption[MLP network output for electrons and the other particle species 
in the momentum range between 300\,\mevc and 5\,\gevc.]
{MLP network output for electrons and the other particle species 
in the momentum range between 300\,\mevc and 5\,\gevc.}
\label{fig:soft:nno}
\end{center}
\end{figure}

Since a lot of partially correlated EMC properties are suitable for electron 
identification, a Multilayer 
Perceptron (MLP) with 10 input nodes, 13 hidden nodes, and one output node has been applied. 
The advantage of a neural network is that it can 
provide a correlation between a set of input variables and one or several output 
variables without any knowledge of how the output formally depends on the input. The
training of the MLP has been done with a data set of 850\,k single tracks for each 
particle species (e, $\mu$, $\pi$, K and p) in the momentum range between 200\,\mevc 
and 10\,\gevc in such a way  that the output values are constrained to be 1 for 
electrons and -1 for all other particle types. 10 input variables in total have been
used, namely $E/p$, $p$, the polar angle $\theta$ of the cluster, and 7 shower 
shape parameters
($E_1/E_9$, $E_9/E_{25}$, the lateral moment of the shower and 4 Zernike moments). The
response of the trained network to a test data set of single particles in the 
momentum range between 300\,\mevc and 5\,\gevc is illustrated in \Reffig{fig:soft:nno}. 
The logarithmically scaled histogram shows that an almost clean electron recognition with a
quite small contamination of muons and hadrons can be obtained by applying a cut on the network output.

\begin{figure}[htb]
\begin{center}
\includegraphics[width=\swidth]{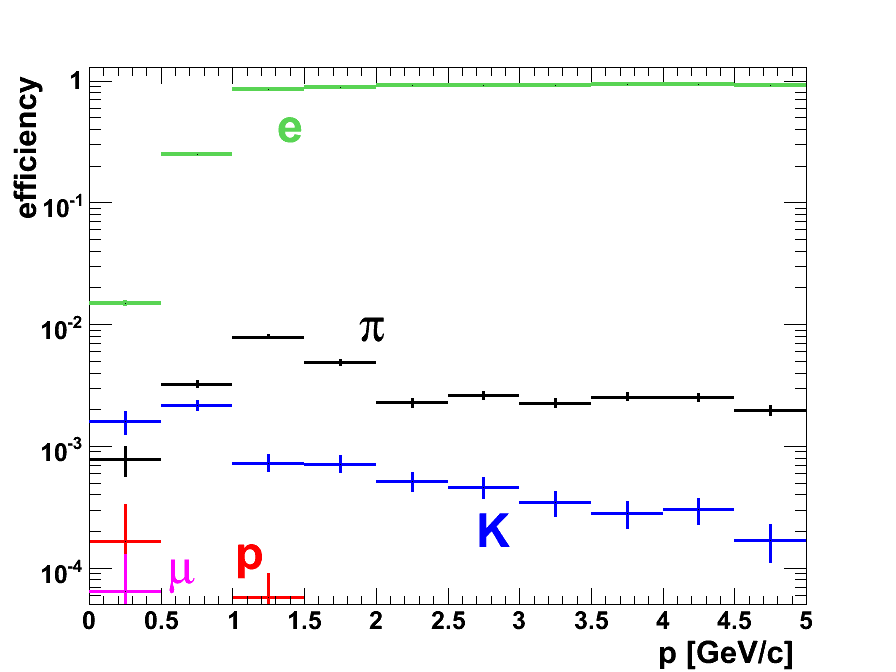}
\caption[Electron efficiency and contamination rate for muons, pions,
 kaons and protons in different momentum ranges by using the EMC information.]
{The electron efficiency and contamination rate for muons, pions,
 kaons and protons in different momentum ranges by using the EMC information.}
\label{fig:soft:lhEMC}
\end{center}
\end{figure}

For the global PID a correlation between the network output and the PID 
likelihood of the EMC has been calculated. \Reffig{fig:soft:lhEMC} shows the electron efficiency 
and contamination rate as a function of momentum achieved  by requiring an 
electron likelihood fraction of the EMC of more than 95\percent.
For momenta above 1\,\gevc one can see that the electron efficiency is greater than 98\percent
while the contamination by other particles is substantially less than 1\percent. 
For momenta below 1\,\gevc, the electron identification based solely on the EMC information
has a poor efficiency and an insufficient purity.

\begin{figure}[htb]
\begin{center}
\includegraphics[width=\swidth]{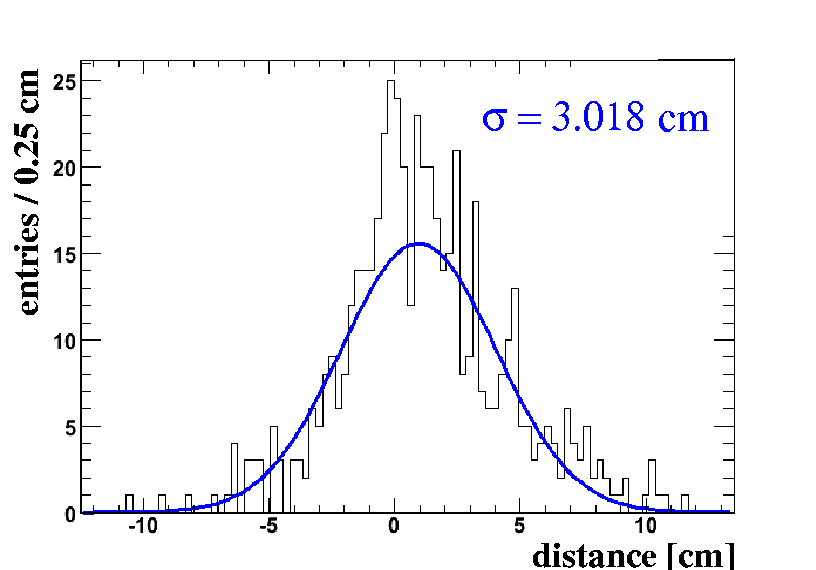}
\caption[The distribution of the distance in x-direction between the expected hits obtained by the 
extrapolation and the 
corresponding detected hits in the muon detector for generated single muon particles.]
{The distribution of the distance in x-direction between the expected hits obtained by the extrapolation 
and the corresponding detected hits in the muon detector. This figure illustrates an example 
for generated single muon particles hitting on specific layer within the barrel yoke.}
\label{fig:soft:resMuoDet}
\end{center}
\end{figure}

\paragraph{PID with the Muon Detector}
\label{sec:soft:muoPID}
The particle ID for the muon detector is based on an algorithm which quantifies for each reconstructed
charged track the compatibility with the muon hypothesis. It propagates the charged particles from
the tracking volume outward through the neighbouring detectors like DIRC and EMC and finally through
the iron, where the scintillator layers of the muon device are located. This procedure takes into 
account the obtained parameters of the global tracking, the magnetic field as well as the muon 
energy loss in the material and the effects of multiple scattering. Then the extrapolated 
intersection points with the scintillators are compared with the detected muon hit positions.  
In case the distance between the expected and the detected hit is smaller than 12\,cm,
according to 4$\sigma$ of the corresponding distribution, the
muon hit will be associated with the corresponding charged track. 
\Reffig{fig:soft:resMuoDet} shows an example for the obtained spatial resolution between 
the expected and the corresponding detected hits. The distribution of the distance in x-direction 
for generated single muons yields to a resolution $\sigma_x\,=\,3.0$\,cm.
Based on the numbers of associated and expected muon hits, the 
likelihoods for each particle type are estimated. 

The procedure results in a good muon identification for momenta above approximately 1~\gevc.
While electrons can be completely suppressed, a contamination rate of only a few percent can be 
achieved for hadrons.  
 
\subsubsection{Global PID}
The global PID, which combines the relevant information of all subdetectors associated with one 
track, has been realised with a standard likelihood method. Based on the likelihoods obtained by each 
individual subdetector the probability for a track originating from a specific particle type $p(k)$ is
evaluated from the likelihoods as follows:
\begin{equation}
      p(k) = \frac{\prod_{i} p_i(k)}{\sum_{j}\prod_{i} p_i(j)} \,\, ,
\label{equ:soft:globlikelihood}
\end{equation}
where the product with index $i$ runs over all considered subdetectors and the sum
with index j over the five particle types $e$, $\mu$, $\pi$, $K$ and p. 

\begin{figure}[htb]
\begin{center}
\includegraphics[width=\swidth]{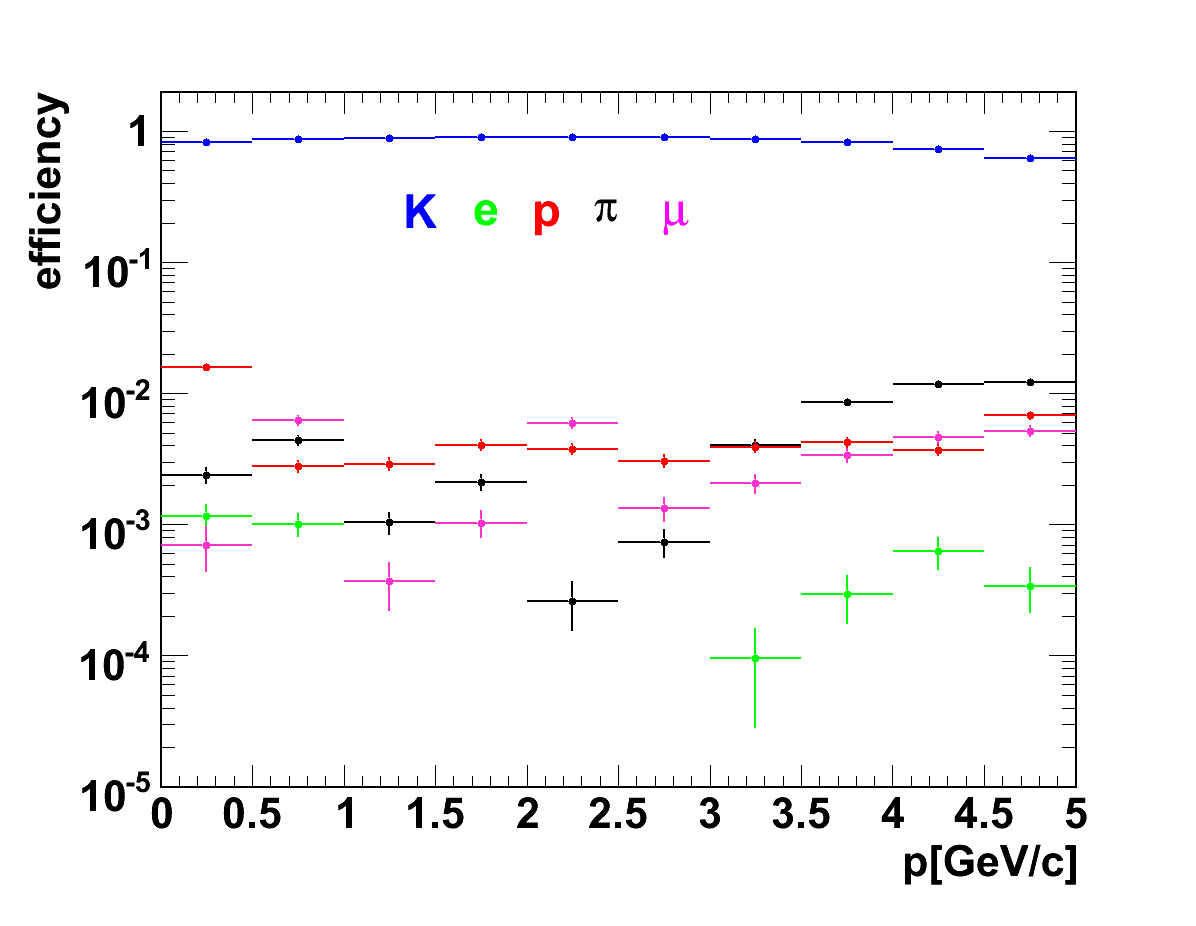}
\includegraphics[width=\swidth]{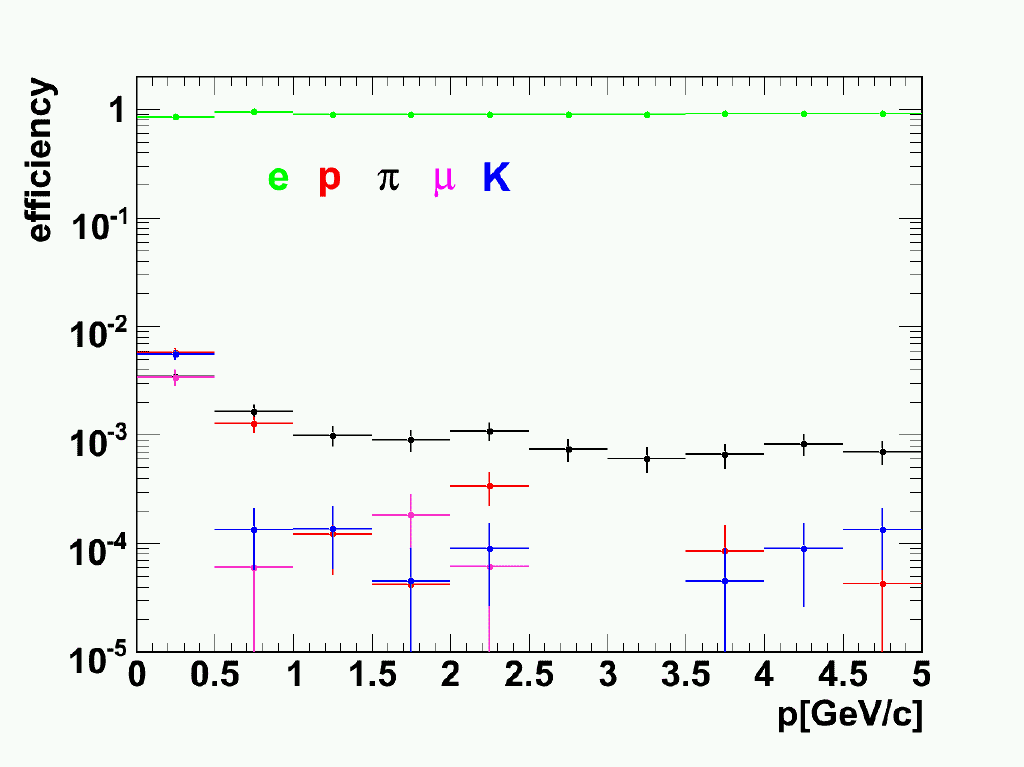}
\caption[Kaon and electron efficiency with the contamination rate of the remaining 
particle species in different momentum ranges by applying \vtig PID cuts]
{The kaon (upper histogram) and electron (lower histogram) efficiency with the contamination rate of the 
 remaining particle species in different momentum ranges by applying \vtig cuts 
 on the global likelihood. The results are based on single particles generated within 
 the $\theta$ range between 25$^\circ$ and 140$^\circ$.}
\label{fig:soft:effGlobalPID}
\end{center}
\end{figure}

Due to the variety of requirements imposed by the different characteristics of the benchmark channels 
various kinds of particle candidate lists depending on different selection criteria on the global 
likelihood are provided for the analysis (see. \Reftbl{tab:soft:emc_pid_cuts}).    
The usage of the so-called \vloo and \loo candidate lists allows to achieve 
good efficiencies, and the \tig and \vtig lists are optimised to obtain 
a good purity with efficient background rejection. \Reffig{fig:soft:effGlobalPID} represents
the performance of the global PID for the \vtig list for kaon and electron candidates,
respectively.

\begin{table}[htb!]
\begin{center}
\begin{tabular}{c|cccc}
\hline\hline
  & \multicolumn{4}{c}{candidate list} \\
particle & {\footnotesize \vloo} & {\footnotesize \loo} & {\footnotesize \tig} & {\footnotesize \vtig} \\
\hline
e     & 20\percent & 85\percent & 99\percent & 99.8\percent \\
$\mu$ & 20\percent & 45\percent & 70\percent & 85\percent \\
$\pi$ & 20\percent & 30\percent & 55\percent & 70\percent \\ 
K     & 20\percent & 30\percent & 55\percent & 70\percent \\
p     & 20\percent & 30\percent & 55\percent & 70\percent \\  
\hline\hline
\end{tabular}

\caption[Selection criteria for the particle candidate lists]
{Selection criteria for the particle candidate lists provided for the analysis. The table 
represents the minimal values for the global likelihood (see. \Refeq{equ:soft:globlikelihood}),
which are required for the corresponding particle types.}
\label{tab:soft:emc_pid_cuts}
\end{center}
\end{table}

%


%% file: soft/soft_ana.tex
%
\section{Physics Analysis}

The event data used for physics analysis is structured in three levels of detail:
\begin{itemize}
\item the small TAG level contains brief event summary data
\item the Analysis Object Data (AOD) mainly consists of PID lists of particle
candidates, and, in the case of MC data, also of MC truth data. Most analysis jobs run
on AOD data.
\item the Event Summary Data (ESD), which holds all reconstruction objects down
to the detector hits which are necessary to redo the track and neutral particle
reconstruction, and to rebuild the PID information. The detailed ESD data is needed
only by very few analysis jobs.
\end{itemize}
This event data design directly supports the typical analysis tasks, which usually can
be subdivided into three steps:
\begin{itemize}
\item a fast event preselection that uses just the TAG data
\item an event reconstruction and refined event selection step using the AOD data. 
In the event reconstruction, decay trees are built up, and geometrical
and kinematic fits are applied on them. Cuts on the fit
probabilities, invariant masses or kinematic properties of the candidates
are common to refine the event selection in this step.  The output of
this analysis step usually are n-tuple like data.
\item in the last analysis step the event selection can be further refined by applying cuts
on the n-tuple data, and histogram fits or partial wave analyses can be performed on the
final set of events.
\end{itemize}

\subsection{Analysis Tools}
The used application framework provides filter modules which allow a TAG based event selection. For
events which do not pass this selection, only the relative small TAG has to be read in. This yields in
a significant speed up of the analysis jobs.

The analysis user has the choice to reconstruct decay trees, perform geometrical and kinematic fits, 
and to refine the event selection by using Beta, BetaTools and the fitters provided by the analysis
software \cite{bib:soft:Beta} directly in an application framework module, or by defining the 
analysis in a more abstract
way using {\it{SimpleComposition}} tools. This TCL based high level analysis tool package provides 
an easy-to-learn user interface for the definition of an analysis task and the production of
n-tuples, and it allows to set up analysis jobs without the need to compile any code.
SimpleComposition as well as Beta, BetaTools and the geometric and kinematic fitters were taken
over as well tested packages from \INST{BaBar}, and adapted and slightly extended for \Panda.

\begin{figure}[htb]
\begin{center}
\includegraphics[width=\swidth]{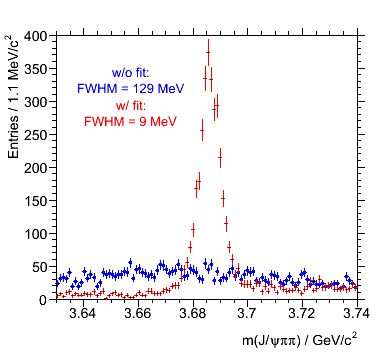}
\caption[Improving the $\jpsi \pip \pim$  mass resolution by applying a kinematic fit]
{$\jpsi\, \pip\, \pim$ mass spectra from 
$\pbarp \,\rightarrow\, \Psi(2S)\, \pip\, \pim \,\rightarrow\, \jpsi\, 2\pip\, 2\pim$ 
before and after applying a 4C-fit to the decay tree with requiring a common vertex for $\jpsi\to\ee$.}
\label{fig:soft:4CFitResult}
\end{center}
\end{figure}

For the exclusive benchmark channels it turned out that especially a 4C-fit of the reconstructed
decay tree is a powerful tool to improve the data quality and to suppress background. 4C-fits 
and cuts on the results can also be defined in SimpleComposition. \Reffig{fig:soft:4CFitResult}
shows an example of a drastic improvement of the mass resolution for a composite particle.

The last n-tuple based analysis steps were carried out utilising the ROOT \cite{bib:soft:ROOT} 
toolbox, which provides powerful, interactively usable instruments among others for cutting, histogramming,
and fitting.

%

%% file: soft/soft_prod.tex
%
\section{Data Production}

\subsection{Bookkeeping}
The Monte Carlo data used for this document were produced in a centrally
organised way. All simulation requests are stored in a MySQL database, and all job scripts
and configuration files are produced automatically by Perl and PHP scripts. Also the
scanning of the log files of the simulation jobs is automated, and the determined job
status is noted in the database, too.

The analysis users could get the actual status of the simulation production for an
individual channel from a web page. With a tool available in the \Panda software they
could configure the input of their analysis jobs and split an analysis task into an 
adequate number of jobs, which could run in parallel on one of the batch farms.

\subsection{Event Production}
Because none of the involved computing sites listed in \Reftbl{tab:soft:comp_resources}
had sufficient computing and storage resources available for \Panda, the data
production for the simulation was distributed over four sites, namely the IPN in Orsay 
\cite{bib:soft:prod:grille}, the CCIN2P3 in Lyon, the Ruhr-Universit\"at Bochum, and the GSI at
Darmstadt. \Reftbl{tab:soft:comp_resources} lists the contributions
of the sites to the simulation production. In total $22\cdot10^6$
signal events for 22 channels, $1002\cdot10^6$ dedicated background
events including the dominant background reactions for the individual
analyses and $280\cdot10^6$ generic background events were produced
in 29 weeks.  The typical event size was 3.5 kByte for AOD and TAG
data, and 4.4 kByte for the ESD component.
\begin{table}
\begin{center}
\begin{tabular}{r c c}
\hline\hline
site & events [$10^6$] & stored data [TB] \\ \hline
Lyon & 551 & 4.4 \\ 
GSI & 308 & 2.4 \\ 
Orsay & 149 & 1.2 \\ 
Bochum & 297 & 2.3 \\ \hline\hline
\end{tabular}
\end{center}
\caption[Contributions of the computing sites to the Physics Book simulation production]
{Contributions of the computing sites to the Physics Book simulation production.}
\label{tab:soft:comp_resources}
\end{table}

The network bandwidth between the production sites was limited and did not allow to copy over
analysis data from one site to another for a large number of events. Therefore, most
analysis jobs ran at the site where the analysis data were produced.

\subsection{Filter on Generator Level}
\label{sec:soft:genfilter}
The most time consuming parts of the data processing chain are the simulation and reconstruction. 
In comparison to that, the first stage, namely the generation of events, is in general few order
of magnitudes faster. In order to achieve the relevant number of required background events, an 
event filter on the generator level has been applied for a certain part of these data. The 
strategy is to consider 
only those events which are possible candidates for passing all the selection criteria in the 
analysis by cutting on the kinematics of the generated particles.

This technique was used especially for the relevant background of those benchmark channels which
contain a \jpsi decaying to a \ee pair. In this 
case only events have been processed completey if the invariant mass of two charged particles is within a 
certain \jpsi mass window.  This method will now be presented and justified in detail for the example 
$\jpsi\eta$.

\subsubsection{Validation of the \jpsi Generator Filter} 

The $\pi\pi\eta$ final state is one of the most important background sources for the $\jpsi\eta$ channel.
It had to be justified that the generated mass of the $\pi\pi$ system, with the pions falsely identified 
as electrons, can be limited to the \jpsi  signal region.

Here a generator-level filter was applied in the following way: the four-vectors of the pions were 
recalculated with an electron mass hypothesis and were combined afterwards. The resulting invariant mass
must lie within [2.8; 3.2] \gevcc, corresponding to the \jpsi  signal region. Another two data samples 
were created with the combined invariant mass between [2.4; 2.8] \gevcc and [3.2; 3.6] \gevcc, 
respectively, corresponding to the sideband regions below and above the \jpsi  signal region.

10 million events have been generated for each of the three mass regions at a beam momentum of 
8.6819 \gevc.

\begin{figure}[htb]
\begin{center}
\includegraphics[width=\swidth]{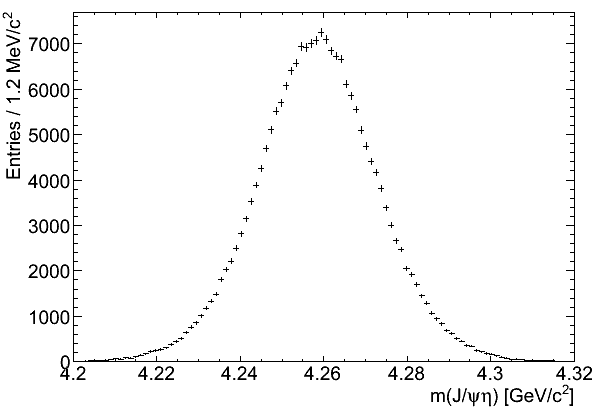}
\caption[Reconstructed $\jpsi \eta$ mass from the background mode $\pi \pi \eta$]
{Reconstructed $\jpsi \eta$ mass from the background mode $\pi \pi \eta$, with the $\pip \pim$ mass (under electron mass hypothesis) within the \jpsi signal region.}
\label{fig:soft:jpsieta_genfilter}
\end{center}
\end{figure}

\begin{figure}[htb]
\begin{center}
\includegraphics[width=\swidth]{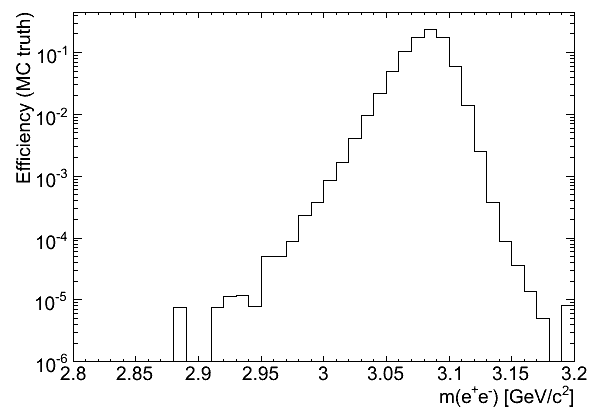}
\caption[Reconstruction efficiency for wrongly identified \ee pairs (truth mass)]
{ Reconstruction efficiency for wrongly identified \ee pairs (Monte Carlo truth mass).}
\label{fig:soft:ee_eff_genfilter}
\end{center}
\end{figure}

The $\pbarp \to \jpsi \eta \to \ee \gamma\gamma$ analysis module was
run on the reconstructed $\pi\pi\eta$ data. The number of reconstructed entries in the signal region in
dependence of the electron PID criteria is shown in \Reftbl{tab:soft:recev}.
In \Reffig{fig:soft:jpsieta_genfilter} the resulting reconstructed $\jpsi \eta$ mass is shown after 
applying all cuts on the data sample corresponding to the \jpsi  signal region. No PID criteria 
have been applied on the false electron candidates. About 200000 candidates are left out of 10 million. 
For events from the 
\jpsi  sideband regions, there are no candidates left after applying the same selection criteria. 
\Reffig{fig:soft:ee_eff_genfilter} shows the reconstruction efficiency on the Monte Carlo truth level.

\begin{table}
\begin{center}
 \begin{tabular}{ccccc}\hline\hline
\multicolumn{2}{c}{$\mathcal{L}_{\ee}$} & \multicolumn{3}{c}{reconstr. events} \\
$e^{\pm}$ & $e^{\mp}$ & $SR$ & $SB_{l}$ & $SB_{r}$ \\ \hline
$>0\%$ & $>0\%$ & $204269$ & $0$ & $0$ \\ 
$>0\%$ & $>20\%$ & $18965$ & \\
$>20\%$ & $>20\%$ & $338$ & \\
$>20\%$ & $>85\%$ & $64$ & \\
$>85\%$ & $>85\%$ & $3$ & \\
$>85\%$ & $>99\%$ & $1$ & \\
$>99\%$ & $>99\%$ & $0$ & \\ \hline \hline
 \end{tabular}
\caption[Number of reconstructed $\jpsi\eta$ candidates from the background mode
 $\pi\pi\eta$, for different electron PID criteria, in three regions of the generated invariant $\pi \pi$ mass under electron mass hypothesis]
{Number of reconstructed $\jpsi\eta$ candidates from the background mode $\pi\pi\eta$, for different electron PID criteria, in three regions of the generated invariant $\pi \pi$ mass under electron mass hypothesis: $\jpsi$ signal region $SR$ and left and right \jpsi sideband regions ($SB_{l}$ and $SB_{r}$, resp.).}
\label{tab:soft:recev}
\end{center}
\end{table}

In summary, one can limit the production to events with a generated $\pi \pi$ mass which is close to
the \jpsi  mass. With a filter efficiency of 14\%, the production time can
be reduced by a factor of 7.


%% file: soft/soft_pandaroot.tex
%

\begin{figure*}[hbt]
\begin{center}
\includegraphics[width=0.8\dwidth]{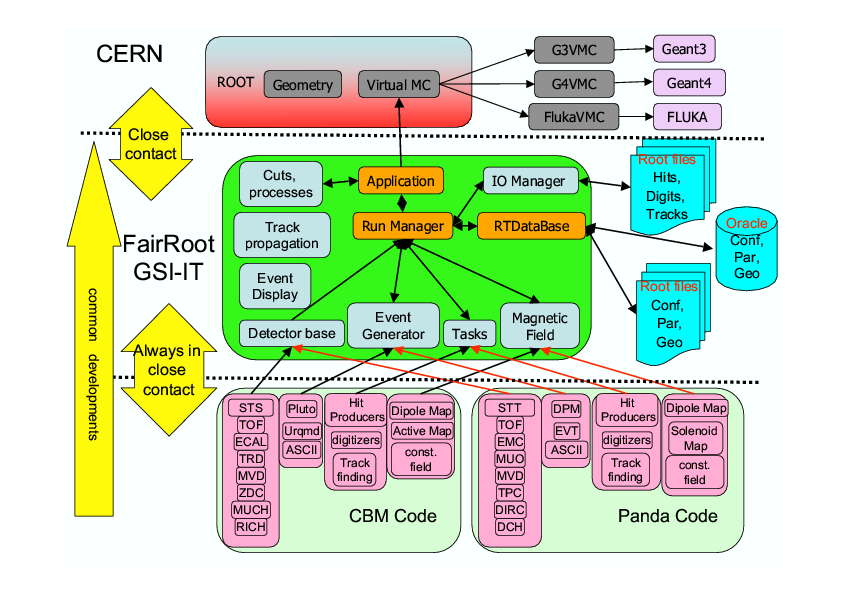}
\caption[Functionality of \INST{FAIRroot}]
{An illustration of the building blocks of \INST{FAIRroot}. The framework inherits the functionalities of
ROOT and Virtual Monte Carlo. The core elements of \INST{FAIRroot} are indicated in the green
box in the middle of the figure. The two boxes at the bottom of the figure represent the ingredients 
of the \INST{CBMroot} and \INST{PANDAroot} branches and both are based on \INST{FAIRroot}.}
\label{fig:soft:FAIRroot}
\end{center}
\end{figure*}

\section{Software Developments}


The development of a next-generation simulation and analysis framework for \PANDA 
was initiated at the end of 2006. The new software 
infrastructure, called \INST{PANDAroot}, is designed to improve the accessibility for beginning users and developers, 
to increase the flexibility to cope with future developments and to enhance synergy with 
other nuclear and high-energy physics experiments. In addition, it will provide a complete 
reconstruction and pattern-recognition chain to overcome the shortcomings of the older
analysis framework. Most of the
functionalities of the software infrastructure as described in the previous sections
have been embedded in \INST{PANDAroot} as well. Below, only the most important new elements
of \INST{PANDAroot} are described.


The core services for the detector simulation and offline analysis are provided
by the \INST{FAIRroot} framework~\cite{bib:soft:FAIRroot}, which is based on the 
object-oriented data analysis framework, ROOT~\cite{bib:soft:ROOT}, and the 
Virtual Monte-Carlo (VMC) interface~\cite{bib:soft:vmc}. 
\INST{FAIRroot} enables the integration of a detailed magnetic field map, advanced parameter handling with
an interface to an Oracle database, the usage of a large set of event generators,
presently EvtGen, DPM, UrQMD, and Pluto, and, since recently, a graphical tool to 
display reconstructed or Monte Carlo based hits and tracks. Furthermore, a modular 
design of the framework 
is guaranteed via a task mechanism. With this, the developer can setup simulation, 
reconstruction, and analysis algorithms in well-separated and exchangeable tasks, which 
the user can exploit to compare different algorithms and methods for his or her 
application. \Reffig{fig:soft:FAIRroot} depicts the various ingredients of \INST{FAIRroot} and
its coupling to ROOT, VMC, and the two available branches, \INST{PANDAroot} and \INST{CBMroot}.
 

The VMC interface allows to perform easily, {\it e.g.} without the need to alter the user code, 
simulations using various transport models, presently Geant3~\cite{bib:soft:geant3}, 
Geant4~\cite{bib:soft:GEANT4}, and Fluka~\cite{bib:soft:FLUKA}. For the modelling of 
the detector geometry, a vitalization scheme, virtual geometry model (VGM), has been 
employed as well. The vitalization concept foresees a transparent transition 
between older and newer transport models, thereby, improving the validity and lifetime 
of the framework significantly. Besides the option to run a simulation using one of the 
Monte-Carlo transport codes, the new framework includes a fast simulation package based on a 
parametrization of the individual detector responses. The parametrization is obtained
by a comparison with the results from experimental data or from simulations using one
of the full transport code. The fast-simulation package is used to generate large 
numbers of background events within an acceptable period of time.

\begin{figure}[htb]
\begin{center}
\includegraphics[width=1.1\swidth]{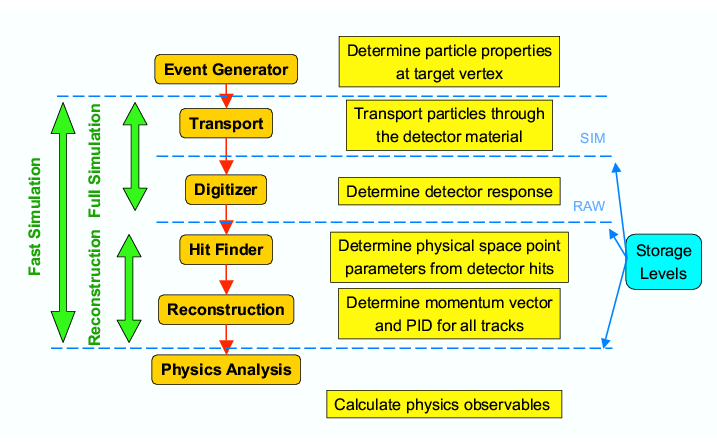}
\caption[Flow diagram of \INST{PANDAroot}]
{A sketch of the simulation and reconstruction chain of \INST{PANDAroot}.}
\label{fig:soft:PandaRootFlow}
\end{center}
\end{figure}

\Reffig{fig:soft:FAIRroot} sketches the various simulation and analysis steps
which are part of the \INST{PANDAroot} framework. 
The new framework enables a complete reconstruction of tracks which does not depend
upon the true origin of hits given by the Monte Carlo transport code.
For this, the information of the transport model is pre-processed to simulate the signals from the
individual detector components. In \Reffig{fig:soft:PandaRootFlow} this procedure is
indicated as {\it digitizer}. The data after this process represent the digitised detector 
response and are, therefore, comparable with experimental data. The following part, 
{\it reconstruction}, takes only this data as input to find tracks and to reconstruct the momenta, 
scattering angles, and the type of particle. Note that such an unbiased reconstruction procedure allows 
to study background due to fake tracks and pileup effects as well.

Efficient and fast algorithms, based on conformal 
mapping~\cite{bib:soft:cm} or extended Riemann~\cite{bib:soft:riemann} techniques, can be 
exploited to find charged tracks, to correlate them with the information of various PID
detectors of \PANDA and to use as a pre-fit for a more precise track-fitter package.
A track follower based on the well-tested GEANE package~\cite{bib:soft:geant3} combined with a generic Kalman 
filter provide a tool set to optimise the momentum reconstruction for charged particles.
With these algorithms and tools, the new framework enables a realistic and complete track finding
and pattern recognition. 

For a complete particle identification, the propagation of Cerenkov photons is simulated and ring-finding 
algorithms for the DIRC are being implemented. Furthermore, for the global particle identification, multi-dimensional
probability distributions for the different particle types are provided using the
k-nearest neighbourhood technique~\cite{bib:soft:knn} together with a very fast multi-variate classification 
method based on self-organising maps~\cite{bib:soft:som}. 


For the higher-level analysis activities, the Rho package~\cite{bib:soft:rho} have been 
embedded in the \INST{PANDAroot} framework. The Rho package is an analysis tool kit which 
is optimised for interactive work and performance. It owes a lot to its predecessors, 
including the BETA package~\cite{bib:soft:Beta}, which has been described briefly in the previous sections. 
Unlike Beta, Rho is based on the solid fundamentals of the ROOT framework 
and runs interactively on all computing platforms supporting ROOT. 
The \INST{PANDAroot} framework is enriched with vertex and kinematic fitting tools, based on the KFitter 
package from the Belle collaboration~\cite{bib:soft:kfitter}. 


The \INST{PANDAroot} framework is designed to run on a large variety of computing platforms. 
This has the advantage that the software can be employed easily on a GRID environment. 
The \PANDA collaboration is presently expanding and maintaining an AliEN$^2$ GRID network~\cite{bib:soft:alien} 
in synergy with the Alice collaboration and with the \INST{PANDAroot} developments. A complete 
simulation and analysis chain has been tested successfully on the GRID which presently 
consists of about ten sites and gradually expanding. In addition, advanced monitoring tools, 
based on MonALISA, are being used in connection to the GRID and framework developments.


The development of the new simulation and analysis framework ran in parallel with the 
physics-benchmark simulations for this report. Hence, nearly all the channels
were so-far studied using the predecessor of the \INST{PANDAroot} framework. However, one of the
electromagnetic channels, namely $\bar p p\rightarrow \gamma\gamma$, was simulated and
analysed using \INST{PANDAroot}. This channel depends primarily on the reconstruction code
of the electro-magnetic calorimeter, which was successfully derived from the old
framework, and, therefore, identical to the code which was used for the studies of all 
other channels. The new framework in combination with the GRID infrastructure will be
used in the near future to perform simulations of an extended set of benchmark 
reactions.
 
%

%% file: panda_pb_phys.tex
%
%
\cleardoublepage
\chapter{Physics Performance}
\label{sec:phys}
\COM{Author(s): D. Bettoni, K. Peters}
%
%
\input{./phys/phys_overview}
\input{./phys/phys_qcd_states}
\input{./phys/phys_qcd_dynamics}
\input{./phys/phys_nuclearmedium}
\input{./phys/phys_hypernuclei}
\input{./phys/phys_nucleonstructure}

\clearpage
\input{./phys/phys_electroweak}
%
%
\newpage
\bibliographystyle{panda_pb_lit}
\bibliography{./phys/lit_phys,./phys/lit_phys_heavylight,./phys/lit_phys_charmonium,./phys/lit_phys_gluonic,./phys/lit_phys_hypernuclei,./phys/lit_phys_nuclearmedium,./phys/lit_phys_baryons,./phys/lit_phys_qcd_dynamics,./phys/lit_phys_nucleonstructure_spinstructure,./phys/lit_phys_nucleonstructure_gda,./phys/lit_phys_nucleonstructure_elmff,./phys/lit_phys_electroweak,./main/lit_main}
%

%% file: phys/phys_overview.tex
%
%
\section{Overview}
\COM{Author(s): D. Bettoni, K. Peters}

The \PANDA experiment will use the
antiproton beam from the \HESR colliding with an
internal proton
target and a general purpose spectrometer to carry out a rich and
diversified hadron physics program.

The experiment is being designed to fully exploit the extraordinary 
physics potential arising from the availability of high-intensity, cooled
antiproton beams.
The aim of the rich experimental program is to improve our knowledge of the 
strong interaction and of hadron structure.
Significant progress beyond the present understanding of the field is expected
thanks to improvements in statistics and precision of the data.

Many measurements are foreseen in \PANDA.
\begin{itemize}
\item The study of {\bf QCD bound states} is of fundamental importance for a better, 
quantitative understanding of QCD. Particle spectra can be computed within the framework
of non-relativistic potential models, effective field theories and Lattice QCD. 
Precision measurements are needed to distinguish between the different approaches and identify
the relevant degrees of freedom. 
The measurements to be carried out in \PANDA include charmonium, D meson and baryon spectroscopy.  
In addition to that \PANDA will look for exotic states such as gluonic hadrons (hybrids and
glueballs), multiquark and molecular states.

\item {\bf Non-perturbative QCD Dynamics.} In the quark picture hyperon pair 
production either involves the
creation of a quark-antiquark pair or the knock out of such pairs out
of the nucleon sea.  Hence, the creation mechanism of quark-antiquark
pairs and their arrangement to hadrons can be studied by measuring the
reactions of the type $\pbarp \to \overline{Y} Y$, where $Y$
denotes a hyperon.  
Furthermore the self-analysing weak decays of most hyperons give
access to spin degrees of freedom for these reactions.
By comparing several\
reactions involving different quark flavours the OZI
rule, and its possible violation, can be tested for different levels of disconnected
quark-line diagrams separately.

\item {\bf Study of Hadrons in Nuclear Matter.} The study of medium 
modifications of hadrons embedded in hadronic matter
is aimed at understanding the origin of hadron masses in the context
of spontaneous chiral symmetry breaking in QCD and its partial restoration
in a hadronic environment. 
So far experiments have been focused on the light quark sector.
The high-intensity $\pbar$ beam of up to 15\,\gevc will allow an extension of
this program to the charm sector both for hadrons with hidden and open charm.
The in-medium masses of these states are expected to be affected primarily
by the gluon condensate.
A complementary study aims to unravel the onset of the regime where the Colour Transparency phenomenon reveals the short distance dominance of some exclusive reactions.

\item {\bf Hypernuclear Physics.} Hypernuclei are systems in which up or 
down quarks are replaced by strange
quarks. In this way a new quantum number, strangeness, is introduced into 
the nucleus.
Although single and double
$\Lambda$-hypernuclei were discovered many decades ago, only 6 double
$\Lambda$-hypernuclei are presently known. 
The availability of $\pbar$ beams at \FAIR will allow efficient production
of hypernuclei with more than one strange hadron, making \PANDA competitive
with planned dedicated facilities. This will open new perspectives for
nuclear structure spectroscopy and for studying the forces between hyperons
and nucleons. 

\item {\bf Electromagnetic Processes.} In addition to the spectroscopic 
studies described above \PANDA will be able to
investigate the structure of the nucleon using electromagnetic processes, such as
Deeply Virtual Compton Scattering (DCVS) and the process $\pbarp \to \ee$, which 
will allow the determination of the electromagnetic form factors of the proton 
in the timelike region over an extended $q^2$ region.
The associated process $\pbarp \to e^+ e^- \pi^0 $ should reveal the shape of the meson cloud in the nucleon through the study of the Transition Distribution Amplitudes describing the proton to pion transition.

\item {\bf Electroweak Physics.} With the high-intensity antiproton beam 
available at HESR a large
number of D-mesons can be produced. This gives the possibility to
observe rare weak decays of these mesons allowing to study electroweak
physics by probing predictions of the Standard Model and searching 
for enhancements introduced by processes beyond the Standard Model.
\end{itemize}

In this chapter we will discuss in detail the various items of the 
\PANDA physics program and we will describe the Monte Carlo simulations
which have been carried out on a number of benchmark channels to study acceptances, 
resolutions, background rejection.
In order to perform these studies a number of tools have been developed, which
are described in detail in the previous chapter, and which include
a full simulation of the detector response and a series of sophisticated
reconstruction and analysis tools. These studies have allowed us to
obtain reliable estimates for the performance of the \PANDA experiment
and the sensitivity of the various measurements.

%

%% file: phys/phys_qcd_states.tex
%
\section{QCD Bound States}
\COM{Author(s): D. Bettoni, K. Peters, A. Gillitzer}
%
\input{./phys/phys_boundstates}
\input{./phys/phys_charmonium}
\input{./phys/phys_exotic}
\input{./phys/phys_heavylight}
\input{./phys/phys_baryons}
%

%% file: phys/phys_boundstates.tex
%
\subsection{The QCD Spectrum}
\begin{figure*}[htb]
\begin{center}
\includegraphics[width=0.85\dwidth]{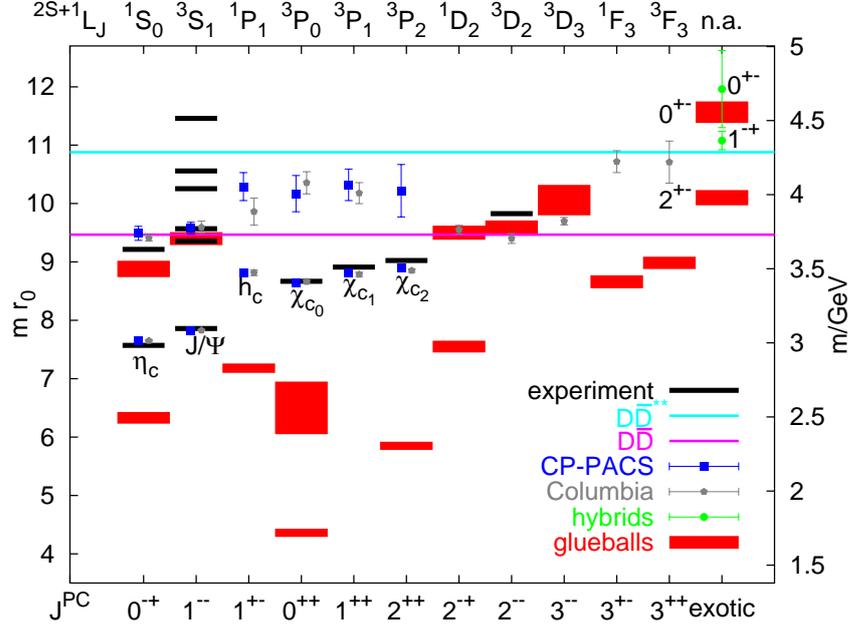}
\caption[Charmonium spectrum from LQCD]{Charmonium spectrum from LQCD.
See~\cite{qwgyr} for details.}
\label{fig:phys:exotic:ccbar_lqcd_spectrum}
\end{center}
\end{figure*}

%
%
The spectrum of charmonium, like bottomonium, is very similar (apart from the scale)
to the spectrum positronium. It is therefore suggestive to assume that 
$\ccbar$ an $\bbbar$ (so called conventional mesons) can be understood
in the strong interaction in an analogue way as positronium in electroweak interactions.
This would imply a Coulomb-like potential and a term which takes care of
linear confinement.
%
%
But this $a/r+br$ potential arising mainly from the exchange of one gluon
is by far not sufficient to explain the spectrum of hadrons.
The coherent exchange of gluons which manifest in a gluon tube
is an important aspect when one wants to understand the binding
among strongly interacting particles. 
Together with mesic excitations and pure gluonic states which
are possible in QCD they are are usually present in the wave function
of hadrons and are often referred
to as Fock-States of the ground state meson.
The Fock-States may decouple from the ground state and thus being individually
observable as individual states. This happens if the lifetime
of the objects allow a partitioning into individual objects
({\it e.g.} width smaller than the mass difference).
We see this usually happening for the first rotational excitations of hadrons.
For higher excitations they tend to end up in a continuum.
Therefore the QCD spectrum is richer than that of the naive quark model (see~\Reffig{fig:phys:exotic:ccbar_lqcd_spectrum}).
We distinguish conventional, gluonic and mesic hadrons. Conventional hadrons
have been already discussed in~\Refsec{sec:qcd:ccb}.
Gluonic hadrons fall into two categories: glueballs and
hybrids. Glueballs are predominantly excited states of glue while
hybrids are resonances consisting largely of a quark, an
antiquark, and excited glue. The properties of
glueballs and hybrids are determined by the long-distance features
of QCD and their study will yield fundamental insight into the
structure of the QCD vacuum. Mesic hadrons are dimesons and/or tetra-quarks,
often referred to as multi-quarks or baryonium.
They may be viewed as a loosely bound meson-anti-meson system or a diquark-antidiquark
ensemble.
\par
The search for glueballs and hybrids has mainly been restricted to
the mass region below 2.2$\,\gevcc$. Experimentally, it would be
very rewarding to go to higher masses because of the unavoidable
problems due to the high density of normal $\qqbar$ mesons below
2.5$\,\gevcc$.
\par
In the search for glueballs, a narrow state at 1500$\,\mevcc$,
discovered in antiproton annihilations by \INST{Crystal
Barrel}~\cite{bib:phy:cb1500a,bib:phy:cb1500b,bib:phy:cb1500c,bib:phy:cb1500d,bib:phy:cb1500e},
is considered the best candidate for the glueball ground state
($\JPC\,=\,\JPCzeropp$). However, the mixing with nearby
conventional scalar $\qqbar$ states makes the unique
interpretation as a glueball difficult.
\par
Both cases indicate the problems of light quark spectroscopy due
to large widths, deteriorated line-shapes and mixing among states.
Thus heavy quarks states are decay modes give a more unbiased view
to the spectrum of QCD states.
\par
In the past decades many resonances have been associated with the quest
of the existence of multi-quarks. The $a_0(980)$ and $f_0(975)$ were always believed
to have a strong $\KKbar$. Now a lot of charmonium states are discussed in
the same framework.
%

%% file: phys/phys_charmonium.tex
%
\subsection{Charmonium}
\label{sec:qcd:ccb}
\COM{Author(s): D. Bettoni, M. Negrini}
\COM{Referee(s): P. Gianotti}
%
\input{./phys/charmonium/charmonium_introduction}
%

%% file: phys/charmonium/charmonium_introduction.tex
%
%
\subsubsection{Introduction}
\label{sec:phys:qcd:ccbar:intro}

Ever since its discovery in 1974 \cite{psidisc1, psidisc2} charmonium
has been a powerful tool for the understanding of the strong
interaction.  The high mass of the $c$ quark (m$_c$ $\approx$ 1.5\,\gevcc) 
makes it plausible to attempt a description of the dynamical
properties of the ($\ccbar$) system in terms of non-relativistic
potential models, in which the functional form of the potential is
chosen to reproduce the asymptotic properties of the strong
interaction. The free parameters in these models are to be determined
from a comparison with the experimental data.
\par
Now, more than thirty years after the $\jpsi$ discovery, charmonium
physics continues to be an exciting and interesting field of research.
The recent discoveries of new states ($\etacprime$, X(3872)), and the
exploitation of the B factories as rich sources of charmonium states
have given rise to renewed interest in heavy quarkonia, and stimulated
a lot of experimental and theoretical activities.  The gross features
of the charmonium spectrum are reasonably well described by potential
models, but these obviously cannot tell the whole story: relativistic
corrections are important and other effects, like coupled-channel
effects, are significant and can considerably affect the properties of
the $\ccbar$ states. To explain the finer features of the charmonium
system, model calculations and predictions are made within various,
complementary theoretical frameworks. Substantial progress in an
effective field theoretical approach, labelled Non-Relativistic QCD
(NRQCD) has been achieved in recent years. This analytical approach
makes it possible to expect significant progress in lattice gauge
theory calculations, which have become increasingly more capable of
dealing quantitatively with non-perturbative dynamics in all its
aspects, starting from the first principles of QCD.
\subsubsection*{Experimental Study of Charmonium}
Experimentally charmonium has been studied mainly in $\ee$ and $\pbarp$ experiments.
\par
In $\ee$ annihilations direct charmonium formation is possible 
only for states with the quantum numbers of the 
photon $\JPC = \JPConemm$, namely  the 
$\jpsi$, $\psiprime$ and $\psi(3770)$ resonances.
Precise measurements of the masses and widths of these states can be
obtained from the energy of the electron and positron beams, which
are known with good accuracy.
All other states can be reached by means of other production mechanisms,
such as photon-photon fusion, initial state radiation, B-meson decay
and double charmonium.
\par
On the other hand all $\ccbar$ states can be directly formed in $\pbarp$
annihilations, through the coherent annihilation of the three quarks in the
proton with the three antiquarks in the antiproton. 
This technique, originally
proposed by P. Dalpiaz in 1979 \cite{dalpiaz79}, 
could be successfully employed a few years 
later at \INST{CERN} and \INST{Fermilab} thanks to the development of stochastic cooling.
With this method the masses and widths of all charmonium states can be measured 
with excellent accuracy, determined by the very precise knowledge of the initial
$\pbarp$ state and not limited by the resolution of the detector.
 \par
The parameters of a given resonance can be extracted by
measuring the formation rate for that resonance as a
function of the c.m. energy $E_{cm}$, as explained in
detail in section 2.4.1.
\begin{figure}[htb]
\begin{center}
\includegraphics[width=\swidth]{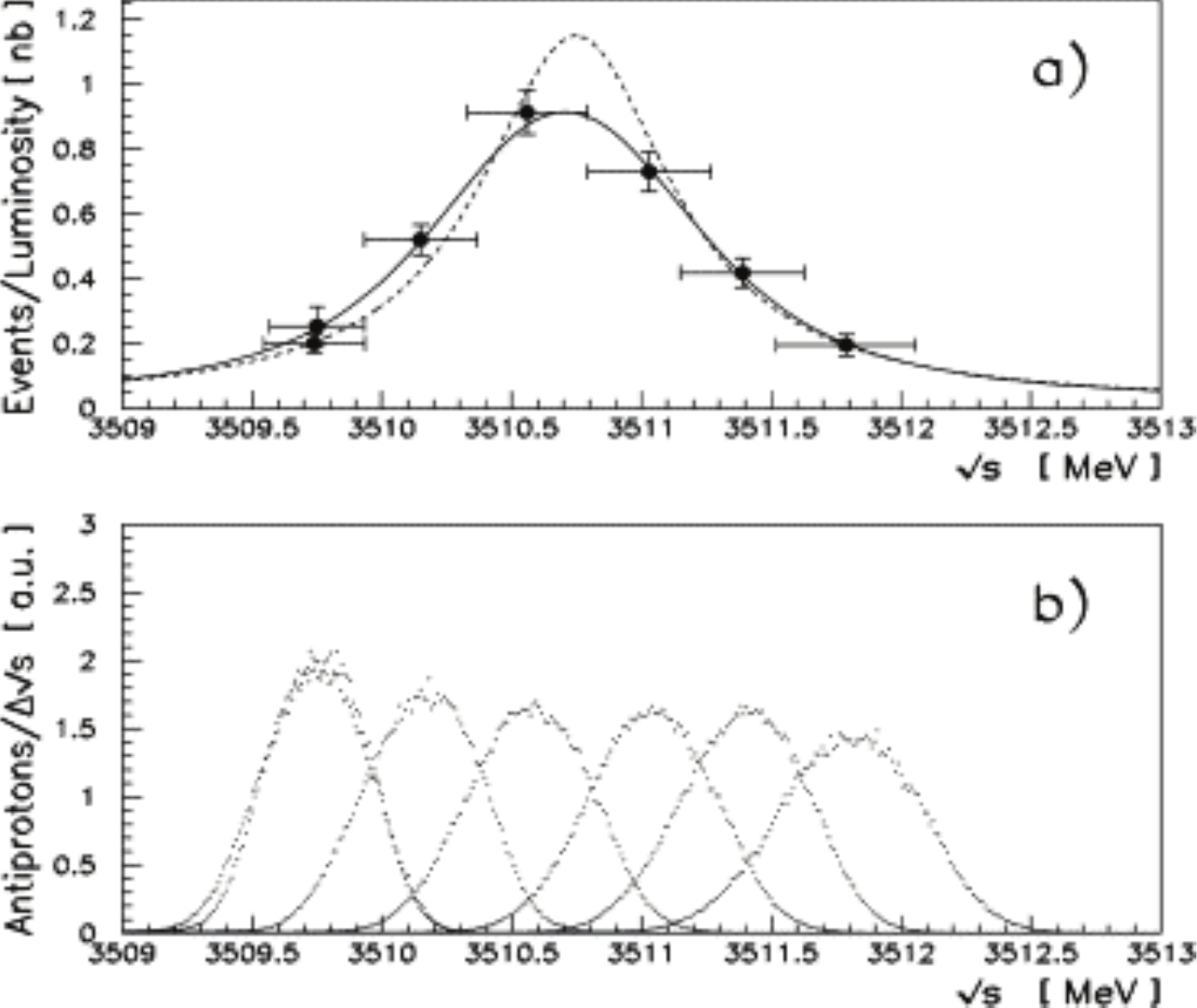}
\caption[Resonance scan at the $\chicone$.]
{Resonance scan at the $\chicone$ carried out at \INST{Fermilab} (a) and beam energy
distribution in each data point (b). 
\label{fig:chi1scan}}
\end{center}
\end{figure}
As an illustration of this technique we show in \Reffig{fig:chi1scan} a scan of the
$\chicone$ resonance carried out at the \INST{Fermilab} antiproton accumulator by the E835
experiment~\cite{chi1E835} using the process $\pbarp \to \chicone \to \jpsi\gamma$. 
For each point of the scan the horizontal error bar in (a)
corresponds to the width of the beam energy distribution. The actual beam energy
distribution is shown in (b). This scan allowed the E835 experiment to carry out the
most precise measurement of the mass ($3510.719 \pm 0.051 \pm 0.019\,\mevcc$) and total
width ($0.876 \pm 0.045 \pm 0.026\,\mev$) of this resonance.
\subsubsection*{The Charmonium Spectrum}
The spectrum of charmonium states is shown in \Reffig{fig:phys:exotic:ccbar_lqcd_spectrum}.
It consists of eight narrow states below the open charm threshold  (3.73\,GeV) and 
several tens of states above the threshold.
\par
All eight states {\bf below $\DDbar$ threshold} are well established, but whereas 
the triplet states are measured with very good accuracy, the same cannot be said 
for the singlet states.
\par
The $\etac$ was discovered almost thirty years ago and many measurements of its
mass and total width exist, with six new measurements in the last four years. Still
the situation is far from satisfactory. 
The Particle Data Group (PDG)~\cite{PDG} value of the mass 
is $2980.4 \pm 1.2\,\mevcc$, an
average of eight measurements with an internal confidence level of 0.026: the
error on the $\etac$ mass is still as large as $1.2\,\mevcc$, to be compared with few
tens of $\kevcc$ for the $\jpsi$ and $\psiprime$ and few hundreds of $\kevcc$ for the $\chi_{cJ}$
states. The situation is even worse for the total width: the PDG average is 
$25.5 \pm 3.4\,\mev$, with an overall confidence level of only 0.001 and individual
measurements ranging from $7\,\mev$ to $34.3\,\mev$. The most recent measurements 
have shown that the $\etac$ width is larger than was previously believed, with
values which are difficult to accommodate in quark models. 
This situation points
to the need for new high-precision measurements of the $\etac$ parameters. 
\par
The first experimental evidence of the $\etac(2S)$ was reported by the \INST{Crystal Ball} 
collaboration~\cite{etacpCB}, but this finding was not confirmed in
subsequent searches in $\pbarp$ or $\ee$ experiments. The $\etac(2S)$ was finally
discovered by the \INST{Belle} collaboration~\cite{etacpBelle}
in the hadronic decay of the B meson
$B \to K + \etac (2S) \to K+ (K_sK^-\pip)$ with a mass which was incompatible with
the \INST{Crystal Ball} candidate. The Belle finding was then confirmed by \INST{CLEO}~\cite{etacpCLEO}
and \INST{BaBar}~\cite{etacpbabar} which observed this state in two-photon fusion.
The PDG value of the mass is $3638 \pm 4\,\mevcc$, corresponding to a surprisingly
small hyperfine splitting of $48 \pm 4\,\mevcc$, whereas the total width is only 
measured with an accuracy of 50\percent. The study of this state has just started and all
its properties need to be measured with good accuracy.
\par
The $^1P_1$ state of charmonium ($\honec$) is of particular importance in the
determination of the spin-dependent component of the $\qqbar$ confinement potential.
The \INST{Fermilab} experiment \INST{E760} reported an $\honec$ candidate in the decay channel
$\jpsi\pi^0$~\cite{1p1E760}, with a mass of $3526.2 \pm 0.15 \pm 0.2\,\mevcc$. 
This finding was not confirmed by the successor experiment
E835, which however observed an enhancement in the $\etac\gamma$~\cite{bib:sim:hc_E835}
final state at a mass
of $3525.8 \pm 0.2 \pm 0.2\,\mevcc$. 
The $\honec$ was finally observed by the \INST{CLEO} collaboration~\cite{1p1CLEO}
in the process
$\ee \to \psiprime \to \honec + \piz$ with $\honec \to \etac + \gamma$, in 
which the $\etac$ was identified via its hadronic decays. They
found a value for the mass of $3524.4 \pm 0.6 \pm 0.4\,\mevcc$.
It is clear that the study of this state has just started and that many more
measurements will be needed to determine its properties, in particular the width.
\par
The region {\bf above $\DDbar$ threshold} is rich in interesting new physics. 
In this region, close to the $\DDbar$ 
threshold, one expects to find the four $1D$ states. 
Of these only the $1^3D_1$, identified with the
$\psi$(3770) resonance, has been found.
The $J = 2$ 
states ($1^1D_2$ and $1^3D_2$) are predicted to be narrow, because
parity conservation forbids their decay to $\DDbar$. 
In addition to the $D$ states, the radial excitations of the $S$ and
$P$ states are predicted to occur above the open charm threshold.
None of these states have been positively identified.
\par
The experimental knowledge of this energy region comes from data taken at the
early $\ee$ experiments at \INST{SLAC} and \INST{DESY} and, more recently, at the $B$-factories, \INST{CLEO-c}
and \INST{BES}.
The structures and the higher vector states observed by the early $\ee$ experiments
have not all been confirmed by the latest much more accurate 
measurements by \INST{BES}~\cite{RBES1, RBES2}. 
A lot of new states have recently been discovered at the $B$-factories, mainly in
the hadronic decays of the $B$ meson: these new states ($X$, $Y$, $Z$ ...) are associated
with charmonium because they decay predominantly into charmonium states such
as the $\jpsi$ or the $\psiprime$, but their interpretation is far from obvious.
The situation can be roughly summarised as follows:
\begin{itemize}
\item the $Z(3931)$~\cite{Z3931}, observed in two-photon fusion and decaying predominantly
 into $\DDbar$, is tentatively identified with the $\chictwo(2P)$;
\item the $X(3940)$~\cite{X3940}, observed in double charmonium events, is 
tentatively identified with the $\etac(3S)$;
\item for all other new states ($X(3872)$, $Y(3940)$, $Y(4260)$, $Y(4320)$ and so on) the
interpretation is not at all clear, with speculations ranging from the missing
$\ccbar$ states, to molecules, tetraquark states, and hybrids. 
It is obvious that further measurements are needed to determine the nature
of these new resonances.
\end{itemize}
\par
The main challenge of the next years will be thus to understand what these new 
states are and to match these experimental findings to the theoretical expectations
for charmonium above threshold. 
\subsubsection*{Charmonium in \PANDA}
Charmonium spectroscopy is one of the main items in the experimental program
of \PANDA, and the design of the detector and of the accelerator are optimised
to be well suited for this kind of physics. 
\PANDA will represent a substantial improvement over the \INST{Fermilab} experiments
\INST{E760} and \INST{E835}:
\begin{itemize}
\item up to ten times higher instantaneous luminosity 
(${L} = 2 \times 10^{32}$\,cm$^{-2}$s$^{-1}$ in high-luminosity mode, compared
to $2 \times 10^{31}$\,cm$^{-2}$s$^{-1}$ at \INST{Fermilab});
\item better beam momentum resolution ($\Delta p/p = 10^{-5}$ in high-resolution
mode, compared with $10^{-4}$ at \INST{Fermilab});
\item a better detector (higher angular coverage, magnetic field, ability to
detect the hadronic decay modes).
\end{itemize}
At full luminosity \PANDA will be able to collect several thousand $\ccbar$ states
per day.
By means of fine scans it will be possible to measure masses with accuracies of the
order of 100\,keV and widths to 10\percent or better.
The entire energy region below and above open charm threshold will be explored.
\subsubsection{Benchmark Channels}
\label{sec:phys:qcd:ccbar:bench}
One of the main problems in the experimental study of charmonium spectroscopy in $\pbarp$
annihilation is the high hadronic background. 
It is therefore necessary to select those decays of charmonium which are less affected 
by background. 
In general we can identify four main classes of charmonium decay:
\begin{itemize}
\item {\bf decays with a} $\boldsymbol{\jpsi}$ {\bf in the final state}: 
$\ccbar \to \jpsi + X$, with $\jpsi \to \ee$ or $\jpsi \to \mumu$.
These channels can be used to identify states such as the $\chicJ$, the $\psiprime$ or the
$X(3872)$.
The presence of the lepton pair in the final state
makes these channels relatively clean, with the main background coming from misidentified $\pip\pim$
pairs.
The analysis of these channels relies on the positive identification of the lepton pair in
the final state, with an invariant mass compatible with the $\jpsi$.
In case of an exclusive analysis (e.g. $\jpsi\pip\pim$) a further improvement on the
signal to background ratio can be obtained by means of a kinematical fit;
\item {\bf two- and three-photon decays} can be used to identify states such as the $\etac$, the
$\etacprime$ or the $\honec$ (via $\honec \to \etac + \gamma \to 3\gamma$).
The main background comes from $\piz\gamma$ and $\piz\piz$ final states in which one or two photons
are lost, either outside the calorimeter acceptance or below the calorimeter low-energy threshold.
This background can be calculated by measuring the $\piz\gamma$ and $\piz\piz$ cross sections and 
then using Monte Carlo techniques to estimate the feed-down to $\gamma\gamma$ and $3\gamma$.
The analysis requires the presence of the required number of photons in the final state, with a veto
on $\piz$s and charged particles. Also in this case a kinematical fit can help to improve the background
situation;
\item {\bf decays to light hadrons}. As stated previously these decay modes are affected by a large
hadronic background. In this case signal/background separation can be obtained by cutting on
discriminating topological variables or angular distributions, whenever possible.
An example of such decay is 
$\honec \to \etac \gamma \to \phi \phi \gamma$;
\item {\bf decays to} $\boldsymbol{\DDbar}$:
charmonium states above open charm threshold will generally be identified
by means of their decay to $\DDbar$, unless forbidden by some conservation rule. 
\end{itemize}
\par
In what follows we will present a discussion of individual benchmark channels.
\subsubsection{$\boldsymbol{\pbarp \to \jpsi \,+ \, X}$}
\begin{figure}[htbp]
  \begin{center}
  \includegraphics[angle=90,width=\columnwidth]{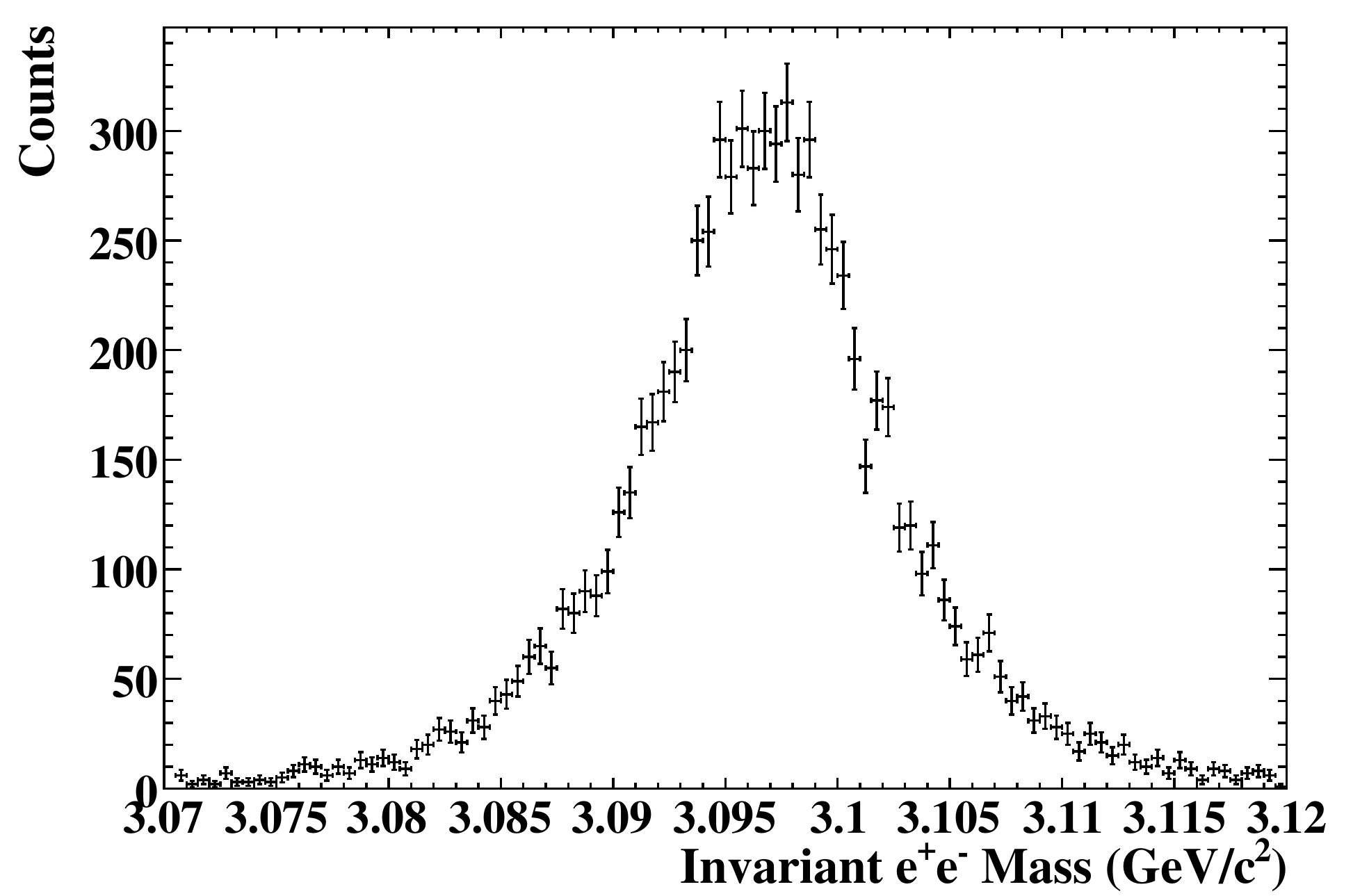}
  \end{center}
\caption{Invariant $\ee$ mass reconstructed at $\sqrt{s}=4.260$\,GeV.}
\label{fig:jpsiMass}
\end{figure}

A class of charmonium decays that will be studied at \PANDA presents a $\jpsi$ in the final state resulting, 
for example, from de-excitation of a higher level charmonia with the emission of hadrons or photons.  
The existence of a $\jpsi$ in the final state represents a clean signature for the signal.\\
In what follows we present the study of the benchmark channels:
\begin{itemize}
\item $\pbarp \to \jpsi\pip\pim\to \ee\pip\pim$;
\item $\pbarp\to \jpsi\piz\piz\to \ee\gamma\gamma\gamma\gamma$;
\item $\pbarp\to\chi_{c1,c2}\gamma\to \jpsi\gamma\gamma\to \ee\gamma\gamma$;
\item $\pbarp\to \jpsi\gamma\to \ee\gamma$;
\item $\pbarp\to \jpsi\eta$.
\end{itemize}
In all these cases the analysis strategy will focus on the detection of the $\jpsi\to \ee$ 
in the final state, allowing an efficient rejection of the hadronic background, 
followed by the full event reconstruction.
%
The first step in the analysis is the reconstruction of the $\jpsi$ candidate starting from the lepton pair, 
following this strategy:
\begin{enumerate}
 \item select one electron candidate from charged tracks with \loo PID criteria, and one electron candidate 
with Tight PID criteria;
\item kinematical fit of both electrons in order to reconstruct the $\jpsi$ candidates with vertex constraint;
\item probability of $\jpsi$ vertex fit: $P_{\jpsi}>0.001$.
\end{enumerate}
This strategy was followed for each benchmark channel.\\
\Reffig{fig:jpsiMass} shows the invariant mass distribution of the $\jpsi$ candidates 
for $\pbarp\to Y(4260)\to \jpsi\pip\pim$.\\
The results of the reconstructed $\jpsi$, at different centre-of-mass energies and for 
several channels, are summarised in \Reftbl{tab:jpsi}
\begin{table*}[htbp]
\begin{center}
\begin{tabular}{c c c c c}
\hline
\hline
Channel   &  Events  &  $\sqrt{s} (GeV)$   &    Mean (GeV) &   RMS (MeV)\\
\hline
$\jpsi\pip\pim$ &  25\,k  &3.526              &      3.097   &     2.5 \\
& 25\,k & 3.686            &    3.097   &     3.9\\
& 25\,k & 3.872            &    3.097   &     4.3\\
& 25\,k & 4.260            &    3.097   &     7.0\\
& 25\,k & 4.600            &    3.097   &     5.7\\
& 25\,k & 5.000            &    3.097   &     6.4\\
\hline
$\jpsi\piz\piz$ &  360\,k  & 4.260 &  3.096   & 8.8\\
\hline
$\chi_{c1}\gamma$ &   20\,k  &3.686              &   3.096      &        6.8  \\
& 20\,k & 3.872              &  3.096      &        7.6  \\
& 20\,k & 4.260              &  3.095      &        8.3  \\
\hline
$\chi_{c2}\gamma$   &  20\,k &3.686       &    3.096    &          6.9\\
& 20\,k & 3.872              &  3.096      &         7.5\\
& 20\,k & 4.260              & 3.096      &          8.3\\
\hline
$\jpsi\gamma$ &        100\,k  &3.510              &   3.097      &          3.1\\
& 100\,k & 3.556              &  3.097      &    3.4\\
& 20\,k & 3.872              &  3.096      &   3.7\\
\hline
$\jpsi\eta$   &       40\,k   & 3.638    &    3.093    &    7   \\
&  40\,k  &   3.686    &     3.094     &     7   \\
&   40\,k    &   3.872    &    3.096    &    6\\
&    80\,k   &   4.260   &   3.096    &    6\\
\hline\hline
\end{tabular}
\caption[Number of simulated events, mean value and RMS of the reconstructed $\jpsi$]
{Number of simulated events, mean value and RMS of the reconstructed $\jpsi$ 
invariant mass distribution for each energy and channel analysed, after 4C fit.}
\label{tab:jpsi}
\end{center}
\end{table*}
\par

We now turn to a detailed discussion of the five benchmark channels.

\subsubsection*{$\boldsymbol{\pbarp\to \jpsi\pip\pim}$}
\par
\begin{figure}[htbp]
  \begin{center}
  \includegraphics[angle=90,width=\columnwidth]{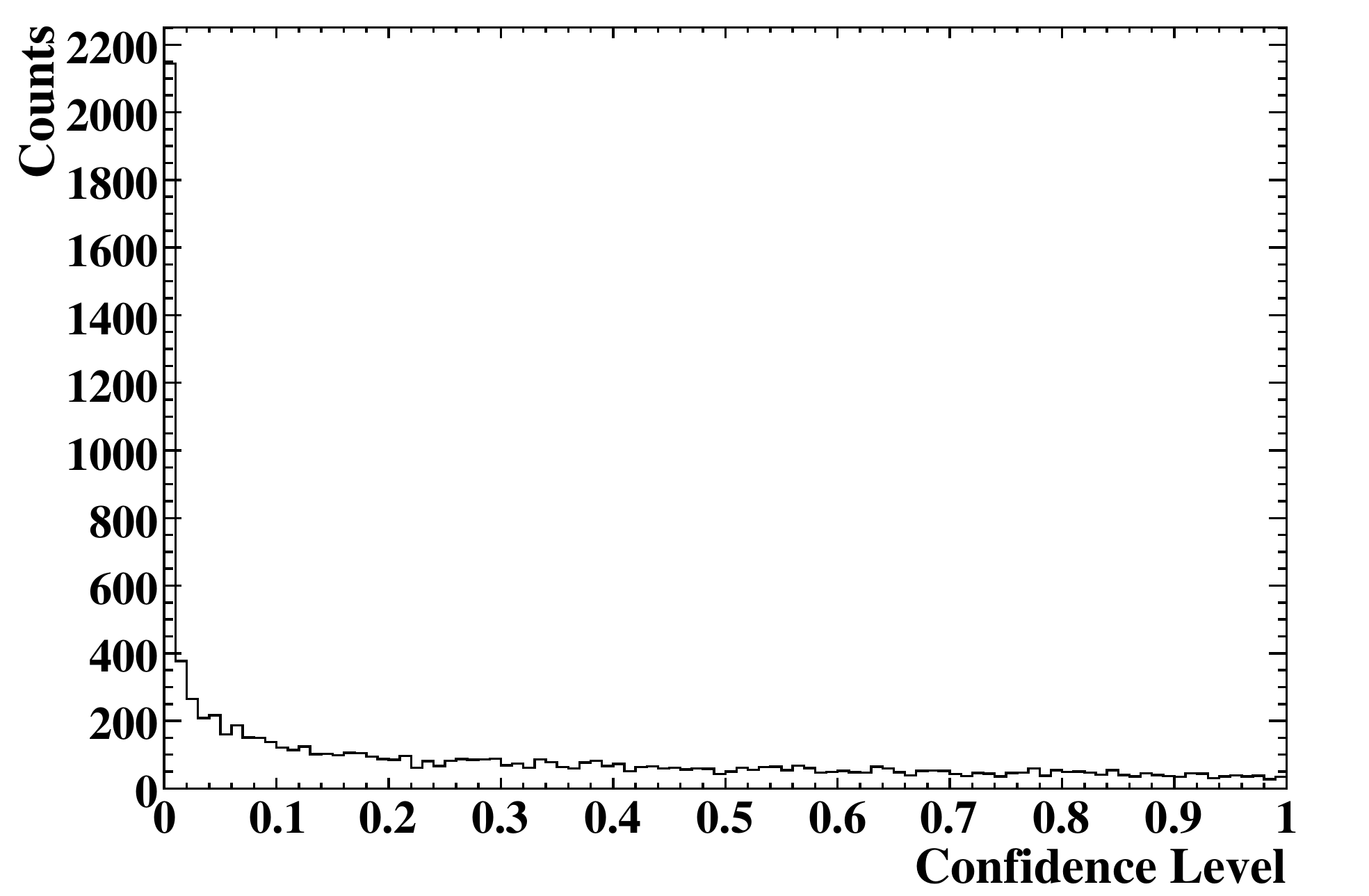}
  \end{center}
\caption[Confidence level of the kinematical fit to $\jpsi\pip\pim$ for the data simulated 
at $\sqrt{s}=4.260$\,GeV energy.]{ The cut applied is $P_{\jpsi\pip\pim}>0.001$.}
\label{fig:ppbprob}
\end{figure}

The reaction $\pbarp\to \jpsi\pip\pim\to \ee\pip\pim$ has been 
simulated for several centre-of-mass energies.\\
Event selection is done in the following steps:
\begin{enumerate}
 \item select a well reconstructed $\jpsi$ in the event;
\item select two pion candidates from charged tracks with \vloo PID criteria;
\item kinematical fit of the $\jpsi\pip\pim$ candidates with vertex constraint;
\item probability of $\jpsi\pip\pim$ vertex fit: $P_{\jpsi\pip\pim}>0.001$.
\end{enumerate}
Fig. \ref{fig:ppbprob} shows the confidence level of the fit for the data simulated at $\sqrt{s}=4.260$\,GeV.
\par
After this selection, the reconstruction efficiency and RMS of the invariant mass distribution 
are reported in \Reftbl{tab:ppb}.
\begin{table}[htbp]
\begin{center}
\begin{tabular}{c c c }
\hline
\hline
$\sqrt{s}$ [GeV]   &    Eff [\%] &   RMS [MeV]\\
\hline
3.526              &      27.52   &     3.7 \\
3.686              &      30.90   &     5.7\\
3.872              &      32.07   &     8.3\\
4.260              &      32.58   &     13.4\\
4.600              &      30.60   &     18.5\\
5.000              &      29.70   &     24.3\\
\hline
\hline
\end{tabular}
\caption[Efficiencies and RMS of the reconstructed $\jpsi\pip\pim$ invariant 
mass distributions]{Efficiencies and RMS of the reconstructed $\jpsi\pip\pim$ invariant 
mass distributions for each energy analysed.}
\label{tab:ppb}
\end{center}
\end{table}
\par
\begin{figure}[htbp]
  \begin{center}
  \includegraphics[angle=90,width=\columnwidth]{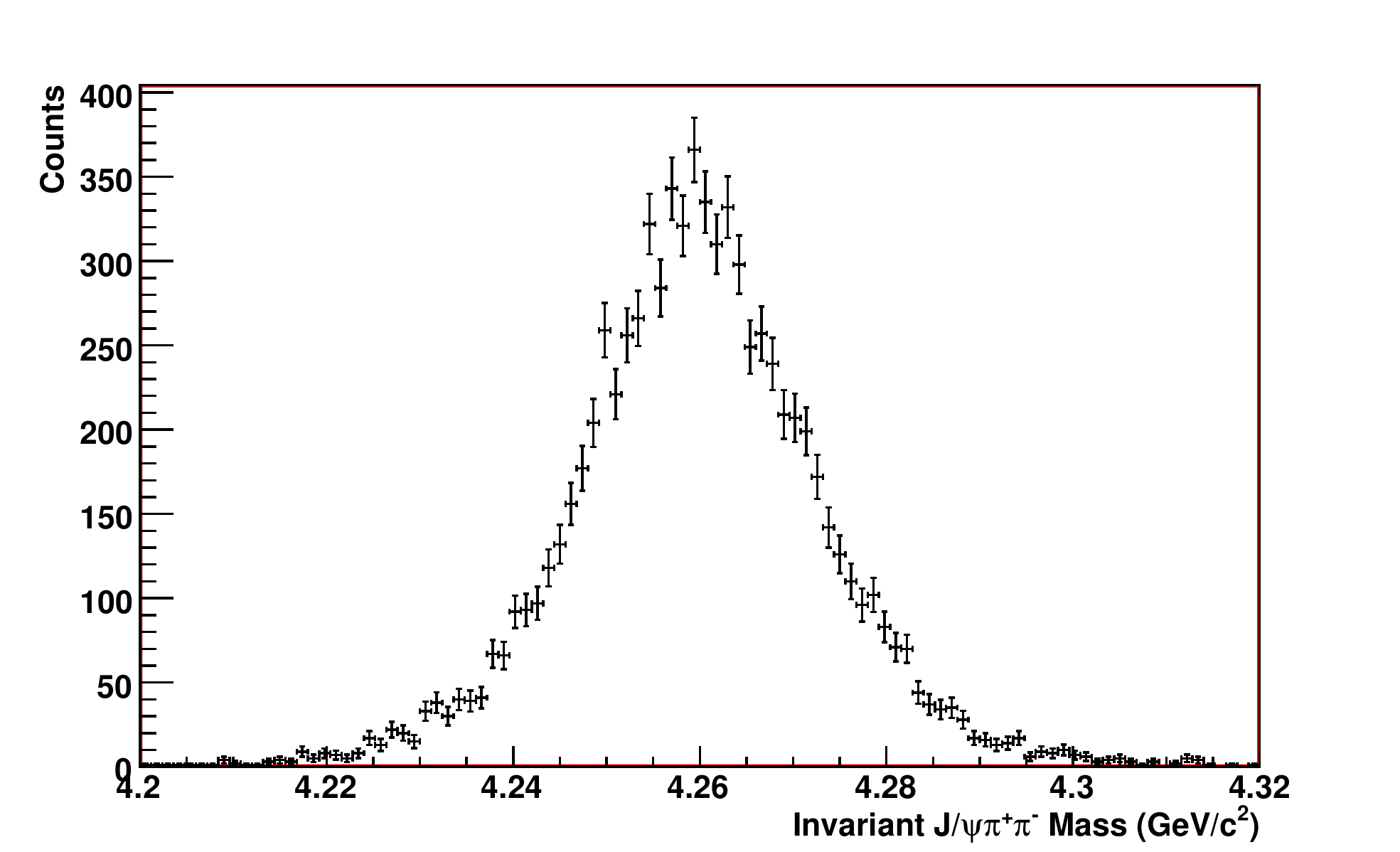}
  \end{center}
\caption{Invariant $\jpsi\pip\pim$ mass, in the case of Y(4260) resonance. }
\label{fig:ppbMass}
\end{figure}

As an example, we will discuss in more detail the results obtained for this channel 
at the energy of the resonance $Y(4260)$. This resonance was observed for the first time by 
BaBar in Initial State Radiation events \cite{Y}, in the decay $Y(4260)\to \jpsi\pip\pim$.  
The natural quantum number assignment for this state is $J^{PC}=1^{--}$ and one of its 
possible interpretation is a hybrid. \\
In \Reffig{fig:ppbMass} the invariant mass distribution for $Y(4260)$ candidates, 
obtained with the described selections, is presented. 
\par
In the simulation, the dipion invariant mass ($m_{\pi\pi}$) distribution was implemented according to the 
following parametrisation \cite{angular}:
\begin{equation}
\frac{d\Gamma}{dm_{\pi\pi}}\propto PHSP\cdot\left(m_{\pi\pi}^2-\lambda m_{\pi}^2\right)^2
\end{equation}
where PHSP is the phase-space factor,
$m_{\pi}$ is the pion mass and $\lambda$ is a parameter that can be obtained 
from the data; in this analysis we used $\lambda = 4.0$ \cite{tesi}. 
This choice is motivated by measurements of $\psi(2S)\to \jpsi\pip\pim$, 
considering the fact that $\psi^{\prime}$ and $Y(4260)$ have the same quantum numbers.\\ 
\begin{figure}[htbp]
  \begin{center}
  \includegraphics[angle=90,width=\columnwidth]{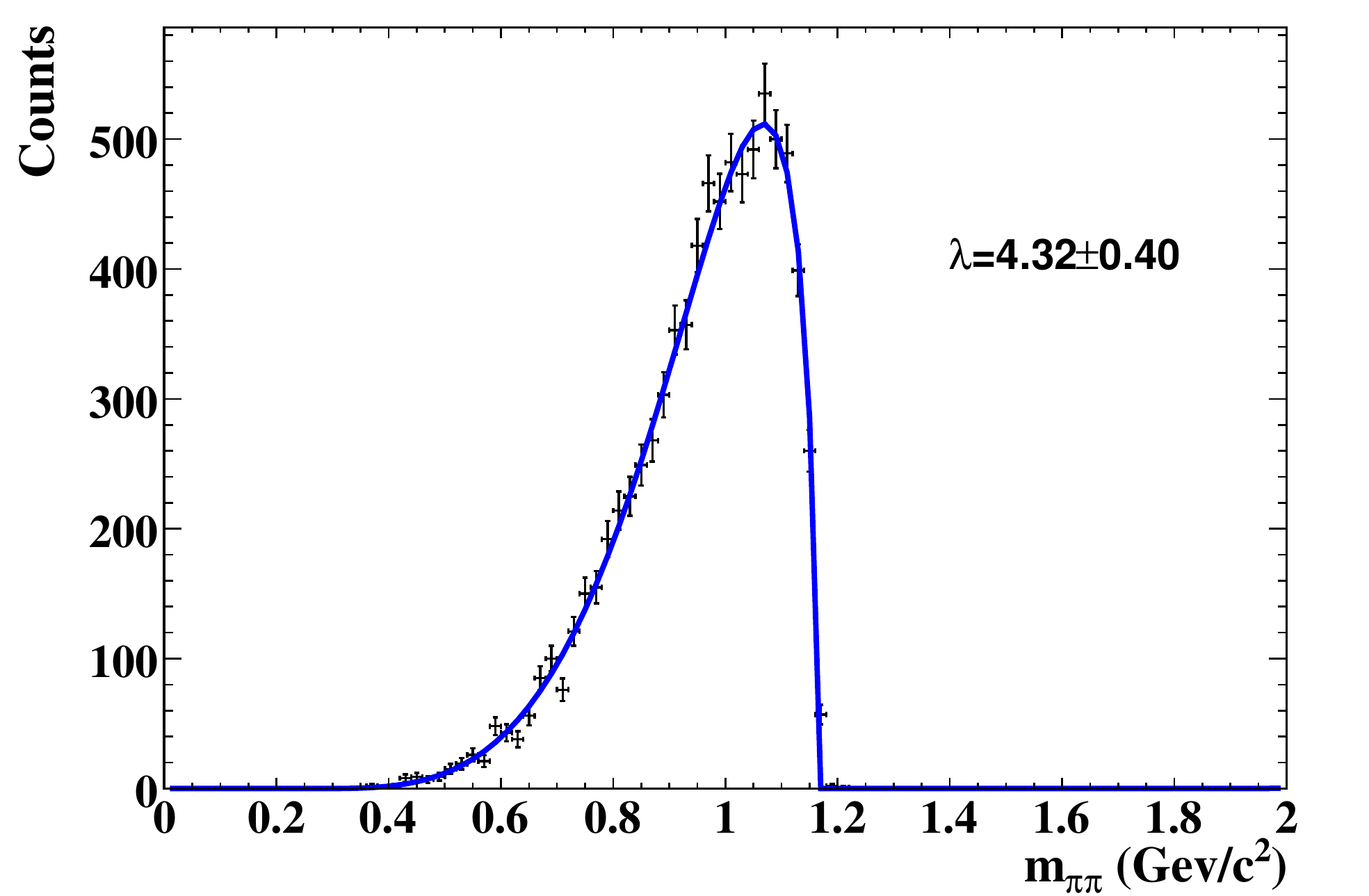}
  \end{center}
\caption[Invariant dipion mass of Y(4260) candidates.]{ The black line is the simulated and reconstructed 
data and the blue line is the fit with the theoretical function. 
The result of $\lambda$ after the fit seems to be consistent with the input data.}
\label{fig:mpipi}
\end{figure} 
\Reffig{fig:mpipi} shows the $m_{\pi\pi}$ distribution after the reconstruction. 
The blue line is the result of the fit with the theoretical formula, which is consistent with the input data. 
The main background for this channel comes from $\pbarp\to\pip\pim\pip\pim$ where two pions 
may be misidentified as electrons and contaminate the signal. \\
The study of background contamination is done only at the $Y(4260)$ energy.
At $\sqrt{s}=4.260$\,GeV the cross section of the background reaction is approximately equal 
to 0.046\,mb \cite{bkg}, while using available data from E835 experiment \cite{tesi}, 
we can estimate the cross section of $\pbarp\to Y(4260)\to \jpsi\pip\pim\to \ee\pip\pim$ 
to be about 60 pb.\\
In order to estimate the signal/noise ratio, 55\,M background filtered events were simulated 
(filter efficiency: 16.66\%). 
Only 60 events satisfy the selection criteria, and present an invariant mass of the reconstructed 
$\jpsi$ in the region between [2.8;3.2]\,\gevcc, and none events show a peak at the $\jpsi$ mass. 
We conclude that the signal/noise ratio is about 2, so this channel could be well identified in \PANDA.
%

\subsubsection*{$\boldsymbol{\pbarp\to \jpsi\piz\piz}$}
\par
With its excellent electromagnetic calorimeter \PANDA will also
be able to study the neutral dipion transition into $\jpsi\piz\piz$ in great detail.
In order to determine the acceptance and background rejection capability of the detector,
Monte Carlo simulations have been done for this channel at $\sqrt{s}=4.26$\,GeV.

\par
The event selection has been done in the following way.
The $\jpsi$ is reconstructed through the decay mode $\ee$
with the same cuts as described in the $\jpsi\pip\pim$ selection.
Photons from \piz candidates must have an energy deposit in the calorimeter larger than 20 MeV.
After the 4C fit with CL$>0.1\percent$, only those events with
m$(\ee)$ within $[3.07;3.12]$\,\gevcc and m$(\gamma\gamma)$ within $[120;150]$\,\mevcc
are accepted.
In order to reduce background, the remaining events are fitted
with $\jpsi\piz\piz$ and $\jpsi\eta\piz$ hypothesis.
Only events with exactly one combination with CL($\jpsi\piz\piz$)$>0.1\percent$
pass the event selection.
Events with at least one $\jpsi\eta\piz$ combination with CL($\jpsi\eta\piz$)$>0.01\percent$
are rejected.
\par
The results are summarised in \Reftbl{tab:perf:jpizpiz_result}.
Assuming a cross section for $\pbarp \to \jpsi\piz\piz \to \ee 4\gamma$ of 30 pb \cite{tesi} at $\sqrt{s}=4.26$\,GeV,
\PANDA will be able to reconstruct about 40 events per day.
The main background channels could be sufficiently suppressed.
Only 1 from 250 million simulated $\pip\pim\piz\piz$ events pass the event selection, which
results in a signal/background ratio S/B$=25$ .
\begin{table*}[htbp]
\begin{center}
\begin{tabular}{l l l l}
\hline
\hline
channel   &    assumed $\sigma$  &  efficiency  &   \\
\hline
$\pbarp \to \jpsi\piz\piz \to \ee 4\gamma$    &  30\,pb   & 16.9\percent & $n_{rec}$= 40 events / day \\
\hline
background reactions: \\
$\pbarp \to \pip\pim\piz\piz \to \pip\pim 4\gamma$    &  50\,$\mu$b  & 1 / 250M  & S/B$=25$ \\
$\pbarp \to \jpsi\eta\piz \to \ee 4\gamma$            &  $<$30\,pb   & 0 / 20K   & S/B$>10^{3}$ \\
$\pbarp \to \jpsi\omega\piz \to \ee 5\gamma$          &  $<$10\,pb   & 4 / 20K   & S/B$>10^{3}$ \\
\hline
\end{tabular}
\caption[results $\jpsi\piz\piz$]{Simulation results for the channel $\pbarp \to \jpsi\piz\piz$. 
To save computing time
$\pip\pim\piz\piz$ events with $m_{\pip\pim} < 2.4$\,\gevcc or $m_{\pip\pim} > 3.4$\,\gevcc are rejected without
detector simulation.
}
\label{tab:perf:jpizpiz_result}
\end{center}
\end{table*}

\begin{figure}[htbp]
 \begin{center}
\includegraphics[angle=0,width=\columnwidth]{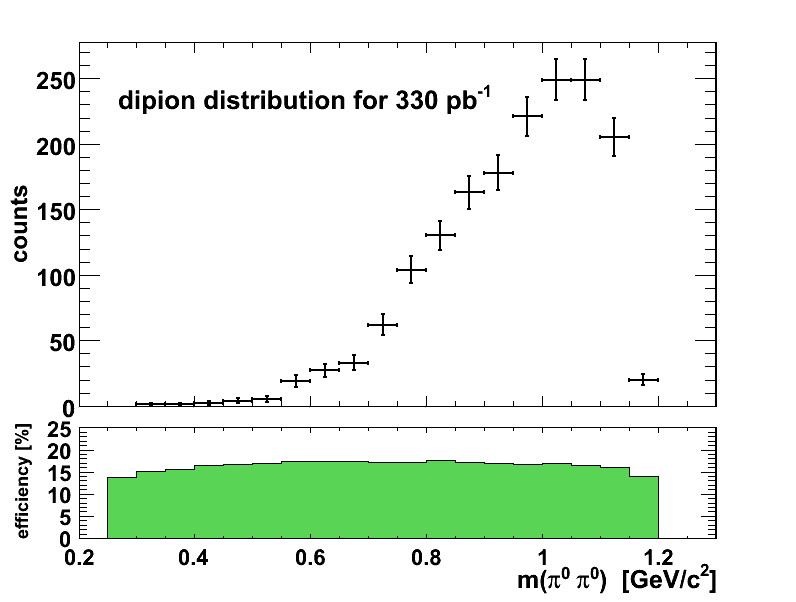}
  \end{center}
\caption{Invariant dipion mass of $\jpsi\piz\piz$ candidates.}
\label{fig:dipion}
\end{figure}

The dipion mass distribution which is simulated with the same shape as that of
the decay of $Y(4260)\to \jpsi\pip\pim$ is shown in \Reffig{fig:dipion}.
No strong efficiency variation in the $m_{\piz\piz}$ spectrum is visible.

%
\subsubsection*{$\boldsymbol{\pbarp\to \chi_{c}\gamma}$}
\par
\begin{figure}[htbp]
  \begin{center}
  \includegraphics[angle=90,width=\columnwidth]{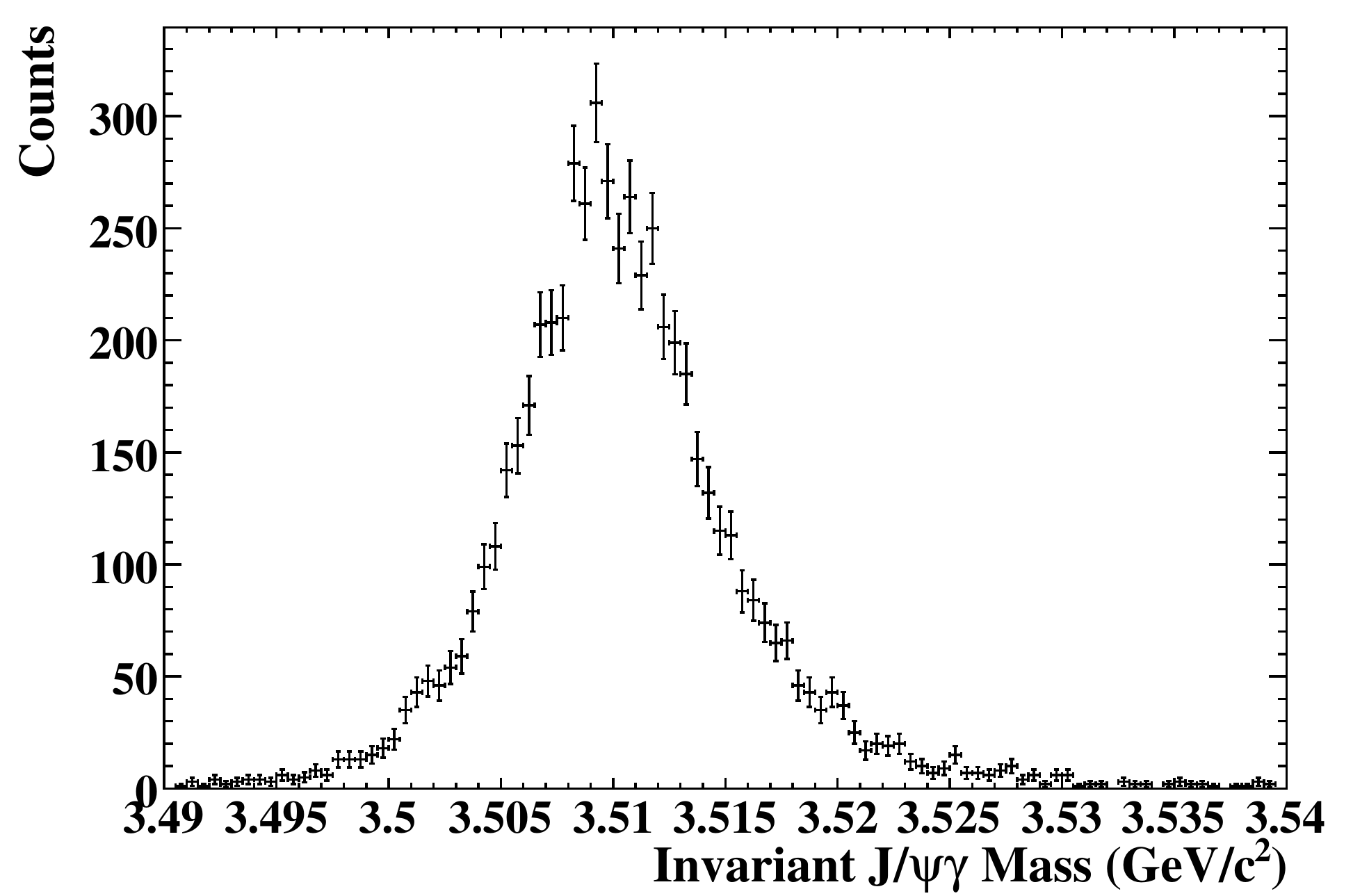}
  \end{center}
\caption{Invariant $\jpsi\gamma$ mass at $\sqrt{s}=3.686$\,GeV.}
\label{fig:chiMass}
\end{figure}

For the study of radiative decays to $\chi_c$, it is possible to make use of the subsequent 
decay $\chi_c\to \jpsi\gamma$. Starting from the $\jpsi$ sample it is possible to add a 
photon to reconstruct the $\chi_c$ candidate; a second photon is then added to fully reconstruct 
the final state.\\
We will consider the following processes:
\begin{equation}
\pbarp\to\chi_{c1,c2}\gamma\to \jpsi\gamma\gamma\to \ee\gamma\gamma\ .
\end{equation}
The event selection is done in the following steps:
\begin{enumerate}
\item select a well reconstructed $\jpsi$ in the event;
\item select reconstructed photon candidates;
\item kinematic fit of the $\chi_{c1,c2}$ candidates with vertex constraint;
\item probability of $\chi_{c1,c2}$ vertex fit: $P_{\chi_{c1,c2}}>0.001$;
\item probability of $\chi_{c1,c2}\gamma$ vertex fit: $P_{\chi_{c1,c2}\gamma}>0.001$;
\item $\chi_{c1,c2}$ mass window: [3.3,3.7]\,GeV.
\end{enumerate}
The results for the reconstructed $\chi_{c1,c2}$ are summarised in \Reftbl{tab:chi1}.
\begin{table*}[htbp]
\begin{center}
\begin{tabular}{c c c c c}
\hline
\hline
           &  \multicolumn{2}{c}{$\chicone$}  &  \multicolumn{2}{c}{$\chictwo$}\\
\hline
$\sqrt{s} [GeV]$   &   Mean [GeV] &    RMS [MeV]  &    Mean [MeV] &    RMS [MeV]  \\
\hline
3.686              &   3.510     &        5.5    &  3.556      &        5.5 \\
3.872              &   3.509      &       6.9    &  3.556      &        6.1 \\
4.260              &   3.510      &       7.0    &  3.556      &        7.4 \\
\hline
\hline
\end{tabular}
\caption[Various properties of reconstructed $\jpsi\gamma$ candidates for the radiative decay of $\chi_{c1,c2}$]
{Mean value and RMS of the reconstructed $\jpsi\gamma$ candidates for each energy analysed for 
the radiative decay of $\chi_{c1,c2}$.}
\label{tab:chi1}
\end{center}
\end{table*}
\Reffig{fig:chiMass} shows the $\chicone$ invariant mass distribution reconstructed at $\sqrt{s}=3.686$\,GeV.
\par
According to the selection cuts described above, a second photon is added to the reconstructed 
$\chi_{c1,c2}$ in order to reconstruct the complete final state, 
where it is possible to perform the kinematical fit to the $\pbarp$  4-momentum. 
The results of the reconstructed $\chi_{c1,c2}\gamma$ are summarised in \Reftbl{tab:chigamma1}:
\begin{table*}[htbp]
\begin{center}
\begin{tabular}{c c c c c c c}
\hline
\hline
 &  \multicolumn{3}{c}{$\chicone$}  &  \multicolumn{3}{c}{$\chictwo$}\\
\hline
$\sqrt{s} [GeV]$   &   Mean [GeV] & Eff [\%]  &  RMS (MeV) & Mean [GeV] & Eff [\%]  &  RMS [MeV] \\
\hline
3.686              &   3.687     &    28.88 &     7.6   &  3.686  &      29.13  &   8.5 \\
3.872              &   3.875      &   29.98 &     14.3  &  3.873  &      28.78  &   10.8\\
4.260              &   4.262      &   28.61 &     15.5  & 4.262   &      29.26  &   15.4\\
\hline
\hline
\end{tabular}
\caption[Various properties of reconstructed $\chi\gamma$ candidates for the radiative decay of $\chi_{c1,c2}$]
{Mean value, efficiencies and RMS of the reconstructed $\chi\gamma$ candidates 
for each energy analysed for the radiative decay of $\chi_{c1,c2}$.}
\label{tab:chigamma1}
\end{center}
\end{table*}

\begin{figure}[htbp]
  \begin{center}
  \includegraphics[angle=90,width=\columnwidth]{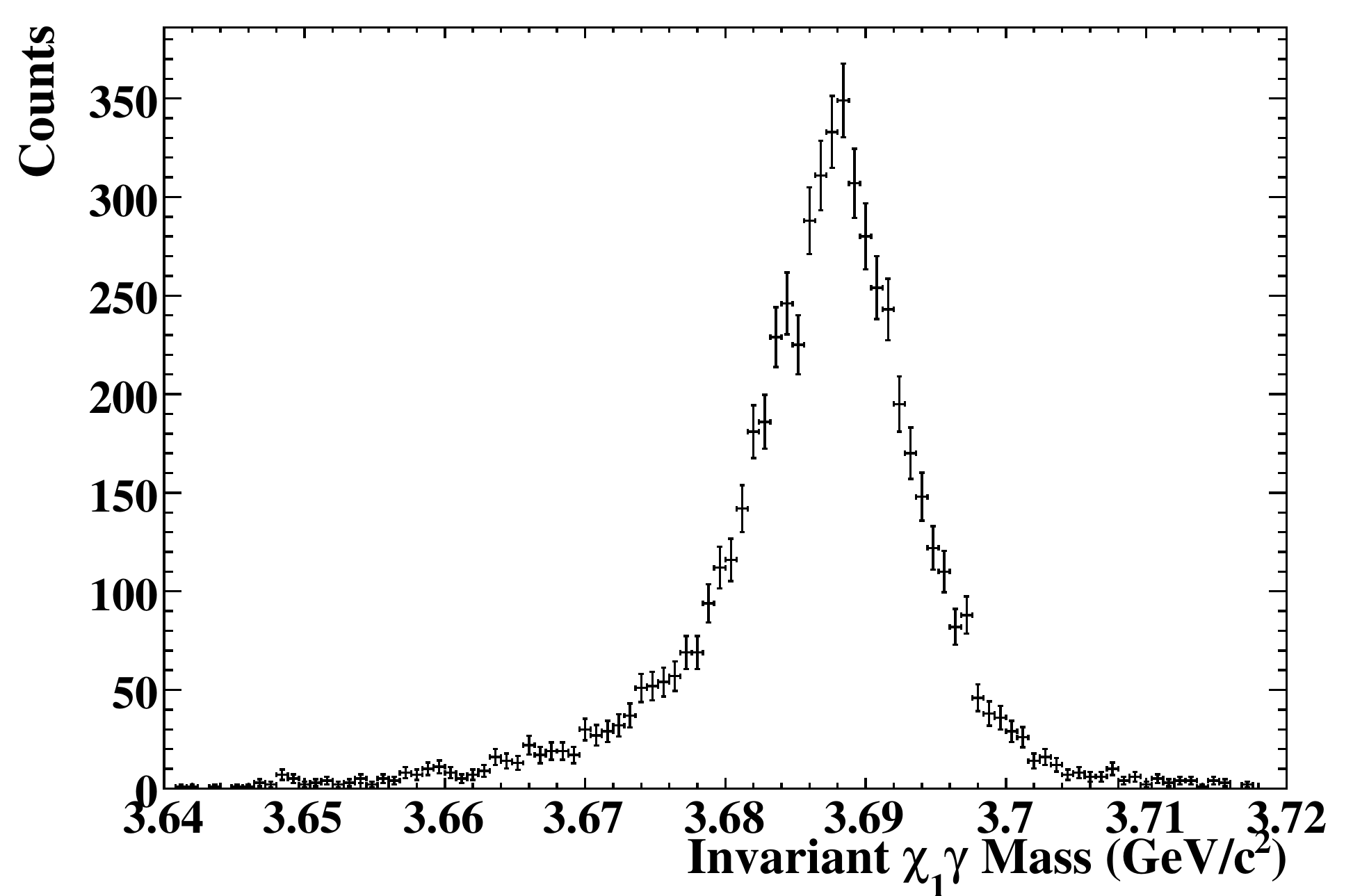}
  \end{center}
\caption{Invariant $\chicone\gamma$ mass at $\sqrt{s}=3.686$\,GeV.}
\label{fig:chigamma}
\end{figure}

\Reffig{fig:chigamma} shows the $\chicone\gamma$ candidates reconstructed at $\sqrt{s}=3.686$\,GeV.
%
\par
The major background comes from $\pbarp\to\pip\pim\piz$. 
This study is done at the $Y(4260)$ and $X(3872)$ energies.\\
The cross section of the background reactions at $\sqrt{s}$\,=\,4.260\,GeV is 
approximately equal to 30\,$\mu b$ \cite{bkg}, while at $\sqrt{s}$=3.872\,GeV the 
cross section is about 0.29\,mb~\cite{bkg}.\\
36 million filtered events (filter efficiency: 13.5\%) were analysed at the $Y(4260)$ energy and, 
for the $\chicone\gamma$ final state, only 7 events pass the selection, while for $\chictwo\gamma$ 
only 8 event satisfy the selection criteria, corresponding to an effective background cross-section
of 0.8\,pb and 0.9\,pb respectively.\\
68 million filtered events (filter efficiency: 12.7\%) were analysed at the X(3872) energy and, 
for the $\chicone\gamma$ final state, only 12 events pass the selection, while for $\chictwo\gamma$ 
only 15 event satisfy the selection criteria, corresponding to an effective background cross-section
of 6.5\,pb and 8.1\,pb respectively.\\
Since the signal cross-section for these reactions is not known it is not possible to calculate
a signal/background ratios. However
these simulations show that for these channels a very good background suppression can be achieved.
\par

\subsubsection*{$\boldsymbol{\pbarp\to \jpsi\gamma}$}
\par

$\pbarp\to \chi_c \to \jpsi\gamma\to \ee\gamma$ is the most important channel to 
study the radiative decay and the angular distribution of $\chi_{c}\to \jpsi\gamma$.\\
The first step is the reconstruction of a $\jpsi$ from electron-positron decay and then, adding one photon, 
one can reconstruct the $\jpsi\gamma$ candidate.\\ 
Event selection is done in the following steps:
\begin{enumerate}
\item select a well reconstructed $\jpsi$ in the event;
\item select reconstructed photon candidates;
\item kinematic fit of the $\jpsi\gamma$ candidates with vertex constraint;
\item probability of $\jpsi\gamma$ vertex fit: $P_{\jpsi\gamma}>0.001$.
\end{enumerate}
\par
According to the selection cuts described above, the results of the reconstructed $\jpsi\gamma$ 
are summarised in \Reftbl{tab:jpsigamma}.
\begin{table}[htbp]
\begin{center}
\begin{tabular}{c c c c}
\hline
\hline
$\sqrt{s} [GeV]$   &   Mean [GeV] &  Eff[\%]  & RMS [MeV]\\
\hline
3.510              &   3.512     &    44.47  &    10.5 \\
3.556              &   3.557      &   45.10  &    11.0 \\
3.872              &   3.874      &   37.96  &    15.3  \\
\hline
\hline
\end{tabular}
\caption [Various properties of reconstructed $\jpsi\gamma$ candidates]
{Mean value, efficiencies and RMS of the reconstructed $\jpsi\gamma$ candidates for each energy analysed.}
\label{tab:jpsigamma}
\end{center}
\end{table}
\par
\begin{figure}[htbp]
  \begin{center}
  \includegraphics[angle=90,width=\columnwidth]{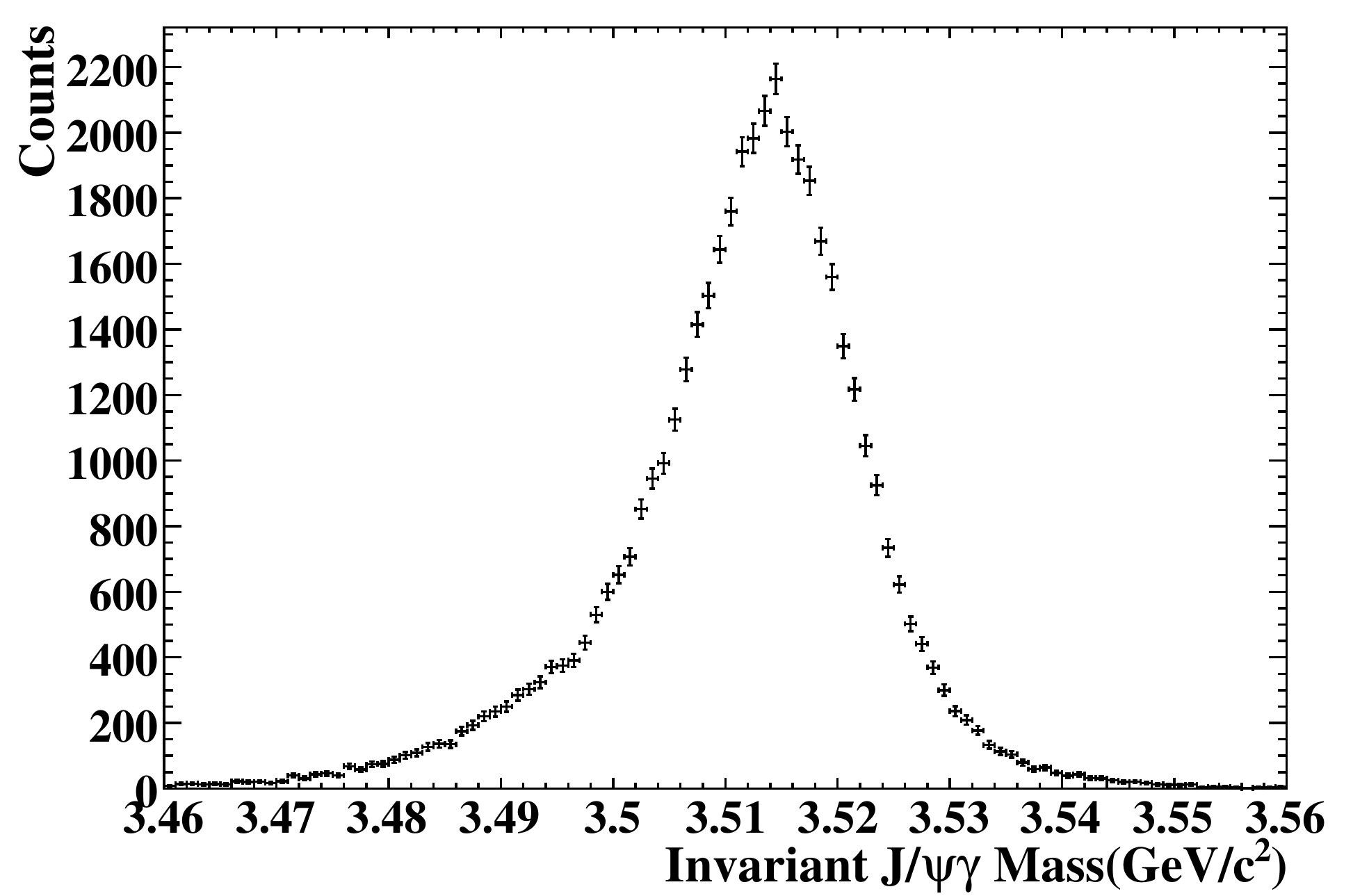}
  \end{center}
\caption{Invariant $\jpsi\gamma$ mass at $\sqrt{s}=3.510$\,GeV.}
\label{fig:jpsigamma}
\end{figure}

\Reffig{fig:jpsigamma} shows the $\jpsi\gamma$ candidates reconstructed at $\sqrt{s}=3.510$\,GeV.

The major background comes from $\pbarp\to\pip\pim\piz$. \\
The study of background contamination is done at the $X(3872)$ and $\chictwo$ energies.\\
The cross section of the background reaction is approximately 
equal to 0.29\,mb at $\sqrt{s}=3.872$\,GeV  and 0.12\,mb at $\sqrt{s}=3.556$\,GeV\,\cite{bkg}.\\
The cross section for $\pbarp \to \chictwo\to \jpsi\gamma$ is, from \INST{E835} measurements, 
about 2\,nb \cite{chi1E835}.
The number of simulated filtered background events was 26\,M and 68\,M 
at $\sqrt{s}=3.556$\,GeV and $\sqrt{s}=3.872$\,GeV, respectively. 
(filter efficiencies are 10.5\% and 12.7\% respectively). 
In the first case only 2 events pass the selection criteria, but if the constraint on the $J/\psi$ 
invariant mass is applied, these events disappear. In the second case 13 events satisfy the selection criteria; 
also in this case, for many background events the 4C-fit does not converge so it is completely suppressed by 
the cut \CL$ > 0.1\percent$.\\ 
Since no background events survive the selection we can set upper limits for the effective background
cross-sections of 1.2\,pb and 1.3\,pb at the $\chictwo$ and $X(3872)$ energies, respectively.
For the $\chictwo$, where the signal cross section is known, this translates into a signal/background 
ratio S/B of about $10^3$. 
%
\subsubsection*{$\boldsymbol{\pbarp\to \jpsi\eta\to \ee\gamma\gamma}$} 

The first step in the analysis of this channel 
is the reconstruction of a $\jpsi$ via its electron decay; 
the second step is the reconstruction of a $\eta$ candidate from two photons decay 
and then reconstruct the $\jpsi\eta$ candidate.\\ 
Event selection is done in the following steps:
\begin{enumerate}
\item select a well reconstructed $\jpsi$ in the event;
\item select a well reconstructed $\eta$ in the event;
\item kinematic fit with beam, $\jpsi$ and $\eta$ mass constraints;
\item $\jpsi$ mass window: [3.07;3.12]\,\gevcc;
\item $\eta$ mass window: [0.535;0.565]\,\gevcc;
\item probability of $\jpsi\eta$ vertex fit: $P_{\jpsi\eta}>0.001$.
\end{enumerate}
\par
According to the selection cuts described above, the results of the reconstructed 
$\jpsi\eta$ are summarised in \Reftbl{tab:jpsieta}.
\begin{table}[htbp]
\begin{center}
\begin{tabular}{c c c c}
\hline
\hline
$\sqrt{s}$    &   Mean       &  Eff    & FWHM\\
$   [GeV]$    &   [GeV]      &  [\%]   & [MeV]\\
\hline
3.638         &   3.645      &   15.7  &    1 \\
3.686         &   3.686      &   18.8  &    5 \\
3.872         &   3.872      &   18.6  &    11  \\
4.260         &   4.260      &   18.8  &    18     \\
\hline
\hline
\end{tabular}
\caption[Various properties of reconstructed $\jpsi\eta$ candidates]
{Mean value, efficiencies and RMS of the reconstructed $\jpsi\eta$ candidates 
for each energy analysed. These results are listed after a fit with mass constraint on $\jpsi$ and $\eta$.}
\label{tab:jpsieta}
\end{center}
\end{table}
\par
\begin{figure}[htbp]
  \begin{center}
  \includegraphics[width=0.45\textwidth,angle=90]{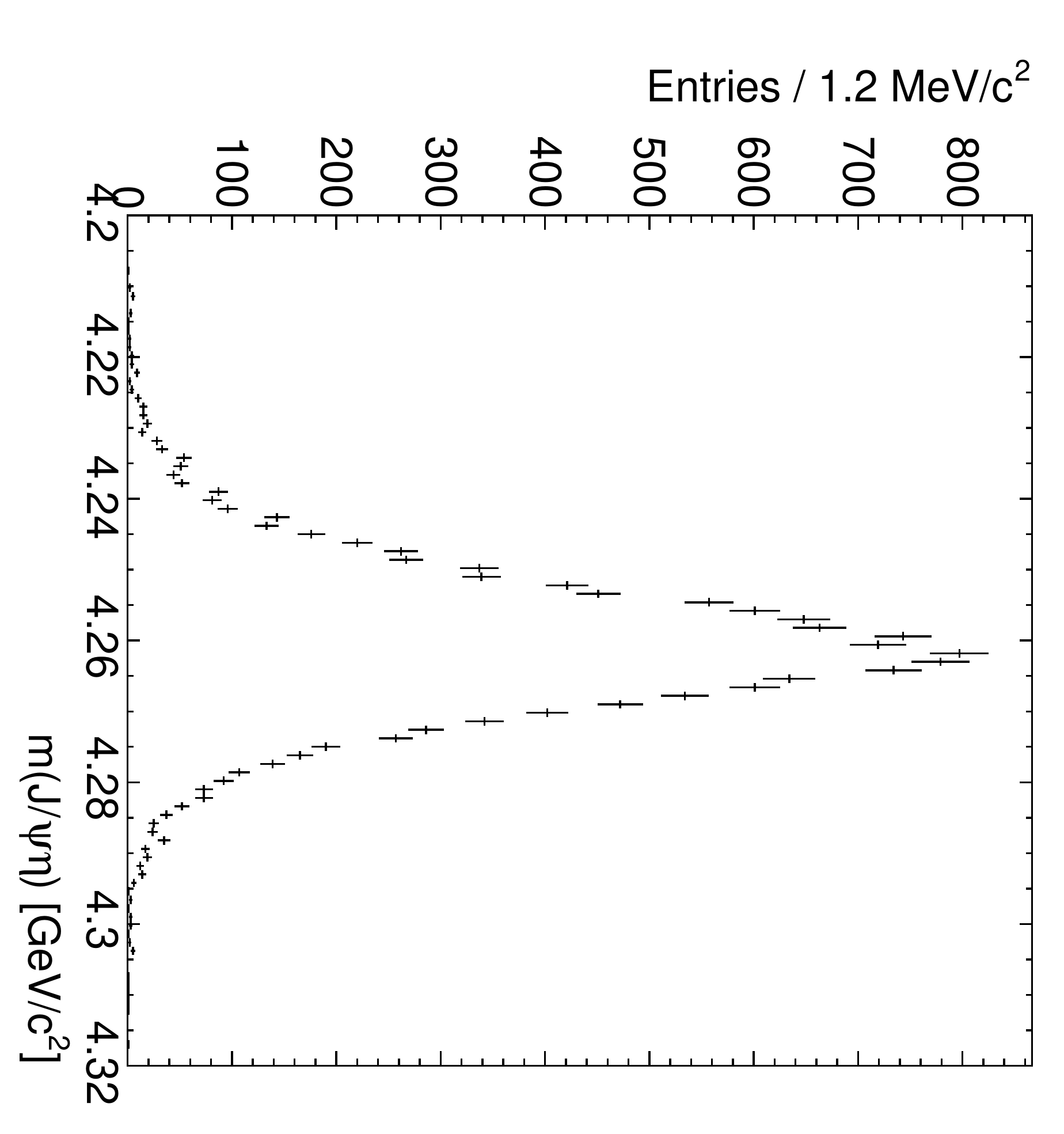}
  \end{center}
\caption{Invariant $\jpsi\eta$ mass at $\sqrt{s}=4.260$\,GeV.}
\label{fig:jpsietay}
\end{figure}

\Reffig{fig:jpsietay} shows the $\jpsi\eta$ candidates reconstructed at $\sqrt{s}=4.260$\,GeV.
%


For this signal channel we investigated the following background reactions:

\begin{itemize}
\item $\pbarp\to \jpsi\pi^0\gamma$
\item $\pbarp\to \jpsi\pi^0\pi^0$
\item $\pbarp\to\jpsi\eta\gamma$
\item $\pbarp\to\jpsi\eta\pi^0$
\item $\pbarp\to\jpsi\eta\eta$
\item $\pbarp\to\pi^+\pi^-\eta$
\item $\pbarp\to\pi^+\pi^-\pi^0$
\end{itemize}

with $\pi^0\to\gamma\gamma$, $\eta\to\gamma\gamma$ and
$\jpsi\to e^+e^-$.  Most cross sections for decays including a
$\jpsi$ have not been measured yet.

The cross sections and branching fractions for the $\jpsi\eta$ signal
and background modes are summarised
in \Reftbl{tab:jpsietabkg}. \Reftbl{tab:jpsietabkgresult} presents the
results of background contamination studies.

\begin{table*}[!thp]
\begin{center}
\begin{tabular}{c c c c}
\hline
\hline
Reaction $\pbarp\to$   &   $\sqrt{s}$ [GeV] &  $\sigma$         &     B\\
\hline
$\jpsi\eta$           &   3.638  &  $\sigma_s$ & 2.34\%$\times$B$(\eta_c(2S)\to\jpsi\eta)$\\
$\jpsi\pi^0\pi^0$      &   3.638  &  $\sigma_b$ & 5.78\%\\
$\jpsi\pi^0\gamma$     &   3.638  &  $\sigma_b$ & 5.84\%\\
$\jpsi\eta\gamma$      &   3.638  &  $\sigma_b$ & 2.32\%\\
\hline
$\jpsi\eta$            &   3.686  &  $\sigma_s$ & 0.07\%\\
$\jpsi\pi^0\pi^0$      &   3.686  &  $\sigma_b$ & 5.78\%\\
$\jpsi\pi^0\gamma$     &   3.686  &  $\sigma_b$ & 5.84\%\\
$\jpsi\eta\gamma$      &   3.686  &  $\sigma_b$ & 2.32\%\\
\hline
$\jpsi\eta$           &   3.872  &  $\sigma_s$ & 2.34\%$\times$B$(X(3872))\to\jpsi\eta)$\\
$\jpsi\eta\pi^0$       &   3.872  &  $\sigma_b$ & 2.30\%\\
$\jpsi\pi^0\pi^0$      &   3.872  &  $\sigma_b$ & 5.78\%\\
$\jpsi\pi^0\gamma$     &   3.872  &  $\sigma_b$ & 5.84\%\\
$\jpsi\eta\gamma$      &   3.872  &  $\sigma_b$ & 2.32\%\\
$\pi^+\pi^-\pi^0$      &   3.872  &  290 $\mu b$ & 98.80\%\\
\hline
$\jpsi\eta$           &   4.260  &  $\sigma_s$ & 2.34\%$\times$B$(Y(4260)\to\jpsi\eta)$\\
$\jpsi\eta\eta$        &   4.260  &  $\sigma_b$ & 0.92\%\\
$\jpsi\eta\pi^0$       &   4.260  &  $\sigma_b$ & 2.30\%\\
$\jpsi\pi^0\pi^0$        &   4.260  &  $\sigma_b$ & 5.78\%\\
$\jpsi\pi^0\gamma$     &   4.260  &  $\sigma_b$ & 5.84\%\\
$\jpsi\eta\gamma$      &   4.260  &  $\sigma_b$ & 2.32\%\\
$\pi^+\pi^-\eta$       &   4.260  &  1.54 $\mu b^1$ & 39.38\%\\
$\pi^+\pi^-\pi^0\pi^0$ &   4.260  &  50 $\mu b$ & 97.61\%\\
$\pi^+\pi^-\pi^0$      &   4.260  & 30$\pm$10 $\mu b^2$ & 98.80\%\\
\hline
\hline
\end{tabular}
\caption[Cross sections and branching fractions for the $\jpsi\eta$ 
 signal and background modes.]  {$\sigma_s$ denotes the cross section
for the formation of a given resonance in $\overline{p}p$ events, B
the branching fraction for the decay tree. $\sigma_b$ is the cross
section for the background mode in $\overline{p}p$ annihilation. 1)
obtained from DPM generator. 2) measured value at 8.8 GeV/c.}
\label{tab:jpsietabkg}
\vspace*{+10mm}
\end{center}
\end{table*}

\begin{table*}[!thp]
\begin{center}
\begin{tabular}{c c c c}
\hline
\hline
Decay $\pbarp\to$   &   $\sqrt{s}$ [GeV] & Suppression      &     Signal to noise\\
\hline
$\jpsi\pi^0\pi^0$      &   3.638  & $>10^5$ & 6300$\tilde{\sigma}/\sigma_b$ \\
$\jpsi\pi^0\gamma$     &   3.638  & $>10^5$ & 6200$\tilde{\sigma}/\sigma_b$ \\
$\jpsi\eta\gamma$      &   3.638  & 5       & 1$\tilde{\sigma}/\sigma_b$ \\
\hline
$\jpsi\pi^0\pi^0$      &   3.686  & $>10^5$ & 7600$\tilde{\sigma}/\sigma_b$ \\
$\jpsi\pi^0\gamma$     &   3.686  & 12500   & 900$\tilde{\sigma}/\sigma_b$ \\
$\jpsi\eta\gamma$      &   3.686  & 400     & 100$\tilde{\sigma}/\sigma_b$ \\
\hline
$\jpsi\eta\pi^0$       &   3.872  & $>10^5$ & 18800$\tilde{\sigma}/\sigma_b$ \\
$\jpsi\pi^0\pi^0$      &   3.872  & $>10^5$ & 75$\tilde{\sigma}/\sigma_b$ \\
$\jpsi\pi^0\gamma$     &   3.872  & 8300    & 600$\tilde{\sigma}/\sigma_b$ \\
$\jpsi\eta\gamma$      &   3.872  & 2000    & 400$\tilde{\sigma}/\sigma_b$ \\
$\pi^+\pi^-\pi^0$      &   3.872  & $>7\cdot 10^7$  & 1$\tilde{\sigma}/nb$ \\
\hline
$\jpsi\eta\eta$        &   4.260  & $>10^5$  & 47700$\tilde{\sigma}/\sigma_b$ \\
$\jpsi\eta\pi^0$       &   4.260  & $>10^5$  & 19000$\tilde{\sigma}/\sigma_b$ \\
$\jpsi\pi^0\pi^0$        &   4.260  & $>10^5$  & 7600$\tilde{\sigma}/\sigma_b$ \\
$\jpsi\pi^0\gamma$     &   4.260  & 3600     & 300$\tilde{\sigma}/\sigma_b$ \\
$\jpsi\eta\gamma$      &   4.260  & 3400     & 600$\tilde{\sigma}/\sigma_b$ \\
$\pi^+\pi^-\eta$       &   4.260  & $>7\cdot 10^7$   & 500$\tilde{\sigma}/nb$ \\
$\pi^+\pi^-\pi^0\pi^0$ &   4.260  & $>1.4\cdot 10^8$ & 20$\tilde{\sigma}/nb$ \\
$\pi^+\pi^-\pi^0$      &   4.260  & $>1.7\cdot 10^8$ & 15$\tilde{\sigma}/nb$ \\
\hline
\hline
\end{tabular}
\caption[Suppression $\eta$ and signal to noise ratio for the 
 background modes of the $\jpsi\eta$ analysis.]  {Signal to noise
ratios are given in terms of the unknown cross section
$\tilde{\sigma}$ or $\tilde{\sigma}/nb$.}
\label{tab:jpsietabkgresult}
\end{center}
\end{table*}

\subsubsection{$\boldsymbol{\pbarp  \to \honec \to \etac\gamma}$}
According to theoretical predictions and previous experimental observations \cite{bib:sim:hc_E835, 
bib:sim:hc_CLEO} one of the most promising decay modes for the observation of the $\honec$ is its 
electromagnetic transition to the ground state of charmonium
\begin{equation}
\honec \to \etac + \gamma
\end{equation}
where the energy of the photon is $E_{\gamma}=503\,\mev$. 
\par
The $\etac$ can be detected through many exclusive decay modes, 
neutral ($\etac \to \gamma \gamma$) or hadronic. 
\par
In order to estimate the signal cross-section we calculate the value of the Breit-Wigner formula at the 
resonance peak $E_R$:
\begin{equation}
\sigma_{p}=\frac{3\pi}{k^{2}}B_{\pbarp}\BR_{\etac\gamma}
\end{equation}
where $k^{2}=(E_{R}^{2}-4m_{p}^{2})/4$ 
and the $B$s represent the branching ratios into the initial and final states.
\par
Using the value measured by \INST{E835} $\Gamma_{\pbarp}\BR_{\etac\gamma}$=10\,eV and assuming a value
of 0.5\,MeV for the $\honec$ width we obtain $\sigma_{p}=33$\,nb.
\par

\subsubsection*{$\boldsymbol{\honec \to \etac + \gamma \to 3 \gamma}$}
We first consider the process $\honec \to 3 \gamma$.
This decay mode was observed at \INST{Fermilab} by \INST{E835} \cite{bib:sim:hc_E835}. 
It is characterised by a fairly clean final state, but the low value of the 
$\etac\to\gamma \gamma$ branching ratio ($4.3\cdot10^{-4}$) results in a 
relatively low event rate in comparison with the hadronic decay modes of $\etac$.
\par
The main contributors to the background for the 3$\gamma$ final state are $\gamma$s 
from the $\piz$, $\eta$, $\eta'$ decay in $\gamma \gamma$ decay modes: the loss of one or more
$\gamma$s outside the detector acceptance or below the energy threshold of the calorimeter 
can result in a 3 $\gamma$ final state. 
The background channels considered in this analysis are presented in \Reftbl{tab:sim:hc_3gamma_bkgr} 
with the corresponding cross-sections measured by \INST{E760} and \INST{E835} over the angular 
range in the CM system $|\cos(\theta_{CM})|<0.6$
\cite{bib:Andreotti:2005zr,bib:Armstrong:1997gv}.
\par
\begin{table}[htbp]
\begin{center}
\begin{tabular}{lcc}
  \hline\hline
  Channel & $\sigma$ (nb) & number of events \\
  \hline
  $\pbarp\to \honec \to 3 \gamma$ & & 20\,k \\
  $\pbarp\to\piz\piz$ & 31.4 & 1.3\,M \\
  $\pbarp\to\piz\gamma$ & 1.4 & 100\,k \\
  $\pbarp\to\piz\eta$ & 33.6 & 1.3\,M \\
  $\pbarp\to\eta\eta$ & 34.0 & 1.3\,M \\
  $\pbarp\to \piz\eta'$ & 50.0 & 100\,k \\
  \hline\hline
\end{tabular}
\caption[The main background contributors to $\honec \to 3 \gamma$]
{The main background contributors to $\honec \to 3 \gamma$ with
corresponding 
cross-section integrated over $|\cos(\theta_{CM})|<0.6$.}
\label{tab:sim:hc_3gamma_bkgr}
\end{center}
\end{table}
\par
\begin{figure}[htb]
\begin{center}
\includegraphics[width=\columnwidth]{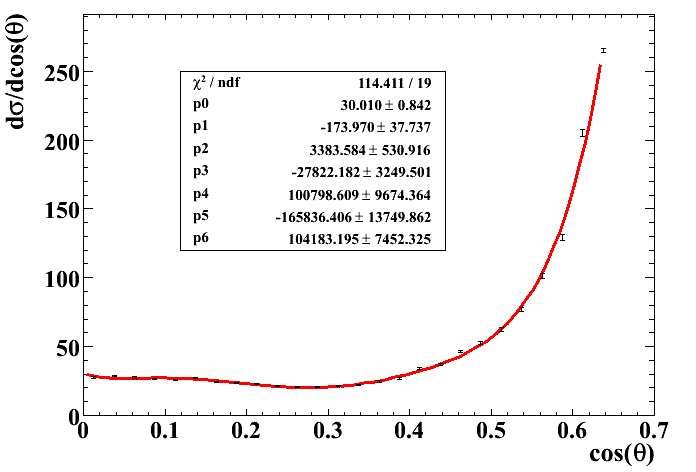}
\caption[Angular dependence of $\piz\piz$ cross-section with parametrisation used in Monte-Carlo simulation.]
{Angular dependence of $\piz\piz$ cross-section in the CM system with parametrisation used in Monte-Carlo simulation.}
\label{fig:sim:hc_pi0pi0_crosssection}
\end{center}
\end{figure}

The angular dependence for all the studied background channels is strongly peaked in the forward and backward 
direction, which is typical of two and three meson production in proton-antiproton annihilation. 
For the Monte Carlo study the angular dependence of the cross-section was parametrised by 6$^{th}$ or 7$^{th}$
 order  polynomials in $\cos(\theta_{CM})$. 
As an example we show the  $\piz\piz$ angular distribution in 
\Reffig{fig:sim:hc_pi0pi0_crosssection}.
\par
\par
The number of Monte Carlo events used for this analysis for signal and all the background channels 
are shown in \Reftbl{tab:sim:hc_3gamma_bkgr}.
\par
\begin{figure}
\begin{center}
\includegraphics[width=\columnwidth]{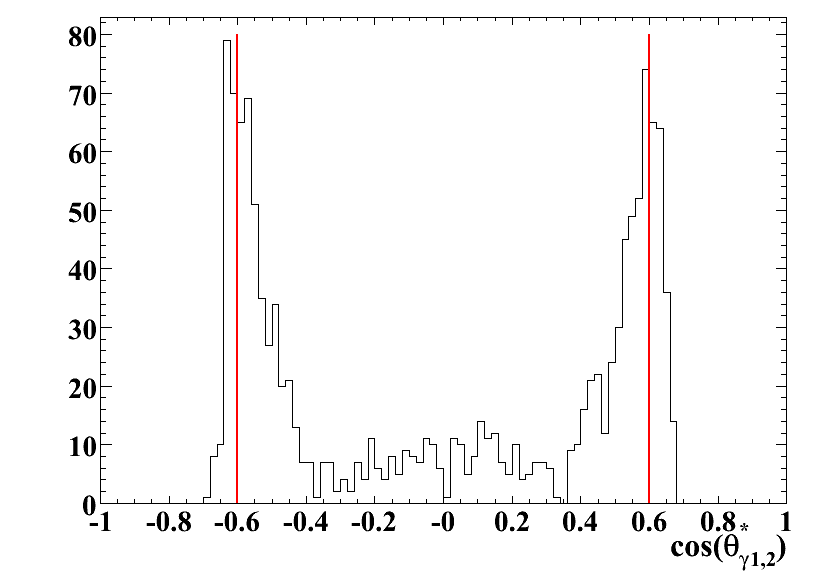}
\caption[Distribution of reconstructed $\cos{\theta}$ of the $\gamma$ in CM system for 
$\pbarp \to \piz\piz$ background.]
{Distribution of reconstructed $\cos{\theta}$ of the 
$\gamma$ in CM system for $\pbarp \to \piz\piz$ background.}
\label{fig:sim:pi0pi0_costheta12}
\end{center}
\end{figure}
\begin{figure}
\begin{center}
\includegraphics[width=\columnwidth]{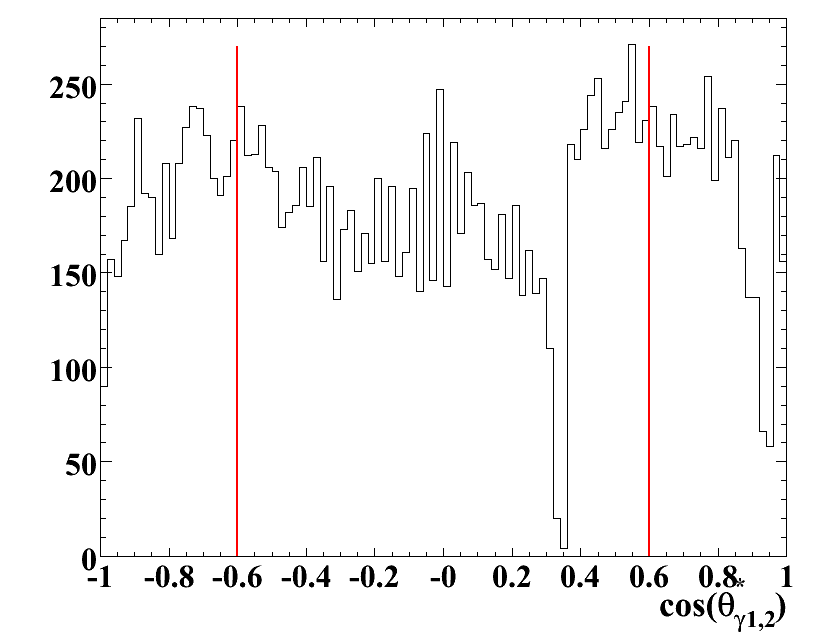}
\caption[Distribution of reconstructed $\cos{\theta}$ of the $\gamma$ in CM system from 
$\honec\to \etac \gamma$.]
{Distribution of reconstructed $\cos{\theta}$ of the $\gamma$ in CM system 
from $\honec\to \etac \gamma$.}
\label{fig:sim:hc_3gamma_costheta12}
\end{center}
\end{figure}

The event selection is done in the following steps:
\begin{enumerate}
\item
An $\etac$ candidate is formed by pairing $\gamma$'s with an invariant mass in the window [2.6; 3.2]\,GeV.
The third $\gamma$ is added to this pair to form the $\honec$ candidate.
\item
A 4C-fit to beam energy-momentum is applied to the $\honec$ candidate and the information on the $\honec$ 
and the updated information on the daughter $\gamma$'s are stored into the root ntuple.
\item
The following cuts are applied at the ntuple level to suppress the background.
\begin{enumerate}
\item
Events with 3 $\gamma$'s were selected. This cut keeps 47\percent of all signal events,
whereas it rejects more
than 90 \percent of all background events (with the exception of the $\piz\gamma$ channel).
\item
Cut on the confidence level of the 4C-fit: $\CL>10^{-4}$.
\item
Cut on the CM energy of the $\gamma$ from the $\honec \to \etac \gamma$ ($E_{\gamma}$)
radiative transition: $0.4\ $GeV$ < E_{\gamma} < 0.6\ $GeV.
\item
Angular cut $|\cos(\theta_{CM})|<0.6$, to reject the background which is strongly peaked in the forward and
backward directions. The $\cos(\theta_{CM})$ distributions for a background channel ($\piz \piz$) and for
the signal are shown in \Reffig{fig:sim:pi0pi0_costheta12} and  \Reffig{fig:sim:hc_3gamma_costheta12},
respectively.
\item
The cut for invariant mass of combination $M(\gamma_1,\gamma_3) > 1.0\,GeV$ and $M(\gamma_2,\gamma_3)>1.0\,GeV$
(the value of the cut is determined by the $\eta'$ mass). 
\end{enumerate}
\end{enumerate}
\par
In \Reftbl{tab:sim:hc_3gamma_eff} the selection efficiencies for different cuts are presented. 
Efficiencies are cumulative, {\it i.e.} applied one after another. 
Taking into account the signal cross-section $\sigma_{\pbarp \to \honec}$\,=\,33\,nb at resonance, 
branching ratio $\BR(\etac\to \gamma \gamma)=4.3 \cdot 10^{-4}$ and background cross-sections it 
results in the expected signal to background ratios presented in \Reftbl{tab:sim:hc_3gamma_sb}. 
The expected event rate for the luminosity in high luminosity mode $L = 2 \cdot 10^{32}\,$cm$^{-2}$s$^{-1}$ is 
20\,events/day, and for high resolution mode with $L =2 \cdot 10^{31}\,$cm$^{-2}$s$^{-1}$ 2.0\,events/day 
correspondingly.
\begin{table*}[htb]
\begin{center}
\begin{tabular}{lcccccc}
  \hline\hline
  Cut & $\honec$ &$\piz\gamma$ & $\piz\piz$ & $\piz\eta$ & $\eta \eta$ & $\piz\eta'$\\
  \hline
  preselection & 0.70 & 0.43 & 0.14 & $8.2\cdot 10^{-2}$ & $4.0 \cdot 10^{-2}$ & $8.5 \cdot 10^{-2}$ \\
  3 $\gamma$ & 0.47 & 0.31 & $1.3 \cdot 10^{-2}$ & $7.5 \cdot 10^{-3}$ & $2.7 \cdot 10^{-3}$ & $8.7 \cdot 10^{-3}$ \\
  $\CL>10^{-4}$& 0.44 & 0.30 & $9.9 \cdot 10^{-3}$ & $4.9 \cdot 10^{-3}$ & $7.2 \cdot 10^{-4}$ & $5.7 \cdot 10^{-3}$ \\
  $E_{\gamma}$ [0.4;0.6]\,GeV& 0.43 & 0.12 & $3.9 \cdot 10^{-3}$ & $2.0 \cdot 10^{-3}$ & $2.8 \cdot 10^{-4}$ & $2.3 \cdot 10^{-3}$ \\
  $|\cos(\theta)|<0.6$ & 0.22 & $9.2 \cdot 10^{-2}$ & $2.7 \cdot 10^{-3}$ & $1.1 \cdot 10^{-3}$ & $7.0 \cdot 10^{-5}$ & $7.5 \cdot 10 ^{-4}$ \\
  $m_{12}^{2}, m_{23}^{2}> 1.0$\,GeV  & $8.1 \cdot 10^{-2}$ & 0 & 0 & 0& 0 & 0 \\
  \hline\hline
\end{tabular}
\caption[Selection efficiencies for $\honec \to 3 \gamma$ and its background channels.]
{Selection efficiencies for $\honec \to 3 \gamma$ and its background channels.}
\label{tab:sim:hc_3gamma_eff}
\end{center}
\end{table*}
\par
\begin{table}
\begin{center}
\begin{tabular}{lc}
  \hline\hline
  Channel & S/B ratio\\
  \hline
  $\pbarp\to\piz\piz$ & $>$ 94 \\
  $\pbarp\to\piz\gamma$ & $>$ 164 \\
  $\pbarp\to\piz\eta$ & $>$ 88 \\
  $\pbarp\to\eta\eta$ & $>$ 87 \\
  $\pbarp\to \piz\eta'$ & $>$ 250 \\
  \hline\hline
\end{tabular}
\caption[Signal to background ratio for $\honec \to 3 \gamma$ and
different background channels.]
{Signal to background ratio for $\honec \to 3 \gamma$ and different
background channels.}
\label{tab:sim:hc_3gamma_sb}
\end{center}
\end{table}
\par
\subsubsection*{$\boldsymbol{\honec \to \etac \gamma \to \phi \phi \gamma}$}
As a benchmark channel with a hadronic decay mode of the $\etac$ we study the $\phi \phi$ final state with 
$\BR=2.6 \cdot 10^{-3}$. We detect the $\phi$ through the decay correspondingly $\phi \to \Kp \Km$, 
with $\BR= 0.49$ \cite{PDG}.
\par
For the exclusive decay mode considered in this study:
\begin{center}
$\pbarp \to \honec \to \etac \gamma \to \phi \phi \gamma \to \Kp \Km \Kp \Km \gamma$
\end{center}
the following 3 reactions are considered as the main contributors to the background:\\
\begin{enumerate}
\item $\pbarp \to \Kp \Km \Kp \Km \piz$,
\item $\pbarp \to \phi \Kp \Km \piz$,
\item $\pbarp \to \phi \phi \piz$.
\end{enumerate}
With one photon from the $\piz$ decay undetected these reactions have the same final state particles 
as the studied $\honec$ decay. 
As in the case of the three photon decay discussed above, it is crucial to have as low an energy threshold
as possible in order to effectively reject this background.\\
Additional possible source of background is $\pbarp \to \Kp \Km
\pip \pim \piz$. This reaction could contribute to background due to pion as
kaon misidentification.


\par
There are no experimental measurements, to our best knowledge, of the cross-sections for the
first three background
reactions, which are supposed to be main contributors to background. 
The only way to estimate their cross-sections was found to use the DPM
(Dual Parton Model) event generator \cite{bib:sim:DPM}. $2 \cdot 10^7$
events were generated with DPM at beam momentum $p_z=5.609 \,\gevc$, which corresponds to the studied $\honec$
resonance.
The corresponding numbers of events are 115 and 12 for the first two
background channels. No events for the $\pbarp \to \phi \phi \piz$ reaction were observed. With the
total $\pbarp$ cross-section at this beam momentum of 60\,mb the
cross-sections for the corresponding background channels are estimated at 345\,nb, 60\,nb
and below 3\,nb, respectively.\\
For $\pbarp \to \Kp \Km \pip \pim \piz$ the cross-section was estimated
by extrapolation from lower energy according to the total inelastic cross-section. 
It gives an estimation $\sigma=30 \mu b$.\\
\par

The number of analysed events are presented in \Reftbl{tab:perf:hc_nevts}.\\
For the $\pbarp \to \Kp \Km \pip \pim \piz$ channel 15 millions out
of 20 millions events were simulated with filter on invariant mass of the
pair of two kaons. The events with $m(\Kp \Km)$ in the range [0.95;1.2] \gev
were selected. The efficiency of the filter is 29.9 \percent, which gives
effective number of simulated events $\sim$ 55 M.\\
\begin{table}
\begin{center}
\begin{tabular}{lc}
  \hline\hline
  Channel & $N$ of events \\
    \hline
  $\pbarp \to \honec \to \phi \phi \gamma$ & 20\,k \\
  $\pbarp \to \Kp \Km \Kp \Km \piz$ & 6.2\,M \\
  $\pbarp \to \phi \Kp \Km \piz$ & 200\,k\\
  $\pbarp \to \phi \phi \piz$ & 4.2\,M\\
  $\pbarp \to \Kp \Km \pip \pim \piz$ & 5\,M + 15\,M \\
& 100 k \\
  \hline\hline
\end{tabular}
\caption[The numbers of analysed events for $\honec$ decay]{The numbers of
analysed events for $\honec$ decay}
\label{tab:perf:hc_nevts}
\end{center}
\end{table}
\par
The following selection criteria were applied:
\begin{enumerate}
\item
$\phi$ candidates were defined as $\Kp$, $\Km$ pairs
with invariant mass in the window [0.8; 1.2]\,GeV.
Two $\phi$ candidates in one event with invariant mass in the window [2.6;3.2]\,$\gev$ 
defined an $\etac$ candidate which, combined with a neutral candidate, formed an $\honec$ candidate.
\item
A 4C-fit to beam energy-momentum was applied to the $\honec$ candidate, which was stored
to a root ntuple together with the
updated information on its decay products.
\item
The following cuts are performed at the ntuple level for additional background suppression:
\begin{enumerate}
\item
cut on the confidence level of the 4C-fit to beam energy-momentum, $\CL> 0.05$,
\item
$\etac$ invariant mass [2.9;3.06]\,\gev,
\item
$E_{\gamma}$ within [0.4;0.6]\,\gev,
\item
$\phi$ invariant mass [0.99;1.05]\,\gev,
\item
no $\piz$ candidates in the event, {\it i.e.} no 2$\gamma$ invariant mass in the range
[0.115;0.15] $\gev$ with 2 different low energy $\gamma$ thresholds: 30\,\mev
and 10\,\mev.
\end{enumerate}
\end{enumerate}
\par
The efficiencies of the various cuts are given in \Reftbl{tab:sim:hc_eff} 
for the signal and three of considered background channels.
\begin{table*}[htb]
\begin{center}
\begin{tabular}{lccccc}
  \hline\hline
  Selection criteria & signal & $4K \piz$ & $\phi  \Kp \Km \piz $ & $\phi
\phi \piz$ & $\Kp \Km \pip \pim \piz$\\
    \hline
  pre-selection & 0.51 & $9.8\cdot 10^{-3}$ & $1.3\cdot 10^{-2}$ & $4.9\cdot
10^{-2}$& $9.0\cdot 10^{-6}$\\
  $\CL>0.05$ & 0.36 & $1.5\cdot 10^{-3}$ & $2.0\cdot 10^{-3}$ & $7.0\cdot
10^{-3}$& $4.0\cdot 10^{-8}$\\
  $m(\etac), E_{\gamma}$ & 0.34 & $4.1\cdot 10^{-4}$ & $5.2\cdot 10^{-4}$ &
$1.8\cdot 10^{-3}$ & 0\\
  $m(\phi)$ & 0.31 & $4.5\cdot 10^{-6}$ & $1.2\cdot 10^{-4}$ & $1.7\cdot
10^{-3}$ & 0\\
  $no \; \piz (30\,MeV)$ & 0.26 & $2.7\cdot 10^{-6}$ & $4.5\cdot 10^{-5}$ &
$9.2\cdot 10^{-4}$ & 0\\
  $no \; \piz (10\,MeV)$ & 0.24 & $1.8\cdot 10^{-6}$ & $3.0\cdot 10^{-5}$ &
$7.1\cdot 10^{-4}$ & 0\\
  \hline\hline
\end{tabular}
\caption[Efficiency of different event selection criteria]{Efficiency of
different event selection criteria.}
\label{tab:sim:hc_eff}
\end{center}
\end{table*}
\par
Assuming the $\honec$ production cross-section of 33\,nb at resonance, one obtains the signal to background
ratios given in \Reftbl{tab:sim:hc_sb_ratio}.
\par
\begin{table}
\begin{center}
\begin{tabular}{lc}
  \hline\hline
  channel & Signal/Background\\
    \hline
  $\pbarp \to \Kp \Km \Kp \Km \piz$ & 8 \\
  $\pbarp \to \phi \Kp \Km \piz$ & 8\\
  $\pbarp \to \phi \phi \piz$ & $>10$\\
  $\pbarp \to \Kp \Km \pip \pim \piz$ & $>12$ \\
  \hline
\end{tabular}
\caption[Signal to background ratio for different $\honec$ background
channels]
{Signal to background ratio for different $\honec$ background channels}
\label{tab:sim:hc_sb_ratio}
\end{center}
\end{table}
\par
For the $\pbarp \to \phi \phi \piz$ background channel the
reduction of low energy $\gamma$-ray threshold from 30\,MeV to 10\,MeV gives a $19\percent$ 
improvement in the signal to background ratio, for the
$\pbarp \to \phi \Kp \Km \piz$ 
the corresponding improvement is $33\percent$.
\par
With the final signal selection efficiency of $25\percent$ and the assumed luminosity 
in high luminosity mode of
$L=2 \cdot 10^{32}$\,cm$^{-2}$s$^{-1}$ the expected signal event rate is 92\,events/day. 
For the high resolution mode with $L=2 \cdot 10^{31}$\,cm$^{-2}$s$^{-1}$ the expected signal event 
rate is 9\,events/day.\\
\subsubsection{$\boldsymbol{\pbarp  \to \DDbar}$}
\label{sec:phys:ddbar}
The main focus of this benchmark study is to assess the ability to
separate the charm signal from the large hadronic background. In
addition to the detection of charmonium states above the \DDbar
threshold this is important for other major parts of the \PANDA physics
program, such as open charm spectroscopy, the search for charmed
hybrids decaying to \DDbar and the investigation of rare decays and CP
violation in the \D meson sector.
\par
In order to study the tracking and PID reconstruction capabilities of
the proposed \PANDA detector, two benchmark channels have been chosen
with decays containing only charged particles (charge conjugated
states included):
\begin{itemize}
 \item \pbarp $\to$ \DpDm with the decay $\ensuremath{D^+\to K^- \pip \pip}\xspace$
 \item \pbarp $\to$ $\ensuremath{D^{*+}D^{*-}}\xspace$ with the
       sequential decays $\ensuremath{D^{*+} \to D^0 \pip}$ and
       $\ensuremath{D^0 \to \Km \pip}$
\end{itemize}
Both channels were simulated at a beam energy corresponding to 
$\sqrt{s}= m_{c\bar{c}}$, the $\Psi(3770)$ for the \DpDm channel and the
$\Psi(4040)$ for the $D^{*+}D^{*-}$ channel, respectively.
The production is done directly into a \DDbar pair, which corresponds to
$\approx~40$\,\mev above the particular \DDbar threshold.

The charm production cross sections close to threshold in \pbarp
annihilations are unknown. To estimate the \DDbar production cross
section a Breit-Wigner approach can be used to calculate the resonant
cross section, where the unknown branching ratios to \pbarp are
estimated by scaling the known ratio $\jpsi \to\pbarp$
\cite{PandaTechRep}.  This method estimates only the strength of the
resonance contribution to the cross section. The strength of the
\DDbar continuum production is unknown and to account for its
contribution, the known decay branchings $\ccbar \to \DDbar$ have been
set to 100\percent, which leads to assumptions for the cross sections
of:
\begin{equation*}
 \sigma(\pbarp \to \Psi(3770)\to\D^+ D^-)= 2.8\, nb
\end{equation*}
for the first channel, and
\begin{equation*}
 \sigma(\pbarp \to \Psi(4040)\to\D^{*+} D^{*-})= 0.9\, nb
\end{equation*}
for the second channel, respectively.
Taking into account the branching ratios of the considered \D meson
decays, the expected  ratio $R$ between the signal and the total \pbarp cross section
within this analysis can be calculated.
The values obtained are listed in \Reftbl{tbl:relDecayBR} together
with the branching ratios of the individual \D meson decays used in this
analysis.
\begin{table}
\begin{center}
 \begin{tabular}{lll}
  \hline
  \hline
  channel  &  $\DpDm$  & $D^{*+}D^{*-}$   \\
  \hline
  decay    &  $D^{\pm} \to K^{\mp}\pi^{\pm} \pi^{\pm}$  &  $D^{*+} \to
  D^0 \pip$ \\
  &  (9.2\,$\percent$)   &    (67.7\,$\percent$) \\
  &  &  $D^0 \to \Km \pip$ \\
  &  &  (3.8 $\percent$) \\
  R &  $4\cdot 10^{-10}$ & $1\cdot 10^{-11}$\\
  \hline
  \hline
 \end{tabular}
\caption[Definition of the \DDbar physics channels, relevant decay
 branching ratios and the expected ratio between signal and total
 \pbarp cross section]{Definition of the \DDbar physics channels, relevant decay
 branching ratios and the expected ratio between signal and total
 \pbarp cross section.}
\label{tbl:relDecayBR}
\end{center}
\end{table}
\par
Using a value of 60\,mb for the total \pbarp cross section at the \DDbar
threshold the relevant production cross section will thus be suppressed
at least by ten orders of magnitude. 
\par
\begin{figure}[htb]
\begin{center}
\includegraphics[width=\swidth]{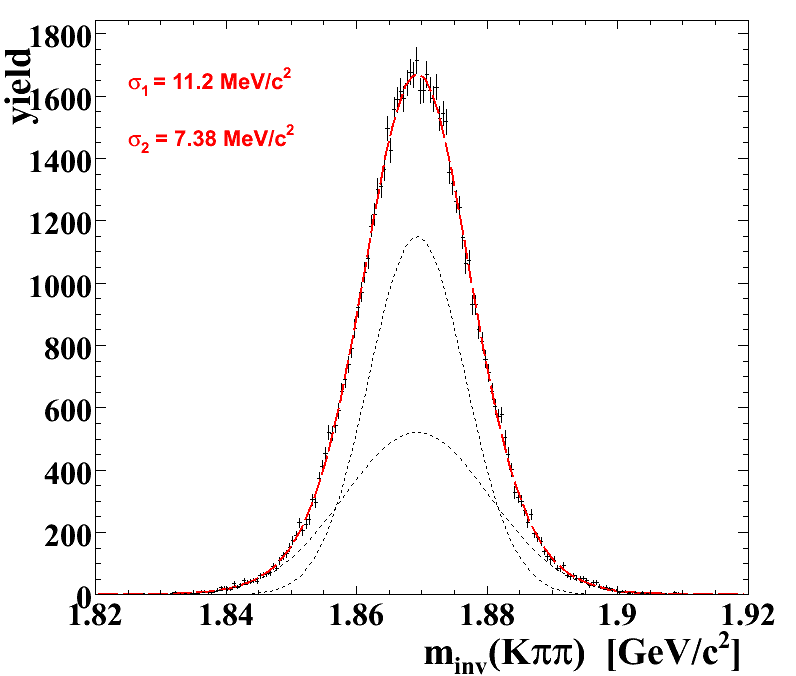}
\caption[Invariant $D^{\pm}$ mass distribution of $\pbarp \to D^{+}
 D^{-}$ signal events]{Invariant $D^{\pm}$ mass distribution of $\pbarp \to
 D^{+} D^{-}$ signal events.} 
\label{fig:DplusMass}
\end{center}
\end{figure}

The event selection requires all six charged tracks to be detected.
\D meson candidates are defined by means of loose mass windows of
$\Delta m = \pm 0.3$~\gevcc
before vertex fitting is done.  
For the \DpDm channel three charged tracks are fitted to a common vertex
and both $D^{\pm}$ 
meson candidates are combined to the initial system, which has to meet
kinematically the beam four-momentum. The confidence level for the
kinematic fit is required to be $CL_{kin}>0.01$. In events
with more than two $D^{\pm}$ candidates the best two are selected according
to the $\chi^2$ value from the vertex fit and the momentum of the candidate,
which should be closer to the kinematically allowed
region. \Reffig{fig:DplusMass} shows the invariant mass distribution of
the $D^{\pm}$ candidates. The signal sample is free of combinatorial
background and 
the distribution is well described by a fit containing a
superposition of two Gaussian components (shown in the plot as
dashed lines) with the individual widths of $\sigma_1=11.2$\,\mevcc and
$\sigma_2=7.4$\,\mevcc. The overall signal efficiency at this stage is
$\epsilon = 39.9$\percent.  
\par
\begin{figure*}[!htb]
\begin{center}
\includegraphics[width=\dwidth]{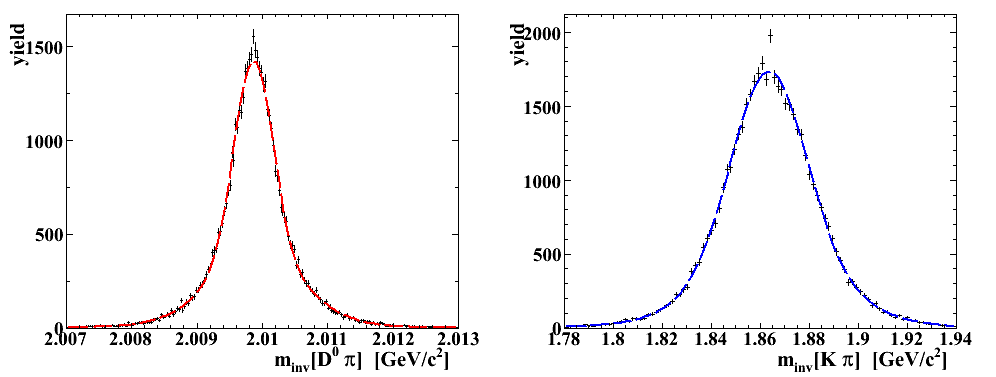}
\caption[Invariant $D^{*\pm}$ and \Dz mass distributions]{Invariant mass
 distributions of the $D^{*\pm}$ candidate (left) and 
 the \Dz candidates (right).}
\label{fig:DstarD0Masses}
\end{center}
\end{figure*}

For the $D^{*+}D^{*-}$ channel, again a loose mass window is set in the
preselection of the \Dz candidate. A kaon and a pion track are combined
and if a common vertex is found, a pion is combined to the previously
selected \Dz candidate
to reconstruct the corresponding $D^{*\pm}$ meson. If both $D^{*\pm}$ are
found, the event is fitted kinematically to the four-momentum of the
beam-target system. A confidence level $CL_{kin}>0.01$
is required.
The invariant masses
of both \D mesons are shown in \Reffig{fig:DstarD0Masses}.  The right
side shows the \Dz invariant mass and the line shape is well described
by a fit containing two Gaussian components with values
$\sigma_1=15.2$\,\mevcc and $\sigma_2=28.3$\,\mevcc.  The left side of
\Reffig{fig:DstarD0Masses} shows the invariant mass distribution of
the $D^{*\pm}$ candidates after a 5C kinematic fit.
\par
This additional mass constraint can be imposed since the 
whole decay tree gets fitted. Together with the beam four-momentum 
constraint, the individual widths of the two Gaussian components are 
$\sigma_1=0.75$\,\mevcc and $\sigma_2=0.29$\,\mevcc. Without the
additional constraint the width of the \Dx mass distribution would be
of the same order as the $D^{\pm}$ mass resolution.
The efficiency for the 5C-fit gets only slightly reduced from
$\epsilon_{signal,4C} = 27.4$~\percent to $\epsilon_{signal,5C} =
24.0$~\percent.  
\par
%
In order to understand the general features of the analysis 
the Dual Parton Model (DPM) was used to produce
background coming from \pbarp annihilations. 
Only inelastic collisions including multi-prong events have been simulated.
Because of the small ratio between the
\DDbar cross section and the total \pbarp cross section, 
a very large number of events needs to be generated. For this reason
this general study was carried out
only for the
\DpDm channel: $8.3\cdot10^7$ events were produced to test the analysis, 
{\it e.g.} for the acceptance of events containing charged track
multiplicities equal or larger than six and to check for possible 
detector-related effects.
Only $11$ events pass the analysis for the \DpDm channel, if no
constraint on the decay vertex position of the two $D^{\pm}$ candidates
is applied . These events contain two charged 
kaons and four pions in the event. Since the DPM model generates the
reaction channel $\pbarp \to K^+K^- 2\pi^+2\pi^-$ as a subset, the
number of remaining events corresponds within the statistics with the
suppression efficiency of the $K^+K^- 2\pi^+2\pi^-$ background channel. 
\par
In a second stage, in order to
evaluate the ability to suppress the background to a sufficient
level, 
a detailed study of specific background reactions was
performed. This includes channels with six charged tracks in the final
state, which could be kinematically interpreted as signal events.
The selected channels and the relative ratio to the total \pbarp
cross section are given in \Reftbl{tab:simDemands}:
%
\par
\begin{table}
 \begin{center}
  \begin{tabular}{lrrr}
   \hline
   \hline
   channel & $\DpDm$ & $D^{*+}D^{*-}$ & ratio to \pbarp\\
   \hline
   DPM                  & 83\,M  &  -  &  - \\
   $3\pip 3\pim\piz$ & 50\,M    &  43\,M & $2.5\cdot 10^{-2}$\\
   $3\pip 3\pim$       & 10\,M   & 14\,M  & $5\cdot 10^{-3}$\\
   $K^+K^- 2\pi^+ 2\pi^-$ & 1\,M    &  10\,M & $5\cdot 10^{-4}$\\
   \hline
   \hline
  \end{tabular}
\caption[Background channels for the \DDbar analysis and amount of
  simulated events]
{Background channels for the \DDbar analysis with the corresponding
  number of simulated events and the ratio to the total \pbarp cross section.}
\label{tab:simDemands}
 \end{center}
\end{table}
\par
\par
\begin{figure*}[t!hb]
\begin{center}
\includegraphics[angle=0,width=0.85\dwidth]{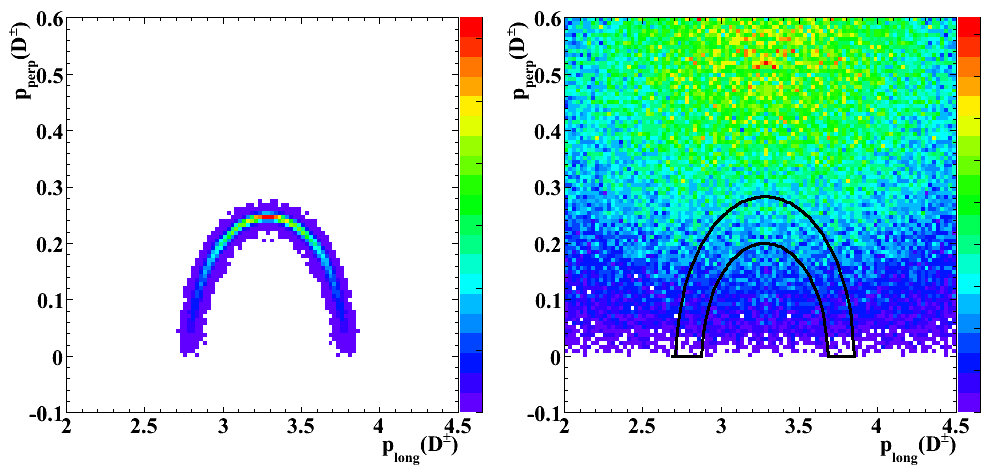}
\caption[Momentum components of $D^{\pm}$ candidates for signal and for background events]
{Momentum components of $D^{\pm}$ candidates for signal events
 (left) and for background events of the reaction  $\pbarp \to K^+K^- 2\pi^+ 2\pi^-$ background
 (right). The surrounded region shows the allow momentum range for
 $D^{\pm}$ candidates.}
\label{fig:backgroundMomentum}
\end{center}
\end{figure*}

To study the background arising from channels with charged kaons and pions
the non-resonant production of $K^{+}K^{-} 2\pi^{+} 2\pi^{-}$
has been analysed, which has at least a $10^6$ times higher cross section
than the \DpDm signal channel close to the  production at
threshold. \Reffig{fig:backgroundMomentum} compares 
the longitudinal vs. the transverse momentum component of the \D meson
candidate for signal (left) and background events (right). 
A constraint on the $D^{\pm}$ momentum  via the definition of a wide
window in the two dimensional
momentum plane gives a reduction factor of $\approx 26$ for background
events. The cut has been chosen to reject mainly those background
events with too large transverse momenta, which would result in lower
values for the invariant $D^{\pm}$ mass. The remaining events 
leave a non-peaking background distribution in the mass region defined
by the loose mass window from the preselection.  Thus, the remaining
background will be kinematically similar to signal events and can
not be separated further by kinematical constraints.
\par
\begin{figure}[!htb]
\begin{center}
\includegraphics[angle=0,width=\swidth]{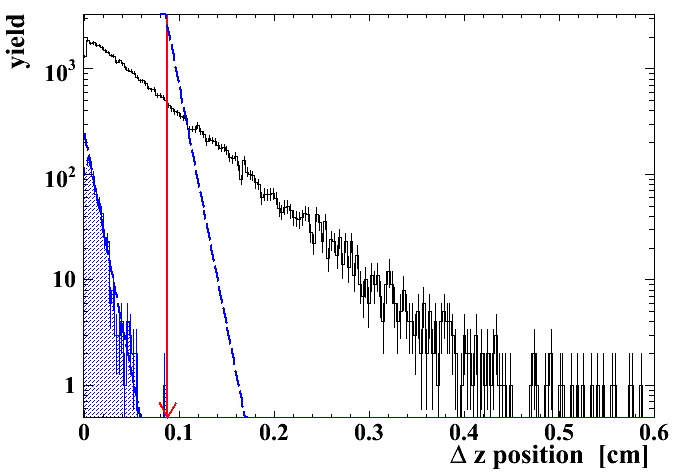}
\caption[]{Distribution of the difference of both reconstructed $D^{\pm}$
 mesons as projection onto the beam axis. Upper curve for signal lower
 for remaining background events (see text for details).}
\label{fig:vertexCut}
\end{center}
\end{figure}

Background events will be produced prompt at the interaction point
and can be in principle separated by finding the $D^{\pm}$ decay vertices
located separated from the primary interaction point. Since the \DDbar
production was studied close to threshold and due to the fixed-target
character of the experiment, the direction of the $D^{\pm}$ mesons will be
close to the beam axis with a small opening angle. 
The uncertainty in the location of the primary vertex is determined by 
the size of the beam spot along the beam which is of the order of
$\sigma_{z,prim} \approx 500\,\mum$. 
On the other hand, using the tracking system, a much better resolution
can be obtained for the $D^{\pm}$ decay
vertex: $\sigma_{x,y}
\approx 35\,\mum$ in the transverse plane and $\sigma_{z}\approx
80\,\mum$ in beam direction, respectively. The reconstructed decay position of both
$D^{\pm}$ candidates can be used to reject background events.
\Reffig{fig:vertexCut} shows the beam axis projection of the difference
vector of the two $D^{\pm}$ meson decay vertices for signal events (black
histogram) and for $K^+K^- 2\pi^+ 2\pi^-$ background events (hatched
histogram) after all kinematic constrains.
Since the simulation requirement for the \DDbar channels is already high, the
amount of simulated background events was chosen to meet the
level of signal generation, under the conservative cross section
assumption given above. The shape
of the background distribution has been scaled by a factor $10^5$, which
is symbolised by an additional blue dashed line in
\Reffig{fig:vertexCut} and would correspond to an equal amount of
produced signal and background events. The   
vertical arrow represents a  $\Delta{z} = 0.088$\,cm cut, where the area
below both distributions is equal. This would
correspond to a signal to background ratio of $1:1$. The remaining 
signal efficiency after the additional vertex cut is $\epsilon_{\Delta z}(\DpDm) =
7.8$\percent. The shape of the background 
distribution can be assumed to be the same like the blue shaded region,
since only vertex constraints determine  the shape of the background 
distribution. An increase of the $\Delta z$ cut can further enhance the signal
to background ratio.
\par
The cross sections for the pionic channels are according to
\Reftbl{tab:simDemands}  a factor of fifty larger
compared to the $\pbarp \to K^+K^- 2\pi^+ 2\pi^-$ background
reaction. For both channels all produced events 
could be suppressed completely, even without demanding a larger kaon
probability in the $D^{\pm}$ reconstruction. To obtain the efficiency
for background events some cuts 
have been relaxed, e.g. the influence of the $D^{\pm}$ momentum cut has
been estimated in an analysis without kinematic
fit.
\begin{table*}[htb]
\begin{center}
 \begin{tabular}{rrrrrr}
  \hline
  \hline
  selection  & \multicolumn{3}{c}{efficiency}  &
  \multicolumn{2}{c}{signal/background}\\
  selection  &  $D^+D^-$  & $3\pi^+ 3\pi^-$ & $3\pi^+ 3\pi^-\pi^0$ &
  $\frac{D^+D^-}{3\pi^+ 3\pi^-}$ & $\frac{D^+D^-}{3\pi^+ 3\pi^-\pi^0}$\\
  \hline
  preselection & 0.43  & $5.4\cdot 10^{-3}$ & $9.6\cdot 10^{-4}$& - &  -\\
  4C-fit & 0.40  & $1.4 \cdot 10^{-6}$ & $4.2\cdot 10^{-7}$  & 0.02 &
  0.015\\
  $D^{\pm}$ momentum & 0.40  & $<1.1 \cdot 10^{-8}$ & $<3.6 \cdot 10^{-9}$ &
  $>2.7$ & $>1.8$\\
  \Kpm \loo PID & 0.23  & $<1.8 \cdot 10^{-9}$ & $<1.7 \cdot 10^{-9}$ &
  $>9.5$ &  $>2.2$ \\
  \hline
  \hline
 \end{tabular}
\end{center}
\caption[Signal to background ratio for the $\pbarp \to \DpDm$
 channel]{Expected signal to background ratios for the \DpDm channel 
to $3\pi^+ 3\pi^-(\piz)$ background events.}
\label{tab:sim3piplus3piminus}
\end{table*}
\par
For the benchmark channel $\pbarp \to 3\pip 3\pim\piz$ the kinematic
suppression of this channel is stronger compared to the $3\pip 3\pim$
channel, although the cross section for this particular channel is a
factor of five larger. The \piz is not reconstructed and in most cases the event does
not fit to the initial beam momentum. For the $3\pip 3\pim$ channel, the
suppression factor caused by the kinematic fit for background events was
estimated to be $\approx 2.6\cdot 10^{-4}$, whereas for the $3\pip 3\pim\piz$
channel it was estimated to $\approx 4.4\cdot 10^{-4}$, respectively.
For both channels no vertex constraints has been used. At this level the
expected signal to background ratio are better than $2:1$.
\Reftbl{tab:sim3piplus3piminus} gives an overview about the
obtained efficiencies and the resulting signal to background ratios.
According to the investigation of the $K^+K^- 2\pi^+ 2\pi^-$
channel and additional cut to the reconstructed $D^{\pm}$ vertex
positions would strongly increase the signal to background ratio.
\par
For the second charmed benchmark channel $\pbarp \to D^{*+} D^{*-}$  in total $10^{7}$
events for the background reaction $K^+K^- 2\pi^+2\pi^-$ have been
analysed. For the generation of the background channels for this physics
channel, a pre-filter has been used, which filtered those events, which
does  not 
have a \Dz and a $D^{*\pm}$ candidate in the event. Only $\approx15$\percent
of all events passed the filter. For this preselection a wide
mass window of $\Delta m = 0.5$~\gevc has been set. All simulated events
could be suppressed by the analysis. This 
corresponds to a signal to background ratio of ${S}/{B} \approx 1:3$.
According to the analysis of this background channel for the \DpDm
signal, an additional suppression factor of $10$ could be obtained by
applying a cut on the difference of the \Dz decay vertices of $\Delta z
= 200\, \mu$m . This constraint would reduce the signal efficiency from
$\epsilon = 24.0$~\percent to $\epsilon = 12.7$~\percent and the signal to
background ratio would improve to $\approx 3:2$. 
\par
To estimate the contributions of the pionic background channels $\pbarp
\to 3\pip 3\pim(\piz)$ to the background of the $D^{*+} D^{*-}$ signal
channel, the total of $5.7\cdot10^7$ background channels have been
simulated (according to \Reftbl{tab:simDemands}). This corresponds to a
total of $3.5\cdot 10^8$ events and no event passes the analysis for
both background channels. To estimate the influence of the suppression
of the kinematic fit, the pionic background channel have been analysed
without kinematic fit and only a few events survive the preselection
including vertex fits. The efficiencies to suppress the background
are given in \Reftbl{tab:Dstarsim3piplus3piminus}. Assuming at least a
factor of $\epsilon_{5C-kin}\approx 10^{-4}$ for the influence of the
kinematic fit  on the background suppression, the expected signal to
background ratio for the pionic channels would than be larger than $S:N
> 12$. This assumption is still very conservative, since the additional
constraint on the \Dz mass in the kinematic fit could not be estimated
and would increase the suppression factor for this background source.
Therefore, the contribution to the expected hadronic background 
from the pionic channels is expected to be small. 
\begin{table*}[htb]
\begin{center}
 \begin{tabular}{rrrrrr}
  \hline
  \hline
  selection  & \multicolumn{3}{c}{efficiency}  &
  \multicolumn{2}{c}{signal/background}\\
  selection  &  $D^{*+}D^{*-}$  & $3\pi^+ 3\pi^-$ & $3\pi^+ 3\pi^-\pi^0$ &
  $\frac{D^{*+}D^{*-}}{3\pi^+ 3\pi^-}$ & $\frac{D^{*+}D^{*-}}{3\pi^+ 3\pi^-\pi^0}$\\
  \hline
  preselection & 0.27  & $5.0\cdot 10^{-7}$ & $7.5\cdot 10^{-8}$& - &  -\\
  5C-fit & 0.24  & $ 5.0 \cdot 10^{-11}$ & $7.5\cdot 10^{-12}$  & 
  $\ge 8.8 $ &  $\ge 12.4$\\
  \hline
  \hline
 \end{tabular}
\end{center}
\caption[Signal to background ratio for the $\pbarp \to D^{*+}D^{*-}$
 channel]{Expected signal to background ratios for the $D^{*+}D^{*-}$ channel 
to $3\pi^+ 3\pi^-(\piz)$ background events.}
\label{tab:Dstarsim3piplus3piminus}
\end{table*}
\par
Assuming the reaction cross sections of
charmonium production above the \DDbar threshold to be in the
order of $3$ nb for the \ensuremath{D^{+}D^{-}} and $0.9$\,nb for the
\ensuremath{D^{*+}D^{*-}} production 
the expected numbers of reconstructed events per year of 
\PANDA operation are at least $1.5\cdot 10^4$ and $1.4\cdot 10^3$,
respectively. For these estimations efficiencies of $\epsilon = 7.8$\percent
for the \DpDm channel and $\epsilon = 24.0$\percent for the
\ensuremath{D^{*+}D^{*-}} channel have been used. These
values are obtained by using only the dominant charged decays of $D$
mesons. Including further decay channels should significantly improve
the signal efficiency. Furthermore, the cross section estimates for the
\DDbar channels given
above are conservative, compared to other estimates using a
quark-gluon string model \cite{bib:phy:Kaidalov94}, the annihilation of a
diquark pair \cite{Kroll89} or the contribution of the \DDbar molecule
hypothesis to higher resonances above the \DDbar threshold
\cite{Braaten08}, which suggest
values up to $10 - 100$~nb in the \panda energy range. These models
usually extrapolate from strange channels, where data exist, to the
charmed region. This allows the study of \DDbar close to threshold and would
strongly increase the expected signal yield for the \DDbar signal
channels.
\par
\subsubsection{$\boldsymbol{\honec}$ Width Measurement}
In order to assess the ability to measure narrow widths we report a study of
the sensitivity of \PANDA to the determination of the $\honec$ width.
For this purpose we performed Monte Carlo simulations of energy scans around the resonance.
Events were generated at 10 different energies around the $\honec$ mass, 
each point corresponding to 5\,days of running the experiment in high resolution mode.
\par
The expected shape of the measured cross-section is the convolution of the Breit-Wigner resonance curve 
with the normalised beam energy distribution and an added background term. 
The expected number of events at the $i$th data point is
\begin{eqnarray}
\nu_{i}=\left[\varepsilon \times \int Ldt\right]_{i} \times \left[\sigma_{bkgd}(E)
+\frac{\sigma_{p}\Gamma^{2}_{R}/4}{(2\pi)^{1/2}\sigma_{i}} \right. \nonumber \\
\left. \times \int  \frac{e^{-(E-E')^{2}/2\sigma_{i}^{2}}}{(E'-M_{R})^{2}+\Gamma_{R}^{2}/4}dE'\right]
\end{eqnarray}
where $\sigma_{i}$ is the beam energy resolution at the $i$th data point, 
$\Gamma_{R}$ and $M_{R}$ the resonance width and mass,
$L$ the luminosity and $\varepsilon $ is an overall efficiency and acceptance factor.
To extract the resonance parameters the likelihood function $-\ln{\cal{\LH}}$ is 
minimised assuming Poisson statistics, where
\par
\begin{equation}
{\LH} =\prod_{j=1}^{N}\frac{\nu_{j}^{n_{j}}e^{-\nu_{j}}}{n_{j}!}
\end{equation}
\par
\begin{figure*}[htb]
\begin{center}
\subfigure[]
{\includegraphics[width=0.9\columnwidth]{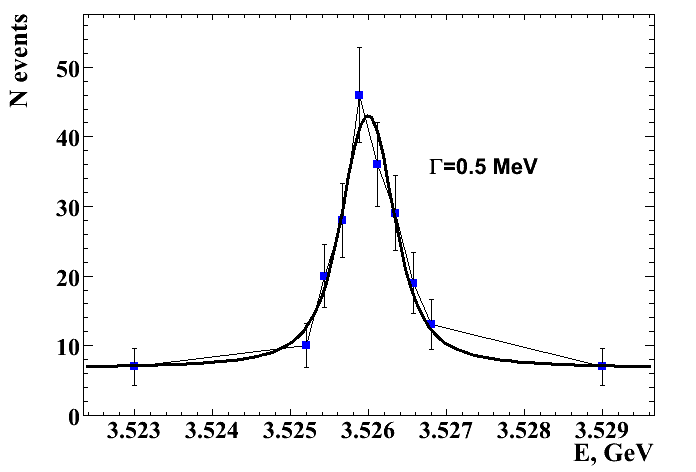}}
\hspace{0.1\columnwidth}
\subfigure[]
{\includegraphics[width=0.9\columnwidth]{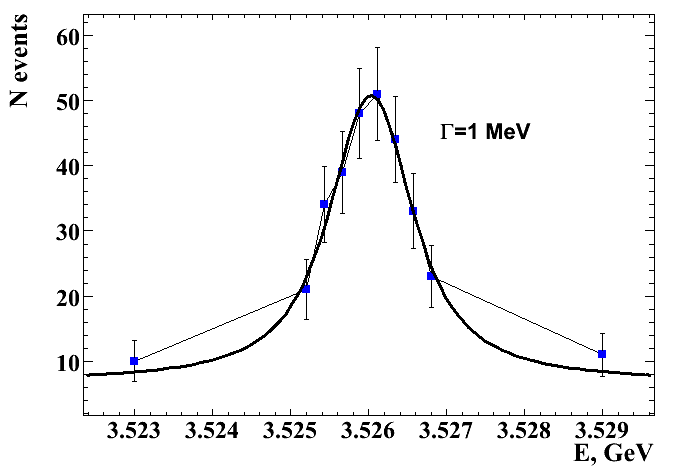}}
\caption{Fit of $\honec$ resonance for $\Gamma$ = 0.5\,MeV (a) and $\Gamma$ = 1\,MeV (b)}
\label{fig:sim:hc_width05MeV}
\end{center}
\end{figure*}

For our simulation we assumed
a signal to background ratio of 8:1 and we used the signal reconstruction 
efficiency of the $\honec \to \etac \gamma \to \phi \phi \gamma$ channel. 
The simulated data were fitted to the expected signal shape with 4 free parameters: 
$E_R$, $\Gamma_{R}$, $\sigma_{bkgd}$, $\sigma_p$. 
The background was assumed to be energy independent.
The study has been repeated for 3 different values of the total width:
$\Gamma_{R}$ = 0.5, 0.75, 1.0\,MeV. 
The results of the fit for 0.5\,MeV and 1.0\,MeV are presented in \Reffig{fig:sim:hc_width05MeV}. 
The extracted $\Gamma_{R}$'s with errors are summarised in \Reftbl{tab:sim:hc_width}.
\par
\begin{table}
\begin{center}
\begin{tabular}{ccc}
  \hline\hline
  $\Gamma_{R, MC}$ [MeV] & $\Gamma_{R, reco}$ [MeV] & $\Delta \Gamma_{R}$ [MeV]\\
  \hline
  1 & 0.92 & 0.24 \\
  0.75 & 0.72 & 0.18\\
  0.5 & 0.52 & 0.14\\
  \hline\hline
\end{tabular}
\caption[Reconstructed $\honec$ width.]{Reconstructed $\honec$ width.}
\label{tab:sim:hc_width}
\end{center}
\end{table}
\par
\par

\subsubsection{Angular Distributions in the Radiative Decay of $\chi_c$}
The measurement of the angular distribution in the radiative decays of the $\chi_{c}$ states provides 
information on the multipole structure of the radiative decay and the properties of the $\overline{c}c$ 
bound state. The process $\pbarp\to \jpsi\gamma\to \ee\gamma$ is dominated 
by the dipole term $E1$. $M2$ and $E3$ terms arise in the relativistic treatment of the interaction between 
the electromagnetic field and the quarkonium system. They contribute to the radiative width at the few 
percent level. \\
The coupling between the set of $\chi$ states and $\pbarp$ is described by four independent 
helicity amplitudes:
\begin{itemize}
\item $\chiczero$ is formed only through the helicity 0 channel;
\item $\chicone$ is formed only through the helicity 1 channel;
\item $\chictwo$ can couple to both.
\end{itemize}
The angular distributions of the $\chicone$ and $\chictwo$ are described by four independent parameters: 
$a_2(\chicone),a_2(\chictwo),B^2_0(\chictwo),a_3(\chictwo)$.\\
The fractional electric octupole amplitude, $a_3\approx E3/E1$ can contribute only to the $\chictwo$ decays, 
and is predicted to vanish in the single quark radiation model if the $\jpsi$ is pure $S$-wave.\\
For the fractional $M2$ amplitude a relativistic calculation yields \cite{teo}:
\begin{equation}
a_2(\chi_{c1})=-\frac{E_{\gamma}}{4m_c}(1+\kappa_c)=-0.065(1+\kappa_c)
\end{equation}
\begin{equation}
a_2(\chi_{c2})=-\frac{3}{\sqrt{5}}\frac{E_{\gamma}}{4m_c}(1+\kappa_c)=-0.096(1+\kappa_c)
\end{equation}
where $\kappa_c$ is the anomalous magnetic moment of the $c$-quark.
\par
\begin{figure}[htbp]
  \begin{center}
  \includegraphics[angle=0,width=\columnwidth]{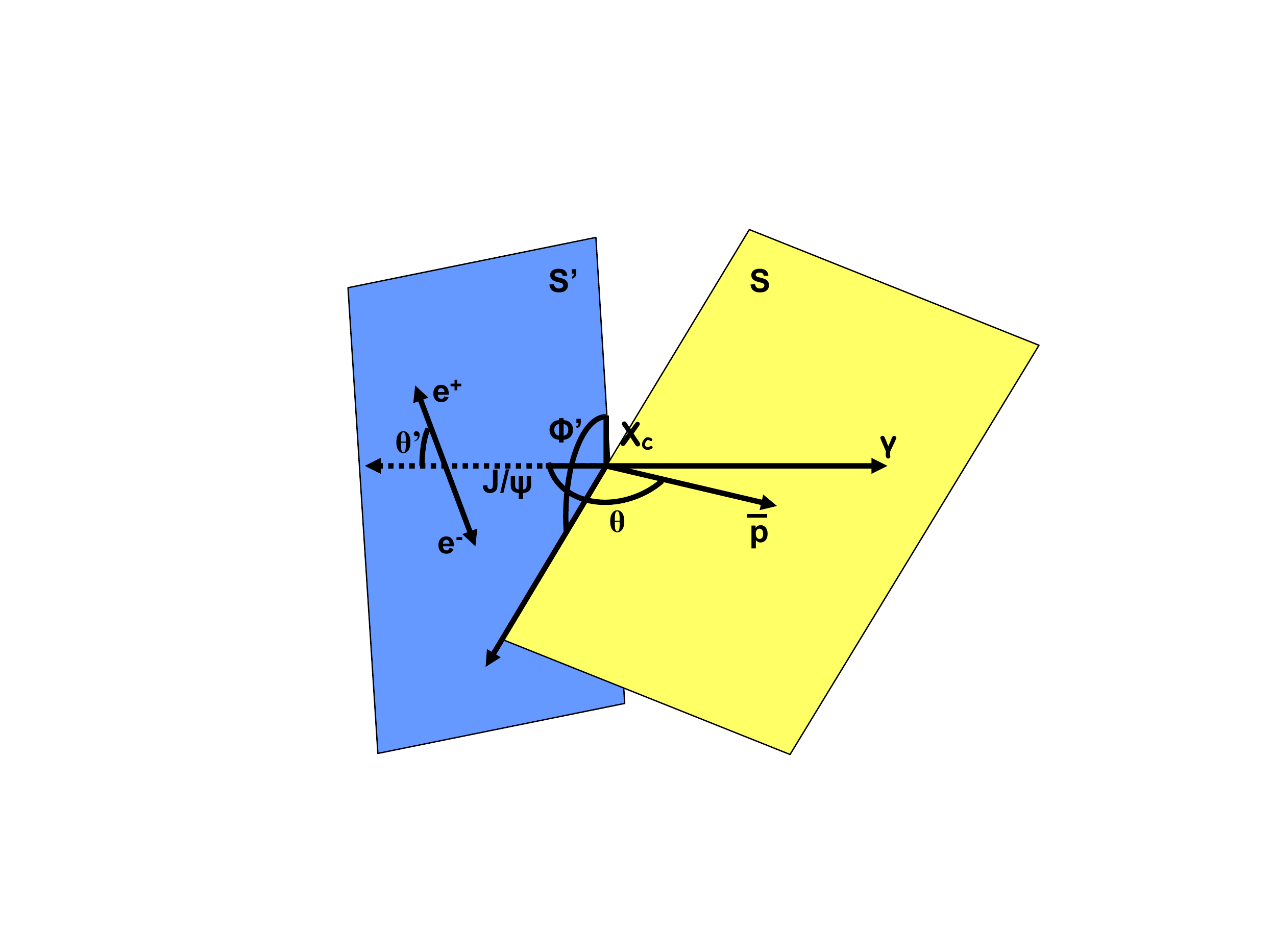}
  \end{center}
\caption{Definition of the angles for the angular distribution of the radiative decay of the $\chi_c$.}
\label{fig:sistrif}
\end{figure}

\Reffig{fig:sistrif} shows the angles used in the description of the angular distribution: 
\begin{itemize}
\item $\theta$ is the polar angle of the $\jpsi$ with respect to the antiproton in the $\pbarp$ 
centre of mass system;
\item $\theta'$ is the polar angle of the positron in the $\jpsi$ rest frame with respect to the $\jpsi$ 
direction in the $\chi_c$ rest system;
\item $\phi'$ is the azimuthal angle between the $\jpsi$ decay plane and the $\chi_c$ plane.
\end{itemize}

The theoretical value of the ratio between $a_2(\chicone)$ and $a_2(\chictwo)$ 
is predicted to be  independent of the c-quark mass and anomalous magnetic moment:
\begin{equation}
 \left(\frac{a_2(\chi_{1})}{a_2(\chi_{2})}\right)_{Th}=\frac{\sqrt{5}}{3}\frac{E_\gamma (\chi_{1}\to \jpsi\gamma)}{E_\gamma (\chi_{2}\to \jpsi\gamma)}=0.676
\end{equation}
E835 measured for the first time this ratio and the result is \cite{fermilab}:
\begin{equation}
\left(\frac{a_2(\chi_{1})}{a_2(\chi_{2})}\right)_{E835}=-0.02\pm 0.34
\end{equation}
While the value of $a_2(\chictwo)$ agrees well with the predictions of a simple theoretical model, 
the value of $a_2(\chicone)$ is lower than expected (for $\kappa_c$=0) and the ratio between the two, 
which is independent of $\kappa_c$ is $\approx 2\sigma$ away from the prediction. 
This could indicate the presence of competing mechanisms, lowering the value of the M2 amplitude 
at the $\chicone$. Further, high statistics measurements of these angular distributions are needed 
to solve this question.\\
In order to do that, following the results of \INST{E835}, a new model of angular distribution was 
implemented using 
the following parameters for the decay of $\chicone$:
\begin{itemize}
\item production amplitudes: $B_0$=0;
\item decay amplitude: $a_2=0.002\pm0.032\pm0.004$;
\end{itemize}
and for the decay of $\chictwo$:
\begin{itemize}
\item production amplitudes: $B_0^2=0.16^{+0.09}_{-0.10}\pm0.01$;
\item decay amplitude: $a_2 = -0.076^{+0.054}_{-0.050}\pm0.009$ and $a_3 = 0.020^{+0.055}_{-0.044}\pm0.009$.
\end{itemize}
The angular distributions for the three angles can be approximately written, for the $\chi_{c1}$, as:\\
\begin{center}
$\overline{W}(cos\theta)\sim 1-\frac{1}{3}\cos^2\theta$\\
$\overline{W}(cos\theta^\prime)\sim 1-\frac{1}{3}\cos^2\theta^\prime$\\
\end{center}
and for the  $\chi_{c2}$:\\
\begin{center}
$\overline{W}(cos\theta)\sim 1-\frac{1}{3}\cos^2\theta$\\
$\overline{W}(cos\theta^\prime)\sim 1+\frac{1}{13}\cos^2\theta^\prime$\\
$\overline{W}(\phi^\prime)\sim 1-\frac{8}{71}\cos 2\phi^\prime$\\
\end{center}
%
\par
\begin{figure*}[htbp]
  \begin{center}
  \includegraphics[angle=90,width=\dwidth]{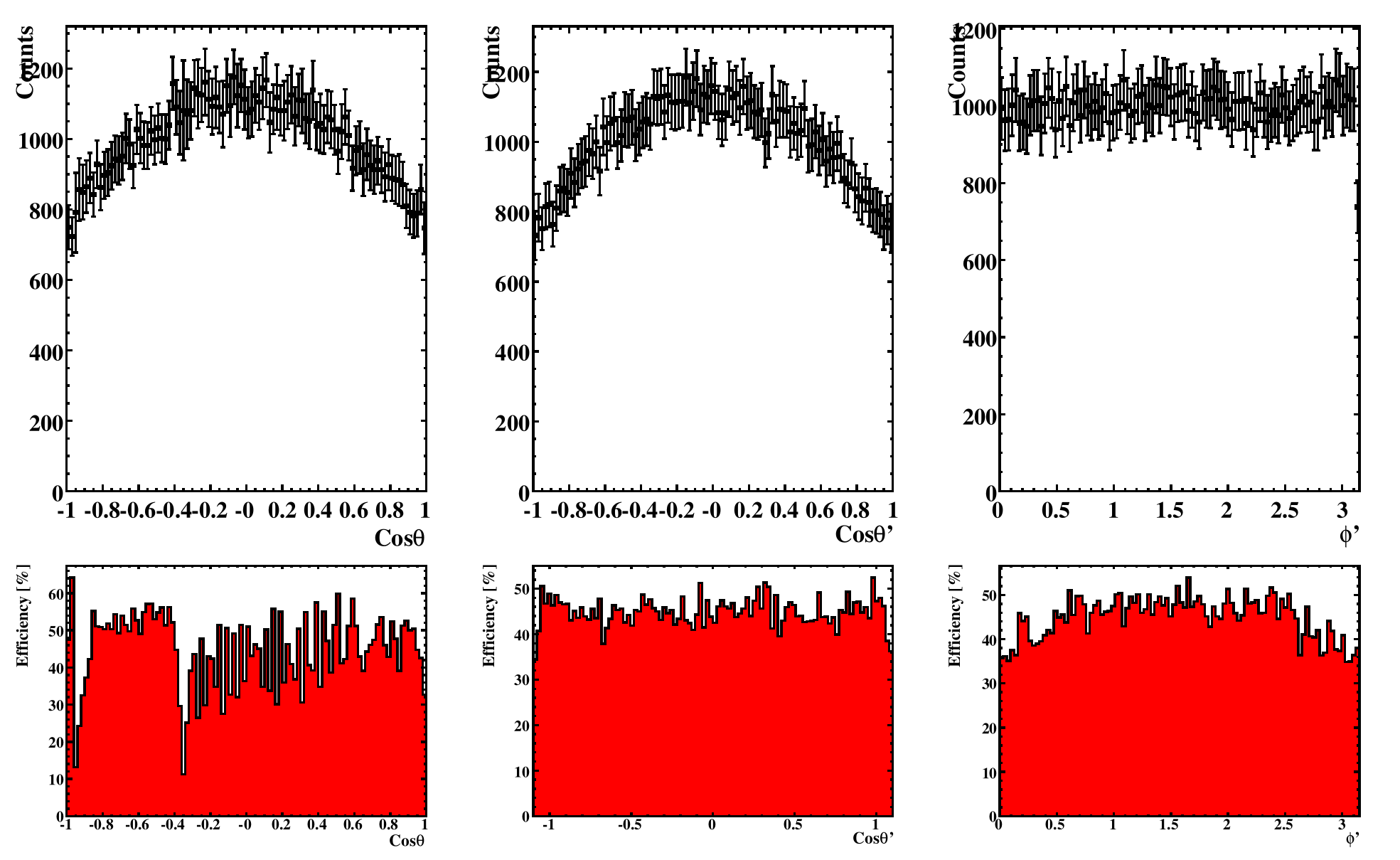}
  \end{center}
\caption{Results for $\cos\theta$, $\cos\theta'$ and $\phi'$ after the generation and reconstruction of 
the events for $\chicone$ decay.}
\label{fig:chi1_eff}
\end{figure*}
\begin{figure*}[htbp]
  \begin{center}
  \includegraphics[angle=90,width=\dwidth]{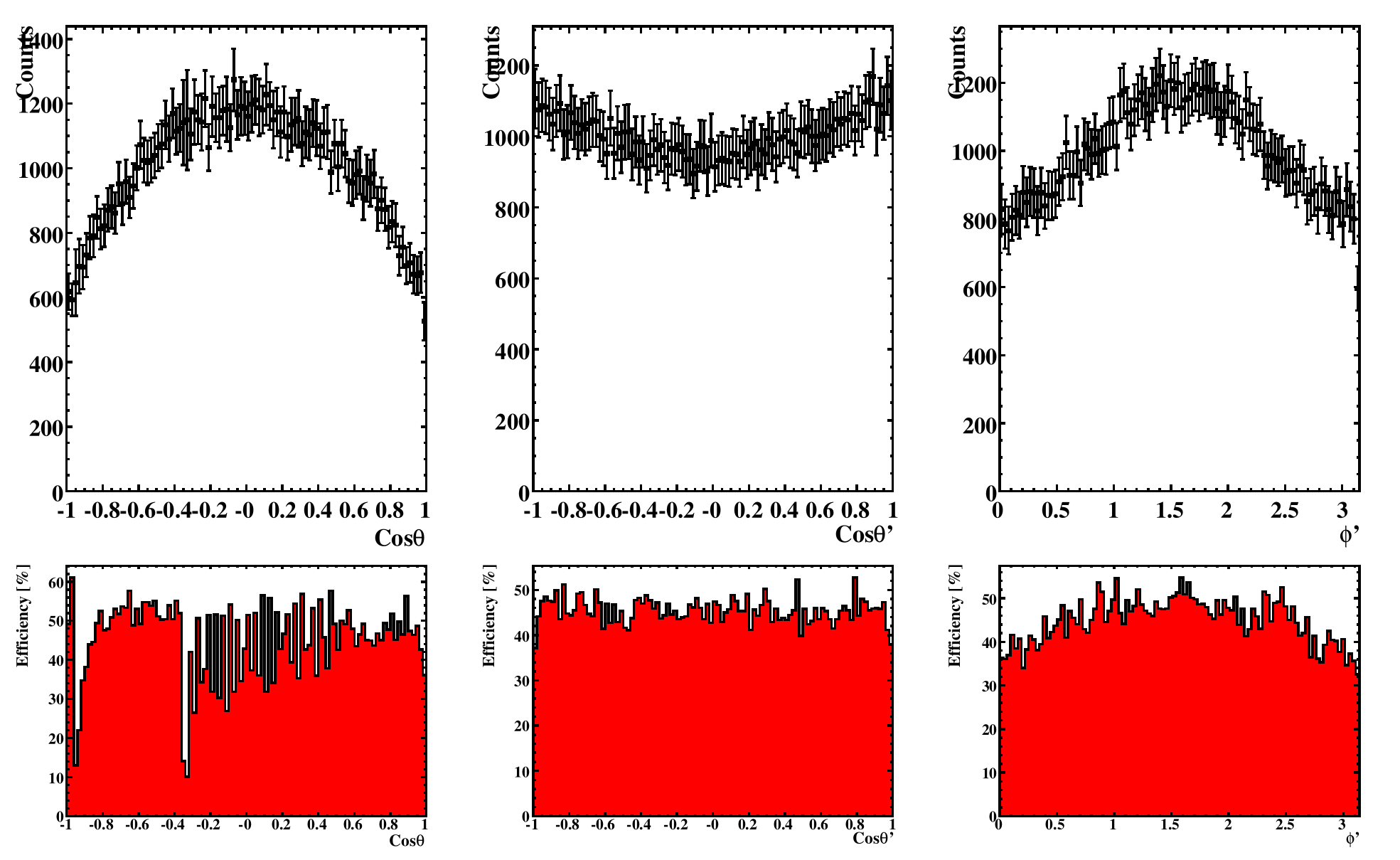}
  \end{center}
\caption{Results for $\cos\theta$, $\cos\theta'$ and $\phi'$ after the generation and reconstruction of 
the events for $\chictwo$ decay.}
\label{fig:chi2_eff}
\end{figure*}

\Reffig{fig:chi1_eff} and \ref{fig:chi2_eff} present the results obtained for the $\cos\theta$, 
$\cos\theta'$ and $\phi'$ distributions after the generation and reconstruction of the events for the 
decay of $\chi_{c1,2}\to \jpsi\gamma$. The top plots show the angle distributions corrected 
with the efficiency, which is presented in the lower part of the plot.
\par
The dip in the efficiency around $\cos(\theta) \sim -0.4$ corresponds to events in which the photons are 
emitted at a polar angle in the laboratory system of $\sim 20\degrees$, in the transition zone between 
the target and forward electromagnetic calorimeters.
The projected angular distributions, after efficiency correction, 
are consistent with the observation from \INST{E835}.
\clearpage

%% file: phys/phys_exotic.tex
%
\clearpage
\subsection{Exotic Excitations}
%
\input{./phys/exotic/phys_exo_hybrids}
\input{./phys/exotic/phys_exo_glueball}
\input{./phys/exotic/phys_exo_molecules}
\subsubsection{Benchmark Channels}
\input{./phys/exotic/phys_exo_hybrids_benchmark.tex}
\input{./phys/exotic/phys_exo_glueballs_benchmark.tex}
%

%% file: phys/exotic/phys_exo_hybrids.tex
%
\subsubsection{Hybrids - Gluonic Excitation of \qqbar{} States}
\begin{figure*}[htb]
\begin{center}
\includegraphics[width=0.85\dwidth]{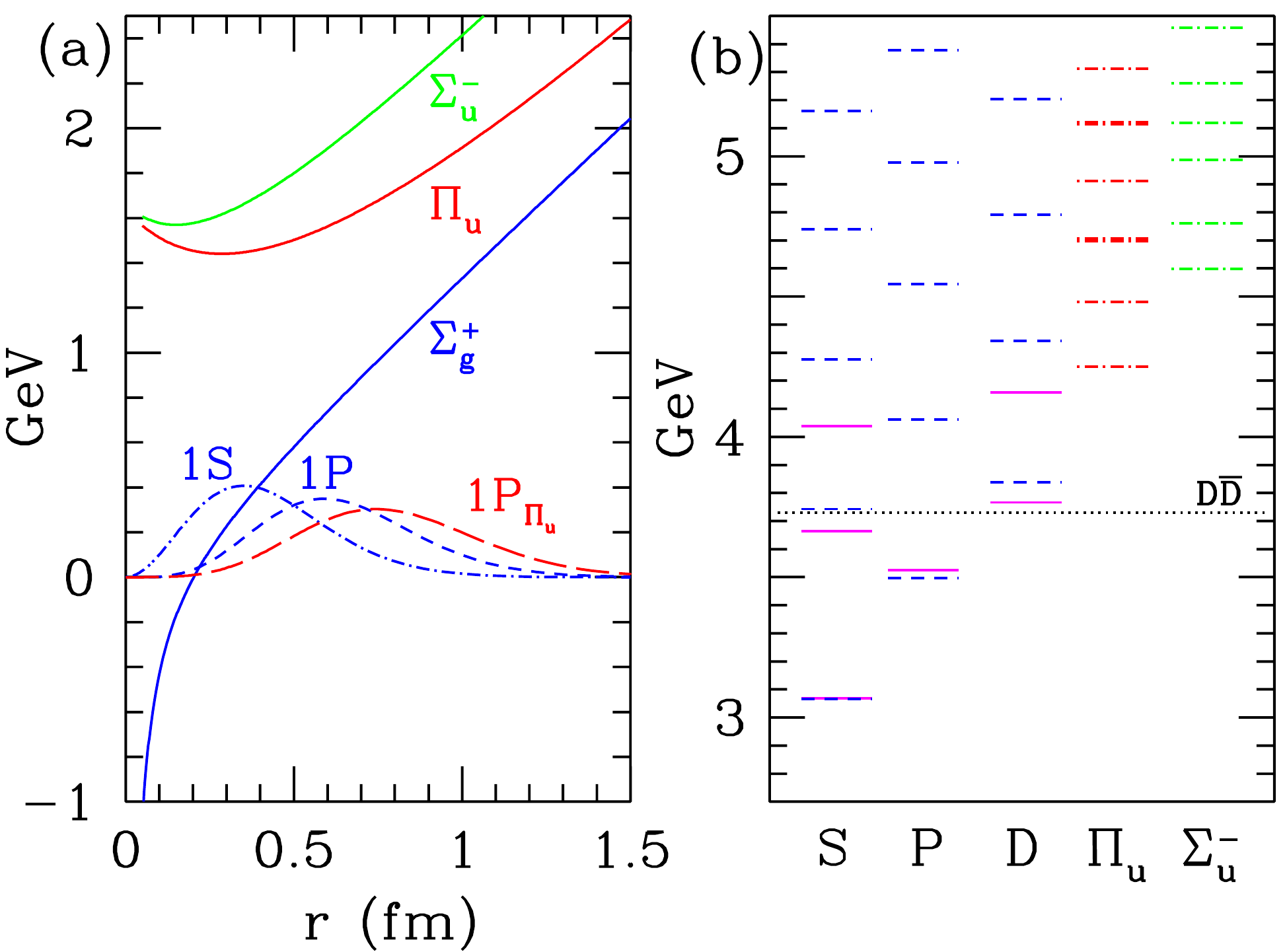}
\caption[Charmonium potentials and hybrid spectrum from LQCD]{(a)
Heavy quarkonium potentials and wave functions for different
excitation levels from LQCD. $\Sigma$ denotes normal one gluon
exchange while the excited $\Pi$ potentials are the lowest lying
hybrid potentials. In that case, the attraction is not mediated by
a single gluon but a string of gluons which carry angular
momentum. (b) shows the charmonium spectrum from LQCD. The
conventional charmonium states are on the right while the hybrids
are found in column $\Pi_u$ and $\Sigma_u^-$.
See~\cite{bib:phy:kutimorningstar2003} for details.}
\label{fig:phys:exotic:hybrid_potential}
\end{center}
\end{figure*}

The glue tube (often referred to as flux-tube) adds degrees of freedom
which may manifest in vibrations of the tube. The higher the excitation
the more units of angular momentum are carried. These different levels
of excitation can be translated into different potentials, one for each
mode. \Reffig{fig:phys:exotic:hybrid_potential} shows this together with the
corresponding wave functions.
\paragraph*{Hybrids with Exotic Quantum Numbers}\ \\
The additional degrees of freedom ({\it e.g.} the vibration of the glue tube) manifest themselves
also in a contribution to the quantum numbers of the topology. In the simplest scenario
this corresponds to adding the quantum numbers of a gluon ($J^P$=$1^+$ or $1^-$ depending
on if it is a colour-electric or colour-magnetic excitation) to a simple $\qqbar$ pair. Therefore
they are often referred to as hybrids. This procedure creates {\it e.g.} for S-wave mesons
8 lowest lying hybrid states (see \Reftbl{tbl:phy:exotic:qqg}).
\par
An important experimental aspect here is that 3 out of these 8 states exhibit quantum numbers
which can not be formed by a normal $\qqbar$ pair. Therefore these quantum numbers are called
exotic.

The most promising results for gluonic hadrons have come from
antiproton annihilation experiments. Two particles, first seen in
$\pi N$ scattering~\cite{bib:phy:thompson97,bib:phy:adams98} with
exotic $\JPC\,=\,\JPConemp$ quantum numbers, $\pione(1400)$
\cite{bib:phy:Abele98} and $\pione(1600)$
\cite{bib:phy:Reinnarth00} are clearly seen in $\pbarp$
annihilation at rest (for a more detailed list of observations see \Reftbl{tbl:phys:exotic:light_hybrids}).
\begin{table}
\begin{center}
\begin{tabular}{lll} \hline\hline
 & \multicolumn{2}{c}{Gluon} \\
($\qqbar$)$_8$ & 1$^{-}$ (TM) & 1$^{+}$ (TE) \\
\hline
$^1$S$_0$, 0$^{-+}$ & 1$^{++}$ & 1$^{--}$ \\
\hline
$^3$S$_1$, 1$^{--}$ & 0$^{+-}$ $\leftarrow$ exotic & 0$^{-+}$ \\
 & 1$^{+-}$ & 1$^{-+}$ $\leftarrow$ exotic\\
 & 2$^{+-}$ $\leftarrow$ exotic & 2$^{-+}$ \\
\hline \hline
\end{tabular}
\caption[The 8 lowest lying hybrid states]{The coupling of spins leads to 8 hybrid states for each pair of pseudoscalar and vector
mesons with equal isospin. Even in this simple case three J$^{PC}$ combination are not allowed for
conventional $\qqbar$ pairs.}
\label{tbl:phy:exotic:qqg}
\end{center}
\end{table}

\begin{table*}[htb]
\begin{center}
\begin{tabular}{llcrlrlcc}
\hline\hline
Experiment & Exotic & $\JPC$ & \multicolumn{2}{c}{Mass [$\mevcc$]} & \multicolumn{2}{c}{Width [$\mevcc$]} & Decay & Refs.\\
\hline
\INST{E852} & $\pi_1(1400)$ & $\JPConemp$ & 1359 &
$^{+16}_{-14}$ $^{+10}_{-24}$ &
314 &$^{+31}_{-29}$ $^{+9}_{-66}$& $\eta\pi$ & \cite{bib:phy:e852a} \\
\INST{Crystal Barrel} & $\pi_1(1400)$ & $\JPConemp$ &
1400 & $\pm$20$\pm$20 &
310 & $\pm$50 $^{+50}_{-30}$ & $\eta\pi$ & \cite{bib:phy:Abele98}\\
\INST{Crystal Barrel} & $\pi_1(1400)$ & $\JPConemp$ &
1360 & $\pm$25 &
220 & $\pm$90 & $\eta\pi$ & \cite{bib:phy:cb1360}\\
\INST{Obelix} & $\pi_1(1400)$ & $\JPConemp$ &
1384 & $\pm$28 &
378 & $\pm$58 & $\rho\pi$ &\cite{bib:phy:ox1384}\\ \hline
\INST{E852} & $\pi_1(1600)$ & $\JPConemp$ &
1593 &$\pm$8 $^{+29}_{-47}$ &
168 &$\pm$20 $^{+150}_{-12}$ & $\rho\pi$ & \cite{bib:phy:e852b}\\
\INST{E852} & $\pi_1(1600)$ & $\JPConemp$ &
1597 &$\pm$10 $^{+45}_{-10}$ &
340& $\pm$40$\pm$50 & $\etaprime\pi$ & \cite{bib:phy:e852b}\\
\INST{Crystal Barrel} & $\pi_1(1600)$ & $\JPConemp$ &
1590 & $\pm$50 &
280 & $\pm$75 & $b_1\pi$ & \cite{bib:phy:cb1590}\\
\INST{Crystal Barrel} & $\pi_1(1600)$ & $\JPConemp$
& 1555& $\pm$50 & 468& $\pm$80 &  $\etaprime\pi$ & \cite{bib:phy:Reinnarth00}\\
\INST{E852} & $\pi_1(1600)$ & $\JPConemp$ &
1709 &$\pm$24$\pm$41 &
403 &$\pm$80$\pm$115 & $f_1\pi$ & \cite{bib:phy:e852e}\\
\INST{E852} & $\pi_1(1600)$ & $\JPConemp$ &
1664&$\pm$8$\pm$10 &
185&$\pm$25$\pm$28 & $\omega\pi\pi$ & \cite{bib:phy:e852d} \\ \hline
\INST{E852} & $\pi_1(2000)$ & $\JPConemp$ &
2001&$\pm$30$\pm$92 &
333&$\pm$52$\pm$49 & $f_1\pi$ & \cite{bib:phy:e852e} \\
\INST{E852} & $\pi_1(2000)$ & $\JPConemp$ &
2014&$\pm$20$\pm$16 &
230&$\pm$32$\pm$73 & $\omega\pi\pi$ & \cite{bib:phy:e852d} \\ \hline
\INST{E852} & $h_2(1950)$ & $2^{+-}$ &
1954 &$\pm$8 &
138&$\pm$3 & $\omega\pi\pi$ & \cite{bib:phy:e852f} \\
\hline\hline
\end{tabular}
\caption[Light states with exotic quantum numbers]{Light states with exotic quantum numbers.
The experiment \INST{E852} at \INST{BNL} was performed with a pion beam on a hydrogen target, 
while \INST{Crystal Barrel} was a $\ppbar$ spectroscopy experiment at \INST{LEAR}.}
\label{tbl:phys:exotic:light_hybrids}
\end{center}
\end{table*}
\paragraph{Charmonium Hybrids}\ \\
Exotic charmonia are expected to exist in the
3--5$\,\gevcc$ mass region where they could be resolved and
identified unambiguously.
\par
Predictions for hybrids come mainly from calculations based on the
bag model, flux tube model, and constituent gluon model and,
recently, with increasing precision, from LQCD
\cite{bib:phy:chen01,bib:phy:michael99}. For hybrids, the
theoretical results qualitatively agree, lending support to the
premise that the predicted properties are realistic. Charmonium
hybrids can be expected since the effect of an extra gluonic
degree of freedom in meson-like systems is evident in the
confining potentials for the $\ccbarg$ system ({\it e.g.}\ as derived
from LQCD calculations in the Born-Oppenheimer approximation
\cite{bib:phy:michael99}).
\par
The discussions have only been centred around the lowest-lying
charmonium hybrids. Four of these states ($\JPC\,=\,\JPConemm$,
\JPCzeromp, \JPConemp, \JPCtwomp) correspond to a $\ccbar$ pair
with $\JPC\,=\,\JPCzeromp$ or \JPConemm, coupled to a gluon in the
lightest mode with $\JPC\,=\,\JPConemm$. The other four states
($\JPC\,=\,\JPConepp$, \JPCzeropm, \JPConepm, \JPCtwopm) with the
gluon mode $\JPC\,=\,\JPConemp$ are probably heavier. 
All models agree that the lightest exotic state would be \JPConemp.
Predictions for the mass are listed in \Reftbl{tbl:phys:exotic:onemp_ccbar}.
In addition to the lightest exotic state there are seven other hidden charmed hybrids to be discovered.
The well accepted picture is that the quartet 1$^{--}$,(0,1,2)$^{-+}$ is lower in mass than 1$^{++}$,(0,1,2)$^{+-}$.
The expected splitting is about 100-250\,MeV from 1$^{-+}$ to 0$^{+-}$~\cite{bib:phy:jkm00a,bib:phy:manke98}.
In addition there is fine-splitting within the
hybrid triplets, so that the levels are spread over a few hundred
MeV ({\it e.g.} 4.14\,\gevcc for 0$^{-+}$ and 4.52\,\gevcc for 2$^{-+}$)~\cite{bib:phy:page,bib:phy:mp} which was verified
by lattice QCD \cite{bib:phy:jkm00b}. The actual signature is therefore not just an individual state, but also the 
pattern of states.
\par
Charmonium hybrids are likely to be narrower since open-charm
decays are forbidden or suppressed below the
$D\overline{D}_{J}^{\ast}+c.c.$ (often referred to as
$DD^{\ast\ast}$) threshold. From experiments at \INST{LEAR} we
know that production rates of such $\qqbar$ states are similar to
those of states with exotic quantum numbers. Thus, we estimate
that the cross sections for the formation and production of
charmonium hybrids will be similar to those of normal charmonium
states which is in the order of 120$\,$pb
($\pbarp\,\to\,\jpsi\piz$ \cite{bib:phy:chester91}), in agreement
with theoretical predictions \cite{bib:phy:gaillard82}.
\begin{table*}[htb]
\begin{center}
\begin{tabular}{lp{4cm}ll}
\hline\hline
(a)& m($\ccbar$g), 1$^{-+}$ & Group & Ref. \\
\hline
& 4\,390$\pm$80$\pm$200 & MILC97 & \cite{bib:phy:milc97} \\
& 4\,317$\pm$150 & MILC99 & \cite{bib:phy:milc99} \\
& 4\,287 & JKM99 & \cite{bib:phy:jkm99} \\
& 4\,369$\pm$37$\pm$99 & ZSU02 & \cite{bib:phy:mei:2002ip} \\
\hline\hline
& \ & \ \\
\hline\hline
(b) & m($\ccbar$g,1$^{-+}$)-m($\ccbar$,1$^{--}$) &  Group &  Ref. \\
\hline
& 1\,340$\pm$80$\pm$200 & MILC97 & \cite{bib:phy:milc97} \\
& 1\,220$\pm$150 & MILC99 & \cite{bib:phy:milc99} \\
& 1\,323$\pm$130 & CP-PACS99 & \cite{bib:phy:cp-pacs99} \\
& 1\,190 & JKM99 & \cite{bib:phy:jkm99} \\
& 1\,302$\pm$37$\pm$99 & ZSU02 & \cite{bib:phy:mei:2002ip} \\
\hline\hline
\end{tabular}
\caption[LQCD predictions for the lightest spin exotic hybrid]{Hybrid masses (a)
and mass differences (b) from quenched LQCD. The results for the lightest
$\JPC$ exotic cluster around the threshold for $D\overline{D}^{\ast\ast}+c.c.$ production.}
\label{tbl:phys:exotic:onemp_ccbar}
\end{center}
\end{table*}
The naming definitions of~\Reftbl{tbl:phy:exotic:ccbhybdef} are used for the subsequent discussions.
\begin{table}
\begin{center}
\begin{tabular}{ll}
\hline \hline
$\tilde{\chi}_{c1} $ 1$^{++}$ & $\tilde{\psi}$ 1$^{--}$ \\
\hline
$\tilde{h}_{c0}$ 0$^{+-}$ $\leftarrow$ exotic & $\tilde{\eta}_{c0}$ 0$^{-+}$ \\
$\tilde{h}_{c1}$ 1$^{+-}$ & $\tilde{\eta}_{c1}$ 1$^{-+}$ $\leftarrow$ exotic\\
$\tilde{h}_{c2}$ 2$^{+-}$ $\leftarrow$ exotic & $\tilde{\eta}_{c2} $2$^{-+}$ \\
\hline \hline
\end{tabular}
\caption{Lowest lying charmonium hybrids corresponding to the definitions as in 
\Reftbl{tbl:phy:exotic:qqg}.}
\label{tbl:phy:exotic:ccbhybdef}
\end{center}
\end{table}
\paragraph*{Production vs. Formation}\ \\
Formation experiments would generate non-exotic charmonium hybrids
with high cross sections while production experiments would yield
a charmonium hybrid together with another particle, such as a
$\pi$ or an $\eta$. In $\pbarp$ annihilation, production
experiments are the only way to obtain charmonium hybrids with
exotic quantum numbers. It is envisaged that the first step of
exploring charmonium hybrids would consist of production
measurements at the highest antiproton energy available
($E_{\pbar}\,=\,15\,\gev$, $\sqrt{s}\,=\,5.46\,\gevcc$) and
studying all possible production channels available to cover
exotics and non-exotic states. The next step would consist of
formation measurements by scanning the antiproton energy in small
steps in the regions where promising hints of hybrids have been
observed in the production measurements, thus having a second
check on the static properties like the $\JPC$ assignment as well
as mass and width.
\par
The discovery of such a reaction would necessitate a very good
charmonium reconstruction efficiency. Thus, apart from the
benchmark channels used for conventional charmonium, the charmed
hybrid production in the mode
$\pbarp\,\rightarrow\,\tilde{\eta}_{c1}\eta\,\rightarrow\,\chicone\pi^0\pi^0\eta$
($\tilde{\eta}_{c1}$ is often referred to as $\psi_g$)
has also been used to study the detector performance in multi-body
charmonium reactions (see~\Refsec{sec:sim:res:hybrid}).
\paragraph*{Proposed Measurements}\ \\
The main goal is to measure all low lying charmonium hybrid states.
From the 8 states, it is possible to measure 7 of them using three
channels with a charmonium final state. The possible reactions are
\begin{eqnarray}
\ppbar & \,\to\, & \tilde{\eta}_{c0,1,2}\eta \,\to\, \chicone\piz\piz\eta\\
\ppbar & \,\to\, & \tilde{h}_{c0,1,2}\eta \,\to\, \jpsi\piz\piz\eta \label{eq:channel2}\\
\ppbar & \,\to\, & \tilde{\psi}\eta \,\to\, \jpsi\omega [\piz\ \mbox{or}\ \eta]
\end{eqnarray}
Since the final states $\chicone\piz\piz\eta$ with $\chicone\,\to\,\jpsi \gamma$ and
$\jpsi\piz\piz\eta$ are very similar in terms of multiplicity and photon energies,
the slightly more complicated channel (\ref{eq:channel2}) with a radiative charmonium decay and a
charmonium hybrid mass of 4.3\,\gevcc is used to
test the sensitivity of the detector.
Although a charmonium final state is likely
an open charm final state may also be possible and 7 of the already mentioned 8
states may be accessed with the final state $D\Dx$. Thus the reactions

\begin{equation}
\ppbar  \,\to\,  [\tilde{\eta}_{c0,1,2},\tilde{h}_{c0,1,2},\tilde{\chi}_{c1}] \eta \,\to\, D\Dx\eta
\end{equation}

should be measured and will be also used as a benchmark channel with a
charmonium hybrid mass of about 4.3\,\gevcc. 
 In order to ensure reasonable event statistics, $D$ decays
with high yields have to be combined. In both cases, the charmonium and the open
charm channels are detailed partial wave decomposition has to be performed
to disentangle the different waves.
Since the experimental findings $Y(3940)$ and $Y(4320)$ are also discussed
in the framework of hybrids, {\it e.g.} as vector charmonium hybrid candidates 
they are also used as benchmark channels.
%

%% file: phys/exotic/phys_exo_glueball.tex
%
\subsubsection{Glueballs - Gluonic Excitation of the QCD Vacuum}
LQCD calculations make rather detailed predictions for the
glueball mass spectrum in the quenched approximation disregarding
light quark loops \cite{bib:phy:mornpeardon99}. For example, the
calculated width of approximately 100$\,\mevcc$
\cite{bib:phy:sexton95} for the ground-state glueball matches the
experimental results. LQCD predicts the presence of about 15
glueballs, some with exotic quantum numbers in the mass range
accessible to the \HESR{}.
\par
Glueballs with exotic quantum numbers are called oddballs which
cannot mix with normal mesons. As a consequence, they are
predicted to be rather narrow and easy to identify experimentally
\cite{bib:phy:page00}. It is conceivable that comparing oddball
properties with those of non-exotic glueballs will reveal deep
insight into the presently unknown glueball structure since the
spin structure of an oddball is different~\cite{bib:phy:page00}.
The lightest oddball, with $\JPC\,=\,\JPCtwopm$  and a predicted
mass of 4.3$\,\gevcc$, would be well within the range of the
proposed experimental program. Like charmonium hybrids, glueballs
can either be formed directly in the $\pbarp$-annihilation
process, or produced together with another particle. In both
cases, the glueball decay into final states like $\phi\phi$ or
$\phi\eta$ would be the most favourable reaction below
3.6$\,\gevcc$ while $\jpsi\eta$ and $\jpsi\phi$ are the first
choice for the more massive states.
\par
The indication for a tensor state around 2.2$\,\gevcc$ was found
in the experiment of \INST{Jetset} collaboration at
\INST{LEAR}~\cite{bib:phy:jones2003}. The acquired statistics was
not sufficient for the complementary reactions to be determined.
We plan to measure the $\pbarp\,\to\,\phi \phi$ channel with
statistics of two orders of magnitude higher than in the previous
experiments. Moreover, other reactions of two vector particle
production, such as $\pbarp\,\to\,\omega \omega, K^\ast
\overline{K^\ast}, \rho \rho$ will be measured. However, the best
candidate for the pseudo-scalar glueball ($\eta_L$(1440)), studied
comprehensively at \INST{LEAR} by the \INST{Obelix}
collaboration~\cite{bib:phy:etaLoblxa,bib:phy:etaLoblxb,bib:phy:etaLoblxc,bib:phy:etaLoblxd,bib:phy:etaLoblxe},
is not widely accepted to be a glueball signal because the
calculations of LQCD predict its mass above 2$\,\gevcc$.
Therefore, new data on many glueball states are needed to make a
profound test of different model predictions.
\par
It is worth stressing again that $\pbarp$-annihilations present a
unique possibility to search for heavier glueballs since
alternative methods have severe limitations. The study of
glueballs is a key to understanding long-distance QCD. Every
effort should be made to identify them uniquely.
\begin{figure}
\begin{center}
\includegraphics[width=0.9\swidth]{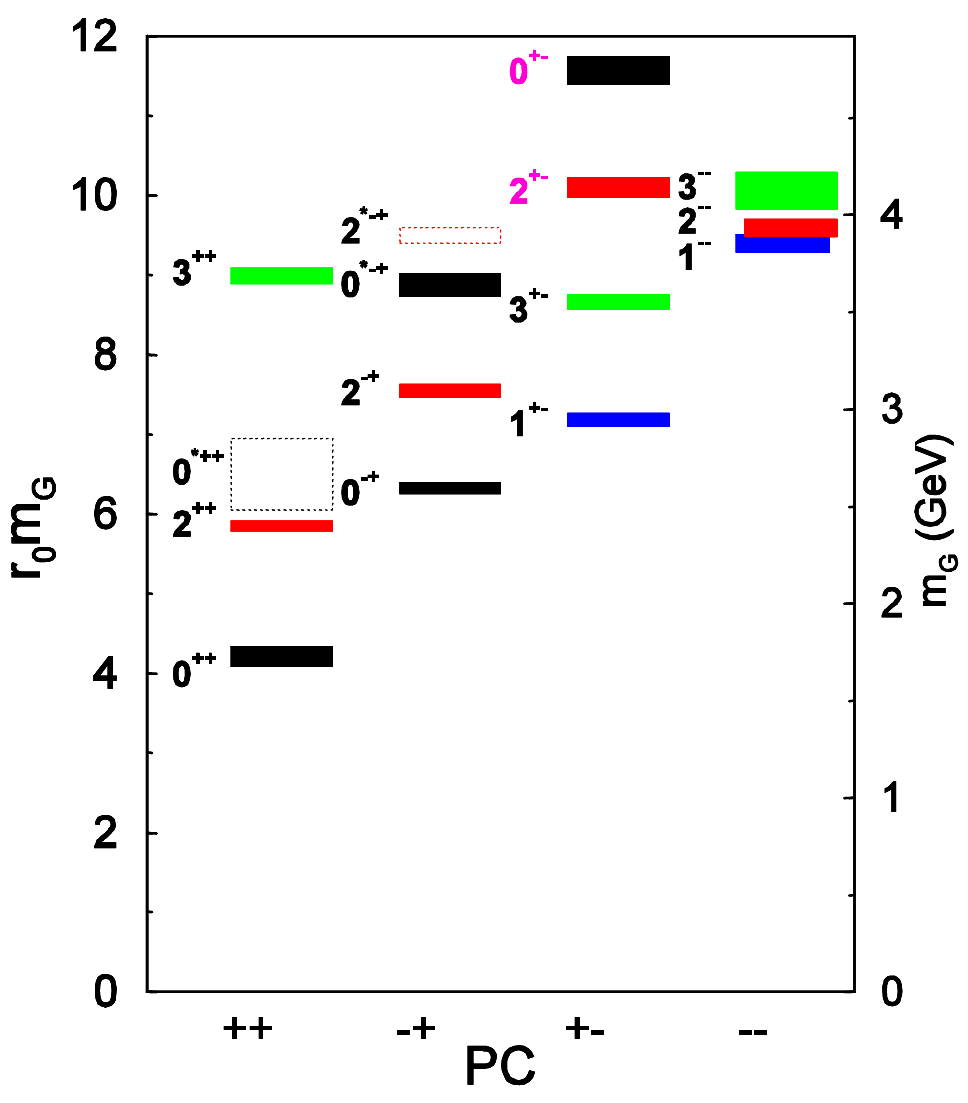}
\caption[Glueball prediction from LQCD calculations]{ Glueball
prediction from LQCD calculations. See
\cite{bib:phy:mornpeardon97,bib:phy:mornpeardon99} for details.
While the region of the ground-state glueball was investigated in
the \INST{LEAR} era (in particular by \INST{Crystal Barrel}) are the
tensor glueball and the spin exotic glueballs with 
$\JPC=0^{+-}$ and $2^{+-}$ important research topics for \Panda.}
\label{fig:phys:exotic:gg:lqcd}
\end{center}
\end{figure}
\par
%
\paragraph*{Light Glueballs}\ \\
Since decades light meson spectroscopy experiments tried to identify the
lowest lying glueball states. Many high statistics experiments have been performed
which delivered excellent information about the scalar and pseudoscalar waves.
Nevertheless, due to the unavoidable mixing problem and the large widths arising from
missing or smooth damping functions
pinning down of the scalar glueballs will very difficult.
\par
In light quark domain the tensor glueball is the best candidate to look at
from experimental means. There is potential mixture from two nonets
($^3P_2$ and $^3F_2$) which sums up to 5 expected isoscalar states, but SU$_F$(3)
forbids $\phi\phi$ decays to first order for the conventional $\qqbar$ states, while
there is no suppression for a potential glueball. The mass for the glueball
is expected in the range from 2.0\,\gevcc to 2.5\,\gevcc. Thus the benchmark reaction is
$\ppbar\,\to\, f_2(2000-2500)\,\to\, \phi\phi$.
\paragraph*{Heavy Oddballs}\ \\
Since glueballs don't have to obey any OZI rule, they may decay in any open channel.
For glueballs above the open charm pair production threshold also decays in to $D$ mesons and its
excitations should be easily possible. The width is completely unknown. Since a lot of
channels are potentially open, the heavy glueballs could be extremely wide. Nevertheless it is known from many other reactions, that nature seems to invest more likely in mass rather than in breakup-momentum,
thus giving the opportunity to look for oddballs in {\it e.g.} $D\Dx$ decays. Decays of this kind are
investigated for the search for charmonium hybrids. The final state to look at is then $D\Dx\eta$
or $D\Dx\piz$. The lightest oddballs are $\JPC=\JPCtwopm$ and $\JPConepm$ (spectroscopic name $b_{0,2}(4000-5000)$). Since they would appear in the same open charm final states as charmonium hybrids,
the conclusions for these hybrid channels apply also to the oddball search.
%

%% file: phys/exotic/phys_exo_molecules.tex
%
\subsubsection{Multiquarks - Mesic Excitation of $\qqbar$ states}
\COM{Author(s): K. Peters}
\COM{Referee(s): F. Iazzi}
It is an widely accepted paradigm, that mesic excitations
are present in the wave functions of QCD bound states, like 
the pion cloud around the nucleons. The mesic excitation is
- if at all - expected to be loosely bound, thus resulting
in extremely large widths. In the vicinity of strong thresholds
this may be different and states with a potentially large
additional mesic component can become substantially more narrow
if they appear sub-threshold. This is for example seen in the $a_0(980)$
and $f_0(975)$ pair, which are believed to be the ground state $I=1$
and $I=0$ scalar mesons, but strongly attracted by the $\KKbar$
threshold and with large (may be dominating) $\KKbar$ component
in the wave function.
\begin{figure*}[htb]
\begin{center}
\includegraphics[width=0.45\dwidth]{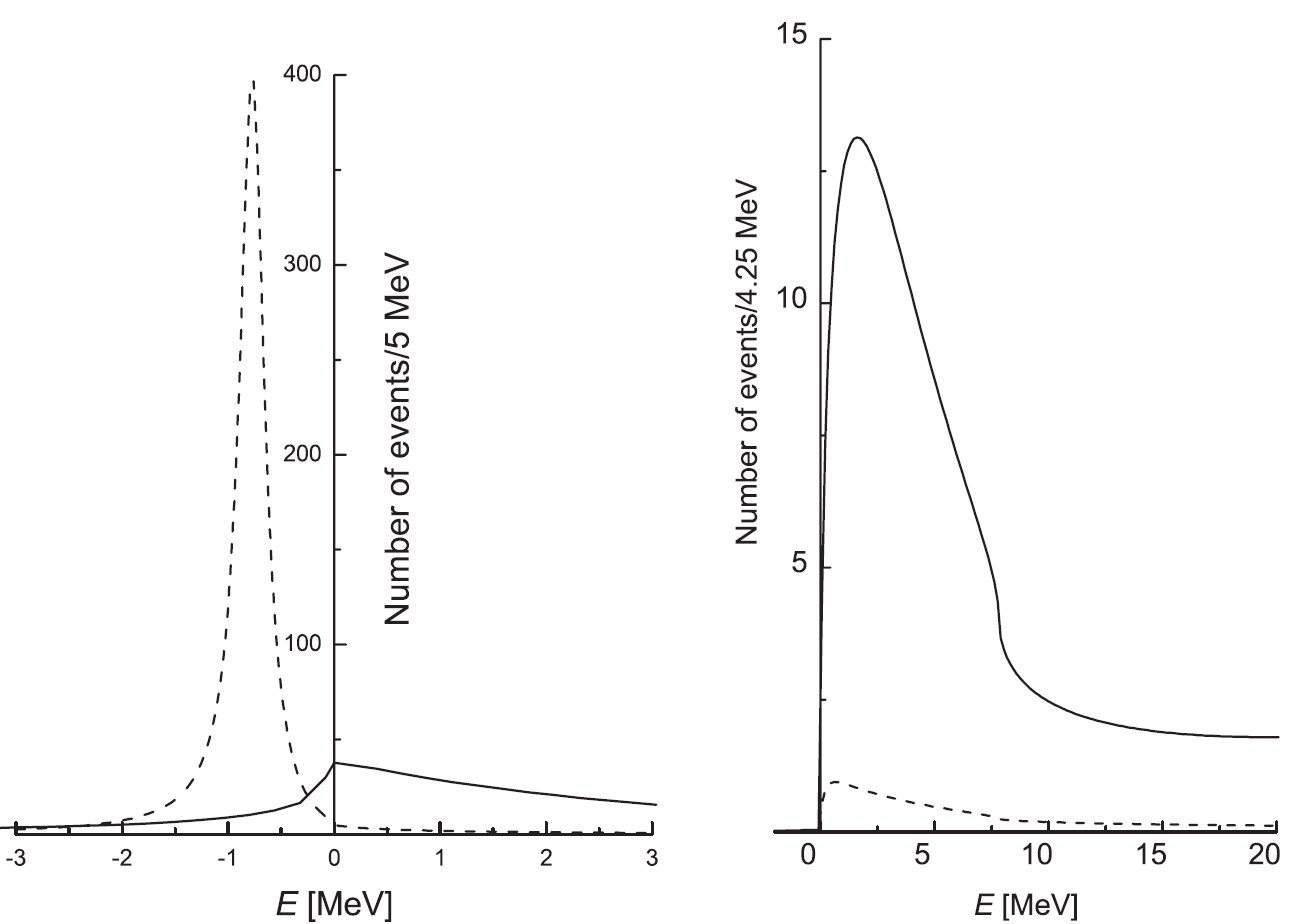}
\vspace{0.05\dwidth}
\includegraphics[width=0.45\dwidth]{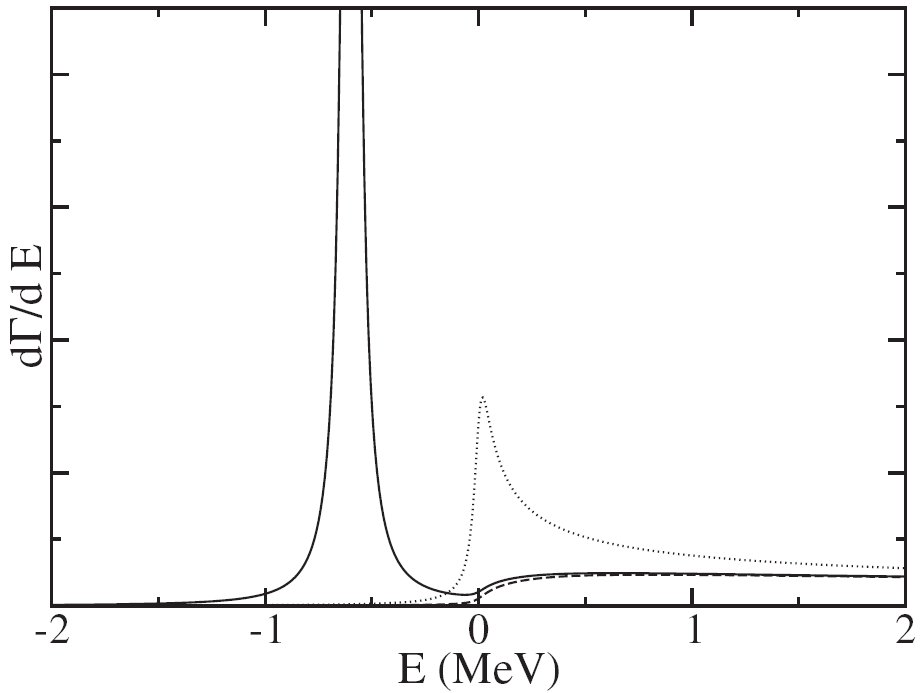}
\caption[Dispersive effects on the $X(3872)$]{Dispersive effects on the $X(3872)$ from various authors: (left) Hanhardt et
al.~\cite{Hanhart:2007yq} and 
(right) Braaten et al.~\cite{Braaten:2007ft}. The left figure
shows differential rates for the $\jpsi\pip\pim$ (first plot) and
\DzDzxbar (second plot) for large $\jpsi\pip\pim$ yield (solid curves) and \DzDzxbar dominance
(dashed curves). The right figure shows the line shapes near the \DzDzxbar threshold for
$X(3872)$ in the \DzDzxbar channel. The line shapes are shown for three different model settings
corresponding to a bound state (solid line), virtual state (dashed line), and
smooth excitation (dotted line).}
\label{fig:phys:exotic:x3872_dispersion}
\end{center}
\end{figure*}
\par
In the case of the extremely narrow $X(3872)$ the $D\Dx$ threshold 
has a dramatic impact on its wave function and dynamics. As discussed
in the previous section it was discovered in typical charmonium reactions,
but it does not really fit very well in any potential model. One solution
could be that it is dominated by a $D\Dx$ bound state. Various calculations
show, that depending on the kind of object, the different dispersive effects
would manifest in different lineshapes~\cite{Braaten:2007ft,Hanhart:2007yq} 
(see also~\Reffig{fig:phys:exotic:x3872_dispersion}).
\par

%% file: phys/exotic/phys_exo_hybrids_benchmark.tex
%
%
\paragraph*{Study of \boldmath $\ppbar\,\to\,\tilde{\eta}_{c1}\eta\,\to\, \chicone\piz\piz\eta$}\ \\
%
%
\label{sec:sim:res:hybrid}
For the production of $\tilde{\eta}_{c1}$ in $\pbarp\to\tilde{\eta}_{c1}\eta$ it is assumed
that the cross section is in the same order of magnitude as for the
process $\pbarp\to\psi(2S)\eta$ including conventional charmonium. The
cross section for this reaction is given in Ref.~\cite{lundborg} to be
$(33\pm8)$\,pb at $\sqrt{s}=5.38\,\gev$ and is calculated from the
crossed process $\psi(2S)\to\eta\pbarp$ observed in $\ee$
annihilation.
%
%
\par
The final state with 7 photons and an \ee lepton pair originating from
\jpsi decays has a distinctive signature and separation from light
hadron background should be feasible.  A source of background
are events with hidden charm, in particular events including a \jpsi
meson. This type of background has been studied by analysing
$\pbarp\,\to\,\chiczero\,\piz\,\piz\,\eta$,
$\pbarp\,\to\,\chicone\,\piz\,\eta\,\eta$,
$\pbarp\,\to\,\chicone\,\piz\,\piz\,\piz\,\eta$ and
$\pbarp\,\to\,\jpsi\,\piz\,\piz\,\piz\,\eta$. The hypothetical hybrid
state is absent in these reactions, but the $\chiczero$ and
$\chicone$ mesons decay via the same decay path as for the
signal. Therefore these events have a similar topology as signal
events and could potentially pollute the $\tilde{\eta}_{c1}$ signal. 
\begin{table}[htbp]
  \begin{center}
    \begin{tabular}{lcc}\hline \hline
      Reaction & $\sigma$ & $\BR$ \\
      $\pbarp\to$ & & \\ \hline
      $\tilde{\eta}_{c1}\eta$               & 33\,pb       & $0.82\percent\times \BR(\tilde{\eta}_{c1}\to\chicone\piz\piz)$ \\
      $\chiczero\piz\piz\eta$    &            & 0.03\percent \\
      $\chicone\piz\eta\eta$    &            & 0.32\percent \\
      $\chicone\piz\piz\piz\eta$ &           & 0.81\percent  \\
      $\jpsi\piz\piz\piz\eta$    &            & 2.26\percent \\ \hline\hline
    \end{tabular}
    \caption[Cross sections for the charmonium hybrid channel]{Cross sections for signal and background reactions.  The
      table lists also the product of branching fractions for the
      subsequent particle decays.  }
    \label{tab:hybridhidden:xsec}
  \end{center}
\end{table}
%
%
\par
The number of analysed signal and background events is summarised in
\Reftbl{tab:hybridhidden:samples}.

\begin{table}[htbp]
  \begin{center}
    \begin{tabular}{lc}\hline \hline
      Reaction & Events \\
      $\pbarp\to$ &  \\ \hline
      $\tilde{\eta}_{c1}\eta$                & $8\cdot10^{4}$ \\
      $\chiczero\piz\piz\eta$     & $8\cdot10^{4}$ \\
      $\chicone\piz\eta\eta$     & $8\cdot10^{4}$ \\
      $\chicone\piz\piz\piz\eta$  & $8\cdot10^{4}$ \\
      $\jpsi\piz\piz\piz\eta$     & $8\cdot10^{4}$ \\ \hline\hline
    \end{tabular}
    \caption[Signal and background events for the charmonium hybrid analysis]{ Number of analysed signal and background events. The
      $\jpsi$ is considered only in the \ee decay mode.}
    \label{tab:hybridhidden:samples}
  \end{center}
\end{table}
%
%
\par
Photon candidates are selected from the clusters found in the EMC with
the reconstruction algorithm explained in
\Refsec{sec:soft:recoalgo}. Two photon candidates are combined and
accepted as \piz and $\eta$ candidates if their invariant mass is
within the interval [115;150]\,\mevcc and [470;610]\,\mevcc, respectively.

From the \jpsi and photon candidates found in an event
$\chicone\,\to\,\jpsi\,\gamma$ candidates are formed, whose invariant
mass is within the range $[3.3$;$3.7]$\,\gevcc.  From these
$\chicone\,\piz\,\piz\,\eta$ candidates are created, where the same
photon candidate does not occur more than once in the final state.
The corresponding tracks and photon candidates of the final state are
kinematically fitted by constraining their momentum and energy sum to
the initial \pbarp system and the invariant lepton candidates mass to
the \jpsi mass. Accepted candidates must have a confidence level of
$\CL>0.1$\percent and the invariant mass of the $\jpsi\gamma$ subsystem
should be within the range $[3.49;3.53]\,\gevcc$, whereas the
invariant mass of the $\eta$ candidates must be within the interval
$[530;565]\,\mevcc$. A FWHM of $13\,\mevcc$ and $9\,\mevcc$ is observed
for the $\eta$ and $\chicone$ signal respectively after the kinematic fit.
\par
For the final event selection the same kinematic fit is repeated with
additionally constraining the invariant $\chicone$, \piz and $\eta$
mass to the corresponding nominal mass values. Candidates having a
confidence level less than 0.1\percent are rejected.
\par
At this stage of the analysis $8.2\%$ of the event are reconstructed,
whereas for a fraction of $5.3\%$ of the reconstructed events more
than one $\chi_{c1}\piz\piz\eta$ combination is found per event. To
ensure an unambiguous reconstruction of the total event, events with a
candidate multiplicity higher than one are rejected.

%
%
\par
\begin{figure}[htbp]
   \centering
   \includegraphics[width=0.45\textwidth,angle=90]{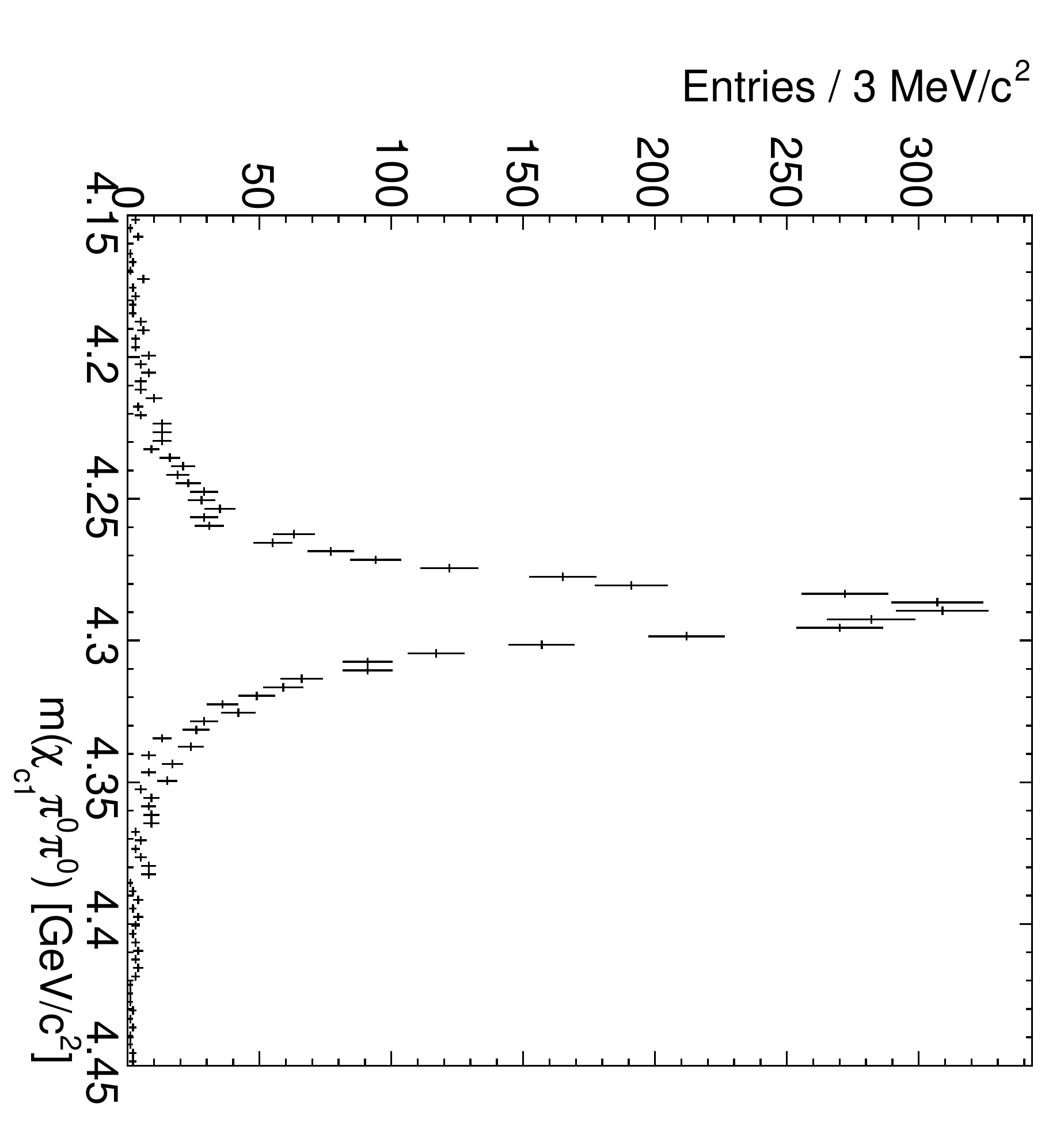}
   \caption[Invariant $\chicone\piz\piz$ mass]{ Invariant
     $\chicone\piz\piz$ mass obtained for the $\jpsi\to\ee$ channel
     after application of all selection criteria. }
\label{fig:hybrichi:hybr}
\end{figure}

The invariant $\chicone\piz\piz$ mass obtained after application of
all selection criteria is shown in \Reffig{fig:hybrichi:hybr}. The
$\tilde{\eta}_{c1}$ signal has a FWHM of $30\,\mevcc$. The reconstruction
efficiency is determined from the number of $\tilde{\eta}_{c1}$ signal entries in
the mass range $4.24-4.33\,\gevcc$ and is found to be 6.83\percent.
\par
The background suppression is estimated from the number of accepted
background events after application of all selection criteria having a
valid $\tilde{\eta}_{c1}$ candidate whose invariant mass is within the
same interval used to determine the reconstruction efficiency for
signal events. In \Reftbl{tab:hybridhidden:suppr} the suppression for
the individual background channels is listed together with the
expected signal to background ratio $S/B$, which is reported in terms
of
\begin{equation}
\mathcal{R}=\frac{\sigma_S\BR(\tilde{\eta}_{c1}\to\chicone\piz\piz)}{\sigma_B}
\label{eq:hybrichi:xsecratio}
\end{equation}
given by the unknown signal (background) cross section $\sigma_S$
($\sigma_B$) and the branching fraction for the
$\tilde{\eta}_{c1}\to\chicone\piz\piz$ decay. Depending on the background
channel and reconstructed $\jpsi$ decay mode $S/B$ is varying between
$250-10100\,\mathcal{R}$. For $\pbarp\to\chicone\piz\piz\piz\eta$ only
a lower limit $>5530\,\mathcal{R}$ is obtained. If the cross
sections $\sigma_B$ for the background processes are not enhanced by
more than an order of magnitude over
$\sigma_S\BR(\tilde{\eta}_{c1}\to\chicone\piz\piz)$ very low
contamination of the signal from these processes is expected.
\begin{table}[htbp]
  \begin{center}
    \begin{tabular}{lcccc}\hline \hline
      Reaction & $\eta$  & $S/B$  \\ 
      $\pbarp\to$ &  [$10^3$]          & [$10^3$]                         \\ \hline
      $\chi_{c0}\piz\piz\eta$     & $5.33$   & $10.1\,\mathcal{R}$   \\ 
      $\chi_{c1}\piz\eta\eta$     & $26.6$    & $4.57\,\mathcal{R}$   \\ 
      $\chi_{c1}\piz\piz\piz\eta$ & $>80$    & $>5.53\,\mathcal{R}$ \\ 
      $\jpsi\piz\piz\piz\eta$     & $9.98$    & $0.25\,\mathcal{R}$ \\ \hline\hline
    \end{tabular}
    \caption[$\tilde{\eta}_{c1}\eta$ signal to background ratio]{ Background
      suppression $\eta$ and the $\tilde{\eta}_{c1}$ signal to background ratio
      $S/B$ for the individual background reactions in terms of
      $\mathcal{R}$ as defined in \Refeq{eq:hybrichi:xsecratio}.}
    \label{tab:hybridhidden:suppr}
  \end{center}
\end{table}
\par
Background reactions including no charm but light mesons in the final
state have not been investigated yet. The studies of the background
types performed for the charmonium states presented in this document
proof that a clean reconstruction via the $\jpsi\to\ee$ decay mode is
possible. Therefore the reconstruction of the $\tilde{\eta}_{c1}$
state via this decay should yield also a sufficient background
suppression.  

The expected number of reconstructed events per day is given by
\begin{equation}
 N=\sigma_S\BR(\tilde{\eta}_{c1}\to\chicone\piz\piz)\times 4.81\,\mbox{nb}^{-1},
\end{equation}
assuming a design luminosity of $2\cdot
10^{32}\mathrm{cm}^{-2}\mathrm{s}^{-1}$ with an efficiency of
50\percent. As before said the cross section for
$\pbarp\to\psi(2S)\eta$ is expected to be $33$\,pb. Assuming the same
cross section for the production of the charmonium hybrid state this
becomes $N=0.16\,\BR(\tilde{\eta}_{c1}\to\chicone\piz\piz)$ events per
day.
\paragraph*{Study of \boldmath $\ppbar\,\to\,\tilde{\eta}_{c1}\eta\,\to\, D\Dx\eta$}\ \\
\label{hybridopencharm}
Two possible background reactions including open charm decays leading
to a similar event topology as signal events have been
investigated. The first is $\pbarp\to\Dz\Dxzbar\piz$, where the recoil
$\eta$ is absent and the D and $\Dx$ mesons decay via the same decay
path as for signal events. Secondly the reaction
$\pbarp\to\Dz\Dxzbar\eta$ is investigated, where the recoil $\eta$ is
present but either the $\Dz$ or the $\Dzbar$ meson (from
$\Dxzbar\to\Dzbar\piz$ decay) is decaying into $\Km\pip\piz\piz$ and
$\Kp\pim\piz\piz$. This $\Dz$ decay mode is listed in PDG as seen and
it is assumed for this study that it's branching fraction is $5\%$,
comparable in the order of magnitude to other $D$ meson decay modes
including a charged kaon.  The product of branching fractions for the
background and signal reactions are shown in
\Reftbl{tab:hybriddds:branching}.
\begin{table}[htbp]
  \begin{center}
    \begin{tabular}{lc}\hline \hline
      Reaction &  \\
      $\pbarp\to$ & $\mathcal{B}$\\ \hline
      $\tilde{\eta}_{c1}\eta$               & $0.47\%\times \mathcal{B}(\tilde{\eta}_{c1}\to\Dz\Dxzbar)$ \\
      $\Dz\Dxzbar\eta$           & $3.2\%\times \mathcal{B}(\Dz\to \Km\pip\piz\piz)=0.16\%^*$ \\
      $\Dz\Dxzbar\piz$           & $1.17\%$ \\ \hline\hline 
     \end{tabular}
    \caption[Decay branching ratios for open charm production]{The product of branching fractions for the subsequent
      particle decays for signal and background reactions. For the
      branching fraction marked with an asterisk ($^\ast$)
      $\mathcal{B}(\Dz\to \Km\pip\piz\piz)=5\%$ is assumed.}
    \label{tab:hybriddds:branching}
  \end{center}
\end{table}
%
\piz and $\eta$ are selected in the standard way discussed before.
Pions and kaons are selected from charged particles
in the event by applying a likelihood based selection algorithm, where
a likelihood value of $\mathcal{L}>0.2$ is required to accept a
candidate as a pion or kaon. All possible $\Kp\pim\piz$ combinations
having an invariant mass in the range $[1.7;2.2]\,\gevcc$ in an event
are formed and fitted by applying a \piz mass constraint and requiring
a common vertex for the tracks of the two charged candidates. A
confidence level $\CL>$0.1\percent is required to accept the candidates as
$\Dz\to \Kp\pim\piz$ candidates. These are then used to form
$\Dz\piz$ candidates having an invariant mass in the interval
$[1.95;2.05]\,\gevcc$, which are kinematically fitted applying a \piz
mass constraint. If the fit yields a confidence level $\CL>$0.1\percent the
$\Dxz\to \Dz\piz$ candidate is accepted for further
selection. Afterwards $\Dz\Dxzbar\eta$ combinations are formed and
fitted by constraining the final state particles' four-vectors to the
initial \pbarp system momentum and energy. Furthermore the invariant
$\gamma\gamma$ mass of the \piz candidates in the decay tree is
constrained to the \piz mass. A confidence level of $\CL>$0.1\percent is
required.
\par
A confidence level of $CL>0.1\%$ is
required. 
For final event selection the fit is repeated but with additional mass
constraints on the \Dz, \Dxz and $\eta$ candidates. Candidates leading
to a confidence level lower than $0.1\%$ are discarded.
\par
A $\chi_{c1}\piz\piz\eta$ candidate multiplicity higher than one is
observed in 9.5\% of the reconstructed events. In order to avoid
unambiguities in later analysis events with a higher multiplicity than
one are are rejected. To estimate the reconstruction efficiency only
the correct combinations are considered. The event yield is determined
as the number of $\Dz\Dxzbar$ signal entries falling in the mass
window $[4.24;4.33]$\,\gevcc.%
%
\par
\begin{figure}[htbp]
   \centering
  \includegraphics[width=0.45\textwidth,angle=90]{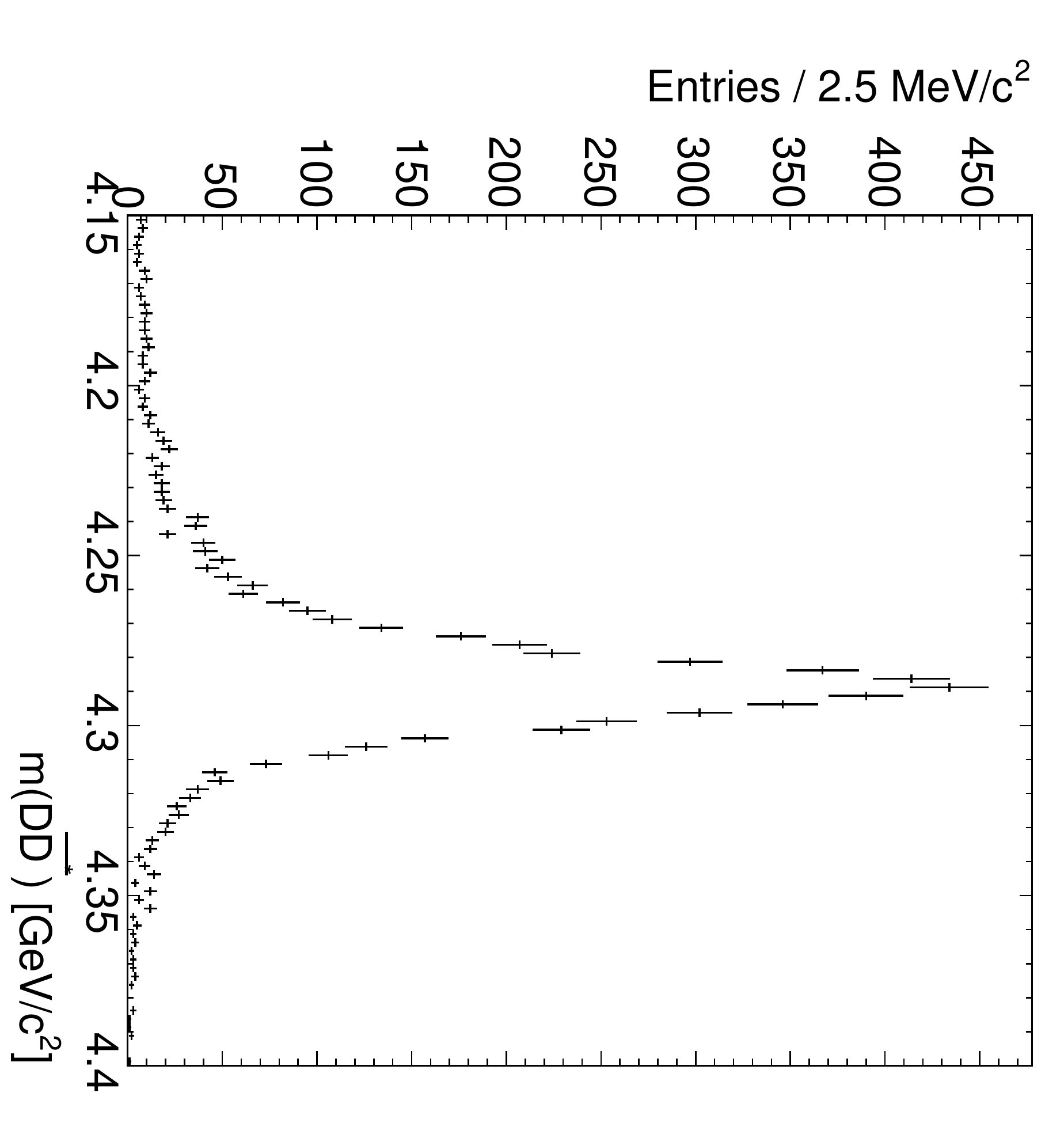}
\caption[Invariant $\Dz\Dxzbar$ mass]{ Invariant
  $\Dz\Dxzbar$ mass obtained after the kinematic fit with
  a momentum and energy constraint on the initial $\pbarp$ system as
  described in the text.}
\label{fig:hybriddds:ddstbar}
\end{figure}

The obtained $\Dz\Dxzbar$ invariant mass distribution is shown
\Reffig{fig:hybriddds:ddstbar}. A signal width (FWHM) of 22.5\,\mevcc is
observed. The reconstruction efficiency is $5.17\%$.
\par
The background reactions $\pbarp\to\Dz\Dxzbar\eta$ (with
$\Dz\to\Kp\pim\piz\piz$) and $\pbarp\to\Dz\Dxzbar\piz$ could be
suppressed by a factor $>1.6\cdot10^5$. Assuming equal cross sections
for these processes and signal events and a branching fraction for
$\Dz\to \Kp\pim\piz\piz$ (which is listed by the PDG as seen) in the
order of 5\percent the expected signal to noise ratio can be expressed by
\begin{eqnarray}
  \frac{S}{N}&>&\frac{\BR(\tilde{\eta}_{c1}\to\Dz\Dxzbar)\times 0.47\,\% \times 5.17\,\%}{ (0.16\,\% + 1.17\,\% )\times 5\cdot10^{-6} } \nonumber \\ 
  &=& \BR(\tilde{\eta}_{c1}\to\Dz\Dxzbar)\times 2.9\cdot10^{3}, 
\end{eqnarray}
where the term $\BR(\tilde{\eta}_{c1}\to\Dz\Dxzbar)$ is the unknown
branching fraction for the decay $\tilde{\eta}_{c1}\to\Dz\Dxzbar$.
\par
With the assumed cross section of $33$\,pb and design luminosity
$\mathcal{L}=2\cdot 10^{32}\mathrm{cm}^{-2}\mathrm{s}^{-1}$ with an
efficiency of 50\percent the expected number of reconstructed events
per day is given by
\begin{equation}
  N=\BR(\tilde{\eta}_{c1}\to\Dz\Dxzbar)\times 0.077.  
\end{equation}
In conclusion this study proofs that the reconstruction of an object
of a mass of $\approx 4\,\gevcc$ decaying to open charm produced in
\pbarp annihilation at 15\,\gevc with a recoiling $\eta$ meson leading
to a final state with high photon multiplicity is feasible. The low
branching fractions of $D$ mesons make the inclusion of other decay
modes necessary to be sensitive for lower $\tilde{\eta}_{c1}$ branching
fractions, but these decays should be detectable with similar
efficiency as the decay mode presented in this study.
\paragraph*{Study of \boldmath $Y(3940)\rightarrow\jpsi\omega$ System in $\ppbar$ Formation}\ \\
In the following an exclusive study of the formation of $Y(3940)$ in
\pbarp annihilation is presented. The \jpsi is reconstructed from its
decay to \ee, whereas the $\omega$ is reconstructed via the
$\pip\pim\piz$ decay mode.
%
%
\par
\label{jpsiomegabg}
As possible sources of background the reactions
\begin{itemize}
\item $\pbarp\to\psi(2S)\piz$ ($\psi(2S)\to\jpsi\pip\pim$)
\item $\pbarp\to\jpsi\rho^0\piz$ ($\rho^0\to\pip\pim$)
\item $\pbarp\to\jpsi\rho^+\pim$ ($\rho^+\to\piz\pip$)
\item $\pbarp\to\pip\pim\piz\rho^0$ ($\rho^0\to\pip\pim$)
\item $\pbarp\to\pip\pim\pim\rho^+$ ($\rho^+\to\piz\pip$)
\item $\pbarp\to\pip\pim\omega$ ($\omega\to\pip\pim\piz$) 
\end{itemize}
have been considered.

None of the cross sections for these reactions have been measured in
the energy range of the $Y(3940)$. The cross sections for the
reactions $\pbarp\to\pip\pim\pi\rho$ and $\pbarp\to\pip\pim\omega$
have been measured in the $\sqrt{s}$ energy range between 2.14 and
3.55\,\gev~\cite{omg:xsec:fivepi}.
\par
For the channels including charmonium states one has to distinguish
between the formation process $\pbarp\to Y(3940)$ followed by the
subsequent $Y(3940)$ decay on the one hand and direct production of a
charmonium state with an associated recoil meson on the other
hand. While the observed decay of the $Y(3940)$ into $\jpsi\omega$ is
considered to be isospin conserving, a decay into $\psi(2S)\piz$ is
isospin violating and thus expected to be suppressed. Here the
non-resonant reaction $\pbarp\to\psi(2S)\piz$ is considered as a
possible source of background. Ref.~\cite{lundborg} quotes the cross
section for this process to be less than 55\,pb. In this study a
value of 55\,pb is assumed for this process as a conservative
estimate. Background could also arise from hypothetical isospin
conserving $Y(3940)$ decays into $\jpsi\rho\pi$, with a possible
intermediate resonance decaying into $\rho\pi$.  Since the
time-scales of the two processes are distinct, less interference
between the final states is expected and the background is considered
to be incoherent to signal. For the branching fractions of the
$Y(3940)$ into $\jpsi\rho^0\piz$ and $\jpsi\rho^+\pim$ a ratio of
$1:2$ is assumed, a decay pattern which is expected for an isoscalar
state.

\Reftbl{tab:omg:xsec} summarises the assumed cross sections and the
branching fractions of the signal and background reactions under
study.

\begin{table}[htbp]
  \begin{center}
    \begin{tabular}{lcc}\hline \hline
      Reaction & $\sigma$ & $\BR$ \\ 
      $\pbarp\to$ & & \\ \hline
      $Y\to\jpsi\omega$    & $\sigma_S$    & $5.2\percent\times \BR(Y\to\jpsi\omega)$ \\
      $\pip\pim\piz\rho^0$ &$149\,\mu\mbox{b}^\ast$  & $100\percent$  \\
      $\pip\pim\pim\rho^+$ &$198\,\mu\mbox{b}^\ast$  & $100\percent$  \\ 
      $\pip\pim\omega$     & $23.9\,\mu\mbox{b}^\ast$& $100\percent$ \\
      $\psi(2S)\piz$       & 55\,pb         & 3.73\percent \\
      $Y\to\jpsi\rho\pi$   & $\sigma$    & $5.9\percent\times \BR(Y\to\jpsi\rho\pi)$ \\ \hline\hline
    \end{tabular}
    \caption[Cross sections for signal and background reactions]
    {Cross sections for signal and background reactions. The
      table lists also the branching ratios for the subsequent
      particle decays. The cross sections marked by an asterisk $(^\ast)$
      take already the branching ratios of subsequent particle decays
      into account and the corresponding branching ratios are listed
      therefore as 100\percent. The reaction $Y\rightarrow \jpsi\rho\pi$, which has
      its own relevance, is treated as background here.}
    \label{tab:omg:xsec}
  \end{center}
\end{table}
%
\par
The number of analysed signal and background events is summarised in
\Reftbl{tab:omg:samples}. Signal events have been generated with phase
space distribution for the reaction $Y(3940)\to\jpsi\omega$. For the
$\omega\to\pip\pim\piz$ a proper angular distribution is
considered.  The background reactions
$\pbarp\to\pip\pim\omega$ and $\pbarp\to\pip\pim\pi\rho$ require a
large amount of data. To simulate the demanded number of events within
a sufficient time a \jpsi mass filter technique 
is applied and the number of events analysed
for these reactions has to be corrected by the filter efficiency,
which is also listed in \Reftbl{tab:omg:samples}.
\begin{table}[htbp]
\begin{center}
 \begin{tabular}{lcc}\hline \hline
Reaction $\pbarp\to$   & Events       & Filter eff. \\ \hline
$\jpsi\omega$          & $2\cdot10^{4}$ & 100\%    \\
$\pip\pim\piz\rho^0$   & $8.49\cdot10^6$   & 0.77\%    \\ 
$\pip\pim\pim\rho^+$   & $8.49\cdot10^6$   & 0.81\%    \\ 
$\pip\pim\omega$       & $9.9\cdot10^6$   & 9.15\%    \\       
$\jpsi\pim\rho^+$      & $2.5\cdot10^5$   & 100\%  \\ 
$\jpsi\piz\rho^0$      & $2.5\cdot 10^5$  & 100\%   \\
$\psi(2S)\piz$         &$8\cdot 10^4$   & 100\%   \\ \hline\hline
\end{tabular}
\caption{Summary of analysed events and the \jpsi mass filter efficiency. 
For channels including charmonium states no filter is applied and the \jpsi is decaying to \ee only.}
\label{tab:omg:samples}
\end{center}
\end{table}
\par
%
The $\jpsi\omega$ system is reconstructed by combining the \jpsi
candidates found in an event with $\omega\to\pip\pim\piz$
candidates. To form the latter, two candidates of opposite charge, both
identified as pions having a likelihood value $\mathcal{L}>0.2$ are
combined together with $\piz\to\gamma\gamma$ candidates, composed from
two photon candidates having an invariant mass in the range
$[115;150]$\,\mevcc.

All combinations of an event are fitted by constraining the sum of the
four-momenta of the final state particles to the initial beam energy
and momentum and constraining the origin of the charged final state
particles to a common vertex.  Combinations where the fit yields a
confidence level less than $0.1\percent$ are not considered for further
analysis. 
The \jpsi and $\omega$ signal
has a FWHM of 9\,\mevcc and 16.5\,\mevcc, respectively.  An $\omega$ mass
window of $[750;810]\,\mevcc$ is applied to cleanly select $\jpsi\omega$
candidates. For the final event selection the accepted candidates are
fitted under the $\jpsi\omega$ hypothesis, where on top of the \pbarp
four-momentum constraint mass constraints are applied to the \jpsi and
\piz candidates. Only candidates where the fit yields a confidence
level $\LH>$0.1\percent are considered further.
\par
\begin{figure}[htbp]
   \centering
  \includegraphics[width=0.45\dwidth,angle=90]{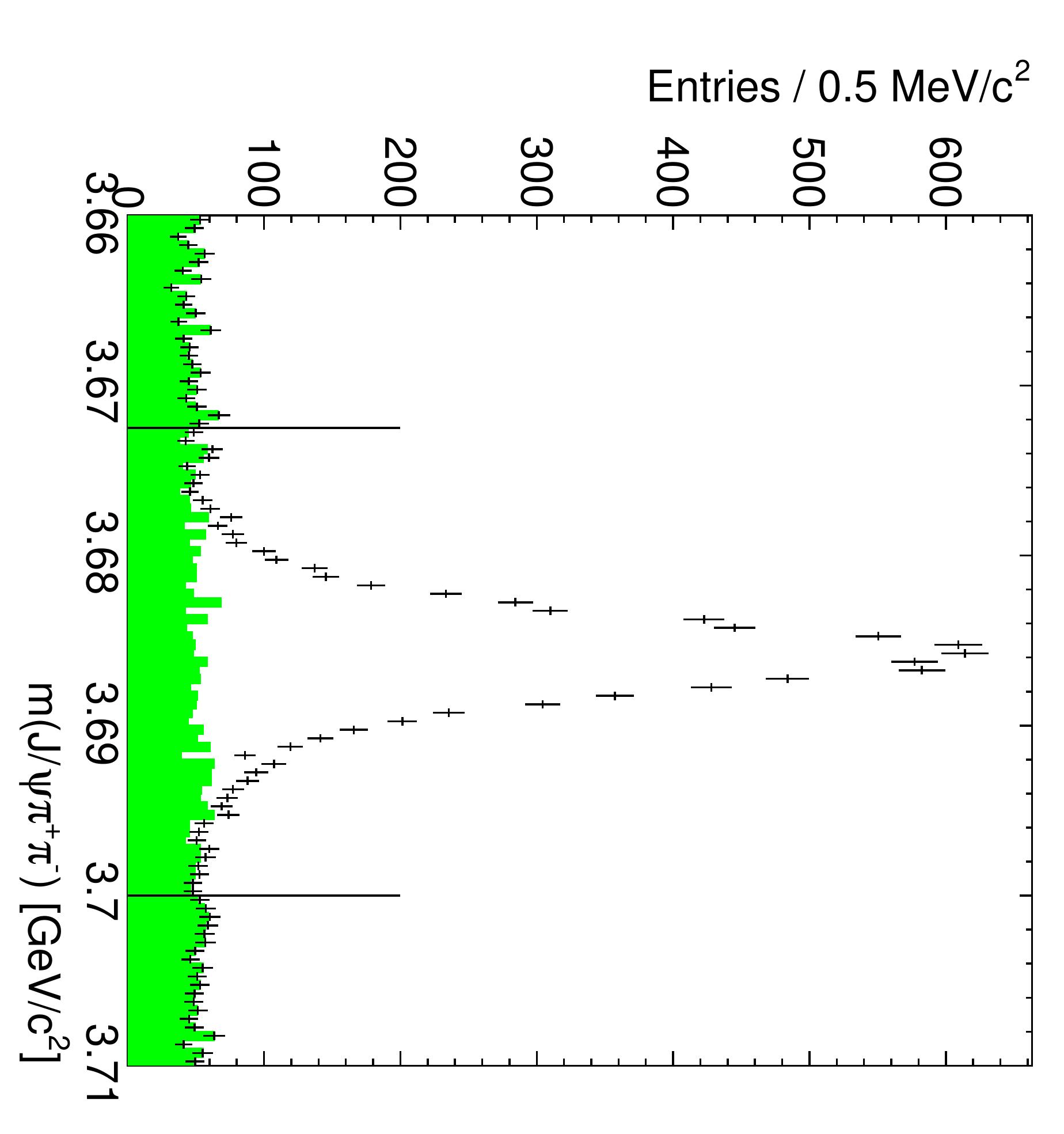}
\caption[Invariant $\jpsi\pip\pim$ ($\jpsi\to\ee$) mass]
{Invariant $\jpsi\pip\pim$ ($\jpsi\to\ee$) mass of the
  accepted $\jpsi\omega$ candidates for signal and $\psi(2S)\piz$
  background events. The background distribution is shown on top of
  the signal distribution (shaded) and is normalised to the signal
  cross section and branching fraction according to
  \Reftbl{tab:omg:xsec}. The vertical lines indicate the mass region
  used for the $\psi(2S)$ veto.}
\label{fig:omg:psi2s}
\end{figure}

At this stage of the analysis in 0.16\percent of the analysed signal events
more than one $\jpsi\omega$ candidate is found. The ambiguity is
solved by selecting the combination in the event which is leading to
the highest confidence level for the fit assuming the $\jpsi\omega$
hypothesis. In 65\percent of the cases the selected combination is
corresponding to the generated combination and thus is the correct
one. In total a negligible fraction of $6\cdot 10^{-4}$ out of the
signal events is reconstructed in the wrong combination.
%
\par
\begin{figure}[htbp]
   \centering
  \includegraphics[width=0.45\dwidth,angle=90]{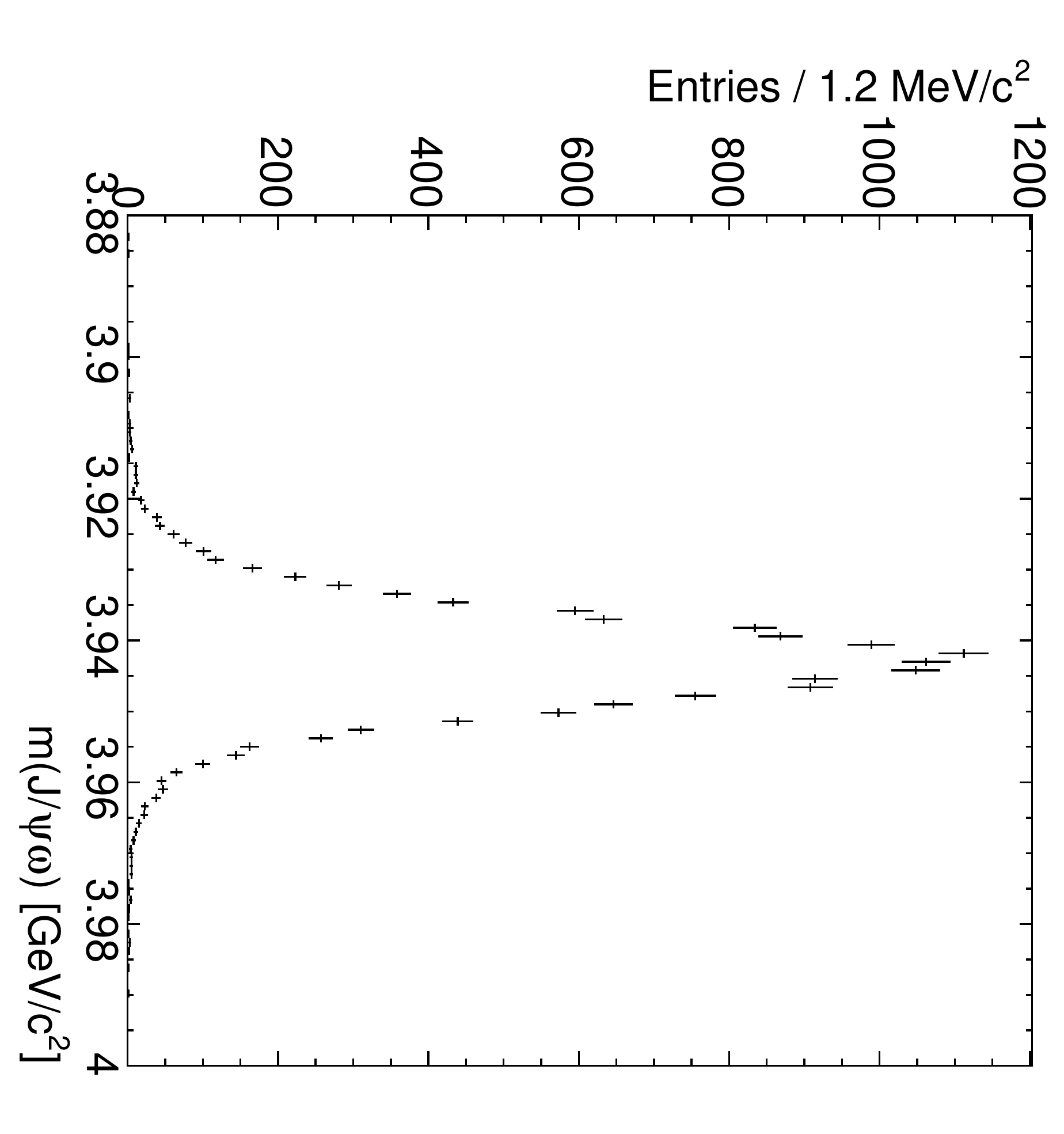}
\caption[ Invariant $\jpsi\omega$ mass distribution ]
{ Invariant $\jpsi\omega$ mass distribution obtained
  after the kinematic fit applying the constraints described in the
  text.}
\label{fig:omg:ysignal}
\end{figure}

The reconstruction efficiency is estimated separately for events
reconstructed via the two different \jpsi decay modes. Only the
correct combinations are considered. The efficiency is found to be
14.7\percent.
The invariant $\jpsi\omega$ mass distribution obtained from a
kinematic fit similar to the final fit applying the $\jpsi$ and $\piz$
mass constraints, but removing the constraint on the beam momentum and
energy is shown in \Reffig{fig:omg:ysignal}. The signal width (FWHM)
is found to be 14.4\,\mevcc.

In order to estimate the pollution of the signal from the considered
background reactions, background events are analysed likewise as
signal events and the number of reconstructed $\jpsi\omega$ candidates
is determined. The suppression $\eta$ for a certain background
reaction is defined as the fraction of generated and accepted events
and is given in \Reftbl{tab:omg:bg} for the individual background
channels. The expected signal to noise ratio is then given by
\begin{equation}
  \frac{S}{N}=\frac{\sigma_S}{\sigma_B}\;\frac{\BR_S}{\BR_B}\;\frac{\epsilon}{\eta^{-1}},
\end{equation}
where $\sigma_s$ ($\sigma_B$) is the cross section and $\BR_S$
($\BR_B$) is the product of branching fractions for signal
(background) reactions, and $\epsilon$ is the signal reconstruction
efficiency. Since the cross section $\sigma_S$ and the branching
fraction $\BR(Y\to\jpsi\omega)$ for the signal process $\pbarp\to
Y\to\jpsi\omega$ are not known the signal to noise ratio is reported
with respect to the product
$\tilde{\sigma}=\sigma_S\BR(Y\to\jpsi\omega)$.
\Reftbl{tab:omg:bg} summarises the expected ratio $S/B$ together
with the suppression for the various background reactions.

\begin{table}[htbp]
\begin{center}
 \begin{tabular}{lcccc}\hline \hline
Reaction                & $\eta$    & ${S/B}$ \\ \hline
$\pip\pim\piz\rho^0$    & $>1.1\cdot10^9$  & $>56.5\,\tilde{\sigma}/\nb$      \\ 
$\pip\pim\pim\rho^+$    & $>1.05\cdot10^9$  & $>40.6\,\tilde{\sigma}/\nb$       \\ 
$\pip\pim\omega$        & $>1.08\cdot10^8$    & $>34.6\,\tilde{\sigma}/\nb$      \\ 
$\psi(2S)\piz$          & $3.33\cdot 10^3$      & $24.8\,\tilde{\sigma}/\pb$    \\ 
$\jpsi\pim\rho^+$       & $25$                     & 4.90$\,\mathcal{BR}$  \\ 
$\jpsi\piz\rho^0$       & $22.1$                      & $7.65\,\mathcal{BR}$ \\ \hline \hline
\end{tabular}

\caption[Background suppression and $S/B$ for $Y(3940)\to\jpsi\omega$]{Observed background suppression $\eta$ and the expected
  signal to noise ratio $S/B$ for the investigated background
  reactions. The ratio $S/B$ is reported for
  $Y\to\jpsi\pi\rho$ with respect to the unknown ratio
  $\BR=\BR(Y\to\jpsi\omega)/\BR(Y\to\jpsi\rho\pi)$
  of the branching fractions for the reactions $\pbarp\to
  Y\to\jpsi\omega$ and $\pbarp\to Y\to\jpsi\rho\pi$. }
\label{tab:omg:bg}
\end{center}
\end{table}

A very good background suppression better than $1\cdot10^9$ and
$1\cdot 10^{8}$ is achieved for the channels
$\pbarp\to\pip\pim\pi\rho$ and $\pbarp\to\pip\pim\omega$,
respectively. Here the expected signal to noise ratio is better than
$34-56\tilde{\sigma}/$nb, depending on the background channel.

For the $\psi(2S)\piz$ background a $S/B$ of $24.8\tilde{\sigma}/$\,pb
is obtained. Thus the expected signal pollution is very low. For the two
$\jpsi\rho\pi$ channels a suppression by a factor of 20--28 is
observed and the expected $S/B$ for the sum of the two $\jpsi\rho\pi$
charge combinations is $5.58\,\BR$, where
$\BR=\BR(Y\to\jpsi\omega)/\BR(Y\to\jpsi\rho\pi)$.

\begin{figure}[htbp]
   \centering
   \includegraphics[width=\swidth]{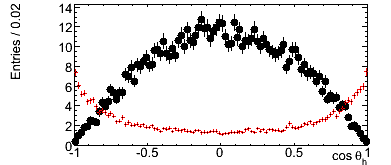}
   \includegraphics[width=\swidth]{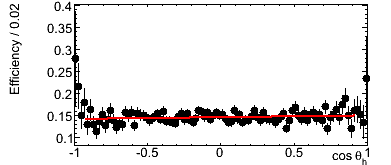}
  \caption[Distribution of the helicity angles in $\jpsi\omega$]
  { Distribution of the helicity angle $\theta_h$ (top) for
  signal (black) and $\pbarp\to\jpsi\rho\pi$ background (red)
  events. The distributions are normalised to the branching fraction
  $\BR(Y\to\jpsi\omega)$ and $\BR(Y\to\jpsi\rho\pi)$,
  respectively. Also shown is the reconstruction efficiency in
  dependence of $\theta_h$ (bottom). The red line is the result of a
  fit using a linear function.}
\label{fig:omg:helang}
\end{figure}

Both reactions can be disentangled performing a partial wave analysis,
which is out of the scope of this document. However, it should be
noted that the $\omega\to\pip\pim\piz$ decay has a distinct angular
distribution from the decay $\rho\pi$ with $\rho\to\pi\pi$. The
$\omega$ helicity angle $\theta_h$ is defined as the angle between the
\pip and \piz momentum computed in the $\pip\pim$ centre of mass
system. For signal events the $\cos\theta_h$ distribution is
$\sim\sin^2\theta_h$. The reconstructed $\theta_h$ distribution for
signal and background events is shown in \Reffig{fig:omg:helang}
together with the reconstruction efficiency in dependence of
$\cos\theta_h$. The efficiency distribution shows only 
structures at extreme forward and backward angles and is homogeneous otherwise
with respect to statistical
uncertainties. A linear fit to the efficiency distribution yields a
gradient consistent with zero within statistical errors. Thus the
angle $\theta_h$ can be cleanly reconstructed without significant
distortion due to an inhomogeneity of the efficiency. The $\cos\theta_h$
distribution reconstructed from signal events shows the expected
$\sim\sin^2\theta_h$ dependence, which is distinct from the
distribution obtained for background events. The result for background
events is derived assuming phase space distribution for the
$Y\to\jpsi\rho\pi$ decay. Depending on the production process of the
$\rho$ meson its helicity and thus the angular distribution for
$\rho\to\pi\pi$ can vary. The two scenarios where the $\rho$ decay
angle is $\sim \sin^2$ and $\sim \cos^2$ have been tested by weighting
the generated events accordingly. The observed $\theta_h$
distributions are independent of the assumed $\rho\to\pi\pi$ angular
distribution. In conclusion it is expected that the $\jpsi\omega$ and
$\jpsi\rho\pi$ decay modes could be disentangled performing a partial
wave analysis of the $\jpsi\pip\pip\piz$ final state.

The number of reconstructed
events per day running the accelerator at the $Y(3940)$ peak position
and design luminosity of $\LH=2\cdot 10^{32}$\,cm$^{-2}$s$^{-1}$
assuming an efficiency of $50\percent$ is given by
\begin{equation}
   N=\epsilon_S\tilde\sigma \int\LH\mathrm{d}t = 66\tilde{\sigma}\,\mbox{nb}^{-1}.
\end{equation}
\paragraph*{Study of \boldmath $Y(4320)\rightarrow\psi(2S)\pi^+\pi^-$ System in Formation}\ \\
In this section the decay $\ppbar \to \psi(2S)\pi^+\pi^-, \psi(2S)\to \jpsi\pi^+\pi^-, \jpsi \to \ee$ at a beam momentum of $p_{\bar{p}} = 8.9578\,\gevc$ (corresponding to the $Y(4320)$ resonance) is examined.
\par
%
The $\jpsi \to \ee$ candidates are reconstructed as described above. Both daughter candidates from a $\jpsi$ decay must be identified as electrons with a likelihood $\LH>$85\percent.  The likelihood value for the particle candidates identified as pions must be $\LH > 0.2$. \\ 
All combinations found in an event are kinematically fitted by constraining the sum of the four-vector of the final state particles to the initial beam energy and momentum and constraining the origin of the charged final state particles to originate from a common vertex. Also a mass constraint is applied on the $\jpsi$ and $\psi(2S)$ candidates (mass and beam constraint - MBC). The fit is repeated with mass constraints only (M) and with beam/energy constraints only (only beam constraint - BC).\\
For accepted candidates the $\psi(2S)$ (BC) is required to be within the interval $[3.67;3.71]~\gevcc$ and the $\jpsi$ mass (BC) to be within $[3.07;3.12]~\gevcc$. The confidence level of the fit (MBC) must be larger than $0.1 \%$. If more than one candidate in an event passes these selection criteria, the candidate with the highest confidence level is chosen and the others are rejected. Reconstructed $\psi(2S)\pi\pi$ candidates from the signal mode which are accepted by the criteria summarised above are checked if they are the correct combination.\\
The background channels are reconstructed in the same way and pass the same selection criteria as the signal modes, without an check for the right combination.

The reconstruction efficiency and the signal to noise ratio are examined only for the decay of $\jpsi \to \ee$.\\
The $\psi(2S)\pi\pi$ signal region is defined as the interval from $[4.29;4.35]\,\gevcc$. The reconstruction efficiency is $14.9\%$. The reconstructed signal has a mean of $4320\,\mevcc$ and a width (FWHM) of $13\,\mevcc$. 
\\
The cross section for the background mode $\pbarp \to 3\pip3\pim$ is $140\,\mu\mbox{b}$ (measured at a beam momentum of $8.8\,\gevc$). The suppression is $\eta_{\ee} = 37\cdot 10^{6}$. The signal to noise ratios is
  $S/B_{ee} = 700\tilde{\sigma}/\mu\mbox{b}$, where $\tilde{\sigma} = \sigma_{S}(\pbarp \to Y(4320))\mathcal{B}(Y(4320)\to \psi(2S)\pip\pim)$ is given in terms of the unknown signal cross section and branching fraction.  

\begin{table}[htbp]
\begin{center}
 \begin{tabular}{ccccc}\hline \hline
   Reaction & Beam mom. & events & Filter eff. \\
   $\pbarp \to$ & [\gevc] &  & $[\%]$ \\ \hline
   $\psi(2S)\pip\pim$ & $8.9578$ & $20000$ & $100$ \\
   $3\pip3\pim$ & $8.9578$ & $10^{6}$ & $0.67$ \\ \hline \hline
 \end{tabular}
\caption{Signal and background modes for the $\psi(2S)\pip\pim$ analysis. For the generator level filter, see \Refsec{sec:soft:genfilter}.}
\label{tab:psi2s}
\end{center}
\end{table}

\begin{table*}[htbp]
\begin{center}
 \begin{tabular}{cccc}\hline\hline
   Reaction & $\sigma$ & $\mathcal{B}$ \\
   $\pbarp \to$ & & \\ \hline
   $\psi(2S)\pip\pim$ & $\sigma_{s}(\pbarp\to Y(4320))$ & $1.9\% \times \mathcal{B}(Y(4320) \to \psi(2S)\pip\pim)$ \\
   $3\pip3\pim$ & $140\,\mu\mbox{b}$ & $100\%$ \\ \hline \hline
 \end{tabular}
\caption[Cross sections and branching fractions for the $\psi(2S)\pip\pim$ signal and background modes]
{Cross sections and branching fractions for the $\psi(2S)\pip\pim$ signal and background modes. For the subsequent $\jpsi$ decay we use to the sole branching fraction to one lepton type. The cross section for the background mode has been measured at a beam momentum of $8.8\,\gevc$.}
\label{tab:psi2scross}
\end{center}
\end{table*}

\begin{table}[htbp]
\begin{center}
 \begin{tabular}{cccccccc}\hline\hline
   Decay & {Eff.} &{Suppr.} & {Sig. to noise} \\
   $\pbarp \to$ & $\epsilon_S$  & $\eta$  & $S/B$ \\ \hline 
   $\psi(2S)\pip\pim$ & $14.9\%$  & $-$  & $-$  \\
   $3\pip3\pim$ & $-$  & $37\cdot10^{6}$  & $700\tilde{\sigma}/\mu\mbox{b}$  \\ \hline \hline
   \end{tabular}
\caption[Suppression $\eta$ and signal to noise ratio for the background modes of the  $\psi(2S)\pip\pim$ analysis]{Suppression $\eta$ and signal to noise ratio for the background modes of the  $\psi(2S)\pip\pim$ analysis. The signal to noise ratios are given in terms of the unknown cross section $\tilde{\sigma}$.}
\label{tab:psi2seff}
\end{center}
\end{table}

\begin{figure}[htbp]
   \centering
  \includegraphics[width=\swidth,angle=90]{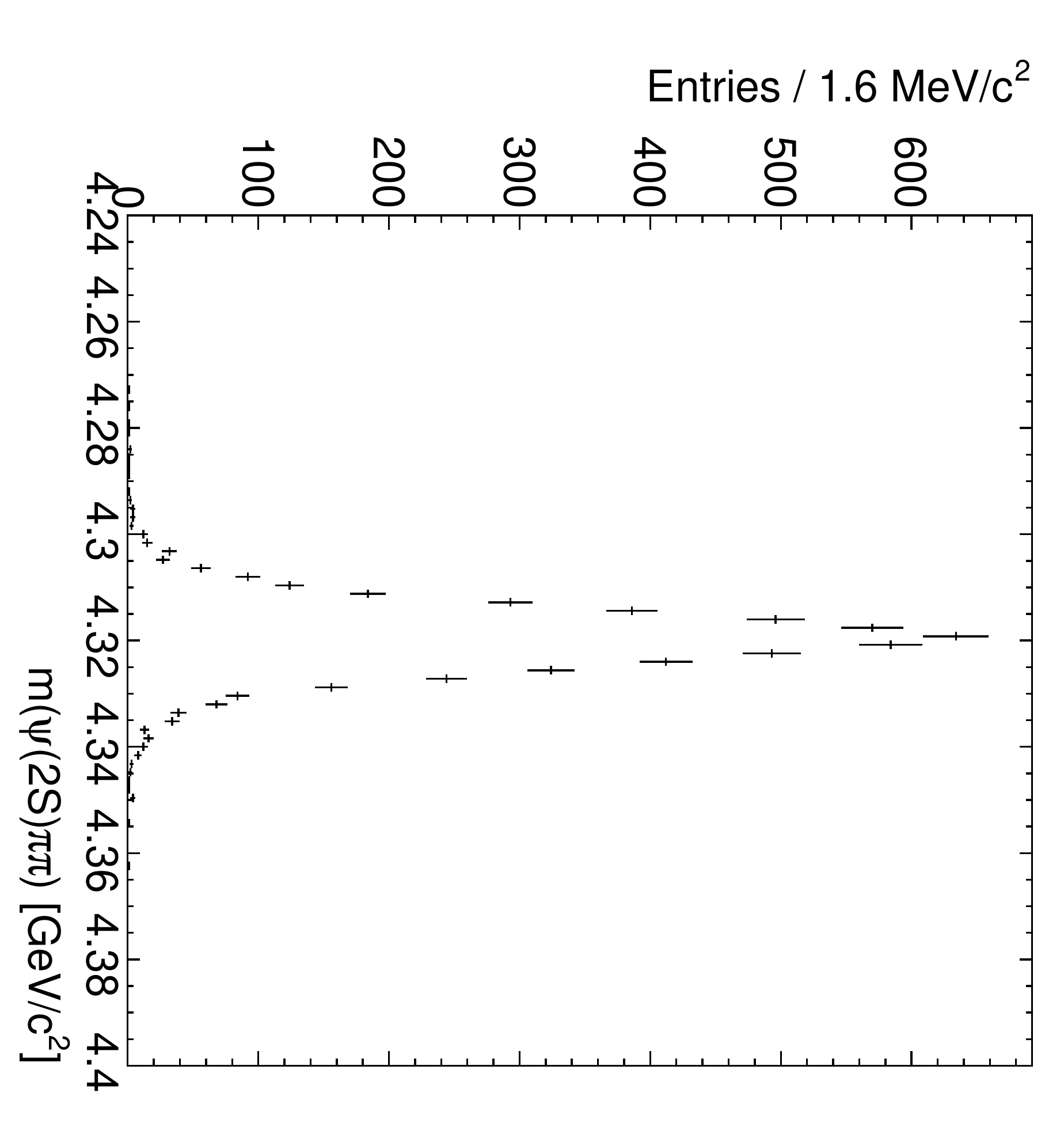}
\caption{Invariant mass distribution for $\psi(2S)\pi\pi$ candidates.}
\label{fig:psi2Spipi}
\end{figure}
%

%% file: phys/exotic/phys_exo_glueballs_benchmark.tex
%
\paragraph*{Study of the Formation Process \boldmath $\ppbar\,\to\, f_2(2000-2500)\,\to\, \phi\phi$}
The primary goal of this study is to proof the feasibility of the reconstruction of exotic states decaying into the favoured decay channel $\phi\phi$ as a feasibility check for the long standing quest about the $\xi(2230)$. Since no explicit assumptions about the exotic particles properties like mass or angular momentum have been made the simulation considered here comprises non resonant reactions of the type $\pbarp\rightarrow \phi \phi$ without an intermediate resonance as a minimum bias approach.
The detection of a possible resonant structure requires an energy scan around the region of interest in order to measure the dynamic behaviour of the cross section $\sigma(\sqrt{s})$. 


Thus this investigation consists of the following two parts:
\begin{enumerate}
\item Reconstruction of the signal
	\bi
	\item determination of efficiency of signal
	\item estimate background level (signal to noise ratio $S/B$)
	\ei
\item Simulation of an energy scan
	\bi
	\item estimate the expected energy dependent cross section with the efficiency measurement from above
	\item estimate the required beam time to detect the signal with a significance of 10\,$\sigma$ for different assumptions for the signal cross section
	\ei
\end{enumerate}
\par
Goal of the reconstruction is to determine the number of reactions of the type
\bn
\pbarp\rightarrow \phi\phi\rightarrow \Kp \Km \Kp \Km
\en
together with the reconstruction efficiency.
 
Under the assumption that the efficiency will not change very much for different energies in the region around 2.5\,GeV signal as well as background events have been generated at $E_{cms}=2.23$\,\gev corresponding to an initial 4-vector
\bn
\label{func:phys:exo:glueball:phiphi:pinit}
P_{\mbox{\scriptsize init}} &=& (p_x , p_y , p_z , E\cdot c) \nonumber \\
 &=& (0, 0, 1432, 2650)\,\mevc,
\en
additionally modified according to an relative jitter $p/dp = 10^{-5}$ accounting for the expected beam uncertainty.  
This is region where is has been found evidence for the tensor ($J^{PC} = 2^{++}$) glueball candidate $\xi(2230)$ by the \INST{BES} experiment~\cite{bib:phy:bai:1996wm}.

Signal events have been generated with the event generator {\tt EvtGen}~\cite{bib:phy:evtgen}. In order to determine the reconstruction efficiency with angular independent accuracy the events have been generated according to phase space resulting in flat angular distributions. 
 
The simulated decay chain was
\bn
\pbarp &\rightarrow& \phi\phi\\
\phi &\rightarrow& \Kp \Km
\en
The particular decay chain leads to a branching ratio related reduction factor of
\begin{eqnarray}
f_{\BR} & = & \BR(\phi\rightarrow\Kp\Km)^2 \cdot \underbrace{\BR( f_2  \rightarrow \phi\phi)}_{\mbox{\scriptsize unknown!}} \nonumber \\
& = & (0.492)^2 \cdot x < 0.242
\end{eqnarray}
As a conservative estimate we will assume the branching ratio to be $\BR(\mbox{Glueball } \rightarrow \phi\phi)= 0.2$ leading to a hypothetical factor $f_{\BR} = 0.05$.


\Reftbl{tab:phys:exo:glueball:phiphi:datasets} summarises the datasets used for these studies.

\begin{table}[!htb]
\centering
\begin{tabular}{lr}
\hline\hline
Channel & Number of events\\
\hline
 $\pbarp\rightarrow \phi\phi$ & 50\,k \\
 DPM generic & 10\,M\\ 
\hline
\end{tabular}
\caption{Datasets for the $\phi\phi$ feasibility study.}
\label{tab:phys:exo:glueball:phiphi:datasets}
\end{table}
\par
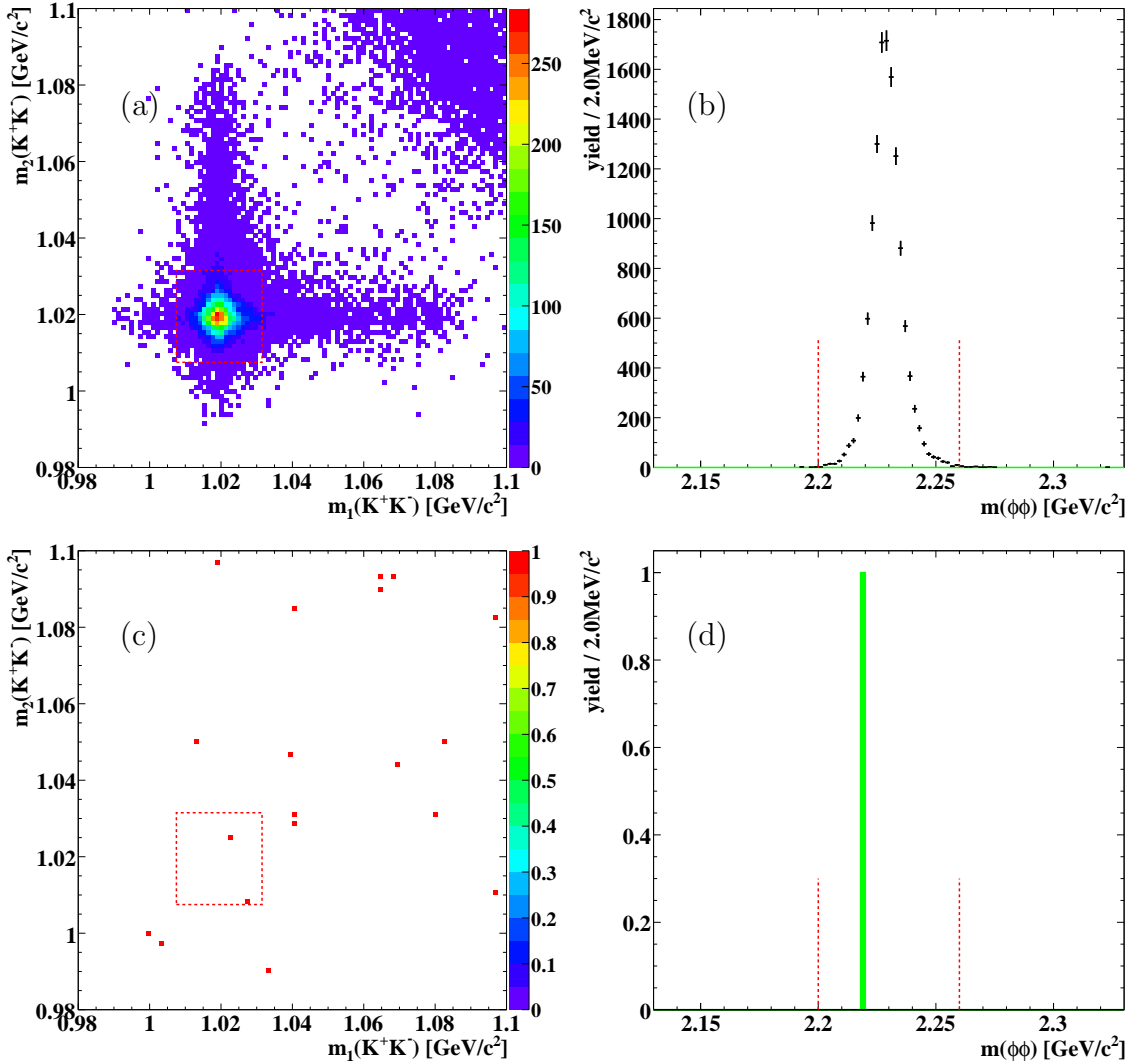
\begin{figure*}[hbtp]
\begin{center} 
\includegraphics[angle=90,width=\swidth]{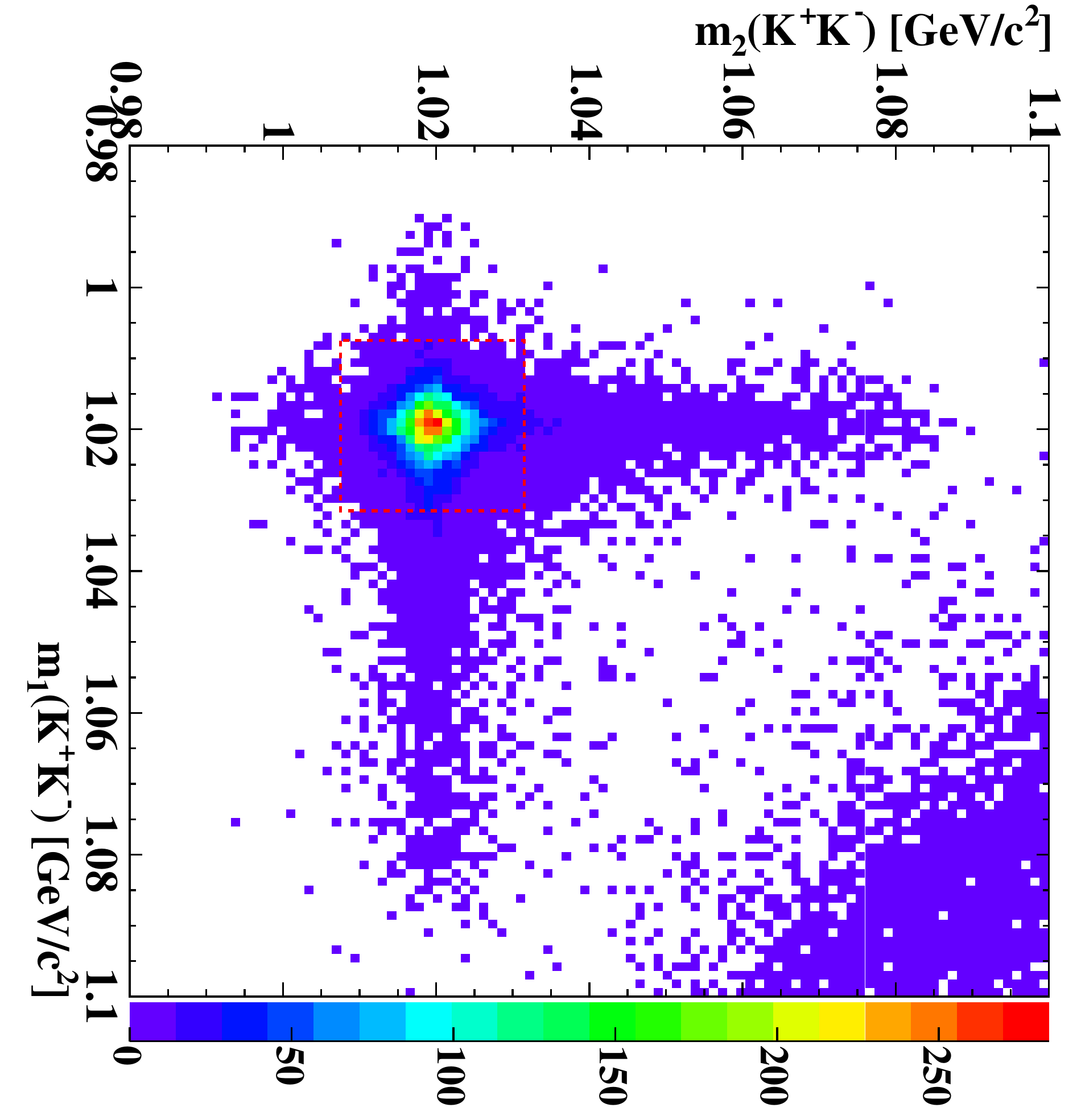}
\includegraphics[angle=90,width=\swidth]{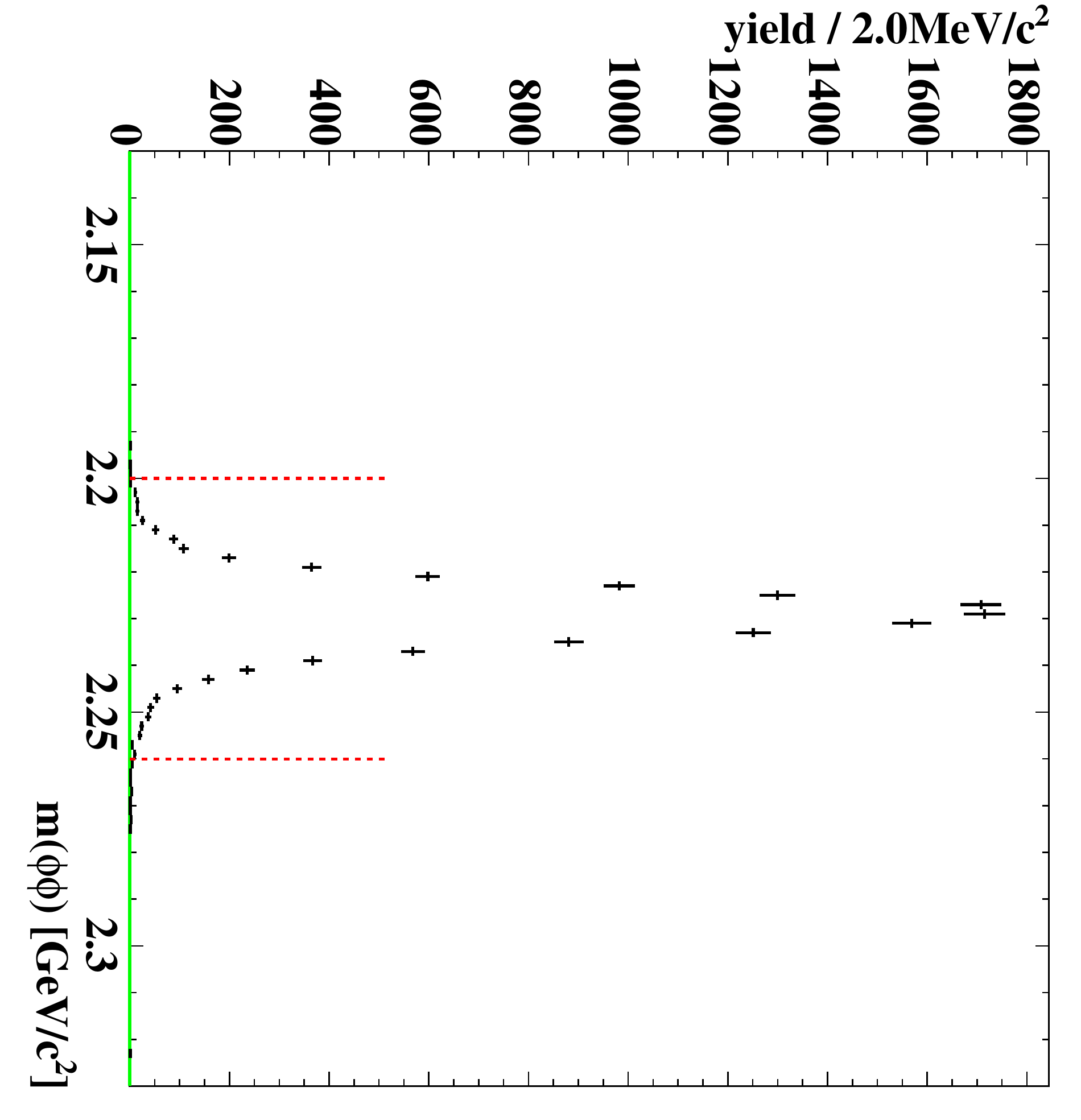}
\includegraphics[angle=90,width=\swidth]{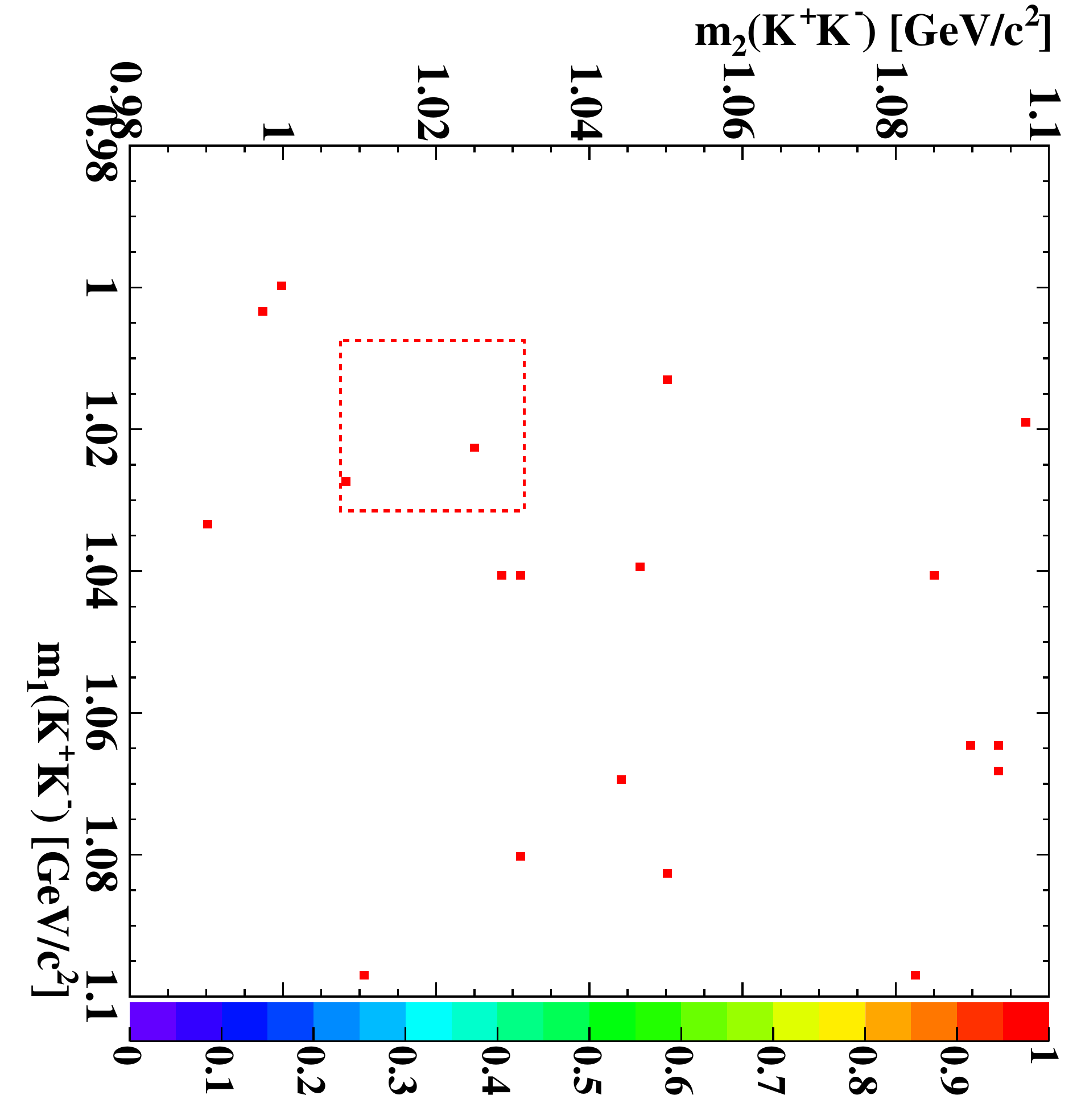}
\includegraphics[angle=90,width=\swidth]{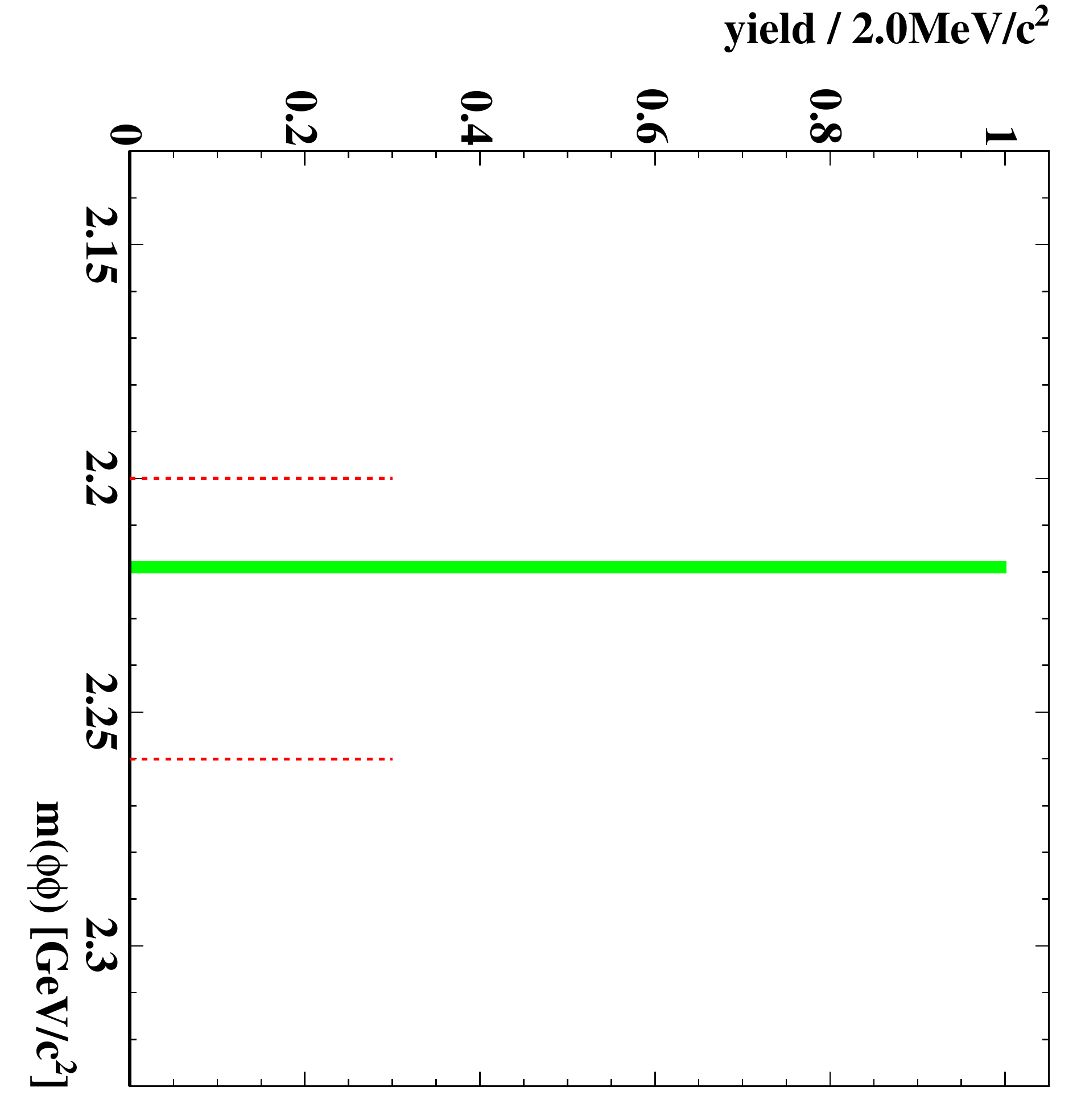}
\caption[Reconstructed events for $\pbarp\rightarrow\phi\phi$]{Distributions for reconstructed events for $\pbarp\rightarrow\phi\phi$. (a) 2D plot of invariant masses $m(\phi_1)$ vs. $m(\phi_2)$ for signal events, (b) invariant mass $m(\phi\phi)$ with MC truth match for signal events.  (c), (d) The same distributions for reconstructed background events generated with the DPM generator. Black histogram corresponds to all reconstructed combinations, the shaded area represents combinations failing the MCT match.}
\label{fig:phys:exo:glueball:phiphi:sigmass}
\end{center}  
 \begin{picture}(0,0)
 \large
  \put(58,438){(a)}
 \put(270,438){(b)}
 \put(58,238){(c)}
 \put(270,238){(d)}
 \end{picture}
\end{figure*}

The procedure for the reconstruction was:
\begin{enumerate}
\item Select kaon candidates from charged tracks with \vloo PID criterion\footnote{See section \ref{sec:soft:recochargedPID} for details.} 
\item Create a list of $\phi$ candidates by forming all combinations of a negative with a positive charged kaon candidate
\item Kinematic fit of the single $\phi$ candidates with vertex constraint
\item Create \pbarp candidates by forming any valid combination of two $\phi$-candidates
\item Kinematic 4-constraint fit of the \pbarp candidates with additional vertex constraint 
\end{enumerate}
Every of the so formed candidates had to fulfil the following requirements:
\begin{enumerate}
\item Probability of $\phi$ vertex fit: $P_\phi > 0.001$
\item Probability of \pbarp kinematic fit: $P_{\scriptsize \pbarp} > 0.001$
\item $\phi$ mass window: $|m(\Kp\Km)-m_{\mbox{\scriptsize PDG}}(\phi)| < 10$\,\mevcc
\item $\phi\phi$ mass window: $|m(\phi\phi)-2.23\mbox{GeV}/c^2|< 30$\,\mevcc
\end{enumerate}
The latter criterion defines the signal region being necessary to determine the efficiency.
\Reffig{fig:phys:exo:glueball:phiphi:sigmass} (a), (b) show the corresponding distributions according to the upper selection criteria for signal Monte Carlo data with kaon selection \vloo. The dashed lines in (b) as well as the box in (a) correspond to the selected mass windows. In plot (b) a superposition is shown of all reconstructed candidates.

In \Reffig{fig:phys:exo:glueball:phiphi:sigmass} (c), (d) the same plots are shown for generic background Monte Carlo data. No $\phi$-signal is seen in the invariant masses $m(\Kp \Km)$ in plot (c) and the signal window in (d) has only one entry, which disappears for tigher PID selection criteria. Therefore the conclusion concerning background level from generic hadronic reactions is limited for the time being. Nevertheless for cases without a single background event a limit is calculated with the assumption of one candidate in the signal region. 

To find an optimum for the PID criterion the selection has been repeated for all available criteria \vloo, \loo, \tig and \vtig.
The results are summarised in \Reftbl{tab:phys:exo:glueball:phiphi:pidsummary}. Since for the PID requirements \vloo and \loo only one event and for \tig and \vtig no background event was observed in the signal region the optimum would be the most loose selection, since all calculations are based on one single background event in the region of interest. The expected signal-to-noise ratios extracted from this study vary between 1:6 and better than 1:9. These values are considered to have large uncertainties due to the direct influence of the relative cross section of background with respect to the signal cross section $\sigma_S$, which has chosen to be $5\cdot 10^6$.

It should be noted that only the order of magnitude of the efficiency, {\it i.e.} $\epsilon\approx$ 15-25\percent, is important, since it only serves as input for the simulation of the energy scan.

\begin{table*}[htb]
\centering
\begin{tabular}{lccccc}
\hline\hline
Channel & rel. X-sec & $\epsilon(\mbox{VL})[\%]$ & $\epsilon(\mbox{L})[\%]$ & $\epsilon(\mbox{T})[\%]$ & $\epsilon(\mbox{VT})[\%]$\\\hline
Signal & 1 & 25.0 & 23.4 & 19.6 & 15.7 \\
 DPM generic & $5\cdot 10^{6}$ & $6.7\cdot 10^{-5}$ & $6.7\cdot 10^{-5}$ & $<6.7\cdot 10^{-5}$ & $< 6.7\cdot 10^{-5}$\\ \hline
 $r=S:N$ & -- & $ 1:6$ & $ 1:6$ & $> 1:7$ & $> 1:9$ \\\hline\hline
\end{tabular}
\caption[PID optimisation summary.]{PID optimisation summary. $\epsilon$ is the efficiency, VL, L, T, VT refer to the PID selection criteria described in the text.}
\label{tab:phys:exo:glueball:phiphi:pidsummary}
\end{table*}

\par
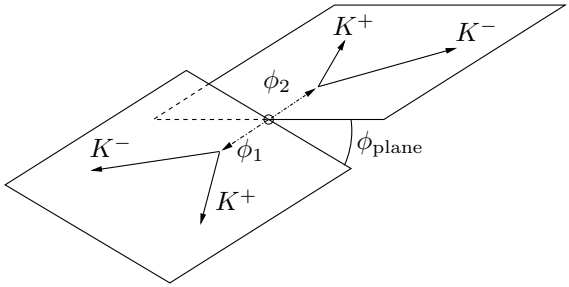
\begin{figure}[hbtp]
\begin{center} 
\includegraphics[width=\swidth]{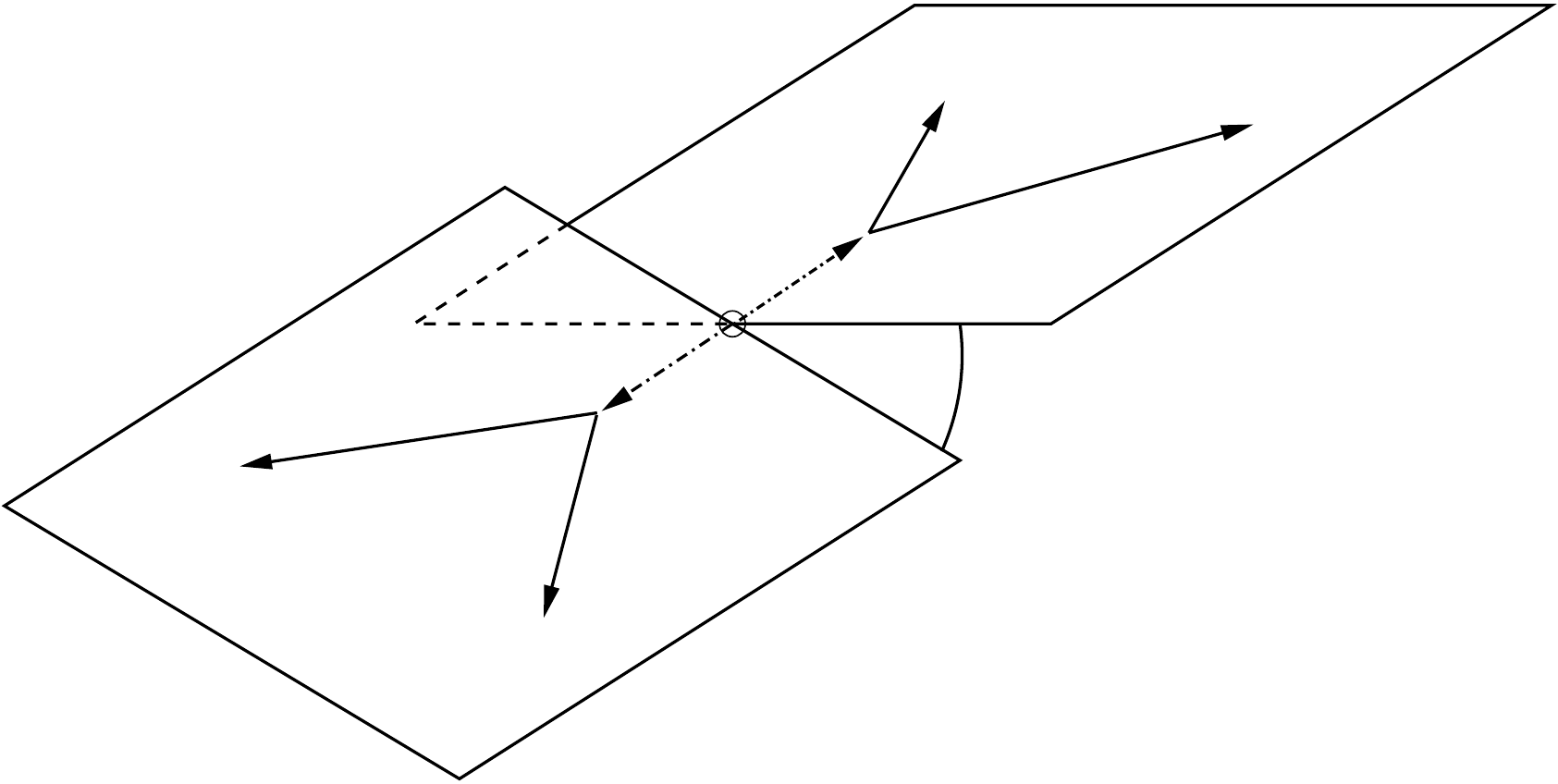}
\caption[Angle between $\phi$ decay planes.]{Angle $\phi_{\mbox{\scriptsize plane}}$ between the decay planes of the two $\phi$'s.}
\label{fig:phys:exo:glueball:phiphi:plane}
\end{center}  
\begin{picture}(0,0)
\put(93,102){$\phi_1$}
\put(103,128){$\phi_2$}
\put(85,82){$\Kp$}
\put(38,102){$\Km$}
\put(129,148){$\Kp$}
\put(175,146){$\Km$}
\put(138,105){$\phi_{\mbox{\scriptsize plane}}$}
\end{picture}
\end{figure}
\begin{figure}[hbtp]
\begin{center} 
\includegraphics[width=\swidth]{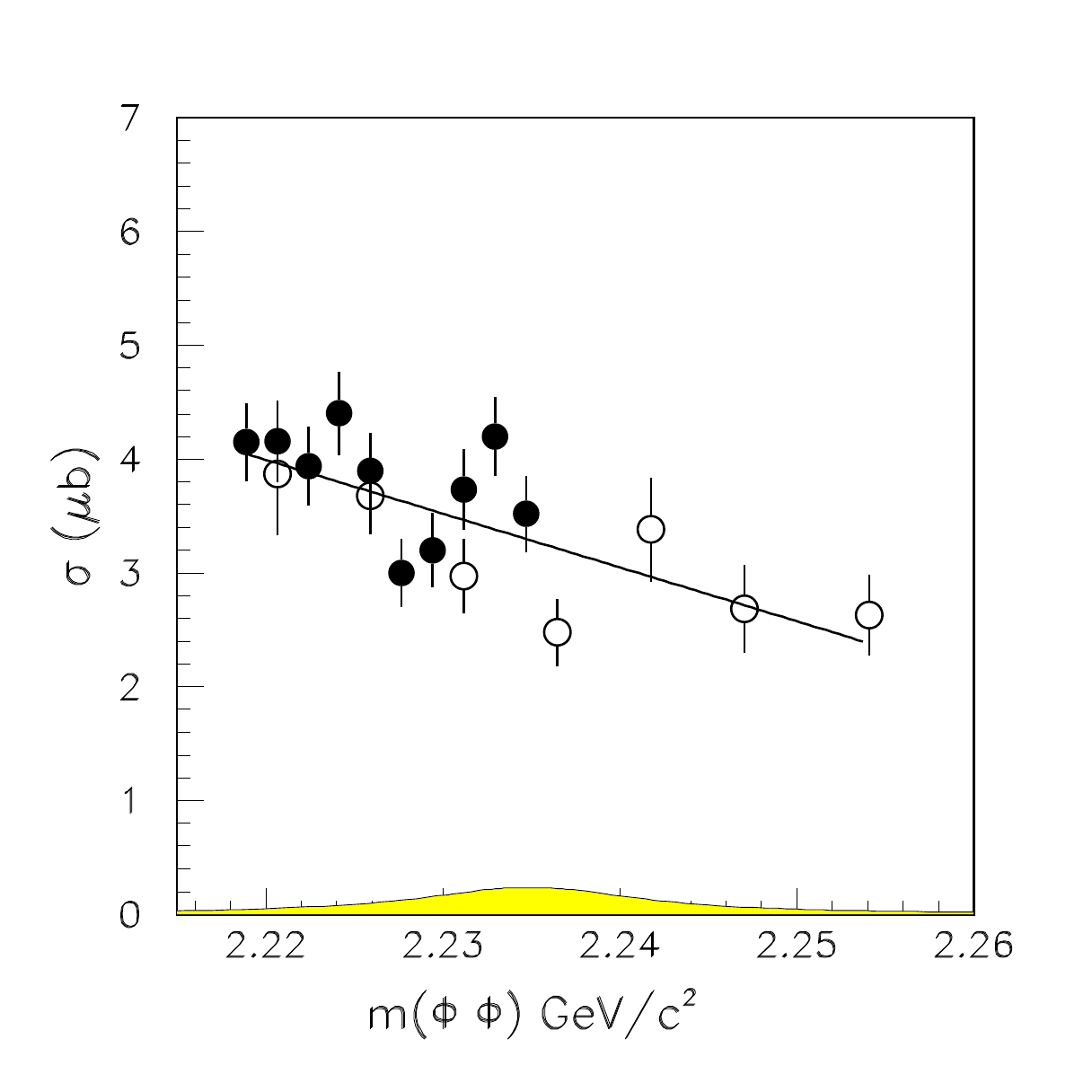}
\caption[Cross section measurement $\pbarp \rightarrow \phi \phi$ by \INST{JETSET}.]{Cross section for the reaction $\pbarp\rightarrow\phi\phi$ measured by the \INST{JETSET} experiment. The yellow curve represents a Breit-Wigner resonance, whose amplitude is at \CL=95\percent upper limit for the production of $f_J(2230)$ at a mass of 2235\,\mev$/c^2$ and a width of 15\,\mevcc~\cite{bib:phy:Evangelista:1998zg}.}
\label{fig:phys:exo:glueball:phiphi:jetset}
\end{center}  
\end{figure}

Besides the fact that there will be uncorrelated hadronic background due to misidentification or secondary particles the main obstacle for a high precision detection of a resonant structure in the cross section will probably be the total cross section of non-resonant 
\bn
\pbarp\rightarrow \phi \phi \rightarrow \Kp \Km \Kp \Km
\en 
reactions being of the order of $\sigma_{\scriptsize \pbarp\rightarrow \phi\phi} \approx $ 3-4\,$\mu$b in that energy region, which has been measured by the \INST{JETSET} experiment as shown in \Reffig{fig:phys:exo:glueball:phiphi:jetset}. These have exactly the same signature as the signal reactions
\bn
\pbarp\rightarrow X \rightarrow \phi \phi \rightarrow \Kp \Km \Kp \Km
\en 
and thus are also kinematically not separable, e. g. by a 4 constraint fit.
The only possibility to disentangle non-resonant from resonant reactions is to perform a spin-parity or partial-wave analysis (PWA). For that purpose it might be crucial to have a 'good' {\it i.e.} flat behaviour of the efficiency dependence with respect to the intrinsic appearing angles of the decay. These are
\bi
\item the K-decay angle $\theta_{\phi_1}$ of the first $\phi$,
\item the K-decay angle $\theta_{\phi_2}$ of the second $\phi$ and
\item the angle $\phi_{\mbox{\scriptsize plane}}$ between the decay planes of the two $\phi$-mesons, as illustrated in \Reffig{fig:phys:exo:glueball:phiphi:plane}.
\ei
Since the initial \pbarp system has been generated phase space distributed the decay angle distributions of the $\phi$'s are expected to be flat. 

\begin{figure*}[hbtp]
\begin{center} 
\includegraphics[angle=90,width=\swidth]{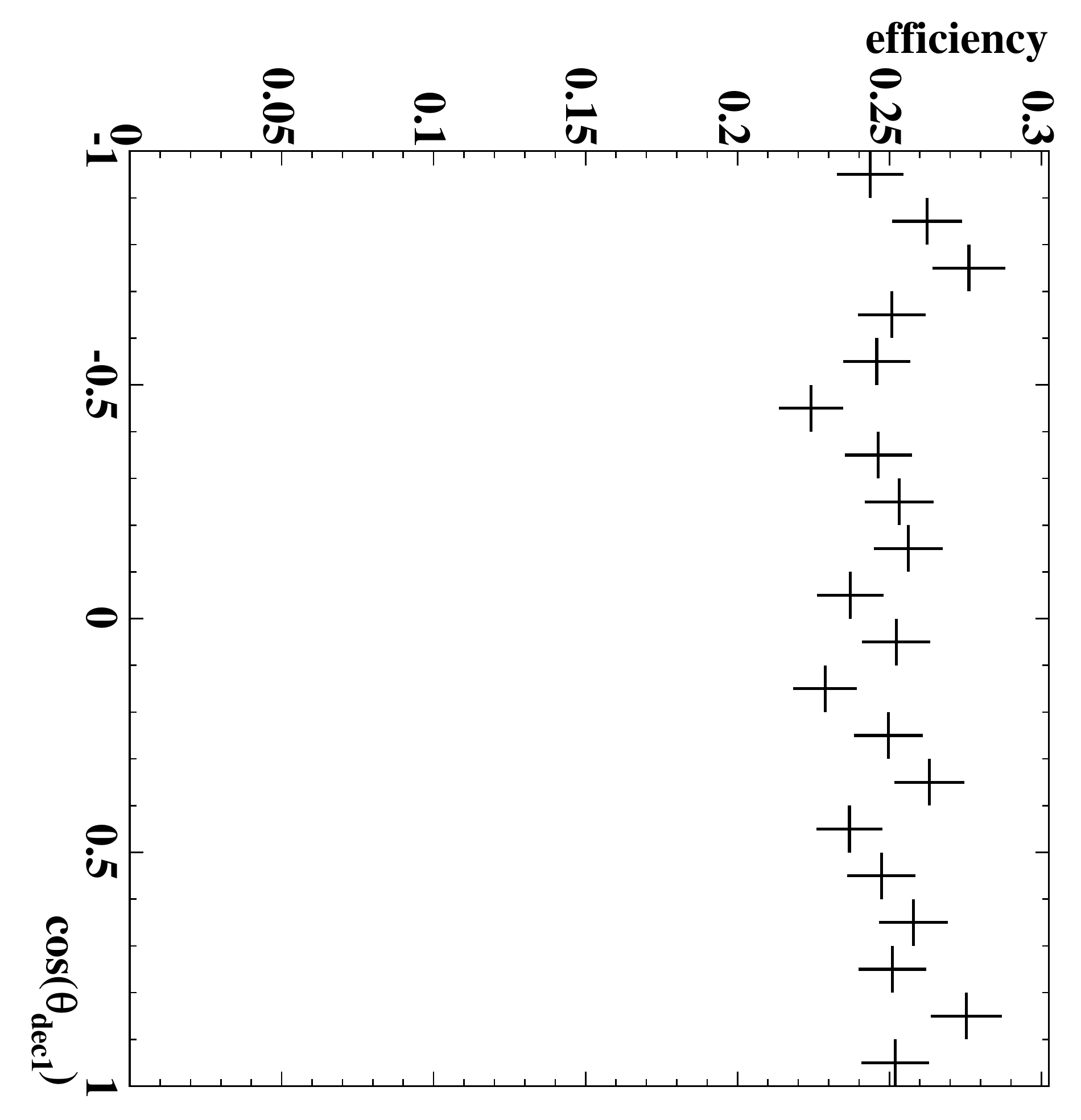}
\includegraphics[angle=90,width=\swidth]{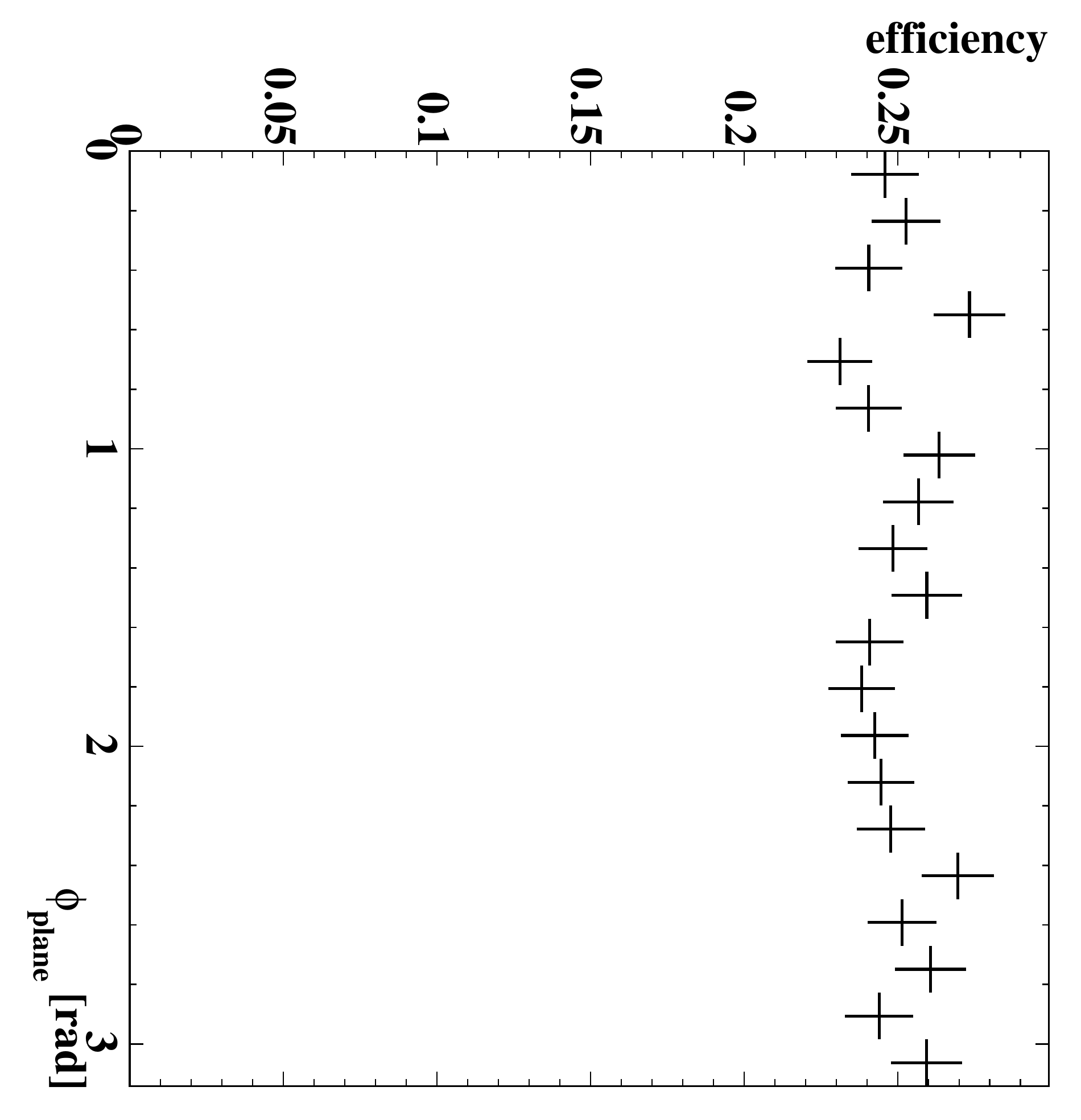}
\caption[Angular dependent efficiencies]{(a) Efficiency as function of the $\phi$ decay angle $\cos\theta_\phi$. (b) Efficiency as function of the angle $\phi_{\mbox{\scriptsize plane}}$ between the two decay planes.}
\label{fig:phys:exo:glueball:phiphi:angeff}
\end{center}  
\begin{picture}(0,0)
\large
\put(58,100){(a)}
\put(270,100){(b)}
\end{picture}
\end{figure*}
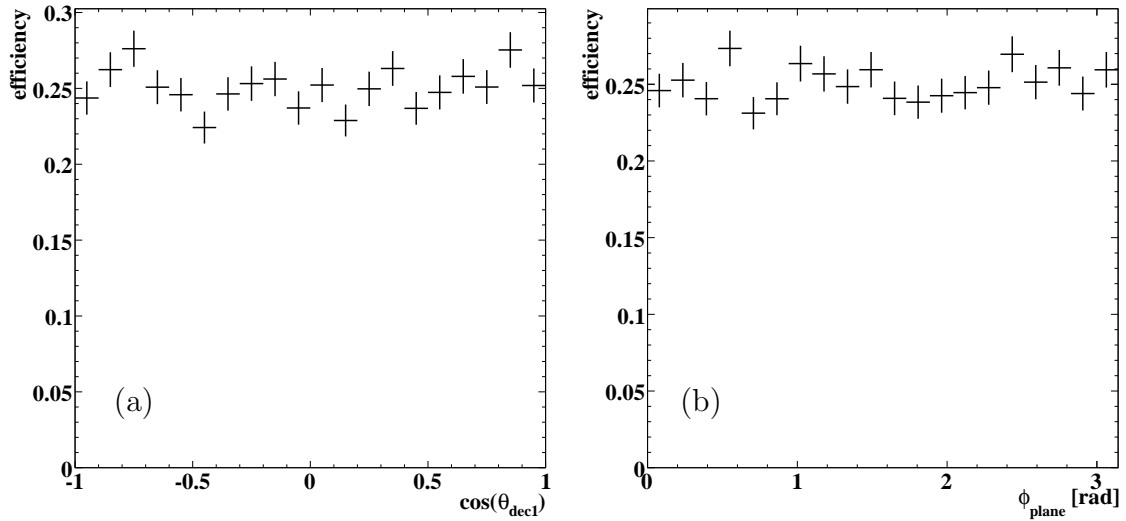

The efficiency as function of these angles has been determined by dividing the distribution of reconstructed candidates by the distribution of the corresponding quantity of generated particles. \Reffig{fig:phys:exo:glueball:phiphi:angeff} shows the results, in (a) for the $\phi$-decay angle $\cos\theta_\phi$ and in (b) for the angle $\phi_{\mbox{\scriptsize plane}}$ between the decay planes. 

It can be seen clearly that the efficiency is independent of any of the involved angles, thus making a potential PWA less difficult.

\par
As previously indicated resonances in formation reactions can only be detected via an energy scan around the potential resonances pole mass. Close to that energy the total cross section will be enhanced according to the line shape of the resonance whose intensity could look like a Breit-Wigner distribution
\bn
\label{func:phys:exo:glueball:phiphi:spectral}
\BW(m) = A\cdot\frac{1}{\pi} \cdot \frac{\Gamma/2}{(m-m_{R})^2 + (\Gamma/2)^2}
\en
with $\Gamma$ and $m_R$ being its total width and pole mass respectively and $A$ being an arbitrary amplitude. This enhancement has to be separated in particular from the non resonant part of the total cross section. The \INST{JETSET} experiment  performed a measurement of exactly this total cross section of the reaction $\pbarp \rightarrow \phi \phi$ leading to a value $\sigma_{\phi\phi} \approx$ 3-4\,$\mu$b, as shown in \Reffig{fig:phys:exo:glueball:phiphi:jetset}. For the following studies the empirical line fit shown has been considered as the background level upon which the signal shape has to be detected. 
The curve explicitly was chosen as
\bn
\label{func:phys:exo:glueball:phiphi:bg}
\sigma_{\mbox{\scriptsize non-res}}(m) = a+b\cdot m
\en
with parameter values $a = 92.8\,\mu$b and $b = -40\,\mu$b/\gevcc.

Figure of merit is the beam time necessary to measure the signal cross section with a significance $S=\sigma/\delta\sigma$ of 10$\sigma$, for now neglecting the background considerations for generic events discussed in the previous chapter. 

\begin{figure*}[hbtp]
\begin{center} 
\includegraphics[angle=90,width=\swidth]{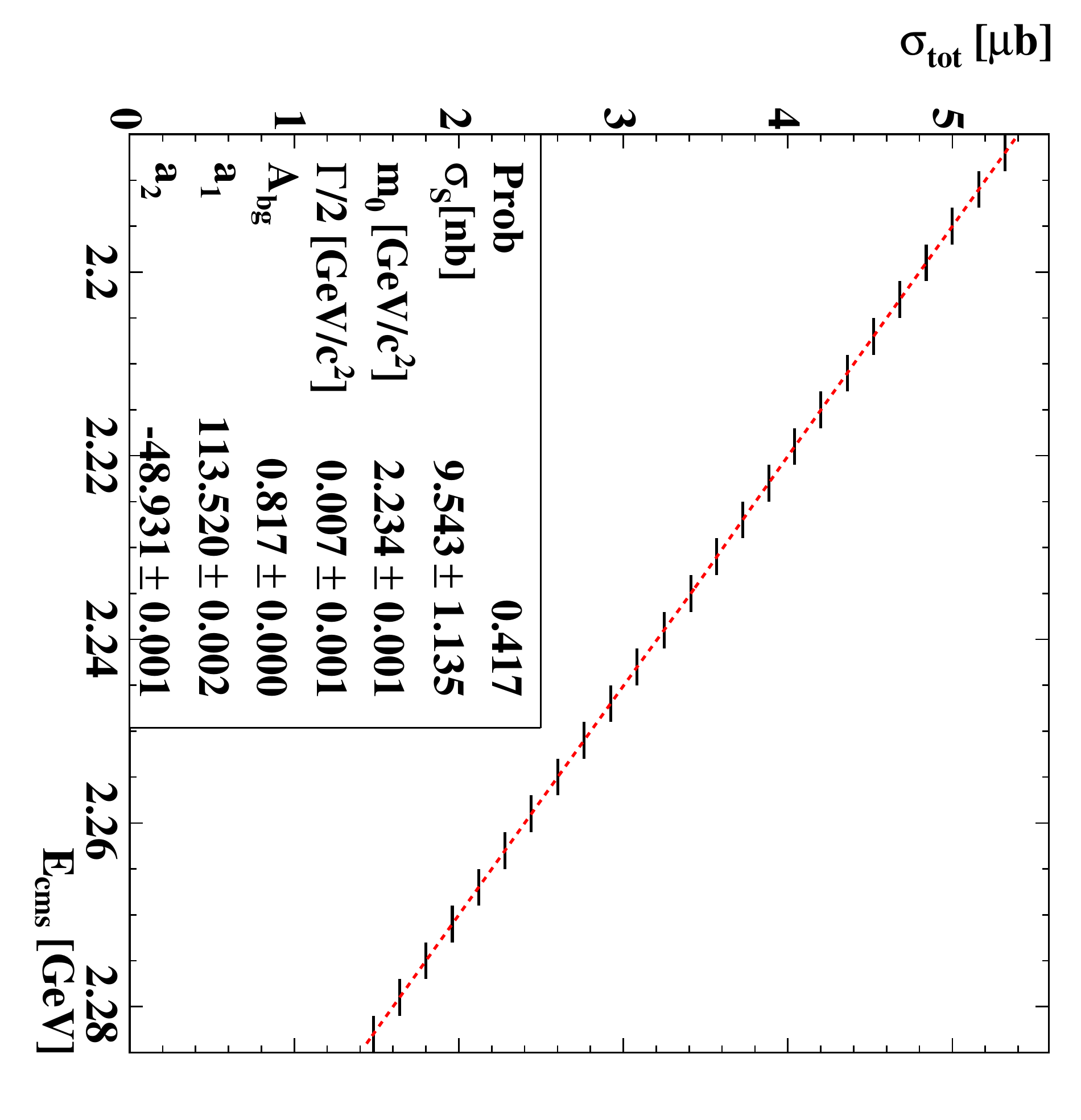}
\includegraphics[angle=90,width=\swidth]{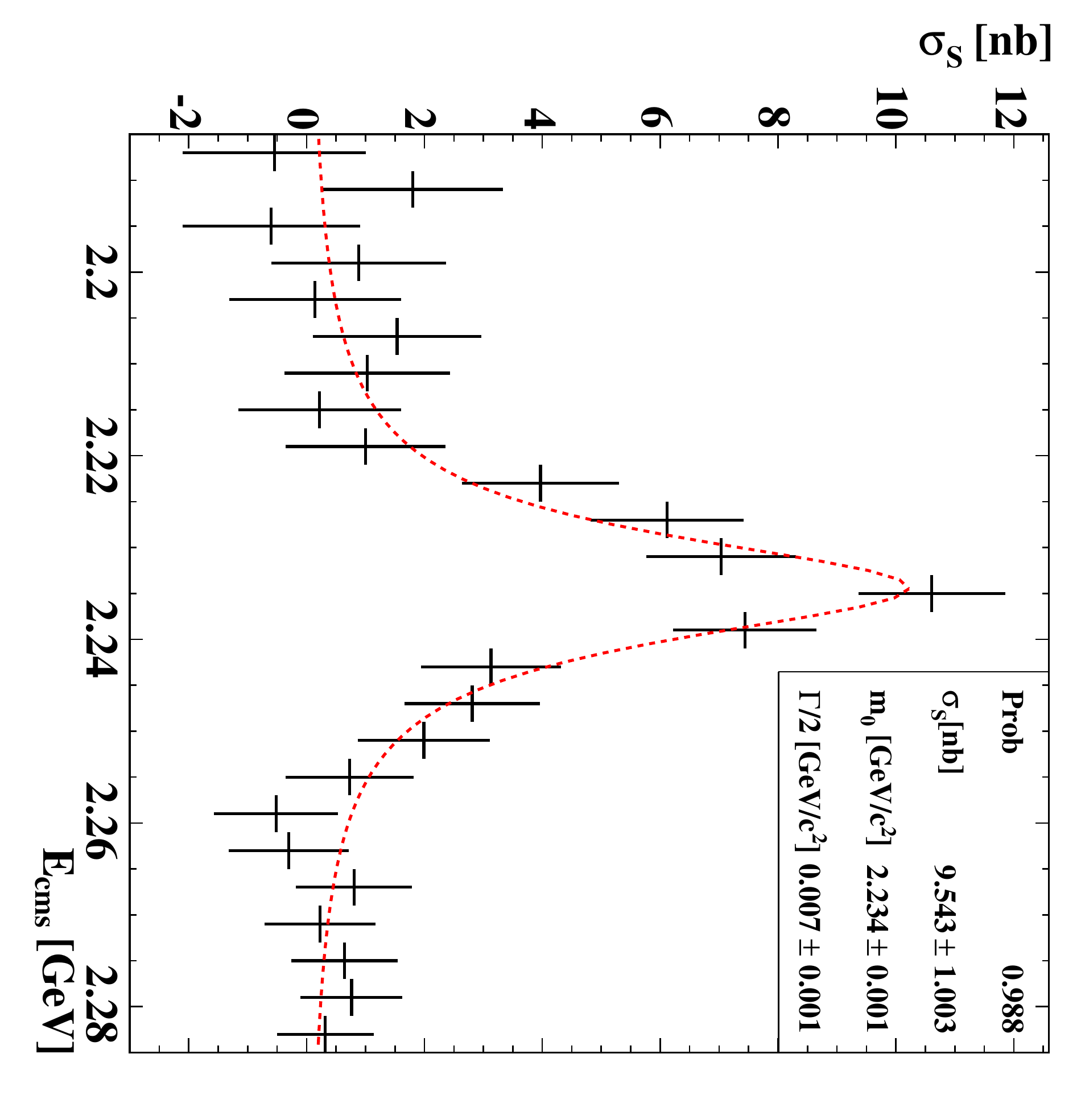}
\caption[Cross section scan examples]{Fits to the total cross section (a) and the derived signal cross section (b) for a scan with $\sigma_S = 10$ nb and beam time according to \Reftbl{tab:phys:exo:glueball:phiphi:scan}.}
\label{fig:phys:exo:glueball:phiphi:xsec1}
\end{center}  
\begin{picture}(0,0)
\large
\put(190,220){(a)}
\put(270,220){(b)}
\end{picture}
\end{figure*}

The procedure for that purpose was:
\begin{enumerate}
\item Assumptions for parameters:
\bi
\item Resonance pole mass: $m_{\mbox{\scriptsize pole}}=2235$\,\mevcc
\item Resonance full width: $\Gamma=15$\,\mevcc
\item Resonance branching ratio to signal channel: $\BR(f_2\rightarrow \phi \phi) = 0.2$
\item Integrated luminosity: ${L} = 8.8$\,pb$^{-1}$/day
\item Energy window: $\pm 50$\,\mev around pole mass
\item Number of equally distributed scan positions: $n=25$ 
\item Reconstruction efficiency of $\phi\phi$ channel: $\epsilon=0.25$
\ei
\item Vary signal cross section $\sigma_S$ at pole mass between 1\,nb and 1\,$\mu b$
\item Determine the approximate total beam time $T_{\mbox{\scriptsize b}}$ for the complete measurement to achieve significance of 10$\sigma$ for the cross section measurement:
\bi
\item Vary  $T_{\mbox{\scriptsize b}}$ arbitrarily an apply following steps until significance is 10$\sigma$
\item Estimate number of expected background entries for each scan energy $E_i = m_i\cdot c^2$ as 
\bn
B_i = \sigma_{\mbox{\scriptsize non-res}}(m_i)\cdot\epsilon\cdot\frac{T{\mbox{\scriptsize b}}[d]\cdot { L}}{n}
\en
\item Estimate number of expected signal entries for $E_i$ as
\bn
S_i &=& \frac{\BW(m_i)}{\BW(m_{\mbox{\scriptsize pole}})} \cdot \sigma_S 
\cdot \epsilon\mbox{\hspace{2.5cm}} \nonumber \\
&& \cdot\BR(f_2\rightarrow \phi \phi)\cdot\frac{T{\mbox{\scriptsize b}}[d]\cdot { L}}{n}
\en
\item Set contents of bin number $i$ of the scan histogram to $c_i = (S_i+B_i) \pm \sqrt{S_i+B_i}$
\item Fit sum of signal function \Refeq{func:phys:exo:glueball:phiphi:spectral} and background function \Refeq{func:phys:exo:glueball:phiphi:bg} to resulting histogram and compute the significance as $A/\delta A$, where $A$ is the fitted amplitude for the resonant part of the fit model.
\ei
\end{enumerate}
It turns out that the results reasonably behave according to statistics expectation, {\it i.e.} twice the beam time results in a precision improved by a factor $\sqrt{2}$. \Reffig{fig:phys:exo:glueball:phiphi:xsec1}  shows some of the corresponding plots with the fits performed to the total cross section (a) as well as to the estimated background subtracted signal cross section determined as the difference of the total cross section and the background expectation computed from \Refeq{func:phys:exo:glueball:phiphi:bg} (b). Both fits agree quite reasonable as expected. It seems surprising that in particular in \Reffig{fig:phys:exo:glueball:phiphi:xsec1} (a) no signal is visible at all whereas the fit result has quite a high significance. This is due to the assumed high precision of the data points reflected in the corresponding difference plots on the right hand side. Of course this evidently depends on the certainty of signal and background line shapes.

\Reftbl{tab:phys:exo:glueball:phiphi:scan} summarises the results for the studies above. The necessary beam times to achieve an accuracy of 10$\sigma$ significance vary from infeasible hundreds to thousands of days with assumed signal cross section of $\sigma_S < 10$\,nb down to comfortable ''far less than a day'' time windows for signal cross sections $\sigma_S > 100$\,nb. 

\begin{table}[htb]
\centering
\begin{tabular}{cc}
\hline\hline
 $\sigma_S$ [nb] & Beam time $T_{\mbox{\scriptsize b}}$ ($\approx$)\\ \hline
1 & 13.7\,y\\
5 & 200\,d\\
10 & 50\,d\\
100 & 12\,h\\
500 & 0.5\,h\\
1000 & 7.2\,min\\\hline\hline
\end{tabular}
\caption{Beam times needed to achieve a significance of 10\,$\sigma$.}
\label{tab:phys:exo:glueball:phiphi:scan}
\end{table}
 
It shall be emphasised at this point that the results might be too optimistic since idealised by the assumption of the correctness of the knowledge about the total cross section measurement. Therefore the given beam time estimates might be considerably sensitive to the uncertainties which clearly can be seen in \Reffig{fig:phys:exo:glueball:phiphi:jetset}.

%

%% file: phys/phys_heavylight.tex
%
\clearpage
\subsection{Heavy-Light Systems}
\COM{Author(s): A. Gillitzer}
\COM{Referee(s): K. Peters}
%
\input{./phys/heavy-light/phys_heavylight_intro}

%% file: phys/heavy-light/phys_heavylight_intro.tex
%
%

\subsubsection*{Introduction}

Consisting of a heavy and a light constituent, the $D$ meson can
be seen as the hydrogen atom of QCD. For the understanding of the
strong interaction $D$ mesons are very interesting objects since
they combine the aspect of the heavy quark as a static colour
source on one side, and the aspect of chiral symmetry breaking and
restoration due to the presence of the light quark on the other
side. In the limit of infinite mass of the heavy quark (heavy
quark limit), the states of heavy-light mesons are degenerate with
respect to the spin degree of freedom of the heavy quark, and the
total angular momentum of the light quark is
conserved~\cite{bib:phy:DeRujula:1976kk,bib:phy:Isgur:1991wq}. In
reality, the charm quark mass is not very much above the hadronic
scale of $\sim{}1$\,GeV, but still, similar to the hyperfine
splitting in the hydrogen atom induced by the proton spin, the
spin orientation of the charm quark has only a small effect on the
mass of the system.

Based on  earlier observations of low-lying $D$ meson states, the
phenomenological quark model was thought to be able to describe
the excitation spectra of heavy-light systems, and thus to predict
also then unobserved $D$ meson states with reasonable
accuracy~\cite{bib:phy:Isgur:1991wq,bib:phy:Godfrey:1986wj,bib:phy:Godfrey:1985xj,bib:phy:DiPierro:2001uu}.
According to the quark model systematics the lowest states are the
$S$-wave states with the spin singlet $J^P=0^-$ as ground state
($D$) and the spin triplet $J^P=1^-$ as first excited state
($D^\ast$), followed by the $P$-wave states with
$J^P=0^+,1^+,1^+,2^+$ ($D_0,D_1,D_1',D_2$). The physical $J^P=1^+$
doublet ($D_1$ with $j_q^P=\frac{1}{2}^+$, $D_1'$ with
$j_q^P=\frac{3}{2}^+$; $j_q$ is the total spin of the light quark)
results from a mixing of $^3P_1$ and $^1P_1$ states, since in
heavy-light systems the total spin $S=s_q+s_Q$ is not a good
quantum number. The experimentally observed non-strange $D$ meson
spectrum~\cite{bib:phy:Hagiwara:2002fs} was consistent with this
pattern of six states, although for some of the states no
spin-parity assignments could be given. In the spectrum of charmed
strange mesons the only states known with established spin-parity
assignments before the recent discoveries were the pseudo-scalar
ground state $D_s$ and the first excited vector state $D_s^\ast$.
Apart from this, a $D_s(2536)$ and a $D_s(2573)$ state had been
observed~\cite{bib:phy:Hagiwara:2002fs}.

\subsubsection*{Recent discoveries}

The series of new observations in the charmonium spectrum at $B$
and charm factories was accompanied by exciting new experimental
results on the spectrum of open charm mesons, starting with the
unexpected discovery of a narrow $D_s(2317)$ state observed in the
decay mode $D_s^+\pi^0$ by \INST{BaBar}~\cite{bib:phy:Aubert:2003fg} in
$\ee$ annihilation data at energies near 10.6\,GeV. Shortly
after, this state was confirmed by
CLEO~\cite{bib:phy:Besson:2003cp} and
Belle~\cite{bib:phy:Krokovny:2003zq}. At the same time, \INST{CLEO} found
a new, also narrow state $D_s(2460)$~\cite{bib:phy:Besson:2003cp}
decaying to $D_s^{*+}\pi^0$. This state was subsequently also seen
by \INST{Belle}~\cite{bib:phy:Krokovny:2003zq}, and confirmed by
\INST{BaBar}~\cite{bib:phy:Aubert:2003pe,bib:phy:Aubert:2004pw}. The
width of both states is small since the $D_s(2317)$ is lying below
the $DK$ threshold, and the $D_s(2460)$ is below the $D^\ast K$
threshold. Thus the $D_s(2317)$ state cannot decay by kaon
emission, and, with $J^P=1^+$, decay with kaon emission is also
forbidden for the $D_s(2460)$ state. Decay to $D_s^{(*)}$ with
single pion emission is isospin forbidden. The properties of the
two states were further studied in
~\cite{bib:phy:Abe:2003jk,bib:phy:Aubert:2006bk}.
\cite{bib:phy:Aubert:2006bk} gives upper limits of the width
$\Gamma{}<3.8$\,MeV for the $D_s(2317)$ state and
$\Gamma{}<3.5$\,MeV for the $D_s(2460)$ state. The observed decay
modes are consistent with spin-parity assignments $J^P=0^+$ for
the $D_s(2317)$ state and $J^P=1^+$ for the $D_s(2460)$ state,
respectively. The so far observed decay modes
are~\cite{bib:phy:PDGlive} $D_s(2317)^+\rightarrow{}D_s^+\pi^0$
and
$D_s(2460)^+\rightarrow{}D_s^{*+}\pi^0,D_s^+\gamma,D_s^+\pi^+\pi^-$.
Very recently, another $D_s$ meson state decaying into $DK$ at a
mass of 2.86\,GeV/c$^2$ and a larger width of $\sim{}47$\,MeV was
observed at \INST{BaBar}~\cite{bib:phy:Aubert:2006mh}.

\subsubsection*{Theoretical Interpretation}

\begin{figure}[htb]
\begin{center}
\includegraphics[width=\swidth]{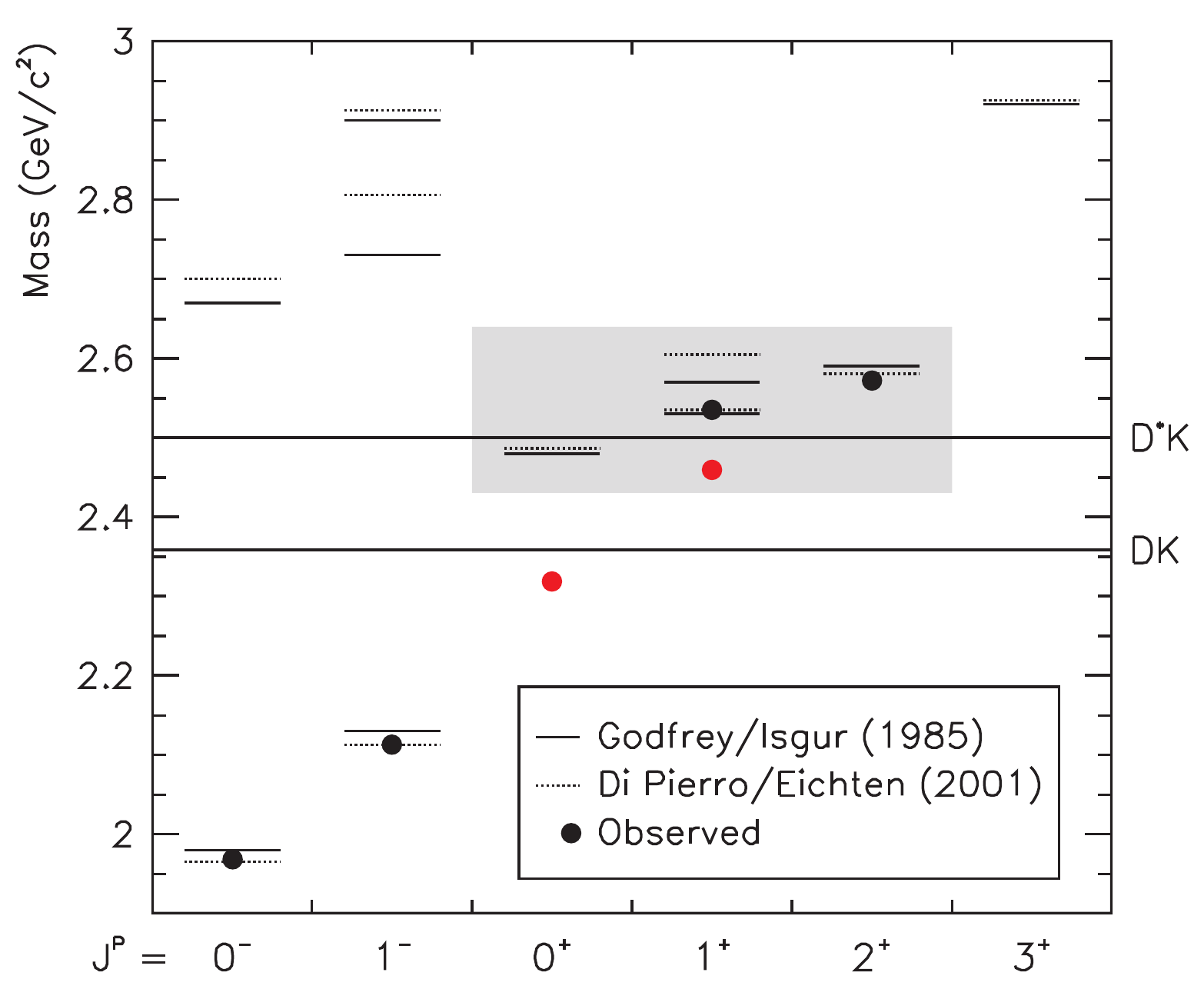}
\caption[The $D_s$ spectrum and quark model predictions]{The $D_s$
meson spectrum as predicted by Godfrey and
Isgur~\cite{bib:phy:Godfrey:1985xj} (solid lines) and Di~Pierro
and Eichten~\cite{bib:phy:DiPierro:2001uu} (dashed lines).
Experimental values are shown by points, the $DK$ and $D^\ast K$
thresholds as horizontal lines. The figure is taken
from~\cite{bib:phy:Aubert:2006bk}. \label{fig:Dspectrum}}
\end{center}
\end{figure}

These observations attracted much interest both in the theoretical
and the experimental hadron physics community, since the new
states don't fit well into the quark model predictions for
heavy-light systems in contrast to the previously known $D$ meson
states. The $D_s(2317)$ state is typically 150\,MeV or more below
the quark model expectation. In particular, it has been considered
very difficult to reproduce the mass difference between the $0^+$
and the $1^+$ state within the quark
model~\cite{bib:phy:Swanson:2006st}. \Reffig{fig:Dspectrum} shows
a comparison between the observed $D_s$ spectrum and quark model
calculations~\cite{bib:phy:Godfrey:1985xj,bib:phy:DiPierro:2001uu}
together with the $DK$ and $D^\ast K$ thresholds.

Long before the discovery of the new states, a model based on
chiral symmetry was developed for heavy-light systems that
predicted the  mass splitting of the  $0^--0^+$ and $1^--1^+$ to
be related to the light  constituent quark
mass~\cite{bib:phy:Nowak:1992um,bib:phy:Bardeen:1993ae}. Since in
that approach the $D$ mesons, like the light quark fields in QCD,
transform linearly under chiral  $SU(3)$ rotations the spectrum
shows a doublet structure. After the discovery of the new states
the scheme was applied to the new $D_s$
states~\cite{bib:phy:Bardeen:2003kt,bib:phy:Nowak:2003ra} and
shown to explain the low $D_s(2317)$ mass and the identical
hyperfine splitting in the $0^--1^--$ and $0^+-1^+$ doublets. Such
results indicate the importance of chiral symmetry for the
understanding of the open-charm meson spectrum.

The discovery of the scalar and axial-vector open-charm states
triggered  a series of theoretical works studying the new states
in various frameworks. For a recent review focusing on approaches
with active quark degrees of freedom see
Ref.~\cite{bib:phy:Swanson:2006st}. The possibility that these
states could have  a simple $c\bar{s}$ structure is discussed in
Refs.~\cite{bib:phy:Cahn:2003cw,bib:phy:Lakhina:2006fy}. The
predicted level scheme of typical quark models may, however,
suggest that the scalar and axial-vector $D_s$ mesons have an
exotic non-$c\bar{s}$ structure. For instance tetra-quark models
for $D_s$ states assume strong
$\left[c\bar{s}\right]\left[u\bar{u}+d\bar{d}\right]$ or
$\left[cq\right]\left[\bar{s}\bar{q}'\right]$ components in the
wave
function~\cite{bib:phy:Terasaki:2003qa,bib:phy:Hayashigaki:2004st,bib:phy:Cheng:2003kg,bib:phy:Maiani:2004uc,bib:phy:Maiani:2004vq,bib:phy:Chen:2004dy,bib:phy:Bracco:2005kt,bib:phy:Nielsen:2006tw}.
However, tetra-quark models usually predict a large number of
states in addition to the $q\bar{q}$ meson states for which there
is little experimental evidence. In
Refs.~\cite{bib:phy:vanBeveren:2003kd,bib:phy:Hwang:2004cd,bib:phy:Hwang:2005tm}
the relevance of higher Fock states involving mesonic degrees of
freedom for the scalar states was investigated applying resonating
group methods to a quark model. The possibility that the scalar
$D_s$  may be a molecular $DK$ state was discussed in
Ref.~\cite{bib:phy:Barnes:2003dj}.

First systematic computations where open-charm resonances are
generated in terms of hadronic degrees of freedom were based on
the chiral Lagrangian written down for the Goldstone bosons and
the pseudo-scalar and vector $D$-meson ground
states~\cite{Kolomeitsev:2003ac,bib:phy:Hofmann:2003je}. The
approach is consistent with the heavy-quark symmetry that arises
in the limit of large charm quark mass. A natural explanation of
the scalar and axial vector spectrum was achieved, where for
instance the scalar $D_s$ states are coupled-channel molecules
with important $DK$ but also $\eta\,D_s$ components. In contrast
to the chiral-doubling approach of
Ref.~\cite{bib:phy:Nowak:1992um,bib:phy:Bardeen:1993ae} the chiral
Lagrangian assumes the $D$ meson fields to transform non-linearly
under chiral transformation. As a consequence the arising spectrum
is not necessarily grouped into chiral doublets. In fact the
leading order chiral interaction predicts weak attraction in
exotic non $c \bar{q}$ channels. The existence of such states
depend2 on the character of sub-leading order terms in the chiral
Lagrangian. A recent study~\cite{Lutz:2007sk} predicts exotic
signals in the $\eta\,D^\ast$ and $\pi\,D$ channels.

\subsubsection*{$\boldsymbol{D}$ Spectroscopy at \PANDA}

An important quantity possibly allowing to distinguish between the
different pictures is the decay width of the two $D_s$
states~\cite{Lutz:2007sk,bib:phy:Sassen:2005ej,bib:phy:Guo:2007up,bib:phy:Sassen05phd,Faessler:2007gv,Faessler:2007us}.
So far their widths are only constrained by upper limits of a few
MeV due to detector resolution, which is not sensitive enough to
draw conclusions on their internal structure. In the following, a
different experimental approach to determine the narrow widths of
these states is discussed. Very close to threshold the energy
dependence of the production cross section for a narrow resonance
can be calculated in a model-independent
way~\cite{bib:phy:Hanhart05pc}, and this function is sensitive to
the resonance width. Provided the beam energy is sufficiently well
known, thus the width of a narrow resonance can be determined in a
measurement of the energy dependence of the production cross
section around the energy threshold, without requiring the
corresponding detector resolution in the reconstruction of the
resonance from the final state. With $\delta{}p/p\simeq{}10^{-5}$
the $\pbar$ momentum spread in the \HESR is sufficiently small to
allow the measurement of the $D_{sJ}$ widths in a threshold scan
of the reaction $\pbarp\rightarrow{}\bar{D}_sD_{sJ}$ at \PANDA
down to values of $\sim{}100$~keV.

As an example, the study of the $D_{s0}$ width is specifically
discussed in the following sections related to simulation studies
which have been performed for this report. Future investigations
of the $D$ and $D_s$ meson spectra at \PANDA have however a wider
scope beyond this specific aspect. If an exotic structure of the
newly found states as molecular or tetra-quark states should be
established, an experimental $D$ meson spectroscopy program should
search for the then missing quark model states with the identical
quantum numbers. Further insight is expected from a comprehensive
study of the decay modes of both $D$ and $D_s$ mesons including
also modes with small branching fraction. Recent theoretical
studies have proposed to measure the partial widths for various
radiative decays of the new $D_s$ states, since these are
predicted to be distinctly different for a molecular or $Q\bar{q}$
structure~\cite{Lutz:2007sk,Faessler:2007gv,Faessler:2007us,bib:phy:Mehen:2004uj,bib:phy:Close:2005se,bib:phy:Colangelo:2003vg,bib:phy:Colangelo:2005hv,bib:phy:Liu:2006jx}.
This requires to measure both the branching fractions for the
radiative transition and the total width of the decaying state
which is certainly an ambitious goal. For not too small total
widths and transition probabilities this may however be possible.

So far only $S$ and $P$ wave states have been observed in both the
$D$ and the $D_s$ spectrum. Information on states with $D$ wave
and higher angular momenta are still missing. This is likely to be
due to the limitation in angular momentum in the production
process of $D$ mesons studied up to now ($\ee$ annihilation and 
$B$ meson decays). In contrast very high partial
waves are available in the $\pbarp$ entrance channel which
should also enhance the population of states with high $L$ values.
As was discussed very recently~\cite{bib:phy:Shifman:2007xn}, the
exploration of the region of states with high angular momenta
could be an important step to clarify the question whether or not
chiral symmetry is restored in hadrons at high excitation
energies.

\subsubsection*{Simulated Signal Channels}
For this report simulation studies focus on the identification of
the $D_{s0}(2317)$ and the determination of its width, based on
the reaction channel
$\pbarp\rightarrow{}D_s^{\pm}D_{s0}^\ast(2317)^{\mp}$. The
recoiling $D_{s0}^\ast(2317)$ is identified inclusively without
specifying its decay mode, as will be discussed below. Signal
events have been generated which the event generator
{\tt{}EvtGen}\cite{bib:phy:evtgen} with the intrinsic width of the
$D_{s0}^\ast(2317)$ set to $\Gamma{}=0.1$\,MeV for the study of signal
reconstruction. A data set for signal events (=Signal 1) was
generated in the following way:
\begin{eqnarray}
\pbarp &\rightarrow& D_s^{\pm} D_{s0}^\ast(2317)^{\mp}\\
D_s^{\pm} &\rightarrow& \phi\pi^{\pm},\quad \phi \rightarrow \Kp\Km\\
D_{s0}^\ast(2317)^{\mp} &\rightarrow& \mbox{ anything}
\end{eqnarray}
A second data set with equal number of events (=Signal 2) was
generated completely inclusive, {\it i.e.}
\begin{eqnarray}
\pbarp &\rightarrow& D_s^{\pm} D_{s0}^\ast(2317)^{\mp}\\
D_s^{\pm} &\rightarrow& \mbox{ anything}\\
D_{s0}^\ast(2317)^{\mp} &\rightarrow& \mbox{ anything}
\end{eqnarray}
in order to allow an independent determination of the
reconstruction efficiency.

Since the $D_{s0}^\ast(2317)$ width with its present upper
experimental limit $\Gamma{}<3.5$\,MeV~\cite{bib:phy:PDGlive} is
smaller than the detector resolution, it cannot be measured
directly. Instead, it can be deduced from a measurement of the
shape of the excitation function of $D_{s0}^\ast(2317)$ production
very close to threshold. The simulation studies thus also address
an energy scan of the signal channel around the threshold energy
and explore how well the shape of the excitation function can be
determined.

\subsubsection*{Background Channels}
In order to estimate the background level several specific
reaction channels with a $D_s$ meson in the final state with
identical decay mode as the signal events have been investigated.
To fill the available phase space at the threshold of the signal
channel of 354\,MeV, background channels with pions or photons in
addition to the second $D_s$ meson have been considered.
Furthermore, a generic hadronic background sample created using
the event generator~{\tt{}DpmGen} based on the dual parton
model~\cite{bib:phy:DPM} has been analysed.
\Reftbl{tab:datasets} lists the investigated data sets. The
data sets were generated at 5\,MeV above the nominal
$D_s^{\pm}D_{s0}^\ast(2317)^{\mp}$ threshold.

\begin{table*}[htb]
\centering
\begin{tabular}{lr}
\hline\hline
Channel & Number of events\\
\hline
 $\pbarp\rightarrow D_s^{\pm}D_{s0}^\ast(2317)^{\mp}$ (Signal 1) & 40\,k \\
 $\pbarp\rightarrow D_s^{\pm}D_{s0}^\ast(2317)^{\mp}$, $D_s^{\pm} \rightarrow$ any (Signal 2) & 40\,k \\ \hline
 $\pbarp\rightarrow D_s^{\pm}D_s^{\mp} \pi^0$  & 40\,k \\
 $\pbarp\rightarrow D_s^{\pm}D_s^{\mp} 2\pi^0$ & 40\,k \\
 $\pbarp\rightarrow D_s^{\pm}D_s^{\mp} \pi^+ \pi^-$ & 40\,k \\
 $\pbarp\rightarrow D_s^{\pm}D_s^{*\mp}$ & 40\,k \\
 $\pbarp\rightarrow D_s^{\pm}D_s^{*\mp} \pi^0$ & 40\,k \\
 $\pbarp\rightarrow D_s^{\pm}D_s^{\mp} \gamma$ & 40\,k \\
 $\pbarp\rightarrow D_s^{\pm}D_s^{*\mp} \gamma$ & 40\,k \\
 DPM generic & 10.5\,M\\
\hline\hline
\end{tabular}
\caption[The data sets to evaluate
signal reconstruction efficiency and signal-to-noise ratio]
{The data sets to evaluate signal reconstruction efficiency and signal-to-noise ratio.}
\label{tab:datasets}
\end{table*}

\subsubsection*{Analysis Strategy}
Corresponding to the goal of the simulation study of
$D_{s0}^\ast(2317)^{\mp}$ production the analysis consists of two
separated steps:
\begin{enumerate}
\item Reconstruct the signal at given energy:
    \begin{itemize}
    \item determine the efficiency of the signal reconstruction
    \item estimate the background level (signal to noise ratio $S/N$)
    \end{itemize}
\item Simulate the energy dependence:
    \begin{itemize}
    \item generate the relevant distributions for the signal events
    based on the results from step 1 at selected energies, and
    determine the number of signal and background events
    \item determine the shape of the excitation function
    \item deduce width and mass of the $D_{s0}^\ast(2317)^{\mp}$ state
    \end{itemize}
\end{enumerate}
These analysis steps will be briefly discussed in the following.

\subsubsection*{Exclusive and inclusive signal reconstruction}
The optimum signal to noise ratio is achieved in a full exclusive
reconstruction of all particles emerging in the decay chains of
the primary particles produced in the reaction. On the other hand,
due to the small branching ratios involved the signal event rate
based on a single decay chain will be very small. In the
simulation studies these small branching ratios reduce the
significance of the achieved background suppression factor at
given size of the analysed background sample. Therefore, as will
be argued below, a different approach using inclusive
$D_{s0}^\ast(2317)$ identification is pursued here.

The $D_s^{\pm}$ meson has many decay branches. As channel with
reasonable efficiency and with characteristic strangeness content
the $D_s^{\pm} \rightarrow \phi \pi^{\pm}$ decay branch with
$\phi\rightarrow{}\Kp\Km$ was selected. The combined branching
ratio is
\begin{eqnarray*}
f_{{\BR},D_s} = {\BR}(D_s^{\pm} \rightarrow \phi
\pi^{\pm})\cdot {\BR}(\phi\rightarrow{}\Kp\Km) =
\\
= 0.044\cdot 0.492 = 0.022.
\end{eqnarray*}
Since the only known decay channel of the $D_{s0}^\ast(2317)$ is the
isospin violating $D_{s0}^\ast(2317)\rightarrow D_s^{\pm} \pi^0$
decay (with unknown branching fraction) one also needs to
reconstruct the $\pi^0\rightarrow 2\gamma$ and the second
$D_s^{\pm}$ in the above channel for a full exclusive
reconstruction resulting in the total branching ratio factor
\begin{eqnarray*}
f_{{\BR},\mbox{\scriptsize excl}} =
\underbrace{{\BR}(D_{s0}^\ast(2317)\rightarrow
\Ds\pi^0)}_{\mbox{\scriptsize unknown}}\cdot {\BR}(D_s^{\pm}
\rightarrow \phi \pi^{\pm})^2\cdot
\\
\cdot {\BR}(\phi\rightarrow{}\Kp\Km)^2\cdot
{\BR}(\pi^0\rightarrow 2\gamma) <
\\
< 4.6\cdot 10^{-4}.
\end{eqnarray*}
The branching ratio ${\BR}(D_{s0}^\ast(2317)\rightarrow\Ds\pi^0)$
is expected to be very close to one. With the assumption of
$\sigma = 1$\,nb signal cross section at threshold, an integrated
luminosity of about ${\cal{}L} = 9000/\mbox{nb}$ per day and
efficiency $\epsilon \approx 0.2$ this results in an expected
number of efficiency and branching ratio corrected signal
reactions of
\begin{eqnarray*}
N_{\mbox{\scriptsize excl}} = \sigma \cdot {\cal{}L} \cdot
\epsilon \cdot f_{{\BR},\mbox{\scriptsize excl}} \approx 9000
\cdot 0.2 \cdot 4.6\cdot 10^{-4} =
\\
= 0.8 \mbox{ detected signals/day}.
\end{eqnarray*}
Such low event rates would require a very long running time for
the measurement of the excitation function. It therefore doesn't
seem promising to only use this specific final state for the
identification of the
$\pbarp\rightarrow{}D_s^{\pm}D_{s0}^\ast(2317)^{\mp}$ reaction,
but to include an as large as possible number of specific channels
in the decay chain in order to increase the event rate in the
exclusive reconstruction.

The task to simulate many different channels is however beyond the
scope of this report. The simulations are therefore focused on the
question whether an inclusive reconstruction of the $D_s^{\pm}$
decay with the identification of the recoiling
$D_{s0}^\ast(2317)^{\mp}$ via the missing mass method, which would
result in higher event rates, yields sufficient background
suppression for signal identification. To enhance the signal to
background ratio kinematic correlations in the event are
exploited, as discussed below. In case of the inclusive
reconstruction the expected number of reactions which can be
detected is estimated to be
\begin{eqnarray*}
N_{\mbox{\scriptsize incl}} = \sigma \cdot {\cal{}L} \cdot
\epsilon \cdot f_{{\BR},D_s} < 9000 \cdot 0.2 \cdot 0.022 =
\\
= 40 \mbox{ detected signals/day}.
\end{eqnarray*}
Hence, to collect a reasonable number like 500 events only around
two weeks of beam time would be required.

\subsubsection*{Event selection criteria}
$D_s^{\pm}$ candidates were created in the following way:
\begin{enumerate}
\item Select kaon candidates from charged tracks with
\vloo PID criterion\footnote{See section \ref{sec:soft:recochargedPID} for details.} (will
be tightened later for better $S/B$ ratio)
\item Create a list of $\phi$ candidates by forming all
combinations of a negative with a positive charged kaon candidate
\item Kinematic fit of the single $\phi$ candidates with vertex
constraint
\item Select pion candidates from charged tracks with
\vloo PID criterion
\item Combine $\phi$ candidates with pion candidates to form
$D_s^{\pm}$ candidates
\item Kinematic fit of the $D_s^{\pm}$ candidates with vertex
constraint
\end{enumerate}

\begin{figure}[htb]
\begin{center}
\includegraphics[angle=90,width=\swidth]{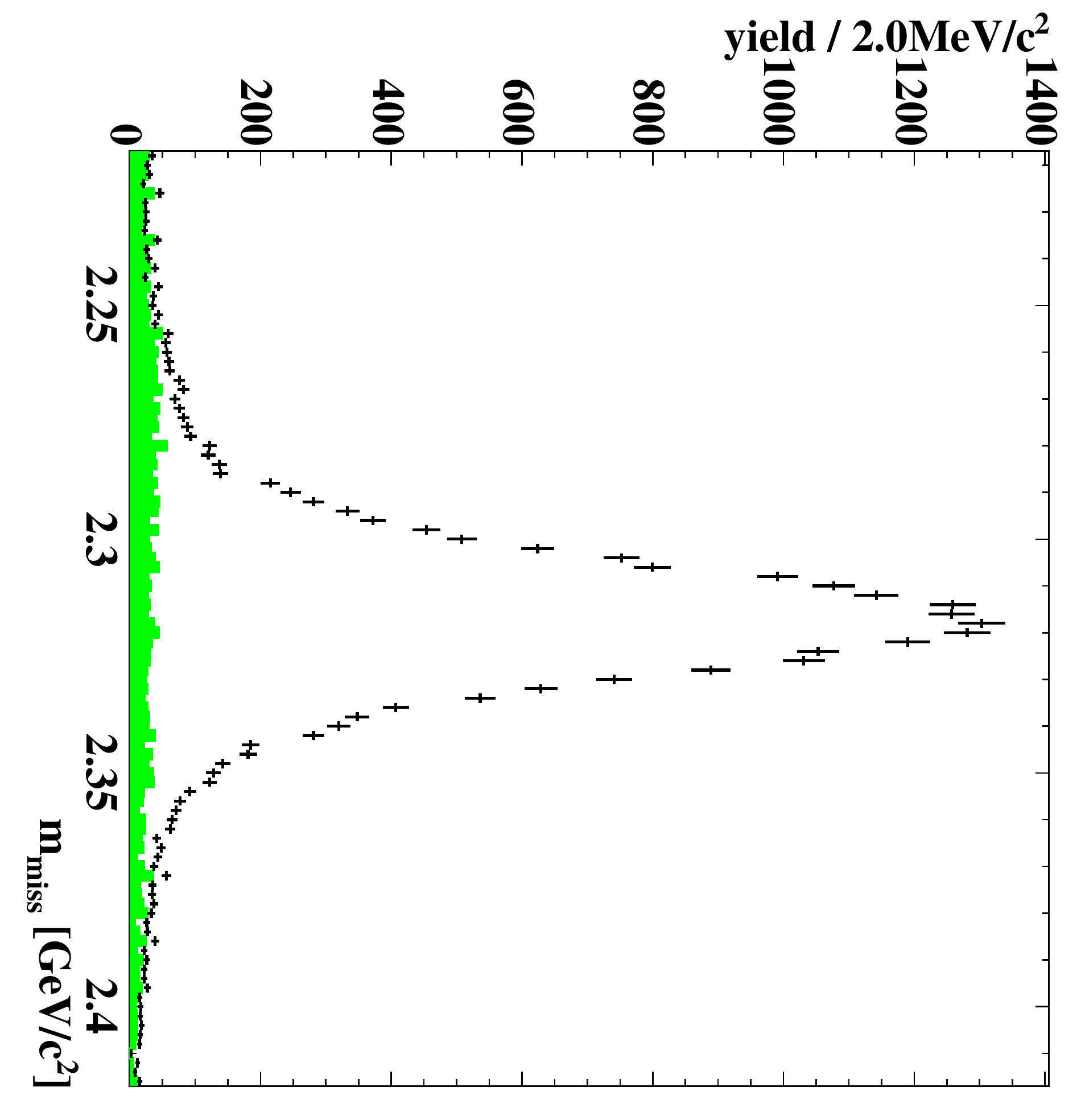}
\caption[Missing mass spectrum obtained for the $D_s^{\pm}$
candidates]{Missing mass spectrum obtained for the $D_s^{\pm}$
candidates based on the known 4-momentum of the initial $\pbarp$
system. The black histogram corresponds to all reconstructed
combinations of the required particles in the final state, the
green filled area represents those combinations which fail the MCT
match criterion (explanation see text).}
\label{fig:Ds-MM}
\end{center}
\end{figure}

On the candidates preselected in this way the following
requirements were applied in addition:
\begin{enumerate}
\item Probability of $\phi$ vertex fit: $P_\phi
> 0.001$
\item Probability of $D_s^{\pm}$ vertex fit: $P_{Ds} > 0.001$
\item $\phi$ mass window: $|m(\Kp\Km)-m_{\mbox{\scriptsize
PDG}}(\phi)| < 10$\,\mevcc
\item $\phi$ decay angle\footnote{The decay angle is defined as
the angle between the direction of motion of the reconstructed
$\phi$ candidate in the laboratory frame and the direction of
motion of one of the kaons in the frame of the $\phi$.}:
$|\cos\theta_{\mbox{\scriptsize dec}}| > 0.5$
\item $D_s^{\pm}$ mass window:
$|m(\phi\pi^{\pm})-m_{\mbox{\scriptsize PDG}}(D_s^{\pm})| <
30$\,\mevcc
\end{enumerate}
\Reffig{fig:Ds-MM} shows the $D_s^{\pm}$ missing mass spectrum
obtained with the selection criteria listed above. The black
histogram represents all reconstructed candidates whereas the
green filled area corresponds to candidates failing the so called
Monte Carlo Truth (MTC) match\footnote{The Monte Carlo Truth match
checks whether or not a decay tree has been exactly reconstructed
the way it was generated.}. Both $D_s^{\pm}$ and $D_{s0}^\ast(2317)^{\mp}$
peaks are reconstructed with a resolution of about
10-15\,\mevcc.

\begin{figure*}[htb]
\begin{center}
\includegraphics[angle=90,width=\swidth]{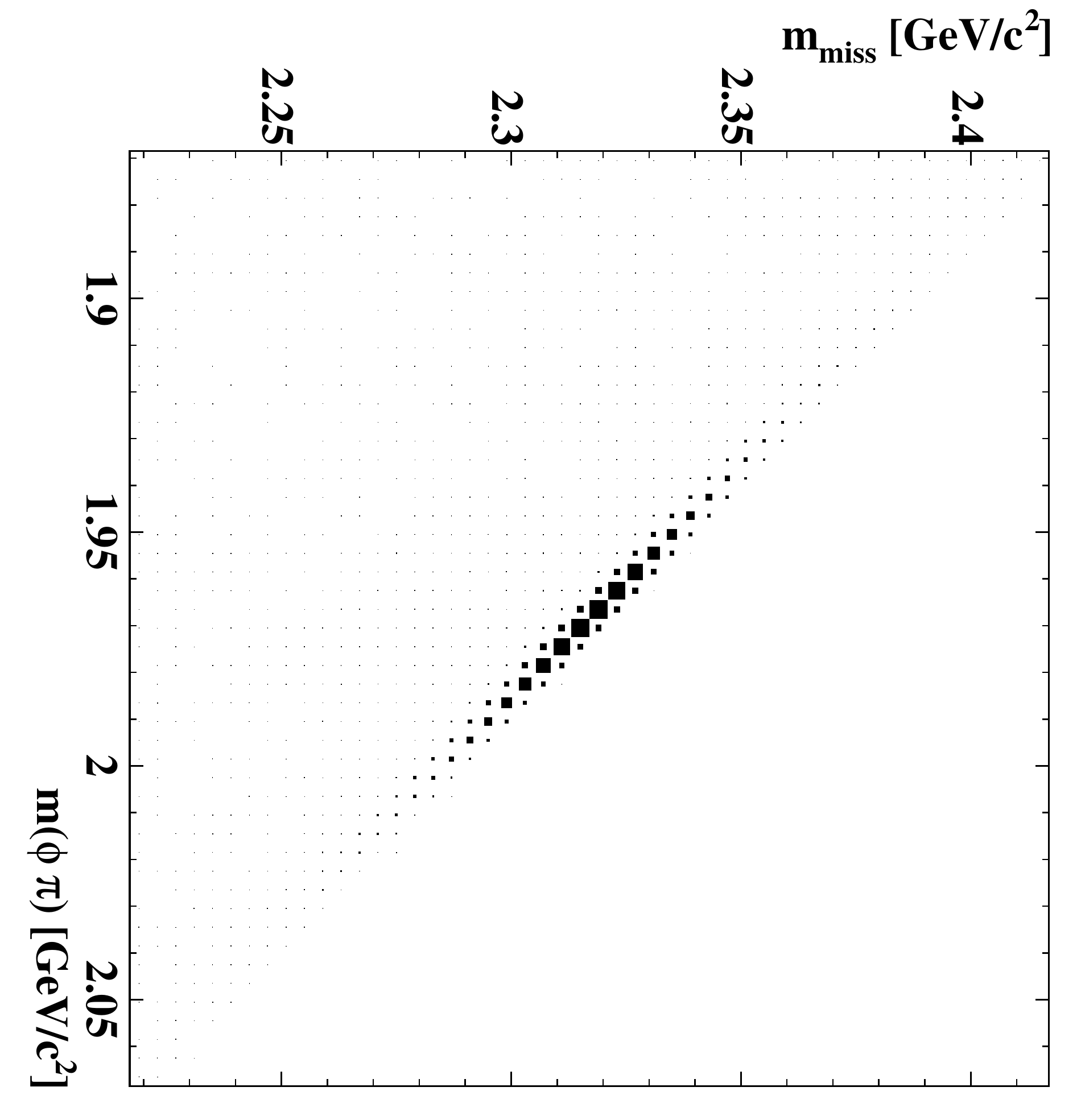}
\hfill
\includegraphics[angle=90,width=\swidth]{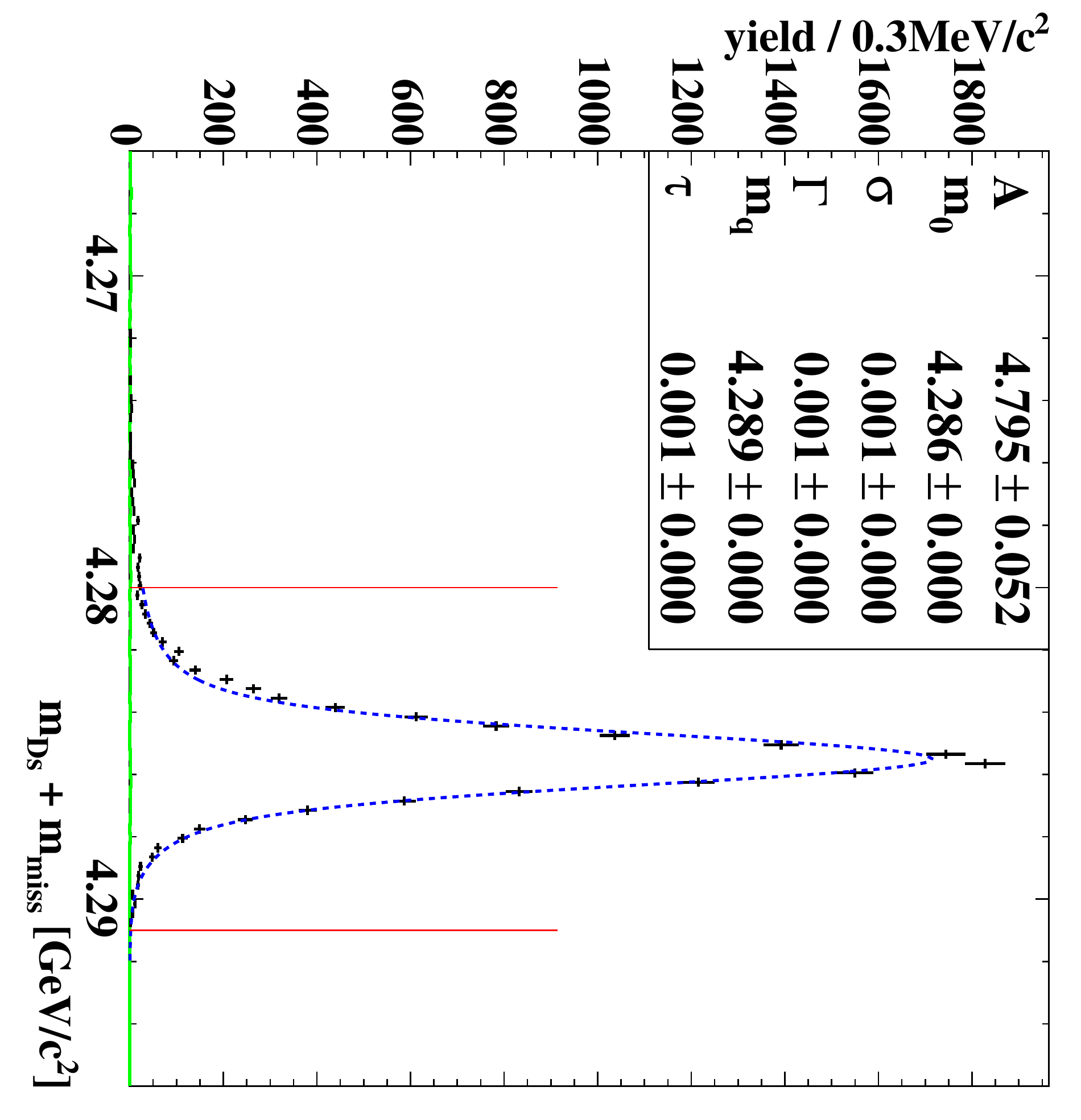}
\caption[Reconstruction of signal events type 1]{Reconstruction of
signal events type 1. (Left) Correlation of missing mass and
invariant $D_s^{\pm}$ mass. (Right) The sum mass
$m_{\mbox{\scriptsize miss}} + m(D_s^{\pm})$. The number of signal
events is evaluated as the number of entries in the interval
between the vertical red lines.}
\label{fig:DsDs0-sum}
\end{center}
\end{figure*}

In a completely exclusive analysis a 4-constraint fit to the sum
of the 4-momenta of the final state particles as given by the
4-momentum vector of the initial $\pbarp$ state in general
improves the signal quality significantly. As the $D_{s0}^\ast(2317)$
decay is not reconstructed, this is not possible here. However,
the reconstructed $D_s^{\pm}$ and $D_{s0}^\ast(2317)^{\mp}$ masses
are kinematically anti-correlated as a consequence of their
production very close to the threshold energy. This correlation,
as shown in \Reffig{fig:DsDs0-sum} (left) is exploited in
order to enhance the signal to noise ratio. The peak of the
$m_{\mbox{\scriptsize sum}} =
m(D_s^{\pm})+m(D_{s0}^\ast(2317)^{\mp})$ sum mass appears to have a
width of about 1\,\mevcc only. Therefore in this analysis the
sum mass $m_{\mbox{\scriptsize sum}}$ is used as quantity to count
the number of signal events. It will be demonstrated later that
background channels exhibit a different behaviour and can be
reasonably well separated from the signal in this projection.

The obtained $m_{\mbox{\scriptsize sum}}$ spectrum is fit with a
Voigt distribution, {\it i.e.} a convolution of a Breit-Wigner with a
Gaussian distribution, with phase space damping\footnote{This
function is given by \\
$f(m)=A\cdot\left[\int_{-\infty}^{+\infty}G(x';m_0,\sigma)\ast
BW(m-x';m_0,\Gamma)\,dx'\right]\times
\frac{1}{1+\exp\left(\frac{m-m_q}{\tau}\right)}$ with a Gaussian
$G(m)$, a non-relativistic Breit-Wigner function $BW(m)$, an
intensity parameter $A$, running variable $m$, resonance pole mass
$m_0$, resonance width $\Gamma$, reconstruction resolution
$\sigma$, phase space limit $m_q$  and decay parameter $\tau$ (all
except $A$ in [\gevcc]).}. However, due to the absence of
background within the signal events the reconstruction efficiency
can simply be determined by counting the number of events in the
range
\begin{equation}
4280 \mbox{ MeV}/c^2 < m_{\mbox{\scriptsize sum}} < 4291 \mbox{
MeV}/c^2 \label{func:signalregion}
\end{equation}
which is marked by the two vertical lines in
\Reffig{fig:DsDs0-sum} (right). For the particular example
here with \vloo kaon identification we find $S=14490$ entries
in the signal region corresponding to an efficiency of $\epsilon =
36.2\%$ An efficiency of the same magnitude is independently found
by the analysis of completely inclusive events (Signal 2), where
also the recoiling $D_s$ decays generically.

\subsubsection*{Background}

\begin{figure*}[htb]
\begin{center}
\includegraphics[angle=90,width=\swidth]{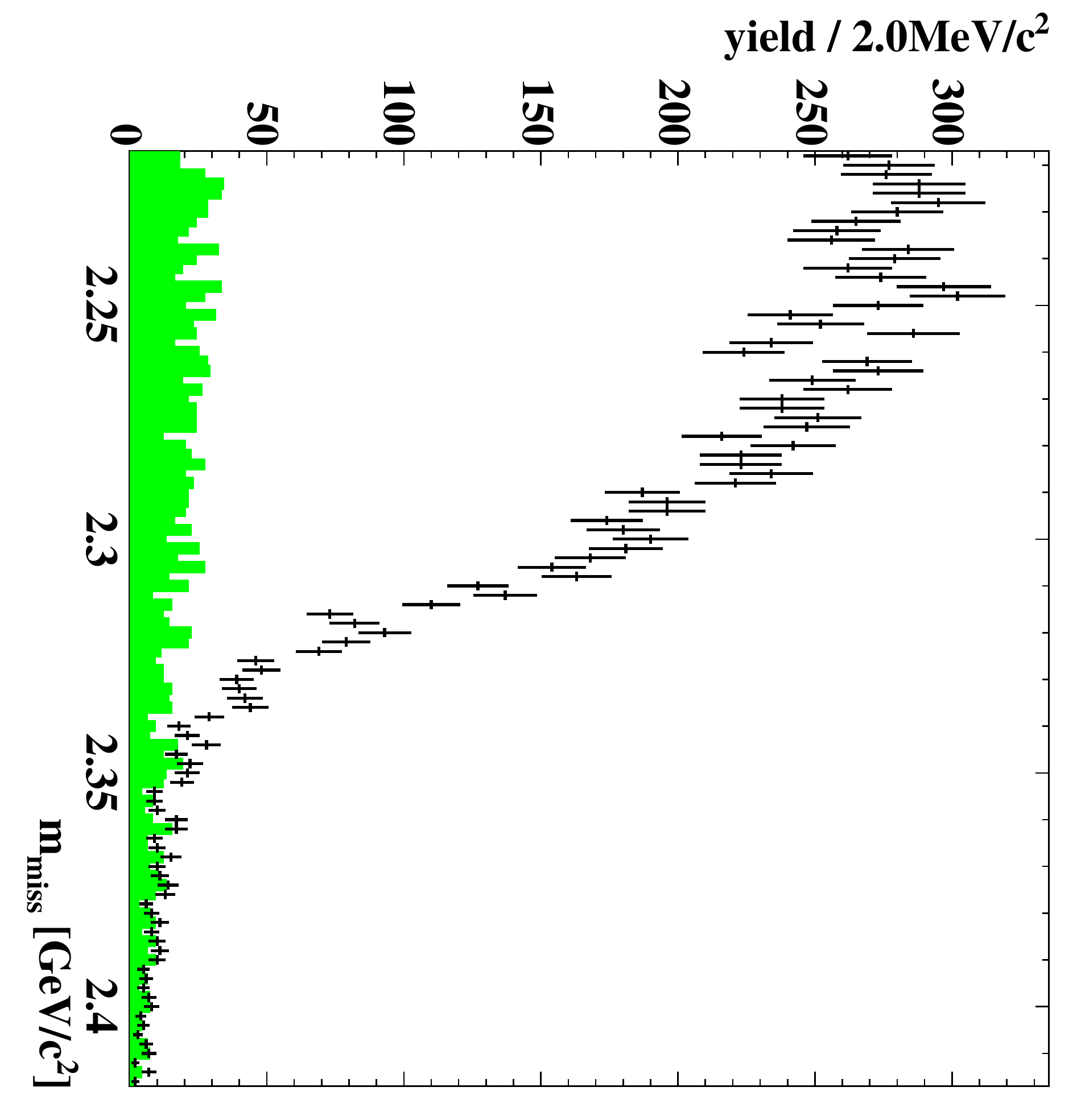}
\hfill
\includegraphics[angle=90,width=\swidth]{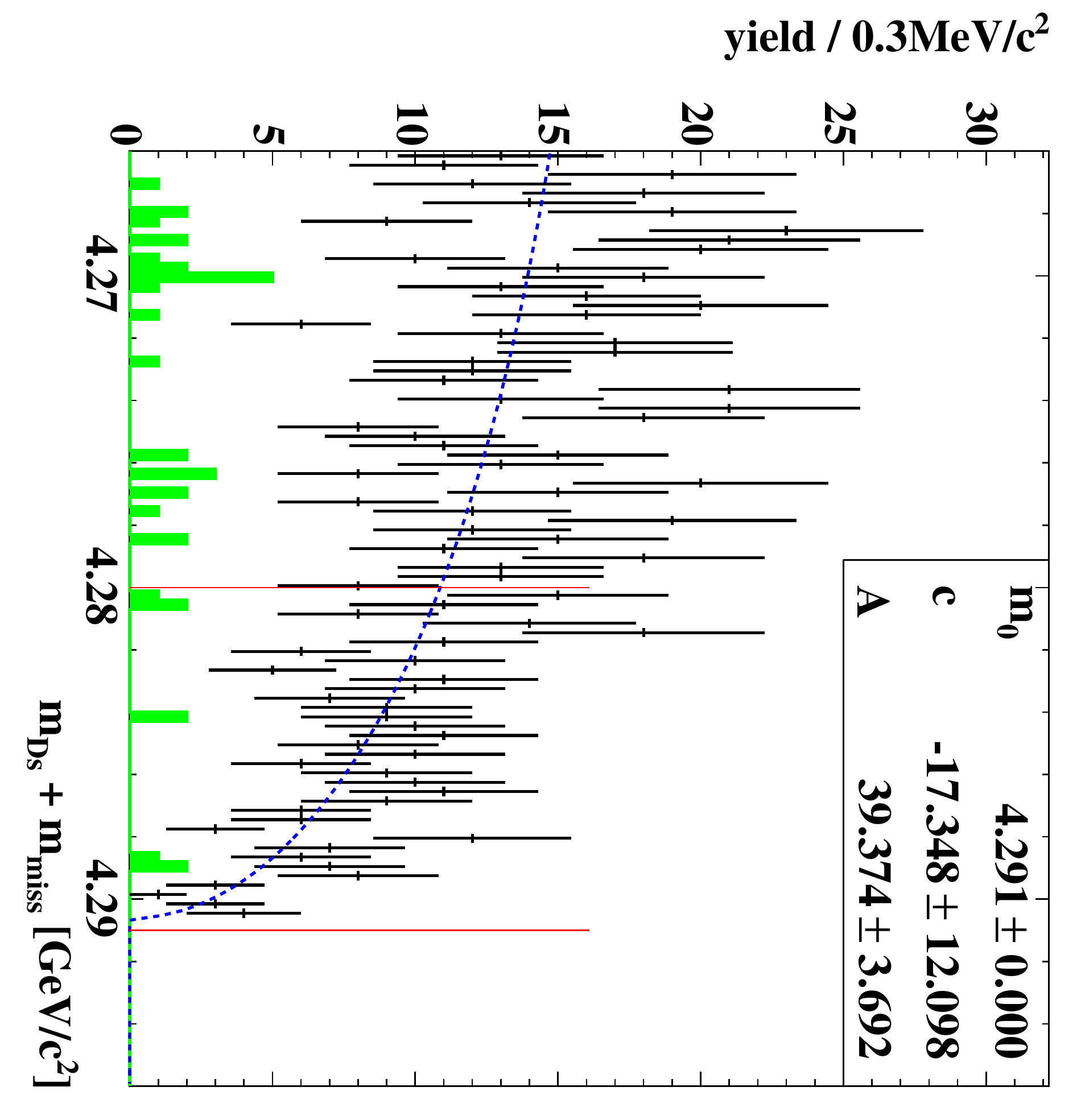}
\caption[Background channel 1:
$\pbarp\rightarrow{}D_s^{\pm}D_s^{\mp}\pi^0$.]{Background channel
1: $\pbarp\rightarrow{}D_s^{\pm}D_s^{\mp}\pi^0$. (Left)
Reconstructed $D_s^{\pm}$ missing mass distribution. (Right)
Obtained sum mass distribution: the number of background events is
evaluated in the interval $4280\,\mevcc<4291\,\mevcc$ as indicated
by the vertical red lines. The black histogram corresponds to all
reconstructed combinations of the required particles in the final
state, the green filled area represents those combinations which
fail the MCT match criterion (explanation see text).}
\label{fig:bg1}
\end{center}
\end{figure*}

Background channels as listed in \Reftbl{tab:datasets} have
been analysed. \Reffig{fig:bg1} shows as an example background
distributions due to the
$\pbarp\rightarrow{}D_s^{\pm}D_s^{\mp}\pi^0$ reaction obtained
for the \vloo kaon candidate selection. Left and right
panels of \Reffig{fig:bg1} show the reconstructed $D_s^{\pm}$
missing mass and the $m_{\mbox{\scriptsize sum}} =
m(D_s^{\pm})+m(D_{s0}^\ast(2317)^{\mp})$ sum mass distributions,
respectively. For this as well as the other specific background
channels the $m_{\mbox{\scriptsize sum}}$ distribution follows a
phase space distribution which is well described by an Argus
function~\footnote{Here the Argus function is defined as
$f_{\mbox{\scriptsize{}bg}}(m)=A_s\cdot{}m\cdot\sqrt{1-(m/m_0)^2}
\cdot\exp\left[c\cdot\left(1-(m/m_0)^2\right)\right]$ with
amplitude parameter $A_s$, phase space limit $m_0$ and shape
parameter $c$.}.

In the background sample generated with DPM - known to be
inadequate for the simulation of charmed hadron production - no
significant amount of $D_s^{\pm}$ mesons are correctly
reconstructed. With the \vloo kaon candidate selection
very few out of 5\,M generated events are found in the
$m_{\mbox{\scriptsize sum}}$ window between 4280\,\mevcc and 4291\,\mevcc.

\subsubsection*{Results}
In order to find the selection criteria for the maximum signal
significance many parameters have to be varied. At the present
stage of the study only the influence of the kaon identification
quality has been systematically analysed.

Four different selection criteria \vloo, \loo, \tig and \vtig based on a global PID likelihood
function $\LH$ are available with requirements according to tab. \ref{tab:soft:emc_pid_cuts}.

\begin{table*}[htb]
\centering
\begin{tabular}{lccccc}
\hline\hline Channel & rel. X-sec & $\epsilon(\mbox{VL})[\%]$ &
$\epsilon(\mbox{L})[\%]$ & $\epsilon(\mbox{T})[\%]$ &
$\epsilon(\mbox{VT})[\%]$\\\hline\hline Signal & 1 & 36.2 & 28.1 &
21.0 & 19.0 \\\hline
 $\pbarp\rightarrow \Dspm\Dsmp \piz$  & 1 &  0.8&   0.6&    0.5&    0.4\\
 $\pbarp\rightarrow \Dspm\Dsmp 2\piz$ & 1 & 6.9&    5.2&    4.0&    3.6\\
 $\pbarp\rightarrow \Dspm\Dsmp \pip \pim$ & 1 & 8.1&    6.1&    4.6&    4.2\\
 $\pbarp\rightarrow \Dspm\Dsxmp$ & 1 & 0.0& 0.0&    0.0&    0.0\\
 $\pbarp\rightarrow \Dspm\Dsxmp \piz$ & 1 & 3.7&    2.8&    2.1&    1.9\\
 $\pbarp\rightarrow \Dspm\Dsmp \gamma$ & 0.1 & 0.6& 0.4&    0.3&    0.3\\
 $\pbarp\rightarrow \Dspm\Dsxmp \gamma$ & 0.1 & 1.1&    0.9&    0.6&    0.6 \\
 DPM generic & $10^{6}$ & $2.5\cdot 10^{-4}$ & $4.5\cdot 10^{-5}$ & $1.9\cdot 10^{-5}$ & $1.9\cdot 10^{-5}$\\\hline\hline
 $r_{SB}$ (w/ DPM) & -- & $1:318$ & $1:74$ & $1:43$ & $1:47$ \\
 $r_{SB}$ (w/o DPM) & -- & 1.86 & 1.90 & 1.89 & 1.88 \\\hline\hline
\end{tabular}
\caption[Results of the simulation studies of signal
reconstruction and background suppression]{Results of the
simulation studies of signal reconstruction and background
suppression. Only relative cross sections are given. $\epsilon$
denotes the signal reconstruction efficiency for the signal
channel and the fake signal finding probability for the studied
background channels, respectively. The resulting values for the
signal-to-noise ratio $r_{SB}$ including or excluding the generic
DPM background are also given (see text).} \label{tab:pidsummary}
\end{table*}

All these four criteria for kaon identification were applied to
the selection procedure, resulting in different numbers of
residual candidates for signal and background in the
$m_{\mbox{\scriptsize sum}}$ signal region. The expected
signal-to-noise ratio $r_{SB}$ depending on the PID selection
criterion can only be given based on relative cross sections of
the background channels as compared to that of the signal. Since
all cross sections are unknown, as reference value for the signal
and for all purely hadronic background channels 50\,nb is assumed,
whereas the reference value is scaled down by a factor 10 to 5\,nb
for channels with photons in the final state. The $S/B$ ratio for
all considered specific background channels based on this
assumption is given in \Reftbl{tab:pidsummary}. The $S/B$ ratio
for the generic background obtained with DPM corresponds to an
assumed ratio of the background to signal cross section of $10^6$.
Using the signal selection criteria described above, 3 events out
of $10.5\cdot{}10^6$ DPM background events are found in the signal
region. For a kaon selection based on criterion \tig this
corresponds to an expected ratio of $r_{SB}\approx 1:43$.

Although this value has a large statistical uncertainty, this
indicates that it may be difficult to measure an excitation
function based on the inclusive reconstruction method described
above. Therefore, an even larger DPM background sample was also
analysed with the exclusive reconstruction of the
$D_s^{\pm}D_{s0}^\ast(2317)^{\mp}\rightarrow{}\phi\pi^{\pm}\piz\phi\pi^{\mp}$
decay chain. None of $4.0\cdot{}10^7$ DPM background events
survived the signal selection cuts using the \vloo kaon selection
criterion. Further background suppression can be expected from
using a tighter kaon selection criterion, and particularly, from a
cut on the $\bar{D}_s^+$ and $D_s^-$ decay vertices.
Unfortunately, due the small branching ratio to the final state,
the non-observation of fake events only corresponds to an
estimated lower limit of signal-to-background ratio $S/B>1:623$.
In order to obtain a more significant result on the achievable
background suppression, much larger background samples have to be
generated and analysed. In addition, the same type of exclusive
analysis needs to be repeated for different decay chains, in order
to acquire a higher signal event rate. Both tasks are beyond the
scope of the present report, and will be pursued in future
studies.


\subsubsection*{Simulation of a Near-Threshold Energy Scan}

\subsubsection*{Excitation function}

At small excess energies above threshold governed by $S$-waves the
energy dependence of the cross section for the reaction
$a+b\rightarrow{}1+2$ where the two final state particles
($i=1,2$) have Breit-Wigner spectral functions
\begin{equation}
\label{func:spectral}
\rho_i(m) = \frac{1}{\pi} \cdot
\frac{\Gamma_i/2}{(m-m_{R_i})^2 + (\Gamma_i/2)^2}
\end{equation}
with resonances pole mass and width $m_{R_i}$ and $\Gamma_i$ is
given by the integral~\cite{bib:phy:Hanhart06}
\begin{eqnarray*}
\sigma(s) = |M|^2 \int_{-\infty}^{+\infty} dm_1
\\
\int_{-\infty}^{+\infty} dm_2\,\,\rho_1(m_1)\rho_2(m_2)\cdot
p\cdot \Theta(\sqrt{s} -m_1-m_2).
\end{eqnarray*}
Here $m_1$ and $m_2$ are the running masses, $\sqrt{s}$ is the
total centre-of-mass energy, $M$ the matrix element of the
process, and $p$ the momentum of the two resonances in the
centre-of-mass frame. Substituting Breit-Wigner spectral functions
with zero width for the $D_s$ meson, and the centre-of-mass
momenta, one obtains a simplified relation for the energy
dependence (with $m_d\equiv m(D_s)$)~\cite{bib:phy:Hanhart06}:
\begin{eqnarray}
\label{func:excitation}
\frac{\sigma(s)}{|M|^2} =
\frac{\Gamma}{4\pi\sqrt{s}}\int_{-\infty}^{\sqrt{s}-m_d}
\\ \nonumber
dm\,\frac{\sqrt{\left[s-(m+m_d)^2\right]\cdot
\left[s-(m-m_d)^2\right]}}{(m-m_R)^2 + (\Gamma/2)^2}.
\end{eqnarray}

\subsubsection*{Scan procedure}
\vspace*{0.5cm}


The simulated energy scan for this write-up is based on several
simplifying assumptions.

Apart from the effects due to energy dependent shifts of the
kinematic limit in the Argus function modelling the background
distributions of the specific channels considered, the background
level is assumed to be energy independent within the small energy
range in which the excitation function is simulated. Also the
signal reconstruction efficiency is assumed to be constant for all
energy steps.

The momentum spread of the beam is only taken into account in a
simplified way, namely by an additional contribution to the energy
selected in the scan procedure. In a fully correct treatment the
effective excitation function instead consists of a convolution of
the energy spread with the theoretical excitation function for
infinitely small momentum spread.

The following parameters are selected:
\begin{itemize}
\item The width $\Gamma_{D_{s0}}$ of the $D_{s0}^\ast(2317)$.
\item The spent beam time $T_{\mbox{\scriptsize beam}}$ for the
complete measurement; a signal cross section of $\sigma_{S}=1$\,nb
at threshold energy $E_{\mbox{\scriptsize thr}}$ and an integrated
luminosity of ${\cal{}L} = 9$\,pb$^{-1}$/day is assumed.
\item The energy region $\Delta E_{\mbox{\scriptsize max}}$ below
and above the threshold to be scanned.
\item The number $n$ of the taken measurements within the energy
scan ($\rightarrow \Delta E = 2\cdot\Delta E_{\mbox{\scriptsize
max}}/n$)
\item The signal-to-noise ratio $r_{SB}$
\item The signal reconstruction efficiency
$\epsilon\cdot{}f_{{\BR},D_s}$
\end{itemize}

The number of signal events at each energy step is computed
according to
\begin{eqnarray*}
S_i = \sigma(E_i/c^2)\cdot \frac{{\cal{L}}_{\mbox{\scriptsize
tot}}}{n} \cdot \epsilon\cdot f_{{\BR},D_s} =
\\
= \sigma_{S}\cdot\frac{f_{\mbox{\scriptsize ex}}(E_i+\delta
E)}{f_{\mbox{\scriptsize ex}}(E_{\mbox{\scriptsize thr}})}
\cdot{\cal{L}}\cdot \frac{T_{\mbox{\scriptsize beam}}}{n} \cdot
\epsilon\cdot f_{{\BR},D_s}
\end{eqnarray*}
where $\delta{}E$ corresponds to the uncertainty of the beam
energy, and is randomly selected within a Gaussian distribution of
150~keV width, and $f_{\rm{}ex}$ is given by the value of the
integral in equation~\ref{func:excitation}.

The signal-to-background ratio $r_{SB}$ listed above is valid at
the highest energy $E_n$ of the scan. According to the lower
number of signal events at the lower energies within the scan the
value of $r_{SB}$ is scaled down appropriately, assuming constant
background apart from a phase space correction.

\subsubsection*{Results}

\begin{figure}[htb]
\begin{center}
\includegraphics[angle=90,width=1.1\swidth]{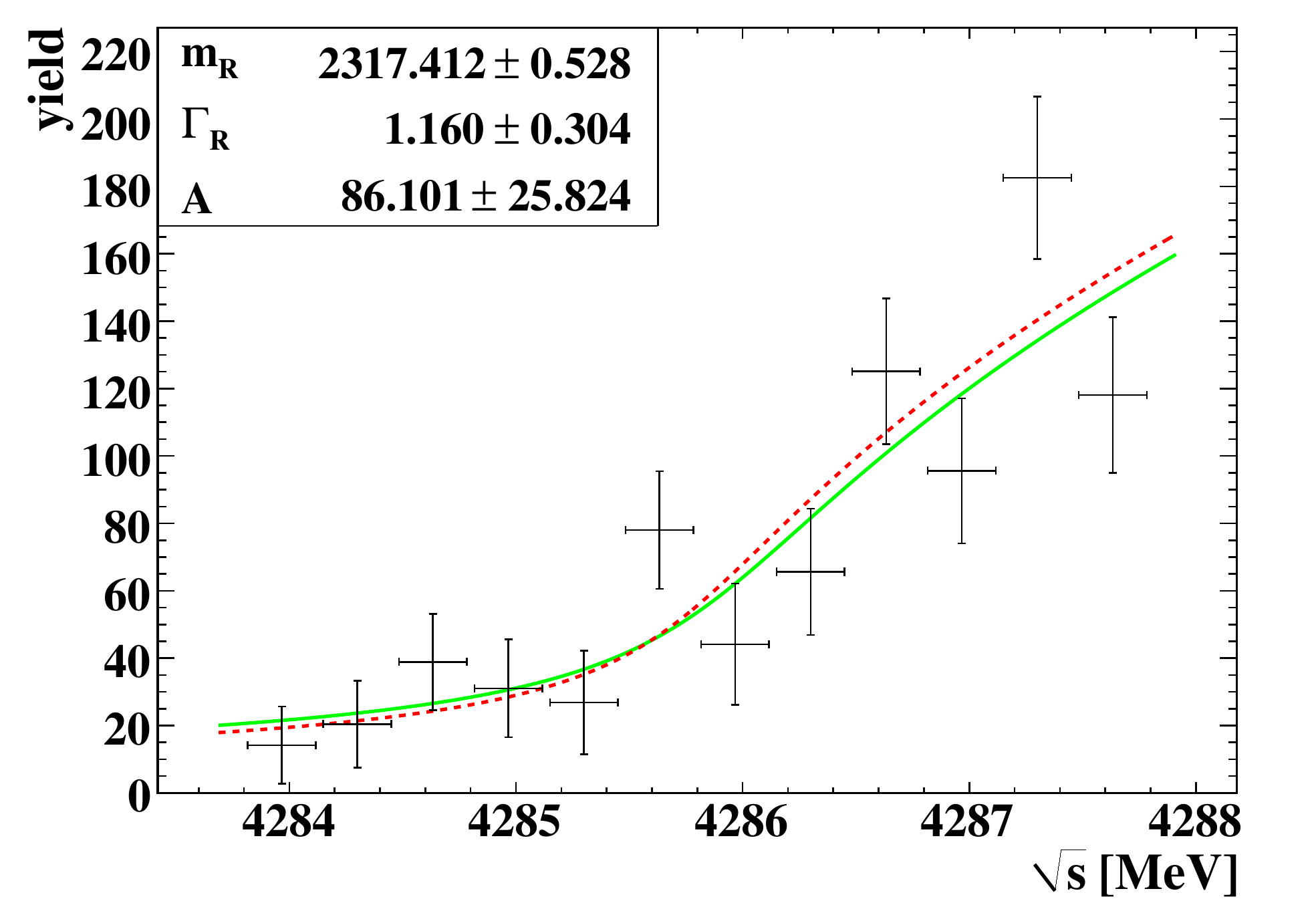}
\caption[Fit of the excitation function obtained from the
reconstructed signal events]{Fit of the excitation function
obtained from the reconstructed signal events, shown by the full
green line. The red dotted line shows the excitation function
corresponding to the generated events.}
\label{fig:scan}
\end{center}
\end{figure}

So far a few combinations of parameters have been explored in the
simulation of the energy scan of the
$\pbarp\rightarrow{}D_s^{\pm}D_{s0}^\ast(2317)^{\mp}$ reaction. No
attempt was made to find the global optimum of the procedure,
which would require a systematic investigation of a
high-dimensional parameter space. The result obtained for a
specific parameter set is shown in \Reffig{fig:scan}. The
selected parameters (scan~1) are:
\begin{itemize}
\item $T=14$ d, $r_{SB}=1:3$, $\Gamma=1$\,\mevcc, $\Delta
E_{\mbox{\scriptsize max}}=2$\,MeV, $n=12$.
\end{itemize}
For this parameter set the fit yields
\begin{equation}
m=2317.41\pm{}0.53\,\mevcc, \quad
\Gamma=1.16\pm{}0.30\,\mevcc,
\end{equation}
to be compared with the input values $m=2317.30\,\mevcc$
and $\Gamma=1.00\,\mevcc$, respectively.

The same procedure with a parameter set corresponding to a much
smaller $S/B$ ratio
\begin{itemize}
\item $T=28$\,d, $r_{SB}=1:30$, $\Gamma=0.5$\,\mevcc, $\Delta
E_{\mbox{\scriptsize max}}=1$\,MeV, $n=12$
\end{itemize}
did not allow to deduce a meaningful fit result for the
$D_{s0}^\ast(2317)$ width.

\clearpage


%% file: phys/phys_baryons.tex
%
\subsection{Strange and Charmed Baryons}
\label{sec:phys:baryons}
\COM{Author(s): A. Gillitzer}
\COM{Referee(s): M. Lutz}
\input{./phys/baryon_spectroscopy/phys_baryons_intro}
%

%% file: phys/baryon_spectroscopy/phys_baryons_intro.tex
%
\subsubsection*{Introduction}
An understanding of the baryon excitation spectrum is one of the
prime goals of non-perturbative QCD. In the nucleon sector, where
most of the experimental information is available and where the
experimental effort is still being concentrated, it is found that
the agreement with quark model predictions is astonishingly small:
some of the low-lying states are not at the energies predicted,
whereas many of the predicted higher lying states have not been
seen experimentally~\cite{PDG}. The latter aspect
has been discussed as the problem of 'missing
resonances'~\cite{bib:phy:Klempt:2002vp}, and different
explanations have been suggested. For example a quark-diquark
structure of baryons would reduce the number of internal degrees
of freedom, and thus the number of states. Another reason could be
a lack of experimental sensitivity, since resonance excitation and
detected final states have so far been largely based on pionic
modes~\cite{PDG}. It is also not clear to which
extent the observed states are three-quark excitations or governed
by meson-baryon dynamics. In the approach of chiral coupled
channel
dynamics~\cite{bib:phy:Lutz:2001yb,bib:phy:Lutz:2003jw,bib:phy:Kolomeitsev:2003kt},
based on the conjecture that the only genuine $qqq$ states are the
octet and decuplet ground states, it is attempted to describe all
excited baryon states as dynamically generated resonances.
\subsubsection*{Strange Baryons}
The question to which extent the excitation spectra of baryons
consisting of light quarks ($u,d,s$) follow the systematics of
$SU(3)$ flavor symmetry, requires knowledge not only on $N^\ast$ and
$\Delta$ spectra but also on those of all species of strange
baryons, {\it i.e.} of $\Lambda$, $\Sigma$, $\Cascade$, and $\Omega$
hyperons. However, as one adds strangeness as an additional degree
of freedom to the baryonic constituents, the experimental data
quality becomes increasingly poor. This is already the case for
the $\Lambda$ and $\Sigma$ spectrum, where recent observations of
new states~\cite{bib:phy:Zychor:2005sj,bib:phy:Adamovich:2007mc}
are waiting for confirmation and interpretation. The data base is
particularly scarce for $S=-2$ and $S=-3$ baryons. $\Cascade$ and
$\Omega$ excited states have in general been seen as bumps in
inclusive experiments only, without determination of spin and
parity quantum numbers. The 2006 edition of the Review of Particle
Physics~\cite{PDG} explicitly mentions that
''nothing of significance on $\Cascade$ resonances has been added since
the 1988 edition''. A large fraction of the data has been obtained
with low statistics in bubble chamber experiments. Apart from the
$\Cascade$ octet and decuplet ($\Cascade(1530)P_{13}$) ground states spin
and parity assignments only exist for the three-star resonances
$\Cascade(1820)D_{13}$ and $\Cascade(2030)$, but their assignment is
labelled as ''merely educated guesses'' in
~\cite{PDG}. More recent information on masses,
widths, and decay modes of the $\Cascade^0(1690)$, $\Cascade^-(1820)$, and
$\Cascade^-(1950)$ states was delivered by the WA89
experiment~\cite{bib:phy:Adamovich:1997ud,bib:phy:Adamovich:1999ic}.
The $\Cascade^0(1690)$ state was also seen in $\Lambdac^+$ decays at
\INST{Belle}~\cite{bib:phy:Abe:2001mb} and at
\INST{BaBar}~\cite{bib:phy:Ziegler07}. The latter study favours a spin 1/2
assignment to this state, and confirms the $J^P=3/2^+$ assignment
for the $\Cascade^0(1530)$ state.
\par
Almost nothing is known on the excitation spectrum of the $\Omega$
baryon: no assignment exists for any of the three seen excited
states (one three-star, two two-star resonances).
\Reftbl{tab:PDG_table} gives an overview of the assignment of
known baryonic states in the light quark
sector~\cite{PDG}.
\par
Recent theoretical work on the $\Cascade$ and $\Omega$ spectrum is
found in Refs~\cite{bib:phy:Loring:2001ky,bib:phy:Guzey:2005vz}.
Due to the lack of experimental data, most of the calculated
states have no experimental counterpart and their existence needs
verification. Ref.~\cite{bib:phy:Guzey:2005vz} also estimates
two-body decay widths, and obtains values of less than 50\,MeV for
some of the states.

\begin{table}[htb]
\begin{center}
\includegraphics[width=\swidth]{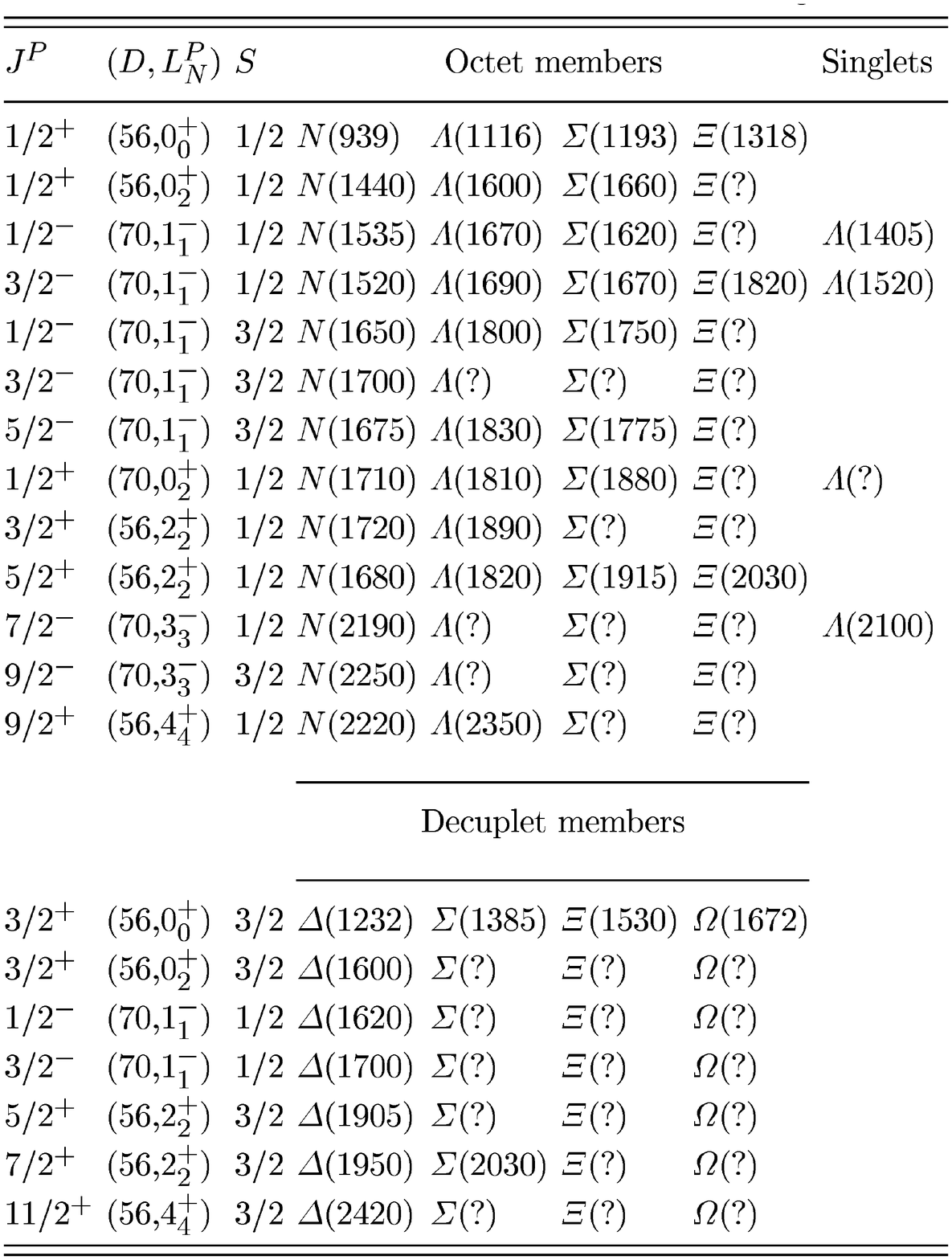}
\caption[Quark-model assignments for some of the known baryon
states]{Quark-model assignments for some of the known baryon
states in flavor-spin SU(6) basis. Part of the spin-parity
assignments are not well established and need confirmation (taken
from ~\cite{PDG}).
\label{tab:PDG_table}}
\end{center}
\end{table}

\subsubsection*{Baryon Spectroscopy with \PANDA}
The \PANDA experiment is well-suited for a comprehensive baryon
spectroscopy program, in particular in the spectroscopy of
(multi-)strange and possibly also charmed baryons. In $\pbarp$
collisions, a large fraction of the inelastic cross section is
associated with channels resulting in a baryon antibaryon pair in
the final state. As an example, at 3\,\gevc $\pbar$ momentum the
total $\pbarp$ cross section is 77\,mb, the inelastic cross
section is 53\,mb, with a one to one ratio of baryonic final states
and of annihilation into mesons. At higher $\pbar$ momenta the
yield of channels with baryonic final states exceeds that of the
mesonic channels, {\it e.g.} at $p_{\pbar}=12$\,\gevc the ratio is
$\sim$2.2. To a large extent reactions with baryonic final states
proceed via excited states giving access to the decay modes of the
populated resonances and to the angular distributions of the decay
particles. A particular benefit of using antiprotons in the study
of (multi-)strange and charmed baryons is that in $\pbarp$
collisions no production of extra kaons or $D$ mesons is required
for strangeness or charm conservation, respectively. This reduces
the energy threshold as {\it e.g.} compared to $pp$ collisions and thus
the number of background channels. In addition, the requirement
that the patterns found in baryon and antibaryon channels have to
be identical reduces the systematic experimental errors. Strange,
multi-strange and charmed baryons are characterised by their or
their daughters' displaced decay vertices, which can be identified
thanks to the good tracking capability of the \PANDA tracking
detectors (MVD, central tracker, tracking detectors of the forward
spectrometer).
\par
Production cross sections for $\Cascade$ resonances are expected to be
of the same order as for ground state $\Cascade$ production, {\it i.e.} for
the reaction $\pbarp\rightarrow{}\CascadeCascadebar$ for which a cross
section up to 2\,$\mu$b has been
measured~\cite{bib:phy:Baltay65,bib:phy:Musgrave65}. The
$\Cascade^\ast(\Cascadebar^\ast)$ yields will thus be sufficiently high to
allow good statistics studies analysing the various $\Cascade^\ast$ decay
modes such as $\Cascade\pi$, $\Cascade\pi\pi$, $\Lambda\bar{K}$,
$\Sigma\bar{K}$, $\Cascade\eta$, and others. Given the extremely scarce
experimental information on the $\Cascade$ excitation spectrum
available, the discovery potential of \PANDA seems to be
particularly large for $\Cascade$ resonances. One should also note that
$\Cascade$ resonances are in general much narrower than nucleon or
$\Delta$ resonances which helps to separate contributions of
different states.
\par
The very poorly known $\Omega$ spectrum can also be studied,
however the creation of an additional $s\bar{s}$ pair has to be
paid by a reduction of the cross section. No experimental data
exist for the reaction $\bar{p}p\rightarrow{}\OmegaOmegabar$,
its predicted maximum cross section is
$\sim$2\,nb~\cite{bib:phy:Kaidalov94}, to our knowledge the
only existing theoretical estimate. Some confidence in this
prediction may be drawn from the consistency of calculated cross
sections with experimental data for other binary reactions in
$\pbarp$ collisions~\cite{bib:phy:Kaidalov94}. At a
luminosity of $10^{32}$\,cm$^{-2}$s$^{-1}$ a cross section of
2\,nb would still correspond to $\sim$700 produced
$\OmegaOmegabar$ pairs per hour which allows to identify
excited $\Omega$ states and their most important decay modes.
\par
For all non-charmed baryons the \INST{HESR} energy range is sufficient to
access excitation energies up to the continuum regime, and thus to
populate the complete discrete part of the spectrum. Depending on
the hyperon resonances and their decay modes to be studied, the
$\pbar$ beam momentum should be chosen such that the excess
energy above the threshold for the respective final state is as
low as possible in order to limit the number of partial waves  and
to facilitate the separation of different resonances.
\par
Also for $\pbarp\rightarrow{}\bar{\Lambda}_c\Lambdac,
\bar{\Sigma}_c\Sigma_c,
\bar{\Lambda}_c\Sigma_c/\Lambdac\bar{\Sigma_c}$ reactions no
experimental data exist. For the $\bar{\Lambda}_c^-\Lambdac^+$
final state ~\cite{bib:phy:Kaidalov94} predicts a cross
section up to 0.2\,$\mu$b which is much larger than for
$\OmegaOmegabar$. However, the decay length of the
$\Lambdac^+$ hyperon is only $c\tau(\Lambdac^+)=60\,\mu$m, which
is too short to be detected by its displaced vertex. It has only
few percent branching for channels with a $\Lambda$ hyperon in the
final state which could be easily identified by its delayed decay.
Thus, in order to estimate the capability of \PANDA to identify
final states with charmed baryons, detailed simulations of these
channels are being planned. One should also note that for charmed
baryon resonances the range of excitation energies accessible is
restricted due to the kinematic limit at the \INST{HESR} of
$\sqrt{s}=5.5$\,GeV, which allows to populate excitation energies
up to 0.93\,GeV and 0.76\,GeV above the $\Lambdac$ and $\Sigma_c$
ground states, respectively.
\par
Recent theoretical studies using a chiral coupled channel
approach~\cite{bib:phy:Hofmann:2005sw,bib:phy:Lutz:2006ya} predict
the existence of narrow crypto-exotic baryon resonances with
hidden charm. In particular, these calculations find a narrow
resonance at 3.52\,\gevcc being a coupled-channel bound state of
$\eta_c{}\Lambda$ and $\bar{D}\Sigma_c$ which should dominantly
decay to $\eta'{}N$. Whereas the exotic or crypto-exotic baryon
resonances for systems with open charm ranging from $C=-1$ to
$C=+3$ also found as dynamically generated states in the same
approach~\cite{bib:phy:Hofmann:2005sw,bib:phy:Lutz:2006ya} are not
accessible within the HESR energy range, we see a good perspective
to confirm or to rule out the existence of narrow crypto-exotic
baryons with hidden charm in the mass range between 3\,\gevcc
and 4\,\gevcc in the \PANDA experiment.
\subsubsection*{Benchmark Channels}
In the context of baryon spectroscopy, the reaction
\begin{equation*}
\bar{p}p\rightarrow\Cascadebar^+\Cascade^-\piz
\end{equation*}
with
\begin{equation*}
\Cascade^-\rightarrow\Lambda\pim, \Lambda\rightarrow{}p\pim
\end{equation*}
(and {\it c.c.}) has been chosen as benchmark channel for the
simulation studies. As a first step, the events have been
generated isotropically over the phase space. The goal of these
simulation studies is to reconstruct $\Cascade^-\piz$ pairs (and
c.c.) as one of the daughter states in the decay of $\Cascade$
resonances, to deduce the acceptance function of the \PANDA
detector - and thus the capability to determine the population of
the three-body final state across the full Dalitz plot, and to
explore the suppression of the presumed dominant background
channels. The capability of identifying specific $\Cascade$
resonances with their quantum numbers in the presence of a
continuum distribution and other $\Cascade$ resonances in the same
final state involves a partial wave analysis, which is beyond the
scope of the studies for this report.

\begin{figure}[htb]
\begin{center}
\includegraphics[width=\swidth]{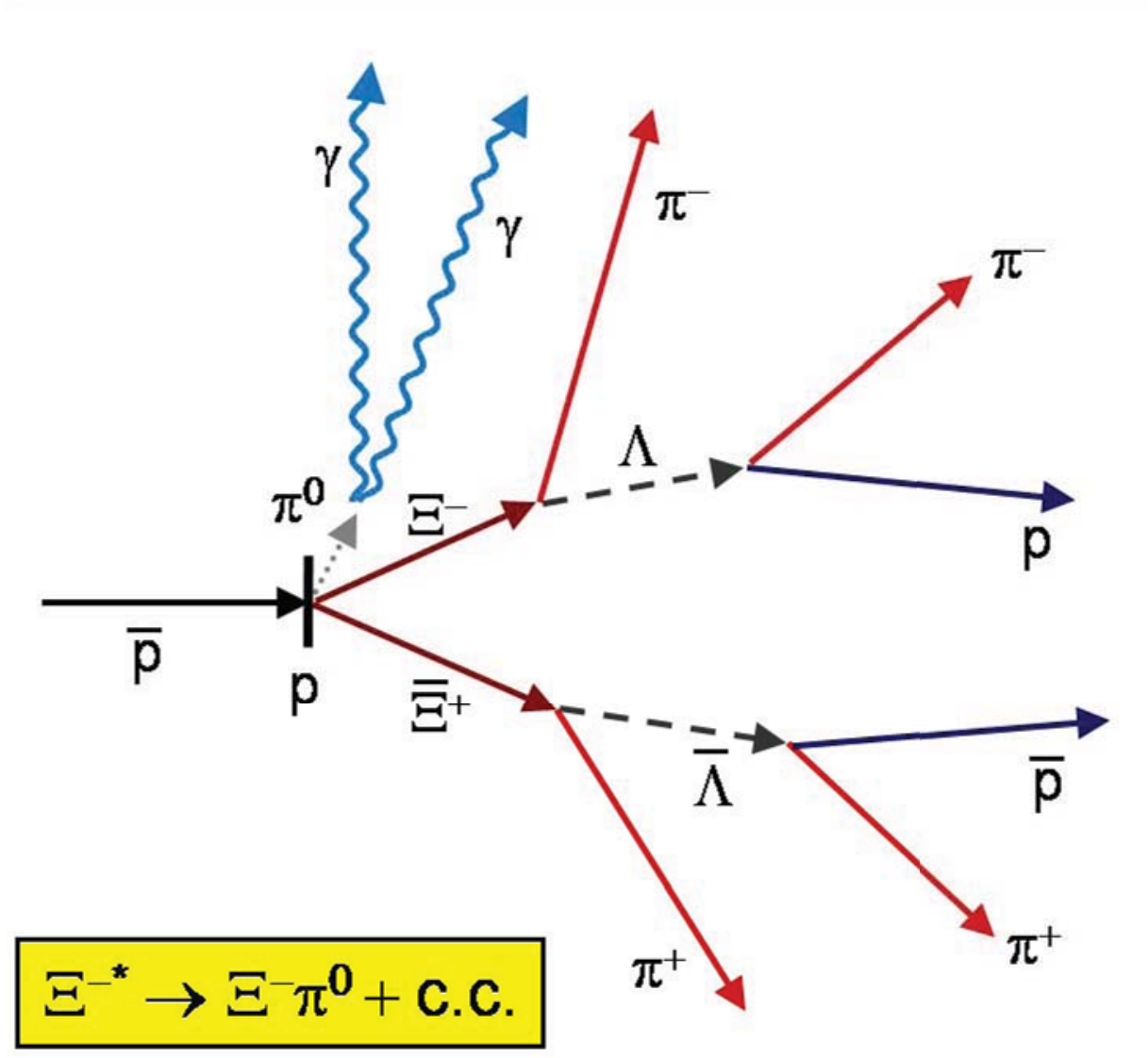}
\caption[Schematic illustration of the investigated
$\bar{p}p\rightarrow{}\Cascadebar^+\Cascade^-\piz$
reaction]{Schematic illustration of the investigated
$\bar{p}p\rightarrow{}\Cascadebar^+\Cascade^-\piz$ reaction with
the considered decay branches. The reaction is characterised by
four delayed decay vertices.}
\label{fig:XiXipi-reaction}
\end{center}
\end{figure}

In order to take into account possible interactions of the charged
$\Cascade$ baryons with the detector material and their bending in the
magnetic field within their propagation before decay
($c\tau=4.9$\,cm), their decays are not generated on event
generator level but within the \INST{GEANT4} detector simulation.
\par
The selected antiproton beam momentum is $p_{\pbar}=6.57\,\gevc$
corresponding to a maximum $\Cascade\pi$ invariant mass of
2.45\,\gevcc. For the $\bar{p}p\rightarrow{}\CascadeCascadebar$
reaction at this incident momentum
Ref.~\cite{bib:phy:Kaidalov94} predicts a cross section of
about $0.3\,\mu$b. In the following this value is also used as
estimate for the $\bar{p}p\rightarrow{}\CascadeCascadebar\piz$
reaction, taking into account that the predicted
$\CascadeCascadebar$ cross section is below the measured value.

\subsubsection*{Background Reactions}

The following background channels are considered:
\begin{itemize}
\item [(a)]
$\bar{p}p\rightarrow\bar{\Lambda}\Lambda\pip\pim\piz\rightarrow\bar{p}p\pip\pim\pip\pim\piz$
\item [(b)]
$\bar{p}p\rightarrow\bar{\Sigma}(1385)^+\Sigma(1385)^-\piz\rightarrow\bar{\Lambda}\Lambda\pip\pim\piz\rightarrow\bar{p}p\pip\pim\pip\pim\piz$
\item [(c)] $\bar{p}p\rightarrow\bar{p}p\pip\pim\pip\pim\piz$
\item [(d)] DPM generic background
\end{itemize}
The background channel (a) is expected to be the main background
source since it has the identical final state as the signal, and
since it also has the same intermediate state characterised by a
$\Lambda\bar{\Lambda}$ pair. This background will be suppressed by
requiring the $\Cascade$ and $\Cascadebar$ delayed decays
according to a decay length $c\tau=4.9$\,cm visible in a kink in
the charged particle track. In addition good $\Lambda\pim$
($\bar{\Lambda}\pip$) invariant mass resolution helps to
distinguish the $\Cascade^-$ ($\Cascadebar^+$) mass peak from the
$\Lambda\pim$ ($\bar{\Lambda}\pip$) continuum. The cross section
for this channel is not known. For the reactions
$\bar{p}p\rightarrow{}\bar{\Lambda}\Lambda\pi^+\pi^-$ and
$\bar{p}p\rightarrow{}\bar{\Lambda}\Lambda{}2\pi^+{}2\pi^-$
Ref.~\cite{bib:phy:Flaminio84} lists a measured cross section of
$(59\pm{}12)\,\mu$b and $(8\pm{}4)\,\mu$b at
$p_{\bar{p}}=6.93$~GeV/$c$, respectively. In the following a cross
section of $70\,\mu$b for process (a) is assumed.
\par
In reality, intermediate and final states of channel (a) are
expected to have some contribution from $\Sigma(1385)^-$ (and
c.c.) production according to channel (b). Even if this fraction
is small, it might be of concern due to the relatively close
masses of $\Sigma(1385)^-$ and $\Cascade^-$ to be reconstructed
from $\Lambda\pim$ pairs (and c.c.). The cross section for this
channel is also not known. For the reaction
$\bar{p}p\rightarrow{}\bar{\Sigma}(1385)^-\Sigma(1385)^+$
Ref.~\cite{bib:phy:Flaminio84} lists a measured cross section of
$(14\pm{}3)\,\mu$b at $p_{\bar{p}}=5.7$~GeV/$c$. In the following
a cross section of $20\,\mu$b for process (b) is assumed.
\par
Channel (c) is expected to have the largest cross section of the
specific background reactions considered here, however it has no
delayed decays and should be suppressed very efficiently by
requiring delayed $\Lambda$ and $\bar{\Lambda}$ vertices. For this
channel Ref.~\cite{bib:phy:Flaminio84} lists a cross section of
$(280\pm{}30)\,\mu$b at $p_{\bar{p}}=6.94$~GeV/$c$. In the
following a cross section of $300\,\mu$b for process (c) is
assumed.

\subsubsection*{Analysis Strategy}

The following selection criteria are chosen in order to
discriminate signal and background events:
\begin{enumerate}
\item $p$ and $\pim$ ($\bar{p}$ and $\pip$) are fitted to a common
vertex with $\chi$ probability $P>0.001$. Proton candidates are
selected from charged tracks with {\tt{}VeryLoose PID} criteria,
pion candidates from all charged tracks.
\item The mass of $\Lambda$, $\bar{\Lambda}$ candidates has to
fulfil the condition $1.105\,\gevcc<m_{\Lambda}<1.125\,\gevcc$.
\item $\Lambda$ and $\pim$ ($\bar{\Lambda}$ and $\pip$) are fitted
to a common vertex with $\chi$ probability $P>0.001$.
\item The mass of $\Cascade^-$, $\Cascadebar^+$ candidates has to
fulfil the condition $1.31\,\gevcc<m_{\Cascade}<1.33\,\gevcc$.
\item $2\gamma$ candidates with $E>25$\,MeV each are combined to a
$\piz$ candidate within a mass window
$110\,\mevcc<m_{\piz}<160\,\mevcc$.
\item $\Cascade^-$ and $\Cascadebar^+$ are fitted to a common vertex with
the assumption that the $\piz$ is emitted from the same vertex.
In addition constraints on the total 4-momentum are set.
Combinations with $\chi$ probability $P<0.001$ are rejected.
\item Only events containing exactly one combination of particles
fulfilling all criteria listed above are further considered.
\item The $\Cascade^-$, $\Cascadebar^+$ decay vertices have to
fulfil the condition that the sum of their distances from the
interaction point is larger than 2\,cm.
\item The complete event is refitted with mass constraints on the
$\Cascade^-$, $\Cascadebar^+$, $\Lambda$, $\bar{\Lambda}$, and
$\piz$.
\end{enumerate}

\subsubsection*{Simulation Results}
\label{sec:phys_baryons_sim}

\vspace*{1cm}
For the signal and all background channels the
$\pbarp\pip\pim\pip\pim\piz$ final state has been investigated.
The size of the analysed signal and background samples is given in
\Reftbl{tab:bg-samples}.
\begin{table}[htb]
\centering
\begin{tabular}{lr}
\hline\hline
Channel & Number of events\\
\hline
 $\pbarp\rightarrow{}\bar{\Xi}^+\Xi^-\pi^0$ & $8.13\cdot{}10^5$ \\
 $\pbarp\rightarrow{}\bar{\Lambda}\Lambda\pip\pim\piz$ & $8\cdot{}10^5$ \\
 $\pbarp\rightarrow{}\bar{\Sigma}(1385)^-\Sigma(1385)^+\piz$ & $8\cdot{}10^5$ \\
 $\pbarp\rightarrow{}\pbarp\pip\pim\pip\pim\piz$ & $3\cdot{}10^5$ \\
 DPM generic & $2\cdot{}10^7$ \\
\hline\hline
\end{tabular}
\caption[Number of events simulated for signal and background
channels]{Number of events simulated for signal and background
channels. For the signal and the specific background channels the
number refers to the $\pbarp\pip\pim\pip\pim\piz$.}
\label{tab:bg-samples}
\end{table}

\Reftbl{tab:Xi-vertex-cut} shows the signal efficiency and the
number of remaining background events depending of the cut
condition on the sum of the distances of the $\Cascade$ and
$\Cascadebar$ decay vertices from the interaction point
$D_{\CascadeCascadebar}^{(IP)}=
{\rm{}dist}(\Cascadebar-{\rm{}IP})+{\rm{}dist}(\Cascade-{\rm{}IP})$.
The selection $D_{\CascadeCascadebar}^{(IP)}>2\,$cm is used for
further analysis. With this selection the signal reconstruction
efficiency is about 16\%. Taking into account a luminosity of
${\cal{}L} = 9000/\mbox{nb}$ per day, a cross section
$\sigma=0.3\,\mu$b, and the branching fraction of the final state,
this corresponds to $1.7\cdot{}10^5$ reconstructed signal events
per day.

Based on the selection criteria, 4 remaining events are found in
the background channel
$\pbarp\rightarrow{}\bar{\Lambda}\Lambda\pip\pim\piz$, whereas no
events survive the selection criteria in the other specific
background channels and in the DPM generic background sample.
Based on the cross sections assumed for signal and background
channels as given above, the resulting values for
signal-to-background ratio are much larger than one in all cases:
\begin{itemize}
\item [(a)]
$S/B=135$ for
$\pbarp\rightarrow{}\bar{\Lambda}\Lambda\pip\pim\piz$
\item [(b)]
$S/B>1896$ for
$\pbarp\rightarrow{}\bar{\Sigma}(1385)^-\Sigma(1385)^+\piz$
\item [(c)]
$S/B>47$ for $\pbarp\rightarrow{}\pbarp\pip\pim\pip\pim\piz$
\item [(d)]
$S/B>19$ for DPM generic.
\end{itemize}

\begin{table*}[!tb]
\centering
\begin{tabular}{lccc}
\hline\hline
Channel & $\epsilon{}(D_{\CascadeCascadebar}^{(IP)}>2\,$\,cm
 & $\epsilon{}(D_{\CascadeCascadebar}^{(IP)}>4\,$\,cm
 & $\epsilon{}(D_{\CascadeCascadebar}^{(IP)}>6\,$\,cm \\
\hline
 $\pbarp\rightarrow{}\CascadeCascadebar\piz$ & 15.8\percent & 15.0\percent & 13.9\percent \\
\hline\hline
 &  $N_B{}(D_{\CascadeCascadebar}^{(IP)}>2\,$\,cm
 &  $N_B{}(D_{\CascadeCascadebar}^{(IP)}>4\,$\,cm
 &  $N_B{}(D_{\CascadeCascadebar}^{(IP)}>6\,$\,cm \\
 \hline
 $\pbarp\rightarrow{}\bar{\Lambda}\Lambda\pip\pim\piz$ & 4
 & 2 & 1 \\
 $\pbarp\rightarrow{}\bar{\Sigma}(1385)^-\Sigma(1385)^+\piz$ &
 0 & 0 & 0 \\
 $\pbarp\rightarrow{}\pbarp\pip\pim\pip\pim\piz$ & 0 & 0 & 0 \\
 DPM generic & 0 & 0 & 0\\
\hline\hline
\end{tabular}
\caption[Signal efficiency and remaining number of background
events]{Signal efficiency and remaining number of background
events depending on the cut condition on the $\Cascade$ and
$\Cascadebar$ decay vertices (explanation see text).}
\label{tab:Xi-vertex-cut}
\end{table*}

\begin{figure}[htb]
\begin{center}
\includegraphics[width=\swidth]{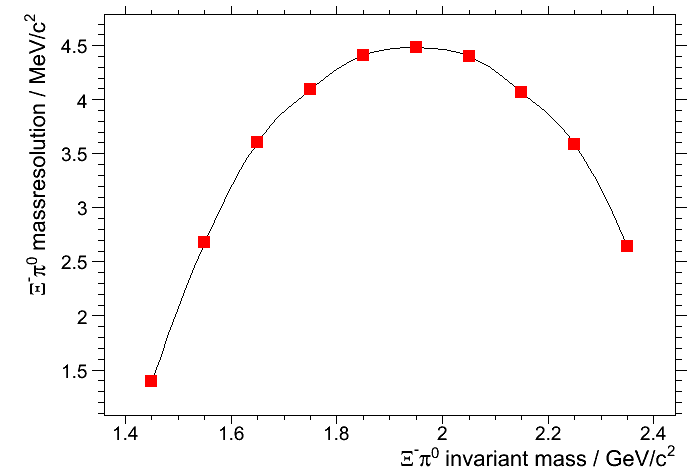}
\caption[$\Cascade^-\piz$ invariant mass resolution]{Resolution
achieved in the invariant mass of reconstructed $\Cascade^-\piz$
pairs as a function of the $\Cascade\piz$ invariant mass.}
\label{fig:Xipi-resol}
\end{center}
\end{figure}

\begin{figure}[htb!]
\begin{center}
\includegraphics[width=\swidth]{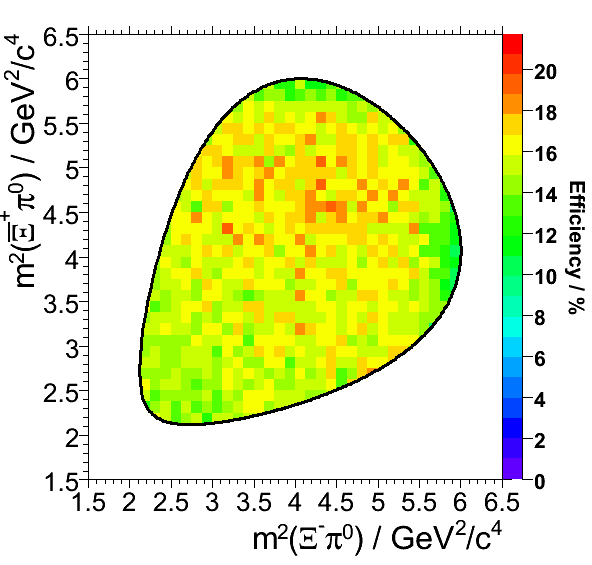}
\caption[Dalitz plot showing the $\Cascadebar^+\Cascade^-\piz$
reconstruction efficiency]{Dalitz plot showing the reconstruction
efficiency for the
$\pbarp\rightarrow{}\Cascadebar^+\Cascade^-\piz$ reaction.}
\label{fig:XiXipi-dalitz}
\end{center}
\end{figure}

\Reffig{fig:Xipi-resol} shows the invariant mass resolution in the
reconstruction of a $\Cascade^-\piz$ pair which is relevant for
the determination of width and pole position of $\Cascade$
resonances as a function of the $\Cascade^-\piz$ invariant mass.
\Reffig{fig:XiXipi-dalitz}, showing the ratio of the number of
reconstructed events relative to that of the generated Monte Carlo
events in a Dalitz plot of the $\Cascadebar^+\Cascade^-\piz$ final
state, demonstrates that the reconstruction efficiency varies
smoothly across the 3-body phase space at average values of
$\sim{}15$\,\%.
\clearpage

%% file: phys/phys_qcd_dynamics.tex
%
\section{Non-perturbative QCD Dynamics}
%

\input{./phys/qcd_dynamics/NP-QCD_intro.tex}

%% file: phys/qcd_dynamics/NP-QCD_intro.tex
%

%
An effective description of reactions in hadron physics relies on the
identification of the relevant degrees of freedom. At highest energies
quark and gluon degrees of freedom seem to describe the observed
reactions very accurately. The energy regime for $\pbarp$
collisions at \HESR is well suited to study the onset of hadron degrees
of freedom. In this regime both {\it ans\"atze} are viable. Thus they
can be experimentally tested separately and compared to each other.

The simplest final state in proton antiproton collisions is two
mesons. Depending on the momentum transfer (Mandelstam variable
$t$), the scattering amplitude is purely hadronic (small $t$) or
is better described in terms of the quark content of the hadrons.
Indeed the success of the dimensional quark counting rules for
large angle exclusive scattering suggests that the amplitude
factorises in a short distance quark subprocess and light cone
wave functions of the lowest Fock state of the participating
hadrons. Whereas experimental data exist for proton proton elastic
scattering, not much is known on all the channels which are open
in antiproton proton scattering. Moreover the coexistence of two
mechanisms as indicated by the proton proton data is an unsettled
question in the other cases. It is very likely that the transition
from one mechanism to another occurs in the Panda energy range.

In the quark picture hyperon pair production either involves the
creation of a quark-antiquark pair or the knock out of such pairs out
of the nucleon sea.  Hence, the creation mechanism of quark-antiquark
pairs and their arrangement to hadrons can be studied by measuring the
reactions of the type $\pbarp \to \overline{Y} Y$, where $Y$
denotes a hyperon.
By comparing several\
reactions involving different quark flavours the OZI
rule~\cite{bib:phy:OZI1,bib:phy:OZI2,bib:phy:OZI3}, and its possible
violation, can be tested for different levels of disconnected
quark-line diagrams separately.

The parity violating weak decay of most ground state hyperons
introduces an asymmetry in the distribution of the decay particles. This
is quantified by the decay asymmetry parameter and gives access to
spin degrees of freedom for these processes, both to the
antihyperon/hyperon polarisation and spin correlations.  One open
question is how these observables relate to the underlying degrees of
freedom.

All strange hyperons, as well as single charmed hyperons are
energetically accessible in $\pbarp$ collisions at \HESR. A
systematic investigation of these reactions will bring new information
on single and multiple strangeness production and its dependence on
spin observables. This is particularly true above 2\,$\gevc$ where
practically nothing is known about the differential distributions and
spin observables. The large amount of observables accessible and high
statistics \PANDA data will allow for a partial wave analysis. Thus it
will be possible to pin down relevant quantum numbers, coupling
constants and possibly find new resonances.

\subsection{Previous Experiments}

\begin{figure*}[htb]
 \includegraphics[width=\dwidth]{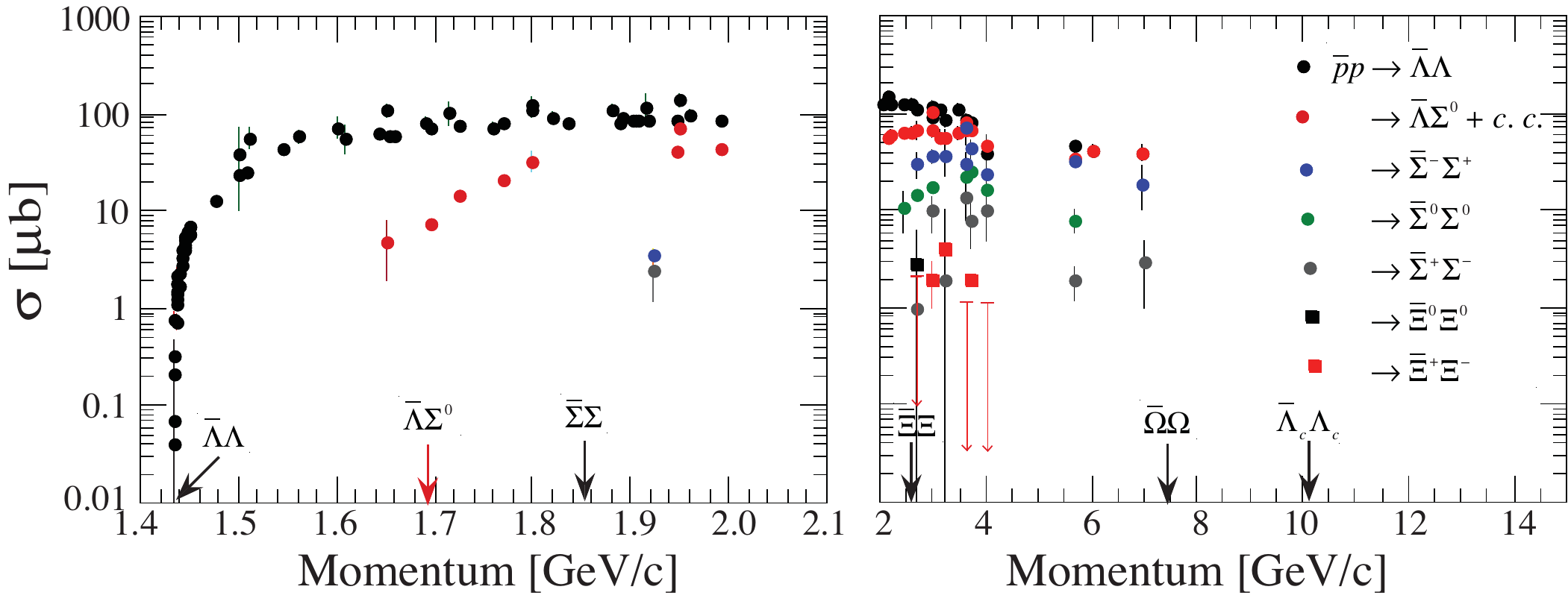}
 \caption[Total cross sections for the $\pbarp \to \overline{Y}
  Y$ reaction.]
  {Total cross sections for the $\pbarp \to \overline{Y}
  Y$ reaction in the momentum range of the \HESR. The figure to the
  left is an expanded view of the threshold region, which reveals
  channels of single strangeness production. The figure to the right
  shows the experimental situation for momenta above
  2\,GeV/c~\cite{bib:phy:PS185, bib:phy:Flaminio84}. The absence of error bars
  for some $\Cascadebar^+\Cascade^-$ points are because they are missing in
  reference~\cite{bib:phy:Baltay65}.
  The upper limits in red also refer to this channel. The arrows pointing to the momentum
  axis indicate the threshold momenta for the different hyperon families.}
 \label{fig:X-sec_all}
\end{figure*}

The $\pbarp \to \Lambdabar \Lambda$ process can be
considered as a prototype reaction in the study of production of
strangeness. This reaction exhibits strong polarisation phenomena and
spin correlation parameters can be extracted when both the
$\Lambdabar$ and $\Lambda$ are reconstructed. The PS185
experiment  at \INST{LEAR}has provided high quality data on the $\pbarp \to
\LambdaLambdabar$ reaction from threshold (1.436\,$\gevc$) up to
2\,$\gevc$~\cite{bib:phy:PS185} which was the maximum momentum of \INST{LEAR}
(see \Reffig{fig:X-sec_all}). The data above 2\,$\gevc$ are dominated
by low statistics bubble chamber experiments and no data exist above
7\,\gevc. Little, if anything, is known about the $\pbarp \to
\overline{Y} Y$ reaction in the multiple strangeness sector.  Only
total cross sections based on a few events have been measured in the
double strangeness channel, {\it {\it i.e.}\ }the production of $\Cascade$
hyperons~\cite{bib:phy:Flaminio84}.  Nothing at all is known about this
reaction for charmed hyperons. All the available data on the total
cross section for the $\pbarp \to \overline{Y} Y$ reaction is
summarised in \Reffig{fig:X-sec_all}. The high statistics data
samples from PS185 comprises 40k completely reconstructed
$\pbarp \to \LambdaLambdabar$
events~\cite{bib:phy:Barnes_164}. Corresponding bubble chamber
experiments have, at most, a few hundred complete
events~\cite{bib:phy:Flaminio84}. There is one counter experiment at 6\,$\gevc$
with comparable statistics to \INST{PS185}~\cite{bib:phy:Becker}, but
normally only one hyperon could be reconstructed for the events
which meant that no spin correlation could be measured.

One example which shows the strength of hyperon pair production in
\pbarp annihilations is the following:  By measuring spin
correlations in the \INST{PS185} experiment, it has been shown that
the $\LambdaLambdabar$ pairs are practically always produced
in a triplet state~\cite{bib:phy:PS185}. It is natural to associate
this with the spin degrees of freedom in the creation process of the
$\ssbar$ pair since the spin of the $\Lambda$ hyperon is
primarily carried by its strange quark. In other words, the
$\ssbar$ pair is predominantly created in a triplet state. The
LEAR data were taken near threshold; and one should verify that this
feature persists as one goes up in momentum transfer into the more perturbative
region. Both meson-exchange models and models based on the constituent
quark model have been applied to the near threshold data from \INST{LEAR} and
both give a relatively good description of the main features of the
data~\cite{bib:phy:Klempt02, bib:phy:Alberg01}. The triplet state is
produced by assuming that the $\ssbar$ pair is created with the
quantum number of the vacuum, $^3P_0$, or with the gluon quantum
number, $^3S_1$, for the quark based models. Alternatively, it has
been suggested that $\ssbar$ pairs may be extracted from the
nucleon or anti\-nucleon sea instead of being created in the
reaction itself~\cite{bib:phy:Alberg95}. In this scenario the triplet state
would reflect the fact that the $\ssbar$ pairs are polarised in
the nucleon sea. In meson-exchange models, the triplet state is
interpreted as being due to a strong tensor force generated by the
exchange of $K$ and $K^\ast$ mesons.

\subsection{Experimental Aims}

As seen from \Reffig{fig:X-sec_all} any measurement of hyperon pair
production in the \PANDA energy regime will significantly improve the
data set.  Even in $\LambdaLambdabar$ production, where a
large data set exists, data points at momenta above 2\,$\gevc$ and a
cross check at low momenta would help understanding the production
mechanism. For all other hyperon pairs \PANDA will provide the first
conclusive insights on the behaviour of the total cross section and
first differential cross sections.  The very first measurements can be
provided for the pair production of charmed hyperons.

Spin observables for the $ \Cascadebar\Cascade$ reaction can be
extracted similarly to the $\LambdaLambdabar$ case. This will
allow for detailed comparisons between the $\ssbar$ and
$\ssbar\ssbar$ creation processes. A comparison between
the $\Lambdabar \Lambda$ channel, which filters isospin $I =
0$, to the $\Lambdabar \Sigma^0$ channel (including its charge
conjugate channel), which forces $I = 1$, gives opportunities to
study the isospin dependence of strangeness production.  In
the naive quark model the spin of the $\Sigma^0$ is opposite to that
of its constituent strange quark . This should lead to differences in the spin
correlations if they are related to the spin state of the created
$\ssbar$-pair.  Studies on the production of charmed hyperons
will allow for detailed comparisons between the $\overline{c}c$ and
the $\ssbar$ creation processes. This may help to disentangle
the perturbative contributions from the non-perturbative ones, as the charm
production will be mainly probing the hard processes while the
strangeness production will be influenced by non-perturbative effects.

\subsection
[Reconstruction of the  $\pbarp \to \overline{Y} Y$ Reaction]
{Reconstruction of the  $\boldsymbol{\pbarp \to \overline{Y} Y}$ Reaction}

Two-body kinematics together with the relatively long lifetime of
strange hyperons makes the identification and reconstruction of
$\overline{Y} Y$ events rather straightforward. The identification
of these reaction channels involve practically always the
reconstruction of a $\Lambdabar \Lambda$ pair as can be seen
from the main decay channels listed in \Reftbl{tab:hyperons}.

\begin{table*}[htb]
  \caption[Properties of strange and charmed ground state
  hyperons]{Properties of strange and charmed ground state
  hyperons~\cite{PDG} that are energetically accessible at
  \PANDA. The hyperon, its valence quark composition, mass, decay
  length $c\tau$, main decay mode, branching ratio \BR\ and the decay
  asymmetry parameter $\alpha_Y$ are listed.}
  \label{tab:hyperons}
  \begin{center}
  \begin{tabular}{ccccccc}
    \hline\hline
      Hyperon  &  Quarks  &  Mass [\mevcc]  &  $c\tau$ [cm]   & Main decay & \BR\ [\%]  &
       $\alpha_Y$\\
    \hline
     $\Lambda$       &  $uds$ & 1116  &  8.0  & $p\pim$    & 64 & +0.64 \\
     $\Sigma^+$      &  $uus$ & 1189  &  2.4  & $p\pi^0$   & 52 &  -0.98 \\
     $\Sigma^0$      &  $uds$ & 1193  &  $2.2\cdot10^{-9}$  & $\Lambda\gamma$  & 100 &  - \\
     $\Sigma^-$      &  $dds$ & 1197  &  2.4  & $n\pim$   & 100 &  -0.07  \\
     $\Cascade^0$         &  $uss$ & 1315  &  8.7  & $\Lambda\pi^0$   & 99 &  -0.41 \\
     $\Cascade^-$         &  $dss$ & 1321  &  4.9  & $\Lambda\pim$   & 100 &  -0.46 \\
     $\Omega^-$      &  $sss$ & 1672  &  2.5  & $\Lambda K^-$   & 68 &  -0.03 \\
     \\
     $\Lambdac^+$   &  $udc$ & 2286  & $6.0\cdot10^{-3}$  & $\Lambda\pip$    & 1 & -0.91(15) \\
     $\Sigma_c^{++}$ &  $uuc$ & 2454  &                    & $\Lambdac^+\pip$   & 100 &   \\
     $\Sigma_c^+$    &  $udc$ & 2453  &                    & $\Lambdac^+\pi^0$ & 100 &   \\
     $\Sigma_c^0$    &  $ddc$ & 2454  &                    & $\Lambdac^+\pim$   & 100 &   \\
     $\Cascade_c^+$       &  $usc$ & 2468  & $1.2\cdot10^{-2}$  & $\Cascade^-\pip\pip$   & seen &   \\
     $\Cascade_c^0$       &  $dsc$ & 2471  & $2.9\cdot10^{-3}$  & $\Cascade^- \pip$   & seen & -0.6(4)  \\
     $\Omega_c^0$    &  $ssc$ & 2697  & $1.9\cdot10^{-3}$  & $\Omega^- \pip$   & seen &   \\

       \hline\hline
  \end{tabular}
  \end{center}
\end{table*}

\begin{figure*}[htb]
 \begin{center}
 \includegraphics[width=0.7\dwidth]{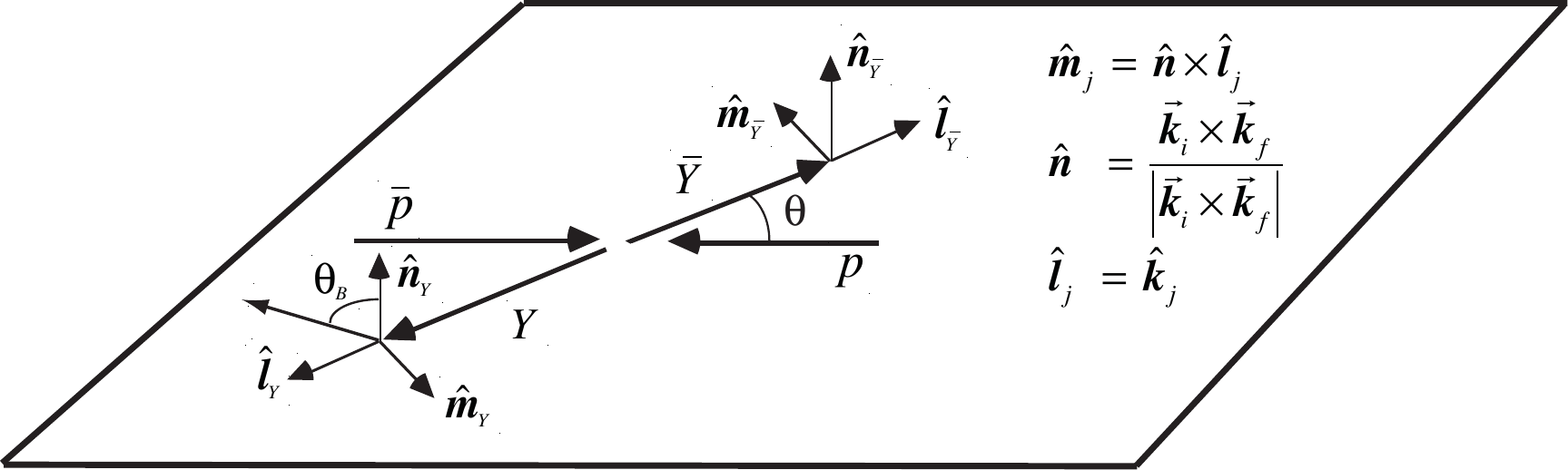}
 \caption[Coordinate system for the $\pbarp \to \overline{Y} Y$
 reaction.]
 {Coordinate system for the $\pbarp \to \overline{Y} Y$
 reaction. $\theta$ is the centre-of-mass (CM) scattering angle,
 $\vec{\bf{k}}_i$ and $\vec{\bf{k}}_f$ are the initial $\pbar$ beam
 and the final antihyperon momentum vectors in the CM system, respectively.
 These two vectors define the scattering plane and also the pseudovector
 $\vec{\mathbf{n}} = \vec{\mathbf{k}}_i \times \vec{\bf{k}}_f$ which
 is normal to the scattering plane. These vectors are then used to
 define a coordinate system for the antihyperon/hyperon rest frame
 with one axis along $\vec{\bf{n}} $ and the two other axes in the
 scattering plane. The direction of the $j^\mathrm{th}$ particle
 momentum $\widehat{\bf k}_j$ (j = $\overline{Y} $ or $Y$) in the CM
 system is taken as the $\widehat{\bf l}_j$ axis. The handiness of the
 coordinate system is taken to be right handed.}
\label{fig:Coord_sys}
\end{center}
\end{figure*}

The \PANDA detector allows for the reconstruction of both neutral and
charged hyperons, due to its capabilities to track charged particles,
to detect photons and discriminate between almost all stable
particles.  For neutral hyperons with charged decay modes, {\it {\it e.g.}\
}the $\Lambda \to p\pim$ decay channel, the decay vertex outside the
interaction region is reconstructed.  Neutral strange hyperons, apart
from the $\Lambda$, are accompanied by photons, which will be detected
in the electromagnetic calorimeter.  The reconstruction of charged
hyperons involve the identification of tracks from the interaction
region that exhibit a ``kink'' that signals the hyperon decay.

A good understanding of the response and reconstruction of tracks that
originate well outside of the interaction region is important for
these studies. This is further emphasised when extracting
spin observables. The parity violating weak decay of hyperons gives a
decay distribution in its own rest frame according to
\begin{equation}
\label{eq:hypol}
 I\left( {\theta_B } \right) =
{1 \over {4\pi }}\left( {1 + \alpha_Y P^Y \cos
\theta_B } \right),
\end{equation}
where $\theta_B$ is the baryon emission angle with respect to the spin
direction of the decaying hyperon, $\alpha_Y$ is the decay asymmetry
parameter (listed in \Reftbl{tab:hyperons}) and $P^Y$ is the hyperon
polarisation. The coordinate system used in the analysis is given in
\Reffig{fig:Coord_sys}. The axes are chosen such that a maximum
use of parity conservation can be made.

Spin observables can be extracted for all strange hyperons, with the
exception of $\Sigma^-$ and $\Omega^-$ where the decay asymmetry
parameter is too small.  The $\Sigma^0$ hyperon decays via the parity
conserving electromagnetic interaction but spin observables can be
extracted from the subsequent $\Lambda \to p\pim$ decay via
\begin{equation}
\label{eq:sigpol}
 I\left( {\theta_p} \right) =
{1 \over {4\pi }}\left( {1 - \frac{1}{3}\alpha_\Lambda P^{\Sigma^0} \cos
\theta_p } \right).
\end{equation}

The measured cross sections for production of single and doubly
strange hyperons range from a $\mu$b to a hundred $\mu$b. These
cross sections are comfortably high with hundred thousands of events produced
per hour already
at nominal luminosity. Nothing is experimentally known of the
cross section for triple strangeness production, {\it i. e.\ }the
reaction $\pbar p \rightarrow{}\overline{\Omega}^+\Omega^-$.
The only existing theoretical estimate predicts maximum cross section
of $\sim 2$\,nb~\cite{bib:phy:Kaidalov94} which would correspond to
$\sim 700$ produced $\overline{\Omega}^+\Omega^-$ pairs per hour at a
luminosity of $10^{32}{\rm{}cm^{-2}s^{-1}}$. This is sufficient for a
measurement of the total cross section and the differential angular
distribution. No spin observables will be directly accessible for this
reaction due to the very small decay asymmetry parameter.

There is only one estimate for the cross section of the
$\pbarp \to \Lambdabar_c^- \Lambdac^+$ reaction
predicting it to be as high as
0.2\,$\mu$b~\cite{bib:phy:Kaidalov94}. The asymmetry parameter in the
$\Lambdac^+ \to \Lambda\pip$ decay is comfortably large to
access spin observables. The challenges in studying this reaction via
this decay channel is that its branching ratio is only 1\%. In fact,
all decay channels for the $\Lambdac^+$ have branching ratios of the
order of 1\percent or less~\cite{PDG}.  The decay length $c\tau$
of the $\Lambdac^+$ cannot be used to identify the reaction as it is
only 60\,$\mu$m.  However, the $\Lambdac^+$ and its antiparticle
decay into $\Lambda$ and $\Lambdabar$, which have a long decay
length.  The reaction may be identified against a large
background as it is possible to additionally pin down the interaction
vertex by the reconstruction of the charged pions from the
$\Lambdac^+$ decay.

It is worthwhile to note that the decay pattern given for all charmed
hyperons listed in \Reftbl{tab:hyperons} will always lead to the
appearance of a $\Lambda$ particle. It is therefore important
to acquire a good understanding of the reconstruction of
$\Lambda$ particles in \PANDA.

\subsubsection{Benchmark Channels}

Among the variety of channels of hyperon pair production accessible at
\PANDA, the channels $\pbarp \to \Lambdabar \Lambda$ and
$\Cascadebar^+ \Cascade^-$ haven been chosen to prove our principle
ability to reconstruct the angular and polarisation
distributions. These channels exhibit the following features which
makes them well suited for a case study.
%
\begin{itemize}

  \item $\boldsymbol{\pbarp \to \Lambdabar \Lambda}$.
   Though well studied close to threshold this basic channel provides
   an essential tool to understand the reconstruction capabilities for
   all hyperon pair production reactions at \PANDA. This is mainly due
   to the fact that most hyperons decay such that a
   $\Lambdabar(\Lambda)$ particle is produced in an intermediate
   state (see \Reftbl{tab:hyperons}).  This is also the case for
   the excited baryons (see sec.~\ref{sec:phys:baryons}). Hence, a
   detailed understanding of the reconstruction and identification of
   $\Lambdabar$ or $\Lambda$ particles in \PANDA detector is
   very important for many aspects of the \PANDA physics programme.
   The reconstruction of the $\Lambda$ decay products (mostly from
   $\Lambdabar \to \pbar\pi^{+}$ and $\Lambda \to p\pim$)
   differs from ordinary charged particle reconstruction in \PANDA as
   the charged particles do not stem from the interaction point. This
   displaced vertex poses a challenge to reconstruction algorithms and
   special attention has to be drawn to background reduction.  The
   extension and comparison of the well measured near-threshold data
   to higher momenta makes the study of this channel also interesting
   in its own right.

 \item $\boldsymbol{\pbarp \to \Cascadebar^+ \Cascade^-}$. This
   channel probes the  tracking capability near the interaction region
   to a much larger extent than the $\Lambda$ pair production.
   This will also be a
   reaction where \PANDA will provide first differential distributions.
   For the simulation we therefore assumed an isotropic distribution
   in the CM system to investigate how well this can be reconstructed with
   \PANDA.
\end{itemize}

\paragraph*{Study of the $\boldsymbol{\pbarp \to
  \Lambdabar \Lambda}$ Reaction.} \  \\
Monte Carlo
data for this reaction have been generated and analysed at three
incoming momenta, 1.64\,\gevc, 4\,\gevc and 15\,\gevc. Approximately
$10^6$ events were generated for each of the three beam momenta. The
lowest momentum corresponds to the near threshold region where high
quality data exist from the \INST{PS185} experiment at \INST{LEAR}. The data for
this momentum were generated according to the experimental angular
distribution~\cite{bib:phy:Barnes_164}. An empirical function composed
of two exponential functions and four parameters in total was used to
generate the corresponding angular distributions at the two higher
momenta as suggested in Ref.~\cite{bib:phy:Becker},
\begin{equation}
\label{eq:twoexp}
d\sigma /dt' = ae^{bt'} + ce^{dt'} \quad .
\end{equation}
Here, $t' = -{1\over{2}}{t'}_{max } (1 - \cos \theta _{CM} )$ is the
reduced four-momentum transfer squared, where mass effects are removed
from the full four momentum transfer
\begin{equation}
\label{eq:fourmomtran}
t = m_p^2  + m_\Lambda ^2  - {s\over{2}} + {1\over{2}}{t'}_{max }\cos \theta _{CM}
\end{equation}
with $t'_{max}= \sqrt {(s - 4m_p^2 )(s - 4m_\Lambda ^2) }$.

\begin{figure}[htbp]
  \begin{center}
  \includegraphics[width=\swidth]{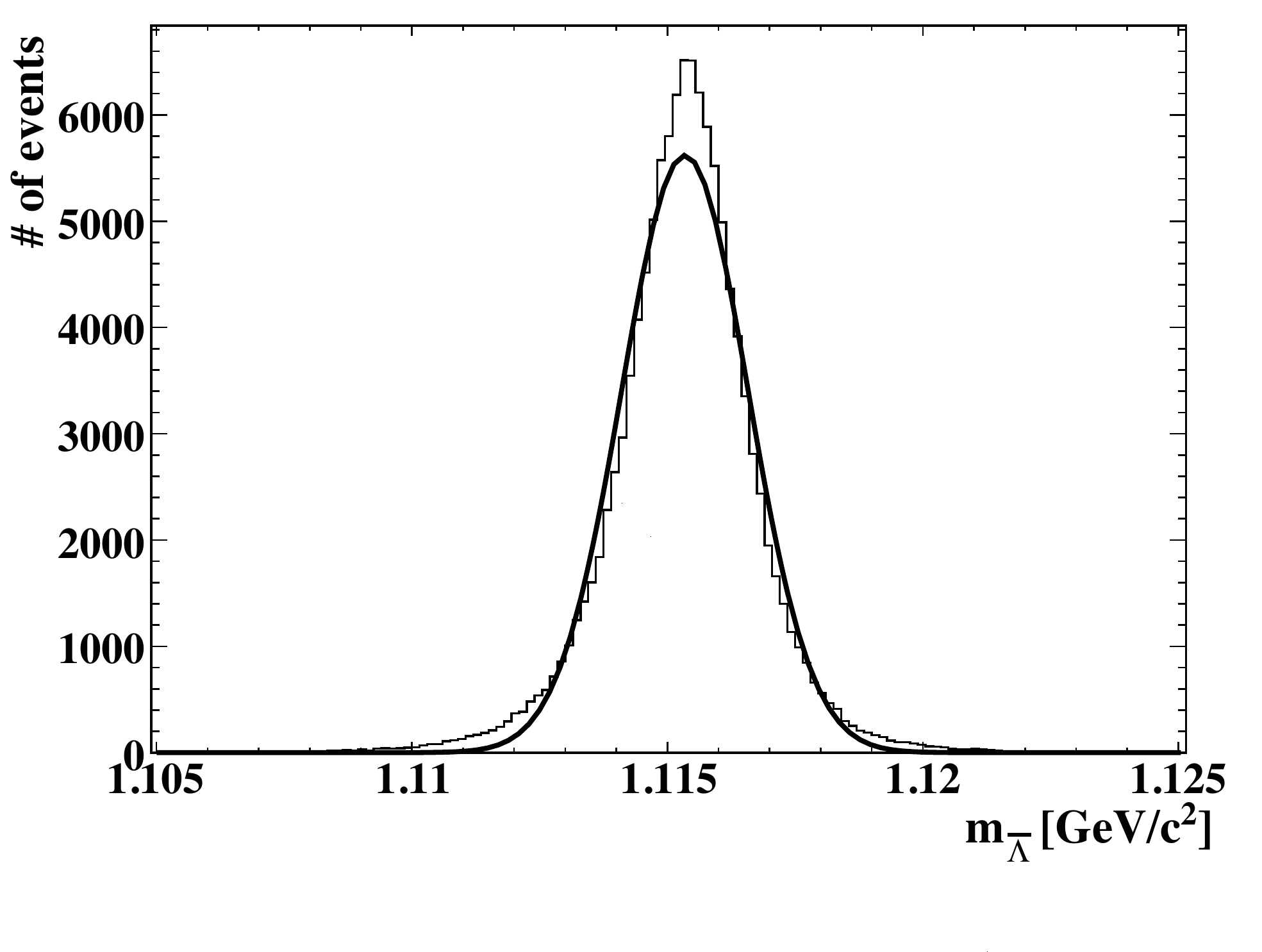}
  \end{center}
  \caption[Reconstructed $\pbar\pip$ invariant mass for
  $\Lambdabar$ candidates from the
  $\pbarp \to \Lambdabar \Lambda$ reaction.]
  {Reconstructed $\pbar\pip$ invariant mass for
  $\Lambdabar$ candidates from the
  $\pbarp \to \Lambdabar \Lambda$ reaction at
  1.64\,$\gevc$ (histogram) and a Gaussian fit to the distribution
  (smooth curve).  The distributions at 4 and 15\,$\gevc$ show a
  similar behaviour.}
\label{fig:Lbar_inv_mass}
\end{figure}

The polarisation was assumed to follow a $\sin{2\theta_\mathrm{CM}}$
dependence for all momenta. The reconstruction of the events was
done consecutively in the following steps:

\begin{enumerate}
 \item Identified pairs of antiprotons(protons) and $\pip(\pim)$
   were fitted to a common vertex under the hypothesis of stemming
   from a $\Lambdabar( \Lambda)$. The $\chi^2$ of the fit was
   then required to be $> 0.001$.

 \item The invariant $\pbar\pip$ and $p\pim$ masses of the
   reconstructed $\Lambdabar$ and $\Lambda$, respectively, are
   required to be within about $4\,\sigma$ of the $\Lambda$ mass, {\it
   {\it i.e.}\ } 1.110 $\gevcc \leq$ M $\leq$ 1.120 $\gevcc$. (See also
   \Reffig{fig:Lbar_inv_mass}.)

 \item The remaining events are fitted to the
   $\pbarp \to \Lambdabar \Lambda$ hypothesis in a
   tree-fit. Again, the $\chi^2$ of the fit was then required to be $>
   0.001$.

 \item Finally, cuts on the $\Lambdabar$ and $\Lambda$
   vertices are applied.  An effective cut on the displaced vertices
   was found to be the requirement that the sum of both path lengths
   is above 2\,cm.
\end{enumerate}

These criteria result in global reconstruction efficiencies of 0.11,
0.24 and 0.14 for the $\pbarp \to\Lambdabar \Lambda$
reaction in the charged decay mode at 1.64\,$\gevc$, 4\,$\gevc$ and
15\,$\gevc$, respectively. The reason for the lower efficiency at
1.64\,$\gev$ is that there are many pions produced with momenta below
50\,$\mevc$ at this momentum. These cannot be reconstructed as will be
shown in the discussion on polarisation below.  The lower efficiency at
15\,$\gevc$ is primarily due to a higher loss of tracks in the
beam pipe. This is a result of the very forward peaked
angular distribution.
The reconstructed $\Lambdabar$ mass at 1.64\,$\gevc$ is shown
in \Reffig{fig:Lbar_inv_mass}. The $\Lambdabar$ mass is
reconstructed obtaining a sigma of 1.2\,$\mevcc$ at this
momentum. Similar mass resolutions are obtained at the higher momenta.

\begin{figure}[htbp]
  \begin{center}
  \includegraphics[width=\swidth]{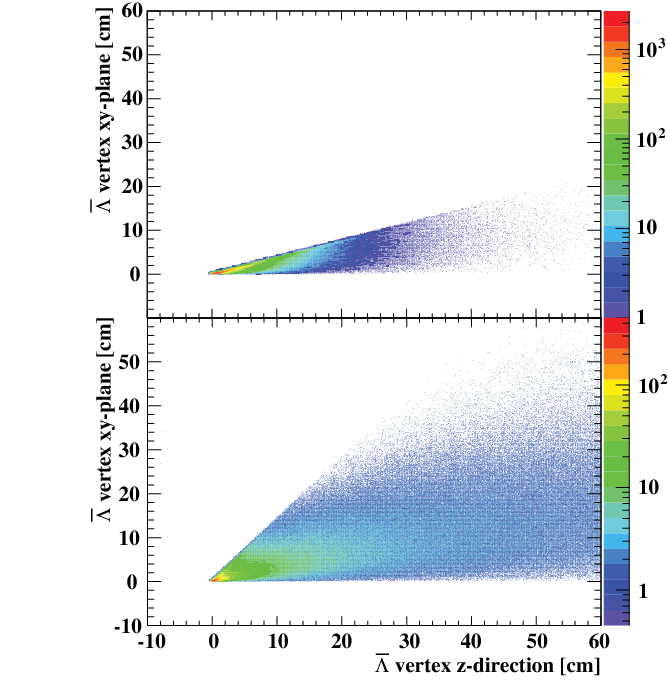}
  \end{center}
  \caption[$\Lambdabar$ decay vertex distributions.]
  { $\Lambdabar$ decay vertex coordinates at beam
  momenta of 1.64\,$\gevc$ (upper panel) and $15\,\gevc$ (lower
  panel) for phase space distributed events. }
\label{fig:Lbar_vertex}
\end{figure}

The spatial distributions of decay vertices for generated
$\Lambdabar$ events are shown in
\Reffig{fig:Lbar_vertex}. The lower figure shows the distribution
at 1.64\,$\gevc$ and the upper figure the corresponding distribution
at 15\,$\gevc$. This illustrates the origin of the charged tracks from
the $\Lambdabar \Lambda$ reaction. No strong dependencies on the
vertex position in the reconstruction efficiencies are found (apart
from obvious geometrical constraints). The decay vertices are
reconstructed with a sigma ranging from 0.6 mm at 1.64 $\gevc$ to 2.8
mm at 15 $\gevc$.  The z component of the vertex is dominating the
uncertainty and this naturally increases with higher beam momenta as
the particles are emitted in smaller angles.

\begin{figure}[htbp]
  \begin{center}
  \includegraphics[width=\swidth]{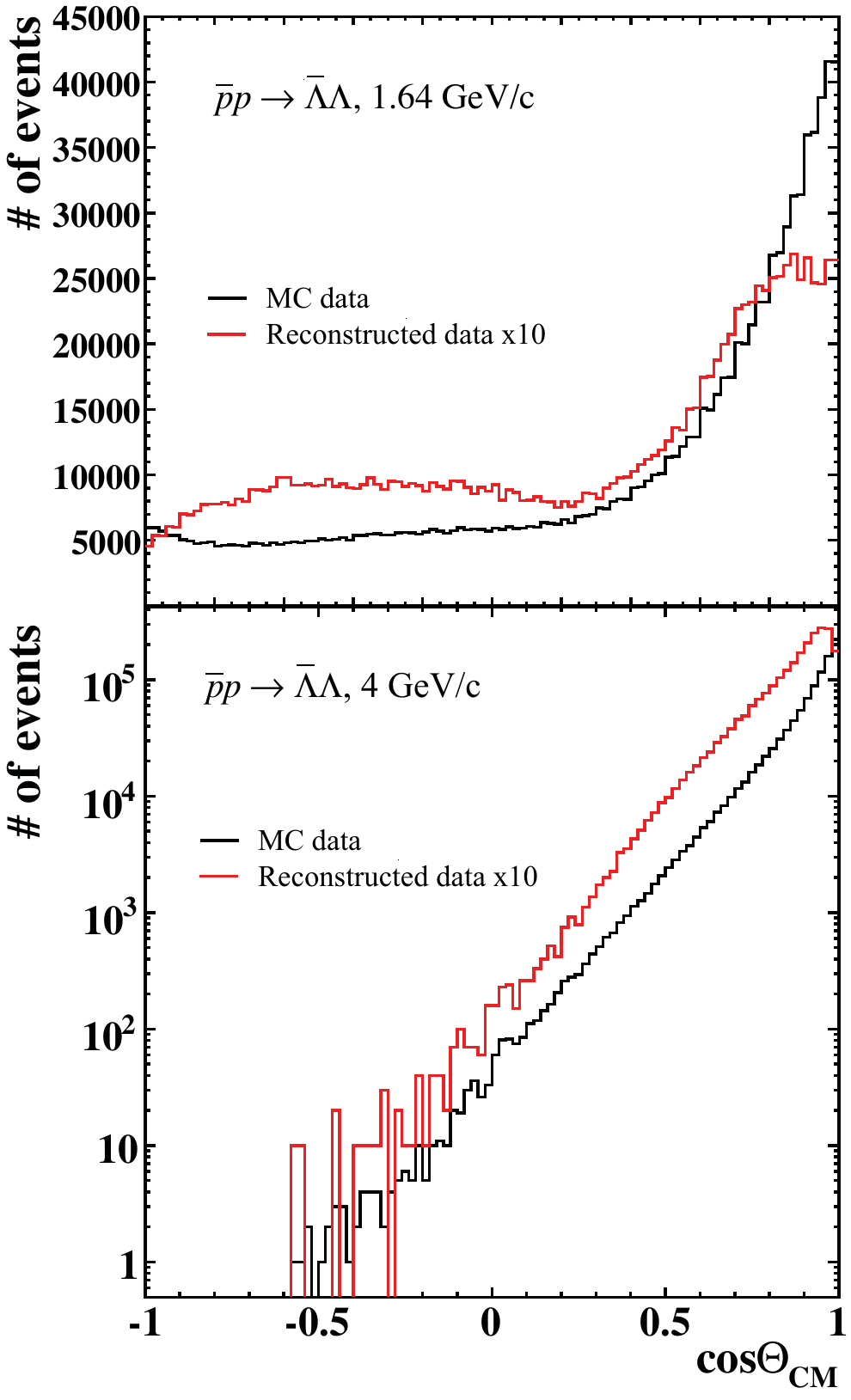}
  \end{center}

  \caption[CM angular distributions for the  $\pbarp \to \Lambdabar \Lambda$ reaction.]
  {Centre-of-mass angular distributions for the
    $\Lambdabar$ hyperon from the
    $\pbarp \to \Lambdabar \Lambda$ reaction at
    1.64\,$\gevc$ (upper figure) and 4\,$\gevc$ (lower figure). The
    Monte Carlo generated angular distributions are shown in black;
    the reconstructed angular distributions (multiplied by a
    factor of ten) are shown in red.}

\label{fig:Lbar_CosCM}
\end{figure}

Centre-of-mass angular distributions for the outgoing
$\Lambdabar$ at 1.64\,$\gevc$ and 4\,$\gevc$ are shown in
\Reffig{fig:Lbar_CosCM}. The Monte Carlo generated events are the
black histograms and the angular dependencies are taken from
references~\cite{bib:phy:Barnes_164} and~\cite{bib:phy:Becker},
respectively. The red histograms show the corresponding reconstructed
angular distributions (multiplied by a factor of 10 to account roughly
for the overall efficiency). It is seen that the experimental
acceptance covers the whole angular region for this reaction at
1.64\,$\gevc$. The lack of events at higher angles at 4\,$\gevc$ is
due to the exponential fall off of the generated angular distribution
(see \Refeq{eq:twoexp}) and this is further emphasised at
15\,$\gevc$. The loss in the very forward direction is due to losses
of events with tracks in the beam pipe region. It has been verified
that the acceptance is covering the full angular region at the two
higher momenta as well, by analysing Monte Carlo event samples with
isotropic CM angular distributions. This means that, after acceptance
correction, the full centre-of-mass angular distributions of the
outgoing hyperons can be deduced for this reaction over the full
momentum range of \HESR.

\begin{figure}[htbp!]
  \begin{center}
\includegraphics[width=\swidth]{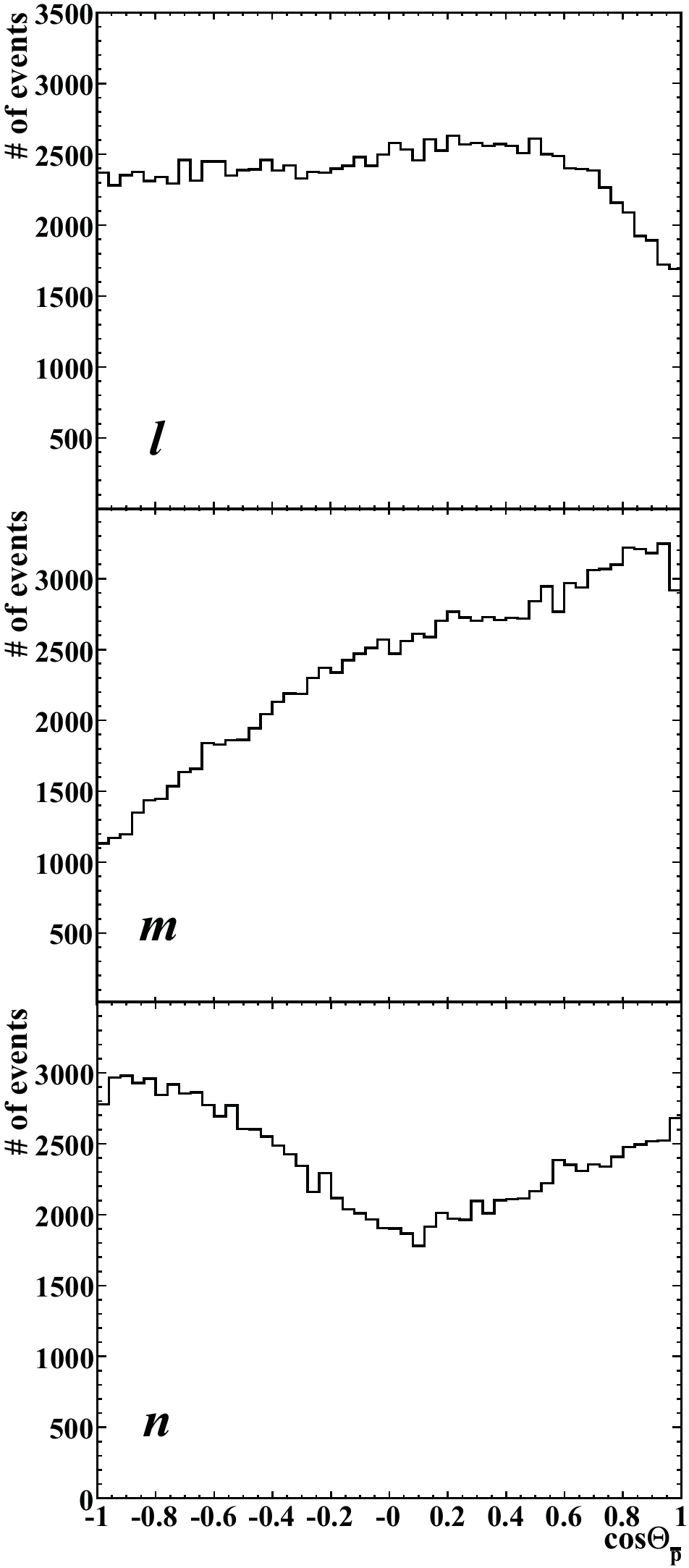}
  \end{center}
\caption[Reconstructed $\pbar$ angular distributions from the $\Lambdabar \to
  \pbar\pip$ decay.]
{Angular distributions
  of $\pbar$ from the $\Lambdabar \to
  \pbar\pip$ decay reconstructed in the \PANDA acceptance in the
  hyperon rest frame according to the coordinate system defined in
  \Reffig{fig:Coord_sys}.}
\label{fig:pbar_Costh}
\end{figure}

The analysis of spin observables requires the knowledge of the angular
distribution of the hyperon decay particles, the proton and pion, in
the rest frame of the hyperon. This angular distribution is isotropic for the
unpolarised case. The coordinate system that is used for
this reconstruction is given in \Reffig{fig:Coord_sys}. It should
be noted that the axes of this coordinate system vary from event to
event. In this coordinate system the projections onto the $\widehat{\bf l}$ and
$\widehat{\bf m}$ axes are isotropic independently of the hyperon polarisation.
The distribution along the $\widehat{\bf n}$ axis will exhibit a slope
equal to $\alpha_{\Lambda} P^{\Lambda}$. Here $\alpha_{\Lambda}$ is
the decay asymmetry parameter and $P^{\Lambda}$ is the polarisation
(see \Refeq{eq:hypol}). CP conservation requires that $P^{\Lambda} =
P^{\Lambdabar}$. \Reffig{fig:pbar_Costh} shows the
projections of the decay antiproton emission vector along the
$\widehat{\bf l}$, $\widehat{\bf n}$ and $\widehat{\bf m}$ axes for
reconstructed non-polarised events at 1.64\,$\gevc$. These
distributions are clearly non-isotropic for all projections which
shows that the \PANDA acceptance is inhomogeneous for these reactions.
Such distributions will therefore be used for acceptance corrections
in the extraction of the polarisation.  The non-isotropies are related
to an inefficiency to reconstruct low-energy pions in \PANDA.

\begin{figure}[htbp!]
  \begin{center}
  \includegraphics[width=\swidth]{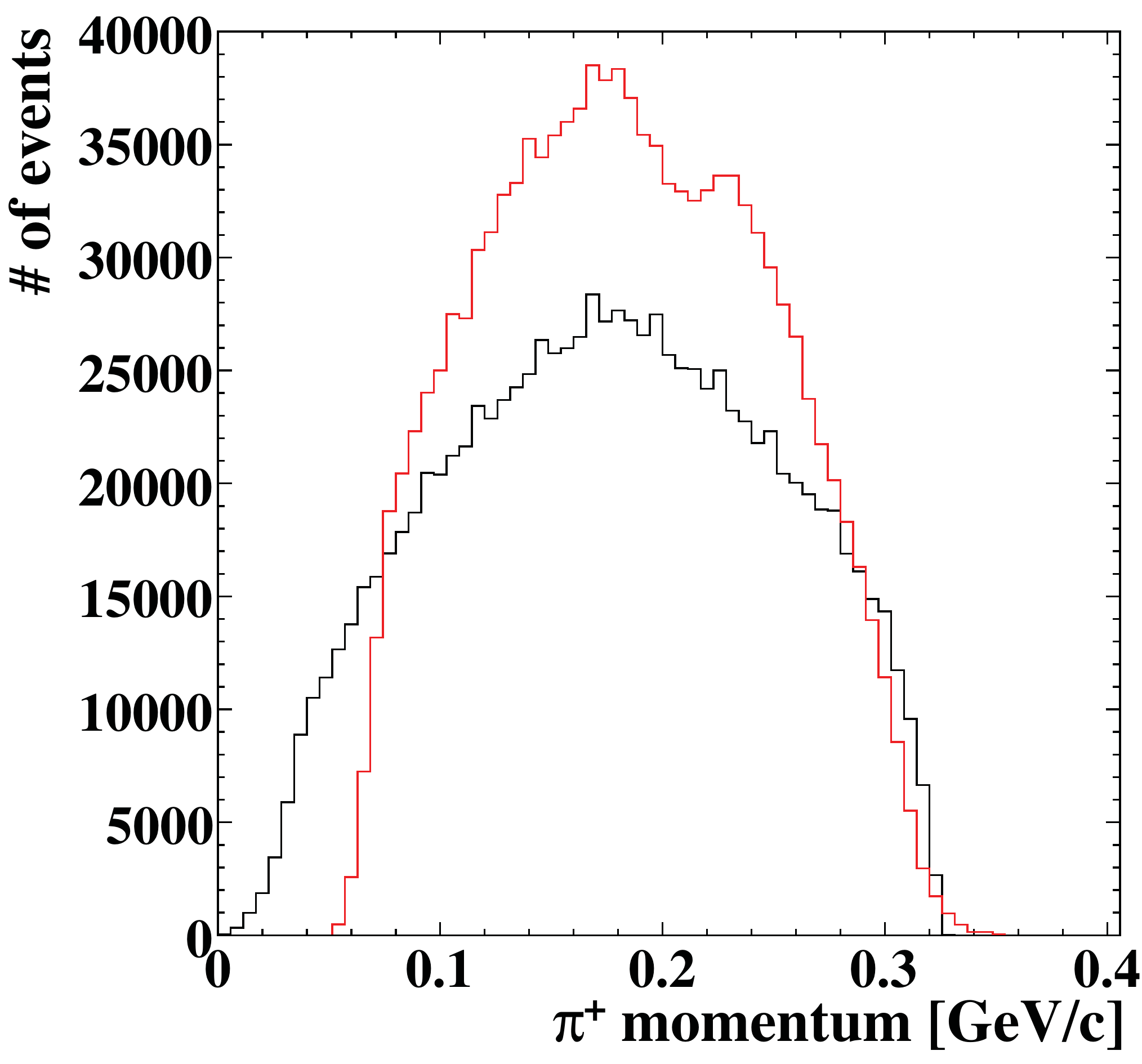}
  \end{center}
\caption[Pion momentum distributions in the laboratory system
  from the $\Lambdabar$ decay 1.64\,$\gevc$.]
{Absolute value of the pion momentum in the laboratory system
  from the $\Lambdabar$ decay 1.64\,$\gevc$. The black
  histogram is Monte Carlo data and the red histogram shows the
  reconstructed events. The histogram for the reconstructed has been
  multiplied by a factor of ten.}
\label{fig:piplusMom}
\end{figure}

That low-energy pions cannot be reconstructed efficiently can be seen
in \Reffig{fig:piplusMom} which shows the absolute value of the
$\pip$ momentum in the laboratory system from the
$\Lambdabar$ decay at 1.64\,$\gevc$ . The black histogram
shows the Monte Carlo generated events and the red histogram shows the
momenta from the reconstructed events multiplied by a factor of
ten. It is clear from this picture that pions with momenta lower than
about 50 $\mevc$ are not reconstructed. This is also the origin of the
non-isotropies in \Reffig{fig:pbar_Costh}. The same pattern is
seen at higher momenta, but the magnitude of the effect decreases
with increasing beam momentum since the amount of low energy pions
decreases with increasing beam momentum.

\begin{figure}[htbp!]
  \begin{center}
  \includegraphics[width=\swidth]{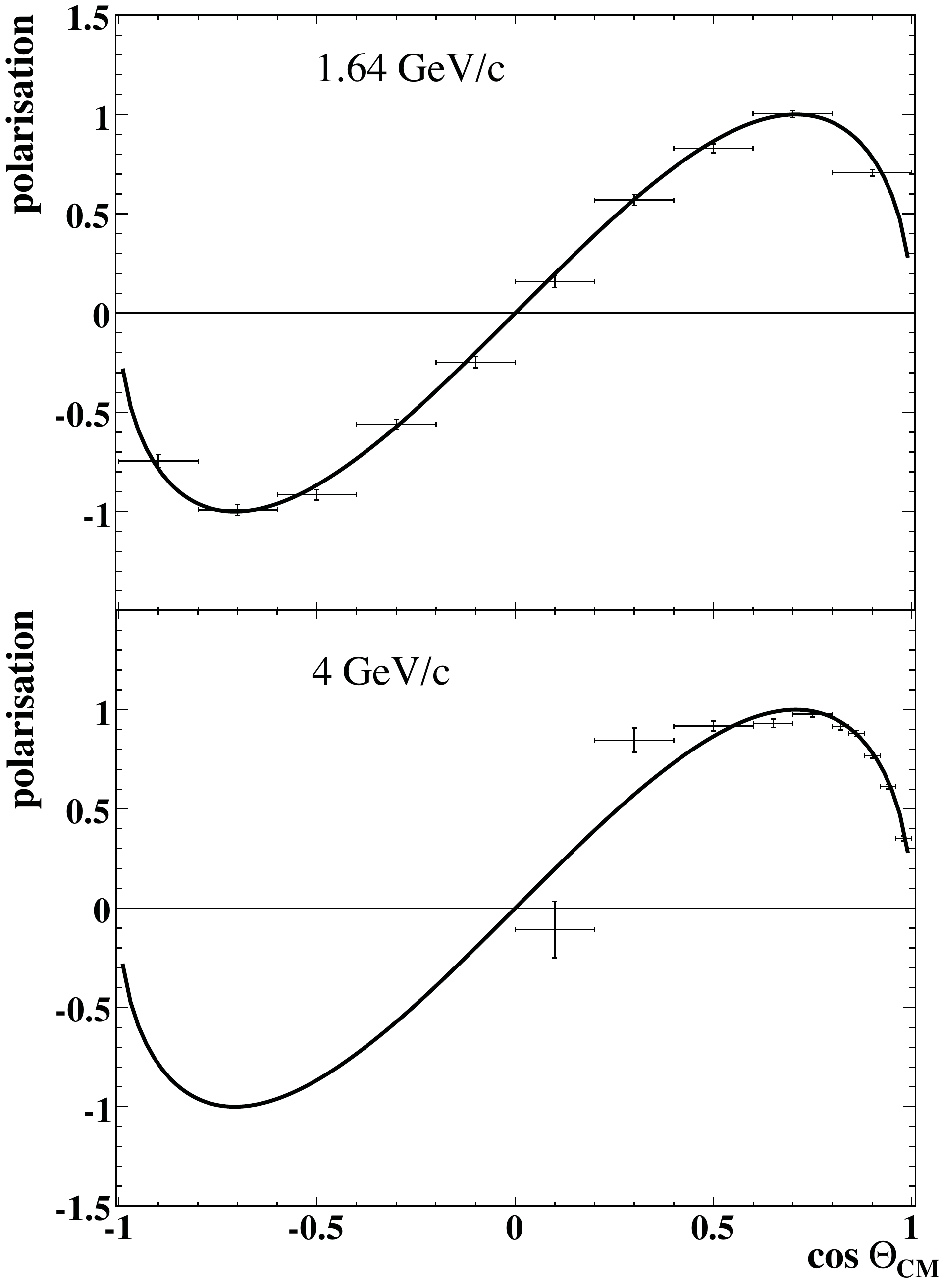}
  \end{center}
\caption[Reconstructed polarisations from the $\pbarp \to
  \Lambdabar \Lambda$ reaction.]
{Reconstructed polarisation from  $\pbarp \to
  \Lambdabar \Lambda$ reaction at 1.64\,$\gevc$ (upper figure)
  and 4\,$\gevc$ (lower figure). The solid line is the
  $\sin2\theta_CM$ function that is used to generate the
  polarisation.}
\label{fig:LBarPol}
\end{figure}

The polarisation distributions are extracted by applying the method of
moments~\cite{bib:phy:Frodesen} with acceptance correction functions from
 non-polarised data in bins of
$\cos{\theta_\mathrm{CM}}$. The polarisation is finally extracted using both
hyperon polarisations, {\it {\it i.e.}}
$(P^{\Lambdabar}+P^\Lambda)/2$.  The extracted polarisations
at 1.64\,$\gevc$ and 4\,$\gevc$ are shown in \Reffig{fig:LBarPol}
together with the superimposed $\sin{ 2\theta_\mathrm{CM}}$ function
which is used to generate the polarisation data. The large error bars
at the higher angles at 4\,$\gevc$ is due to the functional form for
the angular distribution which strongly suppresses events at large CM
emission angles. This is further emphasised at 15\,$\gevc$ where
reasonable statistics is only acquired at the uppermost angular bins.
The background for this channel is treated in section~\ref{Background}.

\begin{figure}[htbp]
  \begin{center}
  \includegraphics[width=\swidth]{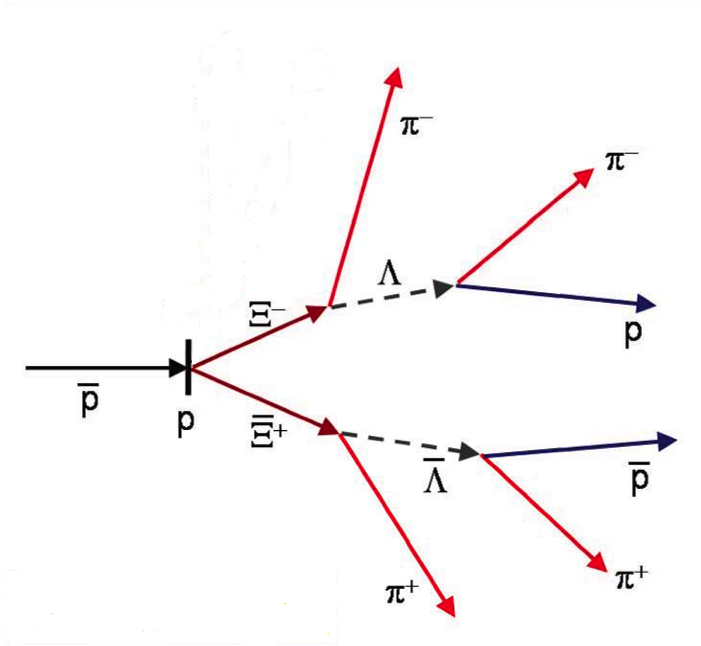}
  \end{center}
\caption[Topology of the $\pbarp \to \Cascadebar^+ \Cascade^-$ reaction.]
{Schematic illustration of the topology for the investigated
  $\pbarp \to \Cascadebar^+ \Cascade^-$ reaction. The
  characteristic pattern of this reaction is four separated decay
  vertices}
\label{fig:XiXi_reaction}
\end{figure}

\paragraph*{Study of the $\boldsymbol{\pbarp \to
\Cascadebar^+ \Cascade^-}$ Reaction.} \  \\
Approximately $2\cdot 10^6$ Monte Carlo events have been generated and
analysed at 4\,$\gevc$ for this reaction. This momentum was chosen
because it is also used for analysing events for the
$\LambdaLambdabar$ channel.  No experimental information on
angular distributions are available for the $\pbarp \to
\Cascadebar^+ \Cascade^-$ channel. The production of two $\ssbar$
quark-pairs in this reaction will most likely lead to a less steep
angular distribution than in the $\LambdaLambdabar$ case. An
isotropic CM angular distribution was therefore chosen. The
reconstructed angular distribution will then directly give the angular
acceptance for the process. A $\sin2\theta_\mathrm{CM}$ function was
used for the polarisation.  The reaction was studied in the $
\Cascadebar^+ \to \Lambdabar\pip \to
\pbar\pip\pip(\Cascade^-\to \Lambda\pim \to p\pim\pim)$ decay
channel which is illustrated in \Reffig{fig:XiXi_reaction}.

The reconstruction was made in the following steps:
\begin{enumerate}
\item Identified pairs of antiprotons(protons) and $\pip(\pim)$ were
  fitted to a common vertex under a $\Lambdabar( \Lambda)$
  hypothesis. The $\chi^2$ of the fit was then required to be $>$ 0.001.

 \item The invariant $\pbar\pip$ masses of the reconstructed
   $\Lambdabar$ and $\Lambda$ are required to be close to the
   $\Lambda$ mass, 1.100\,$\gevcc \leq$ M $\leq$ 1.300\,$\gevcc$.

 \item Pairs of $\Lambdabar(\Lambda)$ and $\pip(\pim)$ are
   fitted to a common vertex under a $\Cascadebar^+(\Cascade^-)$
   hypothesis.  The $\chi^2$ of the fit was then required to be $>
   0.001$.

 \item The invariant mass of the reconstructed $\Cascadebar^+$ and
   $\Cascade^-$ are required to be close to the $\Cascade^-$ mass, {\it {\it i.e.}}
   1.30\,$\gevcc \leq$ M $\leq$ 1.35\,$\gevcc$.

 \item The remaining events are fitted to the $\pbarp \to
   \Cascadebar^+\Cascade^-$ hypothesis in a tree-fit.  The $\chi^2$ of the fit was
   then required to be $> 0.001$.

\end{enumerate}

\begin{figure}[htbp]
  \begin{center}
  \includegraphics[width=\swidth]{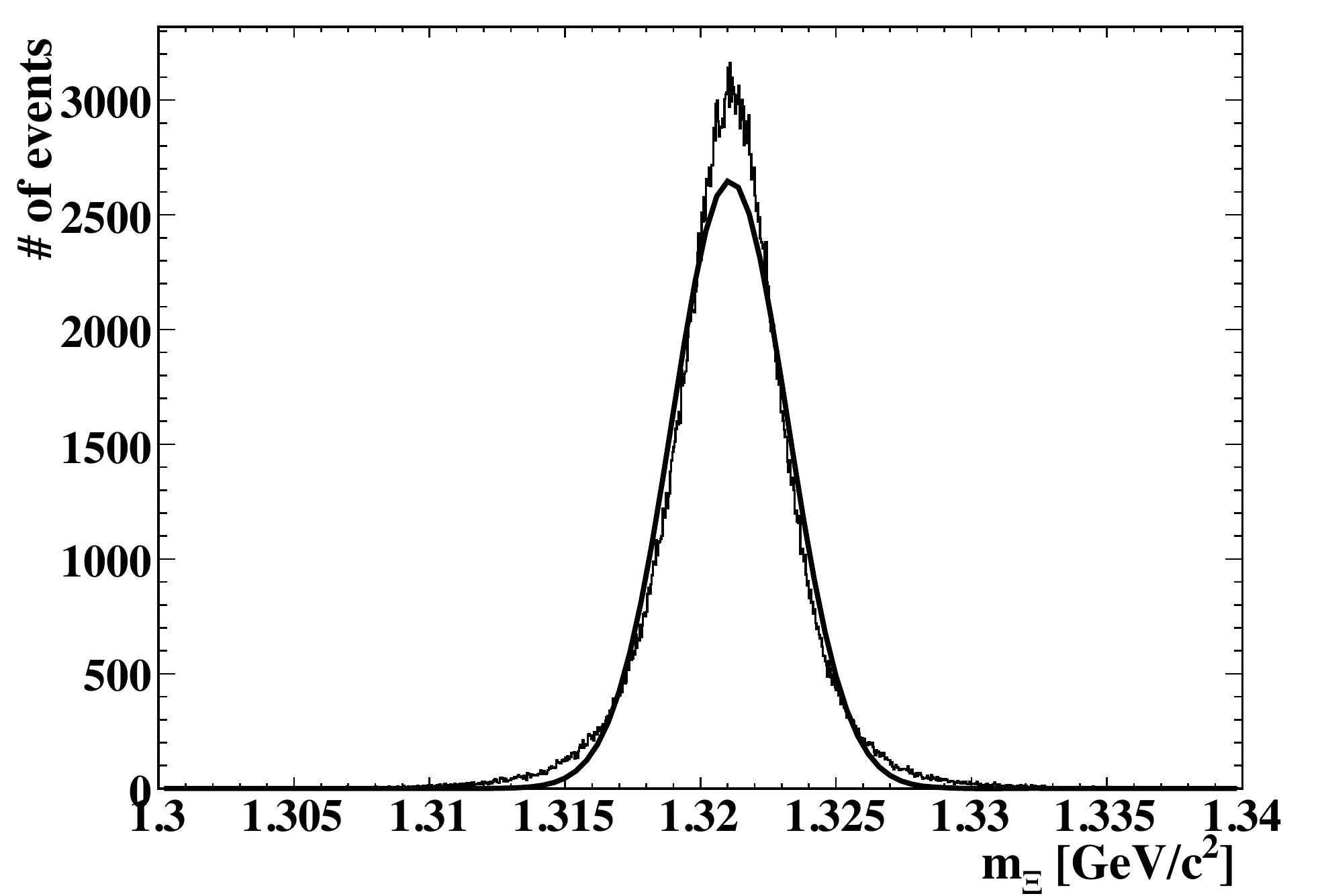}
  \end{center}
\caption[Reconstructed $ \Cascadebar^+$ mass for the
  $\pbarp \to \Cascadebar^+\Cascade$ reaction.]
{Reconstructed $ \Cascadebar^+$ mass for the
  $\pbarp \to \Cascadebar^+\Cascade$ reaction at 4\,$\gevc$.}
\label{fig:Ximass}
\end{figure}

These criteria result in an overall reconstruction efficiency of about
0.19. The reconstructed invariant $\Cascadebar^+$ mass is shown in
\Reffig{fig:Ximass}. The $\Cascadebar^+$ mass is reconstructed
with a sigma of 2.1\,$\mevcc$.  The sigma of reconstructed $\Lambda$
mass is 1.7\,$\mevcc$. The $\Cascade$ and $\Lambda$ decay vertices are
reconstructed with a sigma of 5.2 mm and 4.7 mm, respectively. The
dominating contribution is the resolution in the z direction.

\begin{figure}[htbp]
  \begin{center}
  \includegraphics[width=\swidth]{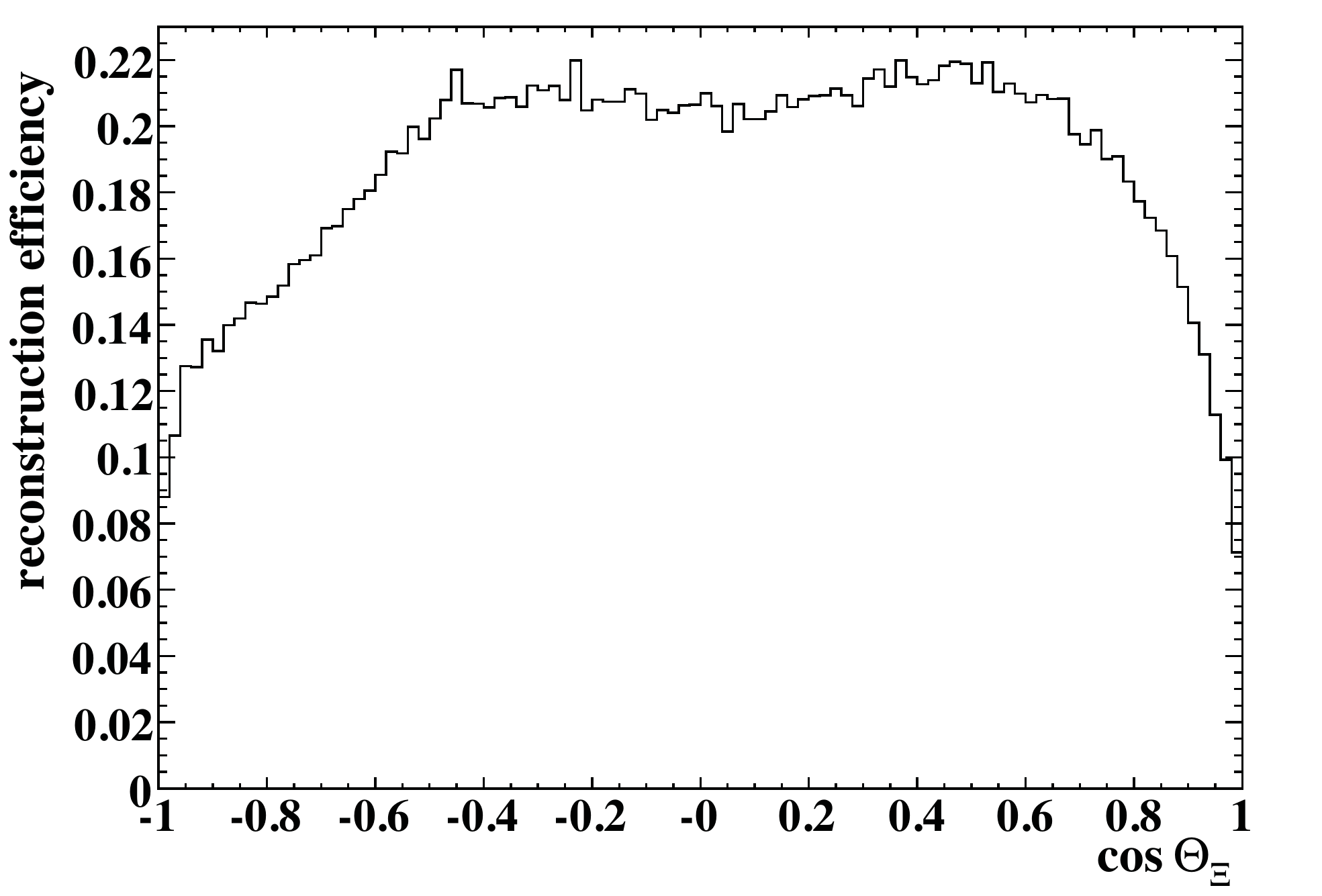}
  \end{center}
\caption[Reconstructed CM angular distribution for the
  $\pbarp \to \Cascadebar^+ \Cascade^-$ reaction.]
{Centre-of-mass angular distribution for the
  $\pbarp \to \Cascadebar^+ \Cascade^-$ reaction at 4\,$\gevc$
  from isotropically generated events reconstructed in the \PANDA
  acceptance.}
\label{fig:XiCosCM}
\end{figure}

The reconstructed $ \Cascadebar^+$ centre-of-mass angular
distribution is shown in \Reffig{fig:XiCosCM}. The acceptance does
not vary more than a factor of two over the full angular range which
means that the distribution can be extracted with small statistical
uncertainties in all angular bins. The dip in the forward and backward
directions are primarily related to events where one track is lost in
the beam pipe region. One must correct for the bending of the charged
$\Cascade$ tracks in the transverse direction to get the correct production
momentum vector. This is done by applying the
formula~\cite{PDG}
\begin{equation}
\label{eq:bending}
R = {{p_ \bot  B} \over {0.3}} \quad ,
\end{equation}
where $R$ is the radius of curvature in meters, $p_ \bot $ the
momentum component in the transverse direction in $\gevc$ and $B$ the
magnetic field in Tesla. This correction improves the resolution in
the x and y components of the production momentum vector of the $\Cascade$
particles from 5.3 $\mevc$ to 1.9 $\mevc$.

\begin{figure}[htbp]
  \begin{center}
  \includegraphics[width=\swidth]{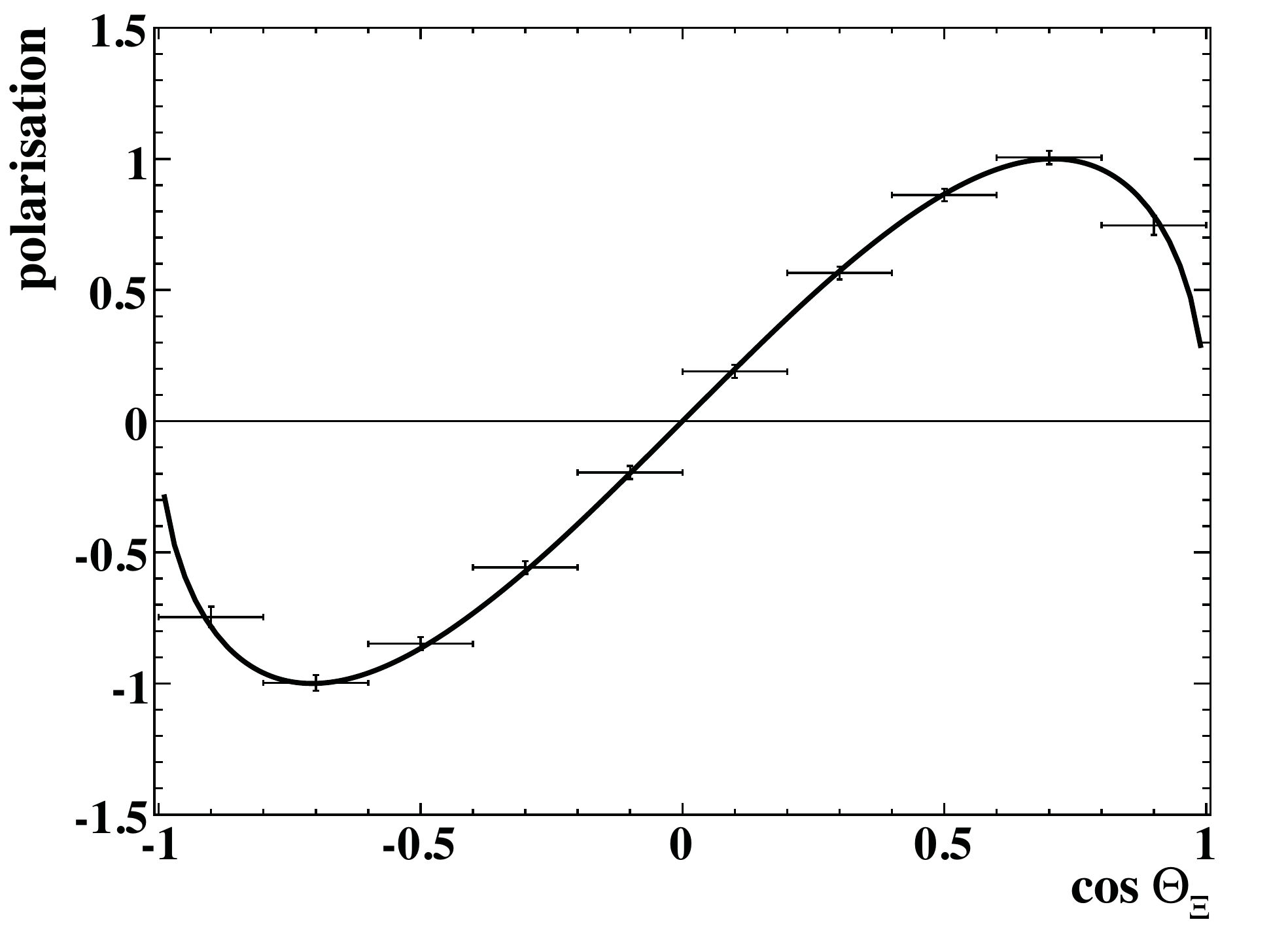}
  \end{center}
\caption[Reconstructed polarisation for the
  $\pbarp \to \Cascadebar^+ \Cascade^-$ reaction.]
{Reconstructed polarisation for the
  $\pbarp \to \Cascadebar^+ \Cascade^-$ reaction at
  4\,$\gevc$. The solid line is the $\sin{2\theta_\mathrm{CM}}$ function used
  to generate the polarisation.}
\label{fig:Xipol}
\end{figure}

The analysis of polarisation is done in the same way as for the
$\LambdaLambdabar$ case described in the previous
section. Here one analyses the distribution of the $\Cascade$ decay
particles, $\Lambda$ and $\pi$, in the $\Cascade$ rest frame. The cosine
distributions of the $\Lambdabar$ distribution look quite
similar to the corresponding $\pbar$ distribution for the
$\LambdaLambdabar$ case. \Reffig{fig:Xipol} shows the
extracted polarisation, using $ P = (P^{\Cascadebar^+} +
P^{\Cascade^-})/2$. The generated polarisation is very well reproduced by
the reconstructed events when corrected for the \PANDA
acceptance.

\subsubsection{Background Reactions\label{Background}}

\begin{table*}[htb]
  \begin{center}
  \begin{tabular}{lcccc}
    \hline\hline
      Channel 1.64\,$\gevc$  & Rec.\ eff.\   &$\sigma$ [$\mu$b] &  Signal \\
    \hline
   $\pbarp \to \LambdaLambdabar$ &  0.11   & 64  & 1  & \\
   $\pbarp \to \pbarp\pi^{+}\pi^{-}$ & $1.2\cdot10^{-5}$  & $\sim10$ & $4.2\cdot10^{-5}$ \\
   \hline
   Channel 4\,$\gevc$ & & & &\\
    \hline
   $\pbarp \to \LambdaLambdabar$ & 0.23 &  $\sim50$  &  1 \\
   $\pbarp \to \pbarp\pi^{+}\pi^{-}$ & $< 3\cdot10^{-6}$ & $3.5\cdot10^{3}$ &
   $< 2.2\cdot10^{-3}$  & \\
   $\pbarp \to \Lambdabar \Sigma^0 $ & $5.1\cdot10^{-4}$ & $\sim50$ & $2.2\cdot10^{-3}$
   \\
   $\pbarp \to \Lambdabar \Sigma(1385) $ & $ < 3\cdot10^{-6}$ &  $\sim50$ &
   $ < 1.3\cdot10^{-5}$
   \\
   $\pbarp \to \Sigmabar^0 \Sigma^0 $ & $ < 3\cdot10^{-6} $ &  $\sim50$ &
   $< 1.3\cdot10^{-5}$
   \\
   \hline
   Channel 15\,$\gevc$ & & & \\
   \hline
   $\pbarp \to \LambdaLambdabar$  &    0.14     & $\sim10$ & 1& \\
   $\pbarp \to \pbarp\pip\pim $  &  $< 1\cdot10^{-6}$  &  $1\cdot10^3$  & $< 2\cdot10^{-3}$
   \\
   $\pbarp \to \Lambdabar \Sigma^0 $ & $2.3\cdot10^{-3}$ & $\sim10$ & $1.6\cdot10^{-2}$
    \\
   $\pbarp \to \Lambdabar \Sigma(1385) $ & $3.3\cdot10^{-5}$  & 60  & $1.4\cdot10^{-3}$
    \\
   $\pbarp \to \Sigmabar^0 \Sigma^0 $ & $3.0\cdot10^{-4}$ & $\sim10$  & $2.1\cdot10^{-3}$
    \\
   DPM    & $< 1\cdot10^{-6}$   & $5\cdot10^{4} $  &$< .09$ \\
   \hline
      Channel 4\,$\gevc$  & Rec.\ eff.\   &$\sigma$ ($\mu$b) &  Signal \\

   \hline
   $\pbarp \to  \Cascadebar^+\Cascade^-$ & 0.19  & $\sim2$ & 1\\
   $\pbarp \to  \Sigmabar^+(1385)\Sigma^-(1385)$& $ < 1\cdot10^{-6}$ & $\sim60$ &
   $ < 2 \cdot10^{-4}$ \\
   \hline
   \hline
  \end{tabular}
   \caption[Background for $\pbarp \to \LambdaLambdabar$
   and $\pbarp \to \Cascadebar^+\Cascade^-$]
   {Background for $\pbarp \to \LambdaLambdabar$
   and $\pbarp \to \Cascadebar^+\Cascade^-$. The
    reconstruction efficiencies give the probability for a generated
    background events to be identified as a physics event. The cross
    sections are taken from refs.~\cite{bib:phy:Barnes_164,
    bib:phy:Flaminio84} or extrapolated from the latter. The cross
    sections and branching ratios into the charged decay mode is taken
    into account in the signal number which gives the normalised
    probability for a background reaction event to be identified as a
    physics event.}
  \label{tab:hyperon_backg}
  \end{center}
\end{table*}

The \INST{PS185} experiment at \INST{LEAR} showed that the $\pbarp \to
\Lambdabar \Lambda$ reaction can be extracted with low
background near the threshold. The background was of the order of
(5-10)\percent which primarily came from quasi-free production on carbon
from the ${\rm{CH}}_{\rm{2}}$ target~\cite{bib:phy:Barnes_164}.
We therefore anticipate a much lower
background in this kinematical region when using pure hydrogen
pellets. The remaining background reactions that have been considered
for the $\pbarp \to \Lambdabar \Lambda$ channel are:

\begin{itemize}
\item[(a)]  $\pbarp \to \pbarp\pi^{+}\pi^{-}$
\item[(b)]  $\pbarp \to \Lambdabar \Sigma^0 $
\item[(c)]  $\pbarp \to \Lambdabar \Sigma(1385) $
\item[(d)] $\pbarp \to \overline{\Sigma^0} \Sigma^0 $
\item[(e)] Dual Parton Model (DPM)
\end{itemize}

These channels could potentially mimic the $\LambdaLambdabar$
channel by producing a $\pbarp\pip\pim$ system in
the final state. It should be noted that channels (b), (c) and (d) are
of interest in their own right, but are treated here as a
background. The only hyperon channel that is energetically accessible
at 1.64\,$\gevc$ is the $\LambdaLambdabar$ channel and
therefore was only reaction (a) considered at this momentum. At
4\,$\gevc$ there are several hyperons channels with a
$\LambdaLambdabar$ pair present as a result of $\Sigma^0$
decays in addition to reaction (a). The same reactions were studied at
15\,$\gevc$ together with the DPM. 1M events were generated and analysed for
reaction (a) at 1.64\,$\gevc$ and 15\,$\gevc$ and the DPM. 300k events
were generated and analysed for the other background channels at all
momenta and reaction (a) at at 4\,$\gev$. The result of this
background study is summarised in \Reftbl{tab:hyperon_backg}.

 At 1.64\,$\gevc$ we can expect a background well below 1\percent, as
anticipated from the \INST{PS185} results. The background will be somewhat
higher at 4\,$\gevc$ due to the increase of the cross section for
reaction (a) and the presence of several neutral hyperon channels. The
result is that we can expect a background of the order of a percent at
this momentum.  Due to the large inelastic cross section the largest
source of background at the highest momentum is assumed to stem from
the DPM process.  This background would not be greater than a few
percent, however.  Hyperon resonances, not considered here, could be
additional sources of background. This background and the $\Sigma^0$
channels would be suppressed by applying a $\chi^2$ test on
these hypotheses.

The $\pbarp \to \Cascadebar^+\Cascade^-$ channel involves four
well separated decay vertices.  Thus any background channel with a
different decay pattern can be suppressed imposing a
constraint on this pattern.  A channel which has the same
$\LambdaLambdabar\pip\pim$ final state is the
\begin{equation*}
  \pbarp \to  \Sigma^+(1385)\Sigma^-(1385)
\end{equation*}
reaction.  We therefore consider it as the main source of
background. This channel has one order of magnitude higher cross
section. However, the contamination from the 1 M events generated
and analysed  is negligible as
can be seen in \Reftbl{tab:hyperon_backg}. The low level of
remaining background is also confirmed in the background studies made
for the $\pbarp \to \Cascadebar^+\Cascade^-\pi^0$ channel at a somewhat
higher momentum in section~\ref{sec:phys_baryons_sim} where more
background channels were studied.

\subsubsection{Simulation Results}
This study shows that the benchmark channels $\pbarp \to
\LambdaLambdabar$ and $\pbarp \to
\Cascadebar^+\Cascade^-$ can be well reconstructed in \PANDA.  There is
acceptance over the full angular range and the whole momentum range of
\HESR for the $\LambdaLambdabar$ channel.  The same will most
likely hold true also for the $\Sigma^0$ channels due to the
kinematical similarities. There is also full CM acceptance for the
$\Cascadebar^+\Cascade^-$ channel at 4 $\gevc$, and most likely over the
full momentum range from threshold.

Acceptance corrections have to be
applied to obtain the final results due to the loss of particles in
the beam pipe direction and the loss of pions below 50 $\mevc$.  The
angular differential cross section and the polarisation can be
extracted to high precision after those corrections.\\ The count rates
will be high for the studied channels. \Reftbl{tab:count_rate}
gives the expected count rates for the benchmark channels in their
charged decay mode channels in \PANDA at a luminosity of $2\cdot10^{32}
\rm{cm^{-2}s^{-1}}$, ranging from a few 10 per second for the
$\Cascadebar^+\Cascade^-$ channel up to a thousand per second for the
$\LambdaLambdabar$ channel.

\begin{table}[htb]
  \begin{center}
  \begin{tabular}{clc}
    \hline\hline
     Momentum [$\gevc$] & Reaction & Rate [s$^{-1}$] \\
     \hline
     1.64 &  $\pbarp \to \LambdaLambdabar$ & 580 \\
     4       &  $\pbarp \to \LambdaLambdabar$ & 980 \\
              &  $\pbarp \to \Cascadebar^+\Cascade^-$                                  &   30 \\
      15   &  $\pbarp \to \LambdaLambdabar$ & 120 \\
    \hline\hline
    \end{tabular}
  \caption[Estimated count rates into their charged decay mode for the
     design luminosity]
  {Estimated count rates into their charged decay mode for the
     benchmark channels at a luminosity of $2\cdot10^{32}
     \rm{cm^{-2}s^{-1}}$}
  \label{tab:count_rate}
    \end{center}
    \end{table}

The high count rate together with the expected low background, not
more than a few percent, makes the study of antihyperon-hyperon pairs
very promising.

\subsection{Two\--Meson Production in \pbarp\--Annihilation at Large Angle}

The scale for the onset of the pQCD regime can only be deduced from
experiments, and this topic has been much discussed recently
especially with regards to the recent electromagnetic form factor
measurements at \INST{JLab}. This onset may well be process-dependent.\\

\begin{figure}[ht]
\begin{center}
\includegraphics[width=\swidth]{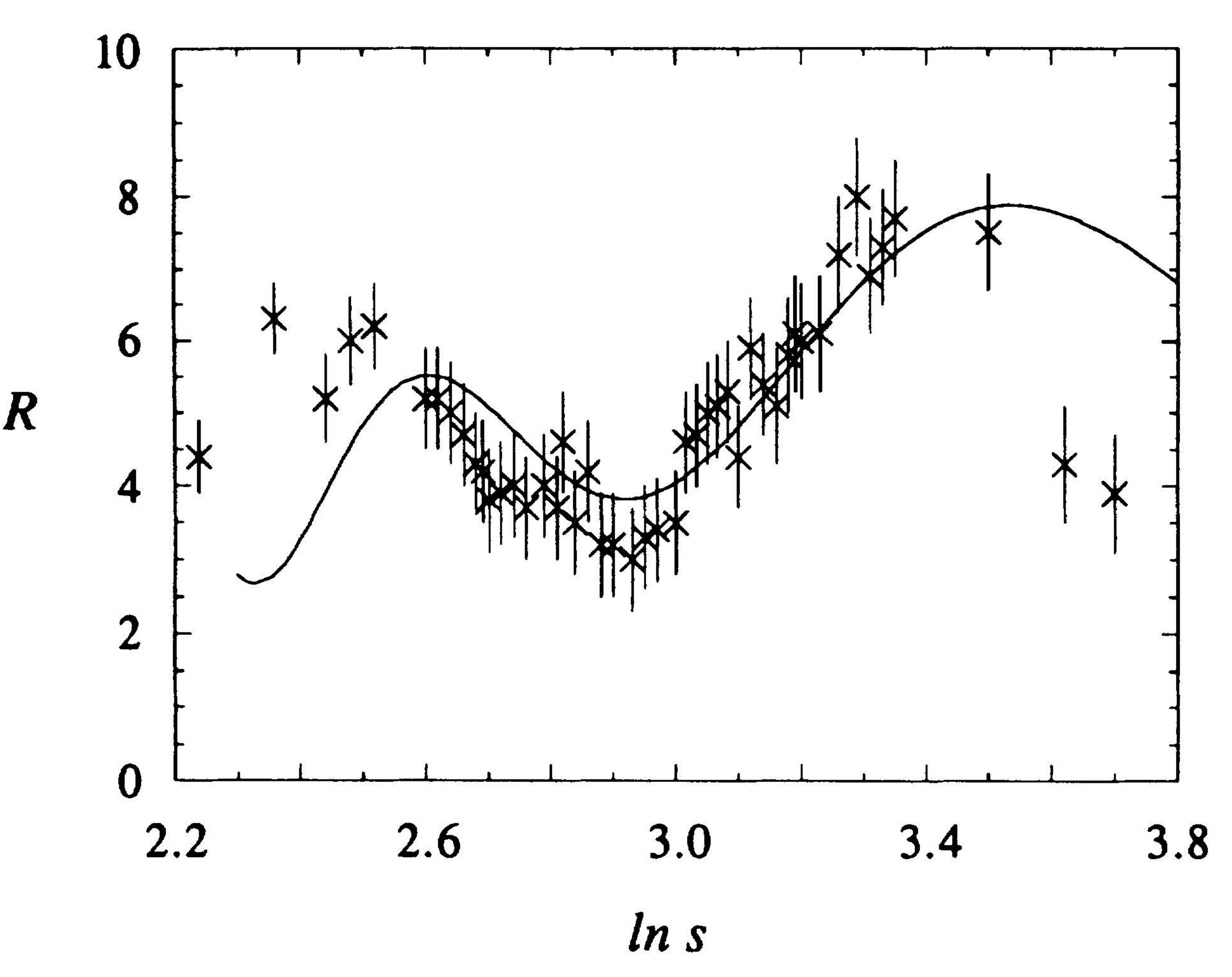}
\caption[oscillatory behaviour of elastic pp]
{$s^{10} d\sigma /dt$ as a function of $ \ln(s)$ for the pp elastic scattering
at $\theta _{CM}= 90^{\circ}$.}
\label{fig:oscill_1.pdf}
\end{center}
\end{figure}

Pions are copiously produced in \pbarp annihilation. At large
angle $(s \sim -t \sim -u \gg \Lambda^2_{QCD})$, The transition to a
perturbative QCD description is expected in the energy range covered
by \PANDA.\\

The scaling power law $s^{-n}$, where $n+2$ is the total number of
elementary constituents in the initial and final state, for exclusive
two-body hard scattering has been in the focus of high energy
scattering theory ever since the first suggestion in the early 70's of
the constituent counting rules~\cite{bib:phy:BroFar,bib:phy:Matveev}.
The subsequent hard pQCD approach to the derivation of the constituent
counting rules has been developed in late 70's-early 80's and is known
as the Efremov-Radyushkin-Brodsky-Lepage (ERBL) evolution technique
~\cite{bib:phy:EfrRad,bib:phy:LepBro}.\\

\begin{figure}[th]
\begin{center}
\includegraphics[width=0.7\swidth]{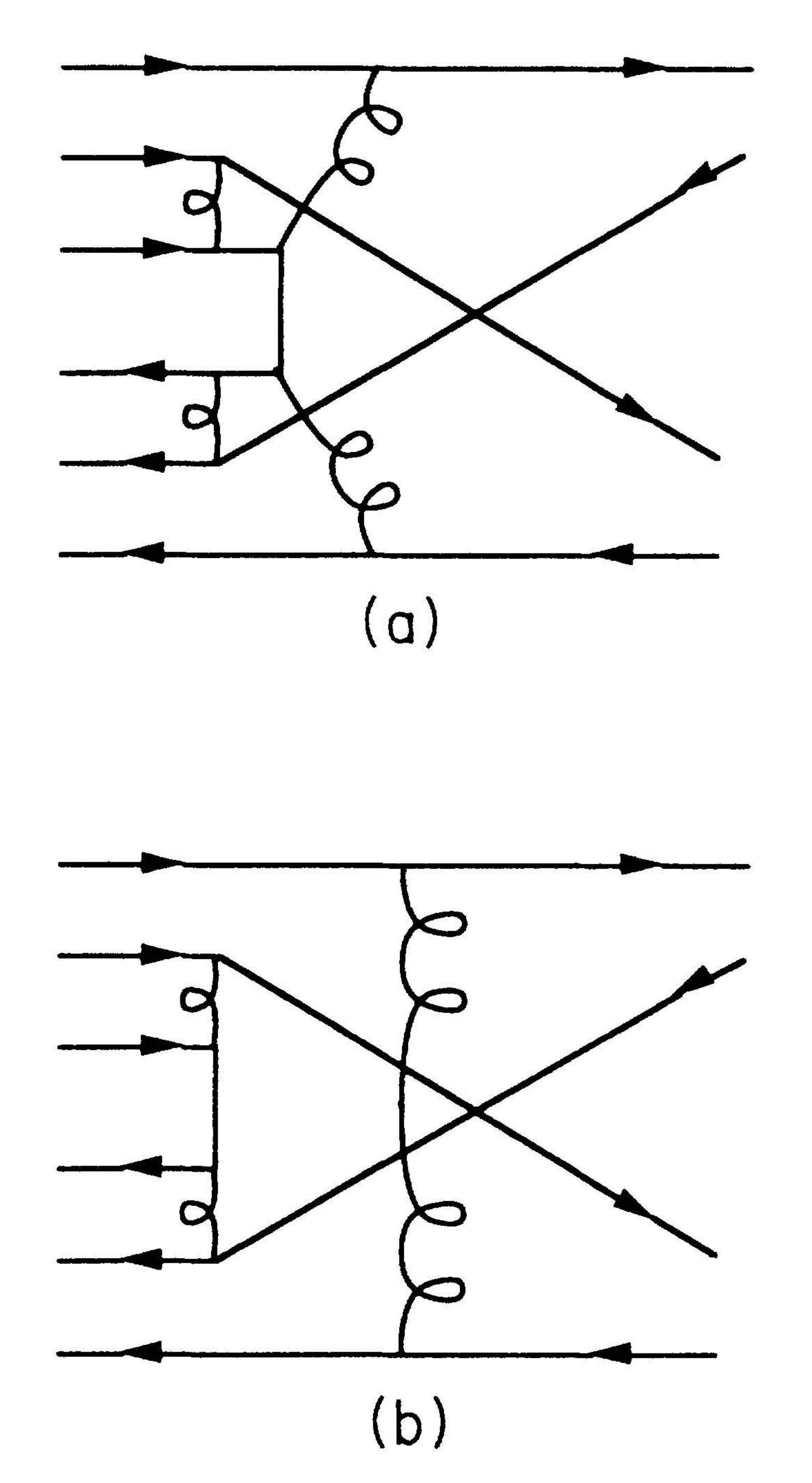}
\caption[Feynman diagram]
{Short-distance QCD diagram for the process $
  p \bar{p} \rightarrow \pi^+ \pi^-$. (a) is the Brodsky-Farrar
  mechanism while (b) represents the Landshoff one.}
\label{fig:landshoff2.pdf}
\end{center}
\end{figure}

\begin{figure}[th]
\begin{center}
\includegraphics[angle=0, width=\swidth]{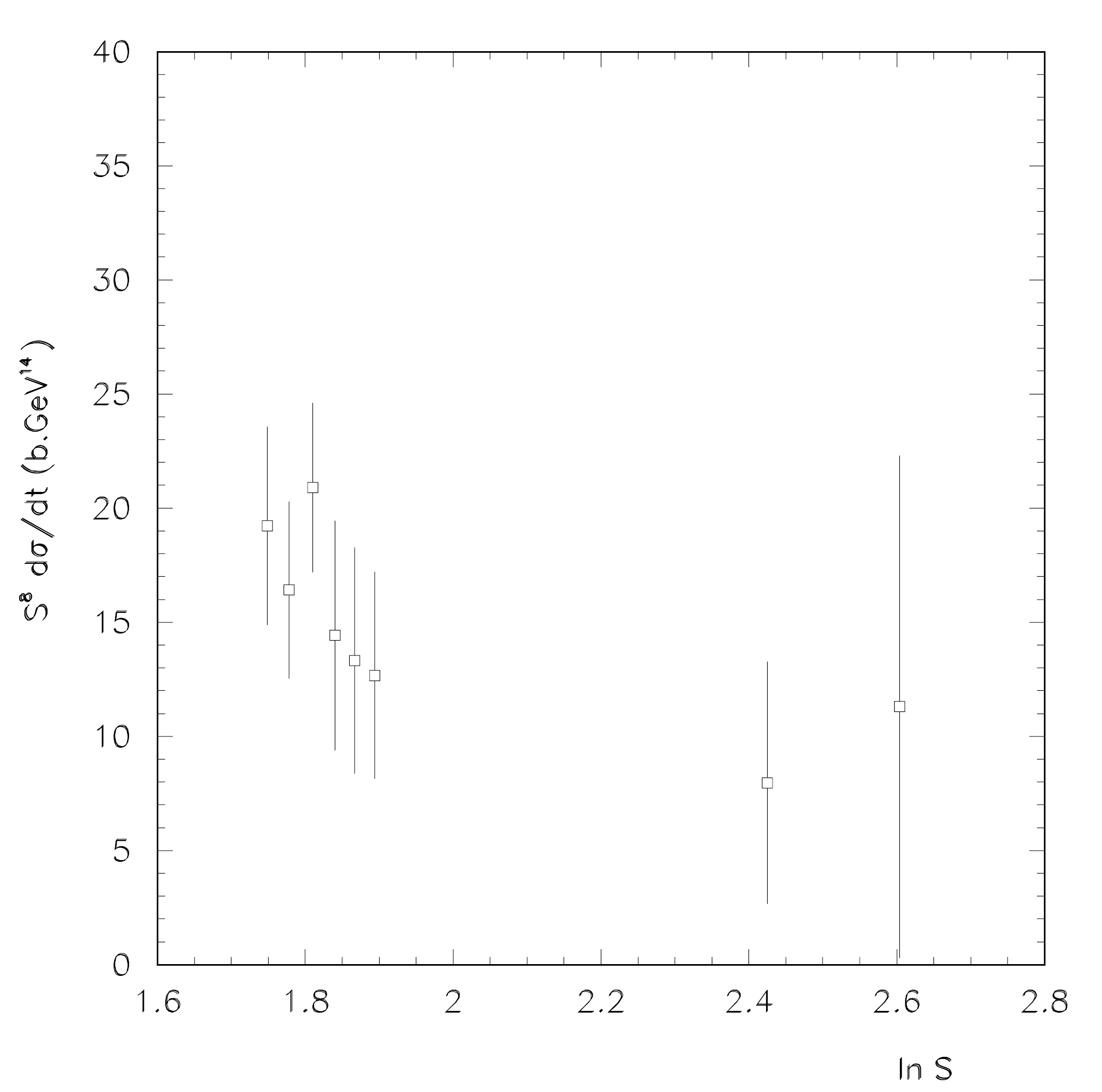}
\caption[oscillation]
{$s^{8} d\sigma /dt$ as a function of $ \ln(s)$ for the
$ p \bar{p} \rightarrow  \pi^+ \pi^-$ reaction
at $\theta _{CM}= 90^{\circ}$.}
\label{fig:pandaoscil.pdf}
\end{center}
\end{figure}

There remains an open and hot issue of the so-called Landshoff
independent scattering mechanism ~\cite{bib:phy:Landshoff} which may
be important at accessible values of $s$ in the \PANDA energy
range. Some indication comes from the experimentally observed
oscillatory s-dependence of $s^{10} d\sigma /dt$ in elastic p p
scattering (see \Reffig{fig:oscill_1.pdf}) which may be understood
as a signal of the interference of the Brodsky-Farrar and Landshoff
mechanisms~\cite{bib:phy:RalPir}. For
$p \bar{p} \rightarrow \pi^+ \pi^-$ a typical short-distance QCD
diagram and the corresponding Landshoff process are shown in
\Reffig{fig:landshoff2.pdf}~\cite{bib:phy:CarCha}. Available data
are shown in
\Reffig{fig:pandaoscil.pdf}~\cite{bib:phy:Eide,bib:phy:Eisenh,bib:phy:Buran}.
Unfortunately, the measurements are not at high enough energy to
clarify the issue.

This oscillation scenario could be investigated in \PANDA in the two
pion channel. Expected counting rates at $ s= 13.5\,\gev^2$ for one
week of beam time at full luminosity are of the order of a few $10^6$
within a $0.1$ wide $cos\theta $ bin at $\theta_{CM}=60^{\circ}$.

Spin and flavour of other two meson channels ($K \bar
K, \phi \phi, \rho \rho$) should help to understand better the
dynamics.

%% file: phys/phys_nuclearmedium.tex
%
\clearpage
\section{Hadrons in the Nuclear Medium}
\COM{Author(s): A. Gillitzer}
\COM{Referee(s): V. Metag}
%
\input{./phys/hadrons_in_medium/phys_nuclearmedium_intro}

%% file: phys/hadrons_in_medium/phys_nuclearmedium_intro.tex
%
The study of hadron properties at finite nuclear densities has a
long history. As the most fundamental aspect, the change of hadron
masses inside the nuclear medium has been proposed to reflect a
modification of the chiral symmetry breaking pattern of QCD due to
the finite density, and thus to be an indicator of changes of
quark condensates. In particular, attractive mass shifts
reflecting the reduced quark pair condensate at finite density
have been predicted for vector
mesons~\cite{bib:phy:Brown91,bib:phy:Hatsuda92}. As the ongoing
theoretical discussion on this issue (see recent review articles
like \cite{bib:phy:Mosel08}) demonstrates, the relation between
the nuclear density dependence of in-medium masses and chiral
condensates is however not a direct one. Furthermore, it is
important to note that in-medium mass shifts of hadrons,
reflecting the real part of a nuclear potential, are driven by low
energy interactions. The most significant medium effects are thus
expected for hadrons which are at rest or have a small momentum
relative to the nuclear environment, whereas the non-observation
of mass shifts at high momenta not necessarily implies the absence
of nuclear potentials. For experimental studies of hadron
in-medium properties therefore appropriate conditions of the
reaction kinematics have to be selected. Due to the released
annihilation energy of almost 2~GeV hadrons produced in
antiproton-nucleon collisions may be implanted in the nuclear
environment at much lower momenta than e.g. in proton induced
reactions.

Numerous experiments using proton-nucleus, photon-nucleus, or nucleus-nucleus
collisions have been, at least partially, devoted to deduce in-medium mass
shifts or nuclear potentials of hadrons in the light quark sector (u,d,s), both
for vector
mesons~\cite{bib:phy:Trnka05,bib:phy:Naruki:2005kd,bib:phy:Muto:2005za,bib:phy:Nasseripour07,bib:phy:Agakishiev95,bib:phy:Arnaldi:2006jq},
and for pseudo-scalar
mesons~\cite{bib:phy:Yamazaki96,bib:phy:Geissel02,bib:phy:Suzuki04,bib:phy:Messchendorp:2002au,bib:phy:Nekipelov02,bib:phy:Rudy02,bib:phy:Barth97,bib:phy:Laue99,bib:phy:Pfeiffer04}.
Only in part of the experiments the mesons were studied at low momenta relative
to the nuclear environment, where possibly significant medium effects may be
expected.

Besides the mass shift also the change of the width of hadrons inside the
nuclear medium is an important observable. In general the width will increase
due to the opening of decay channels which are not accessible in the vacuum. The
measurement of the modification of a hadron's width in the nuclear medium
therefore yields information on its inelastic interactions, which is otherwise
very difficult to access for unstable hadrons. In the case of very short-lived
mesons decaying inside the nucleus, the in-medium width may be measured
directly. As an alternative approach, the measurement of the transparency ratio
determined from the production cross section with different target nuclei allows
to deduce the in-medium width of
mesons~\cite{bib:phy:Kotulla:2008xy,bib:phy:Ishikawa:2004id}.

For some observables, the effect of the nuclear medium becomes
surprisingly small: this is the Colour transparency phenomenon
which is related to the gauge nature of strong interactions. The
study of exclusive hard reactions in a nuclear medium constitutes
a way to test the factorisation of a short distance process
occurring between colour neutral objects whose transverse size is
controlled by the hardness of the collision (measured by the
momentum transfer $t$). Moreover the nuclear medium may even be
able to filter out the non short distance dominated part of the
amplitude. These phenomena and their possible study with \PANDA is
discussed in section \ref{sec:CT}.

\subsection{In-Medium Properties of Charmed Hadrons}

The energy range of the \INST{HESR} and the detection capabilities of the
\PANDA detector in principle allow to extend studies of the
in-medium properties of hadrons into the charm sector by using
antiproton-nucleus collisions as entrance channel. The in-medium
properties of both $D$ mesons and charmonium states have been
studied theoretically in different approaches (for a review on the
earlier work see \cite{bib:phy:Weise01}). In analogy to the
$K/\bar{K}$ splitting in nuclear matter phenomenological estimates
using a quark-meson coupling
model~\cite{bib:phy:Tsushima99,bib:phy:Sibirtsev99} predict an
in-medium mass splitting between $\bar{D}$ and $D$ mesons of about
100\,MeV, and an attractive $D$ meson ($D^+$, $D^0$) potential
with a depth of more than $-100$\,MeV at normal nuclear density
$\rho{}=\rho_0$. A downward shift of the $ \DDbar$ average mass by
about 50\,MeV was obtained in a QCD sum rule
estimate~\cite{bib:phy:Hayashigaki00}. The QCD sum rule analysis
of Refs.~\cite{bib:phy:Weise01,bib:phy:Morath01} predicts a
$D^+-D^-$ mass splitting of more than 50\,MeV with the a downward
shift of the $D^+/D^-$ average mass by about the same order of
magnitude at $\rho{}=\rho_0$. A more involved coupled channel
approach~\cite{bib:phy:Tolos04,bib:phy:Lutz06,bib:phy:Mizutani06}
reveals a complicated structure of $D^+$ mesons (and equivalently
$D^0$ mesons) which is not appropriately described by two simple
parameters denoting in-medium mass and width, respectively. Inside
the nuclear medium the inelastic charm exchange channels
$DNN\rightarrow{}\Lambdac N$ and $DN\rightarrow{}\pi\Lambdac$ are
open at threshold resulting in a low mass component reflecting
$\Lambdac$-hole excitation. Nuclear binding effects for charmonium
have been studied theoretically almost 20 years
ago~\cite{bib:phy:Brodsky90}, but the binding effect was later
found to be rather small, namely between 5 and 10\,MeV for $\jpsi$
and $\etac$ states~\cite{bib:phy:Brodsky97,bib:phy:Klingl99}.
Large attractive mass shifts of $\sim{}100$\,MeV in normal nuclear
matter were predicted for higher-lying charmonium states resulting
from the QCD second order Stark effect due to the change of the
gluon condensate~\cite{bib:phy:Lee03,bib:phy:Lee03a}.

Different experimental methods were discussed to observe signals of the $D$
meson mass modification in nuclear matter. It was
proposed~\cite{bib:phy:Sibirtsev99} to study subthreshold production of $D$ and
$\bar{D}$ mesons in antiproton-nucleus collisions, where reduced in-medium
masses should be visible in an enhancement of the production cross section. A
lowering of the in-medium $D\bar{D}$ threshold has been predicted to result in a
dramatic increase of the $\psi(2S)$ and $\psi(3770)$ width, since in free space
these states are rather narrow due to their vicinity to the $D\bar{D}$
threshold~\cite{bib:phy:Hayashigaki00}. Using a Multiple Scattering Monte Carlo
approach~\cite{bib:phy:Golubeva03} it was however found that the collisional
width of these charmonium states in the nuclear medium is much larger than their
width in free space already for unchanged $D/\bar{D}$ masses. Taking the
collisional width and $D/\bar{D}$ re-scattering into account, no effect of
attractive nuclear potentials survived in the $\DDbar$ channel, and only a very
small effect in the di-lepton channel.

It has also been proposed to study the elementary $DN$ and $\bar{D}N$
interaction more directly in $\pbard$ collisions by using the spectator nucleon
in the deuteron as secondary target, as discussed for the elastic $\DDbar N$
cross section in \cite{bib:phy:Cassing00}. Inelastic channels with charm
exchange like $\pbard\rightarrow{}\Lambdac N(\pi)$ or
$\pbard\rightarrow{}\Sigma_cN(\pi)$ might also be studied. Recently, the
$\bar{D}N$ interaction with its energy dependence was studied in a combined
meson exchange and quark-gluon dynamics approach~\cite{bib:phy:Haidenbauer07}.
It was found that the $\bar{D}N$ cross section is by about a factor two larger
than the $KN$ cross section within the range up to about 150\,MeV above
threshold. In the nucleon rest frame this corresponds to a $\bar{D}$ momentum of
about 1.4\,\gevc.

Due to the high mass of charmed hadrons, it is very difficult to
realise the conditions at which their medium properties are
experimentally accessible. For $D$ mesons produced in direct
annihilation processes in antiproton-nucleus collisions the
condition of low momentum, which is required to be sensitive to
nuclear medium effects, is kinematically not fulfilled. For
example, at threshold the $D$/$\bar{D}$ meson momentum is
3.2\,\gevc. Lower $D$/$\bar{D}$ momenta can be reached by using
backward production at higher beam energies, but at the highest
\INST{HESR} beam momentum of 15\,\gevc the minimum $D$/$\bar{D}$ momentum
is still 1.67\,\gevc. The interesting regime of momenta below
1\,\gevc can only be reached in complicated two- or multi-step
reactions with correspondingly small cross sections.
Qualitatively, the same holds for charmonium states produced in
$\pbarA$ collisions. Presently it is neither known to which extent
the discussed observables are still sensitive to nuclear
potentials at high $D/\bar{D}$ or charmonium momenta relative to
the nuclear medium, or if more complicated processes slowing down
charmed hadrons inside a nucleus can be experimentally identified.
Therefore the study of possible mass modifications of charmed
hadrons in nuclear matter is considered as a long term physics
goal based on further theoretical studies on the reaction
dynamics, and on the exploration of the experimental capability to
identify more complicated processes. It will however not be in the
focus of the physics program during the first years of \PANDA
operation.

\subsection{Charmonium Dissociation}
\label{sec:cc-dissoc}

Apart from the determination of nuclear potentials of charmed
hadrons, specific well-defined problems exist to which the study
of antiproton-nucleus collisions with \PANDA can contribute
valuable information. As an important issue in this respect we see
the still open question of the $\jpsi{}N$ dissociation cross
section. This cross section is as yet experimentally unknown,
except for indirect information deduced from high-energy $\jpsi$
production from nuclear targets. A $\jpsi{}N$ cross section
$\sigma_{\jpsi{}N}=3.5\pm{}0.8$\,mb has been deduced by measuring
$\jpsi$ photo-production from nuclei with a mean photon energy of
17\,\gevc~\cite{bib:phy:Anderson77}. The $\jpsi$ momentum in the
nuclear rest frame was not explicitly determined. In
\cite{bib:phy:Kharzeev97} the authors analysed proton-nucleus
collisions with beam energies between 200 and 800\,\gevc and found
as absorption cross section $\sigma=7.3\pm{}0.6$\,mb, also without
explicitly specifying the $\jpsi$ momentum in the nuclear rest
frame. In an analysis of recent measurements of $\jpsi$ and
$\psiprime$ production in $pA$ collisions at 400 and
450\,\gevc~\cite{bib:phy:Alessandro:2003pc,bib:phy:Alessandro06}
the NA50 Collaboration finds $\sigma_{abs}(\jpsi)=4.6\pm{}0.6$\,mb
and $\sigma_{abs}(\psiprime)=10.0\pm{}1.5$\,mb. The authors of
\cite{bib:phy:Alessandro06} explicitly mention the energy
dependence of $\jpsi$ and $\psiprime$ absorption as a still open
question. Recently, nuclear shadowing effects of $\jpsi$ mesons
were also studied in $d+{\rm{Au}}$ collisions in the \INST{PHENIX}
experiment at \INST{RHIC} at the much higher energy
$\sqrt{s_{NN}}=200$\,\gevc, resulting in a deduced $\jpsi$ breakup
cross section of
$\sigma_{\rm{breakup}}=2.8^{+1.7}_{-1.4}$\,mb~\cite{bib:phy:Adler:2005ph,bib:phy:Adare:2007gn}.

Apart from being a quantity of its own interest, the $\jpsi{}N$
dissociation cross section is closely related to the attempt of
identifying quark-gluon plasma (QGP) formation in
ultra-relativistic nucleus-nucleus collisions. A significant
additional, so-called anomalous suppression of the $\jpsi$ yield
in high-energy nucleus-nucleus collisions had been predicted due
to colour screening of $c\bar{c}$ pairs in a QGP
environment~\cite{bib:phy:Matsui86}. In fact, the \INST{CERN-SPS}
experiments have observed a $\jpsi$ suppression effect increasing
with the size of the interacting nuclear system, and interpreted
this as signature for QGP
formation~\cite{bib:phy:Alessandro05,bib:phy:Arnaldi05,bib:phy:Arnaldi:2007zz}.
The validity of such an interpretation is however based on the
knowledge of the ''normal'' suppression effect due to $\jpsi$
dissociation in a hadronic environment. Nuclear $\jpsi$ absorption
can so far only be deduced from
models~\cite{bib:phy:Sibirtsev:2000aw,bib:phy:Oh07,bib:phy:Hilbert07}
since the available data do not cover the kinematic regime
relevant for the interpretation of the $\jpsi$ suppression effect
seen in the \INST{SPS} heavy ion data, since {\it e.g.} $\jpsi{}N$
dissociation processes will in general occur at higher relative
momenta in 400\,\gevc $pA$ collisions than in 158\,\gevc/u Pb~+~Pb
collisions with partial stopping of the nuclear matter.
Ref.~\cite{bib:phy:Oh07} gives a range from $\sim{}1$\,mb to
$\sim{}7$\,mb for the uncertainty of the estimated values of the
$\jpsi$-nucleon cross section. In cold nuclear matter $\jpsi$
mesons can dissociate via the reaction
$\jpsi{}N\rightarrow{}\bar{D}\Lambdac$ at
$p_{\rm{}lab}\ge{}1.84$\,\gevc, whereas dissociation via
$\jpsi{}N\rightarrow{}\DDbar N$ requires
$p_{\rm{}lab}\ge{}5.17$\,\gevc. Therefore the $\jpsi{}N$
dissociation cross section will be strongly momentum dependent,
and it is important to supply experimental information on this
dependence particularly at lower momenta.

\subsection{\jpsi$N$ Dissociation Cross Section in $\pbar A$ Collisions}

In antiproton-nucleus collisions the $\jpsi{}N$ dissociation cross section can
be determined for momenta around 4\,\gevc with very little model dependence, in
contrast to its values deduced from the previous studies as discussed in section
\ref{sec:cc-dissoc}. The $\jpsi$ momentum inside the nuclear medium is
constrained by the condition that the $\pbarp\rightarrow{}\jpsi$ formation
proceeds 'on resonance' with a target proton ($p_{\pbar}=4.1$\,\gevc). The
determination of the $\jpsi{}N$ dissociation cross section is, in principle,
straight forward: the $\jpsi$ production cross section is measured for different
target nuclei of mass number ranging from light ($d$) to heavy (Xe or Au), by
scanning the $\pbar$ beam momentum across the $\jpsi$ yield profile whose width
is essentially given by the internal target nucleon momentum distribution. The
internal nuclear momentum distribution is sufficiently well known. The $\jpsi$
is identified by its decay branch to $\mumu$ or $\ee$. The attenuation of the
$\jpsi$ yield per effective target proton as a function of the target mass is a
direct measure for the $\jpsi{}N$ dissociation cross section, which can be
deduced by a Glauber type analysis. Note that a study of $\pbard$ collisions 'on
the $\jpsi$ resonance' allows an exclusive measurement of the final state, and
thus with the spectator neutron as secondary target should give direct access to
the cross section for specific $\jpsi{}\mbox{n}$
reactions~\cite{bib:phy:Cassing00}.

\begin{figure}[htb]
\begin{center}
\includegraphics[width=\swidth]{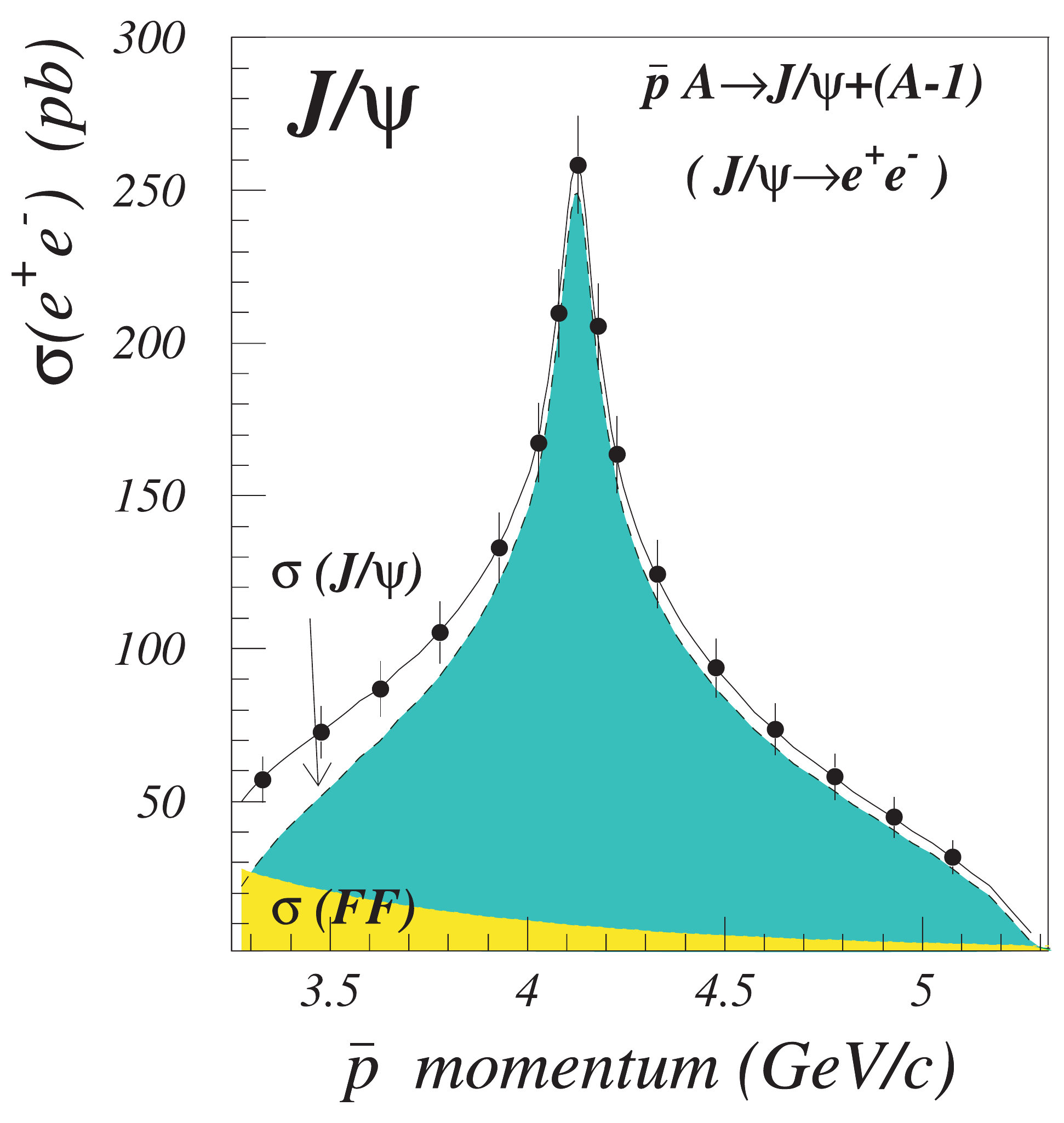}
\caption[Simulated cross section for resonant $\jpsi$ production
on nuclear protons]{Simulated cross section for resonant $\jpsi$
production on nuclear protons with internal Fermi momentum
distribution as a function of the antiproton
momentum~\cite{bib:phy:Seth01}.} \label{fig:jpsi-scan}
\end{center}
\end{figure}

In a second step these studies may be extended to higher charmonium states like
the $\psiprime$ ($\psi(2S)$) which requires $\pbar$ momenta around 6.2\,\gevc
for resonant production. This would also allow to determine the cross section
for the inelastic process
$\psiprime{}N\rightarrow{}\jpsi{}N$~\cite{bib:phy:Gerland05} which is also
relevant for the interpretation of the ultra-relativistic heavy ion data.
Measurement of $\psiprime$ production on nuclear targets is more difficult since
the $\psiprime$ yield will be considerably smaller than that of $\jpsi$.
Neglecting absorption, the estimated ratio of the production cross sections is
$\sigma_{\pbarA\rightarrow{}\psiprime{}X}/
\sigma_{\pbarA\rightarrow{}\jpsi{}X}\simeq{}0.03$, based on the Breit-Wigner
formula with the known~\cite{bib:phy:PDG06} widths and $\pbarp$ branching
ratios. Absorption effects will further reduce this ratio, since due to the
larger size of the $\psiprime$ a larger absorption cross section is expected
than for $\jpsi$. At low momenta in the nuclear rest frame the lower thresholds
for dissociation processes will also play an important role and enhance the
ratio of $\psiprime$ to $\jpsi$ absorption. In contrast to the $\jpsi{}N$ system
as discussed above, the $\psiprime{}N\rightarrow{}\bar{D}\Lambdac$ channel is
already open at threshold, and the dissociation via $\jpsi{}N\rightarrow{}\DDbar
N$ is also open at the $\pbar$ momenta chosen for resonant $\psiprime$
production (in the nucleon rest frame the $\psiprime$ threshold momentum for
$\DDbar$ dissociation is 1.28\,\gevc).

\subsubsection*{Benchmark Channels}

The simulation studies for this report, in the context of antiproton nucleus
collisions, focus on aspects relevant for the determination of the
$\jpsi$-nucleon dissociation cross section. The reaction studied is:
\[
\pbar\,^{40}{\rm{}Ca}\rightarrow{}\jpsi{}\,X\rightarrow{}\ee\,X
\]
The incident $\pbar$ momentum is 4.05\,\gevcc, corresponding to resonant
$\jpsi$ formation with a proton at rest in the target nucleus. Goal of the
simulations is to study the identification of the $\jpsi$ signal in $\ee$pairs,
and to explore the suppression of presumed dominant background channels. No
attempt is made at the present stage to simulate the full experiment required to
determine the $\jpsi{}N$ dissociation cross section, and to estimate its
achievable statistical and systematic errors. The full experiment will comprise
the measurement of absolute cross sections for $\jpsi$ production on a series of
target nuclei ranging from very light to heavy ({\it e.g.} $^2{\rm{}H}$, N, Ne,
Ar, Kr, Xe) including a $\pbar$ momentum scan across the $\jpsi$ excitation
function in each case.

A dedicated event generator for the reaction
$\pbar\,A\rightarrow{}\jpsi{}\,X$~\cite{bib:phy:Sibirtsev08-eg}
has been used to generate 80 thousand events of the signal
channel. The event generator includes realistic Fermi momentum
distributions and average nuclear binding effects of the nuclear
target protons, as well as $\pbar$ and $\jpsi$ absorption in the
nuclear medium.

Various criteria are conceivable to select the
$\jpsi\rightarrow{}\ee$ signal events based on the two-body
kinematics of the process with a slow target nucleon and on the
identification of the \ee pair. The conditions selected in the
simulation studies are given in detail later.

\subsubsection*{Background Reactions}

As compared to resonant $\jpsi$ formation with an antiproton hitting a free
target proton at a cross section of $\sim{}5\,\mu$b, $\jpsi$ production on a
nucleus is reduced roughly by a factor 1000 due to the nuclear Fermi momentum,
and thus the peak cross section is estimated to be a few nb. In contrast, the
total antiproton-nucleus cross section, dominated by annihilation and inelastic
hadronic processes on target nucleons, is approximately given by the geometrical
cross section of the order of 1\,b. Taking into account the $\sim{}6\percent$
branching for the $\ee$ decay, the rate of hadronic background reactions is
almost 10 orders of magnitude larger than that of the $\jpsi$ signal. Obviously,
it is not possible to simulate the detector response for a sample of unspecific
background events which is large enough to test background suppression at a
relative signal level below $10^{-9}$. The background suppression and signal
detection capability can therefore only be estimated by using extrapolations
based on certain assumptions.

Annihilation of antiprotons with one of the nuclear target protons
into \pip\pim pairs without creation of other particles or high
momentum transfer to target nucleons is considered to be the most
dangerous background channel to a di-leptonic $\jpsi$ signal
because of a possible misidentification of \mbox{\pip\pim} pairs
as \ee pairs. A suppression of these background events by
kinematic selection criteria is not possible. Experimental data on
the reaction $\pbarp\rightarrow{}\pip\pim$ at energies above the
\INST{LEAR} energy range are scarce. Ref~\cite{bib:phy:Flaminio84}
lists a total cross section of $7\pm{}5\,\mu$b at
$p_{\bar{p}}=4\,\gevc$. Angular distributions for
$\pbarp\rightarrow{}\pip\pim$ have been measured up to $\pbar$
momentum $p_{\pbar}=0.78$\,\gevc~\cite{bib:phy:Tanimori85} whereas
for $\pbarp\rightarrow{}\pi^0\pi^0$ differential cross sections at
total $\pbarp$ centre-of-mass energies in the charmonium mass
range have been measured by the \INST{Fermilab} \INST{E760} and
\INST{E835} experiments for part of the angular
range~\cite{bib:phy:Armstrong:1997gv,bib:phy:Andreotti:2003sk}.

Samples of $26.4$ million background events for 4.05\,\gevc $\pbar$ on $^{40}$Ca
have been created by using the UrQMD event
generator~\cite{bib:phy:UrQMD1,bib:phy:UrQMD2} in the standard version ({\it
i.e.} statistical de-excitation of the system by emission of low energy
particles is neglected). This sample is representative for unspecific background
in $\pbar\,A$ collisions. Since it is difficult to decide which event patterns
create background in the di-leptonic $\jpsi$ signal at a level below $10^{-9}$
no filter to the UrQMD events has been applied before propagation through the
detector. In addition a sample of $3\cdot{}10^7$ $\pbarp\rightarrow\pip\pim$
background events was generated. Since at the considered energy no experimental
data on \pip\pim angular distributions in $\pbarp$ collisions exist, the
parametrisation of the $\pi^0\pi^0$ angular distribution measured by the
\INST{E835} experiment was used~\cite{bib:phy:Andreotti:2003sk}.

\subsubsection{Simulation Results}

\begin{figure}[htb]
\begin{center}
\includegraphics[height=\swidth,angle=90]{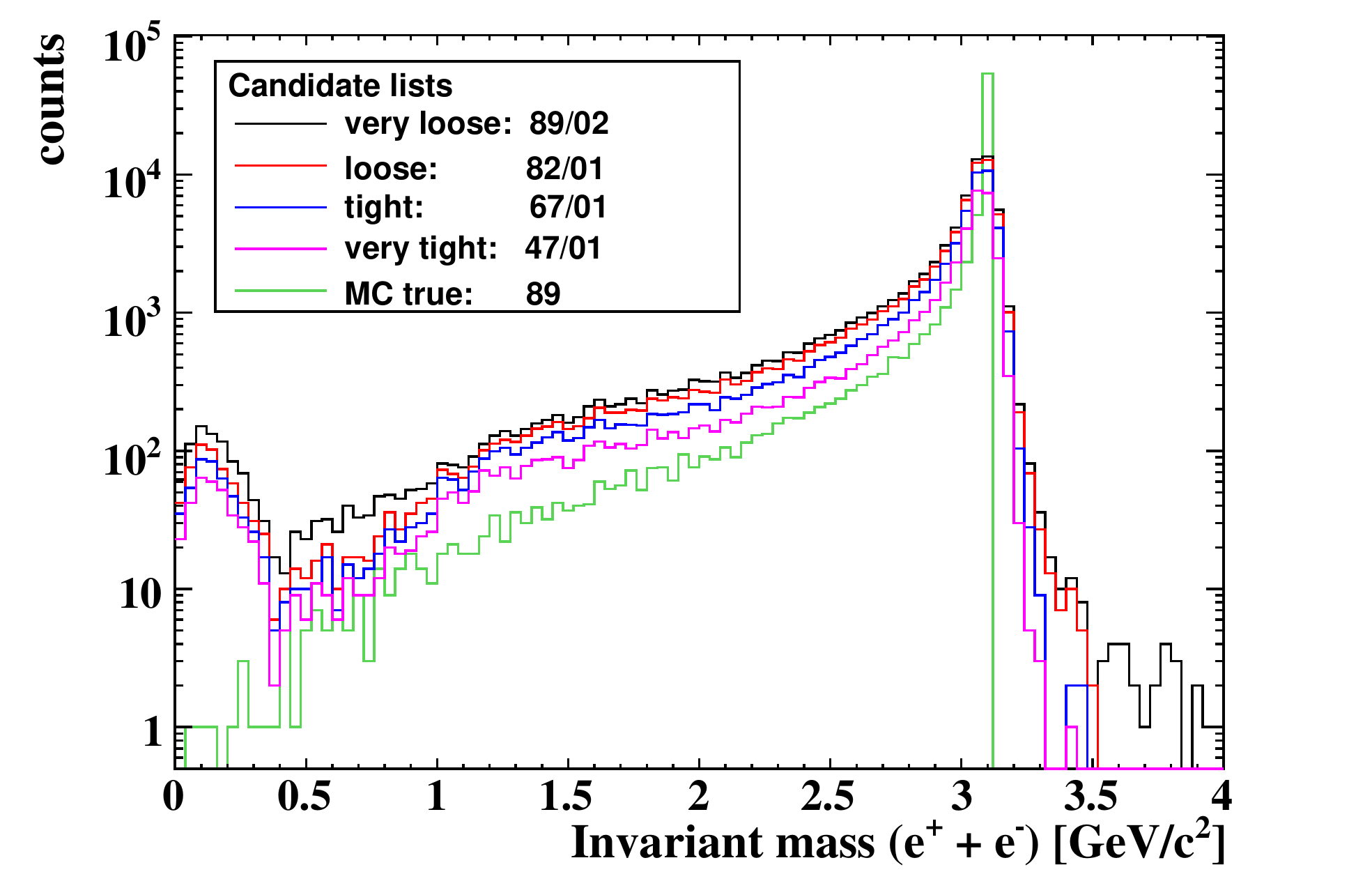}
\includegraphics[height=\swidth,angle=90]{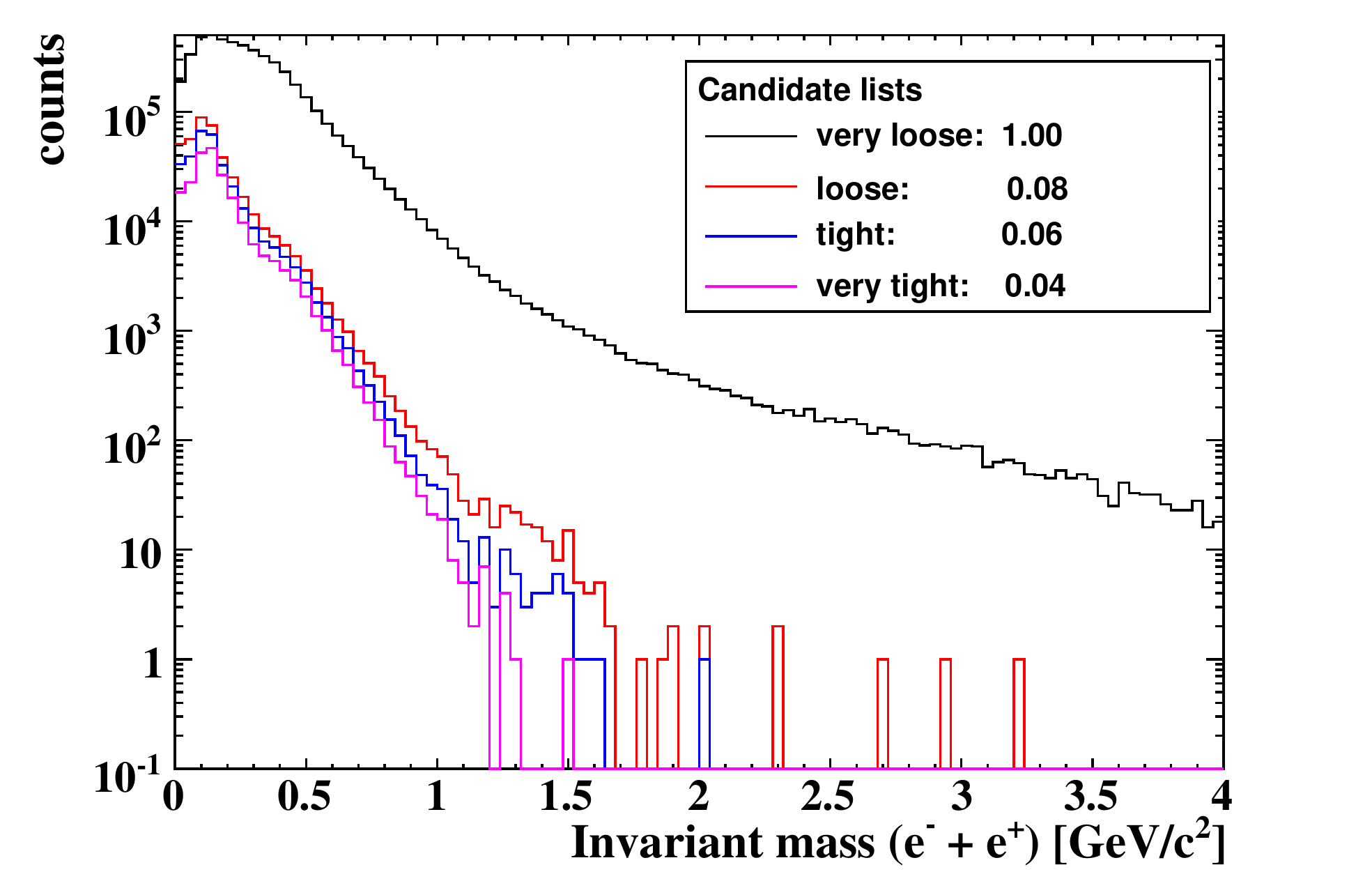}
\caption[Reconstructed invariant mass distribution of $\ee$ pair
candidates for signal and background]{Reconstructed invariant mass
distribution of $\ee$ pair candidates for the signal events (upper
panel) and for the UrQMD background events (lower panel).}
\label{fig:ee-mass-sb}
\end{center}
\end{figure}

The first step in the analysis is the computation of the invariant mass of
selected pairs of $e^-$ and $e^+$ candidates. Four $e^{\pm}$ candidate lists
have been defined - the \vloo, \loo, \tig, and \vtig lists. In
\Reffig{fig:ee-mass-sb} the $\ee$ invariant mass distributions using the
different lists are displayed.

The upper panel contains the signal events, whereas the lower
panel shows the corresponding distributions obtained for the UrQMD
background events. The two numbers given in the legend of the
upper panel are the percentages of reconstructed true and and fake
$\jpsi$ mesons, respectively, at a minimum reconstructed mass of
2.0\,\gevcc.

The green line, labelled with {\em MC true}, represents the true
mass distribution for the \vloo list. The reconstruction
efficiency decreases from $\sim{}89\percent$ with the \vloo list
to $\sim{}47\percent$ with the \vtig list. With the available
statistics the number of fake $\jpsi$ meson candidates in the
considered invariant mass region is zero for all cases. In the
lower panel, showing the background distributions, the number in
the legend is the fraction of reconstructed fake $\jpsi$ mesons
with respect to the number obtained with the \vloo list. The mass
distribution decreases strongly with increasing mass. Except for
the \vloo  list there are only few background events in the mass
range above 2\,\gevcc. With the available number of simulated
background events, using the \loo list $5$ background events
survive in this mass range. With the \tig list $1$ event, with the
\vtig list no background event is found above 2\,\gevcc.

\begin{figure*}[htb]
\begin{center}
\includegraphics[height=\swidth,angle=90]{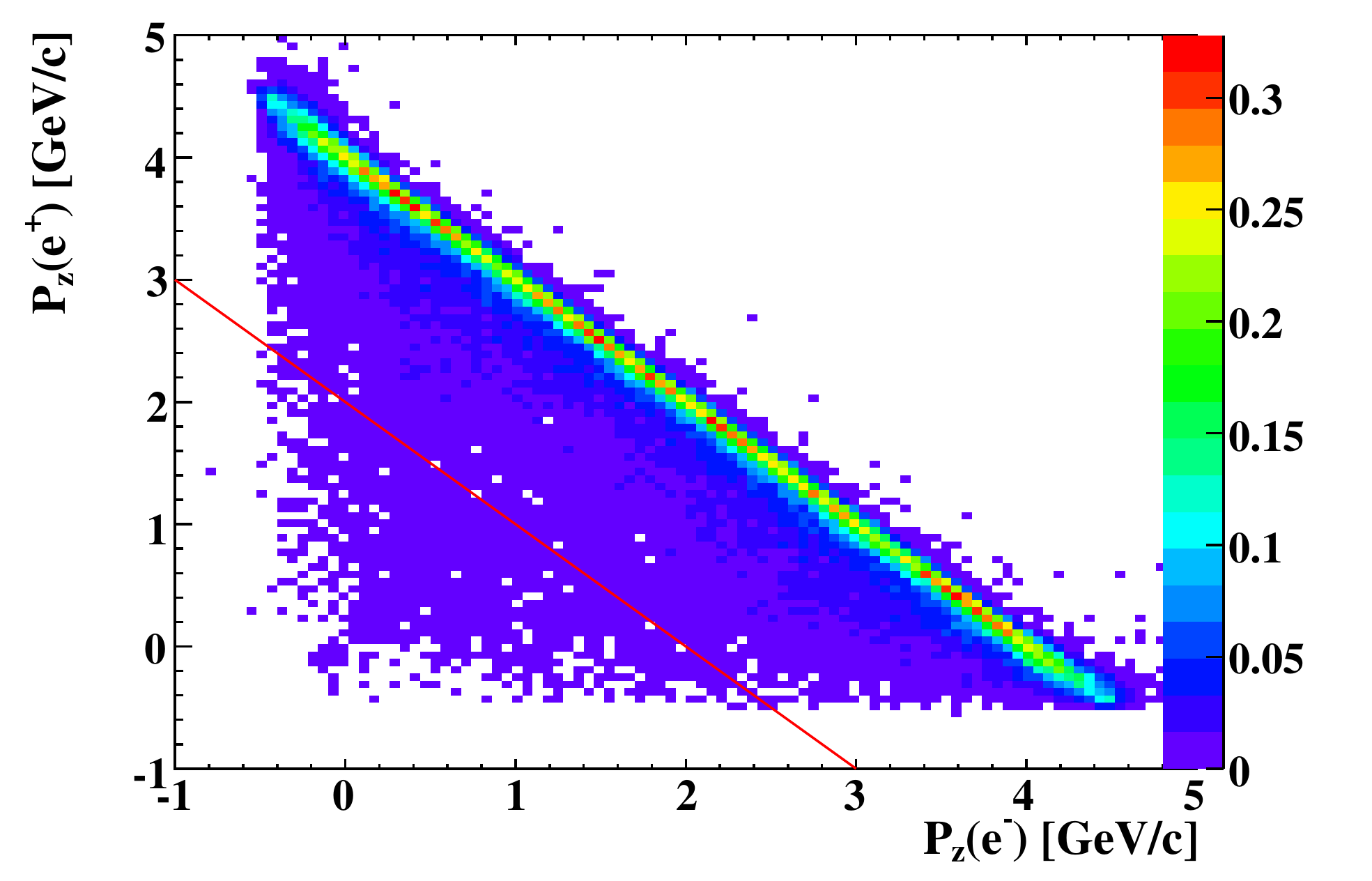}
\hfill
\includegraphics[height=\swidth,angle=90]{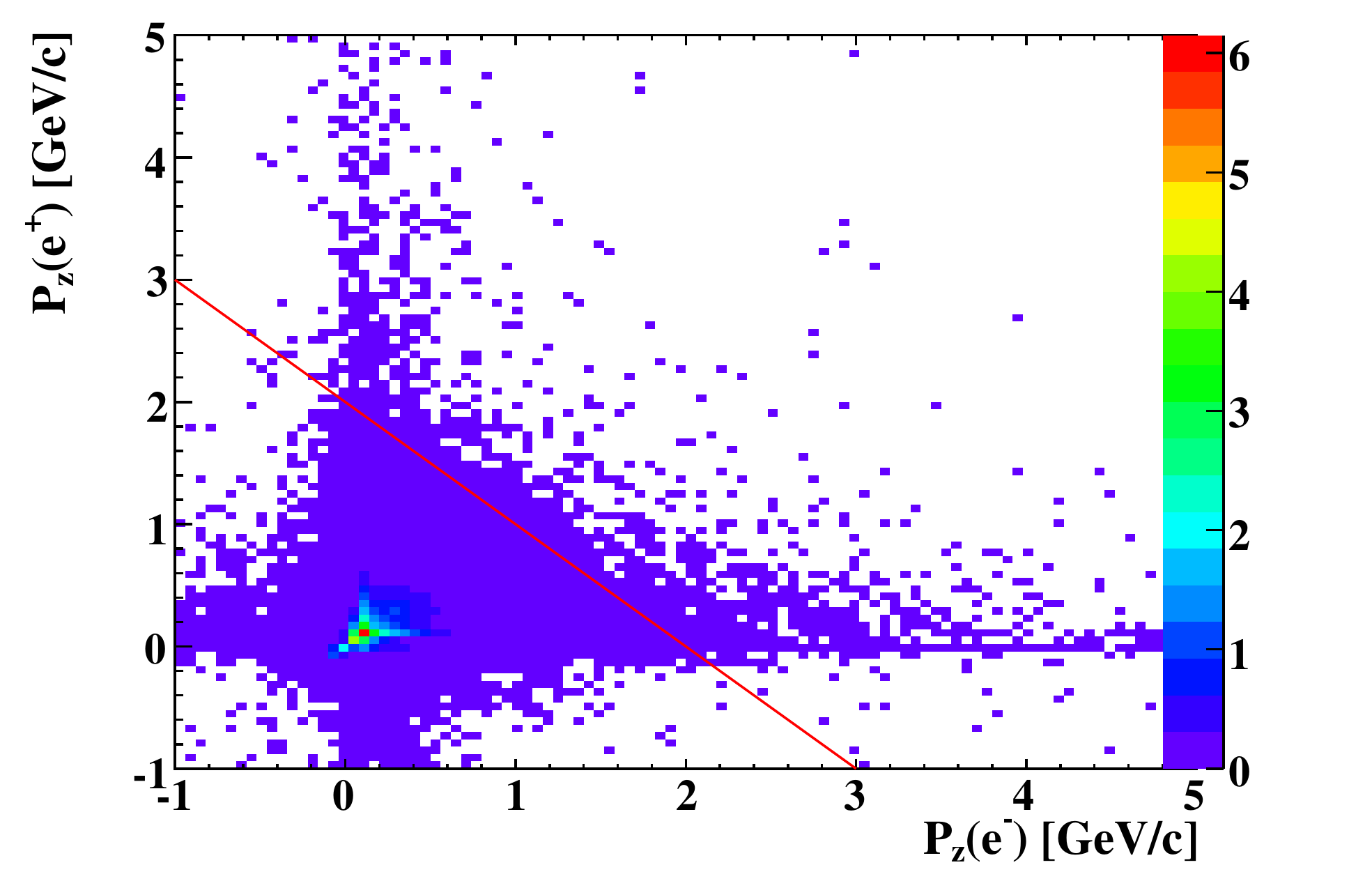}
\includegraphics[height=\swidth,angle=90]{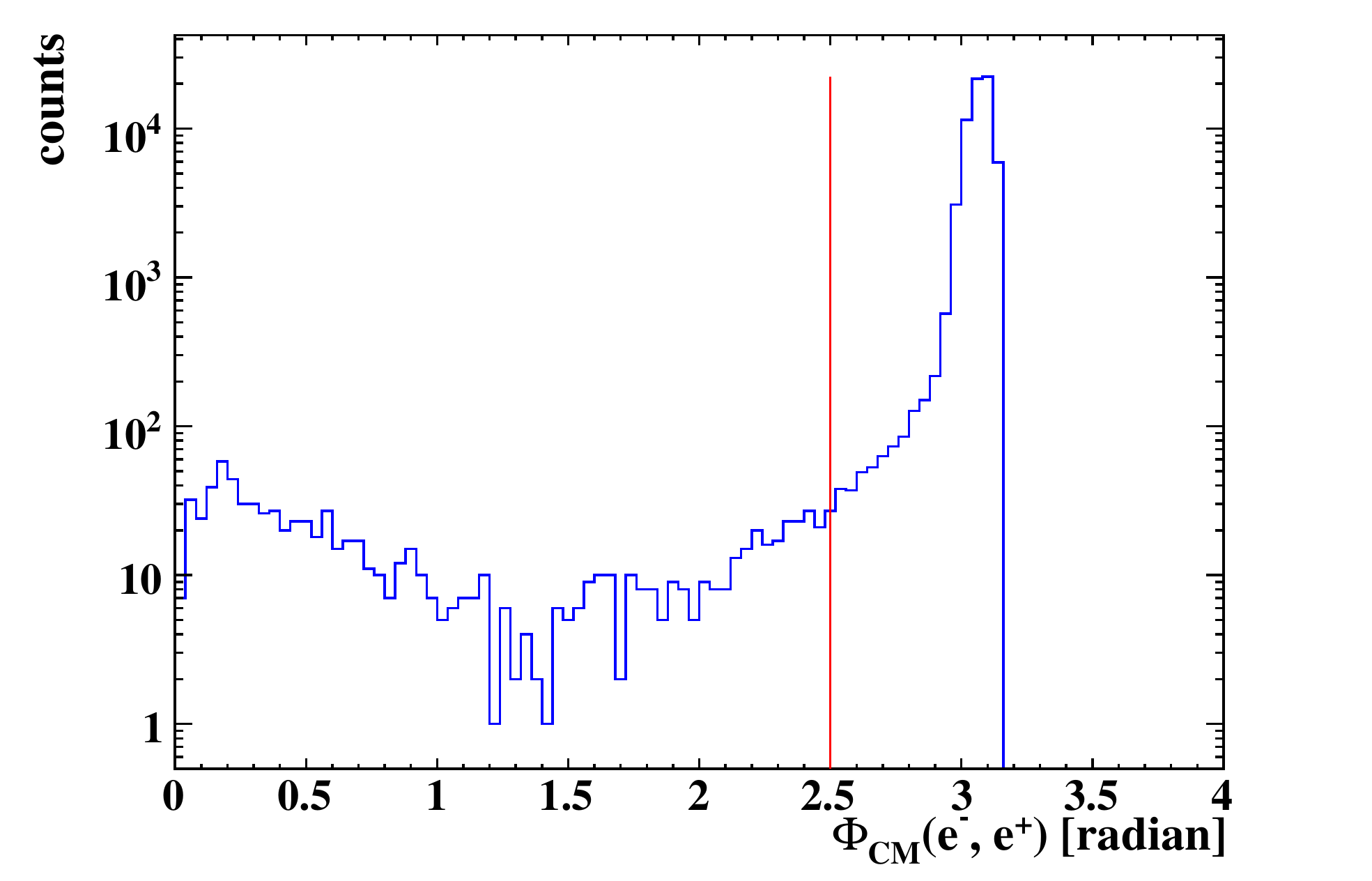}
\hfill
\includegraphics[height=\swidth,angle=90]{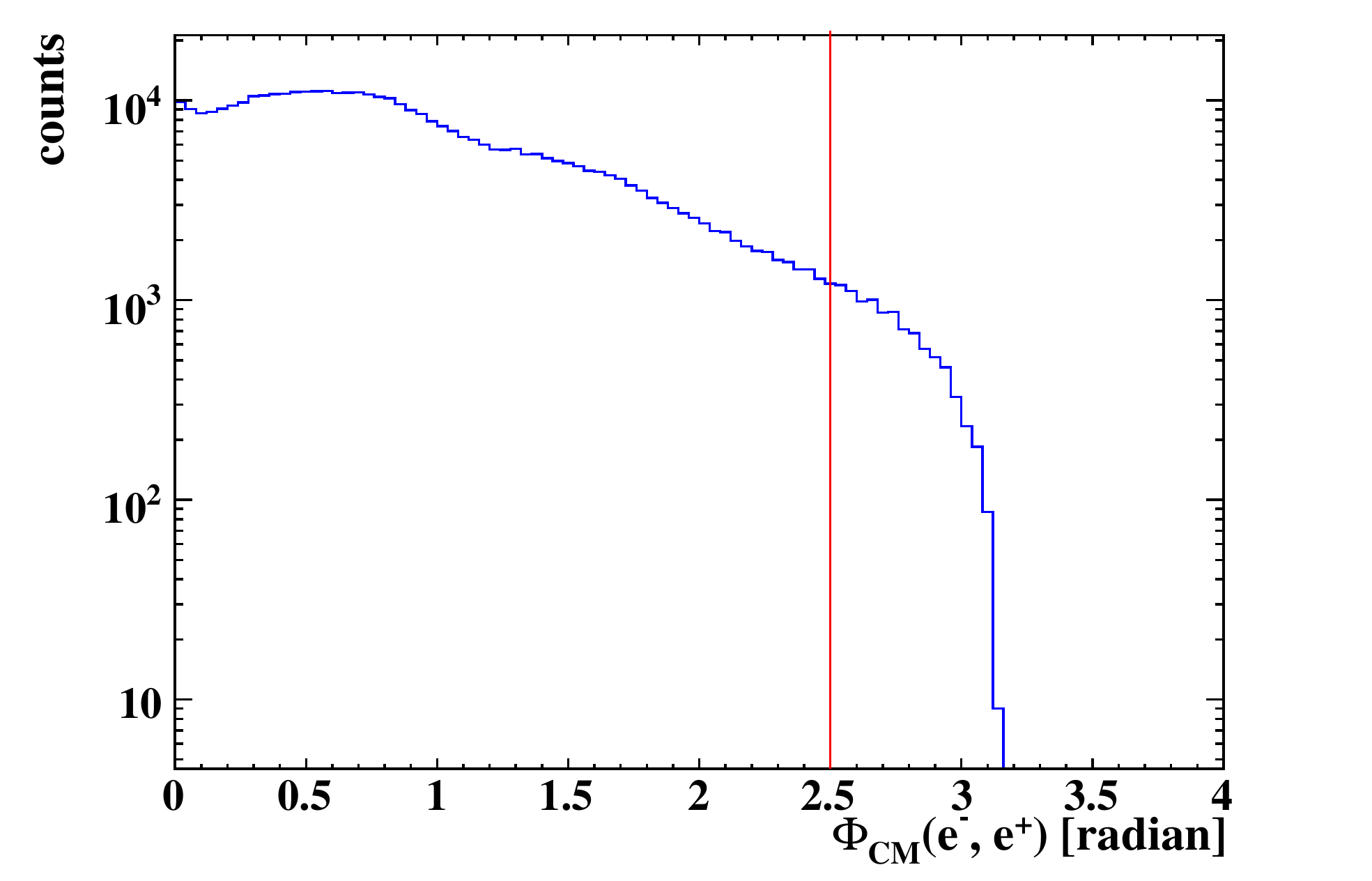}
\includegraphics[height=\swidth,angle=90]{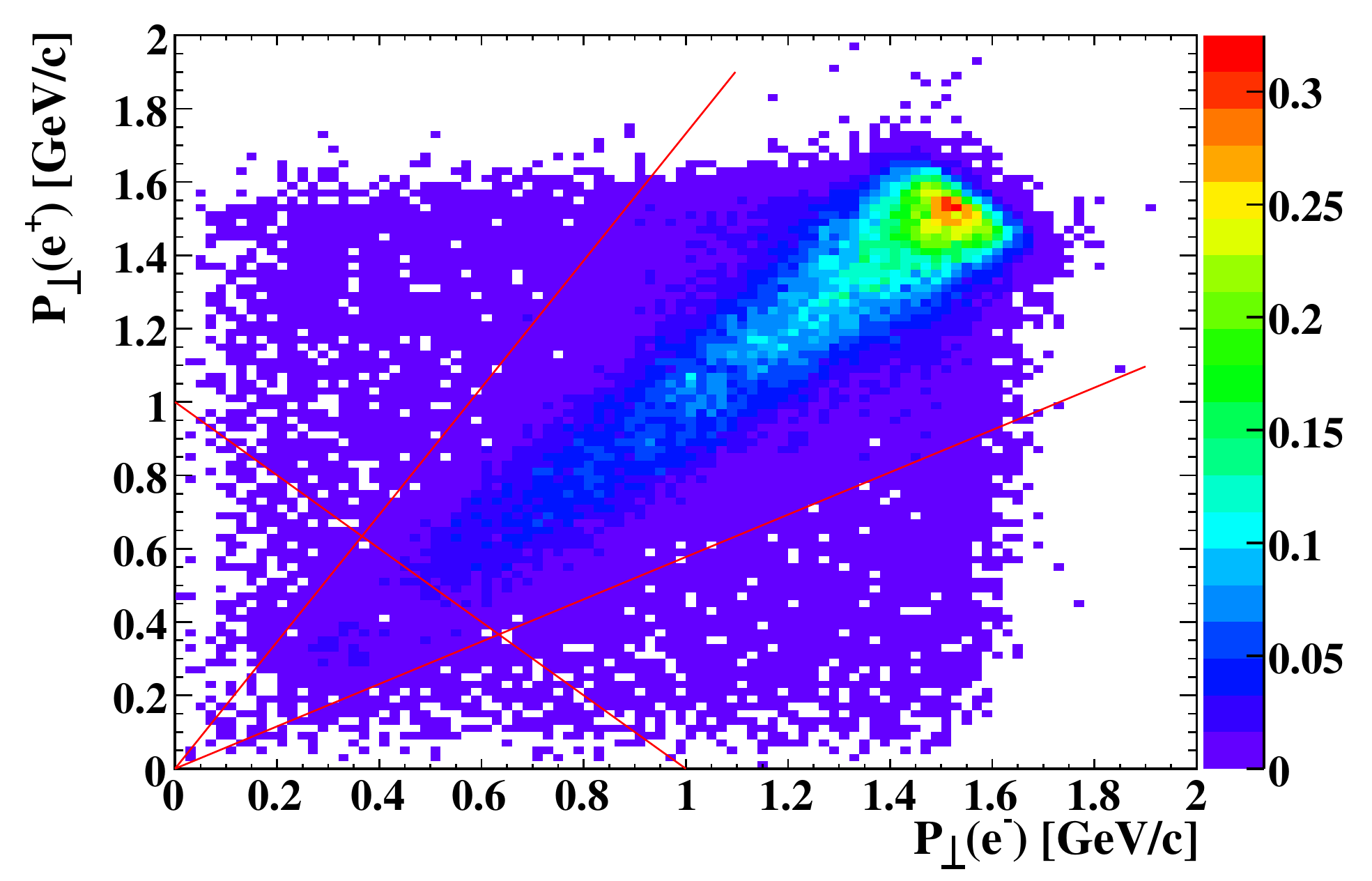}
\hfill
\includegraphics[height=\swidth,angle=90]{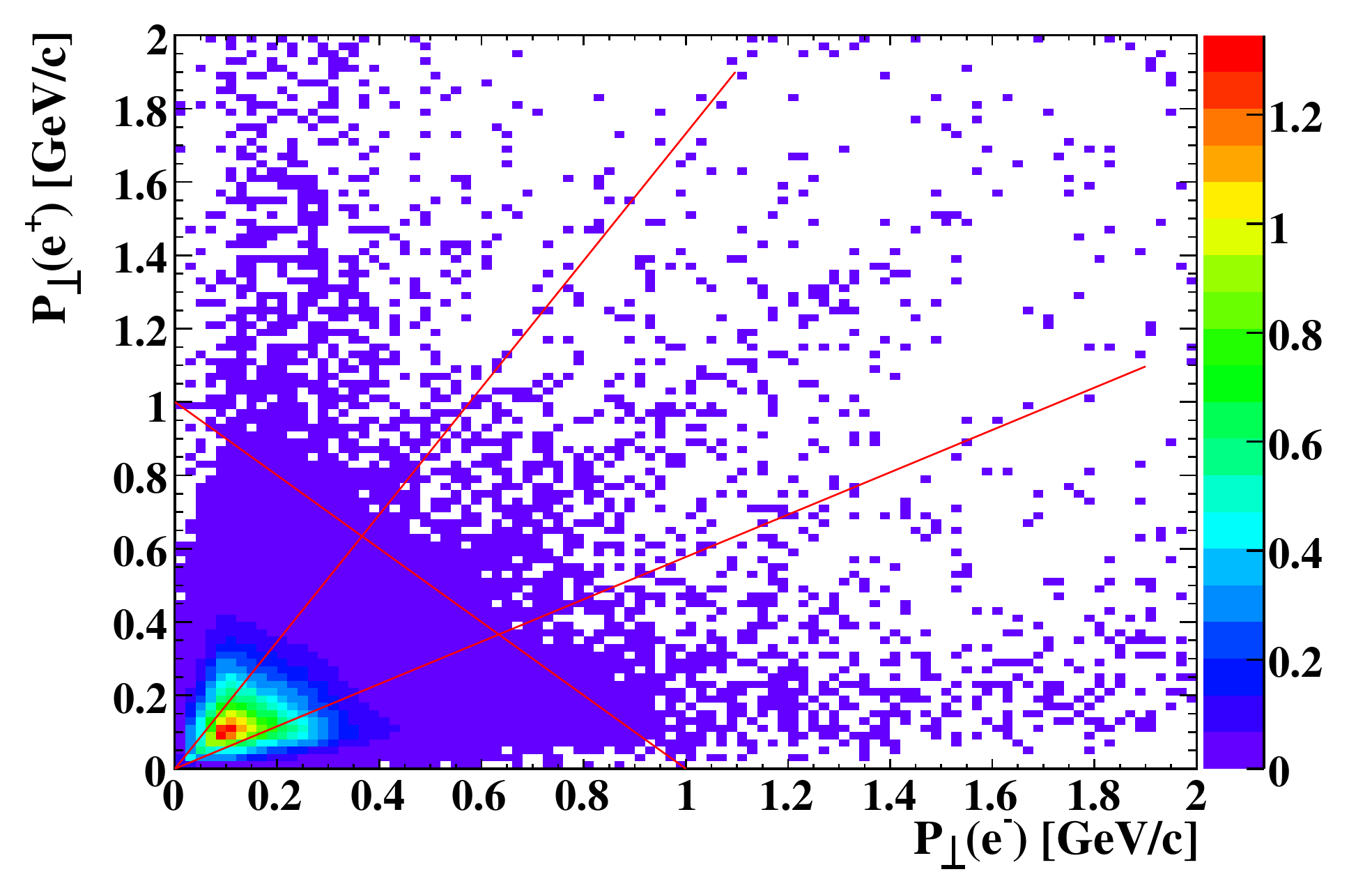}
\caption[Distributions used to suppress the background to the
$\jpsi$ signal]{Distributions used to suppress the background to
the $\jpsi$ signal: Left panels are for signal events, right
panels show the UrQMD background distributions. Upper panels show
the longitudinal momentum of $e^-$ versus that of $e^+$, middle
panels show the azimuthal angle $\phi(e^+, e^-)$ between $e^+$ and
$e^-$ in the centre of mass system, and lower panels show the
perpendicular momentum of e$^-$ versus that of e$^+$. The red
lines represent cuts to enhance the signal to background ratio.
The 2D-histograms are normalised to contain 100 events.}
\label{fig:cuts}
\end{center}
\end{figure*}

A realistic ratio of the numbers of background and signal events
reflecting the cross section ratio of $10^{10}$ may however
require additional cuts for further background suppression. The
simple topology of the signal events helps to select the signal
events from the dominating background. The typical features which
can be exploited to suppress the background are listed below:
\begin{enumerate}
\item[1.] the two leptons emerge from the main vertex
\item[2.] the total momentum approximately equals the incident
$\pbar$ momentum ($P_z(e^+) + P_z(e^-) \approx
P_z(\pbar)$)
\end{enumerate}

The total perpendicular momentum approximately vanishes
($P_{\perp}(e^+) \approx - P_{\perp}(e^-)$) which implies that

\begin{enumerate}
\item[3.] in the centre of mass system the angle between the two
leptons $\Phi(e^+, e^-)$ is  $\sim{}180\degrees$
\item[4.] the absolute values of the perpendicular momenta of the
two leptons is approximately equal ($|P_{\perp}(e^+))| \approx
|P_{\perp}(e^-)|$)
\end{enumerate}

The momentum relations are only approximately valid because of the Fermi-motion
of the protons in the target nuclei.

\begin{table*}[hbt]
\begin{center}
\caption [Definition of cuts for the $\ee$ analysis] {Definition
of cuts and their selection efficiency in combination with the
\loo $e^{\pm}$ candidate list for the $\jpsi\rightarrow{}\ee$
signal, the generic UrQMD, and specific background channel
\pip\pim. The numbers are computed for $\ee$ invariant masses
above a minimum mass of 2\,\gevcc. $v_x,v_y$ are the $x,y$
coordinates of the primary vertex.}
\begin{tabular}{lccc}
\hline\hline
cut & \multicolumn{3}{c}{fraction accepted} \\
 & signal & \multicolumn{2}{c}{background} \\
 &  & UrQMD & \pip\pim \\
\hline
$\sqrt{v_x^2 + v_y^2} < 1$\,mm & 0.77 & $3.8\cdot 10^{-8}$ & $4.1\cdot 10^{-6}$  \\
$P_z(e^+) + P_z(e^-) > 2.0$\,\gevc & 0.77 & $2.3\cdot 10^{-7}$ & $4.9\cdot 10^{-6}$ \\
$\Phi(e^+, e^-) > 2.5$\,rad & 0.77 & $1.5\cdot 10^{-7}$ & $4.9\cdot 10^{-6}$ \\
$[P_{\perp}(e^+) + P_{\perp}(e^-)] > 1$ \gevc \&
    $|\arctan\left( \frac{P_{\perp}(e^+)}{P_{\perp}(e^-)} \right)-45^{\circ}| < 15^{\circ}$ & 0.73 & $3.8\cdot 10^{-8}$ & $2.8\cdot 10^{-6}$ \\
\hline
combined ($IM_{\ee}>2.0$\,\gevcc) & 0.73 & $<3.8\cdot 10^{-8}$ & $2.4\cdot 10^{-6}$ \\
\hline\hline
\end{tabular}
\label{tab:cuts}
\end{center}
\end{table*}

The cuts deduced from these characteristic features are shown in
\Reffig{fig:cuts}. The panels on the left side contain the signal
events, whereas the right side panels show the background events
using the \loo list. The red lines represent possible cuts to
separate signal and background. The used cuts are explicitly given
in \Reftbl{tab:cuts}.
In addition, not shown in the figure, a condition in the $x-y$
plane on the primary interaction vertex at the target can be set.

The efficiencies of these cuts in combination with the \loo $e^{\pm}$ candidate
list are listed in \Reftbl{tab:cuts}. In case of the signal the numbers
represent the fraction of true $\jpsi$ mesons which are reconstructed and
accepted by the given cut in the invariant mass range above 2\,\gevcc. In
case of the background the given fraction is the number of accepted $\ee$ pairs
with a reconstructed invariant mass above 2\,\gevcc relative to the number of
simulated events.

\begin{figure}[htb]
\begin{center}
\includegraphics[height=\swidth,angle=90]{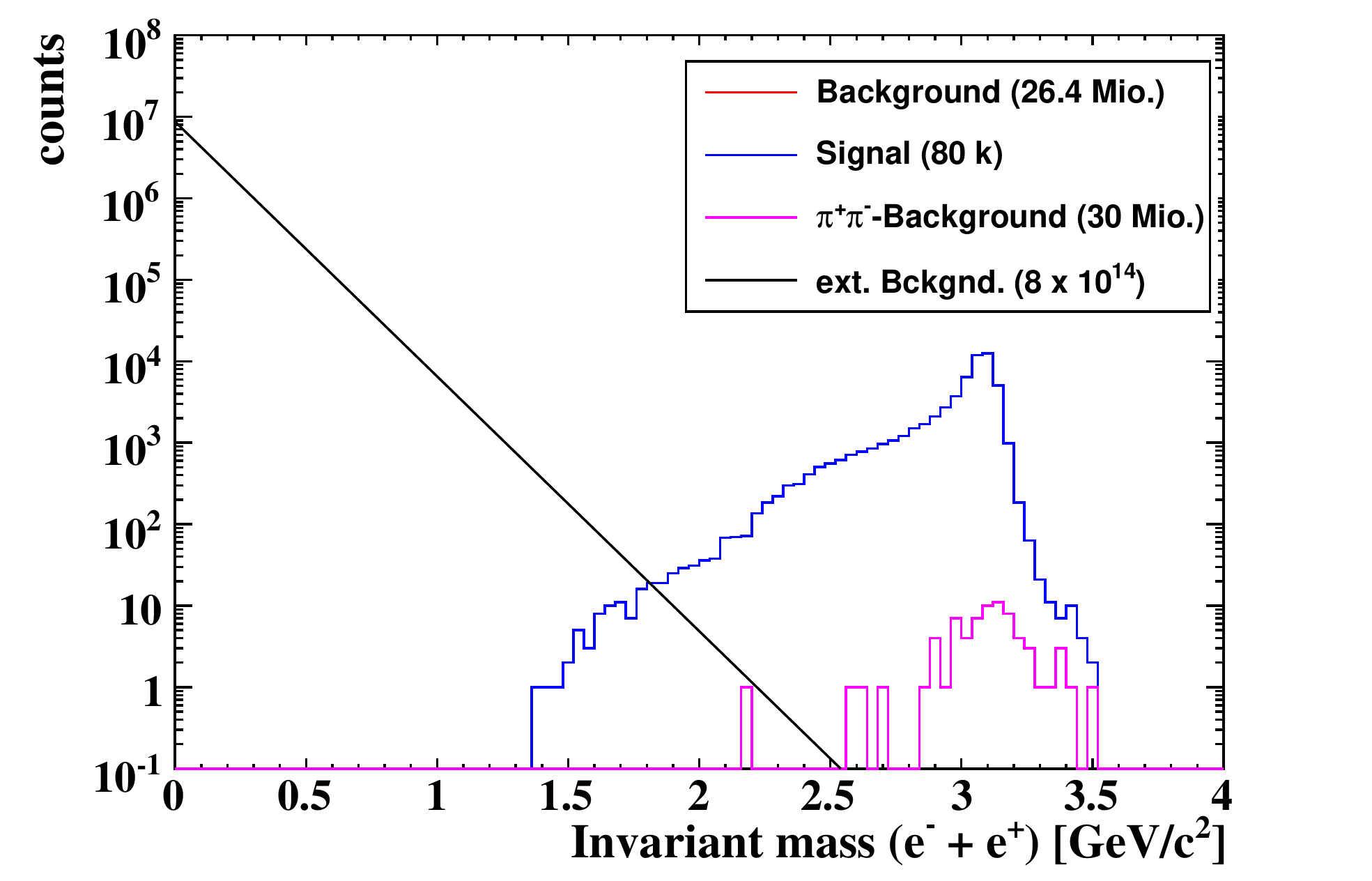}
\caption[Invariant mass distribution of $\ee$ pair candidates for
signal, UrQMD, and \pip\pim events]{Invariant mass distribution of
$\ee$ pair candidates for signal (blue), UrQMD background (red),
and \pip\pim background (magenta) events. The estimated
distribution for an UrQMD background sample scaled up to $10^{10}$
times the number of signal events is shown by the black line.}
\label{fig:ee_mass_sig_wc}
\end{center}
\end{figure}

The combination of all four cuts efficiently enhances the signal to background
ratio. Figure \ref{fig:ee_mass_sig_wc} shows the invariant mass distribution
after application of the combined cut. With the given statistics no UrQMD
background events are left in the investigated mass range. Therefore the
fraction of accepted UrQMD background events shown in \Reftbl{tab:cuts}
represents a lower limit for the achievable background suppression. On the other
hand $73$\percent of the true $\jpsi$ mesons are accepted. Note that this
combination of \loo list and software cuts is considerably more efficient than
simply using a more stringent list to enhance the signal to background relation.

The black line in \Reffig{fig:ee_mass_sig_wc} shows an estimate of the remaining
background with $8\cdot{}10^{14}$ simulated UrQMD background events ($10^{10}$
times the number of signal events). The distribution is assumed to follow an
exponential function which was fit to the \loo list distribution shown in the
lower panel of \Reffig{fig:ee-mass-sb} and was scaled accordingly. This estimate
indicates that also with the realistic relation of signal and background events
it should be possible to measure $\jpsi$ production in antiproton-nucleus
collisions with acceptable signal to background ratio.

In addition, as specific background channel, the reaction
$\pbarp\rightarrow{}\pip\pim$ at the same incident momentum of
4.05\,\gevc has been studied. To mimic the contribution of this
2-charged-pion annihilation channel with a nuclear target, the
momentum of the target proton was smeared isotropically according
to a Gaussian distribution with a width of
$\sigma_p=180\,\mevc$. Assuming
$\sigma_{\bar{p}p\rightarrow{}\pip\pim}=10\,\mu$b at the upper
limit within the experimental
uncertainty~\cite{bib:phy:Flaminio84} together with an $A^{2/3}$
scaling for the nuclear target and a signal cross section of
0.1\,nb, a suppression factor of minimum $10^{-5}$ is required to
keep the signal level above possible background due the \pip\pim
channel. The right most column in \Reftbl{tab:cuts} shows the
fraction of \pip\pim pairs which are erroneously accepted as $\ee$
pairs from $\jpsi$ decay.

The corresponding invariant mass spectrum of \pip\pim pairs accepted out of 30
million generated $\pbar p\rightarrow \pip\pim$ events is represented in
\Reffig{fig:ee-mass-sb} by the magenta distribution.

The suppression factor of $2.4\cdot{}10^{-6}$ obtained with the
combined cut is smaller than the required value and indicates that
a signal-to-background ratio with respect to the \pip\pim channel
above one can be achieved.

\subsection{Antibaryons and Antikaons Produced in $\pbarA$ Collisions}

Although not covered in the simulation section of this write-up,
it is worth mentioning that the special kinematics of $\pbar$
induced reactions combined with the detection capabilities of
\PANDA opens the opportunity to study the in-medium properties of a
number of hadrons in the light quark sector which can be produced
at rest or at very small momenta inside nuclei. This is {\it e.g.} the
case for $\pbar$, $\bar{\Lambda}$, and $\Kbar$, for which the
nuclear potential is a quantity of interest, but could not be
determined experimentally up to now.

The antiproton-nucleon interaction at low energies is dominated by
annihilation.  As a consequence of the strong absorption effects
constraints deduced from the energy levels of antiprotonic atoms
or from low energy antiproton-nucleus scattering on the depth of
the real part of the $\pbar$ nuclear potential have large
uncertainties. In theoretical work G parity transformation has
been proposed as a concept to provide a link between the $NN$ and
$N\overline{N}$ interaction at most for distances where the meson
exchange picture is applicable~\cite{bib:phy:Faessler:1982qt},
whereas at short distances the quark-gluon structure of baryons
may question the validity of this approach. Based on G parity
transformation relativistic mean field models predict a depth of
the $\pbar$ nuclear potential as large as
$-700$\,MeV~\cite{bib:phy:Buervenich02,bib:phy:Mishustin:2004xa,bib:phy:Larionov08}.
The depth of phenomenological potentials deduced from antiprotonic
atoms~\cite{bib:phy:Wong84,bib:phy:Batty97,bib:phy:Friedman05} and
from subthreshold antiproton production in nucleus-nucleus and in
proton-nucleus
collisions~\cite{bib:phy:Teis94,bib:phy:Spieles:1995fs,bib:phy:Sibirtsev:1997mq}.
ranges at lower values, typically between $-100$ and $-350$\,MeV.
Therefore better experimental constraints on the nuclear
antiproton potential, which may help to elucidate the role of the
quark-gluon structure of baryons for the short-range
baryon-antibaryon force, are needed.

The study of antiproton-nucleus collisions at high energy opens
new opportunities: in this case the antiproton can penetrate into
the interior of the nucleus, and probe the nuclear potential quasi
at rest by backward scattering from a nuclear proton. An
attractive $\pbar$ potential will be visible in forward protons
having a higher momentum than the incident antiproton which is a
very sensitive signature. The cross section for
$\pbarA\rightarrow{}\mbox{p}X$ with a recoil proton at
$\theta\simeq{}0\degrees$ is expected to be comfortably large despite
nuclear absorption effects for the incident antiproton and the
outgoing recoil proton. The nuclear antiproton potential reflects
itself in the missing mass distribution of the fast forward
proton.

The same method of low recoil momentum production is applicable
for $\bar{\Lambda}$ antihyperons in the reaction
$\pbarp\rightarrow{}\bar{\Lambda}\Lambda$ on a nuclear proton
with $\Lambda$ emission close to $0\degrees$. This may for the first
time give access to the properties of $\bar{\Lambda}$ antihyperons
inside nuclei, since no experimental information on the nuclear
potential of antihyperons exists so far.

\begin{figure}[htb]
\begin{center}
\includegraphics[width=\swidth]{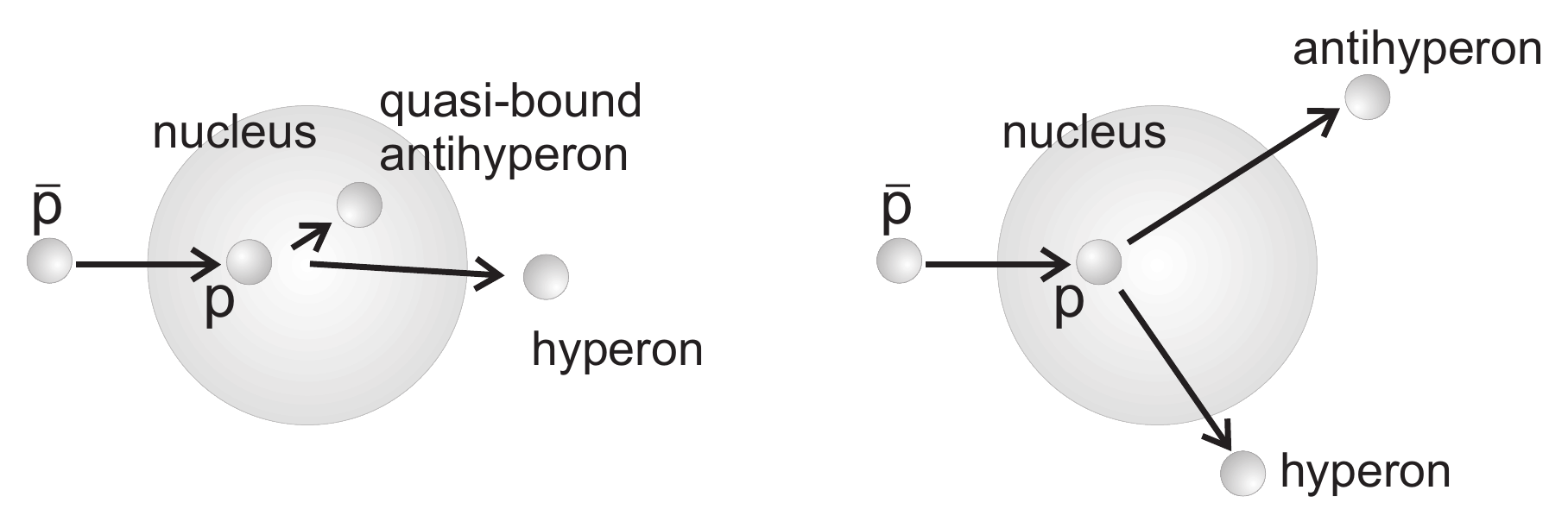}
\caption[Schematic illustration of reactions giving access to the
nuclear potential of $\bar{\Lambda}$ antihyperons]{Schematic
illustration of reactions giving access to the nuclear potential
of $\bar{\Lambda}$ antihyperons: recoilless $\bar{\Lambda}$
production with $\Lambda$ missing mass measurement(left), and
comparison of $\Lambda$ and $\bar{\Lambda}$ transverse momentum
distribution (right).} \label{fig:ppLL}
\end{center}
\end{figure}

In addition, for larger momenta of the produced hadrons,
quantitative information on the 'difference' between baryon and
antibaryon potentials and hence on the potential of antibaryons
may be obtained via exclusive antibaryon-baryon pairs produced
close to threshold after an antiproton-proton annihilation within
a complex nucleus~\cite{bib:phy:Pochodzalla:2008ju}. Once these
hyperons leave the nucleus and are detected, their asymptotic
momentum distributions will reflect the depth of the respective
potentials. A deeply attractive potential for one species could
result in a momentum distribution of antihyperons which is very
different from that of the coincident hyperon. Thus event-by-event
momentum correlation of coincident baryon-antibaryon pairs can
provide a direct and quantitative probe for the nuclear
potentials. \Reffig{fig:ppLL} schematically illustrates both
experimental approaches to access nuclear potentials of
$\bar{\Lambda}$ antihyperons as described above. Both methods may
be used in the same experiment to study baryon and antibaryon
in-medium properties over a larger range of momenta.

Due to its mass being very close to that of the nucleon,
recoilless production is also possible for the $\phi$ meson, and
hence also for the antikaon as one of the decay particles emitted
at very low momentum in the $\phi$ rest frame. This may give
access to the nuclear potential for $\phi$ mesons and for
antikaons. We therefore propose to explore the reaction
$\pbarp\rightarrow{}\phi\phi\rightarrow{}\Kp\Km\Kp\Km$ at
$\theta\simeq{}0\degrees$ on nuclear target protons.

In these reactions the produced slow hadrons are very likely to be
absorbed in the nuclear medium and not to be detected directly.
However a measurement of the $\phi$ or $\phi{}\Kp$ missing mass
may allow to identify the reaction channel, and to deduce the
in-medium properties of $\phi$ or $\Km$ mesons, respectively. In
the latter case it is experimentally challenging to detect and
identify a low momentum $\Kp$ meson.


\subsection{Colour Transparency}
\label{sec:CT}

Colour transparency  (CT)
\cite{bib:phy:Jain:1995dd,bib:phy:Frankfurt:1992dx,bib:phy:Frankfurt:1994hf}
is the predicted phenomenon of reduced strong interactions under
certain conditions, and in particular  when a hard scattering
process occurs, which selects small transverse size components in
hadronic wave functions, and thus triggers a coherent cancellation
of perturbative interactions. Colour transparency is tightly
connected to the property of asymptotic freedom in QCD and is
expected to occur in many kinds of quasi-exclusive reactions with
either electron or hadron beams. A large energy scale is needed
(large quark masses or large transverse momenta) for the coloured
degrees of freedom to be relevant, and an exclusivity condition is
necessary. The concept of colour transparency is related to the
notion of nuclear filtering, which is the conversion of quark wave
functions in hadrons to smaller transverse space dimensions by
interaction with a nuclear medium.

At high energy, colour transparency has been an essential element
of the diffractive physics program at \INST{HERA}  as it forms the basis
of the QCD factorisation theorem for production of vector mesons.
At \INST{FNAL}, perturbative QCD prediction of CT was found to be
consistent with the observation of coherent diffractive
dissociation of  500 GeV pions into di-jets off the nuclei
\cite{bib:phy:Aitala:2000hc}.

At intermediate energies colour transparency of the magnitude
predicted by the colour diffusion model
\cite{bib:phy:Farrar:1988me} was reported recently in the
electroproduction of pions at \INST{JLAB} \cite{bib:phy:Clasie:2007gqa}.
Hence, the  presence of colour transparency for mesons is now
established. Baryons are much more complicated objects than mesons
in particular due to the more important role played by the chiral
degrees of freedom. So looking for onset of colour transparency for
baryon interactions is complementary to the meson case.

Among the various experiments where colour transparency may be
probed, we select here two examples which should be feasible in
the \PANDA set-up.

\subsubsection {Antiproton - Nucleon Annihilation in a Hadron Pair at Large Angle}

The exclusive reactions
$$ \pbarN \to \pi \pi \, , \, K \bar K \, , \, \pbarN $$
and the corresponding ones on a nuclear target
$$ \pbarA \to \pi \pi  (A-1)^* \, , \, K \bar K (A-1)^* \, , \, \pbarN (A-1)^* $$
should be perturbatively calculable at large energies and fixed
large scattering angle. The QCD analysis of these processes
usually distinguishes two competitive mechanisms: the
Brodsky-Farrar-Lepage hard scattering of small-sized objects and
the Landshoff independent scattering process where the scattered
objects have one large transverse size. The different phase
structure of the two interfering amplitudes gives rise to a
pattern of oscillatory cross section
\cite{bib:phy:Pire:1982iv,bib:phy:Ralston:1982pa} which has been
seen at beam energies  in the 5-15 GeV range for p p collisions.

\begin{figure}[t!]
\includegraphics[width=\swidth]{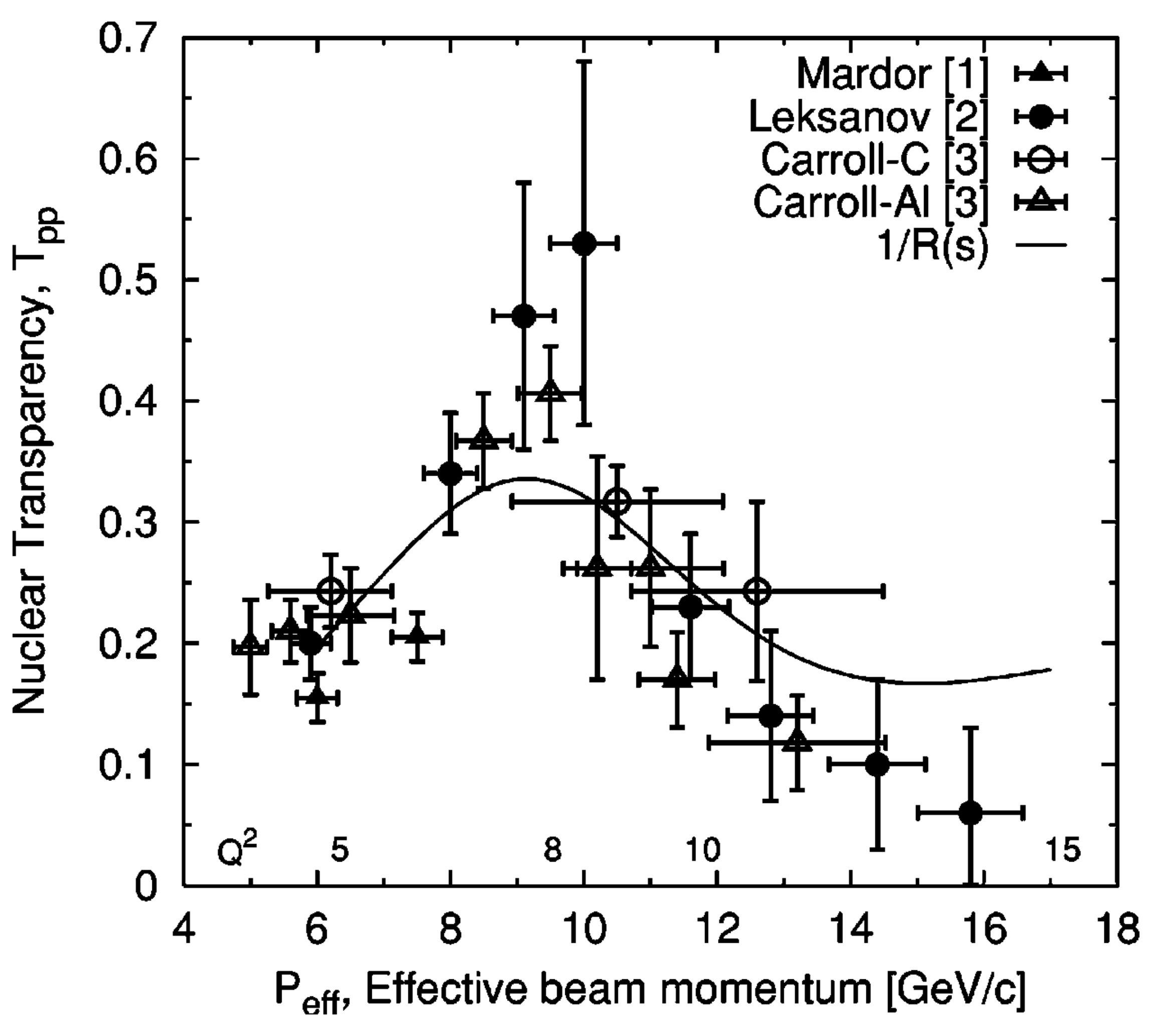}
\caption[The transparency ratio for the elastic proton-proton
scattering]{The transparency ratio for the elastic proton-proton
scattering at large angle, as measured at \INST{BNL}
\cite{bib:phy:Carroll:1988rp,bib:phy:Aclander:2004zm}, compared
with the oscillatory behaviour predicted by
Ref.\cite{bib:phy:Ralston:1988rb,bib:phy:Ralston:1990jj}.}
\label{figCTdata}
\end{figure}

The pioneering colour transparency experiments
\cite{bib:phy:Carroll:1988rp,bib:phy:Aclander:2004zm} involving
hadronic hard scattering was done at BNL by the group of Carroll
et al.. A proton beam was used to study $pA \to p'p''(A - 1)^*$.
Data was taken simultaneously on six targets, namely $^1$H,
$^7$Li, $^{12}$C, $^{27}$Al, $^{64}$Cu and $^{207}$Pb. The
reaction was identified via measurement of the three-momentum of
one proton, the direction of the second proton, and using veto
counters to exclude events with production of extra hadrons but
allowing the residual nucleus to be left in the excited state. In
the subsequent experiment which was focusing on the measurements
with the carbon target momenta of both nucleons were measured. The
experiments introduced and reported the transparency ratio,
$T(p_N, A)$ defined by $T = (d\sigma(pA \to p'p'' (A-1)^*/dt) /
Zd\sigma(pp \to p'p'')/dt$ at 90 degrees in the centre of mass.
The data of the two experiments are consistent in observing a rise
from the value which agrees  with the Glauber theory at
$p_N\sim{}6$~\gevc to the value which is about two times larger
than the Glauber value for $p_N\sim 9.5$~\gevc and falls back to
the Glauber level again at higher momenta all the way up to
14.4~\gevc. This oscillatory energy dependence of the transparency
ratio lead to some debate between different theoretical
explanations
\cite{bib:phy:Farrar:1988me,bib:phy:Ralston:1988rb,bib:phy:Ralston:1990jj,bib:phy:Brodsky:1987xw,bib:phy:Jennings:1989hc}.
Fig. \ref{figCTdata} shows the BNL data together with the
Ralston-Pire interpretation
\cite{bib:phy:Ralston:1988rb,bib:phy:Ralston:1990jj}. These
authors infer the rise and fall pattern in the colour transparency
ratio from the different rates of final state interactions in a
nuclear medium connected to  the different physics underlying the
two possible processes.

The scientific case between the different ways of understanding
the CT phenomenon should be cleared by adding more precise data
from other exclusive reactions, such as the exclusive annihilation
channels of an antiproton with a nucleon bound in a nucleus. The
relative weights of the different mechanisms of production are
process-dependent and it is thus crucial to analyse various
processes before claiming that colour transparency indeed occurs at
intermediate energies for reactions with baryons.

The antiproton beam delivered by FAIR has the right energy and
luminosity to access this controversial question.   The energies
of the produced mesons are at least as high in this case as in the
CT experiment at \INST{JLAB}, so the freezing of the $\bar q q$ system is
strong enough not to mask CT effects. \PANDA has the capacities to
measure in both in vacuum (proton target) and in a nuclear
environment the reactions
$$ \bar p N \to \pi \pi  ~~~~ \bar p N \to K \bar K ~~~~ \bar p N \to \bar p N $$
at various large angles and at various energies in the right
domain where the transition to perturbative QCD may occur and
cross sections are still measurable. Doing experiments at a few
beam energy values and on a few typical nuclear targets ($^{1}$H,
$^{2}$H, $^{12}$C, $^{27}$Al, $^{58}$Ni) should allow to plot the
transparency ratio with a good precision for various centre of
mass angle bins. Simulations already performed in the framework of
the background analysis of the time-like form factor measurement,
namely the reaction $\bar p  p \rightarrow \pi^+ \pi^-$, show that
the hadron identification ($\pi^+ , K^+, p$ or their
antiparticles) is easy to perform. Kinematically, the separation
from the process leading to an additional $\pi$ is as good as in
the case of the reaction on the proton.

\subsubsection {Colour transparency in charmonium production}

Charmonium states will be produced in the nuclear media in the
same resonance reactions as those which will be studied in $\bar p
p$ collisions. A possibility to use these processes for studies of
CT was first suggested in Ref.~\cite{bib:phy:Brodsky:1988xz}. A
subsequent detailed analysis of the effects of CT and Fermi motion
was performed in Ref.~\cite{bib:phy:Farrar:1989vr}. It found that
CT effects due to squeezing of the incoming antiproton are small
since the incident energy is small leading two a short coherence
length for the incident and final particles. The effect of the
Fermi motion is large but well under control. What is unique about
this reaction is that at these energies charmonium is formed very
close ($\le 1 fm$) to the interaction point, hence one gets a
possibility to check the main premise of CT that small objects
interact with nucleons with small cross sections. The ability to
select states of varying size: $J/\psi, \chi_c, \psi'$ will allow
to investigate how the interaction strength depends on the
transverse size of the system. A nontrivial consequence of this
phenomenon is an $A$-dependent polarisation of the $\chi$ states
\cite{bib:phy:Gerland:1998bz} due to filtering of $\bar c c $
configurations of smaller transverse size. The study of this
phenomenon is also important for understanding the dynamics of
charmonium production in heavy ion collisions.

\clearpage

%% file: phys/phys_hypernuclei.tex
%
\section{Hypernuclear Physics}
\label{sec:phys:hyp:summary}
\COM{Author(s): J. Pochodzalla, A. Feliciello, F. Iazzi}
\COM{Referee(s): F. Maas}
%
\input{./phys/hyp/phys_hypernuclei_intro}

%% file: phys/hyp/phys_hypernuclei_intro.tex
%
%
%
Quantum Chromo Dynamics (QCD) is the theory of the force responsible
for the binding of nucleons and nuclei and thus of a significant
fraction of the ordinary matter in our universe. While the internal
structure of hadrons and the spectra of their excited states are
important aspects of QCD, it is at least equally important to
understand how nuclear physics emerges in a more rigorous way out of
QCD and how nuclear structures - nuclei on the small scale and dense
stellar objects on the large scale - are formed
\cite{bib:phy:gre95}. For example, the presence of hyperons in
neutron star cores is expected to lower the maximum mass of neutron
({\it e.g.} ref.\cite{bib:phy:hae07}). Recent measurements of a few
large masses of pulsars in binaries with white dwarfs could be used
to put additional (astrophysical) constraints on the hyperon-nucleon
interaction ({\it e.g.} ref.\cite{bib:phy:hae07}). However, there is
at present no clear picture emerging as to what kind of matter
exists in the cores of neutron stars
\cite{bib:phy:web07,bib:phy:jha08,bib:phy:dap08}.

\begin{figure}[!h]
\begin{center}
 \includegraphics[width=\swidth]{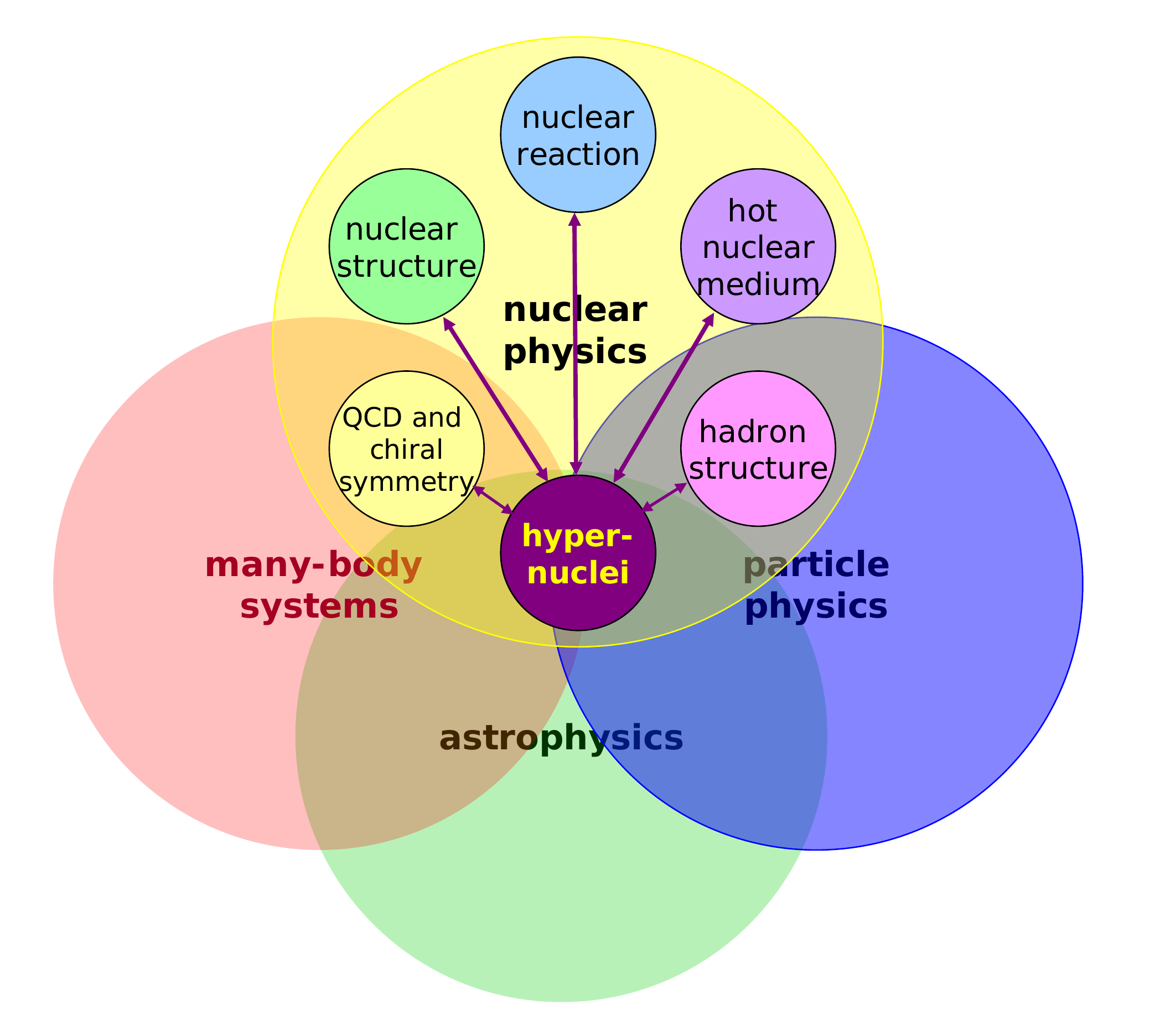}
 \caption{Hypernuclei and their link to other fields of physics.}
 \label{fig:phys:hyp:fig_field}
\end{center}
\end{figure}

A hyperon bound in a nucleus offers a selective probe of the
hadronic many-body problem as it is not restricted by the Pauli
principle in populating all possible nuclear states, in contrast to
neutrons and protons. On one hand a strange baryon embedded in a
nuclear system may serve as a sensitive probe for the nuclear
structure and its possible modification due to the presence of the
hyperon. On the other hand properties of hyperons may change
dramatically if implanted inside a nucleus. Therefore a nucleus may
serve as a laboratory offering a unique possibility to study basic
properties of hyperons and strange exotic objects. Thus hypernuclear
physics represents an interdisciplinary science linking many fields
of particle, nuclear and many-body physics
(\Reffig{fig:phys:hyp:fig_field}).

\subsection{Physics Goals}
\label{sec:hyp:goals}
\subsubsection{Hypernuclei Probing Nuclear Structure} While it is
difficult to study nucleons deeply bound in ordinary nuclei, a
$\Lambda$ hyperon not suffering from Pauli blocking can form deeply
bound hypernuclear states which are directly accessible in
experiments. In turn, the presence of a hyperon inside the nuclear
medium may give rise to new nuclear structures which cannot be seen
in normal nuclei consisting only of nucleons. Furthermore, a
comparison of ordinary nuclei and hypernuclei may reveal new
insights in key questions in nuclear physics like for example the
origin of the nuclear spin-orbit force \cite{bib:phy:kai05}.

An important goal is to measure the level spectra and decay
properties of hypernuclei is in order to test the energies and wave
functions from microscopic structure models. Indeed recent
calculations of light nuclei based on modern nucleon-nucleon
potentials, which also incorporate multi-nucleon interactions, are
able to describe the excitation spectra of light nuclei with a very
high precision of
1-2{\%}~\cite{bib:phy:wir02,bib:phy:pie04,bib:phy:pie07,bib:phy:ste06}.
A challenging new approach to hyperon interactions and structure of
hypernuclei is the relativistic density functional theory. This is a
full quantum field theory enabling an {\em ab initio} description of
strongly interacting many-body system in terms of mesons and baryons
by deriving the in-medium baryon-baryon interactions from free space
interactions by means of Dirac-Brueckner theory
\cite{bib:phy:kei00,bib:phy:hof01}. The field theoretical approach
is also the appropriate starting point for the connection to
QCD-inspired descriptions based for example on chiral effective
field theory (${\chi}EFT$) \cite{bib:phy:bea05,bib:phy:sav07}. At
present, ${\chi}EFT$ is well understood for low-energy meson-meson
\cite{bib:phy:gom02} and meson-baryon dynamics in the vacuum
\cite{bib:phy:Lutz:2001yb,bib:phy:pol06} and infinite nuclear
matter\cite{bib:phy:tol06,bib:phy:lut07}. A task left for the future
is to obtain the same degree of understanding for processes in a
finite nuclear environment. Present nuclear structure calculations
of the light nuclei in ${\chi}EFT$
\cite{bib:phy:ste06,bib:phy:ste07,bib:phy:bor07,bib:phy:nav07}
signal significant progress.

It is this progress made in our theoretical understanding of nuclei
which nurtures the hope that detailed information on excitation
spectra of hypernuclei and their structure will provide unique
information on the hyperon-nucleon and - in case of double
hypernuclei -- on the hyperon-hyperon interactions.

\begin{figure}[!h]
\begin{center}
  \includegraphics[width=\swidth]{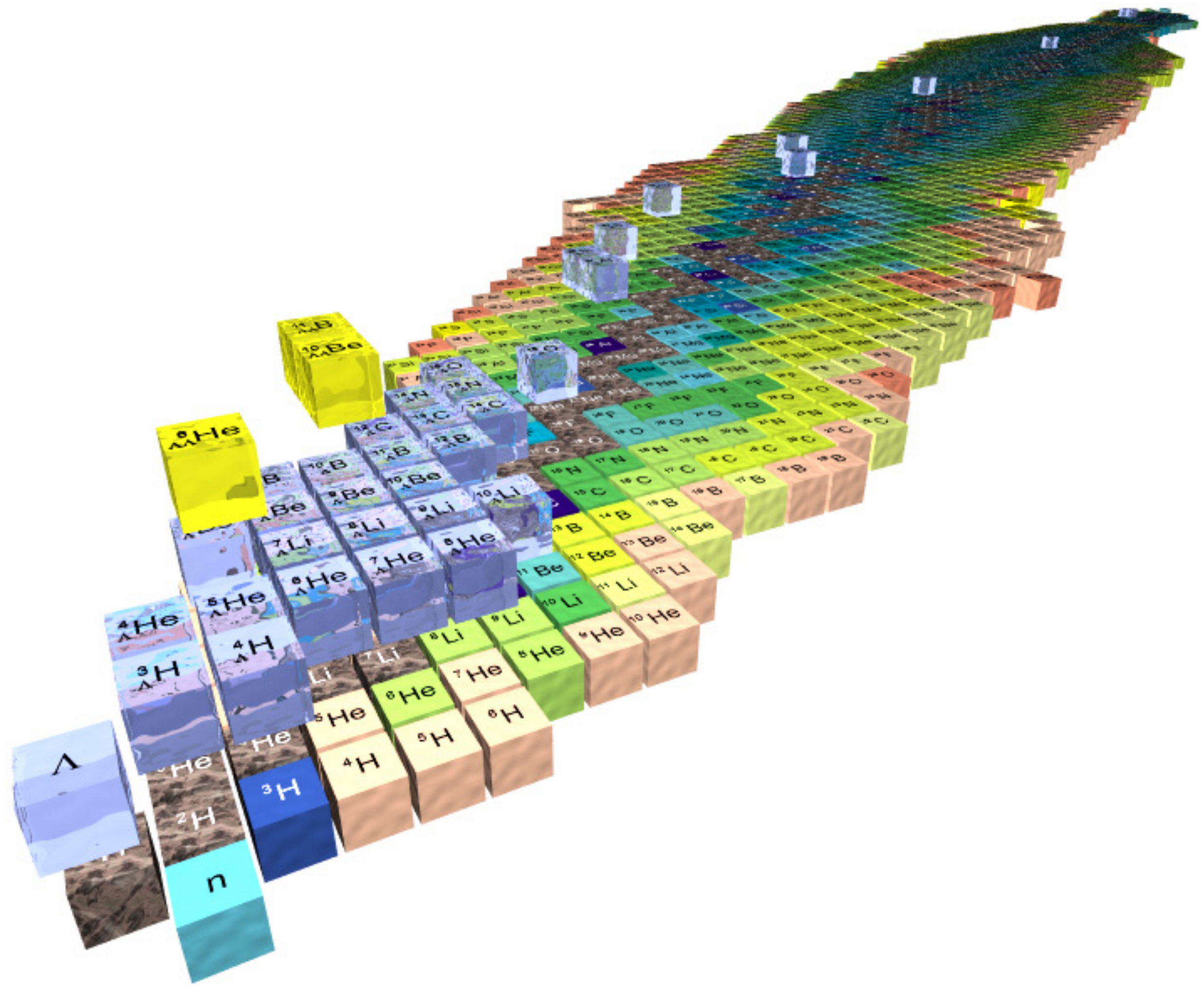}
  \caption{Present knowledge on hypernuclei. Only very
  few individual events of double hypernuclei have been detected and identified so far.}
 \label{fig:phys:hyp:fig_chart}
\end{center}
\end{figure}

\subsubsection{Hypernuclei: Baryon-Baryon Interaction}

While the nucleon-nucleon scattering was extensively studied since
the 50's, direct experimental investigations for the YN interactions
are still very sparse. Because of their short lifetimes, hyperon
targets are not available. Low-momentum hyperons are very difficult
to produce and hyperon-proton scattering is only feasible via the
double-scattering technique~\cite{bib:phy:kon00,bib:phy:ahn06}.
There are only a few hundreds low-momentum $\Lambda$-N and
$\Sigma^{\pm}$-N scattering events available and there is
essentially no data on $\Xi$-$N$ or $\Omega$-$N$ scattering.

In single hypernuclei the description of hyperons occupying the
allowed single-particle states is without the complications
encountered in ordinary nuclei, like pairing interactions and so on.
The strength of the $\Lambda$-$N$ strong interaction may be
extracted with a description of single-particle states by rather
well known wave functions. Furthermore, the decomposition into the
different spin-dependent contributions may be analysed. For these
contributions, significantly different predictions exist for example
from meson exchange-current and quark models.

It is also clear that a detailed and consistent understanding of the
quark aspect of the baryon-baryon forces in the SU(3) space will not
be possible as long as experimental information on the
hyperon-hyperon channel is not at our disposal. Since scattering
experiments between two hyperons are impractical, the precise
spectroscopy of multi-strange hypernuclei at \PANDA will provide a
unique approach to explore the hyperon-hyperon interaction. So far
only very few individual events of double hypernuclei have been
detected and identified (\Reffig{fig:phys:hyp:fig_chart}).

\subsubsection{Hypernuclei: Weak Decays}

\begin{figure}[!h]
\begin{center}
  \includegraphics[width=\swidth]{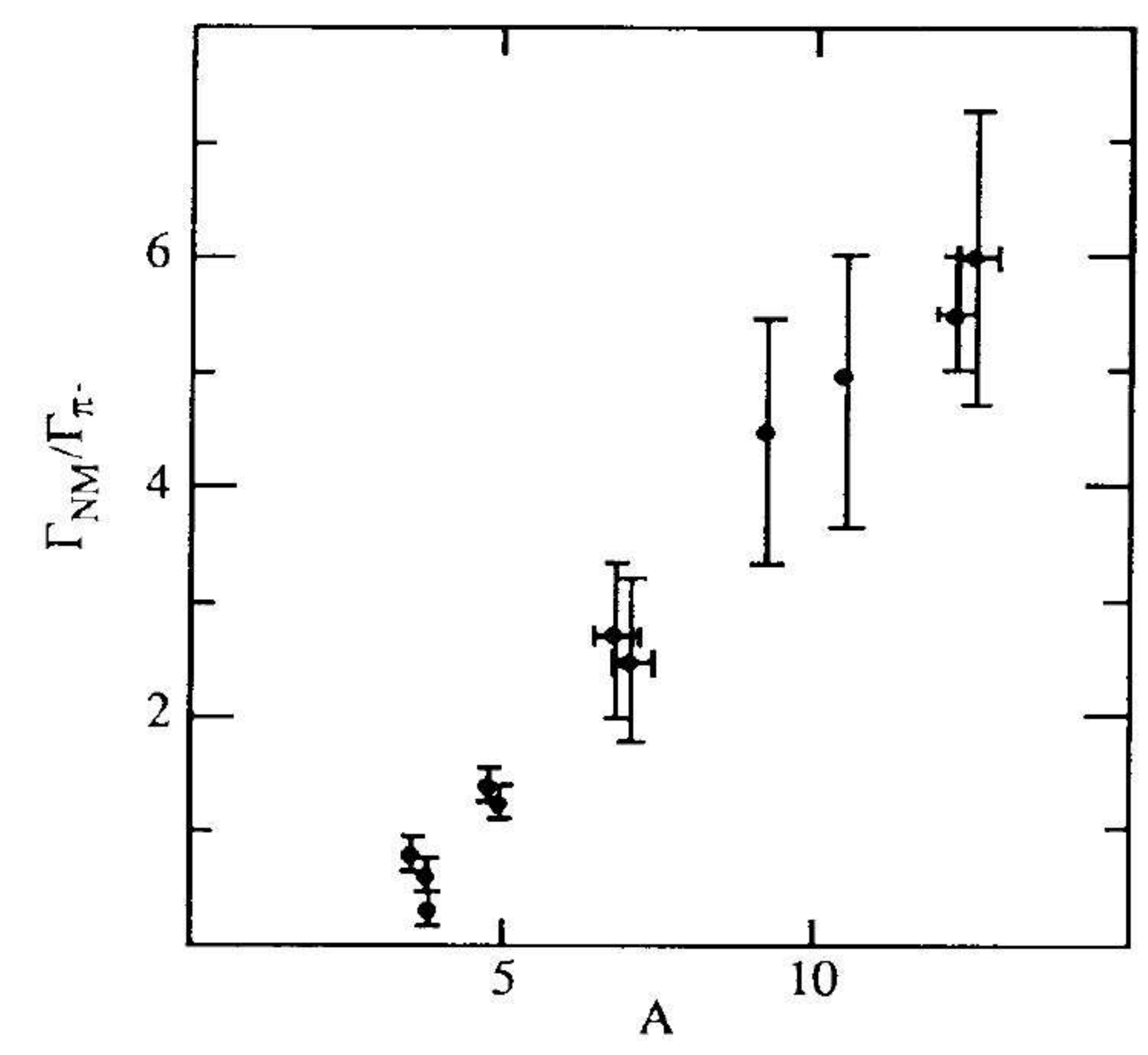}
  \caption{Measured ratio of non-mesonic ($\Lambda N \rightarrow NN$)
  to mesonic ($\Lambda\rightarrow N\pi$) hypernuclear decay widths as a
  function of the nuclear mass \cite{bib:phy:dub96}.}
 \label{fig:phys:hyp:fig_nonmesonic}
\end{center}
\end{figure}

Once a hypernucleus has reached its ground state, it can only decay
via a strangeness-changing weak interaction. Because of the low
Q-value for free-$\Lambda$ mesonic decay at rest of only 40\,MeV,
the mesonic decay of a $\Lambda \rightarrow N \pi$ bound in a
nucleus ($B_{\Lambda} \geq$ -27\,MeV; see e.g.
Ref.~\cite{bib:phy:has06}) is disfavoured by the Pauli principle and
is only for light nuclei still sizable (see
\Reffig{fig:phys:hyp:fig_nonmesonic}). In contrast, processes like
$\Lambda N \rightarrow N\! N$ and $\Lambda \Lambda \rightarrow
\Lambda \! N$ are allowed, opening a unique window for the
four-baryon, strangeness non-conserving interaction. Moreover, in
double hypernuclei hyperon induced non-mesonic weak decays
$\Lambda\Lambda \rightarrow {\Lambda}{N}$ and $\Lambda\Lambda
\rightarrow {\Sigma}{N}$ are
possible~\cite{bib:phy:ito01,bib:phy:par01,bib:phy:sas03} giving
unique access to the $\Lambda{\Lambda}K$ coupling.

In the simulation presented below we consider only the case of two
subsequent pionic decays which amounts to typically 10\% of all weak
decays of the light double hypernuclei (c.f.
\Reffig{fig:phys:hyp:fig_nonmesonic}). Considering in the future
also non-mesonic decays the event statistics may therefore increase
significantly.

\subsubsection{Multi-Strange Atoms}

It is interesting to note that the different S=-2 systems -
$\Xi^-$-atoms and single $\Xi^-$-hypernuclei on one side and double
$\Lambda\Lambda$ hypernuclei on the other side - provide
complementary information on the baryon force: on the one-meson
exchange level strange mesons with I=1/2 do not contribute to the
$\Xi$-$N$ interaction. On the other hand, only strange mesons act in
the ${\Xi}N$-$\Lambda\Lambda$ coupling while only non-strange mesons
contribute in the $\Lambda$-$\Lambda$ interaction
\cite{bib:phy:mot01,bib:phy:rij06}. Indeed, hyperatoms created
during the capture process of the hyperon will supply additional
information on the hyperon-nucleus interaction. X- rays from $\pim$,
$\Km$, $\overline{p}$ and $\Sigma^-$ atoms have already been studied
in several experiments in the past. At J-PARC first precise studies
of $\Xi$-atoms are planned \cite{bib:phy:tan07}. At \PANDA not only
$\Xi^-$ atoms but also $\Omega^-$ atoms can be studied for the first
time thus providing unique information on the nuclear optical
potential of $\Omega^-$ baryons. The $\Omega$ hyperon is
particularly interesting because due to it's long lifetime and it's
spin of 3/2 it is the only 'elementary' baryon with a non-vanishing
spectroscopic quadrupole moment. Since the quadrupole moment of the
$\Omega$ is mainly determined by the one-gluon exchange contribution
to the quark-quark interaction \cite{bib:phy:buc97,bib:phy:buc03}
it's measurement represents a unique benchmark for our understanding
of the quark-quark interaction.

\subsubsection{Hypernuclei: Doorway  to Exotic Quark States}
The claimed observation of pentaquark states places also the
question for other exotic quark states on the agenda. Thus the
possible existence of an S=-2 six quark (uuddss) $H$-dibaryon
\cite{bib:phy:jaf77a,bib:phy:jaf77b} represents another challenging
topic of $\Lambda\Lambda$ hypernuclear physics. Because of their
long lifetimes double $\Lambda$ hypernuclei may serve as 'breeder'
for the $H$-particle. Although some theories predict the
$H$-dibaryon to be stable (\cite{bib:phy:sak00} and references
therein), the observation of several double hypernuclei makes the
existence of a strongly bound free $H$-dibaryon unlikely. However,
since the mass of the H-particle might drop inside a nucleus
\cite{bib:phy:sak00} and due to hyperon mixing
\cite{bib:phy:yam00,bib:phy:myi03,bib:phy:afn03,bib:phy:fil03} it
might be possible to observe traces of a $H$-dibaryon even if it is
unbound in free space by a detailed study of the energy levels in
double hypernuclei.

\subsection{Experimental Integration and Simulation}
\label{sec:hyp:exp} In the \PANDA experiment, bound states of $\Xi$
hypernuclei will be used as a gateway to form double $\Lambda$
hypernuclei \cite{bib:phy:poc05}. The production of low momentum
$\Xi^-$ hyperons and their capture in atomic levels is therefore
essential for the experiment. At \PANDA the reactions
$\overline{p}+p \rightarrow \Xi^- \overline{\Xi}^+$ and
$\overline{p}+n \rightarrow \Xi^- \overline{\Xi}^0$ followed by
re-scattering of the $\Xi^-$ within the primary target nucleus will
be employed (\Reffig{fig:phys:hyp:fig_scheme}). After stopping the
$\Xi^-$ in an external secondary target, the formed  $\Xi$
hypernuclei will be converted into double $\Lambda$ hypernuclei.
This two-step production mechanism requires major additions to the
usual  simulation package PandaRoot as well as the {\PANDA} setup
(\Reffig{fig:phys:hyp:fig_PandaGeSetup}in \Refsec{exp:detector}).
Mandatory for this experiment is a modular and highly flexible setup
of the central {\PANDA} detector:

\begin{figure}[!h]
\begin{center}
 \includegraphics[width=\swidth]{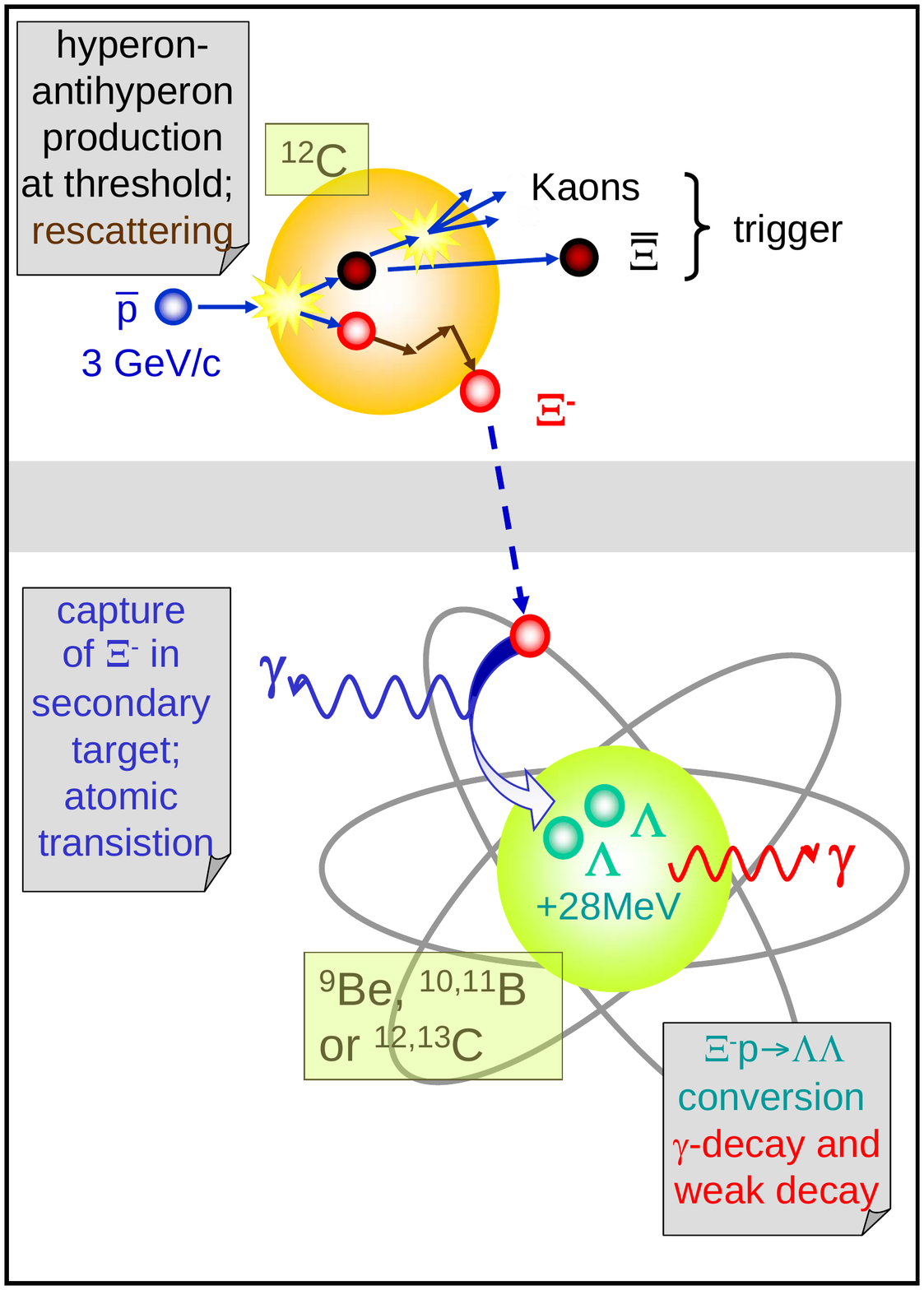}
 \caption{Various steps of the double hypernucleus production in \PANDA.}
 \label{fig:phys:hyp:fig_scheme}
\end{center}
\end{figure}

\begin{itemize}
\item
A primary carbon target at the entrance to the central tracking
detector of \PANDA. To avoid unnecessary radiation damage to the
micro vertex detectors surrounding the nominal target region, these
detectors will be removed during the hypernucleus runs.
\item
A small secondary active sandwich target composed of silicon
detectors and $^9$Be, $^{10,11}$B or $^{12,13}$C absorbers to slow
down and stop the $\Xi^-$ and to identify the weak decay products.
\item
To detect the $\gamma$-rays from the excited double hypernuclei an
array of 15 n-type Germanium triple Cluster-arrays will be added. To
maximise the detection efficiency the $\gamma$-detectors must be
arranged as close as possible to the target. Hereby the main
limitation is the load of particles from $\overline{p}$-nucleus
reactions. Since the $\gamma$-rays from the slowly moving
hypernuclei is emitted nearly isotropic the Ge-detectors will be
arranged at backward angles.
\end{itemize}

\subsubsection{Simulation of Hyperon Production}

At present high statistics production of hyperon-antihyperon pairs
in antiproton-nucleus are not practical within full microscopic
transport calculations like UrQMD. We therefore employed an event
generator \cite{bib:phy:fer07} which is based on an Intra Nuclear
Cascade model and which takes as a main ingredient the re-scattering
of the antihyperons and hyperons in the target nucleus into account.

Target nuclei with larger mass are more efficient for re-scattering
of the produced primary particles and hence for the emission of low
momentum $\Xi^-$ hyperons. However, heavier targets increase the
neutron and x-ray background in the germanium detectors.
Furthermore, Coulomb scattering in heavy primary targets leads to
significant losses of antiprotons. Therefore it is foreseen to use
thin carbon micro-ribbons \cite{bib:phy:loz08} as a primary target
in the HESR ring. For the present simulations 10$^6$ $\Xi^-
\overline{\Xi}$ pairs were generated. At the incident momentum of
3\,\gevc we expect a cross section per nucleon of 2$\mu$b
\cite{bib:phy:kai94}. For comparison, at \PANDA a luminosity of
10$^{32}$cm$^{-2}$s$^{-1}$ for $\overline{p}+^{12}$C reactions
corresponds to about 700000 produced $\Xi^- \overline{\Xi}$ pairs
per hour. Out of the produced 1 million pairs, 50505 contain low
momentum $\Xi^-$ with momentum less than 500\,\mevc.

\subsubsection{Deceleration of $\Xi^-$ Hyperons in a Secondary Target}
In order to limit the number of possible transitions and thus to
increase the possible signal to background ratio, the experiment
will focus on light secondary target nuclei with mass number
$A_0\leq 13$. Since the identification of the double hypernuclei has
to rely on the unique assignment of the detected
$\gamma$-transitions, different isotopically enriched light
absorbers ($^9$Be, $^{10,11}$B, $^{12,13}$C) will be used. In the
following we consider as an example the case of $^{12}$C absorbers
in all four quadrants of the secondary target.

\begin{figure}[!h]
\begin{center}
  \includegraphics[width=\swidth]{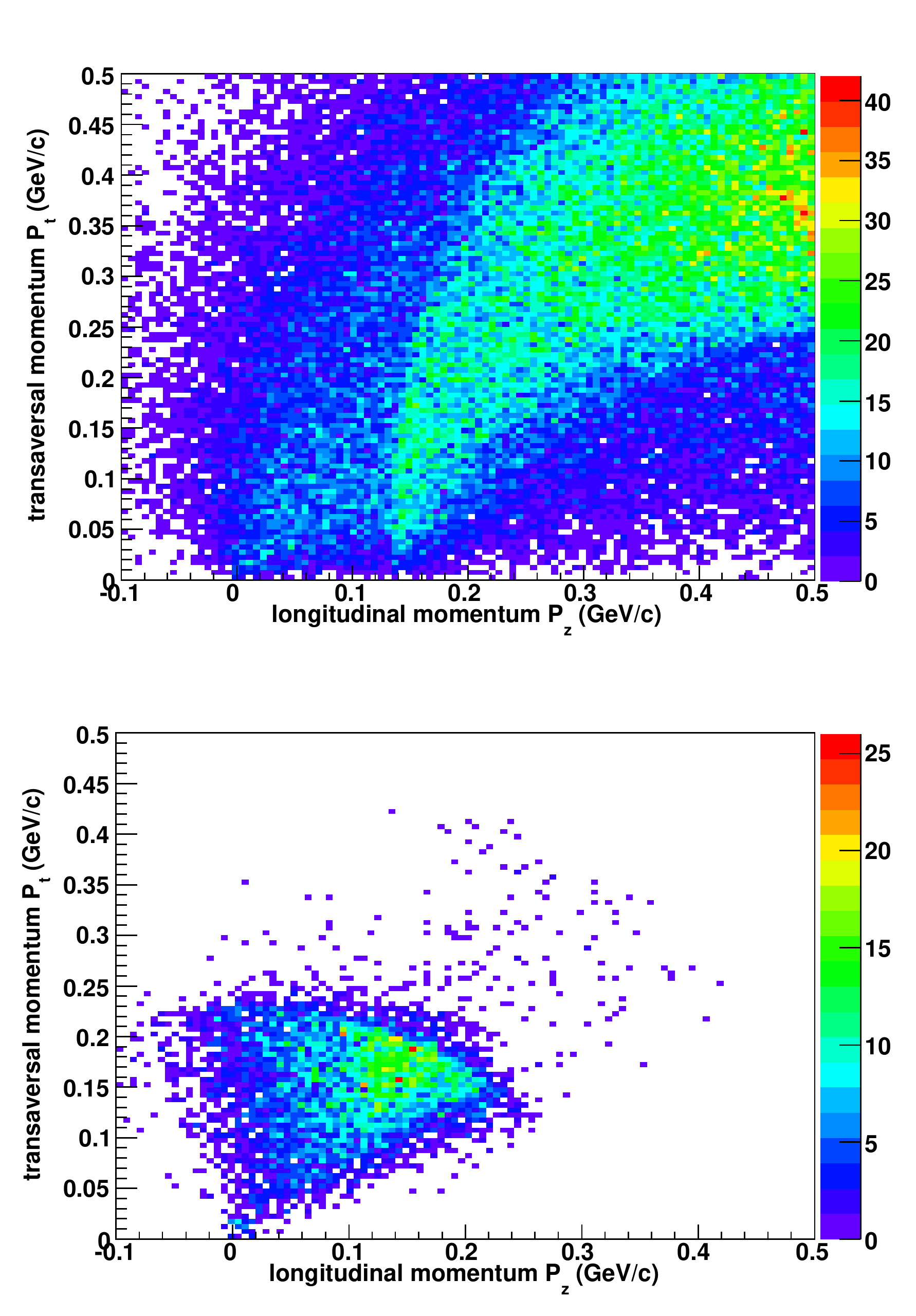}
  \caption{Transverse vs. longitudinal momentum distribution of $\Xi^-$ with
  transverse and longitudinal
  momenta less than 500\,\mevc (upper part) and those stopped within
  the secondary target (lower part).}
 \label{fig:phys:hyp:fig_xiptpl}
\end{center}
\end{figure}

The geometry of the target (see \Reffig{fig:phys:hyp:fig_sectarg1}
in \Refsec{exp:detector}) is essentially determined by the lifetime
of the hyperons and their stopping time in solid material: only
hyperons with momenta smaller than about 500\,\mevc have a
non-negligible chance to be stopped prior to their free decay.
\begin{figure}[!h]
\begin{center}
    \includegraphics[width=\swidth]{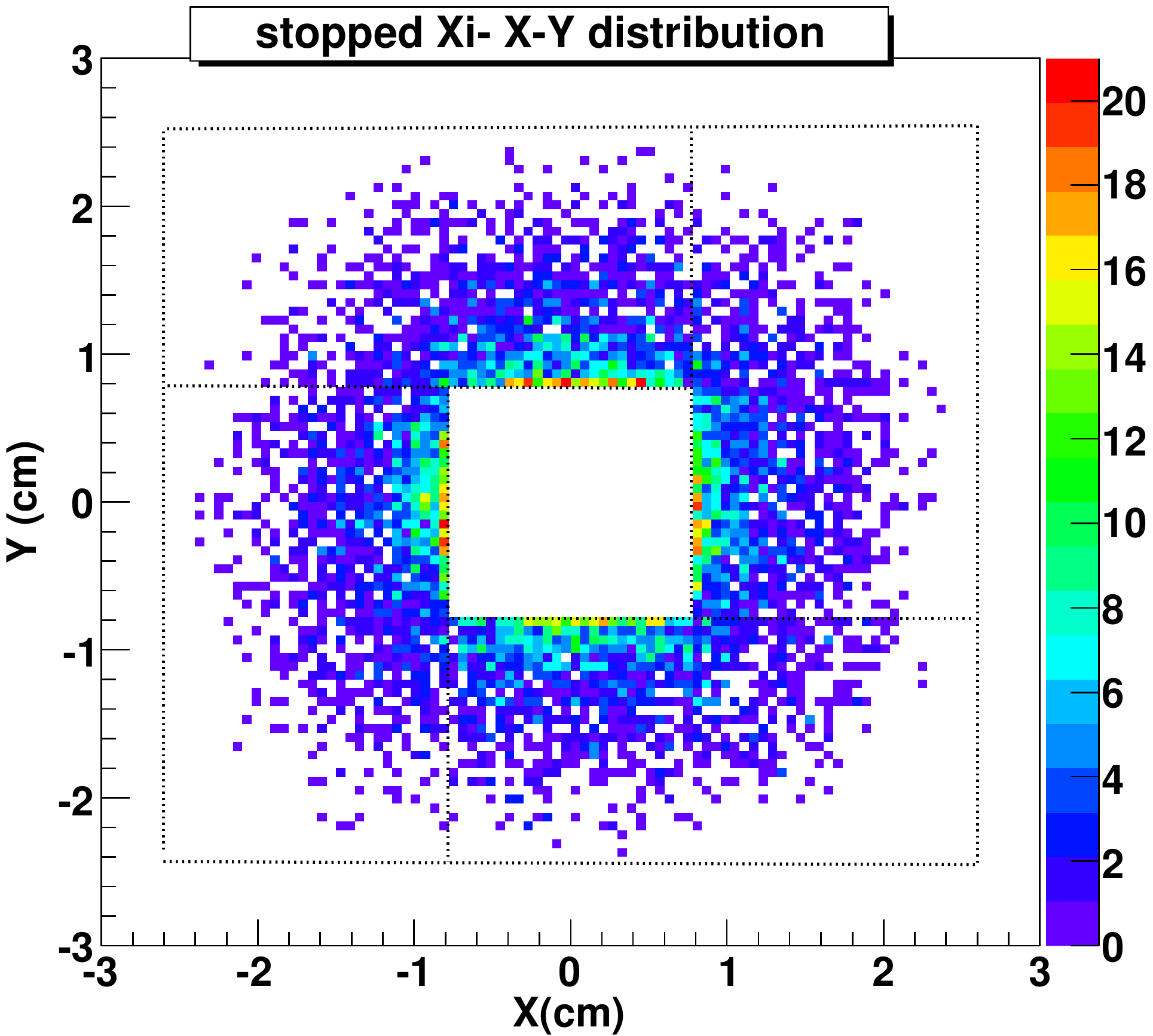}
  \caption[Layout out the secondary sandwich target used in the present simulations.]{
  The figure marks the stopping points of the $\Xi^-$ hyperons
  within the target in the x-y plane transverse to the beam direction. The rectangles indicate the
  outlines of the four target segments. All four segments are equipped with $^{12}$C absorbers.}
 \label{fig:phys:hyp:fig_sectarg2}
\end{center}
\end{figure}
From 50505 produced events which contained a $\Xi^-$ with a
laboratory momentum less than 500\,\mevc, 7396 hyperons are stopped
within the secondary target. The majority of the hyperons are
stopped in the most inner layers of the sandwich structure
(\Reffig{fig:phys:hyp:fig_sectarg2}). The typical momenta of
these stopped $\Xi^-$ are in the range of 200\,\mevc (see lower part
of \Reffig{fig:phys:hyp:fig_xiptpl}).

Recently Yamada and co-workers studied within the framework of the
doorway double-$\Lambda$ hypernuclear picture \cite{bib:phy:Yam95}
the production of double-$\Lambda$ hypernuclei for stopped $\Xi^-$
particles in $^{12}$C \cite{bib:phy:yam97}. Per stopped ${\Xi}^-$
they predict a total double-$\Lambda$ hypernucleus production
probability of 4.7\%. An even larger probability of 11.1\%{} was
recently obtained by Hirata {\it et al.} within the Antisymmetrised
Molecular Dynamics approach \cite{bib:phy:hir97,bib:phy:hir99}.
Since the present studies concentrate on the production of double
hypernuclei, a full microscopic simulation of the atomic cascade and
capture of the $\Xi^-$ hyperons is not performed. For the final rate
estimate we assume a $\Xi^-p \rightarrow \Lambda\Lambda$ conversion
probability of only 5{\%}. Of course for the study of hyperatoms the
simulations need to be complemented in this aspect.

\subsubsection{Population of Excited States in Double Hypernuclei}

For light nuclei even a  relatively  small excitation  energy  may
be comparable with their binding energy. Model calculations
\cite{bib:phy:mil94,bib:phy:ike94,bib:phy:rij06} show that the width
for the conversion of a $\Xi ^{-}$ and a proton into two $\Lambda$'s
is around 2--5\,MeV, i.e. the conversion is rather fast and takes
less than 100 fm/c. In this case we assume that the principal
mechanism of de-excitation is the explosive decay of the excited
nucleus into several smaller clusters. To describe this break-up
process we have developed \cite{bib:phy:lor09} a model which is
similar to the famous Fermi model for particle production in nuclear
reactions \cite{bib:phy:fer50}. In the microcanonical model we
consider all possible break-up channels, which satisfy the mass
number, hyperon number (i.e. strangeness), charge, energy and
momenta conservations, and take into account the competition between
these channels. Previously, this model was applied rather
successfully for the description of break-up of conventional light
nuclei in nuclear reactions initiated by protons, pions, antiprotons
and ions
\cite{bib:phy:bon95,bib:phy:bot90,bib:phy:bot93,bib:phy:sud93}. The
precision was around 20--50\% for the description of experimental
yields of different fragments. This precision is sufficient for the
present analysis of hypernucleus decay, since the main uncertainly
is in unknown masses and energy levels of the produced
hyperfragments.

For double hypernuclei the experimental information is restricted to
a few cases only
\cite{bib:phy:dan63a,bib:phy:pro66,bib:phy:aok91a,bib:phy:ahn01,bib:phy:tak01}.
Except for the $^6_{\Lambda\Lambda}$He nucleus reported in Ref.
\cite{bib:phy:tak01} the interpretation of the observed events is
however not unique
\cite{bib:phy:dan63a,bib:phy:dal89,bib:phy:dov91,bib:phy:hiy02,bib:phy:ran07}.
Furthermore no direct experimental information on possible excited
states is at hand (see e.g. discussion in Ref.
\cite{bib:phy:hiy02}). Therefore, theoretical predictions of bound
and exited states of double hypernuclei by Hiyama and co-workers
\cite{bib:phy:hiy02} were used in the present model calculation for
nuclei with mass number 6$\leq A_0\leq$10. For the mirror nuclei
$^5_{\Lambda\Lambda}$H and $^5_{\Lambda\Lambda}$He there seems to be
a consensus that these nuclei are indeed bound
\cite{bib:phy:myi03,bib:phy:fil03,bib:phy:lan04,bib:phy:sho04,bib:phy:nem05}.
In view of the theoretical uncertainties, we assumed in our
calculations a value for the $\Lambda$-$\Lambda$ bond energy $\Delta
B_{\Lambda\Lambda}$=1\,MeV for both nuclei. In case of
$^4_{\Lambda\Lambda}$H the experimental situation is ambiguous
\cite{bib:phy:ahn01,bib:phy:ran07} and also the various model
calculations predict an unbound \cite{bib:phy:fil02} or only
slightly bound nucleus
\cite{bib:phy:nem03,bib:phy:sho05,bib:phy:nem05}.

Also for heavier nuclei several particle stable excited states are
expected \cite{bib:phy:yam97}. The ground state mass of these double
hypernuclei was estimated from the known masses of single
hypernuclei and adopting a fixed value for $\Delta
B_{\Lambda\Lambda}$ of 1\,MeV. Furthermore, the calculations of
Hiyama and co-workers \cite{bib:phy:hiy02} signal that in the mass
range relevant for this work the level structure of particle stable
double hypernuclei resembles the level scheme of the corresponding
core nucleus. The excitation spectrum of double hypernuclei with
A$\geq$11 was therefore assumed to be given by that of the
corresponding core nucleus. Only states below the lowest particle
decay threshold were considered in the present calculations.

For hypernuclei with a single $\Lambda$ particle, we use the
experimental masses and excited states which are summarised in
various reviews (e.g. Ref.~\cite{bib:phy:ban90,bib:phy:has06}). In
case of the production of conventional nuclear fragments in a
break-up channel, we adopt their experimental ground states masses,
and take into account their excited states, which are stable
respective to emission of nucleons (see nuclear tables, e.g.
\cite{bib:phy:ajz85}). Masses of fragments in excited states were
calculated by adding the corresponding excitation energy to their
ground state masses.

\begin{figure}[!h]
\begin{center}
  \includegraphics[width=\swidth]{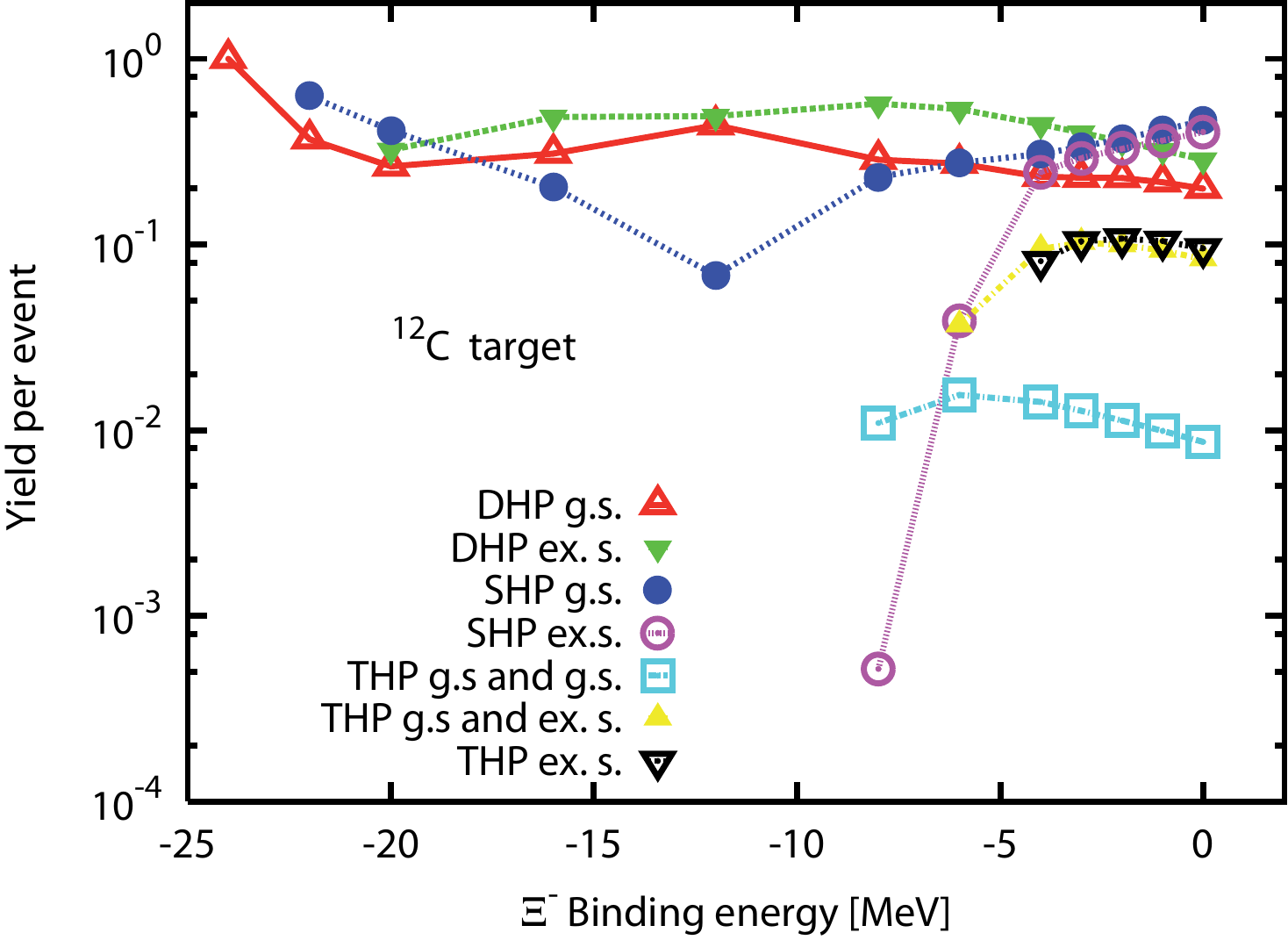}
  \caption{Production probability of ground (g.s.) and excited states (ex.s.) in conventional nuclear fragments
  and in one single (SHP), twin (THP) and double hypernuclei
  (DHP) after the capture of a $\Xi^-$ in a $^{12}$C nucleus and its conversion
into two $\Lambda$ hyperons
  predicted by a statistical decay model.}
 \label{fig:phys:hyp:fig_carbonall}
\end{center}
\end{figure}

\begin{figure}[!h]
\begin{center}
  \includegraphics[width=\swidth]{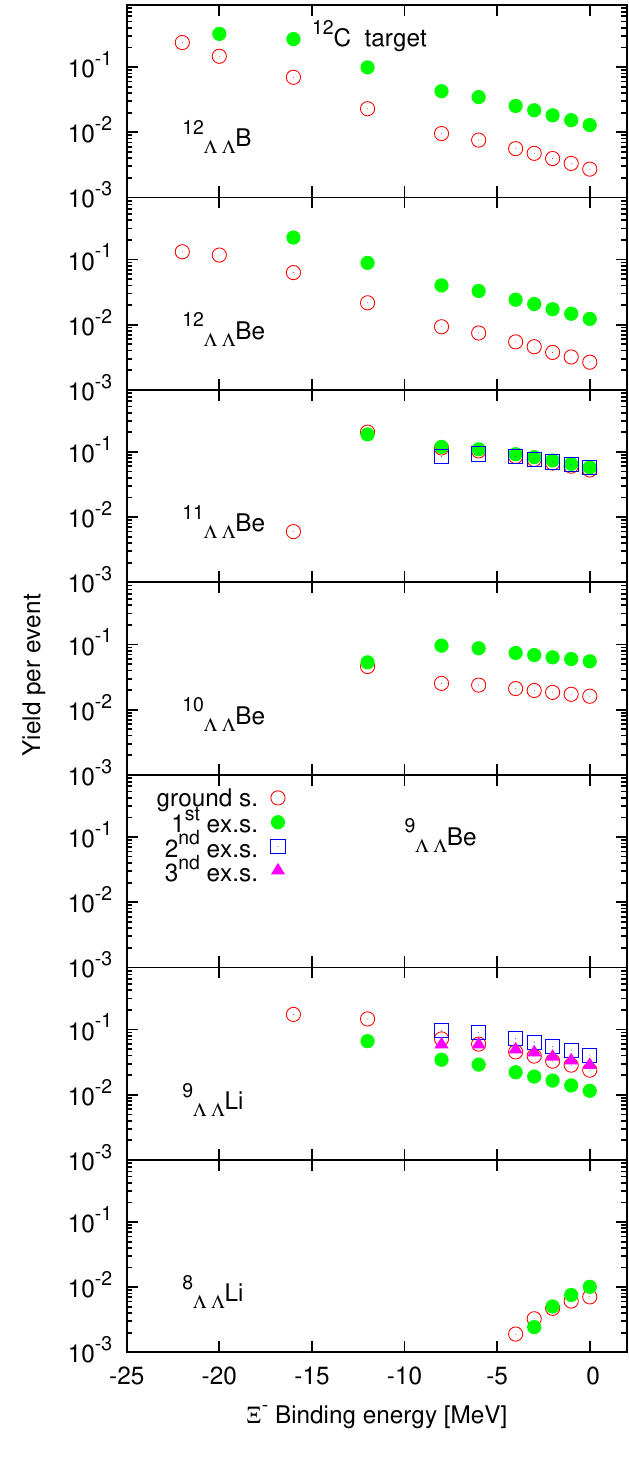}
  \caption[Production probability of ground and excited states of accessible double hypernuclei after
  the capture of a $\Xi^-$ in a $^{12}$C nucleus and the $\Xi^-$ conversion
into two $\Lambda$ hyperons.]{ Excited states in
$^{11}_{\Lambda\Lambda}$Be, $^{10}_{\Lambda\Lambda}$Be and
$^{9}_{\Lambda\Lambda}$Li dominate over a wide range of the $\Xi^-$
binding energy.}
 \label{fig:phys:hyp:fig_carbon12}
\end{center}
\end{figure}

The main reaction which we analyse is the break-up of an excited
hypernucleus with double strangeness produced after absorption of
stopped $\Xi ^{-}$. Unfortunately, the excitation energies of the
produced hypernuclei are not well known, since this conversion may
happen at different energy levels, and a part of the released energy
may be lost. The maximum energy available in this case is
$E_{max}=(M(\Xi^{-})+M_{target})c^2$. In order to take into account
a possible reduction of this energy because of the $\Xi^-$ binding
the calculations were performed for a range of energies less than
$E_{max}$.

The Fermi break-up events were generated by comparing probabilities
of all possible channels with Monte--Carlo methods. The Coulomb
expansion stage was not considered explicitly for such light
systems. The momentum distributions of the final break-up products
were obtained by a random generation over the whole accessible phase
space, determined by the total kinetic energy, taking into account
exact energy and momentum conservation laws. For this purpose we
applied a very effective algorithm proposed by G.I.~Kopylov
\cite{bib:phy:kop70}.

\Reffig{fig:phys:hyp:fig_carbonall} shows as an example the
production of ground (g.s.) and excited (ex.s.) states of
conventional nuclear fragments as well as single (SHP), twin (THP)
and double (DHP) hypernuclei in case of a $^{12}$C target as a
function of the assumed $\Xi^-$ binding energy. According to these
calculations excited states in double hypernuclei (green triangles)
are produced with significant probability.
\Reffig{fig:phys:hyp:fig_carbon12} shows the population of the
different accessible double hypernuclei. For the $^{12}$C target,
excited states in $^{11}_{\Lambda\Lambda}$Be,
$^{10}_{\Lambda\Lambda}$Be and $^{9}_{\Lambda\Lambda}$Li dominate
over a wide range of the assumed $\Xi^-$ binding energy.

Very little is established experimentally on the interaction of
$\Xi$ hyperons with nuclei. Various analyses (see e.g.
\cite{bib:phy:dov83,bib:phy:fri07}) suggest a nuclear potential well
depth around 20\,MeV. Calculations of light $\Xi$ atoms
\cite{bib:phy:bat99} predict that the conversion of the captured
$\Xi^-$ from excited states with correspondingly small binding
energies dominates. In a nuclear emulsion experiment a $\Xi^-$
capture at rest with two single hyperfragments has been observed
\cite{bib:phy:aok95} which was interpreted as $\Xi^- + C \rightarrow
^4_{\Lambda}H + ^9_{\Lambda}Be$ reaction. The deduced binding energy
of the $\Xi^-$ varied between 0.62\,MeV and 3.70\,MeV, depending
whether only one out of the two hyperfragments or both fragments
were produced in an excited particle stable state. Therefore for the
present simulation of the $\gamma$-ray spectra a $\Xi^-$ binding
energy of 4\,MeV was adopted. As can be seen from
\Reffig{fig:phys:hyp:fig_carbon12} this choice of the binding energy
is not crucial for the final $\gamma$-ray yield.

\subsubsection{Gamma Detection}

In the next step the excited particle stable states of double
hypernuclei as well as excited states of conventional nuclei and
single hypernuclei produced during the decay process de-excite via
$\gamma$-ray emission. For the high resolution spectroscopy of
excited hypernuclear states a position sensitive Germanium
$\gamma$--array \cite{bib:phy:san07} has been implemented in the
standard \PANDA framework PandaRoot(see \Reffig{fig:phys:hyp:fig_PandaGeSetup}). To
describe the response of these detectors, processes which are
relevant for the interaction of the emitted photons with matter such
as pair production, Compton scattering and the photoelectric effect
have been taken into account.

\begin{figure}[!h]
\begin{center}
  \includegraphics[width=\swidth]{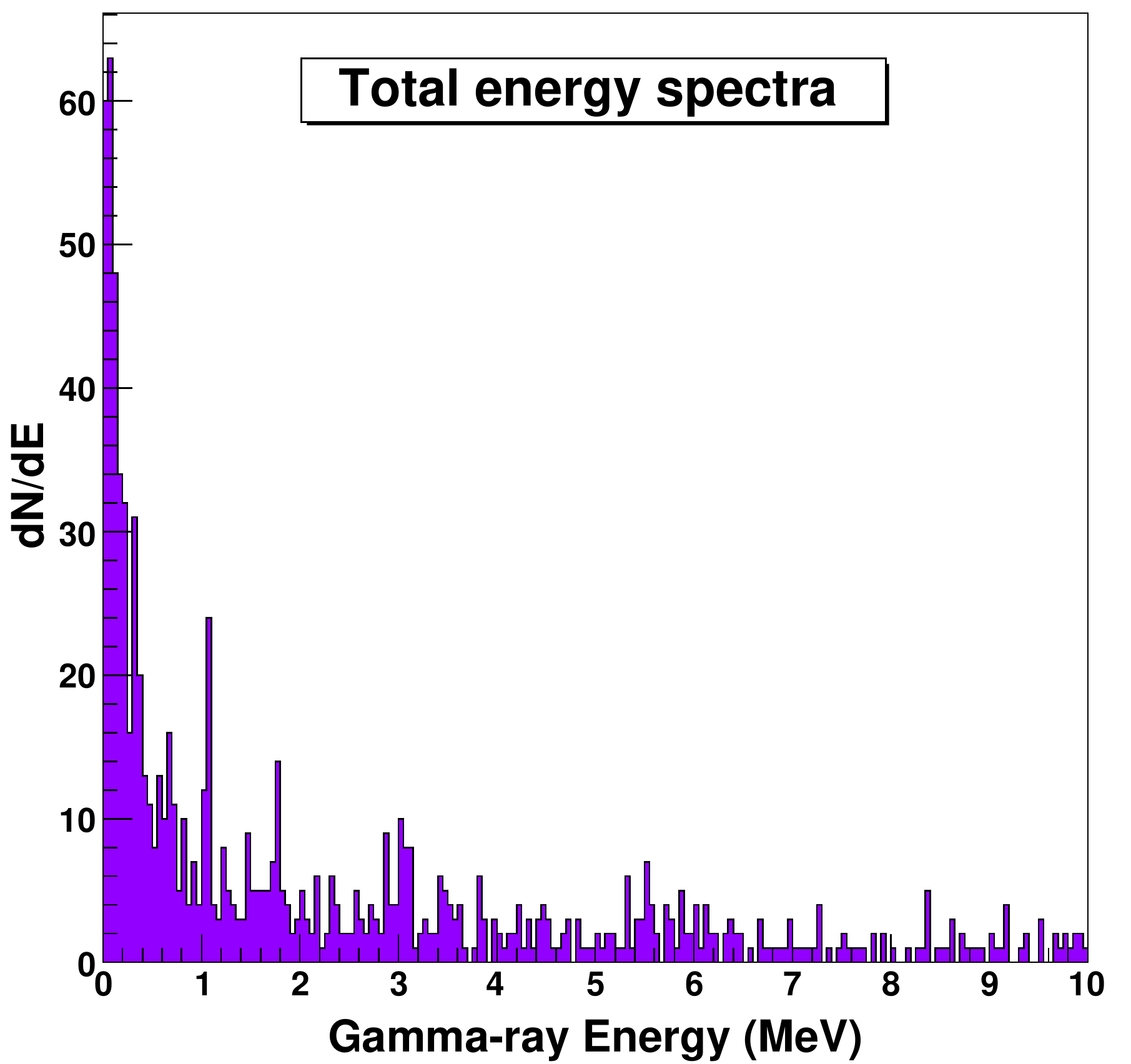}
  \caption[Total $\gamma$-ray spectrum  resulting from the decay of double hypernuclei
  produced in a $^{12}$C target and detected in the germanium array and before additional cuts.]
  {The statistics of the simulations corresponds to a data taking time of about two weeks.}
 \label{fig:phys:hyp:fig_EnerSpec}
\end{center}
\end{figure}

\Reffig{fig:phys:hyp:fig_EnerSpec} shows the total energy spectrum
summed over all germanium detectors for all events where a $\Xi^-$
has been stopped in the secondary target. Note, that the size of the
bins (50keV) in this plot is significantly larger than the
resolution of the germanium detectors expected even for high data
rates at normal conditions (3.4\,keV at 110\,kHz
\cite{bib:phy:kav07}). Even after 100 days of operation at \PANDA
and an integrated neutron fluence of about 6$\cdot$10$^9$
neutrons/cm$^2$, we expect a degradation of the resolution by less
than a factor of 3 to no more than 10\,{keV} \cite{bib:phy:lel03}.
Several peaks seen in the spectrum around 1, 1.68 and 3\,MeV are
associated with $\gamma$-transitions in various hypernuclei.
However, for a clear assignment of these lines obviously additional
experimental information will be needed.

%

\subsubsection{Weak Decays of Hypernuclei}

For the light hypernuclei relevant for the planned experiments the
non-mesonic and mesonic decays are of similar importance. In the
following we will focus on the case of two subsequent mesonic weak
decays of the produced double and single hypernuclei. For the light
nuclei discussed below this amounts to about 10\% of the total decay
width (see \Reffig{fig:phys:hyp:fig_nonmesonic}). Since the momenta
of the two pions are strongly correlated their coincident
measurement provides an effective method to tag the production of a
double hypernucleus. Moreover, the momenta of the two pions are a
fingerprint of the hypernucleus respective its binding energy.

In \PANDA the pions are tracked in the silicon strip detectors of
the secondary target. In the present configuration silicon strip
detectors with a pitch of 100$\mu$m and a two-dimensional readout
are implemented. For the reconstruction the standard software
package of \PANDA was applied. Since most $\Xi^-$ stop in the first
few millimetres of the secondary target, the efficiency for tracking
both pions produced in the subsequent weak decays is rather high.
After the statistical decay of the 7396 produced excited $\Xi^-$
hypernuclei, 14883 charged tracks are reconstructed out of which
8133 tracks are assigned as a $\pi^-$ candidate.

\begin{figure}[!h]
\begin{center}
  \includegraphics[width=\swidth]{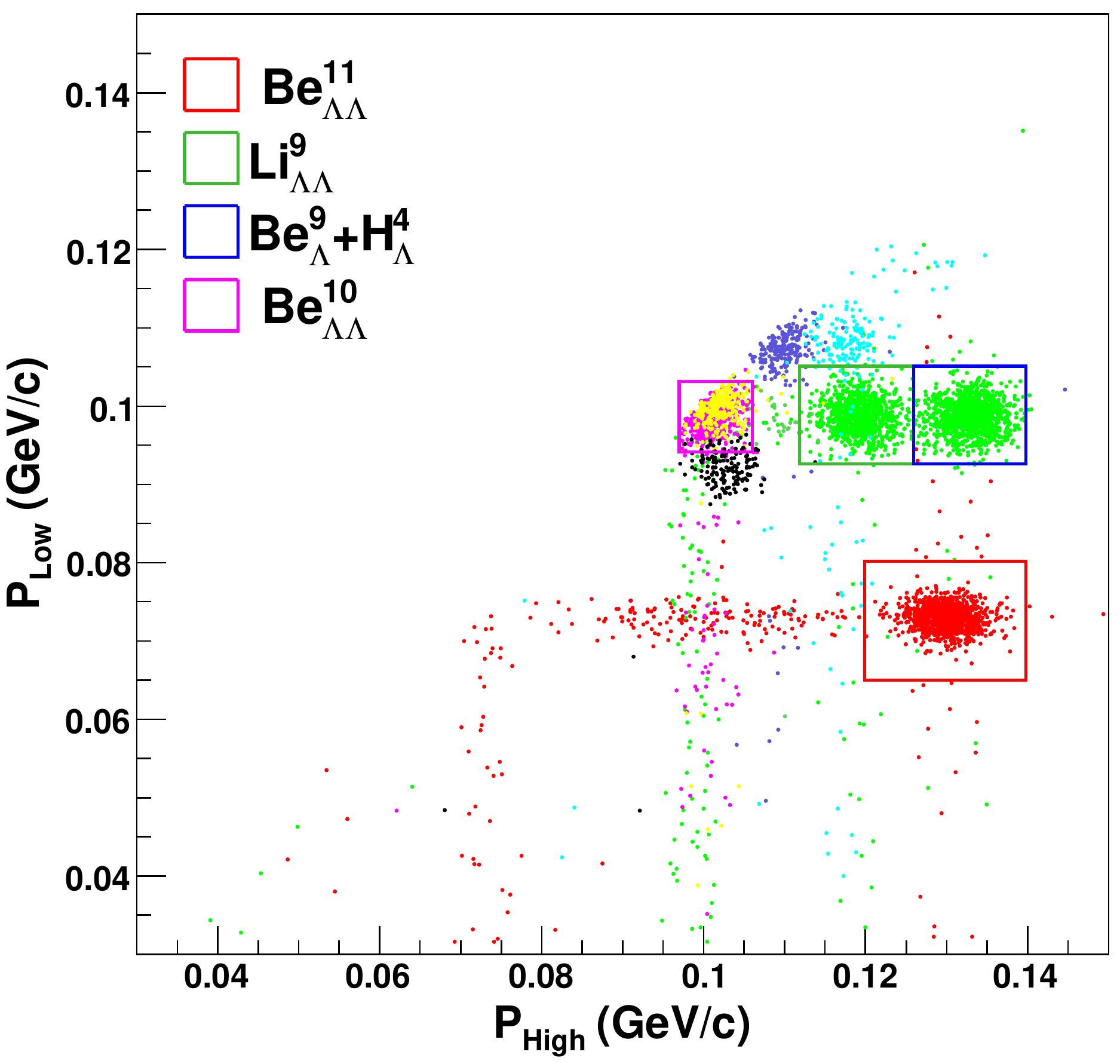}
    \includegraphics[width=\swidth]{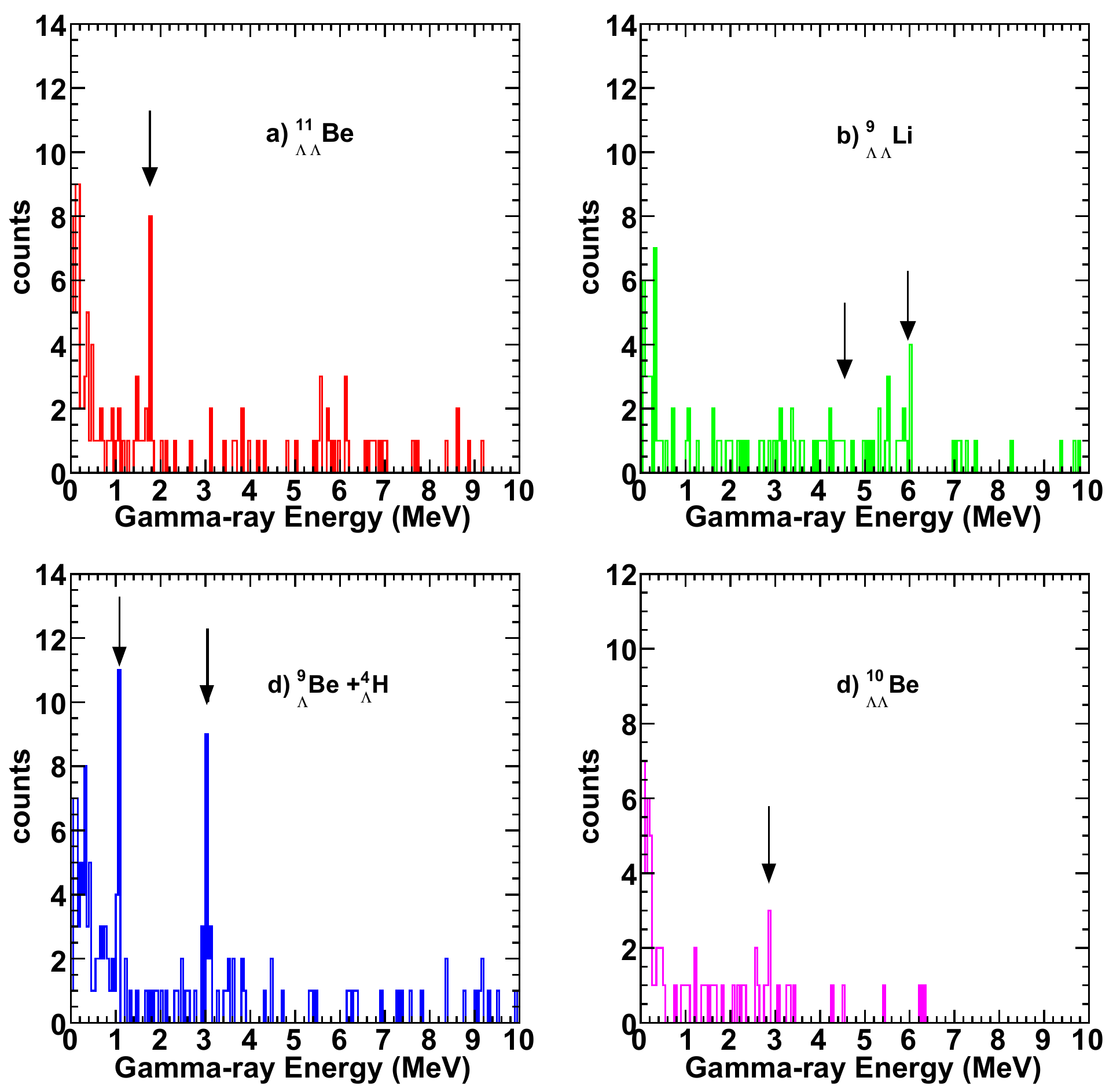}
  \caption[Upper part: Momentum correlation of all negative pion candidates resulting
  from the decay of double hypernuclei in a secondary $^{12}$C target.
  Lower part: $\gamma$-spectrum detected in the Ge-array
  by cutting on the two pion momenta.]{ The expected $\gamma$-transitions energies
  from single and double hypernuclei are marked by the arrows.}
 \label{fig:phys:hyp:fig_decay}
\end{center}
\end{figure}

The upper part of \Reffig{fig:phys:hyp:fig_decay} shows the momentum
correlation of all negative pion candidates from the secondary
$^{12}C$ target. The various bumps corresponding to different double
hypernuclei are marked by different colours. The good separation of
the different double hypernuclei provides an efficient selection
criterion for their decays.


The lower parts of \Reffig{fig:phys:hyp:fig_decay} show the
$\gamma$-ray spectra gated on the four regions indicated in the
two-dimensional scatter plot.  In the plots (a) and (d) the
1.684\,MeV $\frac{1}{2}^+$ and the 2.86\,MeV 2$^+$ states of
$^{11}_{\Lambda\Lambda}$Be and $^{10}_{\Lambda\Lambda}$Be,
respectively, can clearly be identified. Because of the limited
statistics in the present simulations and the decreasing photopeak
efficiency at high photon energies, the strongly populated high
lying states in $^{9}_{\Lambda\Lambda}$Li at 4.55 and 5.96\,MeV
cannot be identified in (b). The two dominant peaks seen in part (c)
result from the decays of excited single hyperfragments produced in
the $\Xi^- + C \rightarrow ^4_{\Lambda}H + ^9_{\Lambda}Be$ reaction,
i.e. $^4_{\Lambda}H$ at an excitation energy of 1.08\,MeV
\cite{bib:phy:bam73, bib:phy:bed76} and $^9_{\Lambda}Be$ at an
excitation energy of 3.029 and 3.060\,MeV \cite{bib:phy:may83,
bib:phy:aki02}.

In the present simulation several intermediate steps, which do not
effect the kinematics and hence the detection of the decay products,
have not been considered on an event-by-event basis. Of course,
these points are relevant for the final expected count rate:
\begin{itemize}
\item
a capture and conversion probability of the $\Xi^-$ of 5-10\%.
\item
typical probability of a double mesonic decay of 10\% (c.f.
\Reffig{fig:phys:hyp:fig_nonmesonic}).
\item
availability of data taking 50\% .
\end{itemize}
With these additional factors taken into account, the spectra shown
in \Reffig{fig:phys:hyp:fig_decay} correspond to a running time at
\PANDA of about two weeks. It is also important to realise that
gating on double non-mesonic weak decays or on mixed weak decays may
significantly improve the final rate by up to a factor 10.

\subsubsection{Background}

\begin{figure}[!h]
\begin{center}
  \includegraphics[width=\swidth]{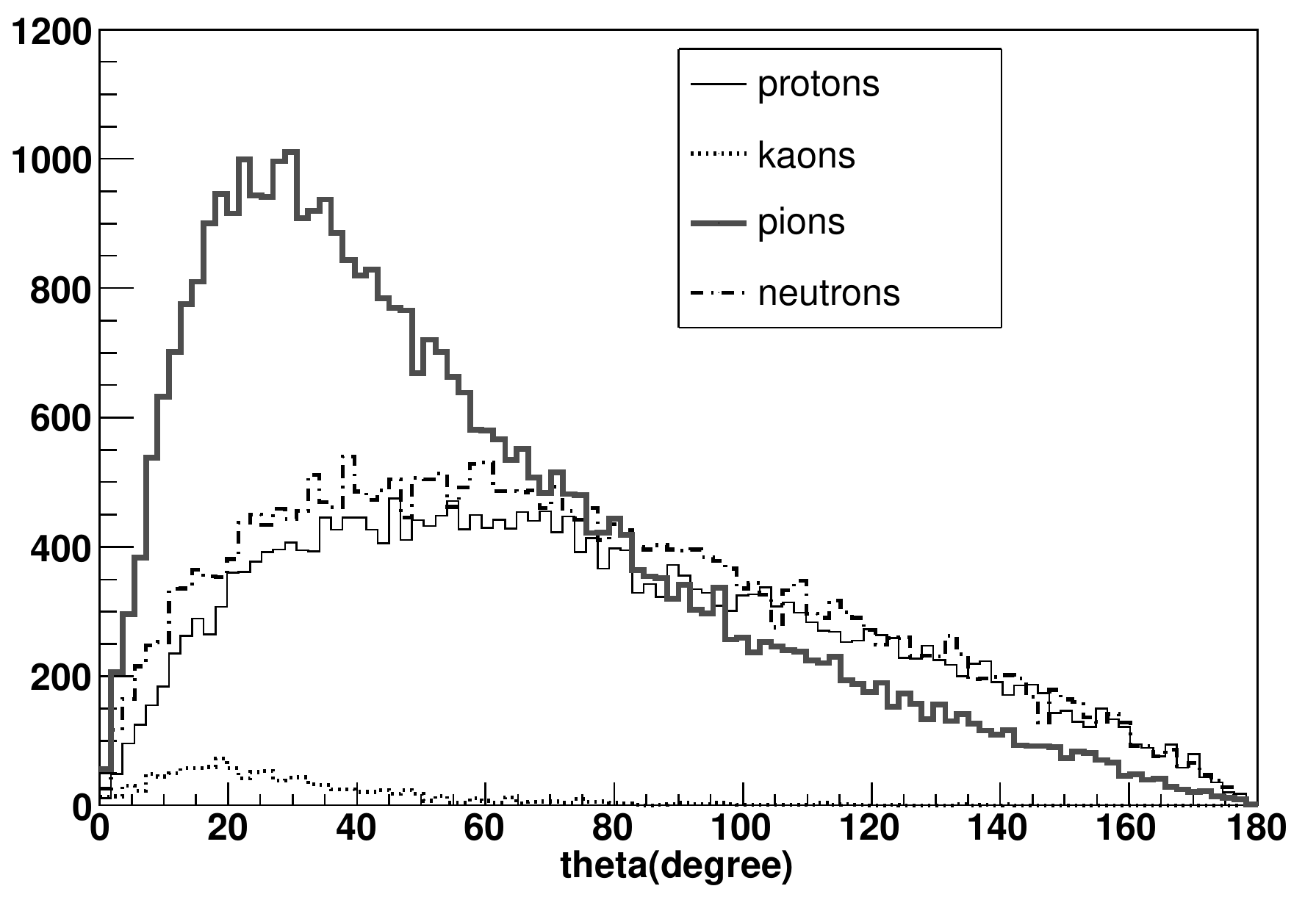}
  \caption[Distribution of produced particles from background
reactions.]{ The Germanium detectors will be affected mainly by
particles emitted at backward axial angle.}
 \label{fig:phys:hyp:fig_urqmd_angle}
\end{center}
\end{figure}

\begin{figure}[!h]
\begin{center}
  \includegraphics[width=\swidth]{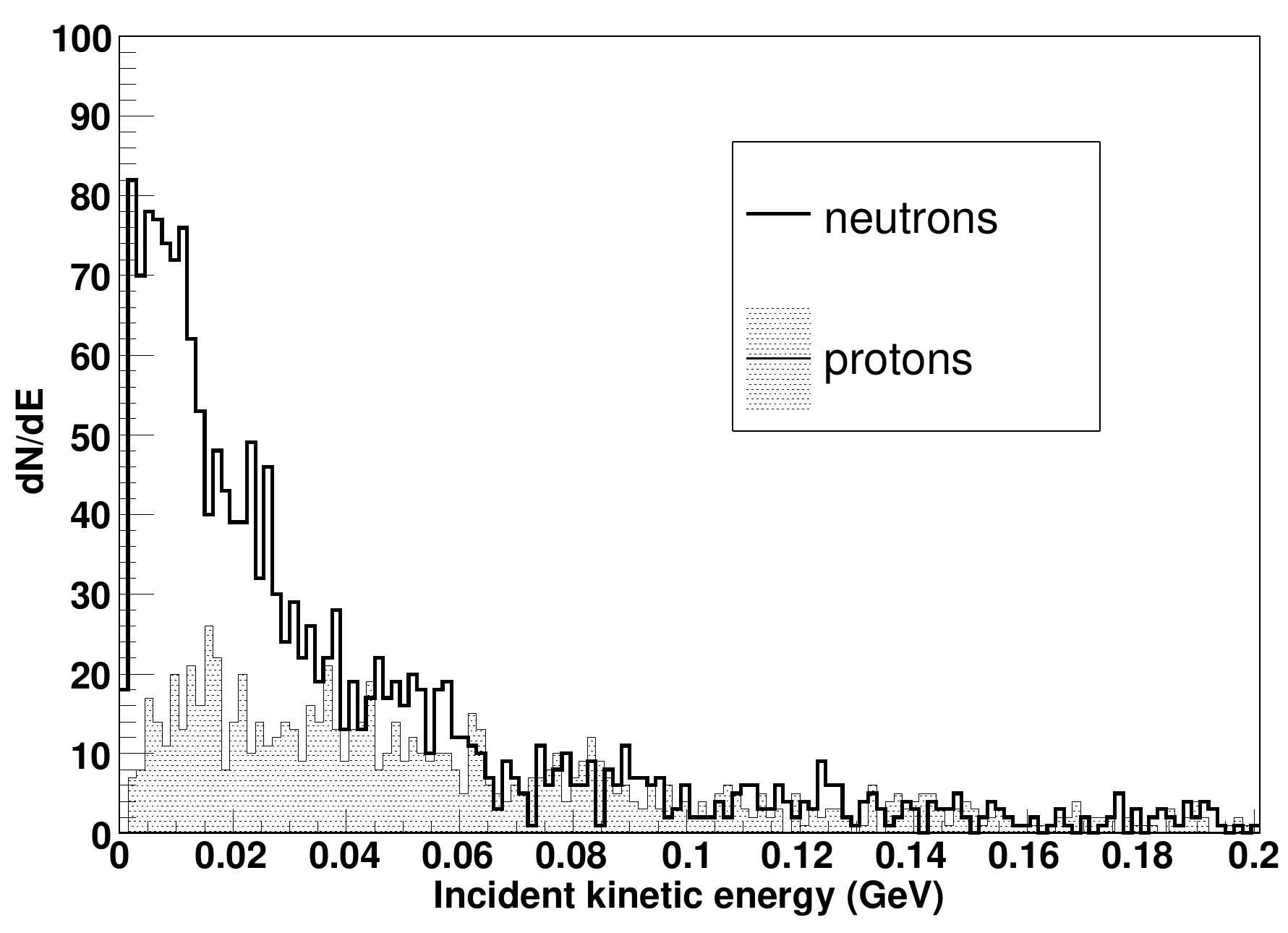}
  \caption[Incident kinetic energy of protons and neutrons
entering the Germanium detector surface.]{ The main contribution to a
possible radiation damage of the detector is provided by neutrons.}
 \label{fig:phys:hyp:fig_gebackg}
\end{center}
\end{figure}

Particles produced simultaneously with the double hypernuclei do not
significantly disturb the $\gamma$-ray detection. The main
limitation is the load of the Cluster--array by the high particles
rate from uncorrelated background reactions. The
$\overline{p}p\rightarrow \Xi^-\overline{\Xi}$ cross section of
2$\mu$b is about a factor 2500 smaller than the inelastic
$p\overline{p}$  cross section of 50mb at 3\,\gevc. Charged
particles and low energy neutrons which are emitted into the region
covered by the Ge detectors will undergo electromagnetic and nuclear
interactions and will thus contribute to the signal of the detector.
The total energy spectra in the crystal has been obtained summing up
event by event the energy contributions of the particles impinging
on the Ge array.

Background reactions have been calculated by using the
UrQMD+SMM~\cite{bib:phy:gal08} event Generator. At present, it is
not possible to simulate the detector response for a sample of
unspecific background events which is large enough to test
background suppression in all details. The background suppression
and signal detection capability can therefore only be estimated by
using extrapolations based on simplified assumptions. For the
present analysis 10000\, $\overline{p}+^{12}C$ interactions at
3\,\gevc were generated. Most of the produced charged and neutrals
particles are emitted into the forward region not covered by the
Germanium array (see \Reffig{fig:phys:hyp:fig_urqmd_angle}). Charged
particles emitted into backward axial angles are very low in kinetic
energy, and will be absorbed to a large fraction in the material
surrounding the primary target. More critical are neutrons emitted
into the backward direction which also contribute to the radiation
damage of the detector. \Reffig{fig:phys:hyp:fig_gebackg} shows the
kinetic energy distribution of protons and neutrons entering the
surface of the Germanium detectors.

The total energy spectra resulting from the background simulation
have been filtered by using the same technique as it was done for
the signal events. Particularly the same cuts on correlated pion
candidates have been applied to the background events, in order to
obtained the corresponding background spectrum for each of the
hypernuclei channels. For $^{11}_{\Lambda\Lambda}$Be as well as
$^{10}_{\Lambda\Lambda}$Be only one single event survived the cuts.
Both of these events had an energy deposition in the germanium
detector exceeding 10 MeV significantly.

Several further improvements of the background suppression are
expected by exploring the topology of the sequential weak decays.
This includes the analysis of tracks not pointing to the primary
target, multiplicity jumps in the detector planes and the energy
deposition in the secondary target. Furthermore kaons detected in
the central detector of \PANDA at forward angles can be used to tag
the $\overline{\Xi}$ production.

\clearpage

%% file: phys/phys_nucleonstructure.tex
%
\section{The Structure of the Nucleon Using Electromagnetic Processes}
\COM{Author(s): M.P. Bussa, M. D\"uren, F. Maas, M. Maggiora} 
%
\input{./phys/nucleonstructure/gda/phys_gda}
\input{./phys/nucleonstructure/spinstructure/phys_spinstructure}
\input{./phys/nucleonstructure/elmff/phys_elmff}

%% file: phys/nucleonstructure/gda/phys_gda.tex
%
\subsection{Partonic Picture of Hard Exclusive $\pbarp$-Annihilation Processes}
%
%
\subsubsection*{Introduction}

A wide area of the physics program of \Panda\  concerns studies of the
non-perturbative region of QCD. However, the experimental setup
foreseen offers the opportunity to study also a certain class of hard 
exclusive processes that give insight into an intermediate region,
which marks the transition towards increasingly important perturbative
QCD effects.  

In the recent years, the theoretical framework of generalised parton
distributions (GPDs) has been developed, which allows treating hard
exclusive processes in lepton scattering experiments on a firm QCD
basis \cite{ref:nuclstruc:gda:Belitsky:2005qn,
ref:nuclstruc:gda:Ji:2004gf, ref:nuclstruc:gda:Diehl:2003ny,
ref:nuclstruc:gda:Goeke:2001tz}. 
This is possible under suitable conditions where one can 
factorise short and long distance contributions to the reaction
mechanism. Being related to non-diagonal matrix elements, GPDs do not
represent any longer a mere probability, but rather the interference
between amplitudes describing different parton configurations of the
nucleon, thus giving access to various momentum correlations. Their
importance was first stressed  in studies of deeply 
virtual Compton scattering (DVCS)\cite{ref:nuclstruc:gda:Ji:1996ek,
ref:nuclstruc:gda:Radyushkin:1996nd, ref:nuclstruc:gda:Ji:1996nm}, for
which it could be rigorously proven that the  
QCD handbag diagram 
\begin{figure}[h]
\begin{center}
  \includegraphics[width=0.5\textwidth]{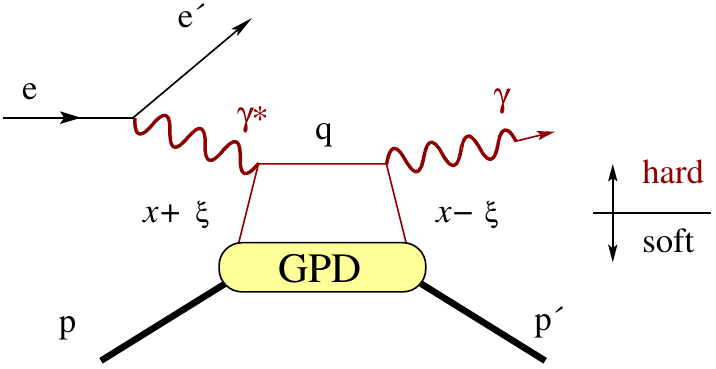}
  \caption{DVCS can be described by the handbag diagram, as there is
factorisation between the upper `hard' part of the diagram which is
described by perturbative QCD and QED,  and a lower `soft' part that is
described by GPDs.}
 \label{fig:nuclstruc:gda:dvcsdiagram}
\end{center}
\end{figure}
(see \Reffig{fig:nuclstruc:gda:dvcsdiagram})
dominates the process in certain kinematical domains and that
factorisation holds, {\it i.e.} that the process is divided into a
hard perturbative QCD process and a soft part of the diagram which is
parametrised by GPDs. The application of perturbative QCD is possible
in DVCS due to the hard scale defined by the large virtuality $Q^2$ of
the exchanged photon. A second example for the application of the
handbag formalism is wide angle Compton scattering (WACS). Here the
hard scale is related to the large transverse momentum of the final
state photons. 

\begin{figure}[h]
\begin{center}
  \includegraphics[width=0.3\textwidth]{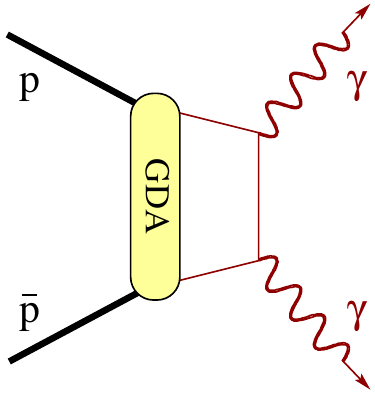} 
  \caption{The handbag diagram may describe the inverted WACS process
$\pbarp\to \gamma\gamma$ at \Panda\  energies. 
 \label{fig:nuclstruc:gda:wacsdiagram}}
\end{center}
\end{figure}

The important question which arises is whether the concepts that are
used in lepton scattering experiments have universal applicability and
can therefore be used in studies of $\pbarp$-annihilation
processes with the crossed kinematics. The crossed diagram of WACS is
the process $\pbarp\to \gamma\gamma$ with emission of the two final
state photons at large 
polar angle in the CM system (see
\Reffig{fig:nuclstruc:gda:wacsdiagram}).
It can be shown that the handbag approach is not appropriate to
describe the crossed channel WACS neither at very small nor at very
large energies
\cite{ref:nuclstruc:gda:Radyushkin:1998rt,
ref:nuclstruc:gda:Diehl:1998kh}. However, there are strong 
arguments and first experimental indications that the handbag approach
is appropriate at the intermediate energy regime where \Panda\  operates 
\cite{ref:nuclstruc:gda:Kroll:2005ni, ref:nuclstruc:gda:Kuo:2005nr},
even though a rigorous proof of factorisation has not been achieved
yet. The corresponding amplitudes 
that parametrise the soft part of the annihilation process ({\it i.e.}
 the counterparts of GPDs) are 
called generalised distribution amplitudes (GDAs). The measurement of
the process $\pbarp\to \gamma\gamma$ as a function of $s$ and $t$ is
an experimental challenge, due to the smallness of the cross
section. The high luminosity and the excellent detector,
especially the $4\pi$ electromagnetic calorimeter, should enable \Panda\
to separate this process from the large hadronic background. 

A second, much more abundant process that can be described in terms of
handbag diagrams is $\pbarp\to \pi^0\gamma$. In contrast to WACS, in
this case one photon is replaced by a pseudo-scalar meson, but
otherwise the theoretical description is similar. First
experimental results from the Fermilab experiment E760 indicate that
the handbag approach is appropriate to accommodate the data in
the range $s\sim 8.5-13.5$\,GeV$^2$ \cite{ref:nuclstruc:gda:Kroll:2005ni}. 

The handbag approach ({\it i.e.} the factorisation assumption) is
suitable for the description of further reactions, like $\pbarp\to
M\gamma$ where $M$ is any neutral meson ({\it e.g.} a $\rho^0$) or
$\pbarp\to \gamma^\ast\gamma$, where $ \gamma^\ast$ decays into an \ee- or
\mumu-pair. The latter process is described by the crossed diagram of 
DVCS. Unfortunately, the factorisation proof in DVCS $\gamma^\ast\mbox{p}
\to \gamma\mbox{p}$ is not applicable for the crossed diagram
$\pbarp\to \gamma^\ast\gamma$, as the virtuality $Q^2$ of the final state
$\gamma^\ast$ is limited to be smaller than $s$, in contradiction to the
assumption made in the proof of factorisation of this diagram in DVCS
kinematics.  

\begin{figure}[h]
\begin{center}
  \includegraphics[width=0.45\textwidth]{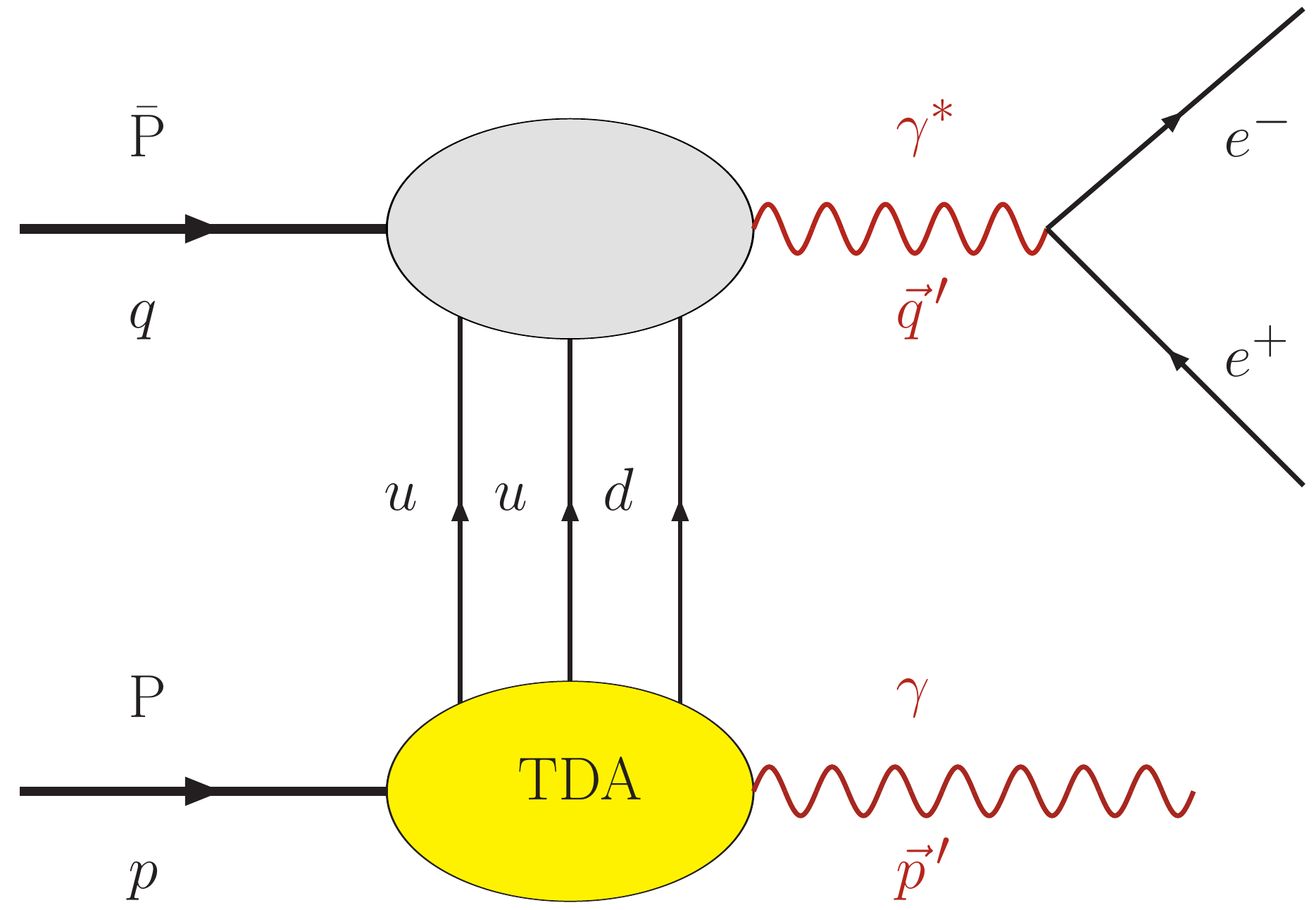} 
  \caption{The production of a hard virtual photon (upper part) is a
hard sub-process that factorises from the lower part,
which can be described by a hadron to photon transition distribution
amplitude (TDA). 
 \label{fig:nuclstruc:gda:tdadiagram}} 
\end{center}
\end{figure}

In a complementary theoretical approach, the process $\pbarp\to
\gamma^\ast\gamma$ is not described by the handbag diagram but by
so-called transition distribution amplitudes (TDAs)
\cite{ref:nuclstruc:gda:Lansberg:2006uh, ref:nuclstruc:gda:Pire:2005ax}
that parametrise the
transition of a proton 
into a (virtual) photon according to the 
diagram in \Reffig{fig:nuclstruc:gda:tdadiagram}.  
In a similar way the exclusive meson production $\pbarp\to
\gamma^\ast\piz$ can be described (see \Reffig{TDAfig1}).
  
The theoretical understanding of GPDs and related unintegrated
distributions is just at its beginning. There is an extended experimental
endeavour by lepton scattering experiments at \INST{DESY}, \INST{CERN} and \INST{JLAB} to get
access to these powerful distributions. \Panda\  has the chance to
join this quest for an improved description of the nucleon structure by
measuring the crossed-channel counterparts of these distributions in
hard exclusive processes with various final states in a new
kinematical region. New insights into the applicability and
universality of these novel QCD approaches can be expected.  

\subsubsection*{Crossed-Channel Compton Scattering}

\begin{figure}[h]
\begin{center}
  \includegraphics[width=0.3\textwidth]{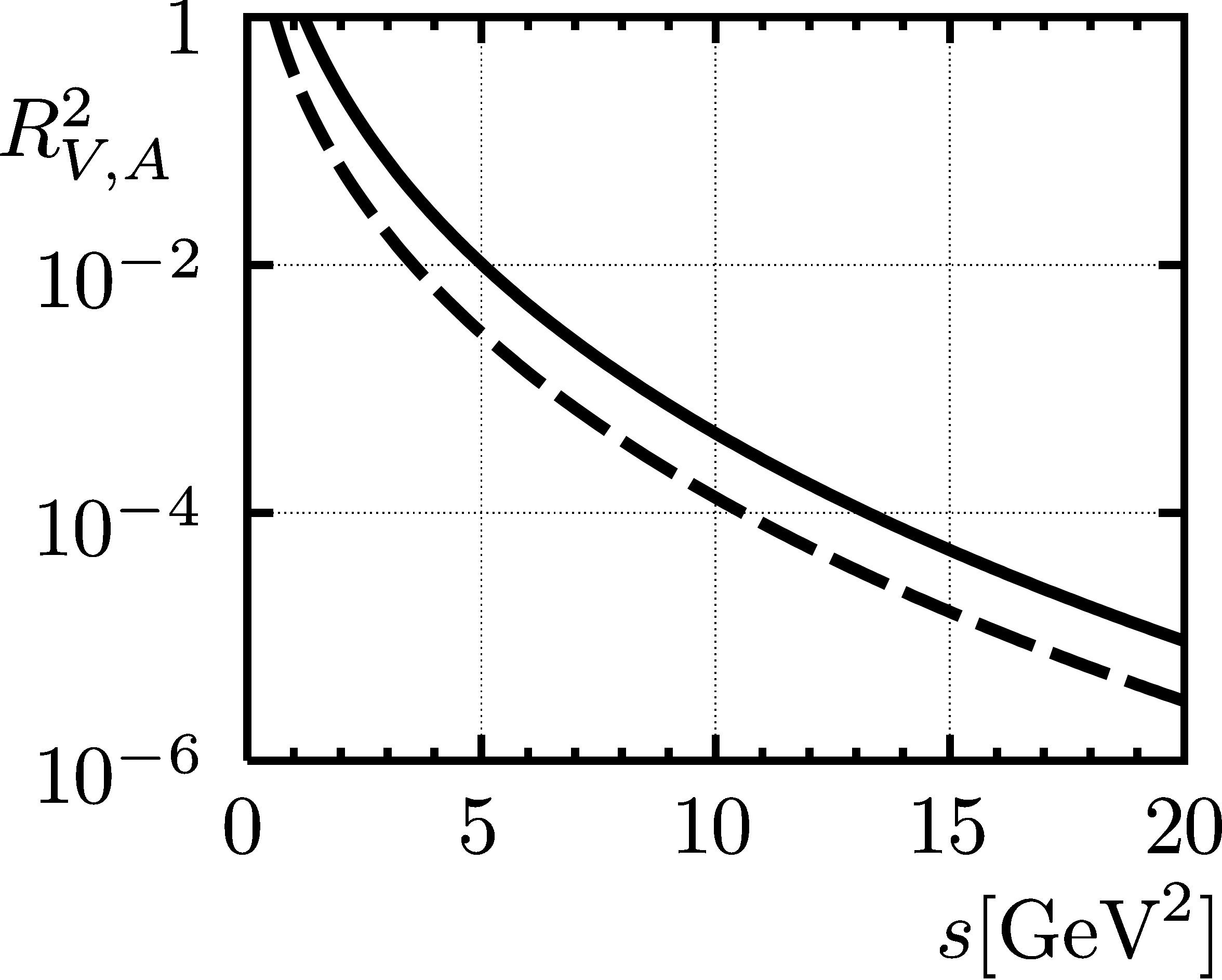}   
\caption{The squared form factors $R_V^2(s)$ (solid line) and
$R_A^2(s)$ (dashed line), as calculated from the double distribution
model (Fig. 2 of ref. \cite{ref:nuclstruc:gda:Freund:2002cq})      
 \label{fig:nuclstruc:gda:rv}} 
\end{center}
\end{figure}

It has been argued \cite{ref:nuclstruc:gda:Freund:2002cq} that the crossed-channel
Compton scattering, namely exclusive proton-antiproton annihilation
into two photons, $\pbarp\to\gamma\gamma$ can also be described in a
generalised parton picture at large $s$ with $|t|,|u|\sim s$. The two
photons are predominantly emitted in the annihilation of a single
``fast'' quark and antiquark originating from the proton and
antiproton. The new double distributions, describing the
transition of the $\pbarp$ system to a $\qqbar$ pair, can be 
related to the timelike nucleon form factors; by crossing symmetry 
they are also connected with the usual quark/antiquark
distributions in the nucleon. With a model for the double partonic
distributions one can compute the $\pbarp\to\gamma\gamma$ amplitude
from the handbag graphs of
\Reffig{fig:nuclstruc:gda:wacsdiagram}. The result for the 
helicity-averaged differential cross section is 
\begin{equation}
\frac{d\sigma}{d\cos \theta}= \frac{2\pi\alpha^2_{\mbox{em}}}{s}
\frac{R_V^2(s) \cos^2\theta + R_A^2(s)}{\sin^2\theta}
\label{nuclstruc:gda:angdiffgamma}
\end{equation}
with the energy dependency of the squared form factors $R_V^2(s)$ and
$R_A^2(s)$ depicted in \Reffig{fig:nuclstruc:gda:rv}. 

Recent measurements of the time-reversed process $\gamma\gamma\to
\pbarp$ by the \INST{BELLE} collaboration \cite{ref:nuclstruc:gda:Kuo:2005nr}
tend to confirm the predicted asymptotic behaviour at higher energies,
however at intermediate energies (2.5 -- 4\,GeV) they can not be
entirely explained by the existing theoretical models.

\subsubsection*{Hard Exclusive Meson Production}

\begin{figure}[h]
\begin{center}
  \includegraphics[width=0.5\textwidth]{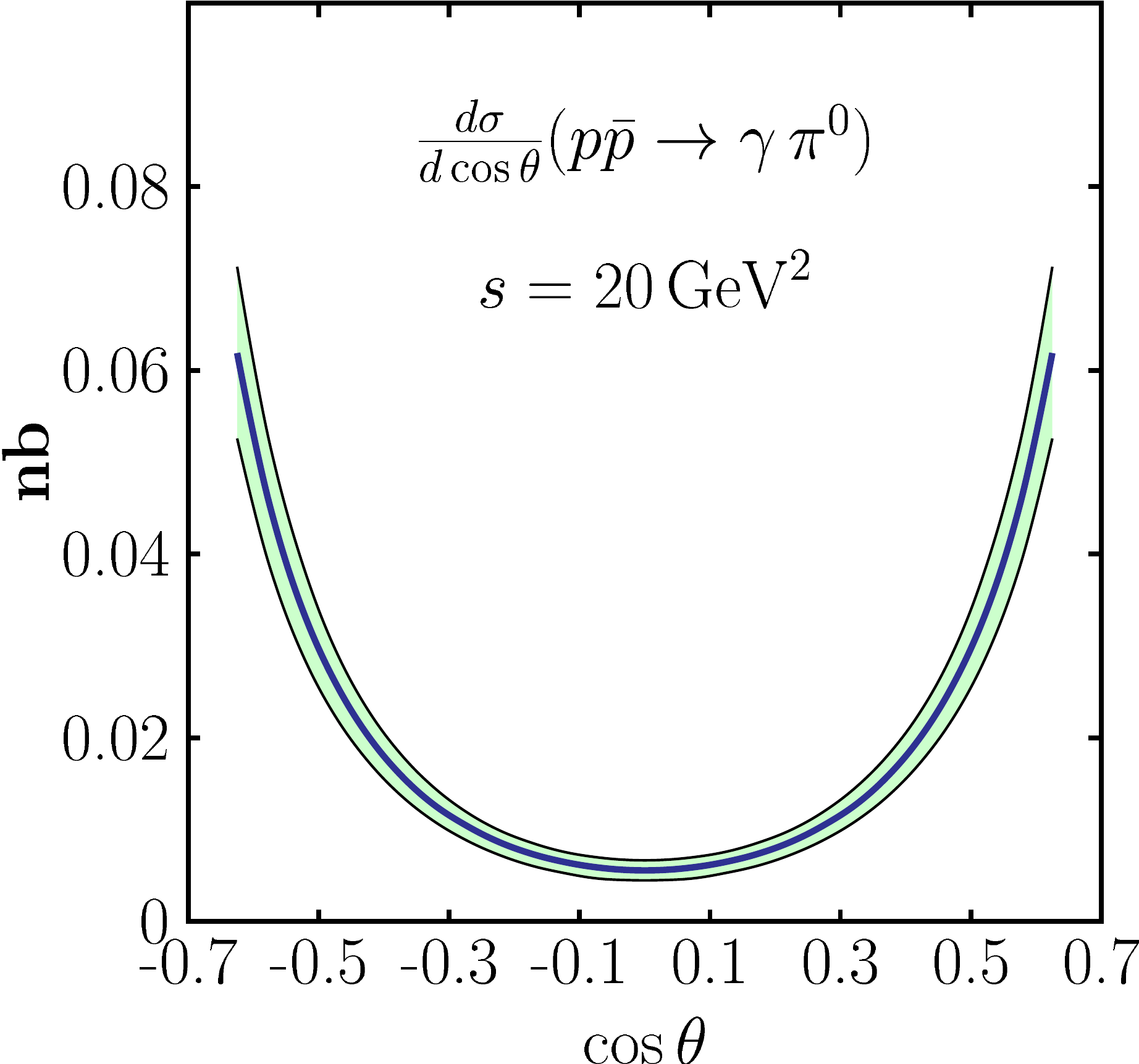}
  \caption{Cross section prediction for the angular distribution at  $s=20$\,GeV$^2$ for \Panda\  
taken from ref.~\cite{ref:nuclstruc:gda:Kroll:2005ni} .
 \label{fig:nuclstruc:gda:pigaangular}}
\end{center}
\end{figure}

Besides detecting $\pbarp\to\gamma\gamma$, hard exclusive meson
production, like  $\pbarp\to\gamma\piz$, 
can also provide valuable
information about the structure of the proton. 
\begin{figure}[h]
\begin{center}
\includegraphics[width=0.45\textwidth]{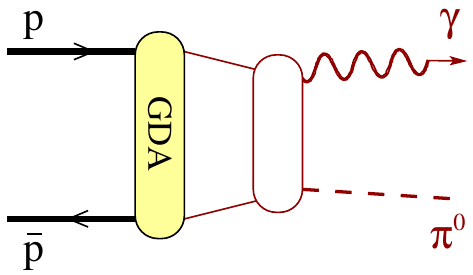}
\caption[The handbag contribution to $\pbarp\to\gamma\piz$ at large
but non-asymptotic $s$.]{ The blob on the right hand side represents the
parametrised general dynamics for $\qqbar \to \gamma\piz$}
\label{fig:nuclstruc:gda:gammapi}
\end{center}
\end{figure}
In treating this
process, one can adopt as starting point the assumption of the handbag
factorisation of the amplitude for the kinematical region
$s,-t,-u\gg\Lambda^2$ (see \Reffig{fig:nuclstruc:gda:gammapi}),
where $\Lambda$ is a typical hadronic scale of the order of 1\,GeV, as
in \cite{ref:nuclstruc:gda:Kroll:2005ni}. A comparison with 
existing \INST{Fermilab} \INST{E760} data \cite{ref:nuclstruc:gda:Armstrong:1997gv}
concerning both the energy dependence of the integrated cross section
and angular  distribution of the differential cross section makes it
possible to have a reliable prediction for the differential cross
section of $\pbarp\to\gamma\gamma$ at \Panda, as shown in
\Reffig{fig:nuclstruc:gda:pigaangular}. 

There is also an approach to describe hard exclusive meson 
production by transition distribution amplitudes for reactions 
like $\pbarp\to\gamma^*\piz\to\ee\piz$. Details are described in ref.~\cite{ref:nuclstruc:gda:Pire:2005xx}.

\subsubsection*{Simulation of $\pbarp \to \gamma\gamma$ and $\pbarp \to
\gamma\pi^0$}

In order to test the possibility of detection of both the $\pbarp \to
\gamma\gamma$ and $\pbarp \to\gamma\pi^0$ reactions, Monte Carlo
simulations were run within the PANDARoot framework. The main goal of
these studies was to estimate the ability of the \Panda detector system
to separate useful physics events from the background.

\begin{figure}[h]
\begin{center}
 \includegraphics[width=0.45\textwidth]{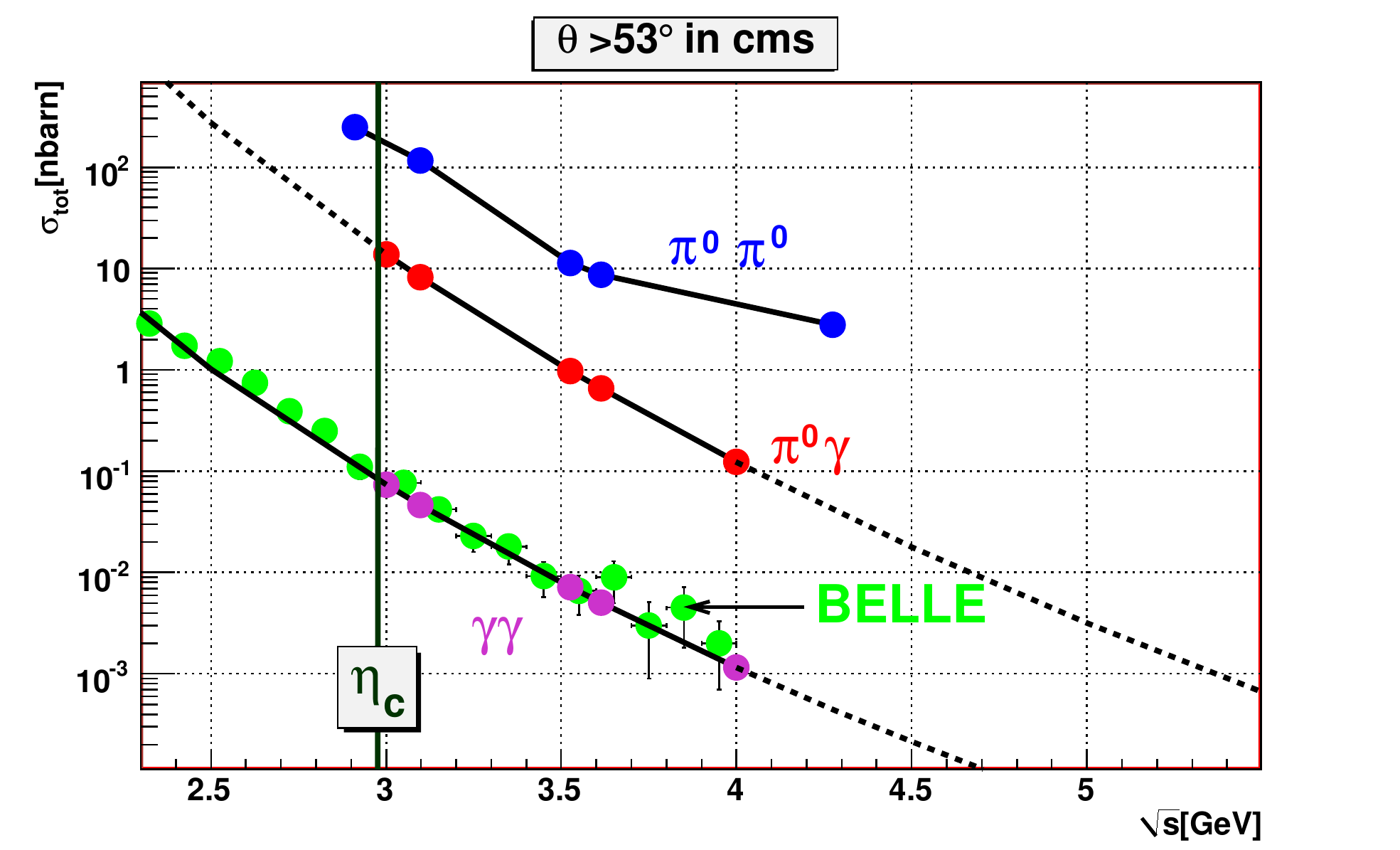}
\caption[Cross sections for processes with $\gamma\gamma$,
$\piz\gamma$ and $\piz\piz$ in the final state, for different
\Panda energies.]{ The results correspond to particles with
$|\cos(\theta)|<0.6$ in the centre of mass system (Refs. \cite{ref:nuclstruc:gda:ONG,
ref:nuclstruc:gda:Kroll:2005ni,ref:nuclstruc:gda:Kuo:2005nr}).}
\label{fig:nuclstruc:gda:crossections}
\end{center}
\end{figure}

The main background for the crossed channel Compton scattering $\pbarp \to
\gamma\gamma$ comes from reactions with neutral hadrons in the final
states, like $\pbarp \to \piz\piz$ or  $\pbarp
\to\gamma\piz$. After a comparison of the various cross sections of
interest at \Panda energies, given in
\Reffig{fig:nuclstruc:gda:crossections},
 it can be concluded that the
number of the exclusive background events considered is roughly three
(respectively two) orders of magnitudes higher than the number of the
events of interest. This result has been taken into account all along
our studies.

The background has a significant constituent originating in reactions
with charged particles in the final states. These events are supposed
to be vetoed by the detector subsystems that are sensitive to charged
particles, however a systematic study of them was not yet performed.

In order to test the detector response,  $\gamma\gamma$,
$\piz\gamma$ and $\piz\piz$ events were generated using {\tt EvtGen}. For
the angular distribution of $\gamma\gamma$ events
\Refeq{nuclstruc:gda:angdiffgamma}  fitted to the \INST{BELLE} data
\cite{ref:nuclstruc:gda:Kuo:2005nr} was used, for $\piz\gamma$
results similar to the one in \Reffig{fig:nuclstruc:gda:pigaangular}
in \cite{ref:nuclstruc:gda:Kroll:2005ni} were considered, the
$\piz\piz$ distribution follows the one given by Ong and Van de
Wiele \cite{ref:nuclstruc:gda:ONG}. The centre of mass angle of the
generated particles was 
limited by  $|\cos(\theta)|<0.6$. 
\begin{figure}[h]
\begin{center}
 \includegraphics[width=0.45\textwidth]{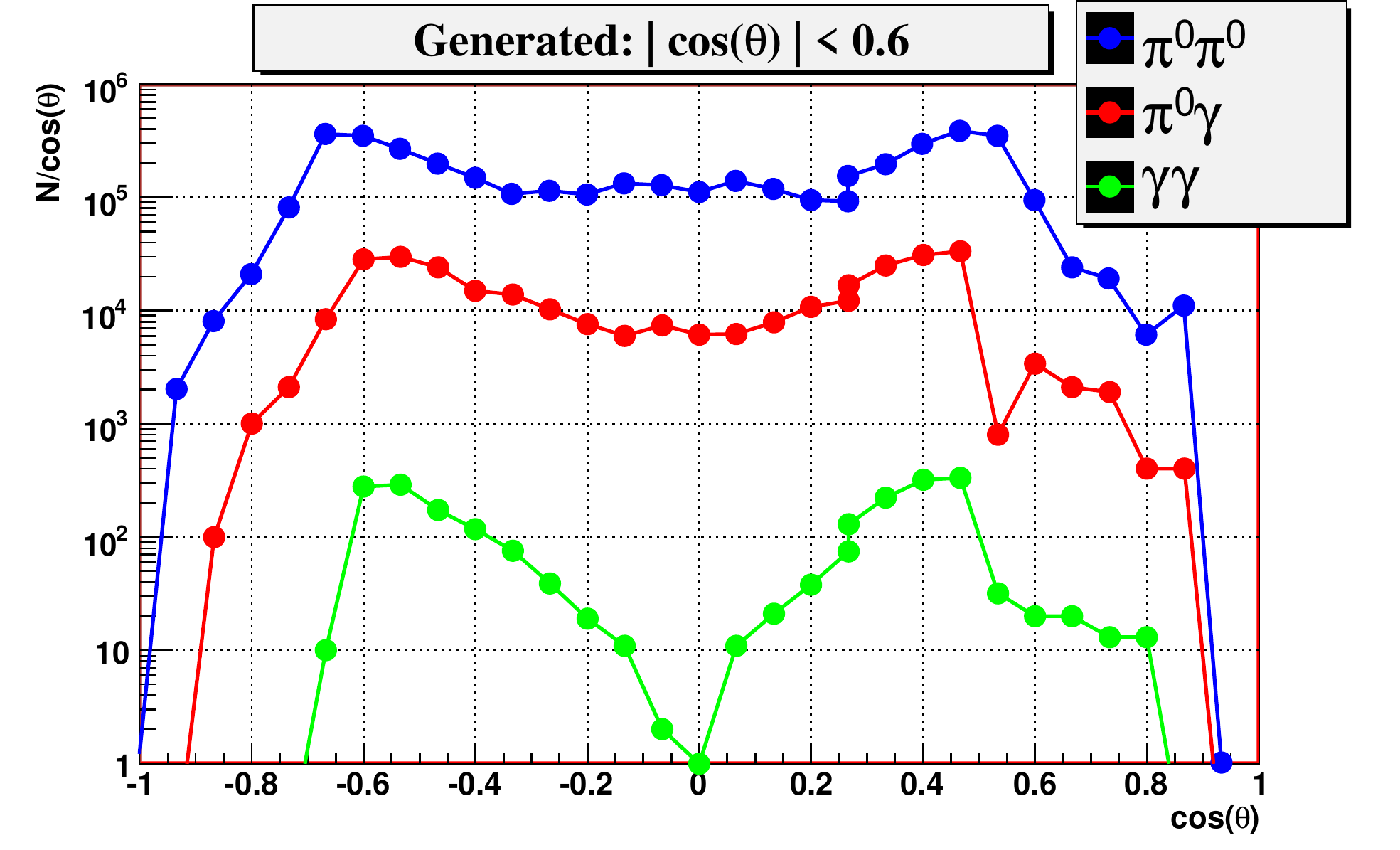}
\caption{Angular distribution of generated particles in the centre of
mass system, as seen by the detector.}
\label{fig:nuclstruc:gda:differential}
\end{center}
\end{figure}

\Reffig{fig:nuclstruc:gda:differential} gives the distribution of
reconstructed events in the electromagnetic calorimeter.
The influence of the asymmetric detector system is clearly
visible, which
introduces a difference between the angular distribution of particles
moving in forward or backward direction. Noticeably, detected
particles seem to be found also outside the $|\cos(\theta)|<0.6$
angular limit, a feature caused by the finite digitisation in the detector.

For the reconstruction of $\gamma\gamma$ events the following
algorithm was used:
\begin{itemize}
\item Identify all the bumps in the electromagnetic calorimeter and
order them according to their energy in the centre of mass system.
\item Associate single photons with the first two bumps of highest
cms-energy.
\item Evaluate the kinematic factor
\begin{eqnarray}
K_{\gamma\gamma}\!\!&\!\!\sim\!\!&\!\!
\sqrt{|\vec{p}_{\overline{\mbox{\tiny p}}}-\vec{p}_{\gamma_1}-\vec{p}_{\gamma_2}|^2
+|E_{\mbox{\tiny L}}-E_{\gamma_1}-E_{\gamma_2}|^2}\nonumber\\
&\times&\left( \frac{1}{\sqrt{
\vec{p}_{\gamma_1}^2+E_{\gamma_1}^2} } + \frac{1}{\sqrt{
\vec{p}_{\gamma_2}^2+E_{\gamma_2}^2} } \right)\label{nuclstruc:gda:Kgamma}
\end{eqnarray}
\end{itemize}

Events with $\gamma\piz$ in the final state are reconstructed in a
slightly different way:
\begin{itemize}
\item Order the bumps in the EMC as above.
\item Associate a single photon to the first one, the photons from
$\piz$ to the next two of them.
\item Evaluate the kinematic factor $K_{\gamma\piz}$ similar to
\Refeq{nuclstruc:gda:Kgamma}
\end{itemize} 
First a comparison of the kinematic factors for the signal and background
is performed,  followed
by a cut on the value of the kinematic factor in order to further
eliminate possible misidentifications. Finally, a cut on the number of
hits in the EMC is applied to differentiate between photons and pions.

\begin{figure}[h]
\begin{center}
 \includegraphics[angle=90,width=0.45\textwidth]{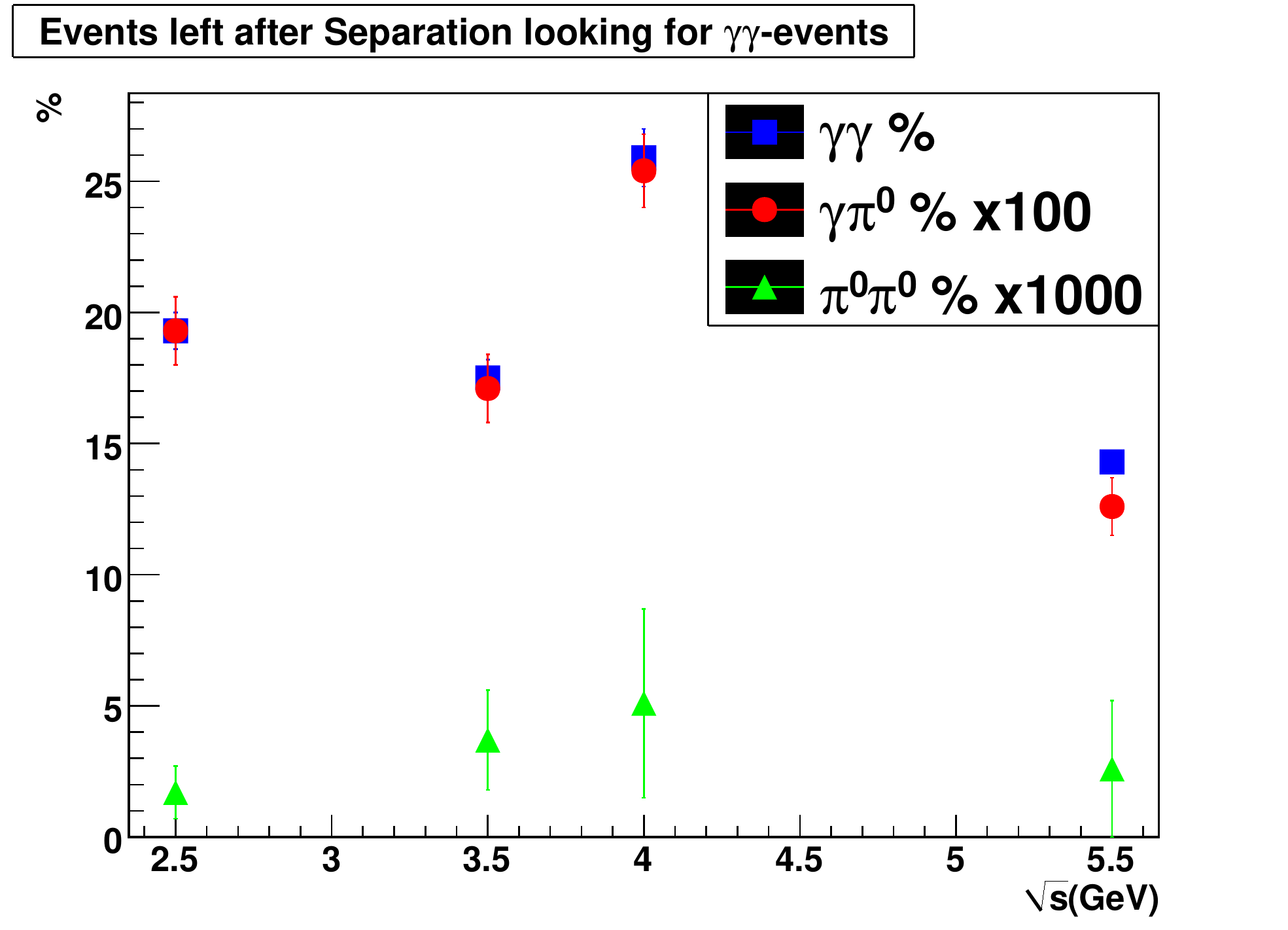}
\caption[Separation of $\gamma\gamma$ events from the neutral
background $\gamma\piz$ and $\piz\piz$ is possible using Monte Carlo corrections.]{Separation of $\gamma\gamma$ events from the neutral
background $\gamma\piz$ and $\piz\piz$ is possible using Monte Carlo corrections. The number of misidentified $\gamma\piz$ ($\piz\piz$)
events and their statistical error were magnified by a factor $10^2$,
($10^3$) to match the limited Monte Carlo statistics to the abundance of the background in the experiment.}
\label{fig:nuclstruc:gda:gammagammasep}
\end{center}
\end{figure}

\begin{figure}[h]
\begin{center}
\includegraphics[angle=90,width=0.45\textwidth]{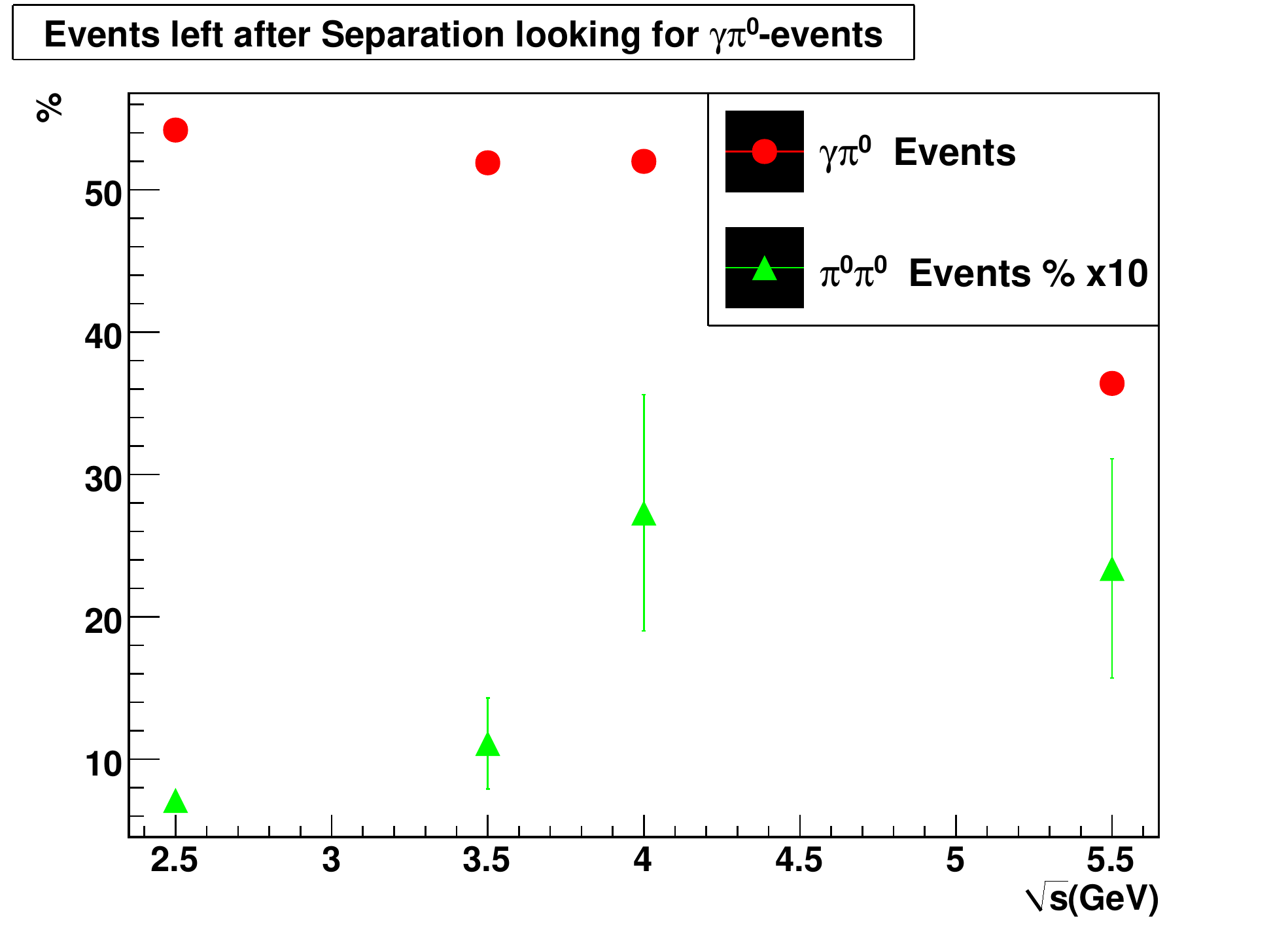}
\caption[Separation of $\gamma\piz$ events from the neutral
background.]{Separation of $\gamma\piz$ events from the neutral
background. The number of misidentified  $\piz\piz$
events and their statistical error were magnified by a factor 10.}
\label{fig:nuclstruc:gda:gammapisep}
\end{center}
\end{figure}

In \Reffig{fig:nuclstruc:gda:gammagammasep} and
\ref{fig:nuclstruc:gda:gammapisep} preliminary results on the separability
 of $\gamma\gamma$, $\gamma\piz$ and $\piz\piz$ events are
presented. Four different 
$\pbarp$ centre of mass energies ($\sqrt{s}=$ 2.5, 3.5, 4 and 5.5\,GeV) 
were considered, and for all energies events of the type $\gamma\gamma$, $\piz\gamma$ and $\piz\piz$  were generated in the
$|\cos(\theta)|<0.6$ angular interval. A full simulation of the
detector system was performed and the event recognition algorithms
with the corresponding cuts were applied. 

To obtain a realistic picture, the ratio of the cross sections 
(see \Reffig{fig:nuclstruc:gda:crossections}) for the various
processes involved must be taken into account, when evaluating the
possibility of detecting an useful physics signal. Therefore, lacking
the necessary Monte Carlo statistics, the numbers of $\piz\gamma$ and
$\piz\piz$ events misidentified as $\gamma\gamma$ have to be
multiplied by factors of $100$ and  $1000$, respectively (see
\Reffig{fig:nuclstruc:gda:gammagammasep}). A similar
approach was taken while identifying $\piz\gamma$ events
(\Reffig{fig:nuclstruc:gda:gammapisep}). In this case the
number of misidentified $\piz\piz$ events  were
multiplied only by a factor of 10.    

After this selection and normalisation, the  $\gamma\gamma$ signal is at the same level as the pion background, which means that the extraction of cross sections for this class of events should be possible by the use of Monte Carlo corrections over the full kinematical range of \Panda. The studies are still very preliminary and only simple cuts were applied, so that we are positive that more sophisticated cuts in future will improve the situation.
 
The separation of exclusive $\gamma\piz$ events is clearly possible over the full energy range after applying the above simple cuts as shown in \Reffig{fig:nuclstruc:gda:gammapisep}. 

\subsubsection*{Conclusions}

First simulation results of the processes $\pbarp \to \gamma\gamma$ and $\pbarp \to \gamma\piz$ suggest that these interesting exclusive reactions can be successfully measured at \Panda.
The performance of the EMC plays a crucial role
for the measurement of these processes. A more detailed study of the
event recognition algorithms and the applied cuts, as well as
considerably increased Monte Carlo statistics are required for more precise predictions.


%

%% file: phys/nucleonstructure/spinstructure/phys_spinstructure.tex
%
\newcommand{\bwt}{
\end{multicols}
\par\noindent\rule{\dimexpr(0.5\textwidth-0.5\columnsep-0.4pt)}{0.4pt}%
\rule{0.4pt}{6pt}
}
\newcommand{\ewt}{
\vspace{\belowdisplayskip}\hfill\rule[-6pt]{0.4pt}{6.4pt}%
\rule{\dimexpr(0.5\textwidth-0.5\columnsep-1pt)}{0.4pt}
\begin{multicols}{2}
}
\newcommand{\half}{\textstyle{\frac{1}{2}}}
\newcommand{\sT}{{\scriptscriptstyle T}}
\newcommand{\ds}{\displaystyle}
\newcommand{\bea}{\begin{eqnarray}}
\newcommand{\eea}{\end{eqnarray}}
\newcommand{\nn}{\nonumber}
\def\eqrule{\hspace*{2cm} \hrulefill \hspace*{2cm}\\}
\onecolumn
\begin{multicols}{2}

\subsection{Transverse Parton Distribution Functions in Drell-Yan Production}
\COM{Author(s): M.P. Bussa, M. Maggiora}
\COM{Referee(s): D. Bettoni}
\subsubsection*{Theoretical Introduction}
\label{sec:nuclstruc:spinstruc:intro}

In a Drell-Yan (DY) process  $\bar{p}^{(\uparrow)} p^{(\uparrow)} \rightarrow \mu^+ \mu^- X$
 two (eventually polarised) hadrons $H_1$ and $H_2$ annihilate into a lepton-antilepton pair $l \bar{l}$; hadrons carry momenta $P_1$ and $P_2$ respectively ($P_{1,2}^2 = M_{1,2}^2$) and spins $S_1, S_2$ ($S	_{1,2}\cdot P_{1,2}=0$), the $l \bar{l}$ momenta being $k_1$, $k_2$ ($k_{1,2}^2 \sim 0$). Relevant kinematic variables are the initial squared energy in the centre-of-mass (CM) frame $s=(P_1+P_2)^2$, and the time-like momentum transfer $q^2 \equiv Q^2 = (k_1+k_2)^2 \geq 0$ directly related to the final state invariant mass ($Q^2 \equiv M^2_{l \bar{l}}$).

In the DIS regime, defined by the limit $Q^2, s \rightarrow \infty$ ($0 \leq \tau = Q^2 / s \leq 1$), a DY process can be described factorising an elementary annihilation process $\bar{q} q \rightarrow l \bar{l}$ with two soft correlation functions describing the annihilating antiparton (1) and parton (2) distributions in the parent hadrons:
\bea
&&\bar{\Phi} (p_1; P_1,S_1) = \nn \\
&& \int \frac{d^4 z}{(2\pi )^4}\, e^{-i p_1 \cdot z}\, 
\langle P_1 S_1 | \psi(z) \, \bar{\psi}(0) |P_1, S_1 \rangle \; ,  \nn \\ 
&&\Phi (p_2; P_2,S_2) =   \label{eq:nuclstruc:spinstruc:correlators} \\
&&\int \frac{d^4 z}{(2\pi )^4}\, e^{i p_2 \cdot z}\, \langle 
P_2, S_2 | \bar{\psi}(0) \, \psi(z) |P_2, S_2 \rangle \; .\nn
\eea
The dominant contribution in leading order is depicted in Fig.~\ref{fig:nuclstruc:spinstruc:dy_hb_lt}~\cite{bib:nuclstruc:spinstruc:Ralston:1979ys}, provided that $M$ is constrained inside a range where the elementary annihilation can be safely 
assumed to proceed through a virtual photon converting into the final $l  \bar{l}$.

At $Q^2 \rightarrow \infty$ the parton 
momenta $p_{1,2}$ are approximately aligned with the corresponding hadron and antihadron momenta $P_{1,2}$, the corresponding light-cone fractions of the parton momenta being $x_{1,2} = \frac{p_{1,2}}{P_{1,2}} \simeq \frac{Q^2}{2\, P_{1,2}\cdot q}$ ($q=p_{1,2}$, by momentum conservation~\cite{bib:nuclstruc:spinstruc:Boer:1999mm}); momenta ${\bf p}_{1,2_T}$ (often addressed in the literature as ${\bf k}_\perp$), the intrinsic transverse-momenta of the partons in the parent hadron w.r.t. the axis defined by the corresponding hadron 3-momentum ${\bf P}s_{1,2}$, are bound by the momentum conservation ${\bf q}_{_T} = {\bf p}_{1_T} + {\bf p}_{2_T}$, where ${\bf q}_{_T}$ is the transverse momentum of the final lepton pair.

\begin{figure}[H]
\begin{center}
  \includegraphics[width=0.8\columnwidth]{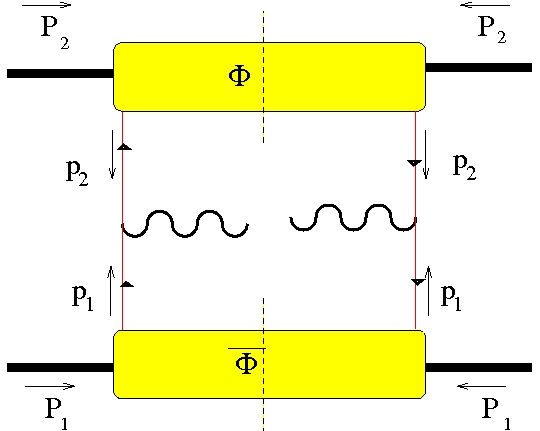}
  \caption[Leading-twist contribution to the Drell-Yan dilepton production]{The leading-twist contribution to the Drell-Yan dilepton production~\cite{bib:nuclstruc:spinstruc:Ralston:1979ys}; the correlation functions for the annihilating hadrons ${\bm \bar{\Phi}}$ and ${\bm \Phi}$ can be parametrised considering explicitly their dependence on the transverse parton momenta ${\bf p}_{1_T}$ and ${\bf p}_{2_T}$~\cite{bib:nuclstruc:spinstruc:Boer:1997nt}, thus leading to Eq. \ref{eq:nuclstruc:spinstruc:unpolxsect}.
 \label{fig:nuclstruc:spinstruc:dy_hb_lt}}
\end{center}
\end{figure}

Since in semi-inclusive processes the annihilation direction is not known and a convenient reference frame is needed; the Gottfried-Jackson (GJ) frame and the $u$-channel frame being other popular choices, the most commonly adopted frame is the so called Collins-Soper frame~\cite{bib:nuclstruc:spinstruc:Collins:1977iv} (see fig.~\ref{fig:nuclstruc:spinstruc:cs_frame}), defined by:
\beq
\hat{t} = \frac{q}{Q}  \; , \;
\hat{z} = \frac{x_1 P_1}{Q} - \frac{x_2 P_2}{Q}  \; , \;
\hat{h} = \frac{q_{_T}}{|{\bf q}_{_T}|}
\label{eq:nuclstruc:spinstruc:colsop-frame}
\eeq
where azimuthal angles lie in the plane perpendicular to $\hat{t}$ and $\hat{z}$: $\phi$ and  $\phi_{S_{1,2}}$ are respectively the angles of $\hat{\bf h}$ and  of the nucleon spin ${\bf S}_{1,2_T}$ 
with respect to the lepton plane. 

\begin{figure}[H]
\begin{center}
  \includegraphics[width=0.8\columnwidth]{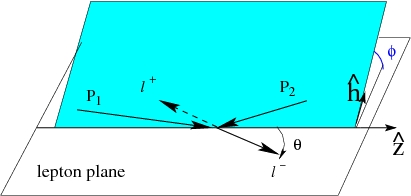}
  \caption[The Collins-Soper frame]{The Collins-Soper (CS) frame~\cite{bib:nuclstruc:spinstruc:Collins:1977iv}.
 \label{fig:nuclstruc:spinstruc:cs_frame}}
\end{center}
\end{figure}

In the so-called collinear kinematics approach (${\bf p}_{1,2_T} \sim 0$, i.e. neglecting dependencies on the ${\bf k}_\perp$), at leading twist (i.e. twist-two) the quark structure of an hadron can be described by three Parton Distribution Functions (PDF): the number density $f_1(x)$ (a.k.a. unpolarised distribution),  the longitudinal polarisation distribution $g_1(x)$ (a.k.a. helicity distribution) and the transverse polarisation $h_1(x)$ (a.k.a. Transversity). 
The density function $f_1(x)$ is the probability of finding a quark with fraction $x$ of the parent hadron longitudinal momentum, regardless of its spins orientation; $g_1(x)$ describes the helicity of a quark in a longitudinally polarised quark, i.e. the asymmetry between the number densities of the quarks with a given momentum fraction $x$ and with spins parallel and antiparallel to the parent hadron spin; the Transversity $h_1(x)$ describes the asymmetry between the number densities of the quarks with a given momentum fraction $x$ and with spins parallel and antiparallel to that of a transversely polarised hadron. The latter PDF has no probabilistic interpretation in the helicity basis, and being chirally odd it could not be observed in the historical DIS experiments. This is the reason why while the former two PDF's have been studied at great length, the Transversity $h_1(x)$, historically introduced right for DY processes~\cite{bib:nuclstruc:spinstruc:Ralston:1979ys}, has been neglected for a very long time and rediscovered only in the beginning of the 90's ~\cite{bib:nuclstruc:spinstruc:Artru:1989zv}, yet being a leading-twist distribution~\cite{bib:nuclstruc:spinstruc:Ralston:1979ys,bib:nuclstruc:spinstruc:Artru:1989zv,bib:nuclstruc:spinstruc:Cortes:1991ja}. 
The current PDF nomenclature has been introduced by Jaffe and Ji ~\cite{bib:nuclstruc:spinstruc:Jaffe:1991kp} and reflects the leading-twist nature of the three considered PDF's (the $1$ subscript). The three PDF's are linked by the well-known Soffer inequality ($2|h_1|\le (f_1+g_1)$) ~\cite{bib:nuclstruc:spinstruc:Soffer:1994ww}.

In collinear kinematics a factorised pQCD approach cannot interpret even the experimental unpolarised cross sections for inclusive particle production in high-energy hadron-hadron production~\cite{bib:nuclstruc:spinstruc:Bourrely:2003bw}; only recently NLO calculations including threshold resummation effects have been developed~\cite{bib:nuclstruc:spinstruc:deFlorian:2005yj}, in reasonable agreement with inclusive cross sections integrated over the hadron rapidity range, but the rapidity-dependent case is still under investigation. The collinear kinematics approach is also not suitable to describe the DY production at low lepton-pair transverse momentum: no transverse momentum can be generated in the collinear LO approximation. 

The role of parton's intrinsic transverse momentum has thus to be explicitly accounted for; since the early interest~\cite{bib:nuclstruc:spinstruc:Field:1976ve,bib:nuclstruc:spinstruc:Feynman:1978dt} huge efforts have been dedicated to provide a full set of Transverse-Momentum Dependent (TMD) PDF's and Fragmentation Functions (FF). The generalisation of the pQCD factorisation theorem to the TMD scenario has been formally proved for the DY processes~\cite{bib:nuclstruc:spinstruc:Collins:1977iv}; at leading twist eight independent TMD PDF distributions are needed to describe the nucleonic structure and the nucleonic correlator can be parametrised as~\cite{bib:nuclstruc:spinstruc:Boer:1997nt}:
\bwt
\bea
\Phi(x,{\bf p}_\sT, S) &=& 
\frac{1}{2}\,\Biggl\{
f_1\,\rlap{/} n_+
+ f_{1T}^\perp\, \frac{\epsilon_{\mu \nu \rho \sigma}
\gamma^\mu n_+^\nu p_\sT^\rho S_\sT^\sigma}{M}
+ \left( S_L g_{1L} +
\frac{{\bf p}_{\perp }\cdot{\bf S}_T}{M} g_{1T} \right)
\gamma^5\rlap{/}n_+
\nonumber \\ & & \qquad
+h_{1T}\,i\sigma_{\mu\nu}\gamma^5 n_+^\mu S_\sT^\nu 
\left( S_L h_{1L}^\perp + \frac{{\bf p}_{\perp }
\cdot{\bf S}_T}{M}h_{1T}^\perp\right)
\frac{i\sigma_{\mu\nu}\gamma^5 
n_+^\mu p_\sT^\nu}{M} 
+ h_1^\perp\,\frac{\sigma_{\mu \nu} p_\sT^\mu n_+^\nu}{M}
\Biggr\},
\label{eq:nuclstruc:spinstruc:correlator}
\eea
\ewt
where $n_{\pm}$ are auxiliary light-like vectors, $M$ is the nucleon mass, and all the TMD PDF's depend on $x$ and $|{\bf p}_{\perp }|$ (e.g. $f_1$ = $f_1(x,{\bf p}_\sT^2)$). This form of the TMD approach is also addressed in the literature as  Generalised Parton Model (GPM). Several notations are used for the TMD's nomenclature ~\cite{bib:nuclstruc:spinstruc:Bacchetta:2004jz,bib:nuclstruc:spinstruc:Barone:2001sp}; in the one ~\cite{bib:nuclstruc:spinstruc:Mulders:1995dh} adopted in Eq.~\ref{eq:nuclstruc:spinstruc:correlator}, $f$, $g$ and $h$ refer respectively to unpolarised, longitudinally polarised and transversely
polarised quarks, to longitudinal and transverse hadron polarisation (subscripts $L$ and $T$), at leading twist (subscript 1), explicitly dependent on intrinsic momenta (apex $\perp$). The three TMD PDF's $f_1(x,{\bf p}_\sT^2)$, $g_{1L}(x,{\bf p}_\sT^2)$ and $h_{1T}(x,{\bf p}_\sT^2)$, upon integration on 
${\bf p}_\sT^2$ yield respectively $f_1(x)$, $g_1(x)$ and $h_1(x)$; the other TMD PDF's cancel out upon integration on ${\bf p}_\sT^2$.

The relevant TMD PDF's for the physics program proposed at \panda~ are: $h_1^{\perp}$ (a.k.a. Boer-Mulders, BM, function), the distribution of transversely polarised partons in unpolarised hadrons; $f_{1T}^\perp$ (a.k.a. Sivers function) and $h_{1T}$ (a.k.a. Transversity), the distributions of respectively unpolarised and transversely polarised partons in a transversely polarised nucleon. These PDF's will be described in details in the next sub-section, while reviewing the experimental data available from the literature and their interpretation.

It was stated before that the DY production at low lepton-pair transverse momentum can not be interpreted in the framework of collinear kinematics; the energy evolution of the spin and TMD distributions had been discussed in~\cite{bib:nuclstruc:spinstruc:Idilbi:2004vb} resumming the large logarithms arising in
perturbative calculations for SIDIS and DY process at low transverse momentum. To extend the validity of the TMD factorisation approach from the very small $q_T$ range to the moderate transverse momentum region (yet with $q_T\ll Q$) soft gluon radiation has to be accounted for by the mean of Sudakov factors included by proper resummation; the outcome is a suppression of the TMD azimuthal asymmetries that becomes more important with rising energy~\cite{bib:nuclstruc:spinstruc:Boer:2001he}.

It should be stressed that theoretical approaches other than the TMD formalism do exist; in particular the one considering twist-three effects in collinear pQCD, whose origin dates back to ~\cite{bib:nuclstruc:spinstruc:Kane:1978nd}, allows to evaluate the azimuthal asymmetries in the framework of pQCD generalised factorisation theorems with the introduction of new twist-three quark-gluon
correlator functions convoluted with ordinary twist-two parton distribution functions and a
short-distance hard scattering part; such functions do not have a simple partonic interpretation, being expectation values between hadronic states of three field operators. See Ref.~\cite{bib:nuclstruc:spinstruc:D'Alesio:2007jt} for a more detailed discussion of such functions and of their relation with SSA.

A comparison of the twist-three and of the TMD
approaches~\cite{bib:nuclstruc:spinstruc:Ji:2006ub,bib:nuclstruc:spinstruc:Ji:2006vf,bib:nuclstruc:spinstruc:Ji:2006br,bib:nuclstruc:spinstruc:Koike:2007dg} to SIDIS and DY processes shows how both mechanisms, yet having different validity domains, describe the same physics in the overlapping kinematic region. In the 
twist-three approach domain (large lepton pair transverse momentum and photon virtuality: ${\bf q}_T , Q  \gg \Lambda_{\rm QCD}$) twist-three quark-gluon correlations can lead to the spin-dependent cross section; at $q_T\simeq \Lambda_{\rm QCD}\ll Q$ single spin asymmetries (SSA) can be generated from spin-dependent TMD quark distributions; in the overlapping domain $ \Lambda_{\rm QCD}\ll q_T \ll
Q$, $q_T$ is large enough for the asymmetry to be a twist-three
effect but at the same time a $q_T \ll Q$ allows for the TMD factorisation formalism.
The connections between such different formalisms lead to strong constraints on those phenomenological studies aiming at the dynamics underlying transverse SSA.

\subsubsection*{Experimental Data and Theoretical Interpretations}
\label{sec:nuclstruc:spinstruc:expdata}

We will focus herewith on the unpolarised and single-polarised DY processes, that can be accessed in the \panda~ scenario respectively since the very beginning and when a polarised target will eventually become available; nevertheless double-polarised DY processes will be shortly addressed as well, being the most promising scenario to access transversity effects in the nucleon dynamics, even if in such a case an antiproton polarisation is needed, and such a polarisation can eventually be provided only in later \INST{FAIR} stages.

\vspace{2mm}
\paragraph*{The Unpolarised Case: $\bar{p} p \rightarrow \l^+ \l^- X$ \\}
\label{sec:nuclstruc:spinstruc:expdata:unpol}

The TMD leading-twist parametrisation of $\bar{\Phi}$ and $\Phi$~\cite{bib:nuclstruc:spinstruc:Boer:1997nt} leads to the following fully differential cross section for the unpolarised DY process~\cite{bib:nuclstruc:spinstruc:Boer:1999mm}:
\bwt
\bea
\frac{d\sigma^o}{d\Omega dx_1 dx_2 d{\bf q}_{_T}} &= &\frac{\alpha^2}{12Q^2}\,
\sum_f\,e_f^2\,\Bigg\{ \left( 1 + \cos^2 \theta \right) \, {\cal F}\left[ \bar{f}_1^f\, f_1^f \right] 
\nn \\
& & + f\, \sin^2 \theta \, \cos 2\phi \, {\cal F}\left[ \left( 2 \hat{\bf h}\cdot {\bf
p}_{1_T} \, \hat{\bf h} \cdot {\bf p}_{2_T} - {\bf p}_{1_T} \cdot {\bf p}_{2_T}
\right) \, \frac{\bar{h}_1^{\perp\,f}\,h_1^{\perp\,f}}{M_1\,M_2}\,\right]\, \Bigg\} 
\; ,
\label{eq:nuclstruc:spinstruc:unpolxsect}
\eea
\ewt
$\alpha$ being the fine structure constant, $e_f$ the charge of a parton with
flavour $f$. The TMD PDF's are convoluted with their antiparton partners according to:
\bwt
\beq
{\cal F} \left[ \bar{f}_1^f \, f_1^f \right] \equiv \int d{\bf p}_{1_T} d{\bf
p}_{2_T} \delta \left( {\bf p}_{1_T} + {\bf p}_{2_T} - {\bf q}_{_T} \right) 
 \left[ \bar{f}_1^f (x_1,{\bf p}_{1_T})\, f_1^f(x_2,{\bf p}_{2_T}) + (1
\leftrightarrow 2) \right] \; .
\label{eq:nuclstruc:spinstruc:convol}
\eeq
\ewt
Cross section (\ref{eq:nuclstruc:spinstruc:unpolxsect}), when considered differential in $\sqrt{\tau}$ and $x_{_F}$ only ($\tau = x_1 x_2$ and the Feynman parameter $x_{_F} = x_1-x_2$) scales as $d^2 \sigma / d \sqrt{\tau} d x_f \sim 1/s$, influencing the choice of the kinematical set-up, 
while after integrating on all the kinematic variables but the angular distribution, at leading order in $\alpha_s$ becomes:
\bea
\frac{1}{\sigma^o}\,\frac{d\sigma^o}{d\Omega} &=& \frac{3}{4\pi}\,\frac{1}{\lambda +3}\, 
\left(  1+\lambda\,\cos^2\theta ~+  \right.  \label{eq:nuclstruc:spinstruc:param}
 \\ 
&&\left.  + \mu\,\sin^2\theta \cos \phi + \frac{\nu}{2}\,\sin^2\theta\cos 2\phi \right) \; \nn
\eea 
In a naive parton model approach the assumption of massless quarks leads to a transversely polarised virtual photon, so that $\lambda = 1$, $\mu = \nu = 0$ and $d\sigma/d\cos\theta\sim 1+\cos^2\theta$; such predictions are confirmed by both LO and NLO perturbative QCD calculations~\cite{bib:nuclstruc:spinstruc:Brandenburg:1993cj}. The so called Lam-Tung rule~\cite{bib:nuclstruc:spinstruc:Lam:1978pu,bib:nuclstruc:spinstruc:Lam:1978zr,bib:nuclstruc:spinstruc:Lam:1980uc} $\lambda=1-2\nu$, analogous to the Callan-Gross relation in DIS, should hold in any reference frame and be unaffected by first-order QCD corrections~\cite{bib:nuclstruc:spinstruc:Lam:1980uc}, even if it could be influenced by parton intrinsic motions and other "soft" effects.

Experimental data from \INST{NA3}~\cite{bib:nuclstruc:spinstruc:Badier:1981ti} and \INST{NA10}~\cite{bib:nuclstruc:spinstruc:Falciano:1986wk,bib:nuclstruc:spinstruc:Guanziroli:1987rp} Collaborations at \INST{CERN} and from the E615~\cite{bib:nuclstruc:spinstruc:Conway:1989fs} Collaboration at \INST{Fermilab} for muon pairs production with $\pi^-$ beams at different momenta ($140\div286~\gevc$) on $^2H$ and $W$ targets have shown no evidence of a CM energy dependence or a nuclear dependence of the angular distribution parameters $\lambda$, $\mu$ and $\nu$.
Data for the former two parameters, leading to $\lambda \sim 1$ and $\mu \sim 0$ as predicted, are mostly independent of the considered kinematic region, except a slight reduction of $\lambda$ at high $x_1$ (the light-cone fraction of the parton in the pion) that has been interpreted with higher-twist effects~\cite{bib:nuclstruc:spinstruc:Berger:1979du,bib:nuclstruc:spinstruc:Berger:1979xz}. But these experimental data show a relevant $\cos2\phi$ dependence with a deviation from zero for $\nu$ at high $q_T$, depending on $\sqrt{s}$, that tops at larger  $q_T$ values $\nu \sim 0.30$ in \INST{CERN} data ~\cite{bib:nuclstruc:spinstruc:Guanziroli:1987rp} and $\nu \sim 0.73$ in \INST{Fermilab} data~\cite{bib:nuclstruc:spinstruc:Conway:1989fs}, clearly departing from pQCD expectations. Other mechanisms, like higher twists or factorisation breaking terms at NLO, are not able to explain such a relevant violation of the Lam-Tung relation~\cite{bib:nuclstruc:spinstruc:Brandenburg:1994wf,bib:nuclstruc:spinstruc:Eskola:1994py,bib:nuclstruc:spinstruc:Berger:1979du}. 

This is not the case if we consider the TMD approach, where the convolution of the two BM functions $h_1^{\perp\,f}$ and $\bar{h}_1^{\perp\,f}$ in the last term of Eq. \ref{eq:nuclstruc:spinstruc:unpolxsect} allows for a leading-twist (hence large) $\cos 2\phi$ azimuthal dependence~\cite{bib:nuclstruc:spinstruc:Boer:1999mm}. Since the BM function describes the transverse polarisation of partons in unpolarised hadrons, it is intimately connected to the orbital motion of the parton inside the hadron; the product $\bar{h}_1^\perp \, h_1^\perp$ brings a change
of two units in the orbital angular momentum, leading to an angular dependence on $2\phi$. The BM function is a chirally-odd PDF, it cannot be extracted from DIS data~\cite{bib:nuclstruc:spinstruc:Jaffe:1993xb}, and can be accessed only through its convolution with another chirally-odd quantity: itself (Eq.~\ref{eq:nuclstruc:spinstruc:unpolxsect}) in the unpolarised DY, or the Transversity $h_{1T}$ in the single-polarised DY (Eq.~\ref{eq:nuclstruc:spinstruc:1polxsect} of the next sub-section) ~\cite{bib:nuclstruc:spinstruc:Boer:1999mm}.

Other mechanisms, like the role of QCD-vacuum structure in hadron-hadron scattering~\cite{bib:nuclstruc:spinstruc:Nachtmann:1983uz,bib:nuclstruc:spinstruc:Brandenburg:1993cj,bib:nuclstruc:spinstruc:Boer:2004mv}, has been considered as well, and more recently the $\cos2\phi$ azimuthal dependence of the unpolarised DY process $p\bar{p}\to\mu^+\mu^- X$ has been studied in the quark-diquark spectator approach~\cite{bib:nuclstruc:spinstruc:Gamberg:2005ip}.

Very recent data from E866/NuSea~\cite{bib:nuclstruc:spinstruc:Zhu:2006gx} Collaboration at \INST{FNAL} for muon pairs production with a $800~\gevc$ proton beam on $^2H$ haven't shown any significant $\cos2\phi$ azimuthal  dependence, constraining thus those theoretical models predicting larger azimuthal dependencies originating from QCD vacuum effects, and pointing toward an almost vanishing sea-quark BM function, much smaller than that related to valence quarks~\cite{bib:nuclstruc:spinstruc:Boer:1997nt}.

Moreover contributions to a  $\cos2\phi$ azimuthal asymmetry for DY dilepton production with (anti)nucleon on nuclear targets could arise from the nuclear distortion of the hadronic projectile wave function, typically a spin-orbit effect occurring on the nuclear surface~\cite{bib:nuclstruc:spinstruc:Bianconi:2005px}; this effect, expected to be on the percent level, should be added to the one originating from the elementary hard event.

\vspace{2mm}
\paragraph*{The Single-Polarised Case: $\bar{p} p^{\uparrow } \rightarrow \mu^+ \mu^- X$ \\} \label{sec:nuclstruc:spinstruc:expdata:singlespin}

When one of the annihilating hadrons is transversely polarised, in a TMD approach at leading-twist a further polarised term shows up in the cross section:
\beq
\frac{d\sigma}{d\Omega dx_1 dx_2 d{\bm q}_\sT} =
\frac{d\sigma^o}{d\Omega dx_1 dx_2 d{\bm q}_\sT} + 
\frac{d\Delta \sigma^\uparrow}{d\Omega dx_1 dx_2 d{\bm q}_\sT} 
\label{eq:nuclstruc:spinstruc:1polxsect_sum}
\eeq
where $d\sigma^o/d\Omega dx_1 dx_2 d{\bm q}_\sT$ is the unpolarised cross section defined in Eq. \ref{eq:nuclstruc:spinstruc:unpolxsect}. The polarised $d\Delta \sigma^\uparrow/d\Omega dx_1 dx_2 d{\bm q}_\sT$~\cite{bib:nuclstruc:spinstruc:Boer:1999mm}:
\end{multicols}
\bea
& &\frac{d\Delta \sigma^\uparrow}{d\Omega dx_1 dx_2 d{\bf q}_{_T}} = 
\frac{\alpha^2}{12sQ^2}\,\sum_f\,e_f^2\,|{\bf S}_{2_T}|\,\Bigg\{ \left( 1 + \cos^2 \theta \right) \, 
\sin (\phi - \phi_{S_2})\, {\cal F}\left[ \hat{\bf h}\cdot {\bf p}_{2_T} \,
\frac{\bar{f}_1^f\, f_{1T}^{\perp\,f}}{M_2}\right] \nn \\
& &\quad - \sin^2 \theta \, \sin  (\phi + \phi_{S_2})\, {\cal F}\left[  \hat{\bf h}\cdot 
{\bf p}_{1_T} \,\frac{\bar{h}_1^{\perp\,f}\, h_{1T}^f}{M_1}\right]   \label{eq:nuclstruc:spinstruc:1polxsect} \\
& &\quad  - \sin^2 \theta \, \sin (3\phi - \phi_{S_2})\, {\cal F}\left[ \left( 
4 \hat{\bf h}\cdot {\bf p}_{1_T} \, (\hat{\bf h} \cdot {\bf p}_{2_T})^2 - 
2 \hat{\bf h} \cdot {\bf p}_{2_T} \, {\bf p}_{1_T} \cdot {\bf p}_{2_T} - 
\hat{\bf h}\cdot {\bf p}_{1_T} \, {\bf p}_{2_T}^2 \right) \, 
\frac{\bar{h}_1^{\perp\,f}\, h_{1T}^{\perp\,f}}{2 M_1\,M_2^2}\,\right]\, \Bigg\} 
\; \nn
\eea
\ewt
depends explicitly on the Sivers function $f_{1T}^{\perp\,f}$, on the Transversity $h_{1T}^{f}$ and on the BM function $h_{1T}^{\perp\,f}$; sum is extended on the parton flavour $f$.

Eq. \ref{eq:nuclstruc:spinstruc:1polxsect} shows how a powerful tool can be the Drell-Yan production of muon pairs, since selecting the proper angular dependence both the BM and the Sivers functions can be accessed. A spin asymmetry weighted by $\sin (\phi - \phi_{S_2})$ leads to the convolution of $f_{1T}^{\perp\,f}$ with the known distribution $f_1^f$ in a mechanism similar to the Sivers effect in DIS with lepton beams~\cite{bib:nuclstruc:spinstruc:Sivers:1990fh}. The asymmetry defined weighting for $\sin (\phi + \phi_{S_2})$ leads to the convolution of $h_{1T}^f$ with $h_1^{\perp\,f}$ in a mechanism similar to the Collins effect~\cite{bib:nuclstruc:spinstruc:Collins:1992kk}; since $h_1^{\perp\,f}$ contribute at leading twist to the unpolarised cross section as well (Eq. \ref{eq:nuclstruc:spinstruc:unpolxsect}), a combined analysis of the $\cos 2\phi$ and $\sin (\phi+\phi_{S_2})$ moments of azimuthal asymmetries respectively in the unpolarised and in the single-polarised Drell-Yan cross sections should allow for the determination of both PDF's at the same time in a single experimental scenario. 

The Transversity distribution $h_{1T}$ is not diagonal in the parton helicity basis, since involves a helicity flipping mechanism at parton level. It is hence chirally-odd, since at leading twist chirality and helicity are identical, and this is reason why it cannot be accessed in
DIS as well: QED and QCD for massless quarks conserve helicity, and thus $h_{1T}$ pertains the "soft" domain where the chiral symmetry of QCD is 
(spontaneously) broken. To be measured in a (chirally-even) cross-section or asymmetry it needs
another chiral-odd partner, in contrast with the unpolarised and the helicity distribution functions, both chiral-even (see for a review 
also Refs.~\cite{bib:nuclstruc:spinstruc:Jaffe:1996zw,bib:nuclstruc:spinstruc:Barone:2001sp}). 
In the transverse spin basis $h_{1T}$ is diagonal and can be interpreted as the difference between the probabilities to find a quark polarised along the transverse
proton polarisation and against it. In a nucleon (and more generally in any spin $\half$ hadron) it has no gluonic counterpart, due to the mismatch in the change of helicity units, and its evolution is hence decoupled from radiative gluons; it also decouples from charge-even $q \bar{q}$ configurations of the Dirac sea, because it is odd also under charge conjugation transformations. The prediction of its weaker evolution ~\cite{bib:nuclstruc:spinstruc:Jaffe:1996zw} could represent a basic test of QCD in the non-perturbative domain. 

The Sivers function $f_{1T}^\perp$ (originally suggested in ~\cite{bib:nuclstruc:spinstruc:Sivers:1989cc,bib:nuclstruc:spinstruc:Sivers:1990fh}) is a T-odd chirally-even  ${\bf p}_\perp$-odd distribution describing how the distribution of unpolarised quarks is affected by the transverse polarisation of the parent proton. 
It appears at leading twist in semi-inclusive DIS (SIDIS) processes like $lp^\uparrow \rightarrow l' \pi X$~\cite{bib:nuclstruc:spinstruc:Efremov:2003tf} or $pp^\uparrow \rightarrow \pi X$~\cite{bib:nuclstruc:spinstruc:Anselmino:2004ky}, where it is responsible of the so-called Sivers effect, the azimuthal asymmetric distribution of the detected pions depending on the 
direction of the target polarisation, since it is proportional ~\cite{bib:nuclstruc:spinstruc:Boer:1997nt,bib:nuclstruc:spinstruc:Bacchetta:2004jz}  to the function $\Delta^N f$ of Refs. ~\cite{bib:nuclstruc:spinstruc:Anselmino:1994tv,  bib:nuclstruc:spinstruc:Anselmino:1997jj} (also addressed in the literature as "Sivers function"). 
Its relevance for the \panda~ physics program is related to the azimuthal asymmetries in single-polarised DY $\bar{p} p^\uparrow \rightarrow l^+ l^- X$ (Eq.~\ref{eq:nuclstruc:spinstruc:1polxsect}): a measurement of a non-vanishing asymmetry would be a direct evidence of the orbital angular momentum of quarks~\cite{bib:nuclstruc:spinstruc:Elschenbroich:2004ba}. 

The Sivers function was originally expected to vanish ~\cite{bib:nuclstruc:spinstruc:Collins:1992kk} in DY, due to parity and time reversal invariance in the light-cone gauge, but the role of the presence of Wilson lines had been reconsidered ~\cite{bib:nuclstruc:spinstruc:Collins:2002kn} since under time-reversal the future-pointing Wilson lines are replaced by past-pointing Wilson lines. The corresponding transverse gauge link is responsible of the gauge invariance of TMD parton distributions ~\cite{bib:nuclstruc:spinstruc:Belitsky:2002sm} and at the same time of the attractive final state interactions (FSI) in SIDIS and of the repulsive initial state interactions (ISI) in DY ~\cite{bib:nuclstruc:spinstruc:Collins:1992kk,bib:nuclstruc:spinstruc:Brodsky:2002cx}. Since in the latter case the past-pointing Wilson lines allow as well an appropriate factorisation of
the Drell-Yan process ~\cite{bib:nuclstruc:spinstruc:Collins:2002kn, bib:nuclstruc:spinstruc:Brodsky:2002rv},
the correct result is not a vanishing Sivers function in DY, but rather a Sivers function showing opposite signs in SIDIS and in DY:
\begin{equation}
  \label{eq:nuclstruc:spinstruc:sidisdy}
f_{1T}^{\perp\,f}|_{\rm SIDIS} = f_{1T}^{\perp\,f}|_{\rm DY} \
\end{equation}
preserving hence the universality of the TMD spin-dependent PDF's. 
Huge theoretical efforts have been aimed to investigate the role of the gauge links in TMD distributions ~\cite{bib:nuclstruc:spinstruc:Ji:2002aa, bib:nuclstruc:spinstruc:Belitsky:2002sm, bib:nuclstruc:spinstruc:Ji:2004wu, bib:nuclstruc:spinstruc:Ji:2004xq,
bib:nuclstruc:spinstruc:Idilbi:2004vb, bib:nuclstruc:spinstruc:Idilbi:2005er}, with a focus on the gauge invariance of the TMD PDF's and on the proper QCD factorisation at leading twist for SIDIS and DY processes; the evaluation of the Sivers function in the DY di-lepton production could allow for a strong test on the universality of the TMD PDF's.

As pointed out by theoretical predictions based on the Sivers effect from SIDIS experimental data, large SSA are expected for DY processes at the large energy scale foreseen in the later phases of the \INST{FAIR} project ~\cite{bib:nuclstruc:spinstruc:Anselmino:2005ea, bib:nuclstruc:spinstruc:Efremov:2004tp, bib:nuclstruc:spinstruc:Vogelsang:2005cs,bib:nuclstruc:spinstruc:Anselmino:2005ea,bib:nuclstruc:spinstruc:Collins:2005rq}, while the Sivers asymmetry at RHIC could be measured only at high rapidity and should be strongly sensitive to the sea-quark Sivers distributions ~\cite{bib:nuclstruc:spinstruc:Collins:2005rq}.

Two TMD mechanisms has been discussed until now, that could lead to transverse SSA's in the DY process $\bar{p} p^{\uparrow } \rightarrow \mu^+ \mu^- X$: the Sivers effect and the Boer-Mulders effect (which involves also the Transversity distribution); in such exclusive process, and this is the main advantage with respect to inclusive processes like $pp\to h + X$, 
the measurement of the lepton-pair angular distribution automatically allows to select one specific effect. A systematic calculation of all leading-twist PDF's in the nucleon has been performed in the framework of a diquark
spectator model~\cite{bib:nuclstruc:spinstruc:Bacchetta:2008af}. But other mechanisms that could also generate SSA's in the DY process had been proposed in the literature~\cite{bib:nuclstruc:spinstruc:Hammon:1996pw,bib:nuclstruc:spinstruc:Boer:1997bw,bib:nuclstruc:spinstruc:Boer:1999si, bib:nuclstruc:spinstruc:Boer:2001tx}, based on higher twist
quark-gluon correlation functions in a generalised pQCD factorisation theorem approach~\cite{bib:nuclstruc:spinstruc:Qiu:1998ia}: in such cases the asymmetries depend on the angle between
the proton polarisation direction and the final lepton pair plane~\cite{bib:nuclstruc:spinstruc:Collins:1977iv}, and vanish upon corresponding angular integrations.

The Sivers function $f_{1T}^\perp$ has recently attracted the deepest interests in the spin physics community. Besides being a T-odd TMD PDF, it describes how the distribution of unpolarised quarks is distorted by the 
transverse polarisation of the parent hadron, and as such it contains
information's on the orbital motion of hidden confined partons and their spatial distribution~\cite{bib:nuclstruc:spinstruc:Burkardt:2003je}. Besides, it offers a natural link between microscopic properties of confined elementary 
constituents and hadronic measurable quantities, such as the nucleon anomalous 
magnetic moment~\cite{bib:nuclstruc:spinstruc:Burkardt:2005km}. And the prediction of Eq.~\ref{eq:nuclstruc:spinstruc:sidisdy} is a strong test of the universality of TMD PDF's.

Recently, very precise data for SSA involving $f_{1T}^\perp$ (the Sivers effect)
have been obtained for the SIDIS process on transversely polarised 
protons~\cite{bib:nuclstruc:spinstruc:Airapetian:2004tw,bib:nuclstruc:spinstruc:Diefenthaler:2005gx,bib:nuclstruc:spinstruc:Diefenthaler:2007rj}. Three different parametrisation of $f_{1T}^\perp$~\cite{bib:nuclstruc:spinstruc:Anselmino:2005ea,bib:nuclstruc:spinstruc:Vogelsang:2005cs,bib:nuclstruc:spinstruc:Collins:2005wb} have been extracted from the HERMES data (see for a comparison among the various approaches Ref.~\cite{bib:nuclstruc:spinstruc:Anselmino:2005an}), showing a relevant non zero effect. On the contrary COMPASS data for non-identified~\cite{bib:nuclstruc:spinstruc:Alexakhin:2005iw,bib:nuclstruc:spinstruc:Ageev:2006da} and identified hadrons~\cite{bib:nuclstruc:spinstruc::2008dn} show small effects, compatible with zero within the statistical errors, interpreted in term of a cancellation between the u- and d-quark contributions~\cite{bib:nuclstruc:spinstruc::2008dn,bib:nuclstruc:spinstruc:Courtoy:2008mj,bib:nuclstruc:spinstruc:Efremov:2007kj,bib:nuclstruc:spinstruc:Efremov:2006qm}.
Although SIDIS data by HERMES and COMPASS do not constrain the large transverse momentum region, at lower transverse momentum they show a serious conflict as long as Sivers effects are concerned, in particular considering the latest data from COMPASS~\cite{bib:nuclstruc:spinstruc:Levorato:2008tv} that show Sivers asymmetries compatible with zero both for positive and negative hadrons. 

New efforts are in progress either to explain the non zero Sivers effects in $p-p$ collisions by the mean of scalar and spin-orbit re-scattering terms~\cite{bib:nuclstruc:spinstruc:Bianconi:2008jj}, either to account for the new data from COMPASS updating the present parametrisations~\cite{bib:nuclstruc:spinstruc:Anselmino:2008uy}. An independent evaluation of the Sivers distribution $f_{1T}^\perp$ in the single-polarised DY processes at \INST{FAIR} would certainly contribute to the present (and probably long lasting) challenge to the spin physics community.

\vspace{2mm}
\paragraph*{The Dream Option: $\bar{p}^{\uparrow} p^{\uparrow} \rightarrow \mu^+ \mu^- X$ \\} \label{sec:nuclstruc:spinstruc:expdata:doublespin}

Although the Transversity $h_{1T}$ can be accessed in the single-polarised DY, convoluted with $h_1^\perp$ (Eq.~\ref{eq:nuclstruc:spinstruc:1polxsect}), the simplest scenario to extract the  Transversity function $h_{1T}$ is the fully polarised DY, as originally proposed in Ref.~\cite{bib:nuclstruc:spinstruc:Ralston:1979ys}: it allows for a direct access to $h_{1T}$, without its convolution with any other PDF, but it requires a reasonable antiproton polarisation; such a dream option could be accessed only in the very last stage of the \INST{FAIR} project, as proposed in ~\cite{bib:nuclstruc:spinstruc:Abazov:2004ih} and in ~\cite{bib:nuclstruc:spinstruc:Pax:2004}. 

In a TMD approach at leading twist the fully polarised cross section, after integrating upon $d{\bf q}_{_T}$, 
becomes~\cite{bib:nuclstruc:spinstruc:Tangerman:1994eh}:
\end{multicols}
\bea
\frac{d\sigma^{\uparrow \uparrow}}{dx_1\, dx_2\, d\Omega} &= &
\frac{\alpha^2}{12 q^2}\, \Bigg[ (1+\cos^2 \theta) \, \sum_f\,e_f^2\, \bar{f}^f_1
(x_1)\,f_1^f(x_2) 	+ \sin^2 \theta \, \cos 2\phi \, \frac{\tilde{\nu} (x_1, x_2)}{2} \nn \\
& &\; + |{\bf S}_{_{T1}}| \, |{\bf S}_{_{T2}}| \, \sin^2\theta \, 
\cos (2\phi - \phi_{_{S_1}} - \phi_{_{S_2}})\, \sum_f \, e_f^2\, \bar{h}^f_{1T}(x_1)
\, h_{1T}^f(x_2) + (1 \leftrightarrow 2) \Bigg] \; ,
\label{eq:nuclstruc:spinstruc:cross2pol}
\eea
\ewt
where $\phi_{_{S_i}}$ is
the azimuthal angle of the transverse spin of hadron $i$ as it is measured with 
respect to the lepton plane in a plane perpendicular to $\hat{z}$ and $\hat{t}$ of the CS frame
(Fig.~\ref{fig:nuclstruc:spinstruc:cs_frame}); the contribution from $h_{1T}^{\perp\,f}$ is hidden in the function $\tilde{\nu}$, while to access the Transversity distribution $h_{1T}^{f}$ a double spin asymmetry can be defined:
\bwt
\beq
A_{_{TT}} = \frac{d\sigma^{\uparrow \uparrow} - d\sigma^{\uparrow \downarrow}}
{d\sigma^{\uparrow \uparrow} + d\sigma^{\uparrow \downarrow}} 
= |{\bf S}_{_{T1}}| \, |{\bf S}_{_{T2}}| \, 
\frac{\sin^2\theta}{1+\cos^2\theta}\, \cos (2\phi - \phi_{_{S_1}} - \phi_{_{S_2}})
\, \frac{\sum_f \, e_f^2\, \bar{h}^f_{1T}(x_1) \, h_{1T}^f(x_2) + (1\leftrightarrow 2)}
{\sum_f\,e_f^2\, \bar{f}^f_1(x_1)\,f_1^f(x_2) + (1\leftrightarrow 2)} 
\; 
\label{eq:nuclstruc:spinstruc:att}
\eeq
\ewt
The asymmetry depends at leading order on $h_{1T}^{f}$ squared, without contribution from sea-quark PDF's nor convolution with fragmentation functions as in SIDIS, providing then the best possible scenario to access Transversity. 

The asymmetry in Eq. \ref{eq:nuclstruc:spinstruc:att} could in principle be accessed at RHIC as well, in the polarised DY process $p^\uparrow p^\uparrow \rightarrow l^+ l^- X$, the first process suggested to access Transversity at leading order ~\cite{bib:nuclstruc:spinstruc:Ralston:1979ys}. But in such a case it would depend on sea-quark PDF's, since it would involve the Transversity of an antiquark in a transversely polarised proton. Moreover, NLO simulations in the RHIC CM energy range ~\cite{bib:nuclstruc:spinstruc:Martin:1997rz,bib:nuclstruc:spinstruc:Barone:1997mj} have shown an  $A_{TT}$ strongly suppressed by QCD evolution and by a Soffer bound on the percent level.
On the contrary making use of the \INST{FAIR}'s antiproton beams the asymmetry would involve Transversities of valence partons only ~\cite{bib:nuclstruc:spinstruc:Efremov:2004qs,bib:nuclstruc:spinstruc:Anselmino:2004ki}.

\vspace{2mm}
\paragraph*{Total Cross Section \\}
\label{sec:nuclstruc:spinstruc:expdata:xs}	

The full expression of the leading-twist differential cross section for the Drell-Yan
 $H_1^{(\uparrow)} H_2^{(\uparrow)} \to l^+ l^- X$ process can be found in the Appendix of Ref.~\cite{bib:nuclstruc:spinstruc:Boer:1999mm}

The clear systematics in the literature showing a production of Drell-Yan pairs distributed with $\langle |{\bf q}_T|\rangle > 1 \gevc$ and depending on $\sqrt{s}$, suggests that
sizeable QCD corrections are needed beyond a simple Quark Parton Model (QPM) approach, since confinement alone induces  much smaller quark intrinsic transverse momenta. The involved higher order Feynman diagrams typically show $q \bar{q}$ annihilations into gluons or quark-gluon scattering ~\cite{bib:nuclstruc:spinstruc:Conway:1989fs}. Two main levels of approximation had been used in the literature ~\cite{bib:nuclstruc:spinstruc:Altarelli:1979ub}. The first one is the so-called Leading-Log Approximation (LLA), where the leading logarithmic corrections to the DY cross section can be re-summed at any order in the strong coupling constant $\alpha_s$, introducing in the PDF's an additional scale dependence on $M^2_{l \bar{l}}$. The so-called DGLAP 
evolution can be obtained by describing the functions with parameters explicitly 
depending on $\log M^2$ (see Ref.~\cite{bib:nuclstruc:spinstruc:Buras:1977yj} and Apps. A, B and D in Ref.~\cite{bib:nuclstruc:spinstruc:Conway:1989fs} for further details). 
The second approximation level in the QCD higher order corrections, the so-called Next-to-Leading-Log Approximation (NLLA), is performed including in the calculation all
processes at first order in $\alpha_s$ involving a quark, an antiquark and a 
gluon~\cite{bib:nuclstruc:spinstruc:Altarelli:1979ub}, and leads to sizeable effects, approximately doubling the pure QPM cross-section. Such a corrections is roughly independent of $x_{_F}$ and $M^2$ (except for the kinematical upper limits) and it is usually indicated as the $K$ factor. $K$-factors depend on the choice of the parametrisation of the distribution functions through their normalisation~\cite{bib:nuclstruc:spinstruc:Anassontzis:1987hk}, and scale as $\sqrt{\tau}$~\cite{bib:nuclstruc:spinstruc:Conway:1989fs}.

The azimuthal asymmetries, which are defined as ratios of cross sections, should be pretty robust w.r.t. such kind of QCD corrections, since the corrections in the numerator and in the denominator should approximately compensate each other~\cite{bib:nuclstruc:spinstruc:Martin:1997rz}.

But this is not the case if we consider the (un)polarised cross-section. DY processes at high CM energy show reduced $K$-factors, but the relevant role of higher-order perturbative QCD corrections ~\cite{bib:nuclstruc:spinstruc:Shimizu:2005fp}, in terms of the available fixed-order contributions as well as of all-order soft-gluon re-summations, leads to large enhancements of the unpolarised DY dilepton production in the \panda~ kinematic regime, due to soft gluon emission near partonic threshold; the unpolarised cross section for DY dilepton production at \panda~ energies is thus matter of investigation itself, and could provide information on
the relation of perturbative and non-perturbative dynamics in hadronic scattering~\cite{bib:nuclstruc:spinstruc:Shimizu:2005fp}.

\subsubsection*{The \panda~ Scenario}
\label{sec:nuclstruc:spinstruc:panda}

The unpolarised and the
single-polarised Drell-Yan $\bar{p} p^{(\uparrow )} \rightarrow \mu^+ \mu^- X$ can be investigated with the \panda~ spectrometer (the former case since the very beginning, the latter {\it if} a polarised target would be developed) and the HESR antiproton beam. In such a scenario a beam energy of 15 \gev on the protons at rest in the fixed target can provide a centre of mass energy up to $s\simeq 30~\gev^2$.

The handbag diagram of Fig.~\ref{fig:nuclstruc:spinstruc:dy_hb_lt} is the dominant contribution only for a CM energy $s$ much bigger than the involved hadron masses. Moreover the di-lepton mass $M_{l \bar{l}}$ should not belong to the hadronic resonance region, in order to select for the elementary process an annihilation into a virtual photon; this is the reason why the DY di-lepton production is usually investigated in the so-called "safe region": $4~\gev \le M_{l \bar{l}} \le 9~\gev$, between the $\psi'$ and the first $\Upsilon$ resonance.

In the DIS regime we can define, in terms of the light-cone momentum fractions, the parameter $\tau = x_1x_2$ and the invariant $x_{_F} = x_1-x_2$ (fraction of the total available longitudinal momentum in the collision CM frame); the unpolarised cross section $d\sigma^0$ of Eq. \ref{eq:nuclstruc:spinstruc:unpolxsect}, kept differential in $M_{l \bar{l}}^2 \equiv Q^2$ and integrated upon $d\tau$:
\bwt
\beq
\frac{d\sigma^o}{dM^2 dx_{_F}} = \frac{4\pi \alpha^2}{9}\,
\frac{1}{M^2 s (x_1+x_2)}\, 
\quad\sum_f \, e_f^2\, \left[ \bar{f}_1^f(x_1) \, f_1^f(x_2) + 
(1\leftrightarrow 2)\right] 
\label{eq:nuclstruc:spinstruc:unpolxsect_mxf}
\eeq
\ewt
shows the scaling in the CM energy $s$ experimentally confirmed~\cite{bib:nuclstruc:spinstruc:Moreno:1990sf} and the enhanced production of di-lepton pairs at lower $M_{l \bar{l}}^2$.

In the \panda~ scenario the upper limit of the "safe region" $M < 9~\gev$ is beyond the accessible kinematic region: since for a $\bar{p}$ beam on a $p$ fixed target is $M^2 = \tau \, s =  \tau\, 2 M_p (M_p + E_{\bar{p}}) \sim  \tau \, 2 M_p E_{\bar{p}}$, even considering the limit $\tau \sim 1$, i.e. the case in which all the available CM energy is transferred to the virtual photon, an $M \sim 9~\gev$ would correspond to an antiproton beam energy $E_{\bar{p}}\sim40~\gev$. The lower cut $M > 4~\gev$ selects then a phase space region $0.5 \lesssim \tau \lesssim 1$ limited to very high values of both $x_1$ and $x_2$. To release the constraint on the lower cut, the $1.5 ~\gev \leq M \leq 2.5 ~\gev$ portion of the di-lepton mass spectrum can be considered as well: a region not overlapping the $\phi$ and $J/\psi$ resonances, that leads to two major benefits in the \panda~ scenario: a wider accessible $\tau$ range (as shown in  Fig.~\ref{fig:nuclstruc:spinstruc:x1x2}) and a larger cross section (see Eq.~\ref{eq:nuclstruc:spinstruc:unpolxsect_mxf} and Ref.~\cite{bib:nuclstruc:spinstruc:Bianconi:2004wu}).

The expected integrated cross section for $\bar{p} p \rightarrow \l^+ \l^- X$ at $s=30~\gev^2$ in the  $1.5 ~\gev \leq M \leq 2.5 ~\gev$ is $\sigma^0_{1.5 \leq M \leq 2.5} \sim 0.8~nb$ ~\cite{bib:nuclstruc:spinstruc:Bianconi:2004wu}; assuming the design luminosity of the High Resolution mode (see Tab. \ref{tab:param} in Sec. \ref{sec:exp:hesr}) $L = 2\cdot 10^{32}~cm^{-2}~s^{-1} $, the expected rate for the DY production of $\mu$-couples would hence be:
\beq
R = 2\cdot 10^{32}cm^{-2}~s^{-1} \times  0.8\cdot 10^{-33}cm^{-2} = 0.16~ s^{-1}
\label{eq:nuclstruc:spinstruc:raw_rate}
\eeq
The cross section at $s=30~\gev^2$ is expected to drop dramatically in the "safe-region" to $\sigma^0_{4 \leq M \leq 9} \sim 0.4~pb$ ~\cite{bib:nuclstruc:spinstruc:Bianconi:2004wu}; since the resulting rate for the DY processes would be incompatible with a detailed investigation in the \panda~ framework, the focus will be herewith on the  $1.5 ~\gev \leq M \leq 2.5 ~\gev$ region only.

\subsubsection*{Simulations}
\label{sec:nuclstruc:spinstruc:sim}

The investigation of the (un)polarised DY $\bar{p} p^{(\uparrow)} \rightarrow \mu^+ \mu^- X$ process in the scenario and with the kinematic conditions described above in this section is certainly a difficult task; to probe its feasibility Monte-Carlo simulations have been performed, based on the event generator kindly provided us by A. Bianconi (a more detailed description of such a generator can be found in Ref. ~\cite{bib:nuclstruc:spinstruc:Bianconi:2004wu,bib:nuclstruc:spinstruc:Bianconi:2008jm}). 
This is the very same event generator involved in the feasibility studies performed for the polarised DY $\pi^\pm p^\uparrow \to \mu^+ \mu^- X$ process at the 
CM energy $\sqrt{s} \sim 14$ \gev reachable at COMPASS ~\cite{bib:nuclstruc:spinstruc:Bianconi:2006hc,bib:nuclstruc:spinstruc:Bianconi:2006mf}, and for the polarised DY 
$pp^\uparrow \to \mu^+ \mu^- X$ process at the CM energy $\sqrt{s}=200$ 
\gev reachable at the Relativistic Heavy-Ion Collider (RHIC) of BNL ~\cite{bib:nuclstruc:spinstruc:Bianconi:2005yj}.

The main goal of the Monte-Carlo simulations reported herewith is to estimate the number of events required for the DY program at \panda,  in order to access unambiguous information on the PDF's of interest, namely the BM function $h_1^{\perp}$ in the $\bar{p} p\rightarrow \mu^+ \mu^- X$ process and the Sivers function $f_{1T}^\perp$, the Transversity $h_{1T}$ and the $h_1^{\perp}$ again in the 
$\bar{p} p^{\uparrow } \rightarrow \mu^+ \mu^- X$ process. The effects of the kinematic cuts above described and of the acceptance introduced by the \panda~ spectrometer, and the possibility to probe the dependence of the experimental asymmetries on the kinematics have to be investigated as well.

An estimation of the fully polarised case is beyond the scope of the present discussion; see Ref.~\cite{bib:nuclstruc:spinstruc:Bianconi:2005bd,bib:nuclstruc:spinstruc:Efremov:2004qs,bib:nuclstruc:spinstruc:Anselmino:2004ki}  for the predictions of double-spin asymmetries in different experimental scenarios at \INST{FAIR} in the case of the fully polarised DY process.

For each one of the two investigated processes $480K$ events have been generated at $s=30~\gev^2$ in order to satisfy the following kinematic cuts: a di-lepton invariant mass $1.5 ~\gev \leq M \leq 2.5 ~\gev$, a di-lepton transverse momentum $q_{_T}>1~\gev/c$, and a polar angle for the $\mu^+$ in the CS frame $60^\circ \leq \theta^{CS}_{\mu^+} \leq 120^\circ$. The second cut, together with the rejection factor introduced by the iron in the magnet, is necessary to select the DY signal events from the hadronic background, as will be discussed later in this section; the latter cut is aimed to select the azimuthal $\theta^{CS}_{\mu^+}$-region where the azimuthal asymmetries are expected to be larger (at $\theta^{CS}_{\mu^+} \sim \frac{\pi}{2}$ \cite{bib:nuclstruc:spinstruc:Bianconi:2004wu}).

If the simulations had included also events in the safe-region ($4 ~\gev \leq M \leq 9 ~\gev$)
the phase space for 
large $\tau = x_1 x_2$ would have been anyway scarcely populated, since the virtual photon 
introduces a $1/M^2 \propto 1/\tau$ factor in the cross section of Eq.~\ref{eq:nuclstruc:spinstruc:unpolxsect_mxf}, which thus decreases for increasing 
$\tau$ (with $0\le \tau \le 1$). The PDF's become hence negligible for $x_1 x_2 \rightarrow 1$ and events  
accumulate in the phase space part corresponding to small $\tau$.

This is indeed the case of Fig.~\ref{fig:nuclstruc:spinstruc:x1x2}, reporting the $x_1~vs~x_2$ scatter-plot for the generated sample of unpolarised DY di-lepton production. The upper
right corner of the figure corresponds to the limit $\tau \rightarrow 1$, when all the available CM energy is transferred to the virtual photon, and is depleted by the cut $1.5 ~\gev \leq M \leq 2.5 ~\gev$; such a cut selects the region $0.075 \le \tau \le 2.1$, and the distribution becomes more and more dense approaching the lower ridge. The line bisecting the plot at $45^\circ$ corresponds to $x_{F} \sim 0$; parallel lines above and below indicate $x_{F}>0$ and $x_{F}<0$, corresponding in the laboratory frame respectively to "forward" (small $\theta^{LAB}_{\mu^+}$) and "backward"  (large $\theta^{LAB}_{\mu^+}$) events.

\begin{figure}[H]
\begin{center}
  \includegraphics[width=0.8\columnwidth]{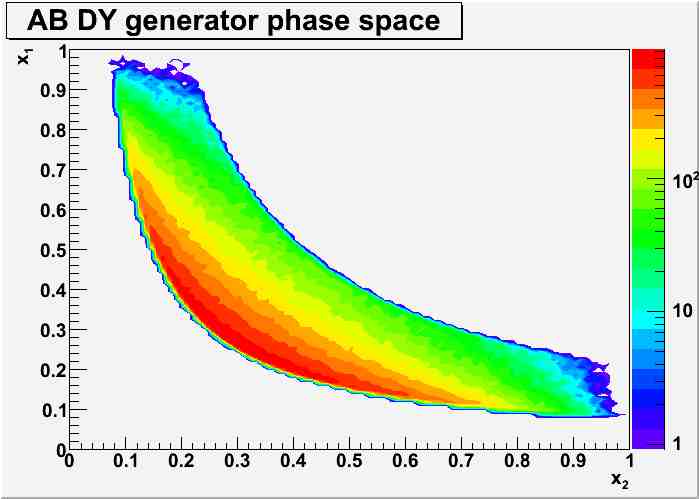}
  \caption[Light-cone momentum fractions scatter plot for unpolarised DY di-lepton production.]{The correlation between the two light-cone momentum fractions of the parton momenta $x_{1,2}$ for the $480K$ DY events generated with the Andrea Bianconi's generator~\cite{bib:nuclstruc:spinstruc:Bianconi:2004wu} for the unpolarised $\bar{p} p \rightarrow \l^+ \l^- X$ processes  at $s=30~\gev^2$ in the following kinematic conditions: a di-lepton invariant mass $1.5 \leq M^2_{l \bar{l}} \leq 2.5~\gev$, a transverse momentum of the lepton pair $q_{_T} > 1~\gevc$ and a polar angle for the $\mu^+$ in the Collins-Soffer frame $60^\circ \leq \theta^{CS}_{\mu^+} \leq 120^\circ$. 
\label{fig:nuclstruc:spinstruc:x1x2}}
\end{center}
\end{figure}

\begin{figure}[H]
\begin{center}
  \includegraphics[width=0.8\columnwidth]{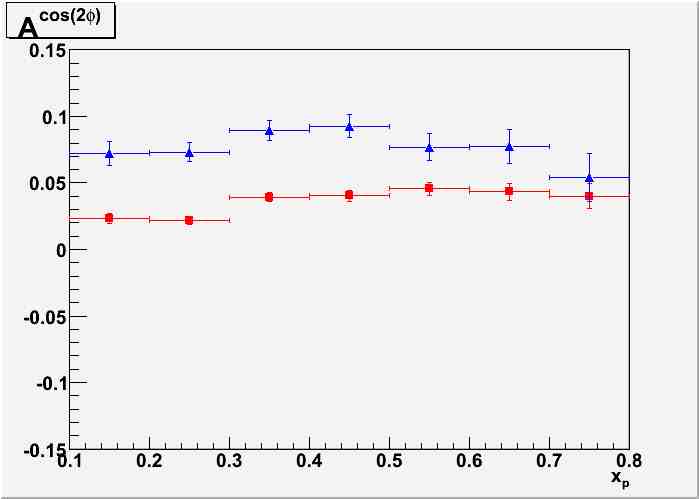}
  \caption[The azimuthal asymmetry for the unpolarised DY process $\bar{p} p \rightarrow \l^+ \l^- X$.]{The azimuthal asymmetry between cross sections related to positive and negative values of the $cos2\phi$ term in Eq.~\ref{eq:nuclstruc:spinstruc:unpolxsect} for $480K$ events of the unpolarised DY process $\bar{p} p \rightarrow \l^+ \l^- X$ in the same kinematic conditions of Fig.~\ref{fig:nuclstruc:spinstruc:x1x2}. The asymmetry is evaluated in $x_p$ bins for a di-lepton pair transverse momentum $1\le q_T \le 2~\gevc$ (squares) or $2\le q_T \le 3~\gevc$ (triangles). Error bars reflect statistical errors only.
\label{fig:nuclstruc:spinstruc:asym_cos2phi}}
\end{center}
\end{figure}

\begin{figure*}[htb]
\begin{center}
  \includegraphics[width=\columnwidth]{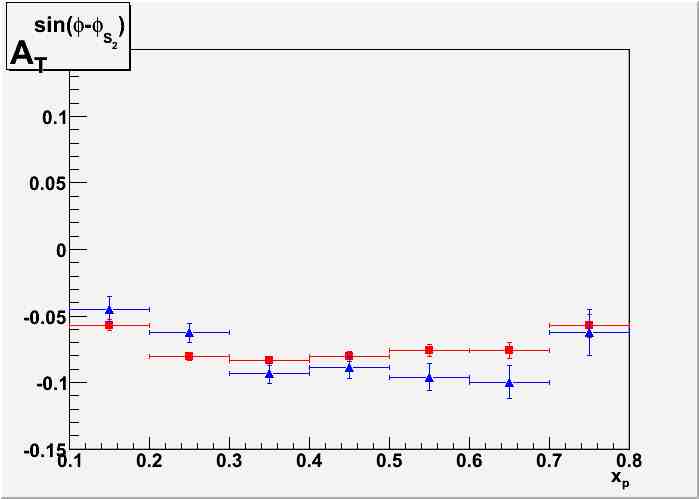}\hfill
  \includegraphics[width=\columnwidth]{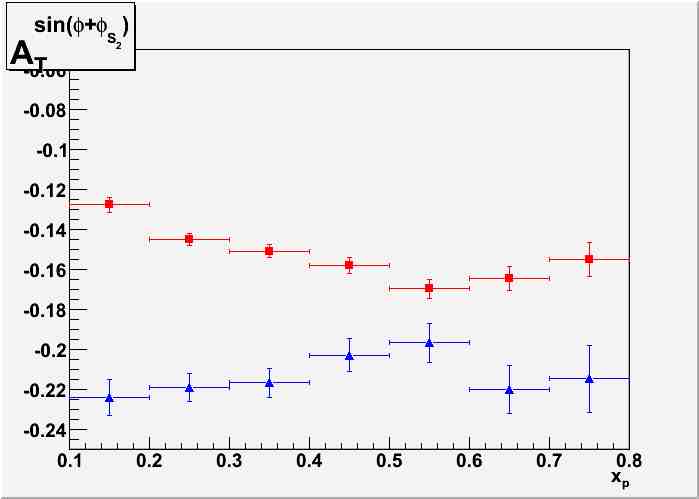}
  \caption[The azimuthal asymmetries for the single-polarised DY process $\bar{p} p^{\uparrow}  \rightarrow \l^+ \l^- X$.]{The single-spin azimuthal asymmetries $A_T^{\sin (\phi - \phi_{S_2})}$ and $A_T^{\sin (\phi + \phi_{S_2})}$ between cross sections related to positive and negative values respectively of the two $\sin (\phi - \phi_{S_2})$ and $\sin (\phi + \phi_{S_2})$ terms in Eq.~\ref{eq:nuclstruc:spinstruc:1polxsect}, for $480K$ events of the single-polarised DY process $\bar{p} p^{\uparrow }  \rightarrow \l^+ \l^- X$ in the same kinematic conditions of Fig.~\ref{fig:nuclstruc:spinstruc:x1x2}. The asymmetries are evaluated in $x_p$ bins for a di-lepton pair transverse momentum $1\le q_T \le 2~\gevc$ (squares) or $2\le q_T \le 3~\gevc$ (triangles). Error bars reflect statistical errors only. 
\label{fig:nuclstruc:spinstruc:asym_ssa}}
\end{center}
\end{figure*}

The azimuthal asymmetry related in the CS frame to the $cos2\phi$ term of Eq.~\ref{eq:nuclstruc:spinstruc:unpolxsect} and Eq.~\ref{eq:nuclstruc:spinstruc:param} (an asymmetry that leads to the BM function $h_1^{\perp}$) has been evaluated for muon pairs with transverse momentum 
$1 <q_T < 2 ~\gevc$ or $2 < q_T < 3 ~\gevc$ in $x_p$ bins ($x_p$ being the light-cone momentum fraction of the parton in the proton) over the range $0.2<x_p<0.8$, where according to the phase space distribution of Fig.~\ref{fig:nuclstruc:spinstruc:x1x2} the statistics is reasonably large. The asymmetry $A^{cos2 \phi}$ has been obtained considering all the $480K$ 
unpolarised $\bar{p} p \rightarrow \l^+ \l^- X$ events generated by the Andrea Bianconi's generator~\cite{bib:nuclstruc:spinstruc:Bianconi:2004wu}, tuned to reproduce the most recent experimental data available in the literature~\cite{bib:nuclstruc:spinstruc:Conway:1989fs}.
For each $x_p$ bin and $q_T$ cut two event samples are stored, corresponding respectively to positive and negative values of $cos2\phi$; the resulting asymmetry $A^{cos2\phi} = (U-D)/(U+D)$ between cross sections with positive ($U$) and negative ($D$) values of $cos2\phi$ is shown in Fig.~\ref{fig:nuclstruc:spinstruc:asym_cos2phi}. Error bars represent statistical errors only; since the asymmetry has been evaluated considering the whole sample of the generated events, both the asymmetry itself and its error bars shown in Fig.~\ref{fig:nuclstruc:spinstruc:asym_cos2phi} are not folded with the acceptance of the \panda~ spectrometer.

The azimuthal asymmetry related to $h_1^{\perp}$ is hence expected to be not negligible, experimentally measurable in the \panda~ energy range, with errors good enough to allow for the investigation of the dependence of the asymmetry on the relevant kinematic variables such as the transverse momentum of the lepton pair $q_T$. Such an investigation is of uttermost importance in order to probe the existence of a possible inversion of the trend in the energy dependence of the asymmetries, to balance soft and hard effects in this kind of processes.

In the case of the single-polarised DY process $\bar{p} p^{\uparrow } \rightarrow \mu^+ \mu^- X$ two different SSA's  can be defined; a first one, $A_T^{\sin (\phi - \phi_{S_2})}$, weighted by the factor $\sin (\phi - \phi_{S_2})$ (Eq.~\ref{eq:nuclstruc:spinstruc:1polxsect}) and related to the Sivers function $f_{1T}^{\perp}$; a second one, $A_T^{\sin (\phi + \phi_{S_2})}$, weighted by the factor $\sin (\phi + \phi_{S_2})$ (Eq.~\ref{eq:nuclstruc:spinstruc:1polxsect}) 	and related to the convolution of the Transversity  $h_{1T}$ with the BM function $h_1^{\perp}$. The single-polarised DY sample have been generated under the assumptions described in Sec. A of~\cite{bib:nuclstruc:spinstruc:Bianconi:2004wu} but assuming the simpler functional hypothesis $\langle h_1(x_p) \rangle / \langle f_1(x_p) \rangle = 1$. The procedure to determine the two asymmetries is the analogous of that used to evaluate the unpolarised asymmetry: for each $x_p$ bin four sample of events are stored, for positive ($U_{\pm}$) and negative ($D_{\pm}$) values of the factors respectively $\sin (\phi + \phi_{S_2})$ and $\sin (\phi - \phi_{S_2})$. The resulting asymmetries $A_T^{\sin (\phi \pm \phi_{S_2})} = (U_{\pm} - D_{\pm})/(U_{\pm} + D_{\pm})$ are plotted in Fig.~\ref{fig:nuclstruc:spinstruc:asym_ssa} in $x_p$ bins in the range $0.2<x_p<0.8$, separately for muon pairs with transverse momentum 
$1 <q_T < 2 ~\gevc$ or $2 < q_T < 3 ~\gevc$.
The asymmetries have been obtained considering the whole sample of $480K$ events generated for the single-polarised DY process and are hence not affected by the spectrometer acceptance; error bars represent statistical errors only.

Under the assumptions above described the asymmetry $A_T^{\sin (\phi + \phi_{S_2})}$ is expected to be relevant, and also in this case the investigation of the dependence of the asymmetry on the di-lepton transverse momentum $q_T$ should be possible. This is not the case for the asymmetry  $A_T^{\sin (\phi - \phi_{S_2})}$, predicted to be much smaller and with a strongly reduced dependence on $q_T$.

\begin{figure*}[t]
\begin{center}
  \includegraphics[width=.66\columnwidth]{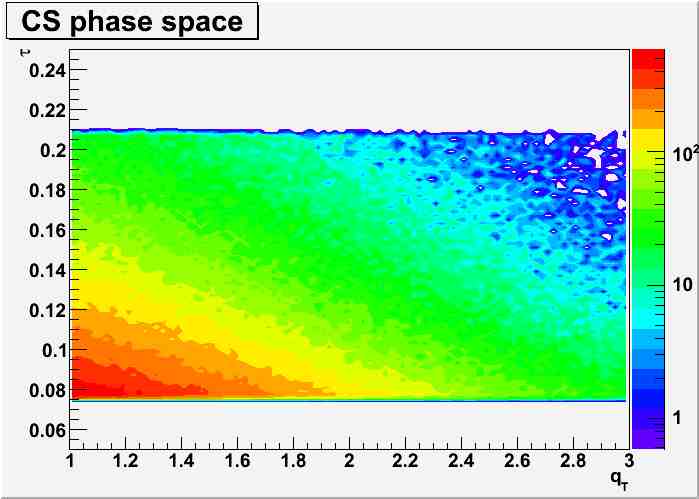}\hfill
  \includegraphics[width=.66\columnwidth]{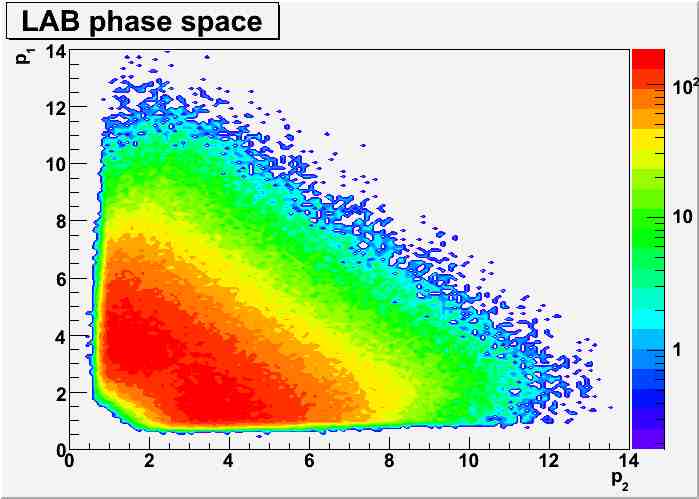}\hfill
  \includegraphics[width=.66\columnwidth]{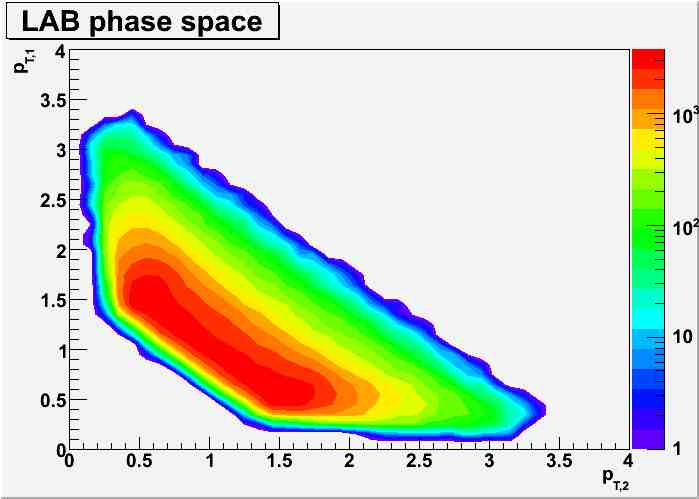}
  \caption[\panda~ acceptance for the unpolarised DY process $\bar{p} p \rightarrow \l^+ \l^- X$.]{Estimated kinematic acceptance for the unpolarised DY process  in the same kinematic conditions of Fig.~\ref{fig:nuclstruc:spinstruc:x1x2}. A "Monte Carlo through" has been performed considering the preliminary estimation of the minimum iron thickness required to resolve the DY signal from the hadronic background. Left panel shows the accessible $\tau$ region and the events accumulating toward low values of the transverse momentum of the lepton pair $q_{T}$ in the Collins-Soper frame. The central and the right panels show the scatter plots respectively of the momenta and of the transverse momenta of the two outgoing muons in the laboratory Frame frame.\label{fig:nuclstruc:spinstruc:kin_acc}}
\end{center}
\end{figure*}

\begin{figure*}[b]
\begin{center}
  \includegraphics[width=\columnwidth]{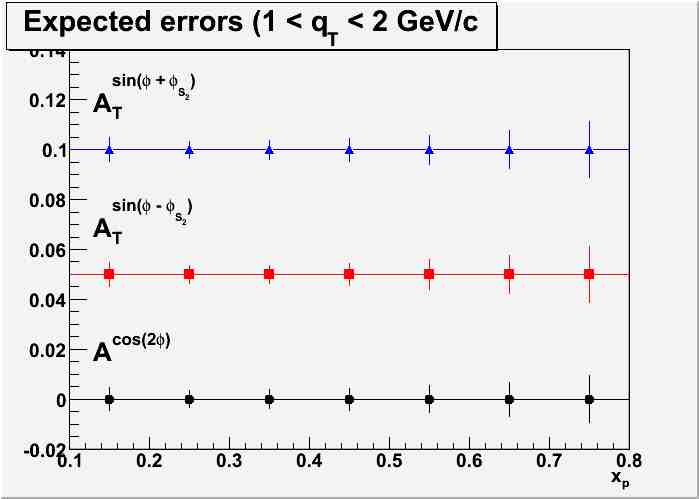}\hfill
  \includegraphics[width=\columnwidth]{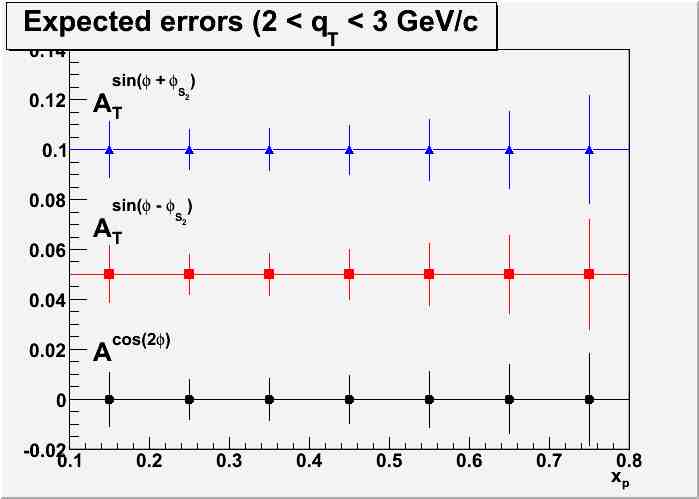}
  \caption[Expected statistical errors for SSA in the (un)polarised DY process $\bar{p} p^{(\uparrow})  \rightarrow \l^+ \l^- X$ folded with \panda~acceptance.]{The statistical errors for the asymmetries of Figures~\ref{fig:nuclstruc:spinstruc:asym_cos2phi} and ~\ref{fig:nuclstruc:spinstruc:asym_ssa} expected once they are folded with the \panda~ global acceptance (see text for a more detailed discussion), for a di-lepton pair transverse momentum $1\le q_T \le 2~\gevc$ (left panel) or $2\le q_T \le 3~\gevc$ (right panel). The "Monte Carlo through" has been performed for each one of the two sets of $480K$ events generated for unpolarised and single-polarised DY processes, in the same kinematic conditions of Fig.~\ref{fig:nuclstruc:spinstruc:x1x2}, considering the preliminary estimation of the minimum iron thickness required to resolve the DY signal from the hadronic background.
\label{fig:nuclstruc:spinstruc:errors_asym_ssa}}
\end{center}
\end{figure*}

For each one of the two simulated DY processes the generated events have then been propagated through the \panda~ spectrometer (Sec.~\ref{exp:detector}) in order to evaluate the global acceptance introduced by the experimental layout, i.e. the geometrical acceptance and the events' loss due to the material budgets, in particular the ones introduced by the Electromagnetic Calorimeter and by the iron of the magnet (in which the muon detectors will be embedded).

The iron shield is mandatory in order to separate the muon couples produced in the investigated DY processes from the pion couples coming from the hadronic background; in fact the required combined rejection factor that has to be provided by the experimental apparatus itself and by the events' reconstruction is very large. The total $\bar{p}p$ annihilation cross section in the \panda~ energy range is $\sigma_{\bar{p}p} \sim 50~mb$; once accounted for its diffractive component, the global cross section for pion production in the final state is of the order of some tens of $mb$. Since the DY expected cross section~\cite{bib:nuclstruc:spinstruc:Bianconi:2004wu} is of the order of the $nb$, the required rejection factor is $\sim 10^7$.

The extended simulations needed to investigate the background rejection are actually still in progress. Nevertheless their preliminary indications point toward a scheme in which the required rejection factor could be achieved by the mean of: the depletion of the pions' sample due to their interaction with the iron in the \panda~ solenoid; a set of kinematic cuts in order to select in the event reconstruction stage the DY processes (the most effective ones being the requirement of a di-muon pair transverse momentum $q_T \ge 1~\gevc$ and of a reaction vertex in the target region); a kinematically constrained refit of the final state in order to reject the residual contamination from the hadronic background. To balance the best possible background rejection and the minimum DY events loss, the geometry of the iron in the barrel and in the endcap of the solenoid has to be segmented, in order to host embedded muon counters. The minimum iron thickness required to achieve such a large rejection factor determines the preliminary global acceptance of the apparatus for the DY processes.

The preliminary indications for such an acceptance from the extended simulation in progress, point to a reduction by a factor of $\sim 2$ of the total cross-section $\sigma^0$ quoted above in this section. That would  rescale Eq.~\ref{eq:nuclstruc:spinstruc:raw_rate} to the corresponding rate expected for those DY muon pairs that can be detected and resolved from the hadronic background in these kinematic conditions by the \panda~spectrometer: $R = 2\cdot 10^{32}\,\mathrm{cm}^{-2}\mathrm{s}^{-1} \times  0.8\cdot 10^{-33}\,\mathrm{cm}^{-2} \times \frac{1}{2} = 0.08\,\mathrm{s}^{-1}$, i.e.  $\sim200\mathrm{K}$ events/month, not accounting for the various experimental efficiencies.

The plots in Fig.~\ref{fig:nuclstruc:spinstruc:kin_acc} show few of the most relevant kinematic distributions for those unpolarised DY events surviving the (preliminarily estimated) minimum iron thickness required to resolve the DY signal from the hadronic background; i.e., those kinematic distributions are folded with the global acceptance of the \panda~spectrometer.	

The most unwanted effect of the \panda~spectrometer acceptance is the introduction of sizeable fake instrumental asymmetries, heavily depending on the geometrical cut introduced in the muons polar angle distribution in the laboratory frame by the hole in the endcap around the beam pipe (varying on the azimuthal angle  $\phi^{LAB}_{\mu^\pm}$ but being in the average $\theta^{LAB}_{\mu^\pm} \ge 7^\circ$). Such an effect will have to be carefully accounted for in the analysis stage and can be partially reduced including in the considered DY sample also those events for which (one of) the outgoing muons can be detected in the Forward Spectrometer (FS) (See Sec.~\ref{exp:detector}). 

The plots in Fig.~\ref{fig:nuclstruc:spinstruc:errors_asym_ssa} show the expected statistical errors for the three considered azimuthal asymmetries once the effects of the preliminary \panda~acceptance, and namely the consequent reduction in the statistics, are accounted for. The estimated errors show how the \panda~spectrometer should be able to detect and experimentally evaluate the above described azimuthal asymmetries, even if they were rather small.

\subsubsection*{Conclusions}
\label{sec:nuclstruc:spinstruc:concl}

The DY production of muon pairs is an excellent tool to access transverse spin effects within the nucleon. As stated before in this Section, the double-polarised case would really be a "dream option", allowing to access the Transversity distribution $h_{1T}$ directly and without any convolution with other PDF's; unfortunately it is beyond the initial scope of the \panda~ physics program. But even considering the unpolarised and the single-polarised cases only, the Boer-Mulders distribution $h_1^{\perp}$ and the Sivers distribution $f_{1T}^{\perp}$ could be accesses as well, and in a single experimental scenario. To access the latter a polarised target is needed, an element not yet included in the present \panda~ layout but by many indicated as an almost necessary upgrade, while the former can be accessed since the very beginning of the \panda~ activities.

A DY program in \panda, besides addressing the present excitement in the spin physics community driven by the discrepancies among the most recent SIDIS data for the Sivers function, would allow for the evaluation of three of the most hunted PDF's in a kinematic region were the valence contributions are expected to be dominant.

A large rejection factor is needed to resolve the DY processes from the hadronic background; a detailed evaluation of the hadronic background in the \panda~ scenario is in progress, in order to complete the feasibility studies relative to the spin physics program.

\end{multicols}
\twocolumn

%

%% file: phys/nucleonstructure/elmff/phys_elmff.tex
%
\subsection{Electromagnetic Form Factors in the Time-like Region}
\COM{Author(s): F. Maas}
\COM{Referee(s): D. Bettoni}
%
%
\input{./phys/nucleonstructure/elmff/phys_elmff_introduction.tex}

\input{./phys/nucleonstructure/elmff/phys_elmff_spacelike_data.tex}
\input{./phys/nucleonstructure/elmff/phys_elmff_timelike_data.tex}
\input{./phys/nucleonstructure/elmff/phys_elmff_simulations_general.tex}
\input{./phys/nucleonstructure/elmff/phys_elmff_simulations_background.tex}

\input{./phys/nucleonstructure/elmff/phys_elmff_simulations_signal.tex}

\input{./phys/nucleonstructure/elmff/phys_elmff_outlook.tex}

%% file: phys/nucleonstructure/elmff/phys_elmff_introduction.tex
\subsubsection{Introduction}
\label{label_emff_intro}

\begin{figure}[h]
\begin{center}
  \includegraphics[width=0.5\textwidth]{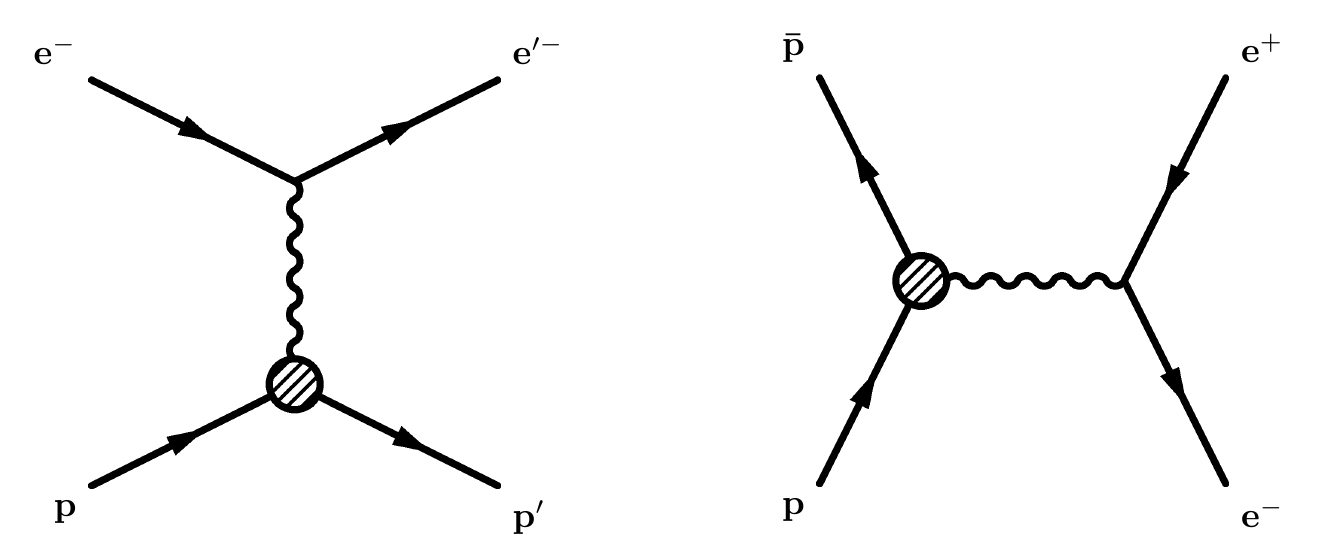}
  \caption[Feynman diagrams for elastic scattering and annihilation]{Feynman diagrams for elastics electron scattering (left) and its crossed channel 
         $\pbarp \rightarrow \ee$  (right) which will be measured with the \Panda-detector. 
 \label{fig:nuclstruc:emff:scattering:annihilation}
         }
\end{center}
\end{figure}

The electromagnetic probe is an excellent tool
to investigate the structure of the nucleon. 
The \Panda-experiment offers the unique possibility to make a precise 
determination of the electromagnetic form factors in the time-like region 
with unprecedented accuracy.
The electric (\GE) and magnetic  (\GM) form factors of the proton parametrise the hadronic current
in the matrix element for elastic electron scattering ($\Pem + \Pp \rightarrow \Pem + \Pp$) and in
its crossed process annihilation ($\pbarp \rightarrow \ee$) as shown in \Reffig{fig:nuclstruc:emff:scattering:annihilation}.
The form factors (FF) measured in electron scattering are intimately connected with those measured 
in the annihilation process. Moreover they are observables that can probe
our understanding of the nucleon structure in the regime of nonperturbative QCD as well 
as at higher energies where perturbative QCD applies.

The interaction of the electron
with the nucleon is described by the exchange of one photon with space-like four momentum transfer \qsq .
The lepton vertex is described completely within QED and on the nucleon vertex,
the structure of the nucleon is parametrised by two real scalar functions depending on one variable
\qsq\ only. These real functions are the Dirac form factor $\Fi^{p,n}$ and the Pauli form factor $\Fii^{p,n}$,
or as a linear combination of F$_{1,2}^{p,n}$ the Sachs form factors $\GE^{p,n}$ and $\GM^{p,n}$.
The standard way of writing the matrix element for elastic electron proton scattering in the framework of
one-photon exchange is:
\begin{eqnarray}
    M = \frac{e^2}{q^2} \: \: \bar{u}(k_2) \: \gamma_\mu \: \: u(k_1) \: \: \: \bar{u}(p_2)\:
    [F_{1}(q^2) \: \gamma_\mu \nonumber \\ + i
                           \: \frac{\sigma_{\mu \nu}q^\nu}{2 m_p}\:  F_{2}(q^2)]\: u(p_1),
\end{eqnarray}
$k_1 (p_1)$ and $k_2 (p_2)$ are the four-momenta of the initial and final
electron (nucleon) represented by the spinors $\bar{u}(k)$ ($\bar{u}(p)$) and
$u(k)$ ($u(p)$), $m_p$ is the nucleon mass, $q = k_1 - k_2$, $q^2 < 0$. Applying 
crossing symmetry yields the matrix element for $\pbarp \rightarrow \ee$
where $k_2 (p_2)$ changes sign so that $q^2 = s$.

The form factors are analytic functions of the four momentum transfer \qsq\ ranging from $\qsq =-\infty$
to $\qsq =+\infty$. While in electron scattering the form factors can be accessed 
in the range of negative \qsq\ (space-like), the annihilation process allows to access
positive \qsq\ (time-like) starting from the threshold of $\qsq = \rm 4 m_\Pp^2$.
Unitarity of the matrix element requires that space-like form factors are real functions 
of \qsq\ while for time-like \qsq\ they are complex functions. 
In the Breit frame, space-like FFs have concrete interpretations, since they are the  
Fourier transforms of the spatial  charge (\GE) and the magnetisation distribution (\GM) of the proton.
Their slope at $\qsq = 0$ directly yields the charge and magnetisation radius of the proton.
In time-like region, FFs reflect the frequency spectrum of the electromagnetic response of the nucleon. That way two 
complementary aspects of nucleon structure can be studied and ask for a 
full and complete description of the electromagnetic form factors over the full kinematical range of 
\qsq .

%% file: phys/nucleonstructure/elmff/phys_elmff_spacelike_data.tex
%
\subsubsection*{Impact from Electron Scattering Data}
The experimental determination of the electromagnetic form factors of the nucleon 
has triggered large experimental programs at all major facilities since they 
have long served as one of the testing grounds for our understanding of nucleon structure
ranging from the low-\qsq\ regime of QCD up to the high energy perturbative regime.
Basically all models of nonperturbative QCD, which are using effective degrees of freedom, 
have been used to estimate the nucleon form factors\cite{bib:nuclstruc:emff:Perdrisat:2006hj}.
For example different constituent quark models, skyrmion type of models, bag models 
and more recently a framework like chiral perturbation theory and lattice gauge theory
have been applied.

Due to their analyticity space-like and time-like form factors are intimately connected
by the application of dispersion relations which are an application of Cauchy's 
integral formula. Perturbative QCD makes predictions for the large \qsq\ behaviour 
of the connection between space-like region and time-like region.
Space-like form factors are connected to the recent developments 
using nonperturbative generalised parton distributions. 

The interest in the time-like form factors of the nucleon has been renewed 
by the recent measurement at \INST{JLAB} using the polarisation transfer and target asymmetry method,
showing that the ratio of $\mu_p \GE/\GM$ ($\mu_p$ magnetic moment of the proton) 
deviates from unity and is in contrast 
to the results derived from Rosenbluth separation technique 
\cite{bib:nuclstruc:emff:poltrans:jones,bib:nuclstruc:emff:exp:poltrans:gayou,
bib:nuclstruc:emff:exp:poltrans:gayou2,bib:nuclstruc:emff:exp:targetasym:Jones,bib:nuclstruc:emff:Punjabi:2005wq}. 
While this discrepancy is most probably connected with radiative corrections, 
it has been shown, that the polarisation transfer method is much less sensitive 
to those effects. It seems that \GE\ is approaching zero around a \qsq\ of 8\,(\gevc)$^2$
while \GM\ follows a dipole form factor indicating that the charge distribution 
has a hard surface in contrast to the magnetisation distribution.
This surprising result has reopened the question on the determination 
of \GE\ and \GM\ in the time-like domain which are complex functions. 
Almost all experiments so far have determined $|\GM|$ in the time-like 
domain using the hypothesis of equality between \GE\ and \GM\ 
which is fulfilled strictly only at threshold. The recent 
\INST{JLAB} results, yielding a different $q^2$ behaviour for 
\GE\ and \GM\, add to the motivation of individual determination 
of time like form factors, which was not obtained up to now and which 
will be possible with the  \Panda-experiment.
Only two experiments had enough statistics to determine the ratio of $|\GE|/|\GM|$ independently
from any hypothesis and which have so far reached contradicting results with large 
experimental uncertainties. The determination of the 
electromagnetic form factors in the time-like domain at low 
to intermediate momentum transfer is therefore regarded to be an open question.

%

%% file: phys/nucleonstructure/elmff/phys_elmff_timelike_data.tex
%
\subsubsection*{Existing Data on Time-like Form Factors}
The \Panda\ experiment offers a unique opportunity to determine 
the moduli of the complex form factors in the time-like domain, 
by measuring the angular distribution of the process $\pbarp \rightarrow \ee$
in a \qsq\ range from about 5\,(\gevc)$^2$ up to 14\,(\gevc)$^2$.
A determination of the magnetic form factor up to 
a \qsq\ of 22\,(\gevc)$^2$ will be possible by measuring the total cross section.

The differential cross section for unpolarised initial and final states
of the process $\pbarp \rightarrow \ee$  is \cite{bib:nuclstruc:emff:exp:zichichi62}:
\begin{eqnarray}
      \frac{\mathrm{d}\sigma}{\mathrm{d}\cos\theta} =
\frac{\pi\alpha^2 (\hbar c)^2}{8m_p^2\sqrt{\tau\left(\tau-1\right)}} 
   \huge[|G_M|^2 \left (1+\cos^2 \theta \right ) \nonumber\\
+ \frac{|G_E|^2}{\tau} \left( 1-\cos^2 \theta \right ) \huge] 
\label{eqn:angdist:crosssection:timelike}
\end{eqnarray}
with $\tau = q^2/4 m_p^2$. A measurement of this differential cross section 
over a wide range of $\cos \theta $ allows an independent determination 
of the moduli $|\GE(q^2)|$ and $ |\GM(q^2)|$ of the electromagnetic form factors
and has been attempted by a number of experiments \cite{bib:nuclstruc:emff:exp:eeE835,bib:nuclstruc:emff:exp:babar}. 
\begin{figure}[h]
\begin{center}
  \includegraphics[width=\swidth,angle=0]{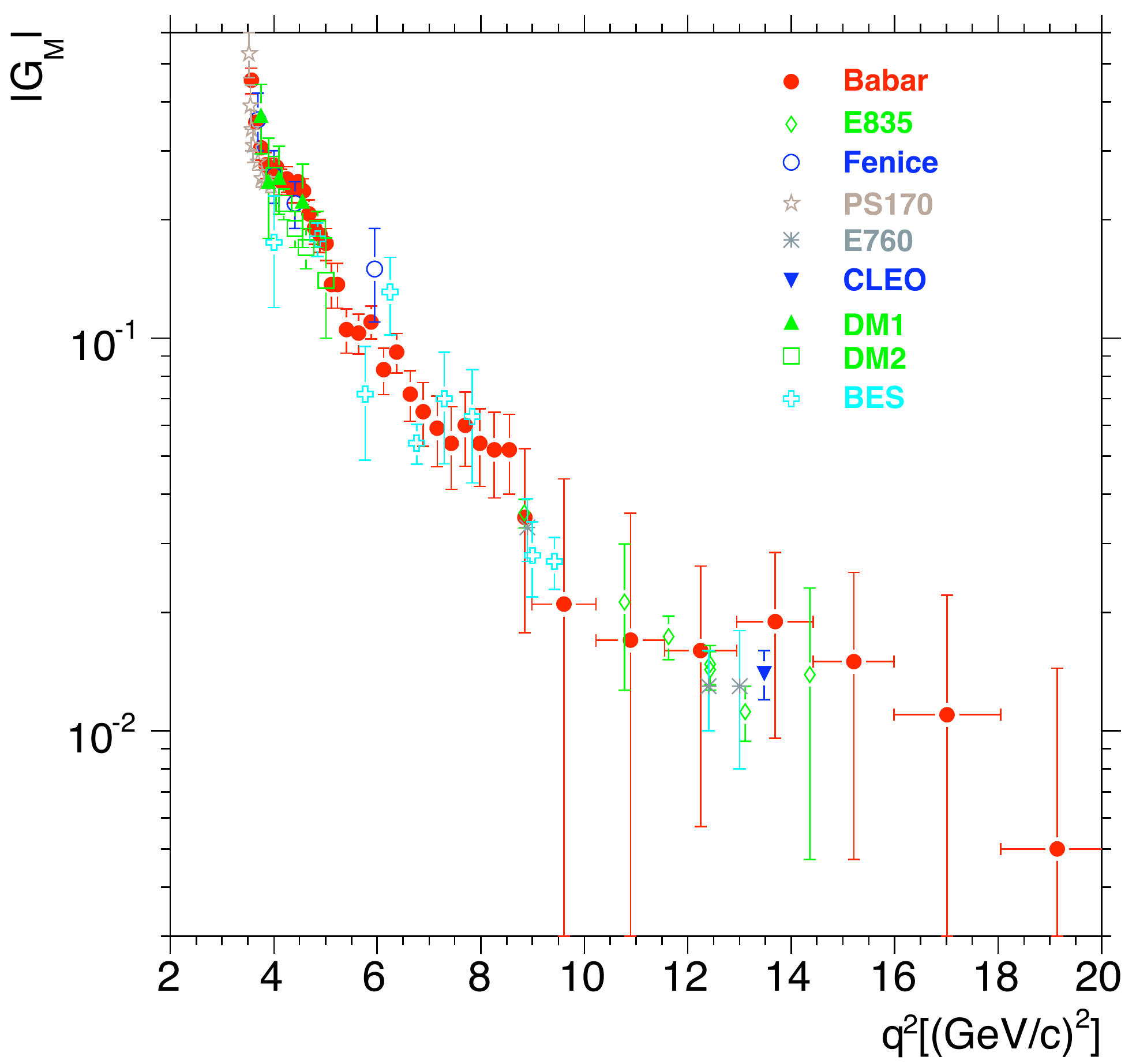}
  \caption[World data on the modulus $|\GM|$ of the time-like magnetic form factor]{ 
         World data on the modulus $|\GM|$ of the time-like magnetic form factor extracted from different 
         experiments using $\pbarp \rightarrow \ee$,$\ee \rightarrow \pbarp$, and $\ee\rightarrow \gamma \pbarp$.
         In all cases, the hypothesis of $R = |\GE|/|\GM| = 1$ has been used to analyse the data.
         \label{fig:nuclstruc:emff:worlddata:gm}
         }
\end{center}
\end{figure}
\Reffig{fig:nuclstruc:emff:worlddata:gm} gives a summary on the world data on the modulus $|\GM|$ of the time-like magnetic 
form factor extracted from different 
experiments using $\pbarp \rightarrow \ee$,$\ee \rightarrow \pbarp$, and $\ee\rightarrow \gamma \pbarp$.
In all cases, the hypothesis of $|\GE| = |\GM|$ has been used to analyse the data using the integrated differential 
cross section.

\begin{figure}[h]
\begin{center}
  \includegraphics[width=\swidth]{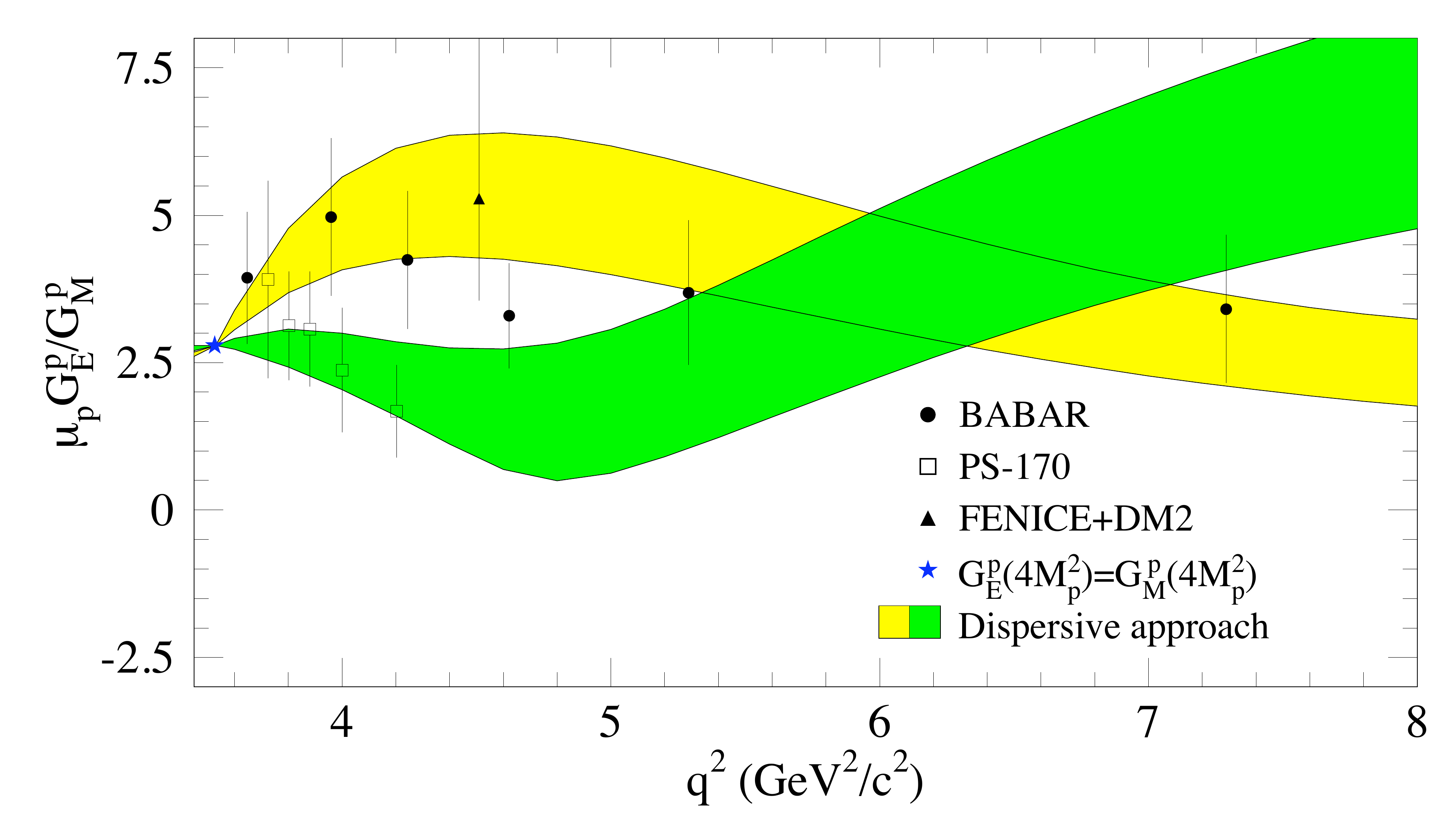}
  \caption[Recent dispersion theory results on $|\GE|/|\GM|$ from \INST{BABAR} data.]{The data from the \INST{LEAR} experiment \INST{PS170} and recent \INST{BABAR} data have been alternatively used as an input to a dispersion relation analysis of the electromagnetic form factors. The recent data from \INST{JLAB} have been used as an input for the space-like region. The green band gives the dispersion relation fit result on $|\GE|/|\GM|$ when using the \INST{PS170} data in the time-like region and the yellow band gives the result for the \INST{BABAR} data.         
The present accuracy in the ratio of  $R = |\GE|/|\GM|$ is of order 50\percent while a future measurement 
using the \Panda\ experiment 
at the design luminosity yields a statistical error of order few \% or better after $10^7$\,s in this region of $q^2$. 
\label{fig:nuclstruc:emff:gegm:pacetti}
}
\end{center}
\end{figure}

So far only two experiments have collected enough statistics in order 
to analyse the angular distribution and extract $|\GE|$ and $|\GM|$ independently
(see \Reffig{fig:nuclstruc:emff:gegm:pacetti}).

The \Panda\ experiment is planned to have unprecedented luminosity 
and rich particle identification capabilities, which are necessary 
in order to discriminate against the very large background of 
$\pbarp \rightarrow \Pgpm\Pgpp$. It is about $10^6$ times higher 
in cross section. We show here the strategy how to reach a
$\Pgpm\Pgpp$ rejection factor of about $10^8$ using 
the particle identification capabilities of each detector.

Any sizeable two-photon exchange contribution in the time-like domain, 
which is regarded to be one of the radiative correction processes responsible for the 
discrepancy between Rosenbluth and polarisation transfer method, 
can be detected in the same measurement since it introduces 
a forward-backward asymmetry in the angular distribution which otherwise 
is symmetric in one-photon exchange \cite{bib:nuclstruc:emff:theory:gakh,bib:nuclstruc:emff:theory:Gakh:2005wa}.

A large experimental activity is coming from $B$-factory $\ee$-colliders
where the energy is fixed to a $b\bar{b}$-resonance. The process of initial state radiation 
$\ee \rightarrow \pbarp\gamma$ (ISR), where the variable energy $\gamma$
from an initial state electron is used in order to "scan" the \qsq\ of the 
virtual photon probing the form factors, is used at those B-factories. 
The \INST{BABAR} experiment has recently published results for the ratio 
$R = |\GE|/|\GM|$, but is penalised by the fact, that the luminosity for 
$\ee \rightarrow \pbarp\gamma$ is then suppressed in average by factors of $10^5$ to $10^6$
as compared to the direct $\ee$-luminosity and cannot so far compete with the proposed 
measurement at \Panda.

An analogous process to ISR would be the emission of a $\pi$ by one of the $\pbarp$
in the initial state which would lower the \qsq\ of the virtual photon at the annihilation vertex. 
That way, one could reach the otherwise unaccessible range below the threshold 
and measure the form factors down to lower \qsq. Vector meson dominance and 
hypothetical baryonium states could be accessed that way.
Another possible extension of the program could be a possibility to access
the axial form factor in time-like domain, by using a neutron (deuteron) target 
\cite{bib:nuclstruc:emff:theory:adamuscinffaxial}.
In analogy to pion electroproduction, the chiral Ward-identities could be used here 
to extract the axial form factor in the time-like domain, for which no data 
exist at all. First estimates of cross sections for both,
subthreshold electromagnetic form factors and an axial form factor measurement have been performed, 
but more theoretical work and simulations is necessary \cite{bib:nuclstruc:emff:theory:adamuscinffaxial}. 

Another way to reach the subthreshold region would be to study
the EM annihilation on a nucleus, using a fast bound proton which
is off-mass-shell. In particular, the reaction could be done
on deuterium ($\overline{p} + d \rightarrow \ee n$) at low beam momentum 
(1.5 \gevc), with the final neutron as a missing particle.
First estimates have been performed regarding the count rate 
and the missing mass resolution. Background rejection remains 
to be studied in detail.

%
%

%% file: phys/nucleonstructure/elmff/phys_elmff_simulations_general.tex
%
\subsubsection{Simulations}
 Time-like form factors (TLFF) measurements through the reaction 
 $ \pbarp \rightarrow \ee $ (or $\mumu$) require the 
 complete identification of the 2 outgoing leptons. The shape of the angular distribution 
 provides a direct access to the moduli of the two proton form factors $|\GM|$ and $|\GE|$ 
 (see \ref{label_emff_intro}). We have studied two aspects concerning the determination of $|\GM|$ and $|\GE|$
 with the \Panda\ detector: the background conditions which will eventually limit the purity of the lepton signal and the
 sensitivity to the shape of the angular distribution after reconstructing the lepton signal in the \Panda\ detector. \ 
 The integrated cross section of the signal reaction $ \pbarp \rightarrow \ee $ was modelled by
 a fit to the world data where the following  Ansatz for $G_M$ had been used (see details in  
 \cite{bib:nuclstruc:emff:theory:eglegm} ):
 \begin{eqnarray}
  \label{fomulagm}
  |G_M|=\displaystyle\frac{a}{\left (1+\displaystyle\frac{q^2}{m_a^2}\right )} G_D\\
  G_D = \displaystyle\frac{1}{\left( 1+\displaystyle\frac{q^2}{m_d^2} \right)^2}, \nonumber \\ 
  m_a^2=3.6 (\gevc)^2, \nonumber \\ 
  m_d^2=0.71 (\gevc)^2 \nonumber  
 \end{eqnarray} 
 $G_D$ denotes the usual dipole-form factor, a = 22.5 is a normalisation constant and $m_{a}$ is an additional parameter 
 describing the deviation from a dipole. $m_{a}^2 = 3.6 \pm 0.9 \mathrm \,\gev^2$, $m_d$ is the usual dipole-mass parameter.

\begin{figure}[htb]
\begin{center}
\includegraphics[angle=90, width=\swidth]{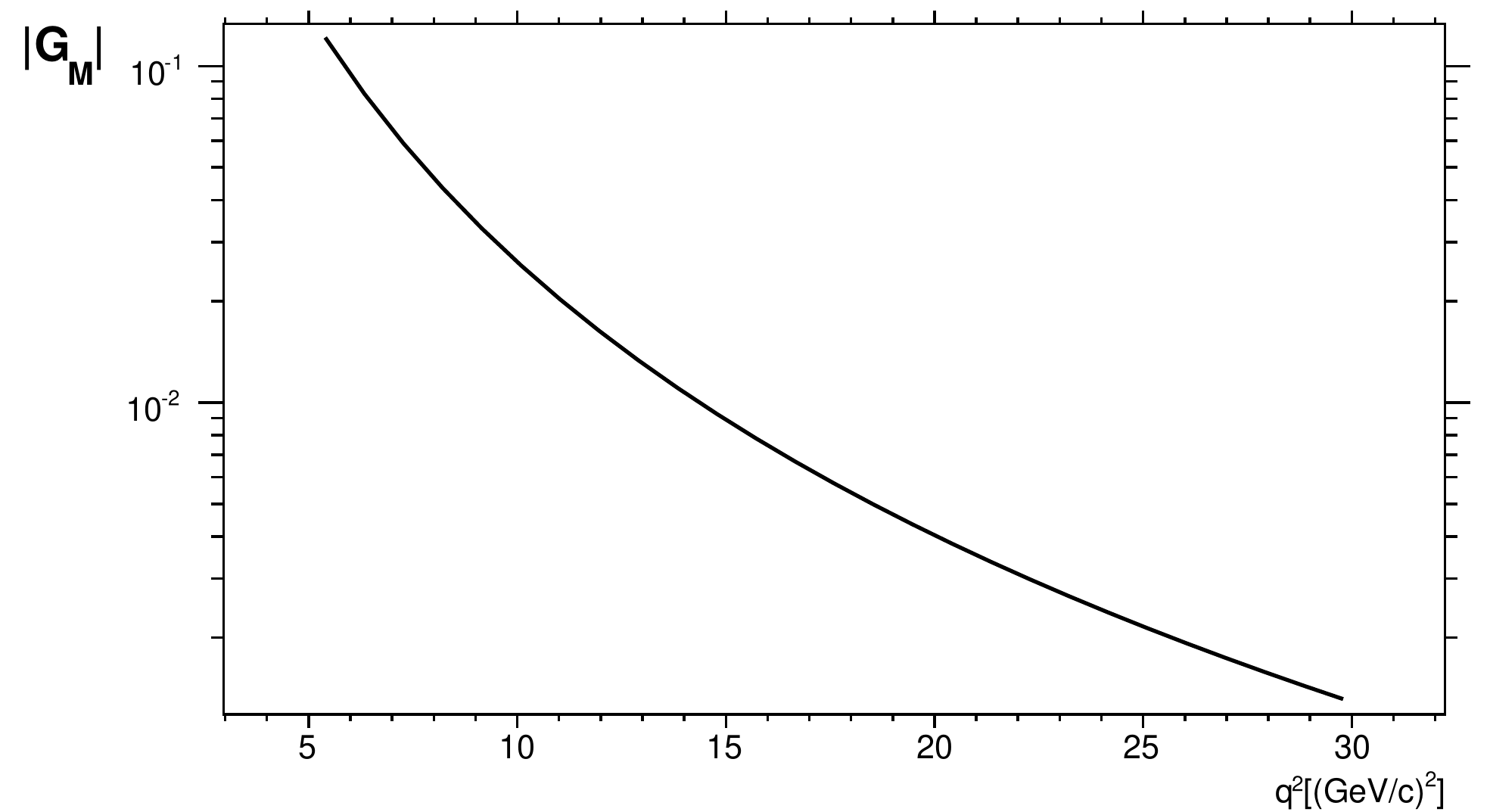}
\includegraphics[angle=90, width=\swidth]{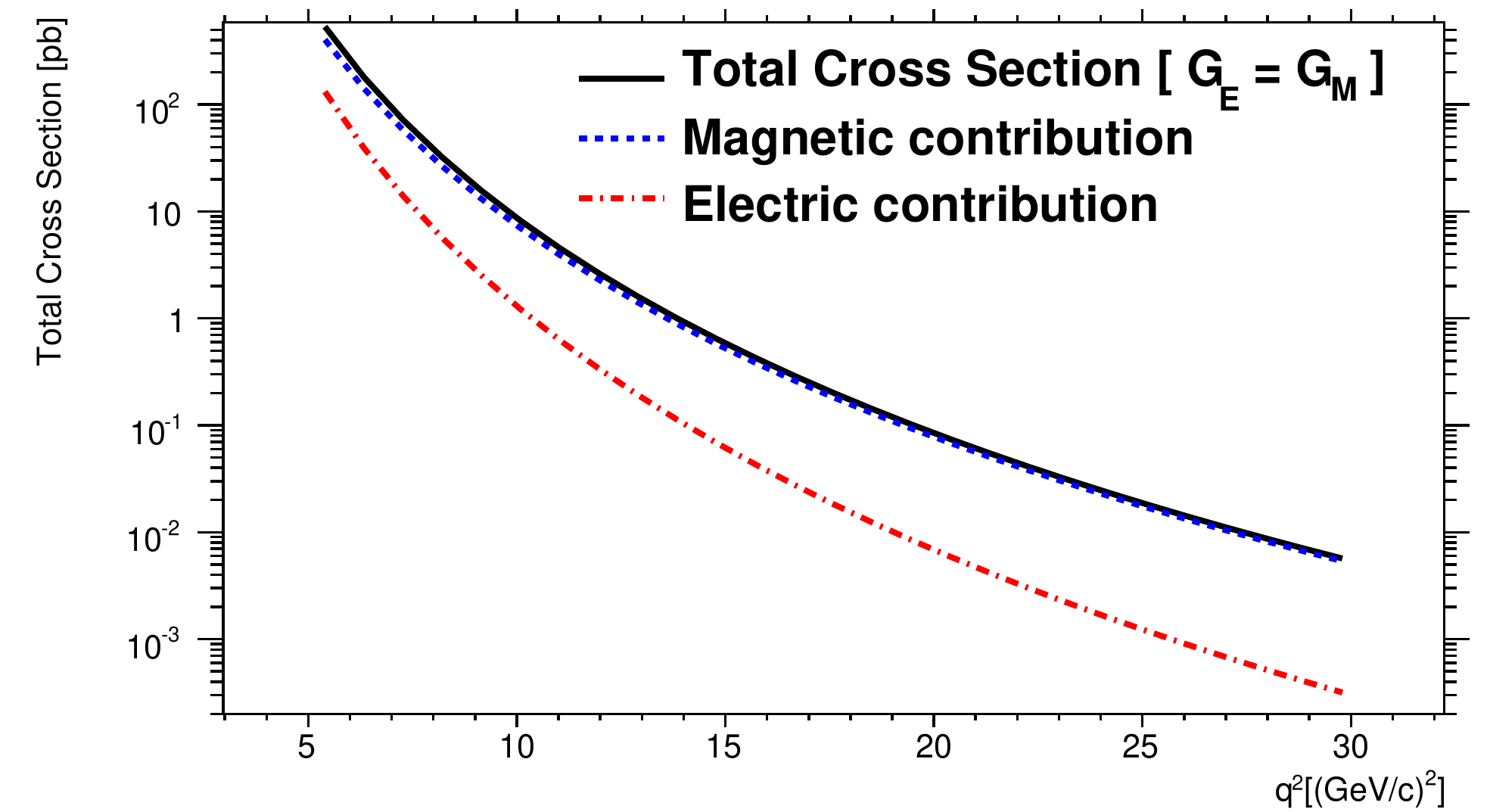}
\caption[\ee cross section as a function of $q^2$.]
{The upper plot shows the modulus of the magnetic time-like proton form factor $|G_M|$
 which has been used in the simulations of the process $ \pbarp \rightarrow \ee $ described here. 
 The lower plot gives the integrated $\pbarp \rightarrow \ee $ cross section 
 as a function of $q^2$.  Also shown is the cross section contribution from magnetic  and electric 
 form factors under the assumption of $|\GE|=|\GM|$.}
\label{fig:phys:cSTot.pdf}
\end{center}
\end{figure}

 The upper plot of \Reffig{fig:phys:cSTot.pdf} shows the magnetic time-like proton form factor $|G_M|$
 which has been used in the signal simulations described here. The lower part of 
 \Reffig{fig:phys:cSTot.pdf} shows the integrated cross section 
 dependence versus $q^2$. The plot shows the individual contributions of $|G_E|$ and $|G_M|$ to the total 
 cross section. One sees that the sensitivity to the electric form factor decreases 
 with increasing $q^2$ due to the kinematic factor $1/\tau$ in front of
 $|G_E|$ ($\tau$ = ${q^{2}} \over {4 m_p^2}$, 
 see Eq. \ref{eqn:angdist:crosssection:timelike}).

\begin{figure*}[htb]
\begin{center}
\includegraphics[angle=90, width=\dwidth]{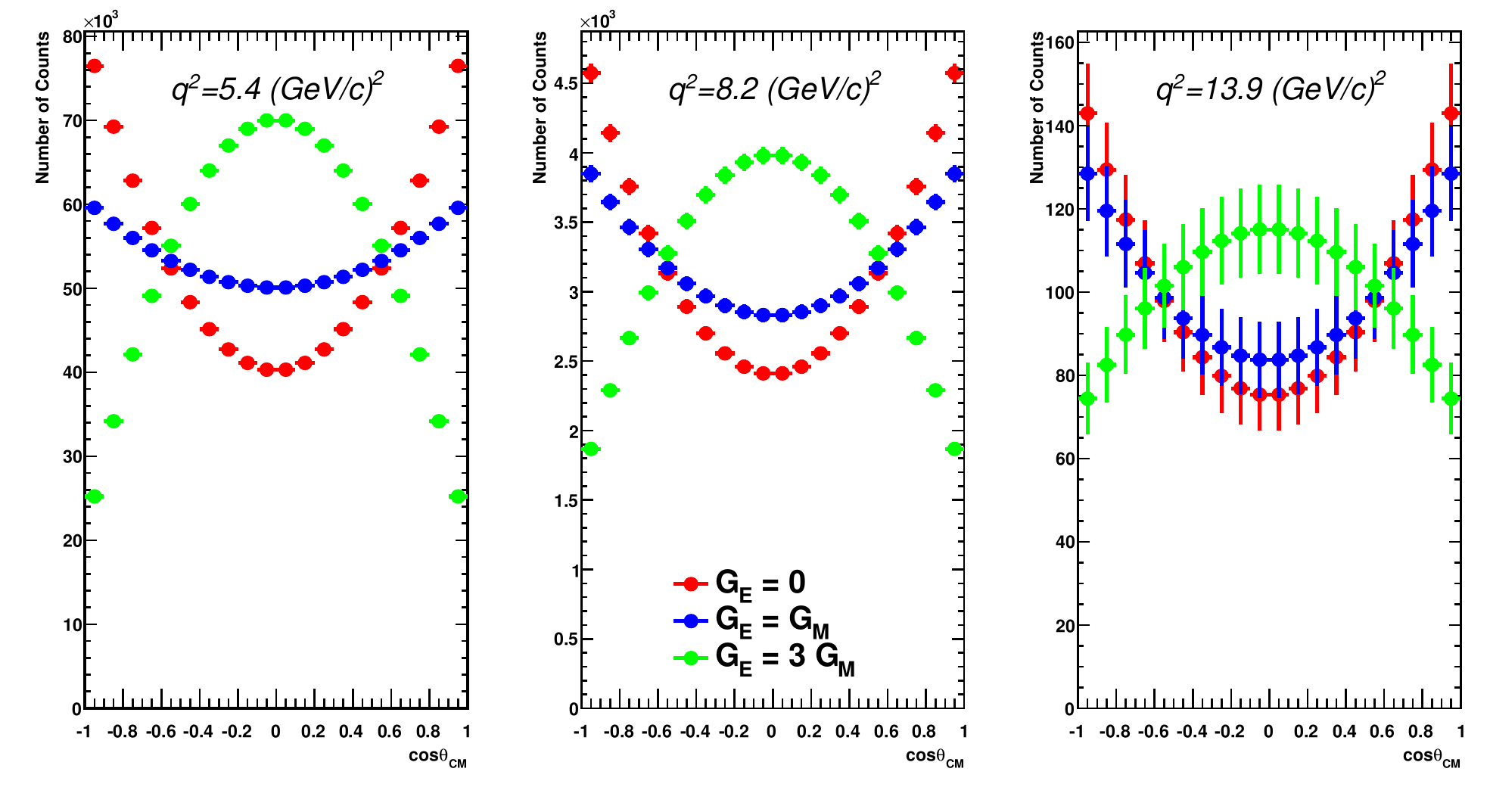}
\caption[\ee cross section as a function of $\theta_{CMS}$.]
{Event generator distributions (events before particle tracking and reconstruction) 
for $\pbarp \rightarrow \ee$ for three different values of $q^2$.
The three different distributions in each plot show the angular distribution in the centre of mass frame (CMS),
for the different models: $|\GE| = 0$ (red), $|\GE| = |\GM|$ (blue) and $|\GE| = 3 |\GM|$ (green). The number 
of expected counts for each model at the same $q^2$ are the same. The error bars
denote the statistical errors where no efficiency correction has been taken into account.}
\label{fig:phys:emff:expectedcountsGEGM}
\end{center}
\end{figure*}

Due to the fact that the world data on time-like form factors have been analysed under the assumption
that $|G_E| = |G_M|$, we can determine and fit the total cross section as shown in \Reffig{fig:phys:cSTot.pdf}.
The knowledge of the ratio $R = |\GE|/|\GM|$ at present is very limited, dispersion theory allows values 
between 0 and 3 for certain $q^2$-values (see \Reffig{fig:nuclstruc:emff:gegm:pacetti}). 
In order to study the sensitivity to the ratio of $|\GE|/|\GM|$ we have used the measured total cross section
to estimate the total number of counts and have simulated the reaction $ \pbarp \rightarrow \ee $ 
with different angular distributions according to $R=0,1,3$ and  the $q^2$-dependence of 
\GM\ according to equation Eq. \ref{fomulagm}.
 \Reffig{fig:phys:emff:expectedcountsGEGM}
shows angular distributions for several assumptions on $|\GE|/|\GM|$.
  \Reftbl{tab:phys:stat_simul_elmff}
 summarises the simulated event numbers reached for the signal simulation ($ \pbarp \rightarrow \ee $
 and $ \pbarp \rightarrow \mumu $)  and the most important background channels
 ($ \pbarp \rightarrow \pip \pim $,  $ \pbarp \rightarrow \piz \piz $).
\begin{table}
\begin{center}
\begin{tabular}{lcccccc}
  \hline\hline
  $q^2$\,[GeV$^2$] & $\ee$            & $\mumu$                     & $\pip \pim$  & $\piz \piz$    \\
    \hline
   $5.4$        & $4 \times 10^6$     &    $4 \times 10^6$        &  -                   &    -            \\
   $7.21$       & $4 \times 10^6$    &    $4 \times 10^6$         &  -                &    -            \\
   $8.21$       & $4 \times 10^6$    &     $4 \times 10^6$        & $10^8$       &   $3 \times 10^6$            \\
   $11.03$      & $4 \times 10^6$     &   $4 \times 10^6$        &  -                 &    -          \\
   $12.9$       & $4 \times 10^6$    &   $4 \times  10^6$        &  $10^8$         &   $3 \times 10^6$             \\
   $16.7$       & $4 \times 10^6$    &     $4 \times 10^6$       & $2. 10^8$     &   $3 \times 10^6$           \\
   $22.3$       & $4 \times  10^6$    &    -                                &  -                &   - 	          \\   
  \hline\hline
\end{tabular}
\caption[Electromagnetic form factor simulation statistics]
{The number of simulated events reached for the simulation of the signal ($\pbarp \rightarrow
\ee $ and $ \pbarp \rightarrow \mumu $) and for the background reactions ($\pbarp \rightarrow
\pip \pim $ and $\pbarp \rightarrow
\piz \piz $). For the signal case we have created $10^6$-events each for the cases
$|\GE| = 0$, $|\GE| = |\GM|$ and $|\GE| = 3 |\GM|$ plus an isotropic $\ee$-distribution for 
acceptance and efficiency corrections. 
We have studied three cases for the final state of  $\piz \piz $: a) both $\piz$ decay into 
2 gammas, b) one of the final state $\piz \piz $ is decaying to 100\% into $\ee \gamma$ (Dalitz decay) and, 
c) both final state $\piz \piz $ do 100 \%  Dalitz decay.}
\label{tab:phys:stat_simul_elmff}
\end{center}
\end{table}

Due to the very low cross sections of the process $ \pbarp \rightarrow \ee $, rejection of the other channels must be very efficient.  The first and most important type of background arises from misidentified hadrons in binary  reactions
($\pip \pim$ and $\Kp \Km$) , 
where both charged hadrons are misidentified as electrons and positrons.
The thickness 
of the electromagnetic calorimeter (about 20 radiation lengths X$_0$) corresponds 
to slightly more the one nuclear interaction length $\lambda_0$, so that we expect 
for about 30\% of the pions nuclear reactions among which charge exchange reactions are 
especially harmfull. Reactions where a $\pip$ converts to a $\piz$ with its subsequent decay into 2 photons
inside the  EMC will deposit energy of the same order as an electron with the same momentum.
This has been completely taken into account in additional simulations used for the determination 
of the PID-likelihood.

The cross sections of the processes with  $\pip \pim$ or $\Kp \Km$ in the final state  
are of the same order of magnitude. However, the kaon mass 
is substantially higher than the pion mass, so rejection through PID and
kinematical  constraints is more efficient for the $\Kp \Km$ channel. 

\begin{figure}[htb]
\begin{center}
\includegraphics[angle=90, width=\swidth]{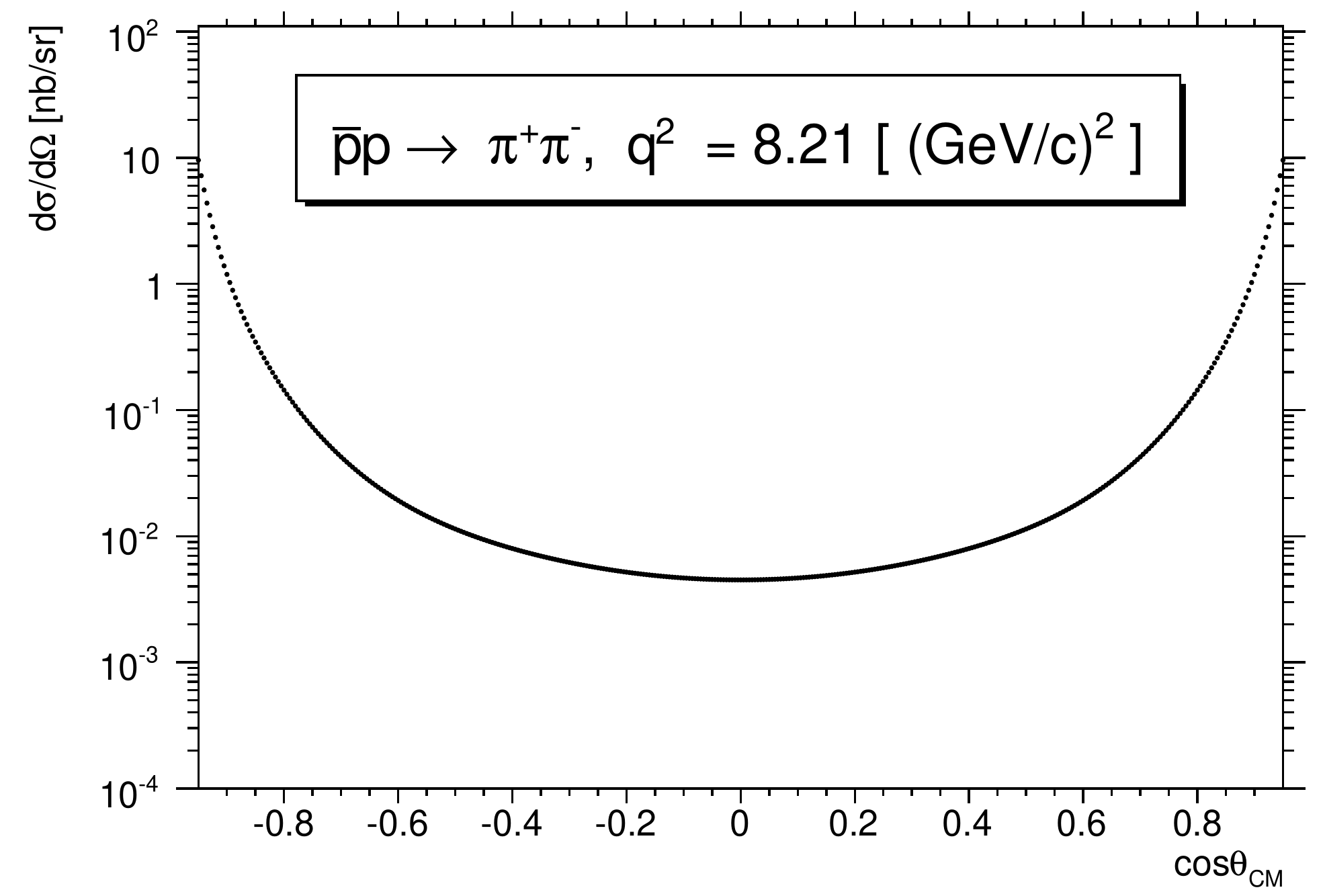}
\caption[Symetrized angular distribution of $\pi^+ \pi^-$ at 8.21.]
{Angular distribution of $\pbarp \rightarrow \pip \pim $ as used in the simulation. The phenomenological fit 
to data has been further symmetrized in $\cos \theta_{CM}$.}
\label{fig:phys:pippim_dSigmadOmega.pdf}
\end{center}
\end{figure}

Consequently, the $\pip\pim$ background channel was simulated as the first step. 
The corresponding angular distributions (see \Reffig{fig:phys:pippim_dSigmadOmega.pdf}) were taken from measured
data and extrapolated where necessary \cite{bib:nuclstruc:emff:exp:pipluspiminus_eisenhandler,
bib:nuclstruc:emff:exp:pipluspiminus_buran,
bib:nuclstruc:emff:exp:pipluspiminus_berglund,
bib:nuclstruc:emff:exp:pipluspiminus_dulude,
bib:Armstrong:1997gv}. 
The ratio of $\pip\pim$ to $\ee$ cross sections, which varies from $10^5$ at $|\cos\theta _{CM}| = 0$ up to 
$3\cdot10^6$ at $|\cos\theta _{CM}| = 0.8$, is then properly taken into account
in the simulation.
 
The second  kind of background arises from exit channels of the type $\ee X$, where $X$ might be a combination of mesons, photons, and lepton pairs. In addition to the direct production from the $\pbarp$ interactions, such final states can also arise when a produced gamma materialises before reaching the tracking detectors. Due to the high resolution of the tracking system, channels, where the missing mass is of the order or larger than the $\pi^0$ mass, are rejected very efficiently. 
Considering the involved cross sections, the main problem therefore arises from 
$\ee$ pairs, each one coming from the decay of one of the two $\pi^0$s produced in  $\pbarp \rightarrow \pi^0 \pi^0$ reaction. This can happen, either from Dalitz decay of $\pi^0$ or after photon conversion following the direct decay. 
The reaction $\pbarp \rightarrow \pi^0 \pi^0$ has therefore been simulated, using the same cross sections and angular distributions as for the $\pbarp \rightarrow \pip \pim $.
Three-body final state background reactions involving 2 opposite charged hadrons and a neutral massive 
particle ($\piz$ or heavier meson) are much easier to
separate, since kinematical considerations can then provide 
additional constraints which can be used to cut very efficiently the corresponding background.
They have not yet been simulated. 

The background events from $\pip\pim$ and $\piz\piz$ were analysed under the hypothesis 
of having an $\ee$-pair. Analogously, the 
$\pip\pim$ events have been reanalysed under the hypothesis of having a $\mumu$-pair. 
The same analysis cuts and kinematical fit constraints have then been applied to the $\ee$ signal 
sample ($\mumu$- sample respectively)  in order to create the signal distributions. 
Different  cuts on the PID were used as described in 
\Refsec{sec:soft:recochargedPID},  corresponding to different thresholds on the global likelihood
(see \Reftbl{tab:soft:emc_pid_cuts}).

Special attention has been paid to the $\piz \piz$ channel. The photons from 
the main $\piz$-decay can eventually convert to $\ee$-pairs in the \PANDA detector, 
notably in the beam pipe before the tracking detectors. Those 
$\ee$-pairs fulfil all PID cuts but can very efficiently be suppressed by the kinematical constraints.
In addition, we studied the case of one (two respectively) $\piz$ decaying via the Dalitz-channel 
($\piz \rightarrow \ee \gamma$). The $\ee$ from Dalitz decay again fulfil 
all PID cuts for electrons, but again, can be efficiently rejected due to 
kinematical constraints. One should note, that the branching ratio for 
Dalitz decay is of order 1\%, i.e. the probability that both $\piz$ decay 
via the Dalitz process is about 10$^{-4}$.

Extensive simulations have been made for the $\ee$ channel and are discussed below
(see \Reftbl{tab:phys:stat_simul_elmff}).
Measuring the $\mumu$ channel could be a very interesting and complementary channel too
which we studied at different $q^2$-values. 
We can not  conclude here on the case of muons, more simulations are needed.

%

%% file: phys/nucleonstructure/elmff/phys_elmff_simulations_background.tex
%
\subsubsection{Background Analysis}

\paragraph*{Separating $\boldsymbol{\ee}$ from $\boldsymbol{\pi^+ \pi^-}$} 
 In the analysis of the process $\pbarp \rightarrow \ee$, the most severe background 
 comes from two pion final states, namely $\pbarp \rightarrow \pip\pim$ and $\pbarp \rightarrow \piz \piz $.
 The large ratio of the cross section for pion final states versus lepton final states
 (of order $10^6$) requires a large event sample of order $10^8$ in order 
 to show that the lepton signal pollution by pion final states is below 1~\%.
 For this reason, we have simulated the background processes for $\pip \pim$ final 
 states with very high statistics only at 3 incident momenta, where we have chosen 
 a low momentum, a medium momentum value and the highest momentum value,
 where we can access form factors:   
 3.3\,\gevc ($q^2 = 8.21\,\gev^2$),
 5.84\,\gevc  ($q^2 = 12.9\,\gev^2$), and
 7.86\,\gevc  ($q^2 = 16.7\,\gev^2$) (see \Reftbl{tab:phys:stat_simul_elmff}) .
 The effect of different cuts in the global PID are shown in
 \Reftbl{tab:phys:background_elmff_cutleft}. For the $\pbarp \rightarrow \ee$-signal processes, 
 particles at CMS-angels of $|\cos\theta _{CM}| > 0.8$, one of the electrons hits the 
 forward spectrometer, which is less powerful concerning the $\ee$ reconstruction. 
 Since we don't use this region of CMS-angle for our simulations of the signal channel, 
 $\pip \pim$ events were simulated only in a
 restricted range ([-0.8,0.8]) of $\cos\theta _{CM}$ values. Outside this
 interval, there are no data on measured $\pbarp \rightarrow \pip \pim$ cross sections and the 
 extrapolation by phenomenological models introduces large uncertainties
 due to the steep rise of the cross section in this $\cos\theta _{CM}$ region. 
 However, as will be shown in the next subsection, the 
 acceptance for the signal is very low close to $|\cos\theta _{CM}|
 =1$ and therefore the uncertainty in the background cross section in 
the interval outside ([-0.8,0.8]) of $\cos\theta _{CM}$ is then not relevant.

\begin{table}[htb]
\begin{center}
\begin{tabular}{lccc}
  \hline\hline
 $q^2 ((\gevc)^2)$    &   8.2         &  12.9        &   16.7       \\
    \hline
    no cut            & $10^8$        & $10^8$       &  $2.\,10^8$   \\
    VL                & 46.8          &  90\,k         &  140\,k       \\
    L                 & 425           &  1.2\,k        &  3\,k         \\
    T                 & 31            &  70        &  120          \\
    VT                & 2             &  5         &  6            \\
    $CL^*$            & $8.\,10^5$    &  $1.\,10^6$  &  $2.5\,10^6$  \\
  \hline\hline
\end{tabular}
\caption[Rejection efficiency for pions]
{Number of $\pip \pim$ events, misidentified as $\ee$, left after 
the different cuts applied for 3 different $q^2$ values. The $CL^*$ cut requires 
2 kinematical fit conditions to be fulfilled, namely \CL $\geq 0.001$ and  $CL_{\ee}\geq 10 CL_{\pip\pim}$.
Please note that the suppression factor of about 100 from the $CL^*$ cut is independent from the cut in the 
PID-probability. By requiring both, PID-cut and $CL^*$ cut, we expect a total suppression factor 
of order $10^{9}$ to $10^{10}$}
\label{tab:phys:background_elmff_cutleft}
\end{center}
\end{table}

\begin{figure}[htb]
\begin{center}
\includegraphics[angle=0, width=\swidth]{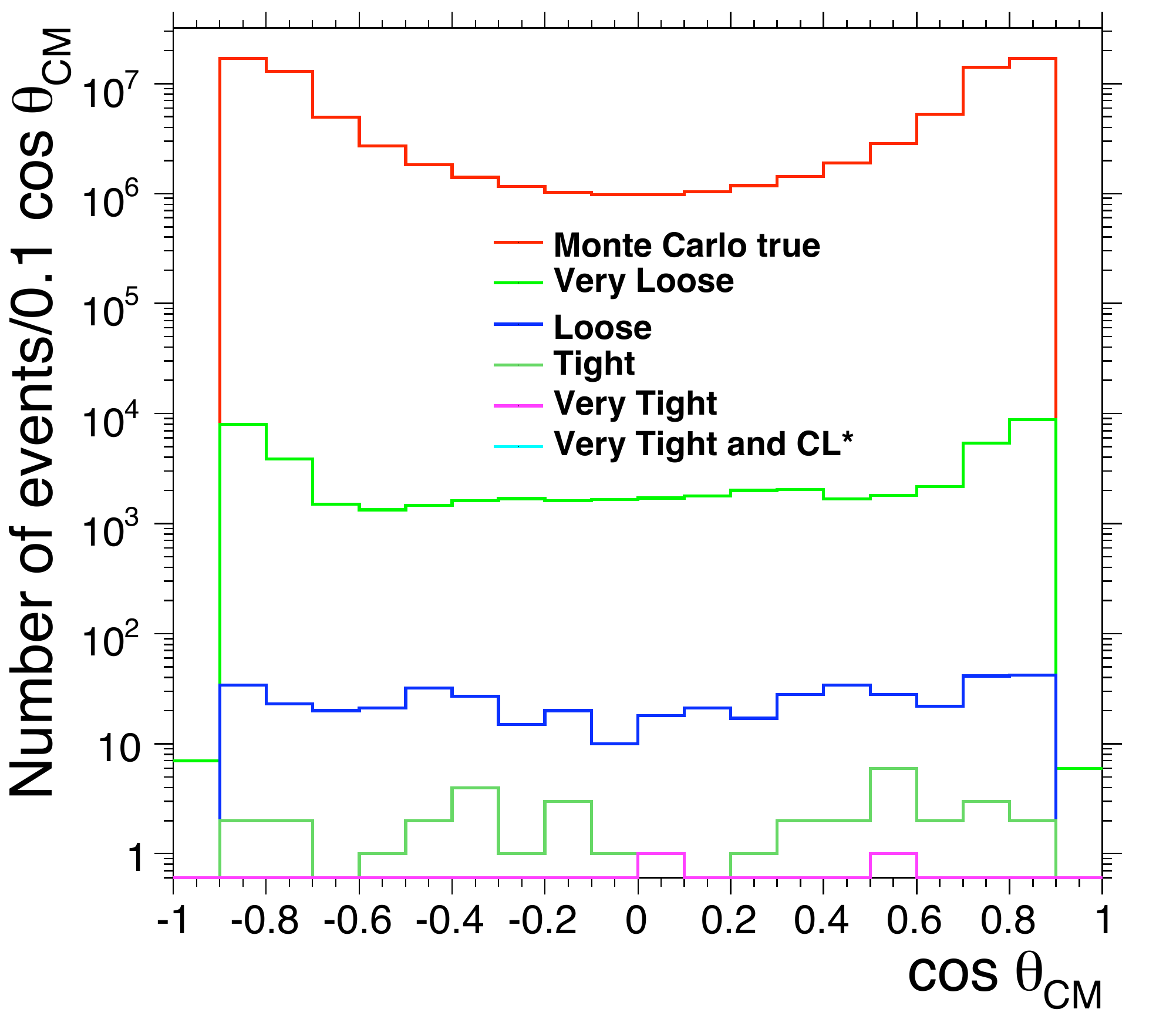}
\caption[Angular distribution of misidentified pions.]
{Centre of mass distribution of pions misidentified as electrons after the different PID cuts at $q^2 = 8.21 (\gevc)^2$.
The $CL^*$ cut provides an independent rejection factor of the order of 100, thus rejecting very efficiently
the 2 remaining events after the VT cut.  }
\label{fig:phys:PionsAsElectrons.pdf}
\end{center}
\end{figure}

\Reffig{fig:phys:PionsAsElectrons.pdf} displays the effect of the different PID cuts.
Only the \vtig cut, corresponding to a global likelihood greater than $99.8\percent$ (and at 
least a $10 \percent$ minimum likelihood on each subdetector)  on both positive
 and negative charged candidates can be used. Combining one \vtig with one \tig candidate is by 
 far not efficient enough to suppress the background.
 Additionally the Confidence Level ($CL^*$) cut resulting from the kinematic refit was used to 
 efficiently reduce the background. This $CL^*$ cut requires 
 2 conditions to be fulfilled, namely \CL $\geq 0.001$ and  $CL_{\ee}\geq 10 CL_{\pip\pim}$. 
 It gives an independent rejection factor of the order of 100, only slightly depending on energy.
 The contamination of signal $\ee$ events by background $\pip\pim$ ones is given for all 3 
 simulated $q^2$ values in  
 \Reftbl{tab:phys:background_elmff_cutleft}. 
 Using \vtig cuts together with the  conditions $\CL^*$ then allows to reach an overall
 rejection factor greater than $10^9$ up to $10^{10}$. Within these conditions, the contamination of 
 the $\ee$ signal by $\pip\pim$ will be well below  $1 \%$ and will therefore not affect the
 precision for extracting the magnetic and the electric proton form factors.

\begin{figure*}[htb]
\begin{center}
\includegraphics[angle=0, width=\dwidth]{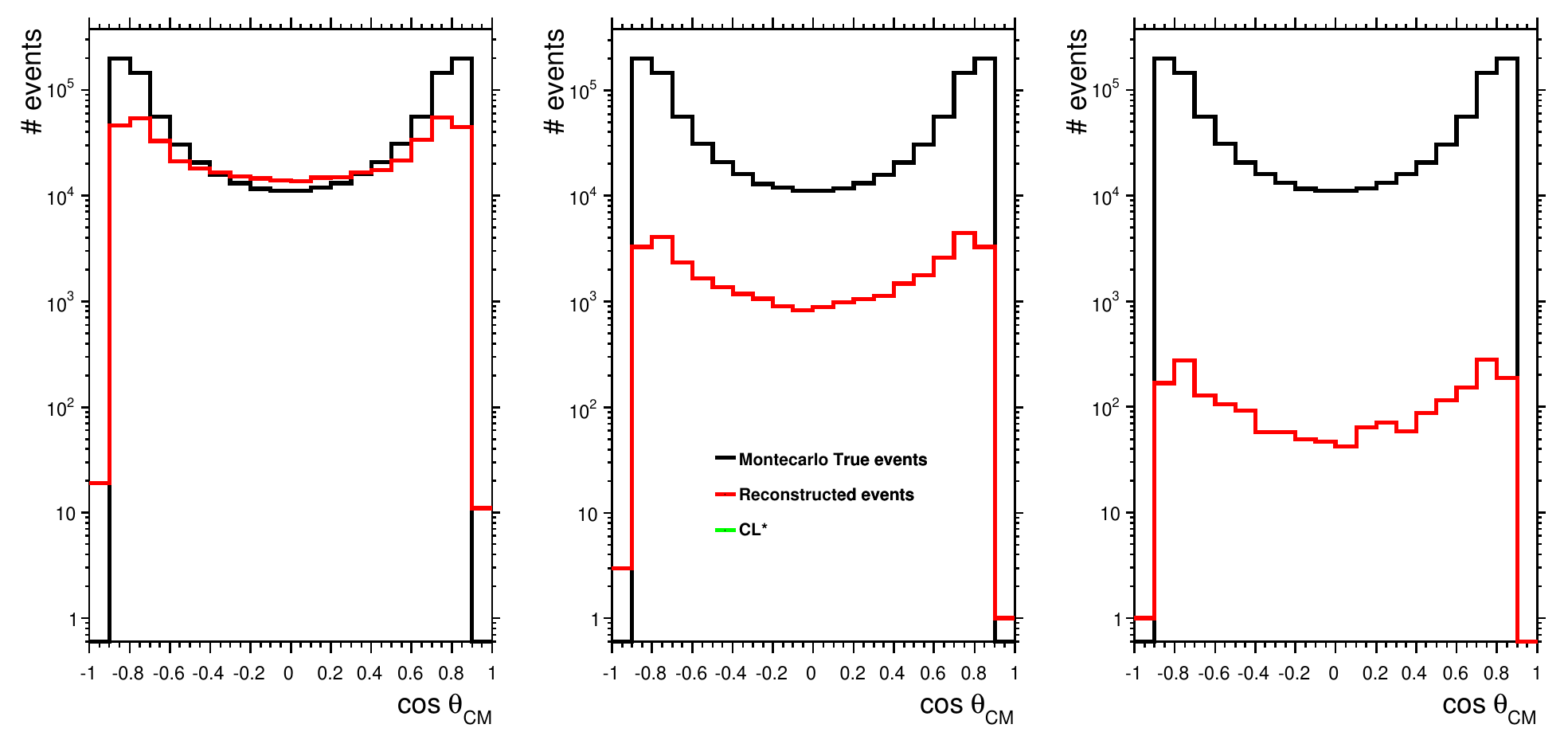}
\caption[Distribution of misidentified  pi0]
{Centre of mass distribution of events after different cuts for the 3 different channels (see text). 
The top curves are the $\Pgpz \Pgpz$ angular distributions
whereas the other curves display the angular distribution of $\ee$. The green line of remaining background after the cuts in kinematical 
fit constraints is not visible since there are zero background events left.}
\label{fig:phys:PiOPiOAsElectrons.pdf}
\end{center}
\end{figure*}

The contamination of signal events by background $\Pgpz\Pgpz$ ones is shown in the figure
 \Reffig{fig:phys:PiOPiOAsElectrons.pdf}.
In this case, electrons are indeed detected and correctly identified as such. They originate from
3 different channels, namely the double $\Pgpz$ Dalitz decay, one $\Pgpz$ Dalitz decay
 associated to a photon conversion from the other $\Pgpz$, or 2 photon conversions from the 2 $\Pgpz$.
 The first one scales as the $\Pgpz$ Dalitz decay branching ratio $\Gamma _{\gamma e^+e^-}$ squared,
  The second one scales as $\Gamma _{\gamma e^+e^-}$ times the conversion probability in the 
  detector material. 
  the third one scales as the conversion probability squared.
  In all cases the final state is a 6-body one ($2e^+,
 2e^-,2\gamma$). From the first case, shown on the left part of the figure, one can deduce
  that kinematical constraints and \PANDA 
 hermeticity provide a rejection of at least a factor $10^4$. This factor applies as well to the case 
 with 2 photon conversion. As a result, we can say that the $\Pgpz\Pgpz$ channel can be rejected by a factor
 close to $10^8$.
 
\paragraph*{Separating $\boldsymbol{\mu^+ \mu^-}$ from $\boldsymbol{\pi^+ \pi^-}$} 

The same full-scale simulations were used to analyse the rejection of $\pi^+ \pi^-$ when using the PID cuts
adapted to the selection of $\mu^+ \mu^-$ pairs.
There are two major effects that limit our ability to discriminate the $\mumu$ signal from the 
background $\pi^+ \pi^-$: the in-flight decay of pions and the unsufficient iron yoke thickness. In the former case
muons from decaying pions in the close vicinity of the target behave like muons from $\ppbar$-annihilation.
In the latter case, the background is due to pions which did not interact strongly with iron and are
detected by the muon counters.  
Since pion and muon masses are not very different, particle identification
through specific energy loss dE/dx or Cerenkov radiation is of very little help. 
The calorimeter only provides a rejection factor for pions, but is however of the order of 2/3.
Preliminary results obtained with simulations done in the present framework 
show that the extraction of the $\mumu$ channel will be difficult.


%% file: phys/nucleonstructure/elmff/phys_elmff_simulations_signal.tex
%
\subsubsection{Signal Analysis}
\paragraph*{$\boldsymbol{\Pep \Pem}$ Channel.}

\begin{figure}[htb]
\begin{center}
\includegraphics[angle=90, width=\swidth]{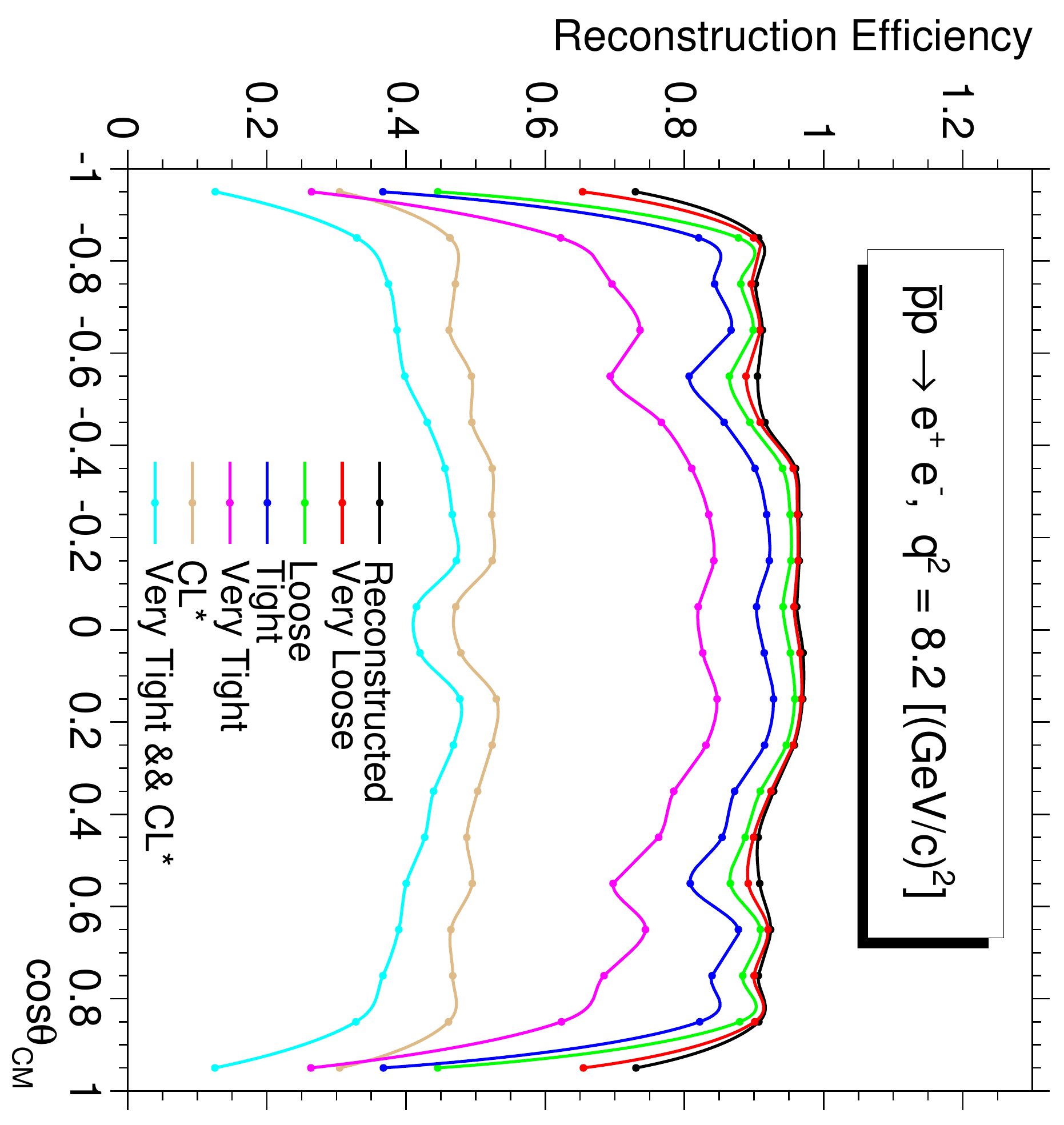}
\caption[\ee reconstruction efficiency at $q^2 = 8.21\, (\gevc)^2$]
{Reconstruction efficiency as a function of $cos\theta _{CM}$ at 
one example of beam energy corresponding to $q^2 = 8.21\, (\gevc)^2$
for the different different PID cuts and kinematical constraints.
The values quoted are averaged over a bin interval of 0.1, but still show 
the drop in efficiency at the transitions between barrel and 
forward end cap detector parts.}
\label{fig:phys:Eff_epem_detail_3300.pdf}
\end{center}
\end{figure}

\begin{figure}[htb]
\begin{center}
\includegraphics[angle=90, width=\swidth]{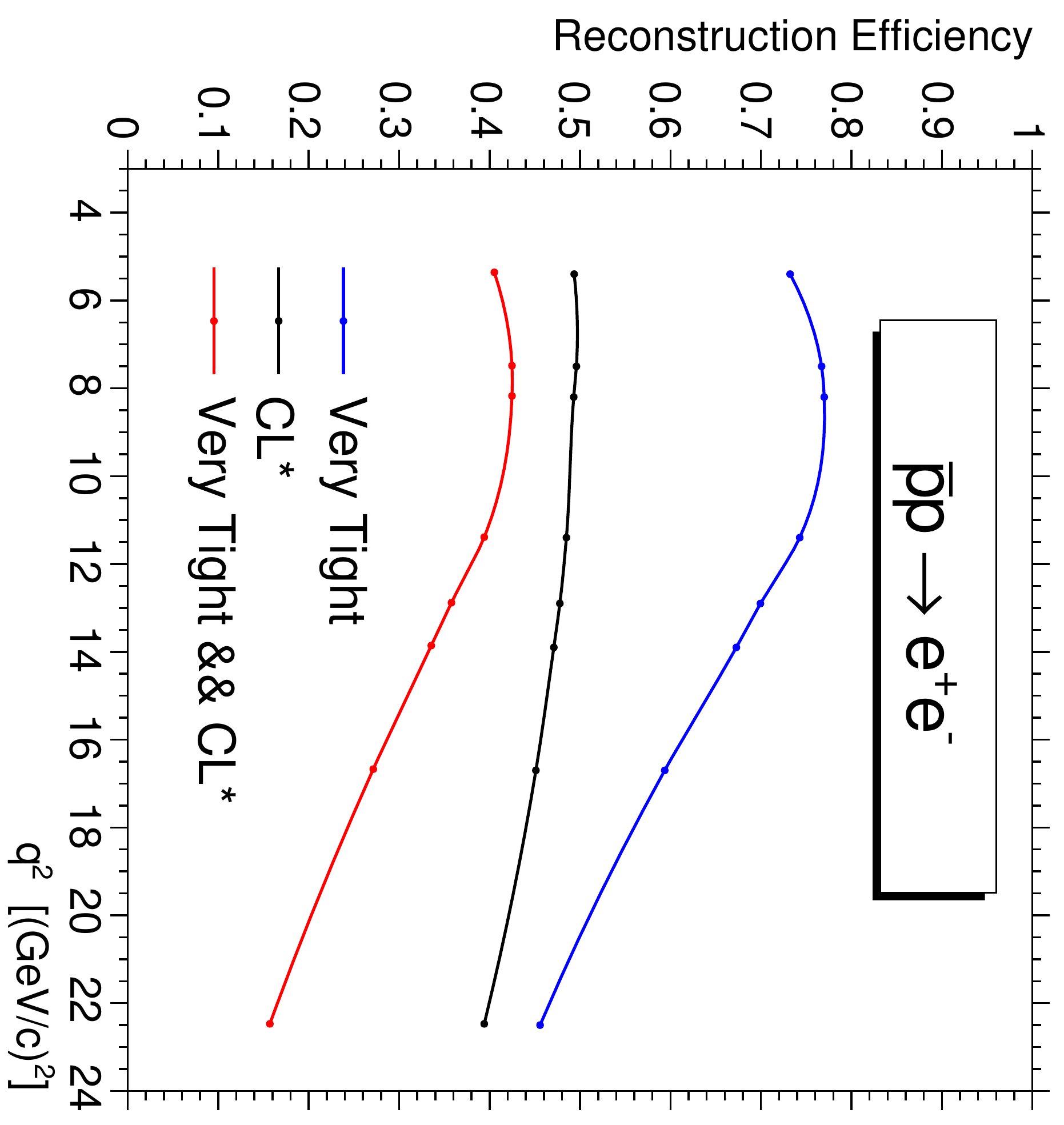}
\caption[Integrated \ee reconstruction efficiency versus $q^2$.]
{Overall integrated reconstruction efficiency as a function of $q^2$
applying kinematical constraints and PID cuts, integrated over 
an interval in $\cos \theta_{CM}$ of  [-0.8,0.8] (red curve). The 
effect of applying the kinematial constraints only ($CL^*$, black curve) 
and PID cuts only (blue curve) are shown separately .}
\label{fig:phys:Eff_vs_q2.pdf}
\end{center}
\end{figure}

\begin{figure}[htb]
\begin{center}
\includegraphics[angle=0, width=\swidth]{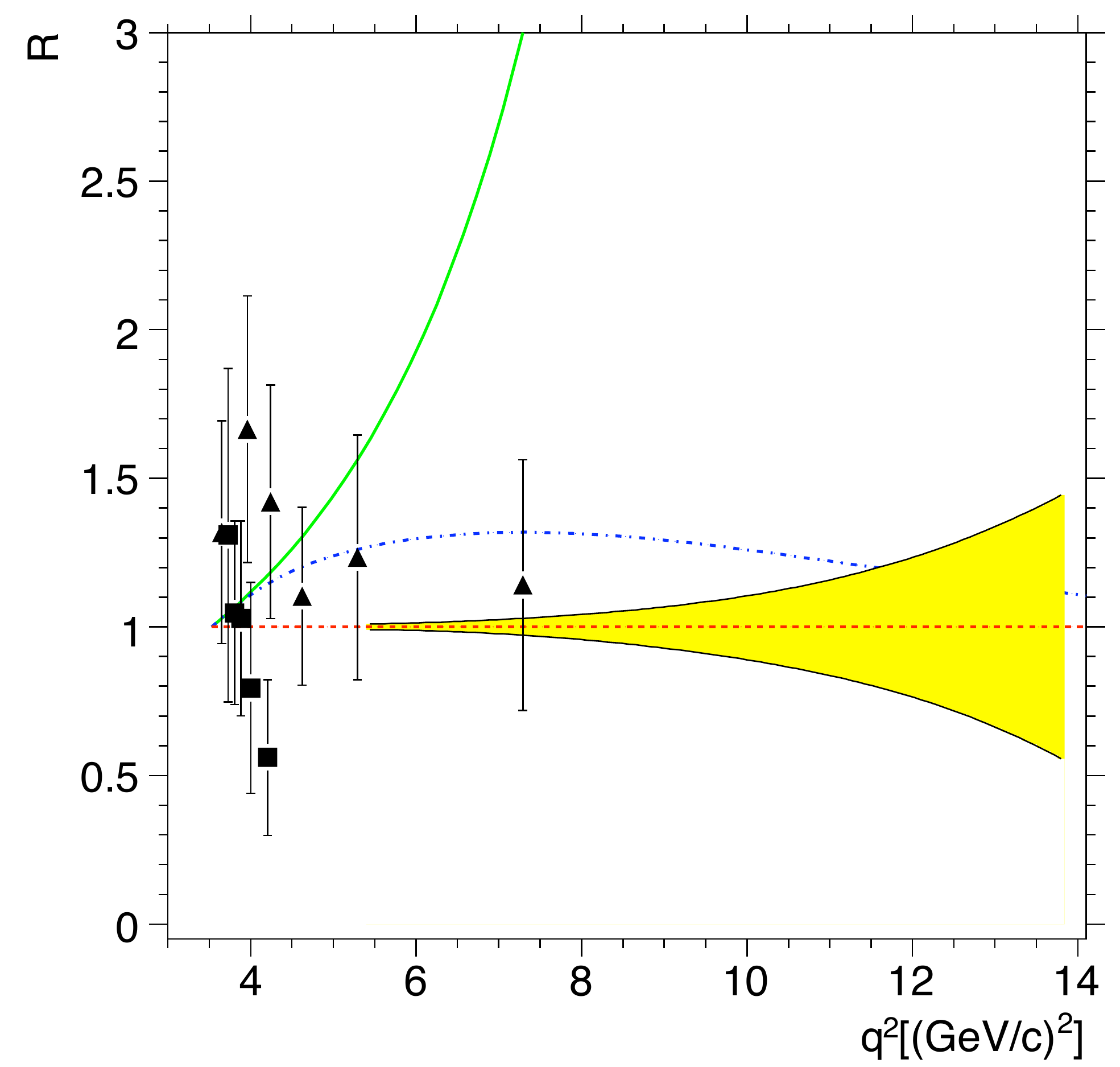}
\caption[precision on ratio GE/GM as a function of $q^2$]
{The expected error of the ratio $R = |\GE|/|\GM|$ is given as a function of $q^2$.
The yellow band represents the errors from the fits to the efficiency 
corrected \ee distributions at the 8 $q^2$ values and is plotted up to $q^2 = 14 (\mathrm{GeV}/c)^2$.
The results for the cases $|G_E|=0$ and $|G_E|=3|G_M|$ look similar.
The data points from \INST{PS170} and \INST{BABAR} are shown as well as three theoretical 
expectations for $R = |\GE|/|\GM|$. In the low $q^2$ region, \PANDA will be able to improve the error bars by an order
of magnitude.}
\label{fig:phys:GE_GM_precision.pdf}
\end{center}
\end{figure}

Signal events for the processes $\pbarp \rightarrow \ee$ and $\pbarp \rightarrow \mumu$
have been simulated. The electromagnetic form factors span the regime 
from nonperturbative QCD to perturbative QCD at higher energies that is why we 
simulated events at 8 different energies. In the simulations presented here, 
we concentrate on the extraction of the ratio $|G_E|/|G_M|$. Due to the kinematical factor 
$1/\tau$ in front of $|G_E|$, the sensitivity to $|G_E|$ from the cross section measurement 
decreases with rising beam momentum (see \Reffig{fig:phys:cSTot.pdf}).
\Reftbl{tab:phys:epem_statistics} gives the expected number of events 
for different antiproton beam momenta. For a given $q^2$ we used the same 
total number of expected events for the different assumptions on the 
ratio $|G_E|/|G_M|$.
\begin{table}[hbt]
\begin{center}
\begin{tabular}{rrr}
  \hline\hline
          $p$  \hspace{5 mm}   & \hspace{5 mm} $q^2$          \hspace{5 mm}  & \hspace{5 mm} number of    \\
	     \gevc \hspace{5 mm}     & \hspace{5 mm} $(\gevc)^2$ \hspace{5 mm} & \hspace{5 mm} events    \\
    \hline
             $1.70$  \hspace{5 mm}  & 5.40  \hspace{5 mm} & 1071\,k     \\
             $2.87$  \hspace{5 mm}  & 7.27  \hspace{5 mm} & 124\,k   \\
             $3.30$  \hspace{5 mm}  & 8.21  \hspace{5 mm} & 64\,k  \\
             $4.85$  \hspace{5 mm}  & 11.03 \hspace{5 mm} & 9094  \\
             $5.86$  \hspace{5 mm}  & 12.90 \hspace{5 mm} & 3198  \\
             $6.37$  \hspace{5 mm}  & 13.84 \hspace{5 mm} & 2003  \\
             $7.88$  \hspace{5 mm}  & 16.66 \hspace{5 mm} & 580  \\
             $10.9$  \hspace{5 mm}  & 22.29 \hspace{5 mm} & 82   \\ 
  \hline\hline
\end{tabular}
\caption[Expected signal counting rates for \ee]{Expected counting rates  for \ee corresponding 
to an integrated luminosity of $2 fb^{-1}$ ( 10$^7$s corresponding to about 
4 months at $L=2\cdot 10^{32} cm^{-2}s^{-1}$)}
\label{tab:phys:epem_statistics}
\end{center}
\end{table}
We have applied the same cuts concerning particle identification and 
kinematical constraints as for the background channels described in
the previous paragraph. \Reffig{fig:phys:Eff_epem_detail_3300.pdf} 
indicates the reconstruction efficiency as a function of $\cos \theta _{CM}$ 
for different cuts for $q^2 = 8.21\, (\gevc)^2$. For the \vtig cut, one
can note the important drop corresponding to the loss of the PID capabilities. 
Holes or drops in the spectrum
correspond to regions where detector transitions occur,{\it e.g.} transition between the barrel
part of the calorimeter and the forward end cap at $\theta _{lab} = 22\degrees$ ($|\cos\theta_{CM}|=
0.33$) or loss of STT dE/dx identification at $\theta _{lab} = 14\degrees$ ($|\cos\theta_{CM}|=
0.67$). 
\Reffig{fig:phys:Eff_vs_q2.pdf} shows the total reconstruction efficiency, integrated over 
an interval in $\cos \theta_{CM}$ of  [-0.8,0.8]. The reconstruction efficiency 
decreases for large $q^2$ values. In addition to the drop in reconstruction efficiency, 
the cross section decreases with rising $q^2$ and the sensitivity for $|G_E|$ 
decreases too. For a $q^2$ above 14~(\gevc)$^2$ a measurement 
of the total cross section will still be possible with unprecedented 
accuracy, yielding a determination of $|G_M|$. 
\begin{figure}[htb]
\begin{center}
\includegraphics[width=\swidth]
{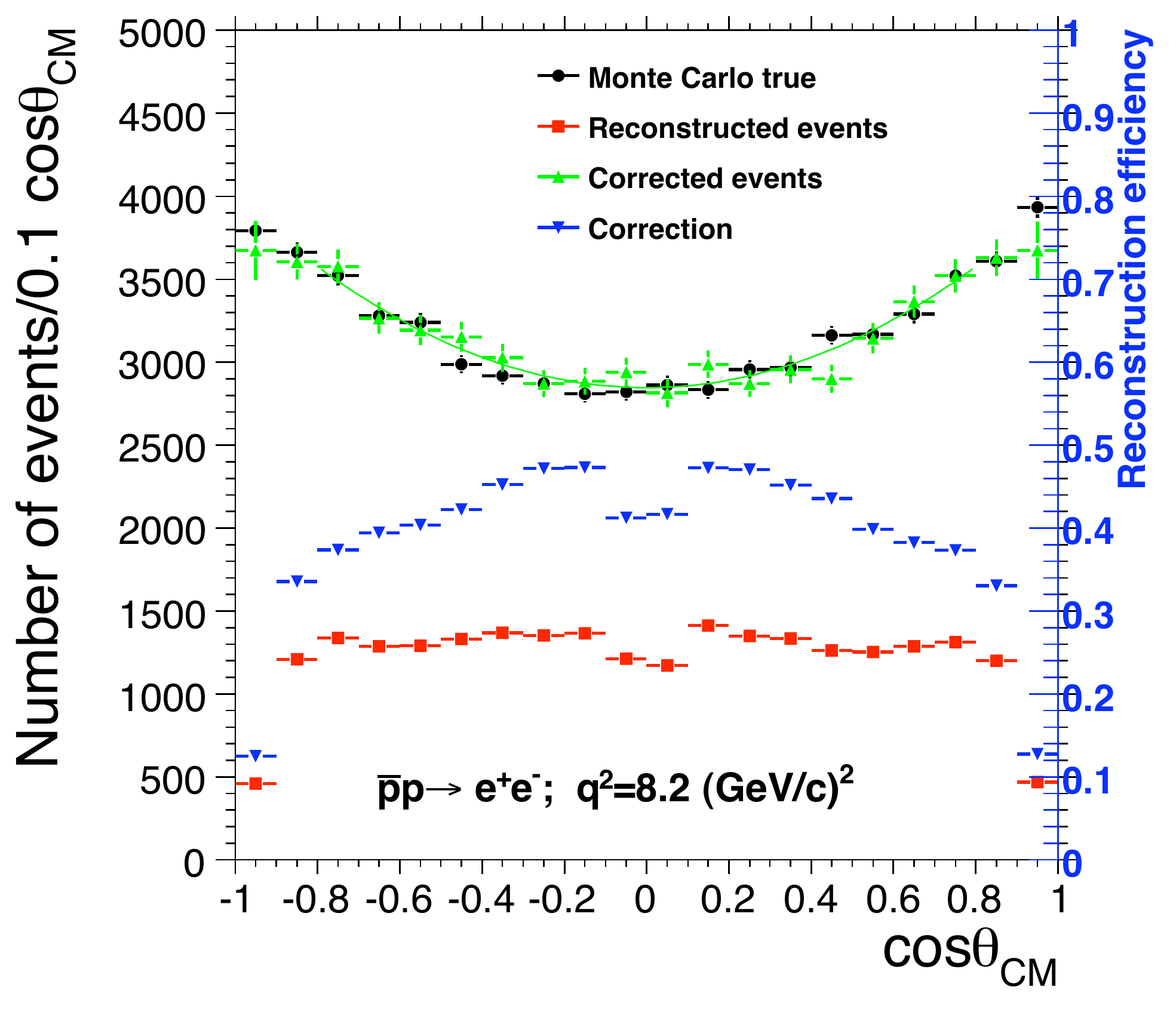}
\caption[Reconstructed angular \ee distribution at $q^2 = 8.21 (\gevc)^2$]
{The figure shows the angular distribution of \ee pairs in the centre of mass system. 
The black symbols denote the distribution at the output of the event generator (Monte Carlo).
The red symbols denote the distribution of the reconstructed \ee-pairs. 
An acceptance and reconstruction efficiency correction (blue points) has been determined from an
\ee data sample which is isotropically distributed in the CM (right scale). The 
angular distribution of the reconstructed and efficiency corrected \ee pairs in the centre of mass system 
at $q^2 = 8.21 (\gevc)^2$ (green points). The corrected angular distribution,
fits nicely to the Monte Carlo one, the ratio $R=|G_E|/|G_M|$ is extracted from
a fit to the angular distribution. Only statistical errors have been taken into account. 
}
\label{fig:phys:Signal_rec_ang_dist_Iso.pdf}
\end{center}
\end{figure}
An \ee\ angular distribution is shown for $q^2 = 8.21 (\gevc)^2$
for one model assumption ($|G_E|=|G_M|$) in \Reffig{fig:phys:Signal_rec_ang_dist_Iso.pdf}.
The event generator output is shown together with the reconstructed event distribution. 
We apply acceptance and reconstruction efficiency corrections, determined 
from an independent isotropical \ee distribution. We have simulated the \ee distributions 
for all $q^2$-values given in table \Reftbl{tab:phys:stat_simul_elmff} and for all 
three assumptions on $|G_E|/|G_M|$ plus isotropic case. 
Reconstruction efficiency corrections have been determined for every $q^2$ 
from the isotropic \ee distribution. We fitted every resulting 
\ee-distribution with a linear 2 parameter function in order to determine the error on $R$. 
\Reffig{fig:phys:GE_GM_precision.pdf} summarises the
results for the case $|G_E|=|G_M|$. The yellow band represents a 
parametrisation of the errors of the fits on R. The results for the cases $|G_E|=0$ and $|G_E|=3|G_M|$
are similar. The different curves correspond to theory estimates \cite{bib:nuclstruc:emff:theory:TomasiGustafsson:2005kc}.
It shows that the separation of the 2 Form Factors can be made almost up to $14\,\gev^2$. 
In the low $q^2$ region, \PANDA will be able to improve the error bars by an order
of magnitude compared to the most recent \INST{BaBar} data, and will consequently severely constrain the
theoretical predictions which today display quite a large dispersion.

\paragraph*{$\boldsymbol{\mu^+ \mu^-}$ Channel.}

Measuring the $\mumu$ channel could be a very interesting and complementary channel to $\ee$.
It was studied at different $q^2$ values and corresponding efficiencies were determined.
Their $\cos\theta_{CM}$ behaviour is different from the $\ee$ channel, since they are strongly
dependent on the geometry of the muon counters and on the thickness of the iron yoke.
At high $q^2$, the average efficiency is only slightly smaller than for $\ee$, but drops dramatically
at low $q^2$, reaching values below $10 \percent$ over an extended $cos\theta_{CM}$ interval
at $q^2=5.4 (\gevc)^2$.

%% file: phys/nucleonstructure/elmff/phys_elmff_outlook.tex
\subsubsection{Conclusion}
\label{label_emff_conclusion}

In extended simulations, we have shown, that it is possible to reject 
the most important background process $\pbarp \rightarrow  \pip \pim$
with a rejection factor of at least $10^9$. The resulting contamination 
of the signal process $\pbarp \rightarrow \ee$ data sample is expected to be 
well below 1~\% and can therefore be safely neglected.

\begin{figure}[h]
\begin{center}
  \includegraphics[width=\swidth,angle=0]{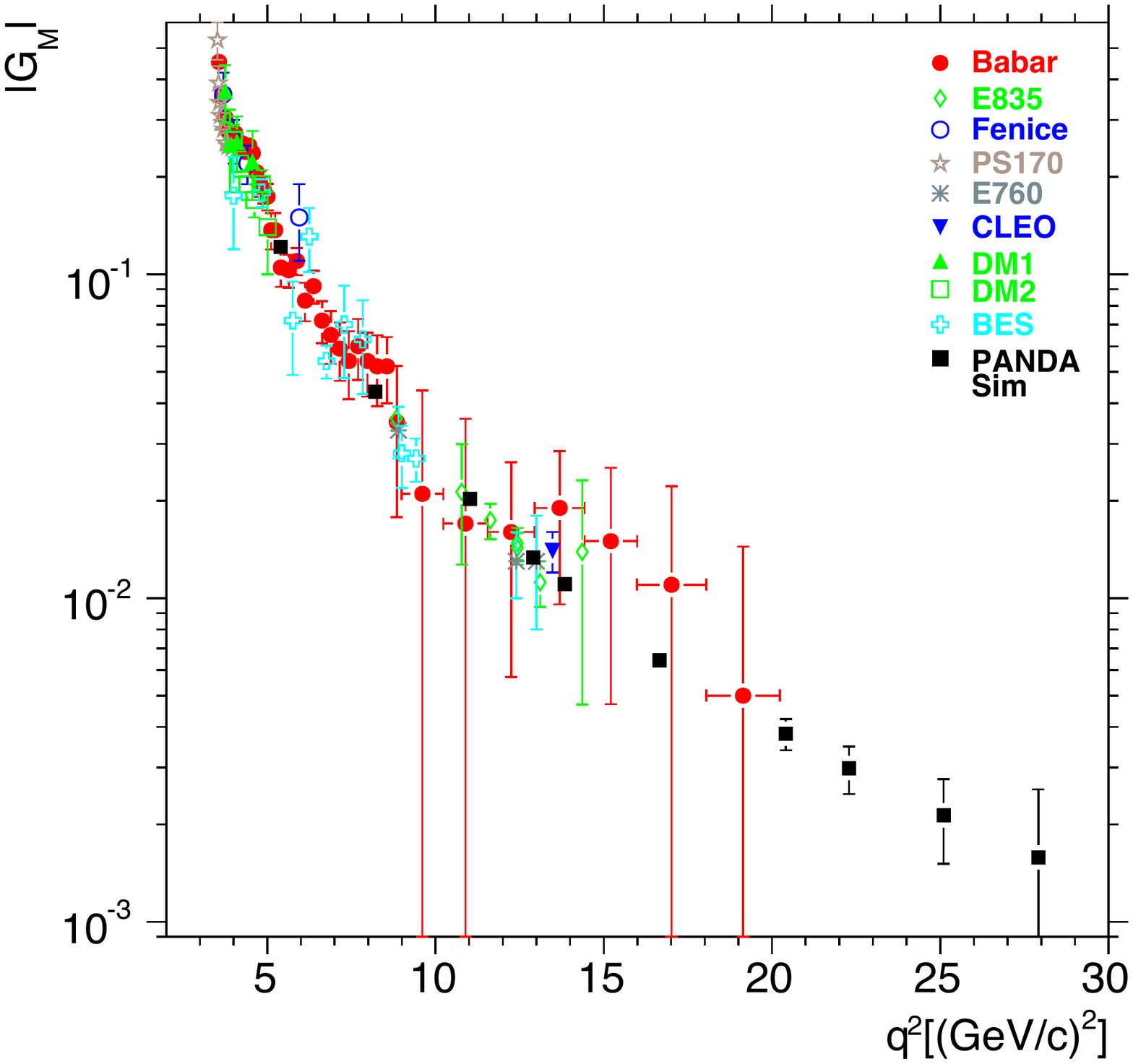}
  \caption[\PANDA \ results]{ Present world data on $ |\GM|$ (extracted using the hypothesis 
          $R = |\GE|/|\GM| = 1$) are shown together with the 
          expected accuracy by measuring $\pbarp \rightarrow \ee$ with the \PANDA 
          experiment at FAIR. Each point corresponds to an integrated luminosity of 2 fb$^{-1}$.
         \label{fig:nuclstruc:emff:worlddata:gm:PANDAsimul}
         }
\end{center}
\end{figure}

Our studies have been performed without explicit assumptions on the 
accuracy of a luminosity measurement. For this we can extract the ratio
$R = |\GE|/|\GM|$ with unprecedented precision up to 14~(\gevc)$^2$.
A factor 10 improved experimental precision is expected with respect to 
present world data. \Reffig{fig:nuclstruc:emff:worlddata:gm:PANDAsimul}
shows the expected accuracy on the \PANDA measurements in comparison 
with the world data, under the assumption that $R = 1$.
With a precise 
luminosity measurement, we can not only determine the ratio $R$ but also 
the absolute and differential cross section up to 22~(\gevc)$^2$.
Moreover separate determination of $|\GE|$ and $|\GM|$ can be made
below 14~(\gevc)$^2$.

In contrast to the \ee case, the situation for the process 
$\pbarp \rightarrow \mumu$ is different. Due to the 
similar mass of muon and pion PID capabilities are 
not sufficient to arrive at a clean separation of pions 
against muons. Our simulations show, that a measurement of 
the electromagnetic form factors using muons is much less
promising. Further studies are required.

Polarisation degree of freedom, either on the target side or with 
transversely polarised $\pbar$-beam would allow to access 
the imaginary part of the complex form factors. For example 
with a transversely polarised target only one could already 
determine the phase difference of the two form factors.

\subsubsection*{Outlook on Transition Distribution Amplitudes (TDA)}

\begin{center}
 \begin{figure}[h!]
\includegraphics[width=\swidth]{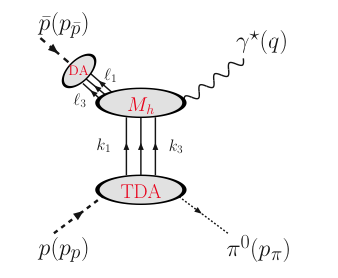}
\caption{The factorisation of the process $\bar p p \to e^+ e^- \pi^0 $
}
\label{TDAfig1}
\end{figure}
 \end{center}

The amplitude of the process 
\begin{equation}
\bar p (p_{\bar p })  p (p_p) \to \gamma^\star(q) \pi(p_\pi)\to \ell^+(p_{\ell^+})  \ell^-(p_{\ell^-})  \pi(p_\pi)
\label{TDAprocess}
\end{equation}
at small $t=(p_\pi-p_p)^2$ (or at small $u= (p_\pi-p_{\bar p})^2$) and large lepton pair invariant mass squared $q^2$
 has been shown to  factorise into a short-distance perturbatively calculable matrix element and 
 long-distance dominated antiproton Distribution Amplitudes (DA)  and proton to pion Transition Distribution Amplitudes (TDA), as shown in \Reffig{TDAfig1}. 
 
 Transition Distribution Amplitudes ~\cite{ref:nuclstruc:gda:Lansberg:2006uh,ref:nuclstruc:gda:Pire:2005ax,Lansberg:2007sr} are universal non-perturbative objects describing the transitions between two different particles (~e.g.~$p\to \pi$, $p\to\gamma$). They are defined from the Fourier transform  of a matrix element of a 
three-quark-light-cone operator between a proton and a meson state.
They obey QCD evolution equations
which follow from the renormalisation-group equation of  the
three-quark operator. Their $Q^2$ dependence is thus completely  under control.
To define the  transition distribution
amplitudes  from a nucleon to a pseudoscalar meson, we introduce light-cone coordinates $v^\pm = (v^0 \pm
v^3) /\sqrt{2}$ and  transverse components $v_T = (v^1,  v^2)$ for any
four-vector $v$.   The skewedness variable  $\xi = -\Delta^+  /2P^+$ with $\Delta = p'-p$ and
$P=(p+p')/2$  describes  the loss  of  plus-momentum  of the  incident
hadron in the proton $\to$ meson transition. The exchanged quarks carry light cone fractions of + 
momenta  labelled by  $x_{1}$, $x_{2}$ and $x_{3}$, 
and their supports are within $[-1+\xi, 1+\xi]$.  Momentum conservation implies that
$\sum_{i}  x_{i} = 2 \xi \, .$

As in the case of generalised parton distributions the simultaneous presence of two transverse 
scales $Q^2$ and $-t$, allows through a Fourier transform
to map the impact parameter dependence of the scattering amplitude. 
In the case under study, the $t-$ dependence
of the $N \to \pi$ transition distribution amplitude allows in 
its ERBL region (namely, when all $x_{i}> 0$) a transverse scan of 
the location of the small sized 
(of the order of $1/Q$ ) hard core made of 
three quarks when a pion carries the rest of the momentum of the nucleon.
This may be phrased alternatively as detecting 
the transverse mean position of a pion inside the proton.

Cross section estimates have been calculated using a TDA ansatz inspired by soft pion theorems in Ref. \cite{ref:nuclstruc:gda:Pire:2005xx}. They show that measurable rates  up to $Q^2$ values of the order a few GeV$^2$ are expected with the designed beam  luminosity. Detailed simulations have not yet been done, 
but the rate estimates show, that the process is accessible with the design luminosity.

\begin{figure}[h]
\includegraphics[width=\swidth]{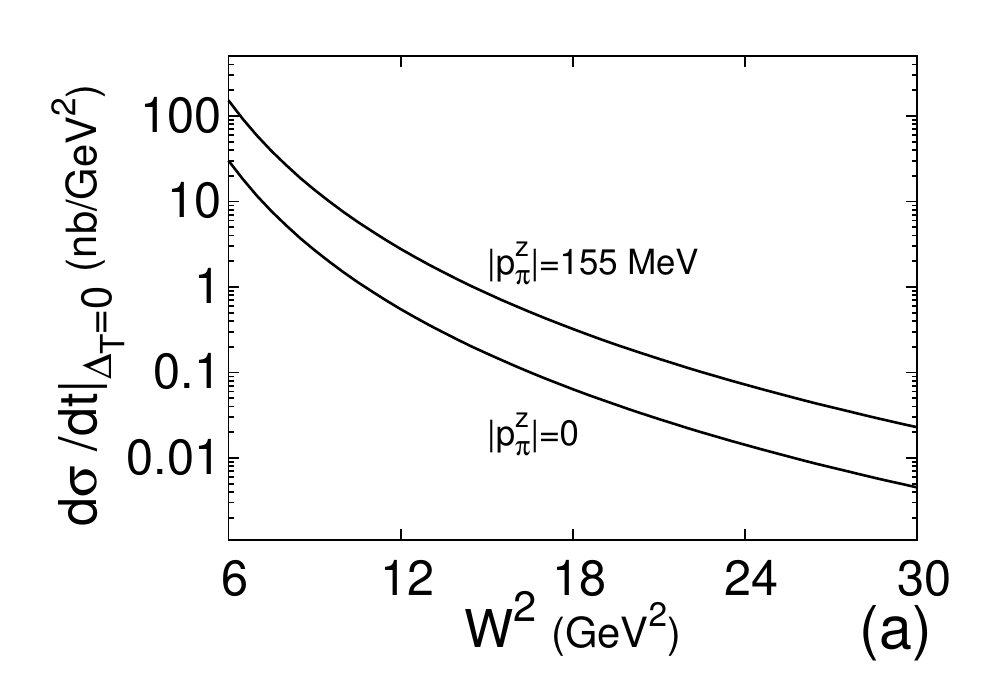}
\includegraphics[width=\swidth]{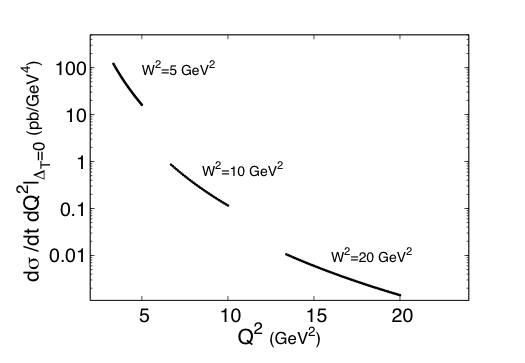}
\caption[Theoretical cross section estimates for accessing TDAs.]{
(a) Differential cross section $d\sigma/dt$ for $\bar p p \to \gamma^\star \pi^0$ 
as a function a $W^2$ for $|p^z_{\pi}| = 0$ (lower curve) and
$|p^z_{\pi}| =M/6 $.
(b) Differential cross section $d\sigma/(dtdQ^2)$ for $\bar p p \to \ell^+ \ell^- \pi^0$ 
as a function of $Q^2$  for various beam energies.}
\label{fig:cross-section}
\end{figure}

%% file: phys/phys_electroweak.tex
%
\section{Electroweak Physics}
\COM{Author(s): L. Schmitt}
\COM{Referee(s): O. Scholten}
With the high-intensity antiproton beam available at HESR a large
number of D-mesons can be produced. This gives the possibility to
observe rare weak decays of these mesons allowing to study electroweak
physics by probing predictions of the Standard Model and searching 
for enhancements introduced by processes beyond the Standard Model.
Since the studied processes are very rare and small deviations are
looked for statistics is the main factor in this class of
measurements. This implies to perform them at highest possible
luminosity, best by even extending the antiproton production rate
beyond the anticipated 2$\times 10^7$/s. However a longterm parasitic
measurement in parallel to spectroscopy and other topics whenever D 
mesons are produced can also provide an interesting statistics over
some years.

\subsection{CP-Violation and Mixing in the Charm-Sector}
$\cal{CP}$ violation \cite{bib:phy:bigi98} has been observed in
neutral kaon and in neutral $B$ meson decays
\cite{bib:phy:aubert01,bib:phy:abe01}. Recently BaBar has seen
evidence for mixing of $\D$ and $\Dbar$ \cite{bib:phy:Aubert:2007wf}. 
The observed lifetime differences are however in no contradiction to
the Standard Model. 
\par
In the standard model,
$\cal{CP}$ violation arises from a single phase entering the
Cabibbo-Kobayashi-Maskawa (CKM) matrix.  As a result, two elements of
this matrix, i.e.\ $V_{ub}$ and $V_{td}$ have large phases.  The
elements have small magnitudes and involve the third generation and
$\cal{CP}$ violation is small in the $K^0$ system.  The violation is
predicted to be even smaller in the $\Dz$ system
\cite{bib:phy:burdman94}.  Thus, a deviation from the small standard
model effect indicating "new physics" can be more easily distinguished
in experiments in the $\D$ meson system. An enhanced mass difference
of mixing $\D$ and $\Dbar$ mesons would constitute a deviation form the
Standard Model. However lifetime difference mostly occur through long 
range interaction of common final states and can be enhanced simply by 
a strong phase.
\par
Working with $\D$ mesons produced at the $\DDbar$ threshold has
advantages arising from the strong correlation of the $\DDbar$ pair
which is kept in the hadronisation process.  Formed near threshold,
asymmetries are not expected in the production process and the
observation of one $\D$ meson reveals the quantum numbers of the other
one when produced in a charge symmetric environment (flavor tagging).
Thus, flavor mixing of $\DDbar$ and $\cal{CP}$ violation can be
searched for in analogy to methods in the $B$-system produced on the
$\Upsilon(4S)$~\cite{bib:phy:babarphysbook}.
\par
The two-body channels
$\pbarp\,\rightarrow\,\psi(3770)\,\rightarrow\,D\overline{D}$ and
$\pbarp\,\rightarrow\,\psi(4040)\,\rightarrow\,D^*\overline{D^*}$ may
serve to investigate the open charm reconstruction abilities of
\PANDA. As shown in section \ref{sec:phys:ddbar} $D$ mesons can be
reconstructed well in \PANDA. The open question is the value of the
production cross-section for $p\bar{p}\rightarrow D\bar{D}$. In
section \ref{sec:phys:ddbar} a very conservative estimate derived from
the decay width $J/\Psi\rightarrow p\bar{p}$ of about 3\,nb was
given. On the other hand Kroll in studies for \INST{SuperLEAR} calculated
a cross section of up to 200\,nb for this process \cite{bib:phy:kroll91}. 
For a measurement of $\alpha_{CP}\sim 10^{-3}$ as predicted by the
Standard Model the production of $10^9$ $D$ meson pairs would be
required, which at a production cross section of 200 nb would
correspond to 3 years of running at ${L}=2\cdot 10^{32}$\,cm$^{-2}$s$^{-1}$, but
would be out of reach at the conservative lower limit of the cross
section.

\subsection{CP-Violation in Hyperon Decays}
In self-analysing two-body decays of hyperons the polarisation of the
mother particle can be obtained directly from the daughters. In these
decays the orbital angular momentum of the final state can be $L = 0$,
$1$, in other words the decay amplitude can be an $S$-wave or a
$P$-wave.  Having two amplitudes, interference can occur and
CP-violating phases can enter. There are two characteristic
parameters, which govern the decay dynamics: The quantity $\alpha$
denotes the asymmetry of the decay angular distribution, the quantity
$\beta$ gives the decay-baryon polarisation. With these quantities and
the decay width $\Gamma$, CP asymmetries can be formed
\cite{bib:phy:hamann91}:
\[
\begin{array}{ccc}
  A = \frac{{\alpha {\rm{ }}\Gamma  + \overline \alpha  {\rm{ }}\overline \Gamma  }}
  {{\alpha {\rm{ }}\Gamma  - \overline \alpha  {\rm{ }}\overline \Gamma  }}
  \approx \frac{{\alpha {\rm{ }} + \overline \alpha  {\rm{ }}}}{{\alpha {\rm{ }} -
      \overline \alpha  {\rm{ }}}} & B = \frac{{\beta {\rm{ }}\Gamma  + \overline \beta
      {\rm{ }}\overline \Gamma  }}{{\beta {\rm{ }}\Gamma  - \overline \beta   {\rm{ }}\overline \Gamma  }}
  \approx \frac{{\beta {\rm{ }} + \overline \beta   {\rm{ }}}}{{\beta {\rm{ }} - \overline \beta   {\rm{ }}}}
  & D = \frac{{{\rm{ }}\Gamma  + {\rm{ }}\overline \Gamma  }}{{{\rm{ }}\Gamma  - {\rm{ }}\overline \Gamma  }}
\end{array}
\]
For the asymmetry $A$, standard model predictions are of the order
$\approx\,$2$\times$10$^{-5}$.  Some models beyond the standard model
predict CP asymmetries of the order of several 10$^{-4}$. To reach the
standard model limit about 10$^{10}$ Hyperon decays have to be
reconstructed, which could be done within one year under ideal
conditions. The detector for this experiment requires good vertex
reconstruction, excellent particle identification both in forward
direction and at large angles and reliable long term stability.

\subsection{Rare Decays}
The study of rare decays can open a window onto physics beyond the
standard model since it probes the violation of fundamental
symmetries. Lepton flavor number violating decays, {\it e.g.}\
$\Dz\,\to\,\mu e$ or $\D^\pm\,\to\,\pi\mu e$, could be searched for.
Flavor changing neutral currents like in the decay $\Dz\,\to\,\mumu$
can occur in the standard model through box graphs or weak penguin
graphs with branching fractions smaller than 10$^{-15}$.  However, the
signatures of the decays are clean leaving hope for their observation,
if processes exist that boost the decay branch.








%

%

%% file: panda_pb_sum.tex
%
%
\cleardoublepage
\chapter{Summary and Outlook}
\label{sec:sum}
\COM{Author(s): D. Bettoni, K. Peters}
%
%
\input{./sum/sum.tex}
%
%
\newpage
%

%% file: sum/sum.tex

\subsubsection*{Software and Data Production}

The aim of the \PANDA Physics Book benchmark studies was to demonstrate the feasibility of the 
planned physics program and moreover to show the expected physics performance for the upcoming 
measurements with the \Panda detector. This requires 
accurate simulation studies of the physics channels of interest and of the relevant 
background by taking into account the complete \Panda detector. A huge number of background events 
were needed because the total \pbarp cross section in this energy regime is between 50~-~100~mb and 
compared to that the cross sections of the most physics channels are expected to be very low and are 
typically in the order of pb or nb. In 
order to meet these requirements an offline software has been devised following an 
object oriented approach and making use of several well-tested software tools and packages from 
other \INST{HEP} experiments. These packages have been adapted to the \PANDA needs. 

The generation and processing of the data is subdivided into a couple of well defined stages:
The event generation, the particle tracking utilising the GEANT4 transport code, the digitisation,
which models the signals and their processing in the front-end electronics of the 
detectors, the reconstruction, which finally creates lists of particle candidates, and at the last stage, 
the physics analysis. For this final step the software provides besides low\--level analysis tools
also  high\--level analysis tools
which allow to reconstruct decay trees, perform geometrical and kinematic fits, and to refine 
the event selection in a very easy and user friendly manner.

The very time consuming data production has been organised in a central way and was distributed 
over four computing farms at \INST{IPNO} in Orsay, at \INST{CCIN2P3} in Lyon, at the 
\INST{Ruhr-Universit\"at Bochum}, and 
at the \INST{GSI} at Darmstadt. Approximately $10^{11}$ events in total have been generated. Sophisticated filters 
have been applied on this event generator level, so that finally $22\cdot10^6$ signal events for 
22 channels, $1002\,\cdot\,10^6$ dedicated background events including the dominant background reactions for 
the individual analyses, and $280\cdot10^6$ generic background events have been simulated and reconstructed 
in 29 weeks. The 
bookkeeping for this data production has been realised using a MySQL database in such a way
that all job scripts and configuration files get produced automatically by Perl and PHP scripts. 

The development of this software has been started after the preparation of the \Panda 
Technical Progress Report in spring 2005. In parallel to these activities a new software project has been 
initiated in autumn 2006, with the aim to realise a next\--generation software for \Panda. 
The objectives for this new software, called \INST{PANDAroot}, are to improve the 
accessibility for beginning users and developers,
to increase the flexibility to cope with future developments and to enhance synergy with other particle
 physics experiments. The core services are provided by the \INST{FAIRroot} code, which is based 
on the object-oriented ROOT software and which makes use of the Virtual Monte-Carlo (VMC) interface. 
Two physics cases have been investigated by making use of the \INST{PANDAroot} software, namely the 
crossed channel compton scattering and the hypernuclei physics. 

\subsubsection*{Physics Case and Outlook}

Thanks to the software tools developed and summarised above
the performance of the detector and the sensitivity
to the various physics channels have been estimated
reliably in terms of geometrical acceptance, 
resolution and signal/background ratios.

For the benchmark channels chosen for this report
the simulations show that the final states of interest
can be detected with good efficiency and that the
background contamination is at an acceptable level.
This demonstrates the feasibility of the planned 
physics programme.

The experimental setup used
in this report is not final: some detector components
are still undergoing R\&D and will be finalised in the
coming months. For this reason the results presented here 
are to be regarded as preliminary.
Further improvements will come from the new software
framework being developed as well as from advances in
the background simulations. 
Progress in theoretical calculations will yield
better estimates of presently unknown cross sections,
which will result in more reliable evaluations of
signal/background ratios.
A new version of the physics book is planned, which will
reflect the progress described above and which will feature
a more complete list of benchmark channels.


%% file: panda_pb_end.tex
%
\cleardoublepage
\onecolumn
\input{./main/thanks}
%
\cleardoublepage
\twocolumn
\input{./main/acronyms}
\cleardoublepage
\addcontentsline{toc}{chapter}{List of Figures}
\listoffigures
\cleardoublepage
\addcontentsline{toc}{chapter}{List of Tables}
\listoftables
%

%% file: main/thanks.tex
%
\begin{center}
\vspace*{2cm}
{\Large\bf Acknowledgments}\addcontentsline{toc}{chapter}{Acknowledgements}
\vskip 2cm
\begin{minipage}[t]{8cm}
\sloppy\large
We acknowledge financial support from
the Bundesministerium f\"ur Bildung und Forschung (bmbf),
the Deutsche Forschungsgemeinschaft (DFG),
the Forschungszentrum J\"ulich GmbH, J\"ulich,
the University of Groningen, Netherlands,
the Gesellschaft f\"ur Schwerionenforschung mbH (GSI), Darmstadt,
the Helmholtz-Gemeinschaft Deutscher Forschungszentren (HGF),
the Schweizerischer Nationalfonds zur F\"orderung der wissenschaftlichen
Forschung (SNF),
the Russian funding agency ``State Corporation for Atomic Energy Rosatom'',
the CNRS/IN2P3 and the Universit\'e Paris-sud,
the British funding agency ``Science and Technology Facilities
Council'' (STFC),
the Instituto Nazionale di Fisica Nucleare (INFN),
the Swedish Research Council,
the Polish Ministry of Science and Higher Education,
the European Community FP6 FAIR Design Study: DIRACsecondary-Beams,
contract number 515873,
the European Community FP6 Integrated Infrastructure Initiative:
HadronPhysics, contract number RII3-CT-2004-506078, 
and the Deutscher Akademischer Austauschdienst (DAAD).
\end{minipage}
\end{center}
\vfill
%
%

%% file: main/acronyms.tex
%
\addcontentsline{toc}{chapter}{List of Acronyms}
\begin{acronym}
\acro{ADC}{Analog to Digital Converter}
\acro{AOD}{Analysis Object Data}
\acro{APD}{Avalanche Photo Diode}
\acro{API}{Application Programming Interface}
\acro{ASIC}{Application Specific Integrated Circuit}
\acro{ATLAS}{A Toroidal LHC ApparatuS}
\acro{BNL}{Brookhaven National Laboratory}
\acro{CAD}{Computer Aided Design}
\acro{CDR}{Conceptual Design Report}
\acro{CERN}{Conseil European pour la Recherche Nucleaire}
\acro{CKM}{Cabbibo-Kobayashi-Maskawa}
\acro{CL}{Confidence Level}
\acro{CLAS}{CEBAF Large Acceptance Spectrometer}
\acro{COMPASS}{Common Muon Proton Apparatus for Structure and Spectroscopy}
\acro{COSY}{Cooler Synchrotron}
\acro{CT}{Central Tracker}
\acro{CVS}{Code Versioning System}
\acro{DAQ}{Data Acquisition}
\acro{DCH}{Drist Chambers}
\acro{DESY}{Deutsches Elektronensynchrotron}
\acro{DIRC}{Detector for Internally Reflected Cherenkov Light}
\acro{DPM}{Dual Parton Model}
\acro{DSP}{Digital Signal Processor}
\acro{DVCS}{Deeply Virtual Compton Scattering}
\acro{DY}{Drell-Yan}
\acro{EMC}{Electromagnetic Calorimeter}
\acro{ESD}{Event Summary Data}
\acro{EU}{European Union}
\acro{FADC}{Flash ADC}
\acro{FAIR}{Facility for Antiproton and Ion Research}
\acro{FE}{Front-End}
\acro{FEE}{Front-End Electronics}
\acro{FNAL}{Fermi National Laboratory}
\acro{FPGA}{Field Programmable Gate Array}
\acro{FS}{Forward Spectrometer}
\acro{GEM}{Gas Electron Multiplier}
\acro{GPD}{Generalized Parton Distribution}
\acro{GSI}{Gesellschaft f\"ur Schwerionenforschnung}
\acro{HC}{Hadron Calorimeter}
\acro{HESR}{High Energy Storage Ring}
\acro{HFG}{Hermann von Helmholtz-Gemeinschaft Deutscher Forschungszentren}
\acro{HV}{High Voltage}
\acro{ISR}{Initial State Radiation}
\acro{IP}{Interaction point}
\acro{IN2P3}{Institut National de Physique Nucleaire et de Physique des Particules}
\acro{INFN}{Istituto Nazionale di Fisica Nucleare}
\acro{KVI}{Kernfysisch Versneller Instituut}
\acro{LAAPD}{Large Area APD}
\acro{LEAR}{Low Energy Antiproton Ring}
\acro{LED}{Light Emission Diode}
\acro{LH}{Likelihood}
\acro{LHC}{Large Hadron Collider}
\acro{LOI}{Letter of Intent}
\acro{LQCD}{Lattice QCD}
\acro{MAPS}{Monolithic Active Pixel Sensor}
\acro{MC}{Monte Carlo}
\acro{MDC}{Mini Drift Chamber}
\acro{MLP}{Multilayer Perzeptron}
\acro{MUD}{Muon Detector}
\acro{MVD}{Micro Vertex Detector}
\acro{NRQCD}{Non-Relativistic QCD}
\acro{PHP}{Hypertext Preprocessor}
\acro{PID}{Particle Identification}
\acro{PMT}{Photomultiplier}
\acro{PSI}{Paul Scherrer Institute}
\acro{PWA}{Partial Wave Analysis}
\acro{PWO}{Lead Tungstate}
\acro{QA}{Quality Assurance}
\acro{QCD}{Quantum Chromo Dynamics}
\acro{QED}{Quantum Electrodynamics}
\acro{RICH}{Ring Imaging Cherenkov Counter}
\acro{SLAC}{Stanford Linear Accelerator Center}
\acro{SQL}{Structured Query Language}
\acro{SSA}{Single Spin Asymmetry}
\acro{SSL}{Secure Socket Layer}
\acro{STT}{Straw Tube Tracker}
\acro{TCL}{Tool Command Language}
\acro{TCS}{Trigger Control System}
\acro{TDA}{Transition Distribution Amplitude}
\acro{TDC}{Time to Digital Converter}
\acro{TLFF}{Time-like Form Factor}
\acro{TOF}{Time-of-Flight Detector}
\acro{TOP}{Time-of-Propagation}
\acro{ToT}{Time-over-Threshold}
\acro{TPC}{Time Projection Chamber}
\acro{TS}{Target Spectrometer}
\acro{TSL}{The Svedberg Laboratory}
\acro{UrQMD}{Ultra-relativistic Quantum Molecular Dynamic}
\acro{VGM}{Virtual Geometry Model}
\acro{VMC}{Virtual Monte Carlo}
\acro{WACS}{Wide Angle Compton Scattering}
\acro{XML}{Extensible Markup Language}
\end{acronym}
\vfill
%
%

%% file: panda_pb.bbl
\begin{flushleft}\begin{thebibliography}{10}\sloppy

\bibitem{bib:tgt:pe:Buescher:2006}
M.~{B{\"u}scher} et~al.,
\newblock {The Moscow-J{\"u}lich Frozen-Pellet Target},
\newblock in {\em Hadron Spectroscopy}, edited by A.~{Reis}, C.~{G{\"o}bel},
  J.~D.~S. {Borges}, and J.~{Magnin}, volume 814 of {\em American Institute of
  Physics Conference Series}, pages 614--620, 2006.

\bibitem{bib:tgt:pe:Ekstroem:1996jt}
C.~Ekstroem et~al.,
\newblock Nucl. Instrum. Meth. {\bf A371}, 572 (1996).

\bibitem{Boukharov:2008vh}
A.~V. Boukharov et~al.,
\newblock Phys. Rev. Lett. {\bf 100}, 174505 (2008).

\bibitem{Atlas:Tdr:11}
Technical report,
\newblock ATLAS Technical Design Report 11, CERN/LHCC 98-13.

\bibitem{Cms:Tdr:5}
Technical report,
\newblock CMS Technical Design Report 5, CERN/LHCC 98-6.

\bibitem{Ketzer:2004jk}
B.~Ketzer, Q.~Weitzel, S.~Paul, F.~Sauli, and L.~Ropelewski,
\newblock Nucl. Instrum. Meth. {\bf A535}, 314 (2004).

\bibitem{Staengle:1997xp}
H.~Staengle et~al.,
\newblock Nucl. Instrum. Meth. {\bf A397}, 261 (1997).

\bibitem{Mengel:1998si}
K.~Mengel et~al.,
\newblock IEEE Trans. Nucl. Sci. {\bf 45}, 681 (1998).

\bibitem{Novotny:2000zg}
R.~Novotny et~al.,
\newblock IEEE Trans. Nucl. Sci. {\bf 47}, 1499 (2000).

\bibitem{Hoek:2002ss}
M.~Hoek et~al.,
\newblock Nucl. Instrum. Meth. {\bf A486}, 136 (2002).

\bibitem{Alice:Tp}
Technical Proposal,
\newblock CERN/LHC 9.71.

\bibitem{Cms:Tp:1994}
Technical Proposal, 1994,
\newblock CERN/LHCC 94-38, LHCC/P1.

\bibitem{Auffray:1999}
E.Auffray et~al.,
\newblock Proceedings of SCINT99,
\newblock Moscow, 1999.

\bibitem{Abbon:2007pq}
P.~Abbon et~al.,
\newblock Nucl. Instrum. Meth. {\bf A577}, 455 (2007).

\bibitem{Akopov:2000qi}
N.~Akopov et~al.,
\newblock Nucl. Instrum. Meth. {\bf A479}, 511 (2002).

\bibitem{bib:emc:E865}
G.~S. Atoyan et~al.,
\newblock Nucl. Instrum. Meth. {\bf A320}, 144 (1992).

\bibitem{bib:emc:PHENIX}
G.~David et~al.,
\newblock Performance of the PHENIX EM calorimeter,
\newblock Technical report, PHENIX Tech. Note 236, 1996.

\bibitem{bib:emc:HERAB}
A.~Golutvin,
\newblock (1994),
\newblock HERA-B Tech. Note 94-073.

\bibitem{bib:emc:LHCb}
LHCb Technical Proposal CERN LHCC 98-4, LHCC/P4, 1998.

\bibitem{bib:emc:KOP99}
I.-H. Chiang et~al.,
\newblock (1999),
\newblock KOPIO Proposal.

\bibitem{bib:emc:KOP1}
H.~Morii,
\newblock (2004),
\newblock Talk at NP04 Workshop at J-PARC.

\bibitem{bib:emc:KOP04}
G.~Atoyan et~al.,
\newblock Test beam study of the KOPIO Shashlyk calorimeter prototype,
\newblock in {\em Proceedings of ``CALOR 2004''}, 2004.

\bibitem{Armstrong:1996np}
T.~A. Armstrong et~al.,
\newblock Phys. Lett. {\bf B385}, 479 (1996).

\bibitem{FAIR:2006}
Baseline Technical Report, subproject HESR,
\newblock Technical report, Gesellschaft f\"ur Schwerionen (GSI), Darmstadt,
  2006.

\bibitem{boine:2006}
O.~Boine-Frankenheim, R.~Hasse, F.~Hinterberger, A.~Lehrach, and P.~Zenkevich,
\newblock Nucl. Instrum. Meth. {\bf A560}, 245 (2006).

\bibitem{parkhomchuk:2000}
V.~Parkhomchuk,
\newblock Nucl. Instrum. Meth. {\bf A441}, 9 (2000).

\bibitem{sorensen:1987}
A.~H. S{\o}rensen,
\newblock Introduction to intrabeam scattering,
\newblock in {\em {CERN Accelerator School: General accelerator physics}},
  number CERN-87-10, page 135, 1987.

\bibitem{hinterberger:1989a}
F.~Hinterberger, T.~Mayer-Kuckuk, and D.~Prasuhn,
\newblock Nucl. Instrum. Meth. {\bf A275}, 239 (1989).

\bibitem{hinterberger:1989b}
F.~Hinterberger and D.~Prasuhn,
\newblock Nucl. Instrum. Meth. {\bf A279}, 413 (1989).

\bibitem{reistad:2007}
D.~Reistad et~al.,
\newblock Calculations on high-energy electron cooling in the {HESR},
\newblock in {\em {Proceedings of COOL 07 --- Workshop on Beam Cooling and
  Related Topics}}, pages 44--48, Bad Kreuznach, Germany, 2007.

\bibitem{stockhorst:2008}
H.~Stockhorst et~al.,
\newblock Stochastic Cooling Developments for the {HESR} at {FAIR},
\newblock in {\em Proc. of the European Accelerator Conference EPAC08}, number
  THP055, page 3491, Genoa, Italy, 2008.

\bibitem{lehrach:2006}
A.~Lehrach, O.~Boine-Frankenheim, F.~Hinterberger, R.~Maier, and D.~Prasuhn,
\newblock Nucl. Instrum. Meth. {\bf A561}, 289 (2006).

\bibitem{hinterberger:2006}
F.~Hinterberger,
\newblock Monte carlo simulations of Thin Internal Target Scattering in
  Ceslius,
\newblock in {\em Beam-Target Interaction and Intra-beam Scattering in the HESR
  Ring: Emittance, Momentum Resolution and Luminosity}, Bericht des
  Forschungszentrum J\"ulich, 2006,
\newblock J\"ul-Report No. 4206.

\bibitem{lehrach:pdb}
W.-M. Yao et~al.,
\newblock Journal of Physics {\bf G33}, 1 (2006).

\bibitem{lehrach:ar}
A.~Lehrach,
\newblock IKP Annual Report {\bf J\"ul-4234}, 138 (2006).

\bibitem{bib:exp:mcginnis:2003}
D.~McGinnis, G.~Stancari, and S.~J. Werkema,
\newblock Nucl.\ Instrum.\ Methods A {\bf 506}, 205 (2003).

\bibitem{bib:exp:andreotti:2007}
M.~Andreotti et~al.,
\newblock Phys.\ Lett.\ B {\bf 654}, 74 (2007).

\bibitem{bib:exp:armstrong:1993}
T.~A. Armstrong et~al.,
\newblock Phys.\ Rev.\ D {\bf 47}, 772 (1993).

\bibitem{bib:exp:kennedy:1992}
D.~C. Kennedy,
\newblock Phys.\ Rev.\ D {\bf 46}, 461 (1992).

\bibitem{bib:exp:aulchenko:2003}
V.~M. Aulchenko et~al.,
\newblock Phys.\ Lett.\ B {\bf 573}, 63 (2003).

\bibitem{bib:exp:pesce:2007}
A.~Pesce,
\newblock \textit{Studio della distribuzione ottimale di luminosit\`a per la
  misura dei parametri di risonanza in fisica delle particelle} (BSc Thesis, in
  Italian),
\newblock 2007.

\end{thebibliography}\end{flushleft}


\begin{flushleft}\begin{thebibliography}{10}\sloppy

\bibitem{PDG06}
W.~M. Yao et~al.,
\newblock J. Phys. {\bf G33}, 1 (2006).

\bibitem{Ber08}
V.~Bernard,
\newblock Prog. Part. Nucl. Phys. {\bf 60}, 82 (2008).

\bibitem{Man00}
A.~V. Manohar and M.~B. Wise,
\newblock Camb. Monogr. Part. Phys. Nucl. Phys. Cosmol. {\bf 10}, 1 (2000).

\bibitem{Hoo74}
G.~'t~Hooft,
\newblock Nucl. Phys. {\bf B72}, 461 (1974).

\bibitem{Pas06}
K.~D. Paschke et~al.,
\newblock Phys. Rev. {\bf C74}, 015206 (2006).

\bibitem{Mor99}
C.~J. Morningstar and M.~Peardon,
\newblock Phys. Rev. D {\bf 60}, 034509 (1999).

\bibitem{Pac07}
S.~Pacetti,
\newblock Eur. Phys. J. {\bf A32}, 421 (2007).

\bibitem{God95}
G.~T. Bodwin, E.~Braaten, and G.~P. Lepage,
\newblock Phys. Rev. D {\bf 51}, 1125 (1995).

\bibitem{Bra04}
N.~Brambilla et~al.,
\newblock hep-ph/0412158  (2004).

\bibitem{Swa06}
E.~S. Swanson,
\newblock Phys. Rept. {\bf 429}, 243 (2006).

\bibitem{Luk97}
M.~Luke and A.~V. Manohar,
\newblock Phys. Rev. D {\bf 55}, 4129 (1997).

\bibitem{Bra05}
N.~Brambilla, A.~Pineda, J.~Soto, and A.~Vairo,
\newblock Rev. Mod. Phys. {\bf 77}, 1423 (2005).

\bibitem{Pol06}
H.~Polinder, J.~Haidenbauer, and U.-G. Meissner,
\newblock Nucl. Phys. {\bf A779}, 244 (2006).

\bibitem{Lut02}
M.~F.~M. Lutz and E.~E. Kolomeitsev,
\newblock Nucl. Phys. {\bf A700}, 193 (2002).

\bibitem{Kra90}
A.~Krause,
\newblock Helv. Phys. Acta {\bf 63}, 3 (1990).

\bibitem{Yan92}
T.-M. Yan et~al.,
\newblock Phys. Rev. D {\bf 46}, 1148 (1992).

\bibitem{Gas85}
J.~Gasser and H.~Leutwyler,
\newblock Nucl. Phys. {\bf B250}, 465 (1985).

\bibitem{Wei90}
S.~Weinberg,
\newblock Nucl. Phys. {\bf B363}, 3 (1991).

\bibitem{Kai97}
N.~Kaiser, T.~Waas, and W.~Weise,
\newblock Nucl. Phys. {\bf A612}, 297 (1997).

\bibitem{Oll98}
J.~A. Oller, E.~Oset, and J.~R. Pelaez,
\newblock Phys. Rev. {\bf D59}, 074001 (1999).

\bibitem{Gar04}
C.~Garcia-Recio, M.~F.~M. Lutz, and J.~Nieves,
\newblock Phys. Lett. {\bf B582}, 49 (2004).

\bibitem{Kol04b}
E.~E. Kolomeitsev and M.~F.~M. Lutz,
\newblock Phys. Lett. {\bf B585}, 243 (2004).

\bibitem{Lut04}
M.~F.~M. Lutz and E.~E. Kolomeitsev,
\newblock Nucl. Phys. {\bf A730}, 392 (2004).

\bibitem{Kol04}
E.~E. Kolomeitsev and M.~F.~M. Lutz,
\newblock Phys. Lett. {\bf B582}, 39 (2004).

\bibitem{Lut07}
M.~F.~M. Lutz and M.~Soyeur,
\newblock Nucl. Phys. {\bf A, in print} (2007).

\bibitem{Guo08}
F.-K. Guo, C.~Hanhart, S.~Krewald, and U.-G. Meissner,
\newblock Phys. Lett. {\bf B666}, 251 (2008).

\bibitem{Hof05}
J.~Hofmann and M.~F.~M. Lutz,
\newblock Nucl. Phys. {\bf A763}, 90 (2005).

\end{thebibliography}\end{flushleft}


\begin{flushleft}\begin{thebibliography}{100}\sloppy

\bibitem{qwgyr}
N.~Brambilla et~al.,
\newblock hep-ph/0412158  (2004).

\bibitem{bib:phy:cb1500a}
C.~Amsler et~al.,
\newblock Phys. Lett. {\bf B342}, 433 (1995).

\bibitem{bib:phy:cb1500b}
C.~Amsler et~al.,
\newblock Phys. Lett. {\bf B353}, 571 (1995).

\bibitem{bib:phy:cb1500c}
A.~Abele et~al.,
\newblock Phys. Lett. {\bf B385}, 425 (1996).

\bibitem{bib:phy:cb1500d}
A.~Abele et~al.,
\newblock Eur. Phys. J. {\bf C19}, 667 (2001).

\bibitem{bib:phy:cb1500e}
C.~Amsler et~al.,
\newblock Phys. Lett. {\bf B340}, 259 (1994).

\bibitem{psidisc1}
J.~Aubert et~al.,
\newblock Phys. Rev. Lett. {\bf 33}, 1404 (1974).

\bibitem{psidisc2}
J.~Augustin et~al.,
\newblock Phys. Rev. Lett. {\bf 33}, 1406 (1974).

\bibitem{dalpiaz79}
P.~Dalpiaz,
\newblock Charmonium and other onia at minimum energy,
\newblock in {\em Karlsruhe 1979, Proceedings, Physics With Cooled Low
  Energetic Antiprotons}, edited by H.~Poth, pages 111--124, 1979.

\bibitem{chi1E835}
M.~Andreotti et~al.,
\newblock Nucl. Phys. B {\bf 717}, 34 (2005).

\bibitem{PDG}
W.-M. Yao et~al.,
\newblock J. of Phys. G {\bf 33}, 1 (2006).

\bibitem{etacpCB}
C.~Edwards et~al.,
\newblock Phys. Rev. Lett. {\bf 48}, 70 (1982).

\bibitem{etacpBelle}
S.~Choi et~al.,
\newblock Phys. Rev. Lett. {\bf 89}, 102001 (2002).

\bibitem{etacpCLEO}
D.~Asner et~al.,
\newblock Phys. Rev. Lett. {\bf 92}, 142001 (2004).

\bibitem{etacpbabar}
B.~Aubert et~al.,
\newblock Phys. Rev. Lett. {\bf 92}, 142002 (2004).

\bibitem{1p1E760}
T.~Armstrong et~al.,
\newblock Phys. Rev. Lett. {\bf 69}, 2337 (1992).

\bibitem{bib:sim:hc_E835}
M.~Andreotti et~al.,
\newblock Phys. Rev. {\bf D72}, 032001 (2005).

\bibitem{1p1CLEO}
J.~Rosner et~al.,
\newblock Phys. Rev. Lett. {\bf 95}, 102003 (2005).

\bibitem{RBES1}
J.~Bai et~al.,
\newblock Phys. Rev. Lett. {\bf 84}, 594200 (2000).

\bibitem{RBES2}
J.~Bai et~al.,
\newblock Phys. Rev. Lett. {\bf 88}, 101802 (2002).

\bibitem{Z3931}
S.~Uehara et~al.,
\newblock Phys. Rev. Lett. {\bf 96}, 082003 (2006).

\bibitem{X3940}
K.~Abe et~al.,
\newblock Phys. Rev. Lett. {\bf 94}, 182002 (2005).

\bibitem{Y}
B.~Aubert et~al.,
\newblock Phys. Rev. Lett. {\bf 95}, 142001 (2005).

\bibitem{angular}
T.N.Pham, B.Pire, and T.~Truong,
\newblock Phys. Lett. {\bf B61}, 183 (1976).

\bibitem{tesi}
M.~Negrini,
\newblock {\em Measurement of the branching ratios $\psi' \rightarrow J\psi X$
  in the experiment E835 at FNAL},
\newblock PhD thesis, University of Ferrara, 2003.

\bibitem{bkg}
V.~Flaminio et~al.,
\newblock CERN-HERA {\bf 70-03} (1970).

\bibitem{bib:sim:hc_CLEO}
J.~L. Rosner et~al.,
\newblock Phys. Rev. Lett. {\bf 95}, 102003 (2005).

\bibitem{bib:Andreotti:2005zr}
M.~Andreotti et~al.,
\newblock Phys. Rev. {\bf D72}, 112002 (2005).

\bibitem{bib:Armstrong:1997gv}
T.~A. Armstrong et~al.,
\newblock Phys. Rev. {\bf D56}, 2509 (1997).

\bibitem{bib:sim:DPM}
A.~Capella et~al.,
\newblock Phys. Rept. {\bf 236}, 225 (1994).

\bibitem{PandaTechRep}
{PANDA} {T}echnical {P}rogress {R}eport,
\newblock Technical report, FAIR-ESAC, 2005.

\bibitem{bib:phy:Kaidalov94}
A.~Kaidalov and P.~Volkovitsky,
\newblock Z. Phys. C {\bf 63}, 517 (1994).

\bibitem{Kroll89}
P.~Kroll, B.~Quadder, and W.~Schweiger,
\newblock Nuclear Physics B {\bf 316}, 373 (1989).

\bibitem{Braaten08}
E.~Braaten,
\newblock Physical Review D (Particles and Fields) {\bf 77}, 034019 (2008).

\bibitem{teo}
K.~Sebastian, H.~Grotch, and F.~L. Ridener,
\newblock Phys. Rev. {\bf D45}, 3163 (1992).

\bibitem{fermilab}
M.~Ambrogiani et~al.,
\newblock Phys. Rev. {\bf D65}, 05002 (2002).

\bibitem{bib:phy:kutimorningstar2003}
K.~Juge, J.~Kuti, and C.~Morningstar,
\newblock Phys. Rev. Lett. {\bf 90}, 161601 (2003).

\bibitem{bib:phy:thompson97}
D.~Thompson et~al.,
\newblock Phys. Rev. Lett. {\bf 79}, 1630 (1997).

\bibitem{bib:phy:adams98}
G.~Adams et~al.,
\newblock Phys. Rev. Lett. {\bf 81}, 5760 (1998).

\bibitem{bib:phy:Abele98}
A.~Abele et~al.,
\newblock Phys. Lett. {\bf B423}, 175 (1998).

\bibitem{bib:phy:Reinnarth00}
J.~Reinnarth,
\newblock Nucl. Phys. {\bf A692}, 268c (2001).

\bibitem{bib:phy:e852a}
S.~U. Chung et~al.,
\newblock Phys. Rev. {\bf D65}, 072001 (2002).

\bibitem{bib:phy:cb1360}
A.~Abele et~al.,
\newblock Phys. Lett. {\bf B446}, 349 (1999).

\bibitem{bib:phy:ox1384}
P.~Salvini et~al.,
\newblock Eur. Phys. J. {\bf C35}, 21 (2004).

\bibitem{bib:phy:e852b}
E.~I. Ivanov et~al.,
\newblock Phys. Rev. Lett. {\bf 86}, 3977 (2001).

\bibitem{bib:phy:cb1590}
S.~U. Chung and E.~Klempt,
\newblock Phys. Lett. {\bf B563}, 83 (2003).

\bibitem{bib:phy:e852e}
J.~Kuhn et~al.,
\newblock Phys. Lett. {\bf B595}, 109 (2004).

\bibitem{bib:phy:e852d}
M.~Lu et~al.,
\newblock hep-ex/0405044.

\bibitem{bib:phy:e852f}
G.~Adams,
\newblock Talk given at the First GHP Meeting of the APS, Chicago, 2004.

\bibitem{bib:phy:chen01}
P.~Chen, X.~Liao, and T.~Manke,
\newblock Nucl. Phys. Proc. Suppl. {\bf 94}, 342 (2001).

\bibitem{bib:phy:michael99}
C.~Michael,
\newblock in {\em Proceedings of ``Heavy Flavours 8''}, 1999.

\bibitem{bib:phy:jkm00a}
K.~Juge, J.~Kuti, and C.~Morningstar,
\newblock Nucl. Phys. (Proc. Suppl.) {\bf B83}, 304 (2000).

\bibitem{bib:phy:manke98}
T.~Manke et~al.,
\newblock Phys. Rev. {\bf D57}, 3829 (1998).

\bibitem{bib:phy:page}
P.~Page,
\newblock PhD thesis, 1995.

\bibitem{bib:phy:mp}
J.~Merlin and J.~Paton,
\newblock Phys. Rev. {\bf D35}, 1668 (1987).

\bibitem{bib:phy:jkm00b}
K.~Juge, J.~Kuti, and C.~Morningstar,
\newblock Nucl. Phys. (Proc. Suppl.) {\bf B86}, 397 (2000),
\newblock Phys. Lett. {\bf B478}, 151 (2000).

\bibitem{bib:phy:chester91}
R.~Cester,
\newblock in {\em Proceedings of the ``Super LEAR Workshop''}, pages 91--103,
  Z\"urich, 1991.

\bibitem{bib:phy:gaillard82}
M.~Gaillard, L.~Maiani, and R.~Petronzio,
\newblock Phys. Lett. {\bf B110}, 489 (1982).

\bibitem{bib:phy:milc97}
C.~Bernard et~al.,
\newblock Phys. Rev. {\bf D56}, 7039 (1997).

\bibitem{bib:phy:milc99}
C.~Bernard et~al.,
\newblock Nucl. Phys. (Proc. Suppl.) {\bf B73}, 264 (1999).

\bibitem{bib:phy:jkm99}
K.~Juge, J.~Kuti, and C.~Morningstar,
\newblock Phys. Rev. Lett. {\bf 82}, 4400 (1999).

\bibitem{bib:phy:mei:2002ip}
Z.-H. Mei and X.-Q. Luo,
\newblock Int. J. Mod. PHys. {\bf A18}, 5713 (2003),
\newblock hep-lat/0206012.

\bibitem{bib:phy:cp-pacs99}
T.~Manke et~al.,
\newblock Nucl. Phys. (Proc. Suppl.) {\bf B82}, 4396 (1999).

\bibitem{bib:phy:mornpeardon99}
C.~Morningstar and M.~Peardon,
\newblock Phys. Rev. {\bf D60}, 34509 (1999).

\bibitem{bib:phy:sexton95}
J.~Sexton, A.~Vaccarino, and D.~Weingarten,
\newblock Phys. Rev. Lett. {\bf 75}, 4563 (1995).

\bibitem{bib:phy:page00}
P.~Page,
\newblock in {\em Proceedings of the ``pbar2000 Workshop''}, edited by
  D.~Kaplan and H.~Rubin, pages 55--64, Chicago, 2001.

\bibitem{bib:phy:jones2003}
R.~Jones,
\newblock in {\em Proceedings of the workshop ``Gluonic excitations''},
  Jefferson Lab, 2003,
\newblock in print.

\bibitem{bib:phy:etaLoblxa}
A.~Bertin et~al.,
\newblock Phys. Lett. {\bf B361}, 187 (1995).

\bibitem{bib:phy:etaLoblxb}
A.~Bertin et~al.,
\newblock Phys. Lett. {\bf B385}, 493 (1996).

\bibitem{bib:phy:etaLoblxc}
A.~Bertin et~al.,
\newblock Phys. Lett. {\bf B400}, 226 (1997).

\bibitem{bib:phy:etaLoblxd}
C.~Cicalo et~al.,
\newblock Phys. Lett. {\bf B462}, 453 (1999).

\bibitem{bib:phy:etaLoblxe}
F.~Nichitiu et~al.,
\newblock Phys. Lett. {\bf B545}, 261 (2002).

\bibitem{bib:phy:mornpeardon97}
C.~Morningstar and M.~Peardon,
\newblock Phys. Rev. {\bf D56}, 4043 (1997).

\bibitem{Hanhart:2007yq}
C.~Hanhart, Y.~S. Kalashnikova, A.~E. Kudryavtsev, and A.~V. Nefediev,
\newblock Phys. Rev. {\bf D76}, 034007 (2007).

\bibitem{Braaten:2007ft}
E.~Braaten and M.~Lu,
\newblock Phys. Rev. {\bf D77}, 014029 (2008).

\bibitem{lundborg}
A.~Lundborg, T.~Barnes, and U.~Wiedner,
\newblock Phys. Rev. {\bf D73}, 096003 (2006).

\bibitem{omg:xsec:fivepi}
J.~Clayton et~al.,
\newblock Nucl. Phys. {\bf B30}, 605 (1971).

\bibitem{bib:phy:bai:1996wm}
J.~Z. Bai et~al.,
\newblock Phys. Rev. Lett. {\bf 76}, 3502 (1996).

\bibitem{bib:phy:evtgen}
D.~J. Lange,
\newblock Nucl. Instrum. Meth. {\bf A462}, 152 (2001).

\bibitem{bib:phy:Evangelista:1998zg}
C.~Evangelista et~al.,
\newblock Phys. Rev. {\bf D57}, 5370 (1998).

\bibitem{bib:phy:DeRujula:1976kk}
A.~De~Rujula, H.~Georgi, and S.~L. Glashow,
\newblock Phys. Rev. Lett. {\bf 37}, 785 (1976).

\bibitem{bib:phy:Isgur:1991wq}
N.~Isgur and M.~B. Wise,
\newblock Phys. Rev. Lett. {\bf 66}, 1130 (1991).

\bibitem{bib:phy:Godfrey:1986wj}
S.~Godfrey and R.~Kokoski,
\newblock Phys. Rev. {\bf D43}, 1679 (1991).

\bibitem{bib:phy:Godfrey:1985xj}
S.~Godfrey and N.~Isgur,
\newblock Phys. Rev. {\bf D32}, 189 (1985).

\bibitem{bib:phy:DiPierro:2001uu}
M.~Di~Pierro and E.~Eichten,
\newblock Phys. Rev. {\bf D64}, 114004 (2001).

\bibitem{bib:phy:Hagiwara:2002fs}
K.~Hagiwara et~al.,
\newblock Phys. Rev. {\bf D66}, 010001 (2002).

\bibitem{bib:phy:Aubert:2003fg}
B.~Aubert et~al.,
\newblock Phys. Rev. Lett. {\bf 90}, 242001 (2003).

\bibitem{bib:phy:Besson:2003cp}
D.~Besson et~al.,
\newblock Phys. Rev. {\bf D68}, 032002 (2003).

\bibitem{bib:phy:Krokovny:2003zq}
P.~Krokovny et~al.,
\newblock Phys. Rev. Lett. {\bf 91}, 262002 (2003).

\bibitem{bib:phy:Aubert:2003pe}
B.~Aubert et~al.,
\newblock Phys. Rev. {\bf D69}, 031101 (2004).

\bibitem{bib:phy:Aubert:2004pw}
B.~Aubert et~al.,
\newblock Phys. Rev. Lett. {\bf 93}, 181801 (2004).

\bibitem{bib:phy:Abe:2003jk}
K.~Abe et~al.,
\newblock Phys. Rev. Lett. {\bf 92}, 012002 (2004).

\bibitem{bib:phy:Aubert:2006bk}
B.~Aubert et~al.,
\newblock Phys. Rev. {\bf D74}, 032007 (2006).

\bibitem{bib:phy:PDGlive}
http://pdglive.lbl.gov.

\bibitem{bib:phy:Aubert:2006mh}
B.~Aubert,
\newblock Phys. Rev. Lett. {\bf 97}, 222001 (2006).

\bibitem{bib:phy:Swanson:2006st}
E.~Swanson,
\newblock Phys. Rept. {\bf 429}, 243 (2006).

\bibitem{bib:phy:Nowak:1992um}
M.~A. Nowak, M.~Rho, and I.~Zahed,
\newblock Phys. Rev. {\bf D48}, 4370 (1993).

\bibitem{bib:phy:Bardeen:1993ae}
W.~A. Bardeen and C.~T. Hill,
\newblock Phys. Rev. {\bf D49}, 409 (1994).

\bibitem{bib:phy:Bardeen:2003kt}
W.~A. Bardeen, E.~J. Eichten, and C.~T. Hill,
\newblock Phys. Rev. {\bf D68}, 054024 (2003).

\bibitem{bib:phy:Nowak:2003ra}
M.~A. Nowak, M.~Rho, and I.~Zahed,
\newblock Acta Phys. Polon. {\bf B35}, 2377 (2004).

\bibitem{bib:phy:Cahn:2003cw}
R.~N. Cahn and J.~D. Jackson,
\newblock Phys. Rev. {\bf D68}, 037502 (2003).

\bibitem{bib:phy:Lakhina:2006fy}
O.~Lakhina and E.~S. Swanson,
\newblock Phys. Lett. {\bf B650}, 159 (2007).

\bibitem{bib:phy:Terasaki:2003qa}
K.~Terasaki,
\newblock Phys. Rev. {\bf D68}, 011501 (2003).

\bibitem{bib:phy:Hayashigaki:2004st}
A.~Hayashigaki and K.~Terasaki,
\newblock Prog. Theor. Phys. {\bf 114}, 1191 (2006).

\bibitem{bib:phy:Cheng:2003kg}
H.-Y. Cheng and W.-S. Hou,
\newblock Phys. Lett. {\bf B566}, 193 (2003).

\bibitem{bib:phy:Maiani:2004uc}
L.~Maiani, F.~Piccinini, A.~D. Polosa, and V.~Riquer,
\newblock Phys. Rev. Lett. {\bf 93}, 212002 (2004).

\bibitem{bib:phy:Maiani:2004vq}
L.~Maiani, F.~Piccinini, A.~D. Polosa, and V.~Riquer,
\newblock Phys. Rev. {\bf D71}, 014028 (2005).

\bibitem{bib:phy:Chen:2004dy}
Y.-Q. Chen and X.-Q. Li,
\newblock Phys. Rev. Lett. {\bf 93}, 232001 (2004).

\bibitem{bib:phy:Bracco:2005kt}
M.~E. Bracco, A.~Lozea, R.~D. Matheus, F.~S. Navarra, and M.~Nielsen,
\newblock Phys. Lett. {\bf B624}, 217 (2005).

\bibitem{bib:phy:Nielsen:2006tw}
M.~Nielsen, R.~D. Matheus, F.~S. Navarra, M.~E. Bracco, and A.~Lozea,
\newblock AIP Conf. Proc. {\bf 814}, 528 (2006).

\bibitem{bib:phy:vanBeveren:2003kd}
E.~van Beveren and G.~Rupp,
\newblock Phys. Rev. Lett. {\bf 91}, 012003 (2003).

\bibitem{bib:phy:Hwang:2004cd}
D.~S. Hwang and D.-W. Kim,
\newblock Phys. Lett. {\bf B601}, 137 (2004).

\bibitem{bib:phy:Hwang:2005tm}
D.~S. Hwang and D.~W. Kim,
\newblock J. Phys. Conf. Ser. {\bf 9}, 63 (2005).

\bibitem{bib:phy:Barnes:2003dj}
T.~Barnes, F.~E. Close, and H.~J. Lipkin,
\newblock Phys. Rev. {\bf D68}, 054006 (2003).

\bibitem{Kolomeitsev:2003ac}
E.~E. Kolomeitsev and M.~F.~M. Lutz,
\newblock Phys. Lett. {\bf B582}, 39 (2004).

\bibitem{bib:phy:Hofmann:2003je}
J.~Hofmann and M.~F.~M. Lutz,
\newblock Nucl. Phys. {\bf A733}, 142 (2004).

\bibitem{Lutz:2007sk}
M.~Lutz and M.~Soyeur,
\newblock arXiv:hep-ph/0710.1545.

\bibitem{bib:phy:Sassen:2005ej}
F.~P. Sassen and S.~Krewald,
\newblock Int. J. Mod. Phys. {\bf A20}, 705 (2005).

\bibitem{bib:phy:Guo:2007up}
F.-K. Guo, S.~Krewald, and U.-G. Mei\ss{}ner,
\newblock arXiv:hep-ph/0712.2953.

\bibitem{bib:phy:Sassen05phd}
F.~Sassen,
\newblock PhD thesis, Universit\"at Bonn, 2005.

\bibitem{Faessler:2007gv}
A.~Faessler, T.~Gutsche, V.~E. Lyubovitskij, and Y.-L. Ma,
\newblock Phys. Rev. {\bf D76}, 014005 (2007).

\bibitem{Faessler:2007us}
A.~Faessler, T.~Gutsche, V.~E. Lyubovitskij, and Y.-L. Ma,
\newblock Phys. Rev. {\bf D76}, 114008 (2007).

\bibitem{bib:phy:Hanhart05pc}
C.~Hanhart,
\newblock priv. comm., 2005.

\bibitem{bib:phy:Mehen:2004uj}
T.~Mehen and R.~P. Springer,
\newblock Phys. Rev. {\bf D70}, 074014 (2004).

\bibitem{bib:phy:Close:2005se}
F.~E. Close and E.~S. Swanson,
\newblock Phys. Rev. {\bf D72}, 094004 (2005).

\bibitem{bib:phy:Colangelo:2003vg}
P.~Colangelo and F.~De~Fazio,
\newblock Phys. Lett. {\bf B570}, 180 (2003).

\bibitem{bib:phy:Colangelo:2005hv}
P.~Colangelo, F.~De~Fazio, and A.~Ozpineci,
\newblock Phys. Rev. {\bf D72}, 074004 (2005).

\bibitem{bib:phy:Liu:2006jx}
X.~Liu, Y.-M. Yu, S.-M. Zhao, and X.-Q. Li,
\newblock Eur. Phys. J. {\bf C47}, 445 (2006).

\bibitem{bib:phy:Shifman:2007xn}
M.~Shifman and A.~Vainshtein,
\newblock Phys. Rev. {\bf D77}, 034002 (2008).

\bibitem{bib:phy:DPM}
A.~Capella, U.~Sukhatme, C.-I. Tan, and J.~Tran Thanh~Van,
\newblock Phys. Rept. {\bf 236}, 225 (1994).

\bibitem{bib:phy:Hanhart06}
C.~Hanhart,
\newblock priv. comm., 2006.

\bibitem{bib:phy:Klempt:2002vp}
E.~Klempt,
\newblock Phys. Rev. {\bf C66}, 058201 (2002).

\bibitem{bib:phy:Lutz:2001yb}
M.~F.~M. Lutz and E.~E. Kolomeitsev,
\newblock Nucl. Phys. {\bf A700}, 193 (2002).

\bibitem{bib:phy:Lutz:2003jw}
M.~F.~M. Lutz and E.~E. Kolomeitsev,
\newblock Nucl. Phys. {\bf A730}, 110 (2004).

\bibitem{bib:phy:Kolomeitsev:2003kt}
E.~E. Kolomeitsev and M.~F.~M. Lutz,
\newblock Phys. Lett. {\bf B585}, 243 (2004).

\bibitem{bib:phy:Zychor:2005sj}
I.~Zychor et~al.,
\newblock Phys. Rev. Lett. {\bf 96}, 012002 (2006).

\bibitem{bib:phy:Adamovich:2007mc}
M.~I. Adamovich et~al.,
\newblock Eur. Phys. J. {\bf C50}, 535 (2007).

\bibitem{bib:phy:Adamovich:1997ud}
M.~I. Adamovich et~al.,
\newblock Eur. Phys. J. {\bf C5}, 621 (1998).

\bibitem{bib:phy:Adamovich:1999ic}
M.~I. Adamovich et~al.,
\newblock Eur. Phys. J. {\bf C11}, 271 (1999).

\bibitem{bib:phy:Abe:2001mb}
K.~Abe et~al.,
\newblock Phys. Lett. {\bf B524}, 33 (2002).

\bibitem{bib:phy:Ziegler07}
V.~Ziegler,
\newblock PhD thesis, University of Iowa, 2007,
\newblock report SLAC-R-868.

\bibitem{bib:phy:Loring:2001ky}
U.~L\"oring, B.~C. Metsch, and H.~R. Petry,
\newblock Eur. Phys. J. {\bf A10}, 447 (2001).

\bibitem{bib:phy:Guzey:2005vz}
V.~Guzey and M.~V. Polyakov,
\newblock arXiv:hep-ph/0512355, 2005.

\bibitem{bib:phy:Baltay65}
C.~Baltay et~al.,
\newblock Phys. Rev. {\bf 140}, B1027 (1965).

\bibitem{bib:phy:Musgrave65}
B.~Musgrave,
\newblock Nuov. Cim. {\bf 35}, 735 (1965).

\bibitem{bib:phy:Hofmann:2005sw}
J.~Hofmann and M.~F.~M. Lutz,
\newblock Nucl. Phys. {\bf A763}, 90 (2005).

\bibitem{bib:phy:Lutz:2006ya}
M.~F.~M. Lutz and J.~Hofmann,
\newblock Int. J. Mod. Phys. {\bf A21}, 5496 (2006).

\bibitem{bib:phy:Flaminio84}
V.~Flaminio, W.~Moorhead, D.~Morrison, and N.~Rivoire,
\newblock Report CERN-HERA 84-01, 1984.

\bibitem{bib:phy:OZI1}
S.~Okubo,
\newblock Phys. Lett. {\bf 5}, 165 (1963).

\bibitem{bib:phy:OZI2}
G.~Zweig,
\newblock CERN report TH-401.

\bibitem{bib:phy:OZI3}
J.~Iizuka,
\newblock Prog. Theor. Phys. Suppl. {\bf 38}, 21 (1966).

\bibitem{bib:phy:PS185}
T.~Johansson,
\newblock Antihyperon-hyperon production in antiproton-proton collisions,
\newblock in {\em AIP Conf. Proc. Eight Int. Conf. on Low Energy Antiproton
  Physics}, page~95, 2003.

\bibitem{bib:phy:Barnes_164}
P.~D. Barnes et~al.,
\newblock Phys. Rev. C {\bf 54}, 1877 (1996).

\bibitem{bib:phy:Becker}
H.~Becker et~al.,
\newblock Nucl. Phys. B {\bf 141}, 48 (1978).

\bibitem{bib:phy:Klempt02}
E.~Klempt et~al.,
\newblock Phys. Rept. {\bf 368}, 119 (2002).

\bibitem{bib:phy:Alberg01}
M.~Alberg,
\newblock Nucl. Phys. A {\bf 692}, 47c (2001).

\bibitem{bib:phy:Alberg95}
M.~Alberg, J.~Ellis, and D.~Kharzeev,
\newblock Phys. Lett. B {\bf 356}, 113 (1995).

\bibitem{bib:phy:Frodesen}
A.~G. Frodesen, O.~Skjeggestad, and H.~T$\o$fte,
\newblock {\em Probability and Statistics in Particle Physics},
\newblock Universitetsf\"{o}rlaget, Bergen, 1979.

\bibitem{bib:phy:BroFar}
S.~Brodsky and G.~Farrar,
\newblock Phys. Rev. Lett. {\bf 31}, 1153 (1973).

\bibitem{bib:phy:Matveev}
V.~Matveev et~al.,
\newblock Lett. Nuovo Cimento {\bf 7}, 719 (1972).

\bibitem{bib:phy:EfrRad}
A.~V. Efremov and A.~V. Radyushki,
\newblock Phys. Lett. {\bf B94}, 245 (1980).

\bibitem{bib:phy:LepBro}
G.~P. Lepage and S.~Brodsky,
\newblock Phys. Rev. {\bf D22}, 2157 (1980).

\bibitem{bib:phy:Landshoff}
P.~Landshoff,
\newblock Phys. Rev. {\bf D10}, 1024 (1974).

\bibitem{bib:phy:RalPir}
J.~P. Ralston and B.~Pire,
\newblock Phys. Rev. Lett. {\bf 49}, 1605 (1982).

\bibitem{bib:phy:CarCha}
C.~E. Carlson et~al.,
\newblock Phys. Rev. D {\bf 46}, 2891 (1992).

\bibitem{bib:phy:Eide}
A.~Eide et~al.,
\newblock Nucl. Phys. B {\bf 60}, 173 (1973).

\bibitem{bib:phy:Eisenh}
E.~Eisenhandler et~al.,
\newblock Nucl. Phys. B {\bf 96}, 109 (1975).

\bibitem{bib:phy:Buran}
T.~Buran et~al.,
\newblock Nucl. Phys. B {\bf 116}, 51 (1976).

\bibitem{bib:phy:Brown91}
G.~Brown and M.~Rho,
\newblock Phys. Rev. Lett. {\bf 66}, 2720 (1991).

\bibitem{bib:phy:Hatsuda92}
T.~Hatsuda and S.~Lee,
\newblock Phys. Rev. C {\bf 46}, 34 (1992).

\bibitem{bib:phy:Mosel08}
U.~Mosel,
\newblock arXiv:0801.49701 [hep-ph], 2008.

\bibitem{bib:phy:Trnka05}
D.~Trnka et~al.,
\newblock Phys. Rev. Lett {\bf 94}, 191303 (2005).

\bibitem{bib:phy:Naruki:2005kd}
M.~Naruki et~al.,
\newblock Phys. Rev. Lett. {\bf 96}, 092301 (2006).

\bibitem{bib:phy:Muto:2005za}
R.~Muto et~al.,
\newblock Phys. Rev. Lett. {\bf 98}, 042501 (2007).

\bibitem{bib:phy:Nasseripour07}
R.~Nasseripour et~al.,
\newblock Phys. Rev. Lett {\bf 99}, 262302 (2007).

\bibitem{bib:phy:Agakishiev95}
G.~Agakishiev et~al.,
\newblock Phys. Rev. Lett {\bf 75}, 1272 (1995).

\bibitem{bib:phy:Arnaldi:2006jq}
R.~Arnaldi et~al.,
\newblock Phys. Rev. Lett. {\bf 96}, 162302 (2006).

\bibitem{bib:phy:Yamazaki96}
T.~Yamazaki et~al.,
\newblock Z. Phys. A {\bf 355}, 219 (1996).

\bibitem{bib:phy:Geissel02}
H.~Geissel et~al.,
\newblock Phys. Rev. Lett. {\bf 88}, 122301 (2002).

\bibitem{bib:phy:Suzuki04}
K.~Suzuki et~al.,
\newblock Phys. Rev. Lett. {\bf 92}, 072302 (2004).

\bibitem{bib:phy:Messchendorp:2002au}
J.~G. Messchendorp et~al.,
\newblock Phys. Rev. Lett. {\bf 89}, 222302 (2002).

\bibitem{bib:phy:Nekipelov02}
M.~Nekipelov et~al.,
\newblock Phys. Lett. B {\bf 540}, 207 (2002).

\bibitem{bib:phy:Rudy02}
Z.~Rudy et~al.,
\newblock Eur. Phys. J. A {\bf 15}, 303 (2002).

\bibitem{bib:phy:Barth97}
R.~Barth et~al.,
\newblock Phys. Rev. Lett. {\bf 78}, 4007 (1997).

\bibitem{bib:phy:Laue99}
F.~Laue et~al.,
\newblock Phys. Rev. Lett. {\bf 82}, 1640 (1999).

\bibitem{bib:phy:Pfeiffer04}
M.~Pfeiffer et~al.,
\newblock Phys. Rev. Lett. {\bf 92}, 252001 (2004).

\bibitem{bib:phy:Kotulla:2008xy}
M.~Kotulla et~al.,
\newblock Phys. Rev. Lett. {\bf 100}, 192302 (2008).

\bibitem{bib:phy:Ishikawa:2004id}
T.~Ishikawa et~al.,
\newblock Phys. Lett. {\bf B608}, 215 (2005).

\bibitem{bib:phy:Weise01}
W.~Weise,
\newblock in {\em Proc. Int. Workshop on the Structure of Hadrons}, Hischegg,
  Austria, 2001.

\bibitem{bib:phy:Tsushima99}
K.~Tsushima, D.~Lu, A.~Thomas, K.~Saito, and R.~Landau,
\newblock Phys. Rev. C {\bf 59}, 2824 (1999).

\bibitem{bib:phy:Sibirtsev99}
A.~Sibirtsev, K.~Tsushima, and A.~Thomas,
\newblock Eur. Phys. J. A {\bf 6}, 351 (1999).

\bibitem{bib:phy:Hayashigaki00}
A.~Hayashigaki et~al.,
\newblock Phys. Lett. B {\bf 487}, 96 (2000).

\bibitem{bib:phy:Morath01}
P.~Morath,
\newblock PhD thesis, TU M\"unchen, 2001.

\bibitem{bib:phy:Tolos04}
L.~Tol\'{o}s, J.~Schaffner-Bielich, and A.~Mishra,
\newblock Phys. Rev. C {\bf 70}, 025203 (2004).

\bibitem{bib:phy:Lutz06}
M.~Lutz and C.~Korpa,
\newblock Phys. Lett. B {\bf 633}, 43 (2006).

\bibitem{bib:phy:Mizutani06}
T.~Mizutani and A.~Ramos,
\newblock Phys. Rev. C {\bf 74}, 065201 (2006).

\bibitem{bib:phy:Brodsky90}
S.~Brodsky, I.~Schmidt, and G.~de~T\'{e}ramond,
\newblock Phys. Rev. Lett. {\bf 64}, 1011 (1990).

\bibitem{bib:phy:Brodsky97}
S.~Brodsky and G.~Miller,
\newblock Phys. Lett. B {\bf 412}, 125 (1997).

\bibitem{bib:phy:Klingl99}
F.~Klingl et~al.,
\newblock Phys. Rev. Lett. {\bf 82}, 3396 (1999).

\bibitem{bib:phy:Lee03}
S.~Lee and C.~Ko,
\newblock Prog. Theor. Phys. Suppl. {\bf 149}, 173 (2003).

\bibitem{bib:phy:Lee03a}
S.~Lee,
\newblock in {\em Proc. $10^{th}$ Int. Conf. on Hadron Spectroscopy
  (Hadron'03)}, Aschaffenburg, 2003,
\newblock [arXiv:nucl-th/0310080].

\bibitem{bib:phy:Golubeva03}
Y.~Golubeva, E.~Bratkovskaya, W.~Cassing, and L.~Kondratyuk,
\newblock Eur. Phys. J. A {\bf 17}, 275 (2003).

\bibitem{bib:phy:Cassing00}
W.~Cassing, Y.~Golubeva, and L.~Kondratyuk,
\newblock Eur. Phys. J. A {\bf 7}, 279 (2000).

\bibitem{bib:phy:Haidenbauer07}
J.~Haidenbauer, G.~Krein, U.-G. Mei\ss{}ner, and A.~Sibirtsev,
\newblock Eur. Phys. J. A {\bf 33}, 107 (2007).

\bibitem{bib:phy:Anderson77}
R.~Anderson et~al.,
\newblock Phys. Rev. Lett. {\bf 38}, 263 (1977).

\bibitem{bib:phy:Kharzeev97}
D.~Kharzeev, C.~Lourenco, M.~Nardi, and H.~Satz,
\newblock Z. Phys. C {\bf 74}, 307 (1997).

\bibitem{bib:phy:Alessandro:2003pc}
B.~Alessandro et~al.,
\newblock Eur. Phys. J. {\bf C33}, 31 (2004).

\bibitem{bib:phy:Alessandro06}
B.~Alessandro et~al.,
\newblock Eur. Phys. J. C {\bf 48}, 329 (2006).

\bibitem{bib:phy:Adler:2005ph}
S.~S. Adler et~al.,
\newblock Phys. Rev. Lett. {\bf 96}, 012304 (2006).

\bibitem{bib:phy:Adare:2007gn}
A.~Adare et~al.,
\newblock Phys. Rev. {\bf C77}, 024912 (2008).

\bibitem{bib:phy:Matsui86}
T.~Matsui and H.~Satz,
\newblock Phys. Lett. B {\bf 178}, 416 (1986).

\bibitem{bib:phy:Alessandro05}
B.~Alessandro et~al.,
\newblock Eur. Phys. J. C {\bf 39}, 335 (2005).

\bibitem{bib:phy:Arnaldi05}
R.~Arnaldi et~al.,
\newblock Eur. Phys. J. C {\bf 43}, 167 (2005).

\bibitem{bib:phy:Arnaldi:2007zz}
R.~Arnaldi et~al.,
\newblock Phys. Rev. Lett. {\bf 99}, 132302 (2007).

\bibitem{bib:phy:Sibirtsev:2000aw}
A.~Sibirtsev, K.~Tsushima, and A.~W. Thomas,
\newblock Phys. Rev. {\bf C63}, 044906 (2001).

\bibitem{bib:phy:Oh07}
A.~Sibirtsev, K.~Tsushima, and A.~Thomas,
\newblock Phys. Rev. C {\bf 75}, 064903 (2007).

\bibitem{bib:phy:Hilbert07}
J.~Hilbert, N.~Black, T.~Barnes, and E.~Swanson,
\newblock Phys. Rev. C {\bf 75}, 064907 (2007).

\bibitem{bib:phy:Seth01}
K.~Seth,
\newblock in {\em Proc. Int. Workshop on the Structure of Hadrons}, Hischegg,
  Austria, 2001.

\bibitem{bib:phy:Gerland05}
L.~Gerland, L.~Frankfurt, and M.~Strikman,
\newblock Phys. Lett. B {\bf 619}, 95 (2005).

\bibitem{bib:phy:PDG06}
W.-M. Yao et~al.,
\newblock J. Phys. G {\bf 33}, 1 (2006).

\bibitem{bib:phy:Sibirtsev08-eg}
A.~Sibirtsev, 2008,
\newblock priv. communication.

\bibitem{bib:phy:Tanimori85}
T.~Tanimori et~al.,
\newblock Phys. Rev. Lett. {\bf 55}, 1835 (1985).

\bibitem{bib:phy:Armstrong:1997gv}
T.~A. Armstrong et~al.,
\newblock Phys. Rev. {\bf D56}, 2509 (1997).

\bibitem{bib:phy:Andreotti:2003sk}
M.~Andreotti et~al.,
\newblock Phys. Rev. Lett. {\bf 91}, 091801 (2003).

\bibitem{bib:phy:UrQMD1}
M.~Bleicher et~al.,
\newblock J. Phys. {\bf G25}, 1859 (1999).

\bibitem{bib:phy:UrQMD2}
S.~A. Bass et~al.,
\newblock Prog. Part. Nucl. Phys. {\bf 41}, 225 (1998).

\bibitem{bib:phy:Faessler:1982qt}
A.~Faessler, G.~Lubeck, and K.~Shimizu,
\newblock Phys. Rev. {\bf D26}, 3280 (1982).

\bibitem{bib:phy:Buervenich02}
T.~B\"u{}rvenich et~al.,
\newblock Phys. Lett. B {\bf 542}, 261 (2002).

\bibitem{bib:phy:Mishustin:2004xa}
I.~N. Mishustin, L.~M. Satarov, T.~J. B\"urvenich, H.~St\"ocker, and
  W.~Greiner,
\newblock Phys. Rev. {\bf C71}, 035201 (2005).

\bibitem{bib:phy:Larionov08}
A.~Larionov, I.~Mishustin, L.~Saratov, and W.~Greiner,
\newblock arXiv:0802.1845 [nucl-th]  (2008).

\bibitem{bib:phy:Wong84}
C.-Y. Wong, A.~Kerman, G.~Satchler, and A.~Mackellar,
\newblock Phys. Rev. C {\bf 29}, 574 (1984).

\bibitem{bib:phy:Batty97}
C.~Batty, E.~Friedman, and A.~Gal,
\newblock Phys. Rept. {\bf 287}, 385 (1997).

\bibitem{bib:phy:Friedman05}
E.~Friedman, A.~Gal, and J.~Mare\v{s},
\newblock Nucl. Phys. A {\bf 761}, 283 (2005).

\bibitem{bib:phy:Teis94}
S.~Teis, W.~Cassing, T.~Maruyama, and U.~Mosel,
\newblock Phys. Rev. C {\bf 50}, 388 (1994).

\bibitem{bib:phy:Spieles:1995fs}
C.~Spieles et~al.,
\newblock Phys. Rev. {\bf C53}, 2011 (1996).

\bibitem{bib:phy:Sibirtsev:1997mq}
A.~Sibirtsev, W.~Cassing, G.~I. Lykasov, and M.~V. Rzyanin,
\newblock Nucl. Phys. {\bf A632}, 131 (1998).

\bibitem{bib:phy:Pochodzalla:2008ju}
J.~Pochodzalla,
\newblock Phys. Lett. {\bf B669}, 306 (2008).

\bibitem{bib:phy:Jain:1995dd}
P.~Jain, B.~Pire, and J.~P. Ralston,
\newblock Phys. Rept. {\bf 271}, 67 (1996).

\bibitem{bib:phy:Frankfurt:1992dx}
L.~Frankfurt, G.~A. Miller, and M.~Strikman,
\newblock Comments Nucl. Part. Phys. {\bf 21}, 1 (1992).

\bibitem{bib:phy:Frankfurt:1994hf}
L.~L. Frankfurt, G.~A. Miller, and M.~Strikman,
\newblock Ann. Rev. Nucl. Part. Sci. {\bf 44}, 501 (1994).

\bibitem{bib:phy:Aitala:2000hc}
E.~M. Aitala et~al.,
\newblock Phys. Rev. Lett. {\bf 86}, 4773 (2001).

\bibitem{bib:phy:Farrar:1988me}
G.~R. Farrar, H.~Liu, L.~L. Frankfurt, and M.~I. Strikman,
\newblock Phys. Rev. Lett. {\bf 61}, 686 (1988).

\bibitem{bib:phy:Clasie:2007gqa}
B.~Clasie et~al.,
\newblock Phys. Rev. Lett. {\bf 99}, 242502 (2007).

\bibitem{bib:phy:Pire:1982iv}
B.~Pire and J.~P. Ralston,
\newblock Phys. Lett. {\bf B117}, 233 (1982).

\bibitem{bib:phy:Ralston:1982pa}
J.~P. Ralston and B.~Pire,
\newblock Phys. Rev. Lett. {\bf 49}, 1605 (1982).

\bibitem{bib:phy:Carroll:1988rp}
A.~S. Carroll et~al.,
\newblock Phys. Rev. Lett. {\bf 61}, 1698 (1988).

\bibitem{bib:phy:Aclander:2004zm}
J.~L.~S. Aclander et~al.,
\newblock Phys. Rev. {\bf C70}, 015208 (2004).

\bibitem{bib:phy:Ralston:1988rb}
J.~P. Ralston and B.~Pire,
\newblock Phys. Rev. Lett. {\bf 61}, 1823 (1988).

\bibitem{bib:phy:Ralston:1990jj}
J.~P. Ralston and B.~Pire,
\newblock Phys. Rev. Lett. {\bf 65}, 2343 (1990).

\bibitem{bib:phy:Brodsky:1987xw}
S.~J. Brodsky and G.~F. de~Teramond,
\newblock Phys. Rev. Lett. {\bf 60}, 1924 (1988).

\bibitem{bib:phy:Jennings:1989hc}
B.~K. Jennings and G.~A. Miller,
\newblock Phys. Lett. {\bf B236}, 209 (1990).

\bibitem{bib:phy:Brodsky:1988xz}
S.~J. Brodsky and A.~H. Mueller,
\newblock Phys. Lett. {\bf B206}, 685 (1988).

\bibitem{bib:phy:Farrar:1989vr}
G.~R. Farrar, L.~L. Frankfurt, M.~I. Strikman, and H.~Liu,
\newblock Nucl. Phys. {\bf B345}, 125 (1990).

\bibitem{bib:phy:Gerland:1998bz}
L.~Gerland, L.~Frankfurt, M.~Strikman, H.~Stoecker, and W.~Greiner,
\newblock Phys. Rev. Lett. {\bf 81}, 762 (1998).

\bibitem{bib:phy:gre95}
W.~Greiner,
\newblock Int. Journal of Modern Physics E {\bf 5}, 1 (1995).

\bibitem{bib:phy:hae07}
P.~Haensel, A.Y.Potekhin, and D.~Yakovlev,
\newblock {\em Neutron Stars 1. Equation of state and structure},
\newblock Springer, 2007.

\bibitem{bib:phy:web07}
F.~Weber, R.~Negreiros, and P.~Rosenfield,
\newblock Springer Lecture Notes ,
\newblock arXiv:arXiv:0705.2708.

\bibitem{bib:phy:jha08}
T.~K. Jha, H.~Mishra, and V.~Sreekanth,
\newblock Phys. Rev. C {\bf 77}, 045801 (2008).

\bibitem{bib:phy:dap08}
H.~Dapo, B.-J. Sch\"afer, and J.~Wambach,
\newblock arXiv:0811.29391 [nucl-th].

\bibitem{bib:phy:kai05}
N.~Kaiser and W.~Weise,
\newblock Phys. Rev. C {\bf 71}, 015203 (2005).

\bibitem{bib:phy:wir02}
R.~B. Wiringa and S.~C. Pieper,
\newblock Phys. Rev. Lett. {\bf 89}, 182501 (2002).

\bibitem{bib:phy:pie04}
S.~C. Pieper, R.~B. Wiringa, and J.~Carlson,
\newblock Phys. Rev. C {\bf 70}, 054325 (2004).

\bibitem{bib:phy:pie07}
S.~C. Pieper,
\newblock Quantum Monte Carlo Calculations of Light Nuclei,
\newblock in {\em Lecture notes for Course CLXIX - "Nuclear Structure far from
  Stability: New Physics and new Technology"}, 2007,
\newblock arXiv:0711.1500v1 [nucl-th].

\bibitem{bib:phy:ste06}
I.~Stetcu, B.~Barrett, and U.~van Kolck,
\newblock Phys. Rev. C {\bf 73}, 037307 (2006).

\bibitem{bib:phy:kei00}
C.~Keil, F.~Hofmann, and H.~Lenske,
\newblock Phys. Rev. C {\bf 61}, 06401 (2000).

\bibitem{bib:phy:hof01}
F.~Hofmann, C.~Keil, and H.~Lenske,
\newblock Phys. Rev. C {\bf 64}, 034314 (2001).

\bibitem{bib:phy:bea05}
S.~Beane, P.~Bedaque, A.~Parreno, and M.~Savage,
\newblock Nucl. Phys. A {\bf 747}, 55 (2005).

\bibitem{bib:phy:sav07}
M.~Savage,
\newblock Baryon-baryon interactions from the lattice,
\newblock in {\em Proc. of the IX International Conference on Hypernuclear and
  Strange Particle Physics}, edited by J.~Pochodzalla and T.~Walcher, page 301,
  SIF and Springer-Verlag Berlin Heidelberg, 2007,
\newblock arXiv: 0612063 [nucl-th].

\bibitem{bib:phy:gom02}
A.~G. Nicola and J.~Pelaez,
\newblock Phys. Rev. D {\bf 65}, 054009 (2002).

\bibitem{bib:phy:pol06}
H.~Polinder, J.~Haidenbauer, and U.-G.~M. ner,
\newblock Nucl. Phys. A {\bf 779}, 244 (2006).

\bibitem{bib:phy:tol06}
L.~Tolos, A.~Ramos, and E.~Oset,
\newblock Phys. Rev. C {\bf 74}, 015203 (2006).

\bibitem{bib:phy:lut07}
M.~Lutz, C.~Korpa, and M.~Moeller,
\newblock Nucl. Phys. A {\bf 808}, 124 (2008).

\bibitem{bib:phy:ste07}
I.~Stetcu, B.~Barrett, and U.~van Kolck,
\newblock Phys. Lett. B {\bf 653}, 358 (2007).

\bibitem{bib:phy:bor07}
B.~Borasoy, E.~Epelbaum, H.~Krebs, D.~Lee, and U.-G. Mei{\ss}ner,
\newblock Eur. Phys. J. A {\bf 31}, 105 (2007).

\bibitem{bib:phy:nav07}
P.~Navratil et~al.,
\newblock Light nuclei from chiral EFT interactions,
\newblock in {\em Proceedings of the 20th European Conference on Few-Body
  Problems in Physics (EFB20)}, 2007.

\bibitem{bib:phy:kon00}
Y.~Kondo et~al.,
\newblock Nucl. Phys. A {\bf 676}, 371 (2000).

\bibitem{bib:phy:ahn06}
J.~Ahn et~al.,
\newblock Phys. Lett. B {\bf 633}, 214 (2006).

\bibitem{bib:phy:has06}
O.~Hashimoto and H.~Tamura,
\newblock Prog. Part. Nucl. Phys. {\bf 57}, 566 (2006).

\bibitem{bib:phy:ito01}
K.~Itonaga, T.~Ueda, and T.~Motoba,
\newblock Nucl. Phys. A {\bf 691}, 197c (2001).

\bibitem{bib:phy:par01}
A.~Parreno, A.~Ramos, and C.~Bennhold,
\newblock Nucl. Phys. A {\bf 65}, 015205 (2001).

\bibitem{bib:phy:sas03}
K.~Sasaki, T.~Inoue, and M.~Oka,
\newblock Nucl. Phys. A {\bf 726}, 349 (2003).

\bibitem{bib:phy:dub96}
J.~Dubach, G.~B. Feldman, B.~R. Holstein, and L.~de~la Torre,
\newblock Ann. Phys. {\bf 249}, 146 (1996).

\bibitem{bib:phy:mot01}
T.~Motoba,
\newblock Nucl. Phys. A {\bf 691}, 231c (2001).

\bibitem{bib:phy:rij06}
T.~Rijken and Y.~Yamamoto,
\newblock arXiv:hep-ph/0207358v1.

\bibitem{bib:phy:tan07}
K.~Tanida et~al.,
\newblock Plan for the measurement of $\Xi^-$-atomic X rays at J-PARC,
\newblock in {\em Proc. of the IX International Conference on Hypernuclear and
  Strange Particle Physics}, edited by J.~Pochodzalla and T.~Walcher, page 145,
  SIF and Springer-Verlag Berlin Heidelberg, 2007.

\bibitem{bib:phy:buc97}
A.~Buchmann,
\newblock Z. Naturforschung {\bf 52}, 877 (1997).

\bibitem{bib:phy:buc03}
A.~Buchmann and R.~F. Lebed,
\newblock Phys. Rev. D {\bf 67}, 016002 (2003).

\bibitem{bib:phy:jaf77a}
R.~Jaffe,
\newblock Phys. Rev. Lett. {\bf 38}, 195 (1977).

\bibitem{bib:phy:jaf77b}
R.~Jaffe,
\newblock Phys. Rev. Lett. {\bf 38}, 617E (1977).

\bibitem{bib:phy:sak00}
T.~Sakai, K.~Shimizu, and K.~Yazaki,
\newblock Prog. Theor. Phys. Suppl. {\bf 137}, 121 (2000).

\bibitem{bib:phy:yam00}
T.~Yamada and C.~Nakamoto,
\newblock Phys. Rev. C {\bf 62}, 034319 (2000).

\bibitem{bib:phy:myi03}
K.~S. Myint, S.~Shinmura, and Y.~Akaishi,
\newblock Eur. Phys. J. A {\bf 16}, 21 (2003).

\bibitem{bib:phy:afn03}
I.~Afnan and B.~Gibson,
\newblock Phys. Rev. C {\bf 67}, 017001 (2003).

\bibitem{bib:phy:fil03}
I.~Filikhin, A.~Gal, and V.~Suslov,
\newblock Phys. Rev. C {\bf 68}, 024002 (2003).

\bibitem{bib:phy:poc05}
J.~Pochodzalla,
\newblock Nucl. Phys. A {\bf 754}, 430c (2005).

\bibitem{bib:phy:fer07}
F.~Ferro, M.~Agnello, F.~Iazzi, and K.~Szymanska,
\newblock Nucl. Phys. A {\bf 789}, 209 (2007).

\bibitem{bib:phy:loz08}
W.~R. Lozowski, D.~Steski, H.~Huang, and C.~Naylor,
\newblock Nucl. Instr. and Meth. in Physics Research A {\bf 590}, 157 (2008).

\bibitem{bib:phy:kai94}
P.~V. A.B.~Kaidalov,
\newblock Z. Phy. C {\bf 63}, 517 (1994).

\bibitem{bib:phy:Yam95}
T.~Yamada and K.~Ikeda,
\newblock Nucl. Phys. {\bf A585}, 79c,
\newblock 1995.

\bibitem{bib:phy:yam97}
T.~Yamada and K.~Ikeda,
\newblock Phys. Rev. C {\bf 56}, 3216 (1997).

\bibitem{bib:phy:hir97}
Y.~Hirata, Y.~Nara, A.~Ohnishi, T.~Harada, and J.~Randrup,
\newblock Nucl. Phys. A {\bf 639}, 389c (1998).

\bibitem{bib:phy:hir99}
Y.~Hirata, Y.~Nara, A.~Ohnishi, T.~Harada, and J.~Randrup,
\newblock Prog. Theor. Phys. {\bf 102}, 89 (1999).

\bibitem{bib:phy:mil94}
D.~Millener, C.~Dover, and A.~Gal,
\newblock Prog. Theor. Phys. Suppl. {\bf 117}, 307 (1994).

\bibitem{bib:phy:ike94}
K.~Ikeda et~al.,
\newblock Prog. Theor. Phys. {\bf 91}, 747 (1994).

\bibitem{bib:phy:lor09}
A.~S. Lorente, A.~Botvina, and J.~Pochodzalla,
\newblock (in preparation).

\bibitem{bib:phy:fer50}
E.~Fermi,
\newblock Progr. Theor. Phys. {\bf 5}, 570 (1950).

\bibitem{bib:phy:bon95}
J.~Bondorf, A.~Botvina, A.~Iljinov, I.~Mishustin, and K.~Sneppen,
\newblock Phys. Rep. {\bf 257}, 133 (1995).

\bibitem{bib:phy:bot90}
A.~Botvina, Y.~Golubeva, and A.~Iljinov,
\newblock INR, P-0657, Moscow  (1990).

\bibitem{bib:phy:bot93}
A.~Botvina et~al.,
\newblock Z. Phys. A {\bf 345}, 413 (1993).

\bibitem{bib:phy:sud93}
A.~Sudov et~al.,
\newblock Nucl. Phys. A {\bf 554}, 223 (1993).

\bibitem{bib:phy:dan63a}
M.~Danysz et~al.,
\newblock Nucl. Phys. {\bf 49}, 121 (1963).

\bibitem{bib:phy:pro66}
D.~Prowse,
\newblock Phys. Rev. Lett {\bf 17}, 782 (1966).

\bibitem{bib:phy:aok91a}
S.~Aoki et~al.,
\newblock Prog. Theor. Phys. {\bf 85}, 1287 (1991).

\bibitem{bib:phy:ahn01}
J.~Ahn et~al.,
\newblock Phys. Rev. Lett. {\bf 87}, 132504 (2001).

\bibitem{bib:phy:tak01}
H.~Takahashi et~al.,
\newblock Phys. Rev. Lett. {\bf 87}, 212501 (2001).

\bibitem{bib:phy:dal89}
R.~Dalitz et~al.,
\newblock Proc. R. Soc. Lond. A {\bf 426}, 1 (1989).

\bibitem{bib:phy:dov91}
C.~Dover, D.~Millener, A.~Gal, and D.~Davis,
\newblock Phys. Rev. C {\bf 44}, 1905 (1991).

\bibitem{bib:phy:hiy02}
E.~Hiyama et~al.,
\newblock Phys. Rev. C {\bf 66}, 024007 (2002).

\bibitem{bib:phy:ran07}
S.~Randeniya and E.~Hungerford,
\newblock Phys. Rev. C {\bf 76}, 064308 (2007).

\bibitem{bib:phy:lan04}
D.~Lanskoy and Y.~Yamamoto,
\newblock Phys. Rev. C {\bf 69}, 014303 (2004).

\bibitem{bib:phy:sho04}
M.~Shoeb,
\newblock Phys. Rev. C {\bf 69}, 054003 (2004).

\bibitem{bib:phy:nem05}
H.~Nemura, S.~Shinmura, Y.~Akaishi, and K.~S. Myint,
\newblock Phys. Rev. Lett. {\bf 94}, 202502 (2005).

\bibitem{bib:phy:fil02}
I.~Filikhin and A.~Gal,
\newblock Phys. Rev. Lett. {\bf 89}, 172502 (2002).

\bibitem{bib:phy:nem03}
H.~Nemura, Y.~Akaishi, and K.~S. Myint,
\newblock Phys. Rev. C {\bf 67}, 051001(R) (2003).

\bibitem{bib:phy:sho05}
M.~Shoeb,
\newblock Phys. Rev. C {\bf 71}, 024004 (2005).

\bibitem{bib:phy:ban90}
H.~Bando, T.~Motoba, and J.~Zofka,
\newblock Int. J. Mod. Phys. {\bf A5}, 4021 (1990).

\bibitem{bib:phy:ajz85}
F.~Ajzenberg-Selove,
\newblock Nucl. Phys. A {\bf 433}, 1 (1985).

\bibitem{bib:phy:kop70}
G.~Kopylov,
\newblock {\em Principles of resonance kinematics},
\newblock Moscow, 1970.

\bibitem{bib:phy:dov83}
C.~Dover and A.~Gal,
\newblock Ann. Phys. {\bf 147}, 309 (1983).

\bibitem{bib:phy:fri07}
E.~Friedman and A.~Gal,
\newblock Phys, Report {\bf 452}, 89 (2007).

\bibitem{bib:phy:bat99}
C.~J. Batty, E.~Friedman, and A.~Gal,
\newblock Phys. Rev. C {\bf 59}, 295 (1999).

\bibitem{bib:phy:aok95}
S.~Aoki et~al.,
\newblock Phys. Lett. B {\bf 355}, 45 (1995).

\bibitem{bib:phy:san07}
A.~S. Lorente et~al.,
\newblock Nucl. Inst. Meth A {\bf 573}, 410 (2007).

\bibitem{bib:phy:kav07}
M.~Kavatsyuk et~al.,
\newblock GSI Annual report  (2007).

\bibitem{bib:phy:lel03}
P.~Leleux et~al.,
\newblock Astronomy and Astrophysics {\bf 411}, L85 (2003).

\bibitem{bib:phy:bam73}
A.~Bamberger et~al.,
\newblock Nucl. Phys. B {\bf 60}, 1 (1973).

\bibitem{bib:phy:bed76}
M.~Bedjidian et~al.,
\newblock Phys. Lett. {\bf 62}, 467 (1976).

\bibitem{bib:phy:may83}
M.~May et~al.,
\newblock Phys. Rev. Lett. {\bf 51}, 2085 (1983).

\bibitem{bib:phy:aki02}
H.~Akikawa et~al.,
\newblock Phys. Rev. Lett. {\bf 88}, 082501 (2002).

\bibitem{bib:phy:gal08}
A.~Galoyan,
\newblock Private Comunication.

\bibitem{ref:nuclstruc:gda:Belitsky:2005qn}
A.~V. Belitsky and A.~V. Radyushkin,
\newblock Phys. Rept. {\bf 418}, 1 (2005).

\bibitem{ref:nuclstruc:gda:Ji:2004gf}
X.~Ji,
\newblock Ann. Rev. Nucl. Part. Sci. {\bf 54}, 413 (2004).

\bibitem{ref:nuclstruc:gda:Diehl:2003ny}
M.~Diehl,
\newblock Phys. Rept. {\bf 388}, 41 (2003).

\bibitem{ref:nuclstruc:gda:Goeke:2001tz}
K.~Goeke, M.~V. Polyakov, and M.~Vanderhaeghen,
\newblock Prog. Part. Nucl. Phys. {\bf 47}, 401 (2001).

\bibitem{ref:nuclstruc:gda:Ji:1996ek}
X.-D. Ji,
\newblock Phys. Rev. Lett. {\bf 78}, 610 (1997).

\bibitem{ref:nuclstruc:gda:Radyushkin:1996nd}
A.~V. Radyushkin,
\newblock Phys. Lett. {\bf B380}, 417 (1996).

\bibitem{ref:nuclstruc:gda:Ji:1996nm}
X.-D. Ji,
\newblock Phys. Rev. {\bf D55}, 7114 (1997).

\bibitem{ref:nuclstruc:gda:Radyushkin:1998rt}
A.~V. Radyushkin,
\newblock Phys. Rev. {\bf D58}, 114008 (1998).

\bibitem{ref:nuclstruc:gda:Diehl:1998kh}
M.~Diehl, T.~Feldmann, R.~Jakob, and P.~Kroll,
\newblock Eur. Phys. J. {\bf C8}, 409 (1999).

\bibitem{ref:nuclstruc:gda:Kroll:2005ni}
P.~Kroll and A.~Schafer,
\newblock Eur. Phys. J. {\bf A26}, 89 (2005).

\bibitem{ref:nuclstruc:gda:Kuo:2005nr}
C.-C. Kuo et~al.,
\newblock Phys. Lett. {\bf B621}, 41 (2005).

\bibitem{ref:nuclstruc:gda:Lansberg:2006uh}
B.~Pire and L.~Szymanowski,
\newblock Phys. Rev. {\bf D71}, 111501 (2005).

\bibitem{ref:nuclstruc:gda:Pire:2005ax}
B.~Pire and L.~Szymanowski,
\newblock Phys. Lett. {\bf B622}, 83 (2005).

\bibitem{ref:nuclstruc:gda:Freund:2002cq}
A.~Freund, A.~V. Radyushkin, A.~Schafer, and C.~Weiss,
\newblock Phys. Rev. Lett. {\bf 90}, 092001 (2003).

\bibitem{ref:nuclstruc:gda:Armstrong:1997gv}
T.~A. Armstrong et~al.,
\newblock Phys. Rev. {\bf D56}, 2509 (1997).

\bibitem{ref:nuclstruc:gda:Pire:2005xx}
J.~Lansberg, B.~Pire, and L.~Szymanowski,
\newblock Phys. Rev. {\bf D76}, 111502 (2007).

\bibitem{ref:nuclstruc:gda:ONG}
S.~Ong and J.~Van~de Wiele,
\newblock HAL:in2p3-00222925 (2008) .

\bibitem{bib:nuclstruc:spinstruc:Ralston:1979ys}
J.~P. Ralston and D.~E. Soper,
\newblock Nucl. Phys. {\bf B152}, 109 (1979).

\bibitem{bib:nuclstruc:spinstruc:Boer:1999mm}
D.~Boer,
\newblock Phys. Rev. {\bf D60}, 014012 (1999).

\bibitem{bib:nuclstruc:spinstruc:Boer:1997nt}
D.~Boer and P.~J. Mulders,
\newblock Phys. Rev. {\bf D57}, 5780 (1998).

\bibitem{bib:nuclstruc:spinstruc:Collins:1977iv}
J.~C. Collins and D.~E. Soper,
\newblock Phys. Rev. {\bf D16}, 2219 (1977).

\bibitem{bib:nuclstruc:spinstruc:Artru:1989zv}
X.~Artru and M.~Mekhfi,
\newblock Z. Phys. {\bf C45}, 669 (1990).

\bibitem{bib:nuclstruc:spinstruc:Cortes:1991ja}
J.~L. Cortes, B.~Pire, and J.~P. Ralston,
\newblock Z. Phys. {\bf C55}, 409 (1992).

\bibitem{bib:nuclstruc:spinstruc:Jaffe:1991kp}
R.~L. Jaffe and X.-D. Ji,
\newblock Phys. Rev. Lett. {\bf 67}, 552 (1991).

\bibitem{bib:nuclstruc:spinstruc:Soffer:1994ww}
J.~Soffer,
\newblock Phys. Rev. Lett. {\bf 74}, 1292 (1995).

\bibitem{bib:nuclstruc:spinstruc:Bourrely:2003bw}
C.~Bourrely and J.~Soffer,
\newblock Eur. Phys. J. {\bf C36}, 371 (2004).

\bibitem{bib:nuclstruc:spinstruc:deFlorian:2005yj}
D.~de~Florian and W.~Vogelsang,
\newblock Phys. Rev. {\bf D71}, 114004 (2005).

\bibitem{bib:nuclstruc:spinstruc:Field:1976ve}
R.~D. Field and R.~P. Feynman,
\newblock Phys. Rev. {\bf D15}, 2590 (1977).

\bibitem{bib:nuclstruc:spinstruc:Feynman:1978dt}
R.~P. Feynman, R.~D. Field, and G.~C. Fox,
\newblock Phys. Rev. {\bf D18}, 3320 (1978).

\bibitem{bib:nuclstruc:spinstruc:Bacchetta:2004jz}
A.~Bacchetta, U.~D'Alesio, M.~Diehl, and C.~A. Miller,
\newblock Phys. Rev. {\bf D70}, 117504 (2004).

\bibitem{bib:nuclstruc:spinstruc:Barone:2001sp}
V.~Barone, A.~Drago, and P.~G. Ratcliffe,
\newblock Phys. Rept. {\bf 359}, 1 (2002).

\bibitem{bib:nuclstruc:spinstruc:Mulders:1995dh}
P.~J. Mulders and R.~D. Tangerman,
\newblock Nucl. Phys. {\bf B461}, 197 (1996).

\bibitem{bib:nuclstruc:spinstruc:Idilbi:2004vb}
A.~Idilbi, X.-d. Ji, J.-P. Ma, and F.~Yuan,
\newblock Phys. Rev. {\bf D70}, 074021 (2004).

\bibitem{bib:nuclstruc:spinstruc:Boer:2001he}
D.~Boer,
\newblock Nucl. Phys. {\bf B603}, 195 (2001).

\bibitem{bib:nuclstruc:spinstruc:Kane:1978nd}
G.~L. Kane, J.~Pumplin, and W.~Repko,
\newblock Phys. Rev. Lett. {\bf 41}, 1689 (1978).

\bibitem{bib:nuclstruc:spinstruc:D'Alesio:2007jt}
U.~D'Alesio and F.~Murgia,
\newblock Prog. Part. Nucl. Phys. 61 {\bf 2008}, 394 (2008).

\bibitem{bib:nuclstruc:spinstruc:Ji:2006ub}
X.~Ji, J.-W. Qiu, W.~Vogelsang, and F.~Yuan,
\newblock Phys. Rev. Lett. {\bf 97}, 082002 (2006).

\bibitem{bib:nuclstruc:spinstruc:Ji:2006vf}
X.~Ji, J.-w. Qiu, W.~Vogelsang, and F.~Yuan,
\newblock Phys. Rev. {\bf D73}, 094017 (2006).

\bibitem{bib:nuclstruc:spinstruc:Ji:2006br}
X.~Ji, J.-W. Qiu, W.~Vogelsang, and F.~Yuan,
\newblock Phys. Lett. {\bf B638}, 178 (2006).

\bibitem{bib:nuclstruc:spinstruc:Koike:2007dg}
Y.~Koike, W.~Vogelsang, and F.~Yuan,
\newblock Phys. Lett. {\bf B659}, 878 (2008).

\bibitem{bib:nuclstruc:spinstruc:Brandenburg:1993cj}
A.~Brandenburg, O.~Nachtmann, and E.~Mirkes,
\newblock Z. Phys. {\bf C60}, 697 (1993).

\bibitem{bib:nuclstruc:spinstruc:Lam:1978pu}
C.~S. Lam and W.-K. Tung,
\newblock Phys. Rev. {\bf D18}, 2447 (1978).

\bibitem{bib:nuclstruc:spinstruc:Lam:1978zr}
C.~S. Lam and W.-K. Tung,
\newblock Phys. Lett. {\bf B80}, 228 (1979).

\bibitem{bib:nuclstruc:spinstruc:Lam:1980uc}
C.~S. Lam and W.-K. Tung,
\newblock Phys. Rev. {\bf D21}, 2712 (1980).

\bibitem{bib:nuclstruc:spinstruc:Badier:1981ti}
J.~Badier et~al.,
\newblock Zeit. Phys. {\bf C11}, 195 (1981).

\bibitem{bib:nuclstruc:spinstruc:Falciano:1986wk}
S.~Falciano et~al.,
\newblock Z. Phys. {\bf C31}, 513 (1986).

\bibitem{bib:nuclstruc:spinstruc:Guanziroli:1987rp}
M.~Guanziroli et~al.,
\newblock Z. Phys. {\bf C37}, 545 (1988).

\bibitem{bib:nuclstruc:spinstruc:Conway:1989fs}
J.~S. Conway et~al.,
\newblock Phys. Rev. {\bf D39}, 92 (1989).

\bibitem{bib:nuclstruc:spinstruc:Berger:1979du}
E.~L. Berger and S.~J. Brodsky,
\newblock Phys. Rev. Lett. {\bf 42}, 940 (1979).

\bibitem{bib:nuclstruc:spinstruc:Berger:1979xz}
E.~L. Berger,
\newblock Z. Phys. {\bf C4}, 289 (1980).

\bibitem{bib:nuclstruc:spinstruc:Brandenburg:1994wf}
A.~Brandenburg, S.~J. Brodsky, V.~V. Khoze, and D.~Mueller,
\newblock Phys. Rev. Lett. {\bf 73}, 939 (1994).

\bibitem{bib:nuclstruc:spinstruc:Eskola:1994py}
K.~J. Eskola, P.~Hoyer, M.~Vanttinen, and R.~Vogt,
\newblock Phys. Lett. {\bf B333}, 526 (1994).

\bibitem{bib:nuclstruc:spinstruc:Jaffe:1993xb}
R.~L. Jaffe and X.-D. Ji,
\newblock Phys. Rev. Lett. {\bf 71}, 2547 (1993).

\bibitem{bib:nuclstruc:spinstruc:Nachtmann:1983uz}
O.~Nachtmann and A.~Reiter,
\newblock Z. Phys. {\bf C24}, 283 (1984).

\bibitem{bib:nuclstruc:spinstruc:Boer:2004mv}
D.~Boer, A.~Brandenburg, O.~Nachtmann, and A.~Utermann,
\newblock Eur. Phys. J. {\bf C40}, 55 (2005).

\bibitem{bib:nuclstruc:spinstruc:Gamberg:2005ip}
L.~P. Gamberg and G.~R. Goldstein,
\newblock Phys. Lett. {\bf B650}, 362 (2007).

\bibitem{bib:nuclstruc:spinstruc:Zhu:2006gx}
L.~Y. Zhu et~al.,
\newblock Phys. Rev. Lett. {\bf 99}, 082301 (2007).

\bibitem{bib:nuclstruc:spinstruc:Bianconi:2005px}
A.~Bianconi and M.~Radici,
\newblock J. Phys. {\bf G31}, 645 (2005).

\bibitem{bib:nuclstruc:spinstruc:Sivers:1990fh}
D.~W. Sivers,
\newblock Phys. Rev. {\bf D43}, 261 (1991).

\bibitem{bib:nuclstruc:spinstruc:Collins:1992kk}
J.~C. Collins,
\newblock Nucl. Phys. {\bf B396}, 161 (1993).

\bibitem{bib:nuclstruc:spinstruc:Jaffe:1996zw}
R.~L. Jaffe,
\newblock hep-ph/9602236  (1996).

\bibitem{bib:nuclstruc:spinstruc:Sivers:1989cc}
D.~W. Sivers,
\newblock Phys. Rev. {\bf D41}, 83 (1990).

\bibitem{bib:nuclstruc:spinstruc:Efremov:2003tf}
A.~V. Efremov, K.~Goeke, and P.~Schweitzer,
\newblock Phys. Lett. {\bf B568}, 63 (2003).

\bibitem{bib:nuclstruc:spinstruc:Anselmino:2004ky}
M.~Anselmino, M.~Boglione, U.~D'Alesio, E.~Leader, and F.~Murgia,
\newblock Phys. Rev. {\bf D71}, 014002 (2005).

\bibitem{bib:nuclstruc:spinstruc:Anselmino:1994tv}
M.~Anselmino, M.~Boglione, and F.~Murgia,
\newblock Phys. Lett. {\bf B362}, 164 (1995).

\bibitem{bib:nuclstruc:spinstruc:Anselmino:1997jj}
M.~Anselmino, A.~Drago, and F.~Murgia,
\newblock hep-ph/9703303  (1997).

\bibitem{bib:nuclstruc:spinstruc:Elschenbroich:2004ba}
U.~Elschenbroich, G.~Schnell, and R.~Seidl,
\newblock hep-ex/0405017  (2004).

\bibitem{bib:nuclstruc:spinstruc:Collins:2002kn}
J.~C. Collins,
\newblock Phys. Lett. {\bf B536}, 43 (2002).

\bibitem{bib:nuclstruc:spinstruc:Belitsky:2002sm}
A.~V. Belitsky, X.~Ji, and F.~Yuan,
\newblock Nucl. Phys. {\bf B656}, 165 (2003).

\bibitem{bib:nuclstruc:spinstruc:Brodsky:2002cx}
S.~J. Brodsky, D.~S. Hwang, and I.~Schmidt,
\newblock Phys. Lett. {\bf B530}, 99 (2002).

\bibitem{bib:nuclstruc:spinstruc:Brodsky:2002rv}
S.~J. Brodsky, D.~S. Hwang, and I.~Schmidt,
\newblock Nucl. Phys. {\bf B642}, 344 (2002).

\bibitem{bib:nuclstruc:spinstruc:Ji:2002aa}
X.-d. Ji and F.~Yuan,
\newblock Phys. Lett. {\bf B543}, 66 (2002).

\bibitem{bib:nuclstruc:spinstruc:Ji:2004wu}
X.-d. Ji, J.-p. Ma, and F.~Yuan,
\newblock Phys. Rev. {\bf D71}, 034005 (2005).

\bibitem{bib:nuclstruc:spinstruc:Ji:2004xq}
X.-d. Ji, J.-P. Ma, and F.~Yuan,
\newblock Phys. Lett. {\bf B597}, 299 (2004).

\bibitem{bib:nuclstruc:spinstruc:Idilbi:2005er}
A.~Idilbi, X.-d. Ji, and F.~Yuan,
\newblock Phys. Lett. {\bf B625}, 253 (2005).

\bibitem{bib:nuclstruc:spinstruc:Anselmino:2005ea}
M.~Anselmino et~al.,
\newblock Phys. Rev. {\bf D72}, 094007 (2005).

\bibitem{bib:nuclstruc:spinstruc:Efremov:2004tp}
A.~V. Efremov, K.~Goeke, S.~Menzel, A.~Metz, and P.~Schweitzer,
\newblock Phys. Lett. {\bf B612}, 233 (2005).

\bibitem{bib:nuclstruc:spinstruc:Vogelsang:2005cs}
W.~Vogelsang and F.~Yuan,
\newblock Phys. Rev. {\bf D72}, 054028 (2005).

\bibitem{bib:nuclstruc:spinstruc:Collins:2005rq}
J.~C. Collins et~al.,
\newblock Phys. Rev. {\bf D73}, 094023 (2006).

\bibitem{bib:nuclstruc:spinstruc:Bacchetta:2008af}
A.~Bacchetta, F.~Conti, and M.~Radici,
\newblock arXiv:0807.0323  (2008).

\bibitem{bib:nuclstruc:spinstruc:Hammon:1996pw}
N.~Hammon, O.~Teryaev, and A.~Schafer,
\newblock Phys. Lett. {\bf B390}, 409 (1997).

\bibitem{bib:nuclstruc:spinstruc:Boer:1997bw}
D.~Boer, P.~J. Mulders, and O.~V. Teryaev,
\newblock Phys. Rev. {\bf D57}, 3057 (1998).

\bibitem{bib:nuclstruc:spinstruc:Boer:1999si}
D.~Boer and P.~J. Mulders,
\newblock Nucl. Phys. {\bf B569}, 505 (2000).

\bibitem{bib:nuclstruc:spinstruc:Boer:2001tx}
D.~Boer and J.-w. Qiu,
\newblock Phys. Rev. {\bf D65}, 034008 (2002).

\bibitem{bib:nuclstruc:spinstruc:Qiu:1998ia}
J.-w. Qiu and G.~Sterman,
\newblock Phys. Rev. {\bf D59}, 014004 (1999).

\bibitem{bib:nuclstruc:spinstruc:Burkardt:2003je}
M.~Burkardt and D.~S. Hwang,
\newblock Phys. Rev. {\bf D69}, 074032 (2004).

\bibitem{bib:nuclstruc:spinstruc:Burkardt:2005km}
M.~Burkardt and G.~Schnell,
\newblock Phys. Rev. {\bf D74}, 013002 (2006).

\bibitem{bib:nuclstruc:spinstruc:Airapetian:2004tw}
A.~Airapetian et~al.,
\newblock Phys. Rev. Lett. {\bf 94}, 012002 (2005).

\bibitem{bib:nuclstruc:spinstruc:Diefenthaler:2005gx}
M.~Diefenthaler,
\newblock AIP Conf. Proc. {\bf 792}, 933 (2005).

\bibitem{bib:nuclstruc:spinstruc:Diefenthaler:2007rj}
M.~Diefenthaler,
\newblock hep-ex 0706.2242, Proc. of Munich 2007 Deep-inelastic scattering ,
  579 (2007).

\bibitem{bib:nuclstruc:spinstruc:Collins:2005wb}
J.~C. Collins et~al.,
\newblock hep-ph/0510342, Proc. of The International Workshop on Transverse
  Polarisation Phenomena in Hard Processes (Transversity 2005)  (2005).

\bibitem{bib:nuclstruc:spinstruc:Anselmino:2005an}
M.~Anselmino et~al.,
\newblock hep-ph/0511017, Proc. of The International Workshop on Transverse
  Polarisation Phenomena in Hard Processes (Transversity 2005)  (2005).

\bibitem{bib:nuclstruc:spinstruc:Alexakhin:2005iw}
V.~Y. Alexakhin et~al.,
\newblock Phys. Rev. Lett. {\bf 94}, 202002 (2005).

\bibitem{bib:nuclstruc:spinstruc:Ageev:2006da}
E.~S. Ageev et~al.,
\newblock Nucl. Phys. {\bf B765}, 31 (2007).

\bibitem{bib:nuclstruc:spinstruc::2008dn}
M.~Alekseev et~al.,
\newblock Phys. Lett. {\bf B673}, 127 (2009).

\bibitem{bib:nuclstruc:spinstruc:Courtoy:2008mj}
A.~Courtoy, S.~Scopetta, and V.~Vento,
\newblock hep-ph/0811.2368, Proc. of International Workshop On Diffraction In
  High Energy Physics (Diffraction 2008)  (2008).

\bibitem{bib:nuclstruc:spinstruc:Efremov:2007kj}
A.~V. Efremov, K.~Goeke, and P.~Schweitzer,
\newblock Czech. J. Phys. {\bf 56}, F181 (2006).

\bibitem{bib:nuclstruc:spinstruc:Efremov:2006qm}
A.~V. Efremov, K.~Goeke, and P.~Schweitzer,
\newblock Phys. Rev. {\bf D73}, 094025 (2006).

\bibitem{bib:nuclstruc:spinstruc:Levorato:2008tv}
S.~Levorato,
\newblock hep-ex/0808.0086, Proc. of II International Workshop on Transverse
  Polarisation Phenomena in Hard Processes (Transversity 2008)  (2008).

\bibitem{bib:nuclstruc:spinstruc:Bianconi:2008jj}
A.~Bianconi,
\newblock J. Phys. {\bf G35}, 115003 (2008).

\bibitem{bib:nuclstruc:spinstruc:Anselmino:2008uy}
M.~Anselmino et~al.,
\newblock hep-ph/0809.3743, Proc. of II International Workshop on Transverse
  Polarisation Phenomena in Hard Processes (Transversity 2008)  (2008).

\bibitem{bib:nuclstruc:spinstruc:Abazov:2004ih}
V.~Abazov et~al.,
\newblock {ASSIA LOI, The structure of the nucleon: A study of spin-dependent
  interactions with antiprotons, hep-ex/0507077},
\newblock 2004.

\bibitem{bib:nuclstruc:spinstruc:Pax:2004}
PAX LOI,
\newblock http://www.fz-juelich.de/ikp/pax.

\bibitem{bib:nuclstruc:spinstruc:Tangerman:1994eh}
R.~D. Tangerman and P.~J. Mulders,
\newblock Phys. Rev. {\bf D51}, 3357 (1995).

\bibitem{bib:nuclstruc:spinstruc:Martin:1997rz}
O.~Martin, A.~Schafer, M.~Stratmann, and W.~Vogelsang,
\newblock Phys. Rev. {\bf D57}, 3084 (1998).

\bibitem{bib:nuclstruc:spinstruc:Barone:1997mj}
V.~Barone, T.~Calarco, and A.~Drago,
\newblock Phys. Rev. {\bf D56}, 527 (1997).

\bibitem{bib:nuclstruc:spinstruc:Efremov:2004qs}
A.~V. Efremov, K.~Goeke, and P.~Schweitzer,
\newblock Eur. Phys. J. {\bf C35}, 207 (2004).

\bibitem{bib:nuclstruc:spinstruc:Anselmino:2004ki}
M.~Anselmino, V.~Barone, A.~Drago, and N.~N. Nikolaev,
\newblock Phys. Lett. {\bf B594}, 97 (2004).

\bibitem{bib:nuclstruc:spinstruc:Altarelli:1979ub}
G.~Altarelli, R.~K. Ellis, and G.~Martinelli,
\newblock Nucl. Phys. {\bf B157}, 461 (1979).

\bibitem{bib:nuclstruc:spinstruc:Buras:1977yj}
A.~J. Buras and K.~J.~F. Gaemers,
\newblock Nucl. Phys. {\bf B132}, 249 (1978).

\bibitem{bib:nuclstruc:spinstruc:Anassontzis:1987hk}
E.~Anassontzis et~al.,
\newblock Phys. Rev. {\bf D38}, 1377 (1988).

\bibitem{bib:nuclstruc:spinstruc:Shimizu:2005fp}
H.~Shimizu, G.~Sterman, W.~Vogelsang, and H.~Yokoya,
\newblock Phys. Rev. {\bf D71}, 114007 (2005).

\bibitem{bib:nuclstruc:spinstruc:Moreno:1990sf}
G.~Moreno et~al.,
\newblock Phys. Rev. {\bf D43}, 2815 (1991).

\bibitem{bib:nuclstruc:spinstruc:Bianconi:2004wu}
A.~Bianconi and M.~Radici,
\newblock Phys. Rev. {\bf D71}, 074014 (2005).

\bibitem{bib:nuclstruc:spinstruc:Bianconi:2008jm}
A.~Bianconi,
\newblock Nucl. Instrum. Meth. {\bf A593}, 562 (2008).

\bibitem{bib:nuclstruc:spinstruc:Bianconi:2006hc}
A.~Bianconi and M.~Radici,
\newblock Phys. Rev. {\bf D73}, 114002 (2006).

\bibitem{bib:nuclstruc:spinstruc:Bianconi:2006mf}
A.~Bianconi and M.~Radici,
\newblock J. Phys. {\bf G34}, 1595 (2007).

\bibitem{bib:nuclstruc:spinstruc:Bianconi:2005yj}
A.~Bianconi and M.~Radici,
\newblock Phys. Rev. {\bf D73}, 034018 (2006).

\bibitem{bib:nuclstruc:spinstruc:Bianconi:2005bd}
A.~Bianconi and M.~Radici,
\newblock Phys. Rev. {\bf D72}, 074013 (2005).

\bibitem{bib:nuclstruc:emff:Perdrisat:2006hj}
C.~F. Perdrisat, V.~Punjabi, and M.~Vanderhaeghen,
\newblock Prog. Part. Nucl. Phys. {\bf 59}, 694 (2007).

\bibitem{bib:nuclstruc:emff:poltrans:jones}
M.~{Jones et al.},
\newblock Phys. Rev. Lett {\bf 84}, 1398 (2000).

\bibitem{bib:nuclstruc:emff:exp:poltrans:gayou}
O.~{Gayou et al.},
\newblock Phys. Rev. Lett {\bf 88}, 092301 (2002).

\bibitem{bib:nuclstruc:emff:exp:poltrans:gayou2}
O.~{Gayou et al.},
\newblock Phys. Rev. C {\bf 71}, 055202 (2005).

\bibitem{bib:nuclstruc:emff:exp:targetasym:Jones}
M.~{Jones et al.},
\newblock Phys. Rev. C {\bf 74}, 035201 (2006).

\bibitem{bib:nuclstruc:emff:Punjabi:2005wq}
V.~Punjabi et~al.,
\newblock Phys. Rev. {\bf C71}, 055202 (2005).

\bibitem{bib:nuclstruc:emff:exp:zichichi62}
A.~Zichichi et~al.,
\newblock Nuovo Cim. {\bf 24}, 170 (1962).

\bibitem{bib:nuclstruc:emff:exp:eeE835}
M.~Ambrogiani et~al.,
\newblock Phys. Rev. {\bf D60}, 032002 (1999),
\newblock and references therein.

\bibitem{bib:nuclstruc:emff:exp:babar}
B.~Aubert et~al.,
\newblock Phys. Rev. {\bf D73}, 012005 (2006).

\bibitem{bib:nuclstruc:emff:theory:gakh}
G.~I. Gakh and E.~Tomasi-Gustafsson,
\newblock Nucl. Phys. {\bf A771}, 169 (2006).

\bibitem{bib:nuclstruc:emff:theory:Gakh:2005wa}
G.~I. Gakh and E.~Tomasi-Gustafsson,
\newblock Nucl. Phys. {\bf A761}, 120 (2005).

\bibitem{bib:nuclstruc:emff:theory:adamuscinffaxial}
C.~Adamuscin, E.~Kuraev, E.~Tomasi-Gustafsson, and F.~Maas,
\newblock Phys.Rev. {\bf C75}, 045205 (2007).

\bibitem{bib:nuclstruc:emff:theory:eglegm}
E.~Tomasi-Gustafsson and M.~P. Rekalo,
\newblock Phys.Lett. {\bf B504}, 291 (2001).

\bibitem{bib:nuclstruc:emff:exp:pipluspiminus_eisenhandler}
E.~Eisenhandler et~al.,
\newblock Nucl. Phys. B {\bf 96}, 109 (1975).

\bibitem{bib:nuclstruc:emff:exp:pipluspiminus_buran}
T.~Buran et~al.,
\newblock Nucl. Phys. B {\bf 116}, 51 (1976).

\bibitem{bib:nuclstruc:emff:exp:pipluspiminus_berglund}
T.~Berglund et~al.,
\newblock Nucl. Phys. B {\bf 137}, 276 (1978).

\bibitem{bib:nuclstruc:emff:exp:pipluspiminus_dulude}
R.~Dulude et~al.,
\newblock Phys. Lett {\bf 79B}, 329 (1978).

\bibitem{bib:nuclstruc:emff:theory:TomasiGustafsson:2005kc}
E.~Tomasi-Gustafsson, F.~Lacroix, C.~Duterte, and G.~I. Gakh,
\newblock Eur. Phys. J. {\bf A24}, 419 (2005).

\bibitem{Lansberg:2007sr}
J.~P. Lansberg, B.~Pire, and L.~Szymanowski,
\newblock Nucl. Phys. Proc. Suppl. {\bf 184}, 239 (2008).

\bibitem{bib:phy:bigi98}
I.~Bigi,
\newblock Surveys High Energy Phys. {\bf 12}, 269 (1998).

\bibitem{bib:phy:aubert01}
B.~Aubert et~al.,
\newblock Phys. Rev. Lett. {\bf 87}, 091801 (2001).

\bibitem{bib:phy:abe01}
K.~Abe et~al.,
\newblock Phys. Rev. Lett. {\bf 87}, 091802 (2001).

\bibitem{bib:phy:Aubert:2007wf}
B.~Aubert et~al.,
\newblock Phys. Rev. Lett. {\bf 98}, 211802 (2007).

\bibitem{bib:phy:burdman94}
G.~Burdman,
\newblock hep-ph/9407378.

\bibitem{bib:phy:babarphysbook}
P.~Harrison and H.~Quinn, editors,
\newblock {\em The Babar Physics Book},
\newblock SLAC, 1998,
\newblock SLAC-R-504.

\bibitem{bib:phy:kroll91}
P.~Kroll,
\newblock Invited talk at SuperLEAR Workshop, Zurich, Switzerland, Oct 9-12,
  1991.

\bibitem{bib:phy:hamann91}
N.~H. Hamann,
\newblock Prepared for SUPERLEAR Workshop, Zurich, Swizterland, 9-12 Oct 1991.

\end{thebibliography}\end{flushleft}


\begin{flushleft}\begin{thebibliography}{10}\sloppy

\bibitem{bib:soft:evtgen}
D.~J. Lange,
\newblock Nucl. Instrum. Meth. {\bf A462}, 152 (2001).

\bibitem{bib:soft:EvtGenLHC}
J.~Catmor, R.~Jones, and M.~Smizanska,
\newblock Development of the EvtGen package for the LHC and ATLAS, Proc. of
  CHEP2006,
\newblock Mumbai, India, September 2006.

\bibitem{bib:soft:DPM}
A.~Capella, U.~Sukhatme, C.-I. Tan, and J.~Tran Thanh~Van,
\newblock Phys. Rept. {\bf 236}, 225 (1994).

\bibitem{bib:soft:Artru}
X.~Artru,
\newblock Nucl. Phys. {\bf B85}, 442 (1975).

\bibitem{bib:soft:Rossi}
G.~C. Rossi and G.~Veneziano,
\newblock Nucl. Phys. {\bf B123}, 507 (1977).

\bibitem{bib:soft:Montanet}
L.~Montanet, G.~C. Rossi, and G.~Veneziano,
\newblock Phys. Rept. {\bf 63}, 149 (1980).

\bibitem{bib:soft:Takahashi}
T.~T. Takahashi, H.~Matsufuru, Y.~Nemoto, and H.~Suganuma,
\newblock Phys. Rev. Lett. {\bf 86}, 18 (2001).

\bibitem{bib:soft:Bali}
G.~S. Bali,
\newblock Phys. Rept. {\bf 343}, 1 (2001).

\bibitem{bib:soft:Kuzmenko1}
D.~S. Kuzmenko and Y.~A. Simonov,
\newblock Phys. Atom. Nucl. {\bf 64}, 107 (2001).

\bibitem{bib:soft:Kuzmenko2}
D.~S. Kuzmenko and Y.~A. Simonov,
\newblock Phys. Atom. Nucl. {\bf 66}, 950 (2003).

\bibitem{bib:soft:Kaidalov_ZPC}
A.~B. Kaidalov and P.~E. Volkovitsky,
\newblock Z. Phys. {\bf C63}, 517 (1994).

\bibitem{bib:soft:Uzhinsky_Galoyan}
V.~V. Uzhinsky and A.~S. Galoyan,
\newblock hep-ph/0212369  (2002).

\bibitem{bib:soft:Andersson1}
B.~Andersson, G.~Gustafson, and C.~Peterson,
\newblock Zeit. Phys. {\bf C1}, 105 (1979).

\bibitem{bib:soft:Andersson2}
B.~Andersson, G.~Gustafson, and T.~Sjostrand,
\newblock Zeit. Phys. {\bf C6}, 235 (1980).

\bibitem{bib:soft:Ritter1}
S.~Ritter,
\newblock Comput. Phys. Commun. {\bf 31}, 393 (1984).

\bibitem{bib:soft:Kaidalov1}
A.~B. Kaidalov,
\newblock Yad. Fiz. {\bf 45}, 1452 (1987).

\bibitem{bib:soft:Kaidalov2}
A.~B. Kaidalov and O.~I. Piskunova,
\newblock Z. Phys. {\bf C30}, 145 (1986).

\bibitem{bib:soft:Ritter2}
K.~Hanssgen and S.~Ritter,
\newblock Comput. Phys. Commun. {\bf 31}, 411 (1984).

\bibitem{bib:soft:UrQMD1}
M.~Bleicher et~al.,
\newblock J. Phys. {\bf G25}, 1859 (1999).

\bibitem{bib:soft:UrQMD2}
S.~A. Bass et~al.,
\newblock Prog. Part. Nucl. Phys. {\bf 41}, 225 (1998).

\bibitem{bib:soft:PDG96}
R.~M. Barnett et~al.,
\newblock Phys. Rev. {\bf D54}, 1 (1996).

\bibitem{bib:soft:donna}
A.~Donnachie and P.~V. Landshoff,
\newblock Phys. Lett. {\bf B296}, 227 (1992).

\bibitem{bib:soft:kochP89a}
P.~Koch, B.~Muller, and J.~Rafelski,
\newblock Phys. Rept. {\bf 142}, 167 (1986).

\bibitem{bib:soft:andersson83a}
B.~Andersson, G.~Gustafson, and B.~Soderberg,
\newblock Z. Phys. {\bf C20}, 317 (1983).

\bibitem{PDBook}
S.~Eidelman et~al.,
\newblock {Phys. Lett.} {\bf B592}, 1+ (2004),
\newblock http://pdg.lbl.gov.

\bibitem{bib:soft:Khaled}
K.~Abdel-Waged,
\newblock Phys. Rev. {\bf C67}, 064610 (2003).

\bibitem{bib:soft:Botvina}
J.~P. Bondorf, A.~S. Botvina, A.~S. Ilinov, I.~N. Mishustin, and K.~Sneppen,
\newblock Phys. Rept. {\bf 257}, 133 (1995).

\bibitem{bib:soft:Cluster}
C.~Hartnack et~al.,
\newblock Eur. Phys. J. {\bf A1}, 151 (1998).

\bibitem{bib:soft:Allison}
J.~Allison et~al.,
\newblock IEEE \ Transactions \ on Nuclear Science {\bf 53}, 270 (2006).

\bibitem{bib:soft:GEANT4}
S.~Agostinelli et~al.,
\newblock Nucl. Instrum. Meth. {\bf A506}, 250 (2003).

\bibitem{bib:soft:DDDcms1}
P.~Arce et~al.,
\newblock Nucl. Instrum. Meth. {\bf A502}, 687 (2003).

\bibitem{bib:soft:PandaEmcTdr}
Technical report,
\newblock Technical Design Report for the PANDA EMC, in progress.

\bibitem{bib:soft:BaBarTracking}
D.~Brown, E.~Charles, and D.~Roberts,
\newblock The BaBar track fitting algorithm, Proc. Computing in High Energy
  Physics Conference,
\newblock Padova, 2000.

\bibitem{bib:soft:Minuit}
F.~James and M.~Roos,
\newblock Comput. Phys. Commun. {\bf 10}, 343 (1975).

\bibitem{bib:soft:BaBarDetector}
B.~Aubert et~al.,
\newblock Nucl. Instrum. Meth. {\bf A479}, 1 (2002).

\bibitem{bib:soft:Strother}
P.~Strother,
\newblock {\em Design and application of the reconstruction software for the
  BaBar calorimeter},
\newblock PhD thesis, 1998,
\newblock University of London and Imperial College, UK.

\bibitem{bib:soft:Beta}
R.~Jacobsen,
\newblock Beta: A High Level Toolkit for BaBar Physics Analysis, 1997,
\newblock presented at Conference on Computing in High Energy Physics, Berlin.

\bibitem{bib:soft:ROOT}
R.~Brun and F.~Rademakers,
\newblock Phys. Res. {\bf A389}, 81 (1996 1997).

\bibitem{bib:soft:prod:grille}
Grille de Recherche d'Ile de France,
\newblock http://www.grif.fr.

\bibitem{bib:soft:FAIRroot}
FairRoot, Simulation and Analysis Framework,
\newblock http://fairroot.gsi.de.

\bibitem{bib:soft:vmc}
Virtual Monte Carlo,
\newblock http://root.cern.ch/root/vmc.

\bibitem{bib:soft:geant3}
CERN Program Library W5013  (1991).

\bibitem{bib:soft:FLUKA}
FLUKA,
\newblock http://www.fluka.org.

\bibitem{bib:soft:cm}
P.~Yepes,
\newblock Nucl. Instr. and Meth. {\bf A380}, 582 (1996).

\bibitem{bib:soft:riemann}
R.~Fr\"uhwirth, A.~Strandlie, and W.~Waltenberger,
\newblock Nucl. Instr. and Meth. in Phys. Res. {\bf A490}, 366 (2002).

\bibitem{bib:soft:knn}
R.~Duda, P.~Hart, and D.~Stork,
\newblock {\em Pattern Classification},
\newblock Wiley Interscience ISBN: 0-471-05669-3.

\bibitem{bib:soft:som}
T.~Kohonen,
\newblock {\em Self-Organizing Maps},
\newblock Springer Series in Information Sciences, Springer, Berlin,
  Heidelberg, New York (1995,1997,2001).

\bibitem{bib:soft:rho}
Rho: A Set of Analysis Tools for ROOT,
\newblock http://savannah.fzk.de/websites/hep/rho/.

\bibitem{bib:soft:kfitter}
Belle Kinematic Fitter,
\newblock http://hep.phys.s.u-tokyo.ac.jp/~jtanaka/BelleSoft/KFitter/.

\bibitem{bib:soft:alien}
Alien$^2$@GRID,
\newblock http://alien.cern.ch.

\end{thebibliography}\end{flushleft}
